\documentclass[prl,aps,amssymb,amsmath]{article}
\usepackage{amssymb,amsmath,tikz}
\usepackage{amsthm}
\usepackage{mathrsfs}
\usepackage{pifont}
\usepackage{slashed}

\showoutput
\DeclareMathOperator*{\p}{p}
\DeclareMathOperator*{\x}{x}

\DeclareMathOperator*{\q}{k}

\DeclareMathOperator*{\Dom}{Dom}

\DeclareMathOperator*{\In}{In}

\renewcommand{\qedsymbol}{$\blacksquare$}

\theoremstyle{plain}

\newtheorem*{twr*}{THEOREM}
\newtheorem*{lem*}{LEMMA}
\newtheorem{twr}{THEOREM}
\newtheorem{lem}{LEMMA}
\newtheorem{defin}{DEFINITION}
\newtheorem*{defin*}{DEFINITION}
\newtheorem{rem}{REMARK}
\newtheorem*{rem*}{REMARK}
\newtheorem{cor}{COROLLARY}
\newtheorem{cor*}{COROLLARY}

\newtheorem*{notn*}{NOTATION}
\newtheorem*{wiener-ito*}{WIENER-IT\^O-SEGAL DECOMPOSITION}
\newtheorem*{prop*}{PROPOSITION}

\begin{document}
\title{{\bf  Solution of the Adiabatic Limit Problem. QED Without Infinities}}
\author{Jaros{\l}aw Wawrzycki \footnote{Electronic address: jaroslaw.wawrzycki@wp.pl or jaroslaw.wawrzycki@ifj.edu.pl}
\\Institute of Nuclear Physics of PAS, ul. Radzikowskiego 152, 
\\31-342 Krak\'ow, Poland}
\maketitle

\newcommand{\ud}{\mathrm{d}}

\vspace{1cm}

\begin{abstract}
In this work we give positive solution to the adiabatic limit problem in causal perturbative QED
on the Minkowski space-time,
as well as give a contribution to the solution of the convergence problem for the perturbative 
series in QED on the Minkowski space-time, by using white noise construction of free fields. The obtained scattering operator is a generalized operator sufficient for the computation of the effective cross section, involving generalized many-particle plane wave states of the fundamental fields. Bound states remain outside the scope of the resulting QED on the Minkowski space-time.
The method is general enough to be applicable to more general causal perturbative QFT,
such as Standard Model with the Higgs field.

\end{abstract}

\tableofcontents

\section{Introduction}\label{intro}

This work is concerned with the causal perturbarive approach to Quantum Field Theories (QFT), initiated by 
St\"uckelberg, Bogoliubov and Shirkov \cite{Bogoliubov_Shirkov}, and developed mainly by
Epstein, Glaser \cite{Epstein-Glaser}, Blanchard, Seneor, Duch \cite{BlaSen} \cite{duch}, D\"utch, Krahe and Scharf and Fredenhagen \cite{DKS1}-\cite{DKS4}, \cite{DutFred}.

In causal perturbative approach to QFT the infra-red-divergence (IR) problem is clearly separated from the ultra-violet-divergence
(UV) problem by using a space-time function $x \mapsto g(x)$  as coupling ``constant''. 
The UV-problem is essentially solved within this approach, \cite{Epstein-Glaser},   -- the origin of infinite counter terms of the renormalization scheme is well understood by now, i. e. using the counter terms (renormalization) 
is equivalent to the causal perturbative construction of the perturbative series due to Bogoliubov-Epstein-Glaser
(scalar massive field), developed further for QED, and other physical theories with non abelian gauge mainly by
D\"utch, Krahe and Scharf, \cite{DKS1}-\cite{DKS4},
where no infinite counter terms appear but instead one uses recurrence rules for the construction of the chronological
product of fields regarded as operator-valued distributions. The renormalization scheme is now incorporated into the following recurrence rules for the chronological product \cite{Epstein-Glaser}, \cite{DKS1}-\cite{DKS4},
\cite{DutFred}, \cite{Scharf}: 
\begin{enumerate}

\item[1)]
causality, 

\item[2)]
symmetricity, 

\item[3)]
unitarity, 

\item[4)]
Translational covariance (Lorentz covariance is not used), 

\item[5)]
Ward identities -- quantum version of gauge invariance (e. g. in case of QED),

\item[6)]
preservation of the Steinmann scaling degree, 
\end{enumerate}
part of the remaining freedom may be reduced by imposing the natural 
field equations for the interacting field (which is always possible for the standard gauge fields)
and the rest of the remaining freedom is pertinent to the St\"uckelberg-Petermann renormalization group . All the recurrence rules should be regarded as important physical laws which incorporate the whole content of the standard pragmatic approach including the renormalization scheme. Causality implies locality for perturbatively constructed
(using the Epstein-Glaser method \cite{Epstein-Glaser}) algebras of localized fields $\mathcal{F}(\mathcal{O})$ regarded as ``smeared out'' operator-valued distributions, where $g$ is constant (equal to the electric charge in case of QED) within the open space-time region $\mathcal{O}$ -- the only step where the UV-problem shows up and is solved by the use of 
Epstein-Glaser method. The IR-problem is solved only partially, i. e. nets 
$\mathcal{O} \mapsto \mathcal{F}(\mathcal{O})$
of algebras  $\mathcal{F}(\mathcal{O})$ of local (unbounded) operator localized fields have likewise been constructed
perturbatively \cite{DutFred}, but in the sense of formal power series only.  
 
The most important and still open problems are the following. 

\begin{enumerate}

\item[(a)] The problem of
existence of the adiabatic limit ($g \mapsto$ \emph{constant function over the the whole space-time})
in each order separately. This is the IR-problem or the Adiabatic Limit Problem.

\item[(b)] 
The convergence of the formal perturbative series for interacting fields (with $g=1$). 

\end{enumerate}

In this work we give a positive solution to the Adiabatic Limit Problem for QED, i.e. the problem (a), 
and give a contribution to the problem (b) for QED. 
The method is based solely on substitution into the casual perturbative series the free fields of the theory which are constructed with the help of white noise calculus. The whole causal perturbative method of Bogliubov-Epstein-Glaser 
remains unchanged. The whole point in constructing the free fields within the white nose set up lies in the fact that
it  allows us to treat them equivalently as integral kernel operators with vector-valued kernels in the sense of 
Obata \cite{obataJFA},
and opens us to the effective theory of such operators worked out by the Japanese School of Hida. 
Using the calculus of such operators we show that the class of integral kernel operators represented (or representing)
free fields allows the operations of differentiation (similarly as Schwartz distributions)
integration, point-wise Wick product, integration of Wick product integral kernel operators
(including spatial integration), convolution of Wick product integral kernel operators with tempered distributions,
and splitting into advanced and retarded parts of integral kernel operators with causal supports.
Thus all operations needed for the causal perturbation series have a well defined mathematical meaning
if understood as operations performed upon integral kernel operators in the sense of Obata.
Therefore the free fields, understood as integral kernel operators with vector-valued kernels in the sense of Obata, can be inserted into the formulas for the higher order contributions to the interacting fields. 
After the insertion we obtain each order term contribution to interacting fields in a form of finite sums of 
well defined integral kernel operators with vector-valued kernels,  similarly as for the free fields themselves or
for the Wick products of free fields. 

But the most essential point is that these formulas do not loose their rigorous mathematical meaning even if we 
put in them the intensity-of-interaction function $g$ equal $1$ everywhere over the whole space-time.
The contributions still preserve their meaning of integral kernel operators with vector valued kernels,
which belong to the same general class of integral kernel operators as the Wick products of free fields.
We therefore arrive at the positive solution of the Problem (a) in QED.
But at the same time we obtain the interacting fields in the form of Fock expansions into integral kernel operators
with vector-valued kernels in the sense of \cite{obataJFA}, with precise estimate of the convergence, which
allows us to give a computationally effective criteria for the convergence of the perturbative series, \emph{i .e}
non-trivial contribution to the solution of the Problem (b). 
  
The method is general enough to be capable of application to other QFT with non abelian gauge. 

In this manner we obtain causal perturbative QED in which there are no infra-red nor ultra-violet divergences
and get insight into problems which were beyond the reach of the conventional approach
involved into renormalization. In particular we hope that have given a step forward on the way in giving
a rigorous construction of a non-trivial (and realistic) quantum interacting field. Some of the prominent 
analyst place this problem also among the most important unsolved problems in the contemporary analysis, 
compare \cite{Segal-ProcStone}.

However we should emphasize, that the causally constructed interactiong fields, of course with the intensity of interaction function $g=1$ everywhere on the space-time, (in fact all the higher order contributions to them) belong to a class of so-called generalized operators. More precisely they are integral kernel operators with vector-valued kernels, transforming continously the nuclear space-time test space $\mathscr{E}$ into the nuclear space $\mathscr{L}((\boldsymbol{E}), (\boldsymbol{E}))$ of continuous maps from the Hida test space $(\boldsymbol{E})$ into its strong dual $(\boldsymbol{E})^*$, \emph{i. e.}
they belong to
\[
\mathscr{L}((\boldsymbol{E}) \otimes \mathscr{E}, \, (\boldsymbol{E})^*) \cong
\mathscr{L}(\mathscr{E}, \, \mathscr{L}((\boldsymbol{E}), (\boldsymbol{E})^*) \, ).
\]
The spaces  $\mathscr{L}((\boldsymbol{E}), (\boldsymbol{E}))$, 
$\mathscr{L}((\boldsymbol{E}) \otimes \mathscr{E}, \, (\boldsymbol{E})^*)$ are endowed with the natural topology of uniform convergence on bounded sets.
In particular each higher order contribution to an interacting field, when evaluated on an element $\phi$ of the space-time test space
$\mathscr{E}$ (\emph{i. e.}``smeared with test function''), is equal to a finite sum of integral kernel operators $\Xi_{l,m}(\kappa_{l,m}(\phi))$ with scalar valued kernels $\kappa_{l,m}(\phi)$. Each such operator $\Xi_{l,m}(\kappa_{l,m}(\phi))$ defines continuous functional
\[
(\boldsymbol{E}) \times (\boldsymbol{E}) \ni \Phi \times \Psi \mapsto
\big \langle \big \langle  \Xi_{l,m}(\kappa_{l,m}(\phi)) \Phi, \Psi  \big \rangle \big \rangle \in \mathbb{C}
\]
with distributional kernel which can be identified with ``matrix elements'' between the many particle 
plane wave states, and which unfortunately are not ordinary numbers, but distributions. Only some of the higher order contributions to interacting fields, when smeared out with test function give generalized operators, which belong to
\[
\mathscr{L}((\boldsymbol{E}) \otimes \mathscr{E}, \, (\boldsymbol{E})) \cong
\mathscr{L}(\mathscr{E}, \, \mathscr{L}((\boldsymbol{E}), (\boldsymbol{E})) \, )
\]
and thus define ordinary operators on the Fock space of free fields, transforming contintinously the
the Hida test space $(\boldsymbol{E})$ into itself. This is in particular the case for the first order contribution to the interacting electromagnetic potential field (of course with the intensity of interaction function $g=1$ everywhere on the space-time).  

Similarily the causal perturbative series for the scattering operator $S(g)$, for $g$ being a space-time test function -- an element of 
a standard nuclear space $\mathscr{E}$,  becomes equal to a Fock expansion
\[
\sum \limits_{l,m} \Xi_{l,m}\Big(\kappa_{l,m}\big(g^{\otimes \, (k+l)}\big)\Big)
\]
into integral kernel operators $\Xi_{l,m}(\kappa_{l,m})$ with scalar-valued distributional kernels 
$\kappa_{l,m}$, and with contribution of each fixed order equal to a finte sum 
\[
\sum \limits_{l,m} \Xi_{l,m}\Big(\kappa_{l,m}\big(g^{\otimes \, (k+l)}\big)\Big)
\]
of integral kernel operators $\Xi_{l,m}\Big(\kappa_{l,m}\big(g^{\otimes \, (k+l)}\big)\Big)$, which in general belong to 
\[
\mathscr{L}((\boldsymbol{E}), \, (\boldsymbol{E})^*).
\]
Only a part of the integral kernel operators entering higher order contributions are regular enough to be elements of
\[
\mathscr{L}((\boldsymbol{E}), \, (\boldsymbol{E})),
\]
and can be interpreted as ordinary operators on the Fock space transforming continously the test Hida space $(\boldsymbol{E})$ into itself,
and this happens if the free fields underlying the theory are all massive. In QED case no higher order integral kernel operators
$\Xi_{l,m}$ behave so regularily. In general however, even for theories including mass less fields as the free fields underlying the theory (e.g. QED), 
each $n$-th order contribution $S_{n}(g)$ to $S(g)$, with $g\in \mathscr{E}$, belongs to $\mathscr{L}((\boldsymbol{E}), \, (\boldsymbol{E})^*)$, and 
each higher order contribution $S_{n}(g)$ to the scattering operator 
$S(g)$, $g\in \mathscr{E}$,  defines a continous functional
\[
(\boldsymbol{E}) \times (\boldsymbol{E}) \ni \Phi \times \Psi \mapsto
\big \langle \big \langle  S_{n}(g) \Phi, \Psi  \big \rangle \big \rangle \in \mathbb{C}, \,\,\, \textrm{for each fixed} \, g\in \mathscr{E},
\]
\emph{i. e.} a \emph{distribution}, with distribution kernel which can be canonically identified with the 
distributional ``matrix elements''
\begin{equation}\label{DistributionSmatrixElements}
\big \langle S_{n}(g) \Phi_{{}_{\ldots s,\boldsymbol{\p} \ldots }}, \Psi_{{}_{\ldots s',\boldsymbol{\p}' \ldots }}\big\rangle, \,\,\, 
\textrm{for each fixed} \, g\in \mathscr{E},
\end{equation}
of the scattering matrix (generalized) operator $S(g)$, $g\in \mathscr{E}$,  in the non-normalizable many particle plane wave
states
\[
\Phi_{{}_{\ldots s,\boldsymbol{\p} \ldots }} =
\cdots a_{s}(\boldsymbol{\p})^{+} \cdots |0\rangle, 
\,\,\,\,
\Phi_{{}_{\ldots s,\boldsymbol{\p} \ldots }} =
\cdots a_{s'}(\boldsymbol{\p}')^{+} \cdots |0\rangle,
\]
with the creation (Hida) operators $a_{s}(\boldsymbol{\p})^{+}$ in the momentum picture and with 
$|0\rangle = \Psi_{{}_{0}}$ being the vacuum in the Fock space of free fields of the theory.
This is the general situation in causal perturbative QFT we arrive at, when using the Hida white noise operators as the annihilation-creation operators. 
Only part of the integral kernel opertors
$\Xi_{l,m}(\kappa_{l,m})$ entering higher order contributions $S_{n}(g)$ to $S(g)$, for each fixed $g\in \mathscr{E}$, belongs to 
$\mathscr{L}((\boldsymbol{E}), \, (\boldsymbol{E}))$, and this happens for theories where all free fields are massive (in particular this is not the case for QED) 
and we can compute for them the ordinary matrix elements
\[
\big\langle\Xi_{l,m}(\kappa_{l,m}) \Phi, \Psi \big\rangle,
\,\,\,
\Phi, \Psi \in (\boldsymbol{E}) \subset \mathcal{H}_{\textrm{Fock}}
\]
in the normalizable states $\Phi, \Psi$ belonging to the test Hida space $(\boldsymbol{E})$
densely included into the total Fock space $\mathcal{H}_{\textrm{Fock}}$ of all free fields of the theory, and composing the Gelfand triple
\[
(\boldsymbol{E}) \subset \mathcal{H}_{\textrm{Fock}} \subset (\boldsymbol{E})^*.
\]

Thus we arrive at the solution of the Adiabatic Limit Problem which, at the first sight, may seem totally unsatisfactory. This may seem so because our scattering matrix $S(g)$, $g\in \mathscr{E}$, and even the separate higher order contributions $S_{n}(g)$, are not ordinary operators, which are in general not well defined on normalizable states. But in fact we should strongly emphasize here, that we do not need the scattering matrix as an operator acting on normalizable states in the computation of the effective cross-section. Paradoxically $S(g)$ acting on a dense domain $\mathscr{D}$ of normalizable states would be even insufficient for the computation of the effective cross-section,
compare \cite{Bogoliubov_Shirkov}, Chap. IV, \S\S 23-25.  In fact in this computation we need an operator
$S(g)$ for each $g\in \mathscr{E}$, defining the distributional matrix elements (\ref{DistributionSmatrixElements}), \emph{i. e.}
we need $S(g)$ as a continuous operator $(\boldsymbol{E}) \rightarrow (\boldsymbol{E})^*$, depending continously on $g \in \mathscr{E}$
in order to compute the limit $g \rightarrow g=1$ in the effective cross section. If the dense domain $\mathscr{D}$ would not be endowed 
with the necessary additional nuclear topology structure, which allows us to understad $S(g)$ as a continous operator defining the distributional kernels
(\ref{DistributionSmatrixElements}), then such $S(g)$ would be useless in computation of the effective
cross section in the limit $g=1$, even if defined on a dense domain $\mathscr{D}$ of normalizable states.  In particular
such operator would be useful in case $\mathscr{D} = (\boldsymbol{E}) \subset \mathcal{H}_{\textrm{Fock}}$, and when it would be a continous oparator $(\boldsymbol{E}) \rightarrow (\boldsymbol{E})$, because such operator is naturally a continuous map $(\boldsymbol{E}) \rightarrow (\boldsymbol{E})^*$, defininig the distributional ``matrix elements'' (\ref{DistributionSmatrixElements}), as its kernel, which moreover defines continous map $\mathscr{E} \ni g \longmapsto S(g) \in 
\mathscr{L}((\boldsymbol{E}), \, (\boldsymbol{E})^*)$. 
This is becase in the scattering 
phenomena we are dealing with non-normalizable many particle plane wave states, 
\emph{i. e.} generalized states belonging to $(\boldsymbol{E})^*$. In order to compute the effective cross section we need the ``matrix elemets'' (\ref{DistributionSmatrixElements}) which are distributions (and not ordinary numbers), because the many particle plane wave states are non-normalizable generalized states. When evaluating the effective cross-section we do not need to know the amplitude of the absolute probability, but only the amplitude for the registration of a particle with given spin and momentum and mass \emph{per unit volume and unit time} (or per unit time in case of scattering by a static classical field). These circumstances allows us to compute
\[
\big| \big \langle S_{n}(g) \Phi_{{}_{\ldots s,\boldsymbol{\p} \ldots }}, \Psi_{{}_{\ldots s',\boldsymbol{\p}' \ldots }}\big\rangle \big|^2,
\,\,\, g \in \mathscr{E},
\]
with the distribution kernel  
\[
\big \langle S_{n}(g) \Phi_{{}_{\ldots s,\boldsymbol{\p} \ldots }}, \Psi_{{}_{\ldots s',\boldsymbol{\p}' \ldots }}\big\rangle,
\,\,\, g \in \mathscr{E},
\]
behaving regularily enough to be represented by ordinary function, for $g\in \mathscr{E}$, which allows the operation of multiplication
\[
\overline{\big \langle S_{n}(g) \Phi_{{}_{\ldots s,\boldsymbol{\p} \ldots }}, \Psi_{{}_{\ldots s',\boldsymbol{\p}' \ldots }}\big\rangle}
\big \langle S_{n}(g) \Phi_{{}_{\ldots s,\boldsymbol{\p} \ldots }}, \Psi_{{}_{\ldots s',\boldsymbol{\p}' \ldots }}\big\rangle
= \big| \big \langle S_{n}(g) \Phi_{{}_{\ldots s,\boldsymbol{\p} \ldots }}, \Psi_{{}_{\ldots s',\boldsymbol{\p}' \ldots }}\big\rangle \big|^2
\] 
with the continous dependence on $g$, which in turn allows for the computation of the ``matrix elements''
\[
 \big \langle S_{n}(g=1) \Phi_{{}_{\ldots s,\boldsymbol{\p} \ldots }}, \Psi_{{}_{\ldots s',\boldsymbol{\p}' \ldots }}\big\rangle,
\,\,\, g \in \mathscr{E},
\]
and extraction of the ``residual part''
\[
\big| \big \langle S_{n}(g=1) \Phi_{{}_{\ldots s,\boldsymbol{\p} \ldots }}, \Psi_{{}_{\ldots s',\boldsymbol{\p}' \ldots }}\big\rangle \big|^2,
\]
in the adiabatic limit $g \rightarrow g=1$, and the computation of the effective cross section in the adabatic limit,
compare e. g. \cite{Bogoliubov_Shirkov}, Chap. IV, \S\S 23-25, even if  
(\ref{DistributionSmatrixElements}) is a distribution. Thus in order to have a theory in a minimal form needed for the computation of the effective cross section, it is sufficient that the interaction defines, together with the causal rules for the construction of the perturbative series, the scattering matrix
$S(g)\in \mathscr{L}((\boldsymbol{E}), \, (\boldsymbol{E})^*)$, \emph{i.e.} with $S(g)$ as a continuous operator $(\boldsymbol{E}) \rightarrow (\boldsymbol{E})^*$, 
for each fixed $g \in \mathscr{E}$, which moreover defines continous map $\mathscr{E} \ni g \longmapsto S(g) \in \mathscr{L}((\boldsymbol{E}), \, (\boldsymbol{E})^*)$. The requirement that the scattering operator $S(g=1)$ should be well defined on normalizable states (intentionally even unitary
in the ordinary sense) is quite unrealistic for realistic interactions of fields on the Minkowski space-time. 

Thus on using the Hida operators as creation-annihilation operators, we arrive at the causal perturbative 
QFT, in particular QED, which can successfully be applied to the class of generalized states, in particular to the high energy scattering phenomena involving the generalized many particle plane wave states. But there is also another class of generalized states, which can also be experimentally extracted
and related to the infrared problem, and which can successfully be treated with the causal perturbative QED we have just constructed with the help of Hida operators. Namely we consider the generalized (belonging to $(\boldsymbol{E})^*$) homogeneous states of homogeneity degree $-1$ in single particle Fock subspace of the free electromagnetic potential field $A$. Correspondingly to these generalized single particle homogeneous states of the free electromagnetic potential field we have the generalized states in the single particle state spaces of the free massive component fields, which are minimally coupled through the minimal coupling interaction term to the electromagnetic potential field. The structure of the generalized single particle states 
of the massive free fields, which correspond to the homogeneous single particle states of $A$, depends on the specific type of the massive field, and in particular for the scalar field they are spanned by the states $\varphi$ whose Fourier transform $\widetilde{\varphi}$ has the general form
\[
\widetilde{\varphi}(p)=(p \cdot k)^{-1 + i \nu}, \,\,\, k \cdot k =0, \,\, p \cdot p = m^2, \,\,\,
\nu \in \mathbb{R} \,\,\, \textrm{fixed}.
\] 
The interacting electromagnetic potential field has well defined restriction (let say the homogeneous part of homogeneity $-1$) to the many particle Fock space of the generalized states over the specified class of generalized single particle states (homogeneous of degree $-1$ on the sigle particle subspace of the free electromagnetic potential field) in the total Fock space. This homogeneous of degree $-1$ part of the interacting electromagnetic potential field provides a special realization of the general quatum theory of the electric charge due to \cite{Staruszkiewicz}. There is a unique relationship between the representation structure of the representation of $SL(2, \mathbb{C})$ acting on the specified class of the generalized states, and the value of the fine structure constant. This relationship, in particular, and in general the reconstruction of the relationship of causal perturbative QED with \cite{Staruszkiewicz} is completely beyond the scope of the theory which uses the renormalization prescription, involved in handling infinite quantities. The last class of generalized states can be experimentally identified through the Bremsstrahlung phenomena: if we look at the particle which radiates the electromagnetic field due to the acceleration from a suitable distance, at which this process of radiation is practically seen as a scattering at a single point with initial and final four-velocities of the particle equal $u$ and $v$, then the registered radiation will degenrate to the homogeneous solution of Maxwell equations, with the Fourier transform of the correponding four-potential of the radiation equal
\[
\frac{e}{2\pi} \Big(\frac{u}{u \cdot p} - \frac{v}{v \cdot p} \Big),
\]
and homogeneous of degree $-1$, with $e$ equal to the electric charge of the particle.

However it is not the computational aspect, that now we can compute the effective cross sections
for high energy scattering processes involving many particle plane wave states of the elementary fields without any use of renormalization and without infrared infinities, which is of primary importance here. 

The most important thing about the elimination of the infrared and ultraviolet infinities altogether from causal perturbative QFT, in particular from QED, lies in the fact that now we have a mathematical theory, with the basic principles formulated in well defined mathematical terms. These principles are: 1) the Hamiltonian formulation of classical theory subject to quantization (e.g. classical QED) together with the canonical commutation rules on which the relation between free classical fields and their quantum counterparts is based, 2) the causal rules for the causal perturbative construction of the scattering (generalized) operator $S(g=1)$, 3) causal geometry of space-time (here the Minkowski spacetime). The interacting quantum fields are obtained from the scattering opertor $S(g=1) = S(\mathcal{L})$, due to the general Bogoliubov rule
\[
A_{{}_{\textrm{int}}}(g=1; x) = \frac{i\delta}{\delta h(x)} 
S(\mathcal{L} +hA)^{-1}S(\mathcal{L})\Big|_{{}_{h=0}}
\]
relating the interacting field $A_{{}_{\textrm{int}}}(g=1)$ to the corresponding free field $A$, compare \cite{Bogoliubov_Shirkov}, \cite{DKS1}, \cite{DutFred}. Here $S(g=1) = S(g\mathcal{L}) =
S(\mathcal{L})$ is the scattering matrix in which the first order term is equal
\[
S_1(g) = \int i\mathcal{L}(x) \, g(x) \, \ud^4 x
\]
undersood as an integral kernel operator with the free fields in the interaction lagrangian density 
$\mathcal{L}(x)$ uderstood as the integral kernel operators with vector-valued kernels.
Similarily $S(\mathcal{L} +hA)$ is the scattering (generalized) operator $S(g=1,h)$ in which the first induction step in the causal perturbative construction is equal
\[
S_1(g,h) = \int i\big(g(x) \, \mathcal{L}(x) + h(x) \, A(x)\big) \, \ud^4 x,
\]
also understood as integral kernel operator.
 Here we only must remember that the canonical commutation rules in case of the system of free fields with infinite number of degrees of freedom are rigorously realized through the annihilation-creation operators, which mathematically are understood as white noise Hida operators. Because the Hida operators indeed do fulfill the canonical commutation rules they fits naturally in as the realization of the annihilation-creation operators of quantum mechanics (of the system with infinite number of degrees of freedom). 
 This theory has well defined rage of experimental applicability: it can be applied at least to the two classes of generalized states. The first class involves the many particle plane wave generalized states in the high energy scattering phenomena and provides the effective cross sections. The second class of generalized states embraces the many particle generalized states constructed by symmetrized/anitisymetrized projective tensor product of homogeneous of degree $-1$ states (of the free field $A$)
 and the correspodnig generalized single partice states of the massive fields coupled to $A$.

The fact that we have a causal perturbative QFT with principles expressed in well-defined mathematical terms cannot be overestimated. Now the rules are under mathematical control and now we can find the sources of difficulties in explaining specific physical problems. This would be impossible within the technique of renormalization which involves handling infinite quantities simply because such a handling is not really a mathematically logical process, and we cannot proceed along any logical-mathematical line from a physical problem to its source identified within the principles. 
We illustrate this by giving two examples of analysis which would be impossible to go through using the renormalization technique. The first is (I) the problem with bound states and the spectrum of stable or meta-stable particles and the second, related to the first, concerns (II) relation of QFT to space-time geometry. At the same time they may serve as an outlook, which is customary to be placed in research publications. 

\begin{center}
{\small (I)}
\end{center}
The Emmy Noether integrals corresponding to the one parameter subgroups of space-time symmetries or to the phase transformations, do exist as ordinary self-adjoint operators in the Fock space for free quantum fields, which we prove using the integral kernel operator analysis. But the same integrals for the interacting fields do not exist as ordinary operators, but only as generalized operators (at least this is so for the separate higher order contributions). They are generalized integral kernel operators transforming continously the Hida space $(\boldsymbol{E})$ into its strong dual $(\boldsymbol{E})^*$. This is the mathematical consequence of the principles 1) -- 3) stated above. In particuar the Noether generator of time translations, the Hamiltonian, is not well defined as an ordinary self-adjoint operator in the Fock space for the system of interacting fields, but it is only a generalized integral kernel operator belonging to $\mathscr{L}((\boldsymbol{E}), \, (\boldsymbol{E})^*)$. In particular the problem of analysis of normaliziable bound eigen-states of the Hamiltonian (say stable particles), or their 
superpositions with small energy uncertainty (say meta-stable particles) cannot be grashped within the principles 1) -- 3), and at least one of these principles will have to be changed in order to account for the existence of stable and metastable particles and generally in order to account for phenomena involved into bound states. We note here that 1) and 2) make sense on general globally causal space-time, at least when there exist four one-parameter groups of space-time symmetries with the corresponding vector fields which span everywhere the the tangent space, playing the role analogous to translations (they do not have to commute). Because 1) and 2) make sense on general globally causal space-time, and because the high energy scattering experiments  confirm 1) and 2) and because these experiments are less sensitive to the global structure of space-time we arrive at the conclusion that the geometry of space-time will have to be changed into some other globally causal. 
This throws some light on the problem why we are rather successful in understanding Rutherford-type experiments with deeply inelastic scattering of electron by a nucleon (e. g. the series of the famous experiments of SLAC-MIT cooperation) whenever we confine ourselves to the scattering at the level of many-particle plane wave states of elementary fields for high energies (say inside the nucleon), but at the same time we cannot account for the lower energy scattering involved into bound state production, using the same elementary fields with the same gauge interactions.

\begin{center}
{\small (II)}
\end{center}
The interacting fields are so much singular, that even after ``smearing'' with test functions do not give any ordinary operators in the Fock space, but only generalized operators transforming continously 
$(\boldsymbol{E}) \rightarrow (\boldsymbol{E})^*$. Even the Wick polynomials of free fields (if they contain zero mass free field factors) behave in the similar singular manner, and even for free fields averages of some physical local quantities, entering e.g. energy-momentum tensor, cannot be sensibly computed even after ``smearing out'' over compact domains (in case we are using Hida operators as annihilation-creation operators). In particular no sensible quasi-classical limit 
exists for interacting fields, and even for free fields no sensible computation of the averages of the analogues of important classical measurable local quantities, can be performed. Thus no sensible quasi-classical limit exists for interacting quantum fields, and even for free fields. These strange theorems are mathematical conclusions of the above principles 1) -- 3). Put otherwise: from 1) and 2) it follows that no interacting (and even free) quantum fields can be constructed on the flat Minkowski space-time which have the classical limit, or which have (in Bohr's parlance) correspondence to classical fields. Thus we arrive at the conclusion, that it follows from the principles 1) and 2) that the quantum fields which have correspondence to their classical counterparts and have classical limits do possess non trivial weight, and cannot be constructed on the flat Minkowski space-time. Thus we arrive at the conclusion that the numerical value of the Newton's gravitational constant $G$, and the Einstein equations of gravitation for the classical limit of the quantum fields should follow from the principles 1), 2) and the assumed causal  geometry of space-time,
by the very rule of construction of the states in which the quantum fields behave as their classical counterparts.  We are lead to the conclusion that no extra law joining gravitation to quantum fields is needed. Each system of (interacting) quantum fields can coexsist with the given spacetime
geometry if the (interacting) quantum fields constructed according to 1) and 2) with the assumed globally causal gemetry of the space-time, admit classical limit. This is all we can know at present about the relationship betweeen the matter and space-time geometry. Of course in order to make this mathematical conclusion to be not empty we need to give at least one non-trivial example of a globally causal space-time, on which the free and the causally constructed interacting fields behave so regularily as to admit classical limits in which the average values of the quanum analogues of the quantities  entering the Hilbert energy-momentum tensor components can sensibly be computed and compared to the classical values. Moreover if we are about to preserve the first two priciples 1) and 2), except in changing the globally causal spacetime geometry of the Minkowski space-time into some other globally causal space-time,
we should make this change so as to preserve agreement with high energy scattering experiments. 
This is strong limitation because theory based on 1) and 2) with the assumption that the many particle plane wave states as \emph{in} and \emph{out} states gives the effective cross sections which are in agreement with experiment although the plane wave states are assumed to live on the flat Minkowski space-time. This is very non trivial and valuable limitation, which in particular limits the set of allowable ``plane wave'' packets as the sensible single particle states in the Fock spaces of local free quantum fields on the chosen space-time. Unfortunately almost all works concerned with construction of local free quantum fields on globally causal space-times other than the Minkowski space-time ignore this condition. But fortunately there is one exception: in the series of works \cite{SegalZhouQED}, \cite{SegalZhouPhi4}, \cite{PaneitzSegalI}-\cite{PaneitzSegalIII}, in which free quantum fields are constructed on the static Einstein Universe, relation to the scattering phenomena on the Minkowski space-time is seriously accounted for in the form sufficent for our purposes. Unfortunately these authors do not use the Hida operators nor the white noise analysis of integral kernel operators, which is very effective in the investigation of mass-less fields. Nonetheless results obtained by them already show that all local massive free fields on the  Einstein Universe as well as the QED interaction Lagrange density behave so regularily as expected: these fields are well defined (unbounded) operators on the Fock space, even when evaluated at single specified space-time point.

\vspace*{1cm}

Presented work is the separated part of the work \cite{wawrzycki2018} which is focused on the 
solution of the ``Adiabatic Limit Problem''. 
The whole work \cite{wawrzycki2018} also contains exploration of
the problems which were beyond the conventional method: 1) analysis of the structure of infra-red states, 2) relationship between the interacting fields and the classical gravitational field. Because both of them 
require a considerable amount of harmonic analysis on $SL(2, \mathbb{C})$ or a non-trivial extension of the harmonic analysis on $T_4 \circledS SL(2, \mathbb{C})$ over to Krein-isometric representations in the Krein-Hilbert space, we have decided to separate off the part devoted to the ``Adiabatic Limit Problem'', not immediately involved into the representation theory. 

From the purely mathematical point of view the present work may be considered as an immediate extension of the works: \cite{hida}, \cite{obataJFA}, \cite{obata-book}, of Hida and his school, on the so called integral kernel operators and Fock expansions into integral kernel operators.

The following Subsection of Introduction gives a more detailed formulation of our result.

\subsection{Adiabatic Limit Problem and its solution. Short account}\label{AdiabaticLimit}

We keep the causal method of St\"uckelbeg-Bogoliubov-Epstein-Glaser
unchanged, with the only proviso: we insert into the formulas the free fields of the theory 
which are constructed with the help of white noise Hida operators -- construction
of free fields which goes back to Berezin and later improved by the Japanese school of 
Hida. This allows us to interpret the free fields as integral kernel operators
with vector-valued distribution kernels in the sense of Obata.
The rest part of the work is reduced to application of the white noise calculus of integral kernel 
operators, which essentially is reduced to the proof that the operations involved in the causal perturbative 
construction of the higher order contributions are well defined when applied to the integral 
kernel operators defined by free fields. The main difficulty lies in the white noise construction of the 
free fields, namely the free Dirac and electromagnetic fields $\boldsymbol{\psi}$, $A$, as finite sums
\[
\boldsymbol{\psi} = \Xi_{0,1}(\kappa_{0,1}) + \Xi_{1,0}(\kappa_{1,0}), \,\,\,
A= \Xi_{0,1}(\kappa'_{0,1}) + \Xi_{1,0}(\kappa'_{1,0}) 
\in \mathscr{L}\big((E) \otimes \mathscr{E}, \, (E)^* \big)
\]
(of two) well defined integral kernel operators, in the sense of Obata \cite{obataJFA},
with vector valued distributional kernels $\kappa, \kappa'$ which belong respectively to
\[
\mathscr{L}\big(E, \mathscr{E}^* \big),
\] 
Here $E$ is the respective nuclear space  of restrictions of the Fourier transforms $\widetilde{\varphi}$ of all 
space-time test functions $\varphi \in \mathscr{E}$ to the respective orbit $\mathscr{O}$
in the momentum space determining the representation of the $T_4 \circledS SL(2, \mathbb{C})$ 
acting in the single particle Hilbert space of the respective  field, $\boldsymbol{\psi}$ or $A$.
$\mathscr{L}\big(E, \mathscr{E}^* \big)$ denotes the space of all linear continuous operators 
$E \rightarrow \mathscr{E}^*$,
\emph{i. e.} $\mathscr{E}^*$-valued distributions over the corresponding orbit $\mathscr{O}$ 
in the momentum space (recall that $\mathscr{O}$ is equal to the positive energy sheet of the hyperboloid
$p \cdot p = m^2$ in the momentum space in case of field of mass $m$). 
We endow $\mathscr{L}\big(E, \mathscr{E}^* \big)$ with the natural topology of uniform convergence 
on bounded sets. $(E), (E)^*$ is the nuclear Hida subspace
 of the Fock space of the corresponding free field, and its strong dual space.

Moreover in order to construct the useful commutative algebra of operators to which the 
perturbative expansion can naturally be applied, we need a construction of the free fields, 
$\boldsymbol{\psi}$, $A$, with as explicit representation of the Poincar\'e group in their Fock spaces as 
possible. Unfortunately no construction of these two most important fields in the whole of 
QFT, namely $\boldsymbol{\psi}$ and  $A$, based on the theory of representations
of $T_4 \circledS SL(2, \mathbb{C})$, has been achieved, which is a well known fact, compare
\cite{Haag}, p. 48, \cite{lop1}, \cite{lop2}. This is because this problem cannot be solved within the ordinary
unitary representations of the $T_4 \circledS SL(2, \mathbb{C})$ group. 
We have been forced to extend the Mackey theory of induced representations over to a more general class 
of representations in order to solve this unsolved problem, compare Section 12 of \cite{wawrzycki2018} 
for this extension. 
But this is not the whole problem, because we additionally need
a white noise constructions of these two free fields $\boldsymbol{\psi}$ and  $A$.       
This construction is essentially worked out for the simplest massive free scalar field by mathematicians
\cite{HKPS}, and its generalization to other massive fields (if the group theoretical aspect is ignored)
presents no essential difficulties. But concerning the mass less fields, such e. g. as $A$,
the white noise construction is far not so obvious and in fact (as to the author's knowledge) 
has not been done before. This is because the white noise construction of the mass less fields requires the modification
of the space-time test space $\mathscr{E}$ which cannot be equal $\mathcal{S}(\mathbb{R}^4; \mathbb{C}^4)$
but instead it has to be equal to the space $\mathscr{E} = \mathcal{S}^{00}(\mathbb{R}^{4};\mathbb{C}^4)$.
Namely $\varphi  \in \mathcal{S}^{00}(\mathbb{R}^{4};\mathbb{C}^4)$ if and only if its Fourier transform
$\widetilde{\varphi} \in \mathcal{S}^{0}(\mathbb{R}^{4};\mathbb{C}^4)$, and 
$\mathcal{S}^{0}(\mathbb{R}^{4};\mathbb{C}^4)$ is the subspace of $\mathcal{S}(\mathbb{R}^{4};\mathbb{C}^4)$
of all those functions which have all derivatives 
vanishing at zero. Correspondingly we have the nuclear algebra $E$
of all restrictions of Fourier transforms to the corresponding orbit $\mathscr{O}$ (positive energy sheet of the cone) 
of the elements of the test space 
$\mathscr{E} = \mathcal{S}^{00}(\mathbb{R}^{4};\mathbb{C}^4)$, equal to
$E= \mathcal{S}^{0}(\mathbb{R}^3; \mathbb{C}^4)$ (of $\mathbb{C}^4$-valued functions in case of the field $A$,
but for the $r$-component mass less fields we will have $\mathbb{C}^r$-valued functions here).
This is related to the singularity of the cone orbit $\mathscr{O}$ at the apex -- the orbit pertinent to the representation associated with mass less fields, i.e. the positive sheet of the cone in the momentum space 
(note that each sheet of the massive hyperboloid $\mathscr{O}_{m,0,0,0} = \{p \cdot p = m^2\}$ in the momentum space is everywhere smooth only for the massive orbit of the point $\bar{p} = (m,0,0,0)$ with $m \neq 0$, the zero mass orbits of
$\bar{p} = (1,0,0,1)$ or $(-1,0,0,1)$, i.e. the positive and negative energy sheets of the cone are singular at the apex).
The need for the modification of the space-time test space $\mathscr{E}$, when passing to mass less fields,
may seem  unexpected for those readers which compare it with the construction of mass less fields in the sense of Wightman,
which allows the ordinary Schwartz test space also for the mass less fields.
We nonetheless choose the white noise construction of free fields as much more adequate mathematical interpretation
of the (free) quantum field. Among other things the white noise construction provides a much deeper insight
into the Wick product construction of free fields  at the same space-time point, which moreover fits well with the needs of the causal perturbative approach. ``Wick product'' construction due to Wighman and G{\aa}rding 
(although also rigorous) is not very much useful for the realistic causal perturbative QFT, such as QED. 
Again that the Wightman-G{\aa}rding ``Wick product'' is not useful in practical computations
such as the causal perturbative approach, or in construction of conserved currents corresponding to the Noether 
theorem (which in fact is the basis for the Canonical Quantization Postulate) has been recognized by Segal
\cite{Segal-NFWP.I}, a prominent analyst
who devoted much part of his research to the mathematical analysis of the Wick product construction.

Thus we give here white noise construction of the free field $\boldsymbol{\psi}$ with the explicit
construction of the representation of $T_4 \circledS SL(2, \mathbb{C})$, compare Sections \ref{e+e-},
\ref{electron}, \ref{positron}. The white noise construction of the electromagnetic potential field
$A$ in the Gupta-Bleuler gauge with explicit construction of the Krein-isometric representation of
$T_4 \circledS SL(2, \mathbb{C})$ acting in the Krein-Fock space of the free field $A$
is given in Sections 4 and 5 of \cite{wawrzycki2018}.
As to the author's knowledge it has not been done before.
In Subsections \ref{A=Xi0,1+Xi1,0} and \ref{equivalentA-s} there are summarized some of the results 
obtained in \cite{wawrzycki2018} concerning the free field $A$ which are used in this work.

In fact the white noise construction of the free fields is not a knew idea and goes back to Berezin.
Subsequently  it was developed mainly by Hida and his school.   

The fact that the test space $\mathcal{S}^{00}(\mathbb{R}^{4};\mathbb{C}^4)$ contains no non-zero
elements with compact support does not destroy splitting of causal homogeneous 
distributions into retarded and advanced parts, because the pairing functions of mass less fields, such as $A$, are homogeneous distributions. 
The test space $\mathcal{S}^{00}(\mathbb{R}^{4};\mathbb{C}^4)$
is flexible enough to contain non zero element for each conic-type set, supported on this set. This allows splitting
of causal homogeneous distributions (Subsection 5.7 of \cite{wawrzycki2018}). 

Having given the free fields,  $\boldsymbol{\psi}$ and  $A$, constructed as (finite sums of) integral kernel 
operators with vector-valued kernels, we show that the operations of differentiation, Wick product
at the same space-time point, integration of the Wick product and its convolution with tempered distribution
are well defined within the class of integral kernel operators to which the free fields and Wick product 
belongs (Subsection \ref{OperationsOnXi}). In particular the formulas for each $n$-th order contributions, with the intensity of the interaction function
$g=1$, are equal to finite sums 
\[
\begin{split}
\boldsymbol{\psi}_{{}_{\textrm{int}}}^{(n)}(g=1, x) = \sum \limits_{l,m} \Xi_{l,m}\big(\kappa_{l,m}(x)\big), \\
A_{{}_{\textrm{int}}}^{(n)}(g=1, x) = \sum \limits_{l,m} \Xi_{l,m}\big(\kappa'_{l,m}(x)\big), \\
\end{split}
\] 
of integral kernel operators (similarly we have for $\Xi_{l,m}\big(\kappa'_{l,m}(x)\big)$)
\begin{multline*}
\Xi_{l,m}\big(\kappa_{l,m}(x)\big) = \\
\sum \limits_{s_1, \ldots, s_{l+m}} \int \limits_{\mathbb{R}^{3(l+m)}}
\kappa_{l,m}(s_1, \boldsymbol{\p}_{1}, \ldots, s_{l+m}, \boldsymbol{\p}_{l+m}; x) \,
a_{s_{1}}(\boldsymbol{\p}_{1})^{+} \cdots a_{s_{l+m}}(\boldsymbol{\p}_{l+m}) \,
\ud^3 \boldsymbol{\p}_{1} \cdots \ud^3 \boldsymbol{\p}_{l+m}, 
\end{multline*}
where $a_s(\boldsymbol{\p})^{+}, a_{s}(\boldsymbol{\p})$ are the creation and annihilation 
operators, constructed here as Hida operators in the tensor product of the Fock spaces of the 
free fields $\boldsymbol{\psi}, A$, in the normal order, with the first $l$ factors
equal to the creation operators and the last $m$ equal to the annihilation operators. Here
\[
\begin{split}
\kappa_{l,m} \in \mathscr{L}\big(E^{\otimes (l+m)}, \mathscr{E}_{1}^{*} \big), 
\,\,\,\, \mathscr{E}_{1} = \mathcal{S}(\mathbb{R}^4; \mathbb{C}^4)  \\
\kappa'_{l,m} \in \mathscr{L}\big(E^{\otimes (l+m)}, \mathscr{E}_{2}^{*} \big), 
\,\,\,\, \mathscr{E}_{2} = \mathcal{S}^{00}(\mathbb{R}^4; \mathbb{C}^4)
\end{split}
\]
with each factor $E$ in the tensor product $E^{\otimes(l+m)}$ equal
\[
E= \mathcal{S}(\mathbb{R}^3; \mathbb{C}^4)\,\,\, \textrm{or} \,\,\, 
E= \mathcal{S}^{0}(\mathbb{R}^3; \mathbb{C}^4).
\]
Each of the operators $\Xi_{l,m}\big(\kappa_{l,m}(x)\big)$, $\Xi_{l,m}\big(\kappa'_{l,m}(x)\big)$ determines
a well defined integral kernel operator 
\[
\begin{split}
\Xi_{l,m}\big(\kappa_{l,m}\big) 
\in \mathscr{L}\big((\boldsymbol{E}) \otimes \mathscr{E}_{1} , \, (\boldsymbol{E})^*\big)
\cong \mathscr{L}\big(\mathscr{E}_{1},  \mathscr{L}((\boldsymbol{E}), (\boldsymbol{E})^*) \big),  \\
\Xi_{l,m}\big(\kappa'_{l,m}\big) 
\in \mathscr{L}\big((\boldsymbol{E}) \otimes \mathscr{E}_{2} , \, (\boldsymbol{E})^*\big) 
\cong \mathscr{L}\big(\mathscr{E}_{2},  \mathscr{L}((\boldsymbol{E}), (\boldsymbol{E})^*) \big)
\end{split}
\]
with vector-valued distribution kernel $\kappa_{l,m}$ , respectively, $\kappa'_{l,m}$, in the sense 
of Obata \cite{obataJFA}, where $(\boldsymbol{E})$ is the nuclear Hida subspace in the tensor product of the Fock spaces 
of the fields $\boldsymbol{\psi}$ and $A$. The integral kernel operators 
$\Xi_{l,m}\big(\kappa_{l,m}(x)\big)$, $\Xi_{l,m}\big(\kappa'_{l,m}(x)\big)$
are uniquely determined by the condition 
\[
\begin{split}
\big\langle \big\langle \Xi_{l,m}(\kappa_{l,m})(\Phi \otimes \phi), \Psi  \big \rangle \big \rangle
= \langle \kappa_{l,m}(\eta_{\Phi, \Psi}), \phi \rangle,
\,\,\,
\Phi, \Psi \in (\boldsymbol{E}), \phi \in \mathscr{E}_{1}, \\
\big\langle \big\langle \Xi_{l,m}(\kappa'_{l,m})(\Phi \otimes \phi), \Psi  \big \rangle \big \rangle
= \langle \kappa'_{l,m}(\eta_{\Phi, \Psi}), \phi \rangle,
\,\,\,
\Phi, \Psi \in (\boldsymbol{E}), \phi \in \mathscr{E}_{2}, 
\end{split}
\]
where 
\[
\eta_{\Phi, \Psi}(s_1, \boldsymbol{\p}_{1}, \ldots, s_{l+m}, \boldsymbol{\p}_{l+m}) =
\Big\langle \Big\langle a_{s_{1}}(\boldsymbol{\p}_{1})^{+} \cdots a_{s_{l+m}}(\boldsymbol{\p}_{l+m}) \,
\Phi, \, \Psi  \Big\rangle \Big\rangle.
\]
Note that
\[
\eta_{\Phi, \Psi} \in E^{\otimes (l+m)}, \,\,\, \Phi, \Psi \in (\boldsymbol{E}),
\]
with the canonical pairing $\langle\langle \cdot, \cdot \rangle\rangle$ on $(\boldsymbol{E})^* \times (\boldsymbol{E})$.
These results are contained as a particular case of Theorem \ref{g=1InteractingFieldsQED} of 
Subsection \ref{OperationsOnXi}, compare also Section \ref{A(1)psi(1)}.

Moreover the interacting fields, in the adiabatic limit $g=1$, can be understood as Fock expansions
\[
\begin{split}
\boldsymbol{\psi}_{{}_{\textrm{int}}}(g=1) = \sum \limits_{l,m} \Xi_{l,m}\big(\kappa_{l,m}\big), \\
A_{{}_{\textrm{int}}}(g=1) = \sum \limits_{l,m} \Xi_{l,m}\big(\kappa'_{l,m}\big), 
\end{split}
\]
into  integral kernel operators in the sense of \cite{obataJFA} with all terms $\Xi_{l,m}\big(\kappa_{l,m}\big)$, 
$\Xi_{l,m}\big(\kappa'_{l,m}\big)$ equal to integral kernel operators with vector-valued kernels, and all 
belonging to the class indicated above. Even more, most of the terms 
$\Xi_{l,m}\big(\kappa_{l,m}\big)$, $\Xi_{l,m}\big(\kappa'_{l,m}\big)$ behave even much more ``smoothly''
(although it is not necessary for the theory to work) and 
\[
\begin{split}
\Xi_{l,m}\big(\kappa_{l,m}\big) 
\in \mathscr{L}\big((\boldsymbol{E}) \otimes \mathscr{E}_{1} , \, (\boldsymbol{E})\big)
\cong \mathscr{L}\big(\mathscr{E}_{1},  \mathscr{L}((\boldsymbol{E}), (\boldsymbol{E})) \big),  \\
\Xi_{l,m}\big(\kappa'_{l,m}\big) 
\in \mathscr{L}\big((\boldsymbol{E}) \otimes \mathscr{E}_{2} , \, (\boldsymbol{E})\big) 
\cong \mathscr{L}\big(\mathscr{E}_{2},  \mathscr{L}((\boldsymbol{E}), (\boldsymbol{E})) \big).
\end{split}
\]
In particular the first order contribution $A_{{}_{\textrm{int}}}^{\mu \,(1)}(g=1)$, given by
\[
A_{{}_{\textrm{int}}}^{\mu \,(1)}(g=1,x) =
-\frac{e}{4 \pi} \int \ud^3 \boldsymbol{x_{1}} 
\frac{1}{|\boldsymbol{x_1} - \boldsymbol{x}|}
\,
: \overline{\psi} \gamma^\mu \psi : (x_0 - |\boldsymbol{x_1} - \boldsymbol{x}|, \boldsymbol{x_1}),
\]
to the interacting potential field, belongs to
\[
\mathscr{L}\big((\boldsymbol{E}) \otimes \mathscr{E}_{2} , \, (\boldsymbol{E})\big) 
\cong \mathscr{L}\big(\mathscr{E}_{2},  \mathscr{L}((\boldsymbol{E}), (\boldsymbol{E})) \big).
\]

\section{White noise construction of the Dirac and electromagnetic potential fields 
as integral kernel operators. Fundamental operations performed upon integral kernel operators}\label{e+e-}

Here we present the white noise construction of the free quantized Dirac field
within the white noise set-up of Hida, Obata and
Sait\^o \cite{hida}, \cite{obata-book},
and which is a rigorous realization of the field along the lines suggested (partially heuristically) by 
Berezin \cite{Berezin}.
This construction can be regarded as a far reaching 
extension of the definition due to Wightman  
\cite{wig} of the (free) field, and enters into the analysis of the distributional
(generalized) states.
We should emphasise here that the definition of Wightman is operationally and computationally 
much weaker. In general the two definitions are not equivalent.  
 The main advantage we gain when constructing free fields within the white noise formalism  
is that we can give a rigorous meaning to the (free) quantum field
 of the so called \emph{integral kernel operator
with vector-valued distributional kernel} (in the sense 
\cite{obataJFA} or \cite{obata-book}, Chap. 6.3), which would be impossible within Wightman set-up. 
This allows to give the meaning of integral kernel operators (with vector-valued kernels) 
to the (generalized) operators under the formula (17.1) in \cite{Bogoliubov_Shirkov}, p. 154, or
equivalently to the (generalized) operators (43) of \cite{Epstein-Glaser}, Sect. 4, p. 229.  
In particular when constructing free fields according to Berezin-Hida we obtain 
Theorem 0 of \cite{Epstein-Glaser} as a corollary to theorems 2.2 and 2.6 of 
\cite{hida} and Thm. 3.13 of \cite{obataJFA} with the domain $\mathcal{D}_0$ 
replaced with the so called Hida test space of white noise functionals. 
Moreover using the Berezin-Hida construction of free fields we gain a rigorous formulation and proof of 
the so called ``Wick theorem'', 
as stated in \cite{Bogoliubov_Shirkov}, Chap. III. It should be emphasized that Wightman's 
definition of the (free) field \cite{wig}, does not provide sufficient computational basis for any rigorous formulation and  proof of the ``Wick theorem'' for free fields as stated in \cite{Bogoliubov_Shirkov}, Chap. III.
Note also that the (free) field constructed within the white noise calculus is well defined at
space-time point as a generalized operator transforming the so called Hida space into its strong dual.

One should note that although the definition of the ``Wick product'' of Wightman and G{\aa}rding
\cite{WightmanGarding} based on the Wightman's definition \cite{wig} of the field, is mathematically rigorous,
it suffers at several crucial points from being computationally ineffective in computations which are important 
from the physical point of view:
\begin{enumerate}
\item[1)] 
The space-time averaging limits
in  Wightman and G{\aa}rding's
\cite{WightmanGarding} definition of the ``Wick product'' are by no means 
canonical and involve a considerable amount
of arbitrariness. 

\item[2)] 
Although Wightman and G{\aa}rding
\cite{WightmanGarding} are able to construct their own ``Wick products'' which, after smearing out over space-time domains becomes well defined densely defined unbounded operators, it would be difficult to investigate the closability questions
for these operators, their eventual self-adjointness,
as well as averaging over space-like (equal-time) surfaces, within the method of Wightman and G{\aa}rding.  
But the equal-time averagings are involved through conserved currents when
we consider Noether theorem for free fields -- fundamental from the more conventional, and used by physicists, approach to commutation rules and the more traditional proof of the Pauli theorem for free fields 
(compare \cite{Bogoliubov_Shirkov}). 

\item[3)] 
 Wightman and
G{\aa}rding definition of the ``Wick product'' \cite{WightmanGarding} is not a sufficient basis for 
the strict formulation
and proof of the ``Wick theorem'' as stated in  \cite{Bogoliubov_Shirkov}, Chap. III, so fundamental for the causal approach to QFT which avoids ultraviolet divergences. Note in particular that Theorem 0 of 
\cite{Epstein-Glaser} is formulated and proved on the basis of partially heuristic (but solid)
arguments of the more traditional approach presented in  \cite{Bogoliubov_Shirkov}, Chap. III, which 
uses the free fields at specified space-time points in the intermediate stage, and which are not merely
symbolic in their character (contrary to what we encounter in the Wightman-G{\aa}rding's approach).
White noise construction of free fields on the other hand do provide a sufficient basis for the 
rigorous formulation and proof of ``Wick theorem'' for free fields of \cite{Bogoliubov_Shirkov}, Chap. III.
  
\item[4)]
But most of all when constructing free fields using the white noise formalism, as integral kernel operators with vector-valued kernels, we are able to give a rigorous
meaning to each order term contribution to interacting fields in QED (within the  causal perturbative approach), of an  integral kernel operator with vector-valued distribution kernel (in the sense \cite{obataJFA}), 
which defines a well defined operator valued distribution on the 
space-time test space -- a continuous map from the space-time test space
to the linear space of continuous linear operators on the Hida space into its dual 
(with the standard topology of uniform convergence on bounded sets). 
Each such contribution can be averaged in the states of the Hida subspace and defines a scalar distribution
as a functional of space-time test function.
The crucial point is that these contributions do not loose this rigorous 
sense even for the ``coupling space-time function $g$'' put everywhere equal to unity, which allows to avoid both: ultraviolet and infra-red infinities in the perturbative (causal) approach to QED. 
For a detailed proof of this assertion and analysis of the all higher order contributions to the Dirac and 
electromagnetic potential interacting fields, compare  Subsection \ref{OperationsOnXi},
Sect. \ref{A(1)psi(1)}. In particular we can
reach in this way a positive solution to the existence problem for the adiabatic limit in QED using a method which is applicable to interactions and fields of more general character, e.g. to the Standard Model. 
   
\end{enumerate}

For these reasons we regard the white noise construction of (free) fields of Berezin-Hida
as integral kernel operators (with vector-valued distributional kernels) as more adequate
mathematical interpretation of the (free) quantum field than the one proposed by Wightman \cite{wig}.

In this Section we present white noise Berezin-Hida construction of the free Dirac field
as an integral kernel operator with  vector-valued distributional kernel in the sense of Obata
\cite{obataJFA}. In the work \cite{wawrzycki2018}
we give the white noise construction of the free electromagnetic potential field $A$,
which again may be interpreted as integral kernel operator 
with vector-valued distributional kernel in the sense of Obata \cite{obataJFA}, compare
Subsections \ref{A=Xi0,1+Xi1,0} and \ref{equivalentA-s} where some of the results of \cite{wawrzycki2018}
concerning the free field $A$ are summarized (in fact these Subsections 
are borrowed from \cite{wawrzycki2018}).
  
   We present the construction of the Dirac field $\boldsymbol{\psi}$
in several steps, keeping the presentation as general as possible, in order to make it to serve as 
an introduction to the construction of (free) local fields within the white noise formalism. 

Firstly, we give definition of the Hilbert space which is subject to second 
quantization functor, and then in the remaining four steps quantize it.
The steps are realized in the following Subsections: \ref{FirstStepH},
\ref{electron}, \ref{positron}, \ref{electron+positron}, \ref{psiBerezin-Hida}. 
Subsection \ref{psiBerezin-Hida} is the longest, but it contains an introduction to the papers
\cite{hida}, \cite{obataJFA} on integral kernel operators with scalar-valued and respectively 
vector-valued distributional kernels in fermi and bose Fock spaces (note that \cite{hida}, \cite{obataJFA}
give detailed analysis for the bose case), which is of use in the remaining part
of the whole work, and which is not so much pertinent to the specific Dirac field
$\boldsymbol{\psi}$, but which is
important for general local fields constructed within the white noise calculus.
In particular we are using the cited theorems of \cite{hida}, \cite{obataJFA} on integral kernel operators 
in the proof of \emph{Bogoliubov-Shirkov Hypothesis} (equivalently the classic Pauli theorem)
for the Dirac field $\boldsymbol{\psi}$ (Subsection \ref{StandardDiracPsiField}) and 
for the electromagnetic potential field (Subsection 5.9 of \cite{wawrzycki2018});
and finally in the analysis of contributions to interacting fields in QED 
(Subsection \ref{OperationsOnXi}).

Subsection \ref{OperationsOnXi} is devoted to the proof that the contributions to interacting
fields in causal perturbative spinor QED are well defined integral kernel operators with vector-valued kernels
in the sense of Obata \cite{obataJFA} whenever we are using in the causal construction of interacting fields
the free fields which themselves are well defined integral kernel 
operators in the sense of Obata. Nonetheless Subsection \ref{OperationsOnXi} is of more general 
character not pertinent to the special case of spinor QED. It is devoted to
the fundamental operations performed upon the free fields, understood as integral kernel operators with vector-valued kernels, which serve as fundamental computational rules in construction of the theory, in particular in construction 
of the perturbative series for interacting fields such as: Wick product of free fields, derivation and  integration
operations. These operations have general character and can be extended over other causal perturbative QFT. 

We add three additional Subsections \ref{psiWightman}, \ref{MotivationForHida}
and \ref{StandardDiracPsiField}.  Subsections \ref{psiWightman} and \ref{MotivationForHida} give a motivation for using white noise 
calculus and for using the construction of fields due to Berezin-Hida, as integral kernel operators with vector-valued kernels. In other words Subsections \ref{psiWightman} and \ref{MotivationForHida}
give motivation for introduction of Hida operators 
as the annihlation and creation operators of free quantum fields of the theory and white
noise analysis into the perturbative casual formulation of QFT.

Subsection \ref{StandardDiracPsiField} contains comparison with the standard realization of the free Dirac field and is devoted to the Bogoliubov-Shirkov Postulate (first Noether theorem for free fields
and the classic Pauli theorem on spin-statistics relation).  

In this Section $m > 0$ has the constant value equal to the electron mass.

\subsection{Definition of the Hilbert space $\mathcal{H}$ which is then subject to the second quantization
functor $\Gamma$}\label{FirstStepH}

This is the Hilbert space $\mathcal{H}$ of bispinor solutions $\phi$ (regular function-like distributions
on the Schwartz space $\mathcal{S}(\mathbb{R}^4; \mathbb{C}^4)$ of testing bispinors transforming according to the 
law (27) of Subsection 2.1 of \cite{wawrzycki2018})  of the Dirac equation
\[
(i \gamma^\mu \partial_\mu ) \phi = m \phi,
\] 
with the inner product 
\begin{equation}\label{Inn-Prod-Single-Dirac}
(\widetilde{\phi}, \widetilde{\phi'}) = m \int \limits_{x^0 = const.} \Big(\phi(x), \phi'(x) \Big)_{{}_{\mathbb{C}^4}}
\, \ud^3 x,
\end{equation}
and transformation law (27) of Subsection 2.1 of \cite{wawrzycki2018}, 
compare e.g. \cite{Scharf} or \cite{Bogoliubov_Shirkov}.
This means that the Fourier transform $\widetilde{\phi}$ of the bispinor $\phi \in \mathcal{H}$ (regular distribution)
is concentrated on the disjoint sum of the positive and negative energy orbits $\mathscr{O}_{m,0,0,0} \sqcup
\mathscr{O}_{-m,0,0,0}$ and $\widetilde{\phi}$ cannot be regarded as ordinary function on the full range of 
$p \in \mathbb{R}^4$ of the momentum space. Nonetheless $\widetilde{\phi}$ is a well defined 
(singular, i.e. non-function-like) distribution in the
Schwartz space 
\[
\mathcal{S}(\mathbb{R}^4; \mathbb{C}^4) = \mathcal{S}(\mathbb{R}^4; \mathbb{C})
\oplus \mathcal{S}(\mathbb{R}^4; \mathbb{C}) \oplus \mathcal{S}(\mathbb{R}^4; \mathbb{C}) \oplus 
\mathcal{S}(\mathbb{R}^4; \mathbb{C})
\]  
of bispinors on $\mathbb{R}^4$ (transforming according to (24) and (25), Subsect. 2.1 of \cite{wawrzycki2018}).
It defines an ordinary bispinor-function $p \mapsto \widetilde{\phi}(p)$ on the disjoint sum $\mathscr{O}_{m,0,0,0} \sqcup \mathscr{O}_{-m,0,0,0}$ of the positive and resp. negative energy orbits, which we denote likewise by the
symbol $\widetilde{\phi}$ (although it makes sense as a function only on the disjoint sum of the respective orbits and not on the whole $\mathbb{R}^4$ space), and which is square integrable with respect to the inner product (compare
(28), Subsect. 2.1 of \cite{wawrzycki2018}) induced by the above inner product 
(\ref{Inn-Prod-Single-Dirac}) in $\mathcal{H}$.   
Namely for $\phi \in \mathcal{H}$, the action of the Fourier transform $\widetilde{\phi}$
on $\widetilde{f} \in \mathcal{S}(\mathbb{R}^4; \mathbb{C}^4)$ is by definition equal to the integration
of the product of the mentioned function $p \mapsto \widetilde{\phi}(p)$ by the restriction of $\widetilde{f}$ to the disjoint sum $\mathscr{O}_{m,0,0,0} \sqcup \mathscr{O}_{-m,0,0,0}$
along $\mathscr{O}_{m,0,0,0} \sqcup \mathscr{O}_{-m,0,0,0}$ with respect to the invariant measure on
$\mathscr{O}_{m,0,0,0} \sqcup \mathscr{O}_{-m,0,0,0} \subset \mathbb{R}^4$ induced by the 
invariant measure $\ud^4 p$ on $\mathbb{R}^4$. Thus, by definition of the singular distribution 
$\delta(P=0)$, where $P$ is a smooth function on $\mathbb{R}^4$ such that $\textrm{grad} \, P \neq 0$
on the surface $P=0$ (compare \cite{GelfandI}, Chap. III), we have
\begin{multline*}
\int \phi(x) \, f(x) \, \ud^4 x = \langle \widetilde{\phi}, \widetilde{f} \rangle 
= \int \widetilde{\phi}(p) \widetilde{f}(p) \, \ud^4p \\
= \int \delta(p\cdot p -m^2) \, \widetilde{\phi}(p) \widetilde{f}(p) \, \ud^4 p \\
=\int \delta(p\cdot p -m^2) \Theta(p_0) \, \widetilde{\phi}(p) \widetilde{f}(p) \, \ud^4 p
+ \int \delta(p\cdot p -m^2) \Theta(-p_0) \, \widetilde{\phi}(p) \widetilde{f}(p) \, \ud^4 p \\
=\int \limits_{\mathscr{O}_{m,0,0,0}} \, \widetilde{\phi}(p) 
\widetilde{f}|_{{}_{\mathscr{O}_{m,0,0,0}}}(p) \, \ud \mu_{{}_{m,0}}(p) 
+ \int \limits_{\mathscr{O}_{-m,0,0,0}} \, \widetilde{\phi}(p) 
\widetilde{f}|_{{}_{\mathscr{O}_{-m,0,0,0}}}(p) \, \ud \mu_{{}_{-m,0}}(p).
\end{multline*}

From now on we agree to denote the ordinary bispinor function $\widetilde{\phi}$ on the disjoint sum
$\mathscr{O}_{m,0,0,0} \sqcup \mathscr{O}_{-m,0,0,0}$ (equal to the distributional Fourier support of the distribution $\widetilde{\phi}$)
by the same symbol $\widetilde{\phi}$ as the distributional Fourier transform $\widetilde{\phi}$
of $\phi \in \mathcal{H}$ (although $\widetilde{\phi}$
makes sense as the ordinary function only on the support of the distribution $\widetilde{\phi}$,
which as a ``function'' is intentionally equal zero outside the support, which makes a precise sense
when $\widetilde{\phi}$ is regarded as distribution defined as above).

In short for $\phi \in \mathcal{H}$ we can write
\[
\phi(x) = \int \limits_{\mathscr{O}_{m,0,0,0}} \, \widetilde{\phi}(p) \, 
e^{-ip\cdot x} \, \ud \mu_{{}_{m,0}}(p) 
+ \int \limits_{\mathscr{O}_{-m,0,0,0}} \, \widetilde{\phi}(p) \, 
e^{-ip\cdot x} \, \ud \mu_{{}_{-m,0}}(p);
\] 
or
\begin{multline}\label{DistributionalSolDiracEq}
\phi(x) = \int \limits_{\mathscr{O}_{m,0,0,0}} \, \widetilde{\phi}(p) \, 
e^{-ip\cdot x} \, \ud \mu_{{}_{m,0}}(p) 
+ \int \limits_{\mathscr{O}_{-m,0,0,0}} \, \widetilde{\phi}(p) \, 
e^{-ip\cdot x} \, \ud \mu_{{}_{-m,0}}(p) \\
= \int \limits_{\mathbb{R}^3} \widetilde{\phi}(\vec{p},|p_0(\vec{p})|) \, e^{-(i|p_0(\vec{p})|t -i\vec{p}\cdot \vec{x}) }
\, \frac{\ud^3 \vec{p}}{2 |p_0(\vec{p})|} - 
\int \limits_{\mathbb{R}^3} \widetilde{\phi}(-\vec{p},-|p_0(\vec{p})|) \,  e^{i|p_0(\vec{p})|t -i\vec{p}\cdot \vec{x}}
\, \frac{\ud^3 \vec{p}}{2 |p_0(\vec{p})|}, \\
 p_0(\vec{p}) = \pm \sqrt{\vec{p} \cdot \vec{p} + m^2}.
\end{multline}
Here of course $p = (p_0(\vec{p}),\vec{p}) = (\sqrt{\vec{p} \cdot \vec{p} + m^2},\vec{p})$ on $\mathscr{O}_{m,0,0,0}$
and $p = (p_0(\vec{p}),\vec{p}) = (-\sqrt{\vec{p} \cdot \vec{p} + m^2},\vec{p})$ on $\mathscr{O}_{-m,0,0,0}$

In particular for the 
solution $\phi \in \mathcal{H}$ whose Fourier transform $\widetilde{\phi}$ is concentrated on 
the positive energy orbit $\mathscr{O}_{m,0,0,0}$
we have
\begin{multline*}
\phi(x)= \phi(\vec{x},t) = \int \limits_{\mathscr{O}_{m,0,0,0}} \widetilde{\phi}(p) \, 
e^{-ip\cdot x} \, \ud \mu_{{}_{m,0}}(p) \\
= \int \limits_{\mathbb{R}^3} \widetilde{\phi}(\vec{p},p_0(\vec{p})) \, e^{-(ip_0(\vec{p})t -i\vec{p}\cdot \vec{x}) }
\, \frac{\ud^3 \vec{p}}{2 p_0(\vec{p})}, \,\,\,\,\, p_0(\vec{p}) = \sqrt{\vec{p} \cdot \vec{p} + m^2}.
\end{multline*}
Similarly we have for the solution $\phi \in \mathcal{H}$ whose Fourier transform is concentrated on the negative energy orbit $\mathscr{O}_{-m,0,0,0}$:
\begin{multline*}
\phi(x)= \phi(\vec{x},t) = \int \limits_{\mathscr{O}_{-m,0,0,0}} \widetilde{\phi}(p) \, 
e^{-ip\cdot x} \, \ud \mu_{{}_{-m,0}}(p) \\
= \int \limits_{\mathbb{R}^3} \widetilde{\phi}(-\vec{p},-|p_0(\vec{p})|) \,  e^{i|p_0(\vec{p})|t -i\vec{p}\cdot \vec{x}}
\, \frac{\ud^3 \vec{p}}{2 p_0(\vec{p})}, \,\,\,\,\, p_0(\vec{p}) = -\sqrt{\vec{p} \cdot \vec{p} + m^2}.
\end{multline*}

We have the following equality for the solutions $\phi,\phi' \in \mathcal{H}$ whose Fourier transforms 
$\widetilde{\phi}, \widetilde{\phi}'$ are 
concentrated on the positive energy orbit $\mathscr{O}_{m,0,0,0}$:
\begin{multline*}
\int \limits_{x^0 =  t = const.} \Big(\phi(\vec{x}, t), \phi'(\vec{x}, t) \Big)_{{}_{\mathbb{C}^4}}
\, \ud^3 x =  \int \limits_{\mathscr{O}_{m,0,0,0}} \Big(\widetilde{\phi}(p), \phi'(p) \Big)_{{}_{\mathbb{C}^4}}
\, \frac{\ud \mu_{{}_{m,0}}(p)}{2 p_0} = \\
\int \limits_{\mathbb{R}^3} \Big(\widetilde{\phi}(\vec{p}, p_0(\vec{p})), 
\phi'(\vec{p}, p_0(\vec{p})) \Big)_{{}_{\mathbb{C}^4}}
\, \frac{\ud^3}{2 p_0(\vec{p})}, \,\,\,\, p_0(\vec{p}) = \sqrt{\vec{p} \cdot \vec{p} + m^2}.
\end{multline*}
Similarly we have for the solutions $\phi, \phi' \in \mathcal{H}$ whose Fourier transforms
$\widetilde{\phi}, \widetilde{\phi}'$ are concetrated on the negative energy 
orbit $\mathscr{O}_{-m,0,0,0}$: 
\begin{multline*}
\int \limits_{x^0 =  t = const.} \Big(\phi(\vec{x}, t), \phi'(\vec{x}, t) \Big)_{{}_{\mathbb{C}^4}}
\, \ud^3 x \\
= \int \limits_{\mathbb{R}^3} \Big(\widetilde{\phi}(-\vec{p}, -|p_0(\vec{p})|), 
\widetilde{\phi}'(-\vec{p}, -|p_0(\vec{p})|) \Big)_{{}_{\mathbb{C}^4}}
\, \frac{\ud^3 \vec{p}}{(2 p_0)^2} \\=
 -\int \limits_{\mathbb{R}^3} \Big(\widetilde{\phi}(\vec{p}, p_0(\vec{p})), 
\widetilde{\phi}'(\vec{p}, p_0(\vec{p})) \Big)_{{}_{\mathbb{C}^4}}
\, \frac{\ud^3 \vec{p}}{(2 p_0)^2} \\ =
-\int \limits_{\mathscr{O}_{-m,0,0,0}} \Big(\widetilde{\phi}(p), 
\phi'(p) \Big)_{{}_{\mathbb{C}^4}}
\, \frac{\ud \mu_{{}_{m,0}}(p)}{2 |p_0|},
 \,\,\,\, p_0(\vec{p}) = -\sqrt{\vec{p} \cdot \vec{p} + m^2}.
\end{multline*}
Note that the last expression is equal to \emph{minus} the inner product (33)
of Subsection 2.1 of \cite{wawrzycki2018}
of the (Fourier transforms of) bispinors $\phi, \phi'$ 
on the Hilbert space of Fourier transforms of bispinors, concentrated on $\mathscr{O}_{-m,0,0,0}$ (up to the irrelevant
constant factor $m>0$), introduced in Subsection 2.1 of \cite{wawrzycki2018}.

Consider now the induced representation
\begin{equation}\label{m}
U^{{}_{(m,0,0,0)} L^{{}^{1/2}}}
\end{equation}
of $T_4 \circledS SL(2, \mathbb{C})$, concentrated on the orbit $\mathscr{O}_{(m,0,0,0)}$.
Now we apply the isometric map $V^\oplus$ to the space of this representation followed by the
Fourier transform (20) (of Introduction to Sect. 2 of \cite{wawrzycki2018} 
with the orbit $\mathscr{O}_{\bar{p}}= \mathscr{O}_{(m,0,0,0)}$),
where $V^\oplus$ is the map defined in Example 1 (Subsection 2.1 of \cite{wawrzycki2018}). Let us denote the composed map
just by $\widetilde{V^\oplus}$. The image of $\widetilde{V^\oplus}$ lies in $\mathcal{H}$. 
Indeed because of eq. (28) of Subsection 2.1 of \cite{wawrzycki2018}
it is even isometric.

Similarly consider the representation  
\begin{equation}\label{-m}
U^{{}_{(-m,0,0,0)} L^{{}^{1/2}}}
\end{equation}
of $T_4 \circledS SL(2, \mathbb{C})$, concentrated on the orbit $\mathscr{O}_{(-m,0,0,0)}$.
To the space of this representation we apply the map $\widetilde{V^\ominus}$ equal to $V^\ominus$ followed by the
Fourier transform (20) (Introduction to Section 2 of \cite{wawrzycki2018} with the orbit 
$\mathscr{O}_{\bar{p}}= \mathscr{O}_{(-m,0,0,0)}$), where
$V^\ominus$ is the map defined in Example 1,
Subsection 2.1 of \cite{wawrzycki2018}. Its image likewise lies in 
$\mathcal{H}$ and by the same (28) of Subsection 2.1 of \cite{wawrzycki2018} -- 
which is also valid for $\widetilde{V^\ominus}$ --
it is isometric too. Now the image $\mathcal{H}_{m,0}^{\oplus}$ of the representation 
space of the representation  (\ref{m}) under
the map $\widetilde{V^\oplus}$ lies in the positve eigenspace subspace 
$E_+ \mathcal{H}$ of the essentially self adjoint Dirac hamiltonian operator 
$H = -i \gamma^0\gamma^k \partial_k + m \gamma^0 = -i \alpha^k \partial_k + m \gamma^0$ acting on 
$\mathcal{H}$, where $E_+$ is the spectral projection corresponding to all positive spectral values
of $H$. Similarly the image $\mathcal{H}_{-m,0}^{\ominus}$ of the space of the representation (\ref{-m}) 
under the map $V^\ominus$ lies in the negative eigenspace subspace $E_-\mathcal{H}$ of the
operator $H$. We have $E_+ + E_- = \bold{1}_\mathcal{H}$ and $E_+ E_- = 0$, i. e.
$E_+ \mathcal{H}$ and $E_- \mathcal{H}$ are orthogonal. Therefore the operator
$\widetilde{V^\oplus} \oplus \widetilde{V^\ominus}$ maps the representation space of the representation 
\begin{equation}\label{m+-m}
U^{{}_{(m,0,0,0)} L^{{}^{1/2}}} \oplus U^{{}_{(-m,0,0,0)} L^{{}^{1/2}}},
\end{equation}
concentrated on the sum theoretic set $\mathscr{O}_{(m,0,0,0)} \cup \mathscr{O}_{(-m,0,0,0)}$
of the orbits $\mathscr{O}_{(m,0,0,0)}$ and $\mathscr{O}_{(-m,0,0,0)}$, isometrically
into $\mathcal{H}$. 

On the the other hand the only eigenvalues of the matrix $\gamma^0$ are 1 and -1, so it follows from the 
theorem of Section 10.1, Part II, 
Chapter II of \cite{Geland-Minlos-Shapiro} (compare also \cite{GelfandYaglom1}-\cite{GelfandYaglom3}), 
that the ordinary Fourier transform 
\[
\widetilde{\phi}(p) = \int \phi(x) \, e^{ip\cdot x} \, \ud^4x
\]
of any element of $\mathcal{H}$ is
concentrated on the set theoretical sum $\mathscr{O}_{(m,0,0,0)} \cup \mathscr{O}_{(-m,0,0,0)}$
of the orbits $\mathscr{O}_{(m,0,0,0)}$ and $\mathscr{O}_{(-m,0,0,0)}$. Thus the operator
$\widetilde{V^\oplus} \oplus \widetilde{V^\ominus}$ regarded as operator on the space of 
the representation (\ref{m+-m})
is onto $\mathcal{H}$, and therefore it is unitary, so that 
\[
E_+ \mathcal{H} = \mathcal{H}_{m,0}^{\oplus} \,\,\, \textrm{and} \,\,\,
E_- \mathcal{H} = \mathcal{H}_{-m,0}^{\ominus}. 
\] 
Therefore in the Hilbert space $\mathcal{H} = \mathcal{H}_{m,0}^{\oplus} \oplus \mathcal{H}_{-m,0}^{\ominus}$
there acts the unitary\footnote{Please, note also that the representation 
\[
V^\oplus \, U^{{}_{(m,0,0,0)} L^{{}^{1/2}}} \, (V^\oplus )^{-1} \, \oplus \,
V^\ominus U^{{}_{(-m,0,0,0)} L^{{}^{1/2}}} \, (V^\ominus)^{-1},
\]
concentrated on $\mathscr{O}_{m,0,0,0} \sqcup \mathscr{O}_{-m,0,0,0}$ is unitary, similarly as the representation
\[
V^{\oplus \ominus} \, \big( U^{{}_{(m,0,0,0)} L^{{}^{1/2}}} \oplus 
 U^{{}_{(m,0,0,0)} L^{{}^{1/2}}} \big) \, (V^{\oplus \ominus})^{-1}
\]
(compare Example 1, Subsection 2.1 of \cite{wawrzycki2018}) concentrated on $\mathscr{O}_{(m,0,0,0)}$. }
representation
\begin{equation}\label{rep-on-H}
\widetilde{V^\oplus} \, U^{{}_{(m,0,0,0)} L^{{}^{1/2}}} \, (\widetilde{V^\oplus} )^{-1} \, \oplus \,
\widetilde{V^\ominus} U^{{}_{(-m,0,0,0)} L^{{}^{1/2}}} \, (\widetilde{V^\ominus})^{-1}
\end{equation}
concentrated on $\mathscr{O}_{(m,0,0,0)} \cup \mathscr{O}_{(-m,0,0,0)}$, with 
\begin{equation}\label{rep-on-H+}
\widetilde{V^\oplus} \, U^{{}_{(m,0,0,0)} L^{{}^{1/2}}} \, (\widetilde{V^\oplus} )^{-1} 
\end{equation}
acting on $\mathcal{H}_{m,0}^{\oplus}$ and with
\begin{equation}\label{rep-on-H-}
\widetilde{V^\ominus} U^{{}_{(-m,0,0,0)} L^{{}^{1/2}}} \, (\widetilde{V^\ominus})^{-1}
\end{equation}
acting on $\mathcal{H}_{-m,0}^{\ominus}$.

To the Hilbert space $\mathcal{H}$ treated as if it was the single particle space we apply the fermionic 
functor of second quantization
$\Gamma$, and obtain the standard absorption and emission operators. Next we split them (i. e. we consider 
their restrictions resp. to  $\mathcal{H}_{m,0}^{\oplus}$ or $\mathcal{H}_{-m,0}^{\ominus}$)
according to the splitting $\mathcal{H} = \mathcal{H}_{m,0}^{\oplus} \oplus \mathcal{H}_{-m,0}^{\ominus}
= E_+ \mathcal{H} \oplus E_- \mathcal{H}$ of the space $\mathcal{H}$,
compare e.g. \cite{Scharf}. We observe then that the absorption and emission operators restricted to 
 $\mathcal{H}_{m,0}^{\oplus}$ compose a fermionic free field and similarly the restrictions
of the absorption and emission operators restricted to $\mathcal{H}_{-m,0}^{\ominus}$ and that the the two sets 
of operators commute and are independent in consequence of the orthogonality of the subspaces
$\mathcal{H}_{m,0}^{\oplus}$ and $\mathcal{H}_{-m,0}^{\ominus}$ (e. g. \cite{Scharf}). 
That is we have two independent fermionic quantizations: the functor $\Gamma$ applied to
$\mathcal{H}_{m,0}^{\oplus}$ and the functor $\Gamma$ applied to $\mathcal{H}_{-m,0}^{\ominus}$
with the tensor product of the two independent sets of annihilation and creation operators acting in 
the tensor product of fermionic Fock spaces
$\Gamma\big(\mathcal{H}_{m,0}^{\oplus} \big) \otimes \Gamma\big(\mathcal{H}_{-m,0}^{\ominus} \big)
= \Gamma(\big( \mathcal{H}_{m,0}^{\oplus} \oplus \mathcal{H}_{-m,0}^{\ominus} \big)$.
In order to repair the energy sign
of the free Dirac field on $\Gamma\big(\mathcal{H}_{m,0}^{\oplus} \big) \otimes 
\Gamma\big(\mathcal{H}_{-m,0}^{\ominus} \big)$ we interchange the absorption 
and emission operators in $\Gamma\big(\mathcal{H}_{-m,0}^{\ominus} \big)$. In this manner we obtain
the following construction which may be described in the following four steps.

\subsection{Application of the Segal second quantization 
functor to the subspace $\mathcal{H}_{m,0}^{\oplus}$}\label{electron}

To the subspace  $\mathcal{H}_{m,0}^{\oplus}$ we apply the Segal's functor $\Gamma$ of fermionic
quantization and obtain the fermionic Fock space
\[ 
\mathcal{H}^{\oplus}_{F} = \Gamma(\mathcal{H}_{m,0}^{\oplus}) = \mathbb{C} \oplus \mathcal{H}_{m,0}^{\oplus} \oplus
\big( \mathcal{H}_{m,0}^{\oplus} \big)^{\widehat{\otimes} 2} \oplus 
\big( \mathcal{H}_{m,0}^{\oplus} \big)^{\widehat{\otimes} 3}
\oplus \ldots;
\] 
with the unitary representation
\begin{multline*} 
\Gamma \Big(\widetilde{V^\oplus} \, U^{{}_{(m,0,0,0)} L^{{}^{1/2}}} \, (\widetilde{V^\oplus} )^{-1} \Big)
= \bigoplus \limits_{n = 0,1,2 \ldots}
\Big(\widetilde{V^\oplus} \, U^{{}_{(m,0,0,0)} L^{{}^{1/2}}} \, (\widetilde{V^\oplus})^{-1} \Big)^{\widehat{\otimes} n},
\end{multline*}
where in the formulas $(\cdot)^{\widehat{\otimes} n}$ stands for $n$-fold anti-symmetrized tensor product,
and $(\cdot )^{\widehat{\otimes} n}$ with $n = 0$ applied to the representation gives the trivial representation on 
$\mathbb{C}$ with each representor acting on 
$\mathbb{C}$ as multiplication by 1.

In this and in the following Sections, we will encounter essentially two types of topological vector spaces
and operators acting upon them: 1) \emph{Hilbert spaces} and 2) \emph{nuclear spaces} 
(the Schwartz $\mathcal{S}(\mathbb{R}^n)$ space of test functions on 
$\mathbb{R}^n$ is an example of a nuclear space). Correspondingly we will
use respectively 1) the \emph{Hilbert space tensor product} $\otimes$ 
(if applied to Hilbert spaces, elements of Hilbert spaces 
and operators upon them) and respectively \emph{projective tensor product} $\otimes$ 
(if applied to nuclear spaces, their elements
and operators acting upon them); for definition, and properties of these standard constructions we refer e.g.
to \cite{Murray_von_Neumann}, \cite{treves}, \cite{Schaefer}.

The linear spaces we encounter (Hilbert spaces and nuclear spaces) will be always over $\mathbb{R}$ or over $\mathbb{C}$,
but whenever they are over $\mathbb{C}$ they will be equal to complexifications of real (Hilbert or nuclear)
spaces with naturally defined complex conjugation $\overline{(\cdot)}$ in them.  

Note that by Riesz representation theorem for such Hilbert spaces $\mathcal{H}'$ we have natural identification 
of linear continuous functionals on $\mathcal{H}'$ with the elements of the adjoint Hilbert space
$\overline{\mathcal{H}'}$, which in fact becomes an isomorphism of Hilbert spaces if we appropriately introduce the multiplication by a number and the inner product into the space of linear functionals
on $\mathcal{H}'$. Recall that the adjoint space
$\overline{\mathcal{H}'}$ have the same set of elements as $\mathcal{H}'$, but with scalar multiplication
by a number $\alpha \in \mathbb{C}$ and inner
product defined by 
\[
\begin{split}
\alpha u \,\,\, \textrm{in}  \,\,\,\overline{\mathcal{H}'} \,\,\, = \,\,\,\overline{\alpha}u \,\,\, \textrm{in} \,\,\,
\mathcal{H}', \\
(u,v) \,\,\, \textrm{in} \,\,\,\overline{\mathcal{H}'} \,\,\, = \,\,\, (v,u) \,\,\, \textrm{in} \,\,\,
\mathcal{H}'.
\end{split}
\]
With such a Hilbert space structure on $\overline{\mathcal{H}'}$ the map 
$\mathcal{H}' \ni u \mapsto \overline{u} \in \overline{\mathcal{H}'}$
defines a canonical \emph{linear} isomorphism.
In the sequel we will regard the dual space $\mathcal{H}'^*$ as the adjoint space  $\overline{\mathcal{H}'}$
with elements the same as elements of $\mathcal{H}'$ (Riesz isomorphism).

For operators on Hilbert spaces we are using the standard notation for 
the ordinary adjoint operation with the superscript $*$,
with the exception of the annihilation operators, denoting the operators which are adjoint to 
them with the superscript $+$ instead $*$ (which is customary in physical literature). 
If working with operators $A$ transforming (continuously) one nuclear space into another $E_1 \rightarrow E_2$, we use the superscript $*$ to denote the linear dual (transposed) operator $A^*$: 
$E_{2}^* \rightarrow E_{1}^*$, transforming continuously  the strong dual space 
$E_{2}^*$ into the strong dual space $E_{1}^*$, for definition and general properties 
of transposition we again refer to \cite{treves}. For operator $A$ transforming
(continuously) nuclear space into nuclear space we denote by 
$A^+$ the operator $\overline{(\cdot)} \circ A^* \circ \overline{(\cdot)}$, i.e. the linear dual
of $A$ composed with complex conjugation (say Hermitean adjoint $=$ linear transposition $+$ complex conjugation).

In the standard way we obtain the map from 
$\mathcal{H}_{m,0}^{\oplus} \ni \widetilde{\phi}$ to the families
$a_{\oplus}(\widetilde{\phi}), 
a_{\oplus}^+(\widetilde{\phi}) = {a_{\oplus}(\widetilde{\phi})}^+$ of ordinary annihilation and creation operators in the fermionic Fock space  
$\Gamma\big(\mathcal{H}_{m,0}^{\oplus}\big)$ fulfilling the canonical anticommutation relations:
\begin{multline*}
\Big\{a_{\oplus}(\widetilde{\phi}), \,\,\,
{a_{\oplus}(\widetilde{\phi}')}^+  \Big\}
= \Big(\widetilde{\phi}, \widetilde{\phi}'\Big)_{{}_{\mathcal{H}_{m,0}^{\oplus}}} \\
= \Big(\widetilde{\phi}, \widetilde{\phi}'\Big) \\
= \int \limits_{x^0 =  t = const.} \Big(\phi(\vec{x}, t), \phi'(\vec{x},t) \Big)_{{}_{\mathbb{C}^4}}
\, \ud^3 x  \\ 
=  \int \limits_{\mathscr{O}_{m,0,0,0}} 
\Big(\widetilde{\phi}(p), 
\widetilde{\phi}'(p) \Big)_{{}_{\mathbb{C}^4}}
\, \frac{\ud \mu_{{}_{m,0}}(p)}{2 p_0} \\
= \int \limits_{\mathbb{R}^3} \Big(\widetilde{\phi}(\vec{p}, p_0(\vec{p})), \,\,
\phi'(\vec{p}, p_0(\vec{p})) \Big)_{{}_{\mathbb{C}^4}}
\, \frac{\ud^3 \vec{p}}{(2 p_0(\vec{p}))^2}, \\
 p_0(\vec{p}) = \sqrt{\vec{p} \cdot \vec{p} + m^2}.
\end{multline*}
Here and in the rest part of this Section we identify the Hilbert space $\mathcal{H}_{m,0}^{\oplus}
= E_+\mathcal{H}$
of positive energy distributional solutions $\phi$ of the Dirac equation with the ordinary functions 
$\widetilde{\phi}$ on the orbit $\mathscr{O}_{m,0,0,0}$
which they induce on the orbit in the manner described above. 
Correspondingly we identify the Hilbert space $\mathcal{H}$ of distributional solutions 
$\phi$ of Dirac equation with the ordinary functions $\widetilde{\phi}$
on the disjoint sum of orbits $\mathscr{O}_{m,0,0,0} \sqcup \mathscr{O}_{-m,0,0,0}$ ($= \textrm{supp} \, \widetilde{\phi}$
of $\widetilde{\phi}$ regarded as distribution). Similarly we identify
the  Hilbert space $\mathcal{H}_{-m,0}^{\ominus}
= E_-\mathcal{H}$ of negative energy distributional solutions $\phi$ of Dirac equation
with the corresponding ordinary functions on $\mathscr{O}_{m,0,0,0} \sqcup \mathscr{O}_{-m,0,0,0}$
having the support on $\mathscr{O}_{-m,0,0,0}$.

In the later stage of the construction of the free Dirac field we will need a unitary
involutive (and thus self-adjoint) operator
$\In$, which we call \emph{parity number operator}, canonically related to the Fock space construction. 
In order to indicate the relation of the parity number operator $\In$ to the corresponding Fock space
$\Gamma\big(\mathcal{H}_{m,0}^{\oplus}\big)$, we use the subscript $\oplus$: ${\In}_\oplus$.

In order to define ${\In}_\oplus$ recall that 
every element $\Phi \in \Gamma\big(\mathcal{H}_{m,0}^{\oplus}\big)$
may be uniquely represented as the sum 
\begin{equation}\label{GeneralPsiInGamma(H)}
\Phi = \sum \limits_{n \geq 0} \Phi_n  
\end{equation}
over all $n= 0, 1, 2, \ldots $ of the orthogonal components 
$\Phi_n \in \big(\mathcal{H'}\big)^{\widehat{\otimes} n}$
-- the so called $n$-particle states, with 
\begin{equation}\label{NormGeneralPsiInGamma(H)}
\|\Phi\|^2 = \sum \limits_{n \geq 0} \| \Phi_n \|^2 < +\infty. 
\end{equation}
We define on the Fock space a bounded self-adjoint operator ${\In}_\oplus$ -- parity number operator --
which maps a general state $\Phi \in \Gamma\big(\mathcal{H}_{m,0}^{\oplus}\big)$ 
defined by (\ref{GeneralPsiInGamma(H)}) into the following state
\[
{\In}_\oplus \Phi = \sum \limits_{n \geq 0} \, (-1)^n \, \Phi_n. 
\]
It is evident that ${\In}_\oplus$ is unitary and involutive  (thus self-adjoint) 
\[
{\In}_\oplus^2 = \boldsymbol{1}, \,\,\, {\In}_\oplus^* = {\In}_\oplus
\] 
and that ${\In}_\oplus$ anti-commutes with the annihilation (and creation) operators:
\[
a_{\oplus}(\widetilde{\phi}) \, {\In}_\oplus =
- {\In}_\oplus \,  a_{\oplus}(\widetilde{\phi}).
\]

Note that the unitary involution $\In$ on general Fock space, and in particular ${\In}_\oplus$,
commutes with any (bounded or even unbounded) operator $B$ which transforms the closed subspaces of fixed 
particle number into themselves
(in case $B$ is unbounded we assume $\Dom \, B$ to be a linear subspace
or still more generally with $\Dom \, B$ to be closed under operation 
of multiplication by $-1$). In particular $\In$ (or ${\In}_\oplus$) commutes
with any operator of the form 
\[
B = \Gamma(A) = \sum \limits_{n=0}^{\infty} A^{\otimes n},
\]
namely:
\[
\big[\Gamma(A), {\In}_\oplus \big] = 0 \,\,\, \textrm{on} \,\,\,
\Dom \Gamma(A),
\]
irrespectively if $A$ is bounded or not, but with linear $\Dom A$
and $\Dom \Gamma(A)$. This in particular means that the operator 
${\In}_\oplus$ commutes:
\[
\Bigg[\,\,\,
\Gamma \Big(\widetilde{V^\oplus} \, U^{{}_{(m,0,0,0)} L^{{}^{1/2}}} \, (\widetilde{V^\oplus} )^{-1} \Big) \,\,\,,
\,\,\,
{\In}_\oplus \,\,\, \Bigg] = 0
\]
with the representation of $T_4 \circledS SL(2, \mathbb{C})$ acting in the Fock space
$\Gamma\big(\mathcal{H}_{m,0}^{\oplus}\big)$.

\vspace*{1cm}

\begin{rem}\label{TwoRepOfaa^+InFermiFock}
Note that in literature, e.g. \cite{Bratteli-Robinson},  there is frequently used the following
construction of annihilation and creation operators, in a general Fock space (here we concentrate on
the fermionic Fock space) $\Gamma(\mathcal{H}')$. For each $u \in \mathcal{H}'$ of the single particle space
$\mathcal{H}'$  we define the operators $a(u), a^+(u)=a(u)^+$
which by definition act on general element 
\begin{equation}\label{GeneralPsiInGamma(H')}
\Phi = \sum \limits_{n \geq 0} \Phi_n, \,\,\,
\Phi_n \in \mathcal{H}'^{\widehat{\otimes} \, n}
\end{equation}
with
\begin{equation}\label{NormGeneralPsiInGamma(H')}
\|\Phi\|^2 = \sum \limits_{n \geq 0} \| \Phi_n \|^2 < +\infty, 
\end{equation}
of the Fock space
$\Gamma(\mathcal{H}')$, in the following manner
\[
\begin{split}
1) \,\,\,\,\, a(u) \big(\Phi = \Phi_0 \big) = 0, \\
2) \,\,\,\,\, a(u) \Phi =  \sum \limits_{n \geq 0} \, n^{1/2} \, \overline{u} \, \widehat{\otimes}_1 \, \Phi_n, \\
3) \,\,\,\,\, a(u)^+ \Phi =  \sum \limits_{n \geq 0} \,(n+1)^{1/2} \, u \, \widehat{\otimes} \, \Phi_n.  
\end{split}
\]
Here $\widehat{\otimes}$ and
$\widehat{\otimes}_1$ denote respectively the anti-symmetrized $n$-fold tensor product and the 
anti-symmetrized $1$-contraction, uniquely determined by the formulae
\[
v_{{}_{1}} \, \widehat{\otimes} \, \cdots \widehat{\otimes} \, v_{{}_{n}} =
(n!)^{-1} \sum \limits_{\pi} \textrm{\emph{sign}} \, (\pi) \, v_{{}_{\pi(1)}} \otimes
\cdots \otimes v_{{}_{\pi(n)}} \,\,\,\,
v_{{}_{i}} \in  \in \mathcal{H}',
\]
\[
u \, \widehat{\otimes}_1 v_{{}_{1}} \, \widehat{\otimes} \, \cdots \widehat{\otimes} \, v_{{}_{n}}
= (n!)^{-1} \sum \limits_{\pi} \textrm{\emph{sign}} \, (\pi) \, \langle u,v_{{}_{\pi(1)}} \rangle \, 
 v_{{}_{\pi(2)}} \otimes
\cdots \otimes v_{{}_{\pi(n)}},
 \,\,\,
u \in \mathcal{H}'^{*}, v_{{}_{i}} \in \mathcal{H}',
\]
with the sums ranging over all permutations $\pi$ of the natural numbers $1, \ldots, n$, and with
the evaluation $\langle u,v_{{}_{\pi(1)}} \rangle$ of $u$, understood as a linear functional $\mathcal{H}'^{*}$,
on $v_{{}_{\pi(1)}} \in \mathcal{H}'$ equal
\[
\langle u,v_{{}_{\pi(1)}} \rangle = (\overline{u},v_{{}_{\pi(n)}})
\]
to the inner product of the elements $\overline{u}, v_{{}_{\pi(n)}} \in \mathcal{H}'$. 
Note that in all the relevant physical situations the single particle Hilbert spaces and the corresponding Fock spaces have natural real structure and are equal to complexifications of real Hilbert spaces with naturally defined
complex conjugations $\overline{(\cdot)}$ in them. Recall also that the map 
$\mathcal{H}' \ni u \mapsto \overline{u}$ defines a linear isomorphism of the Hilbert space   
$\mathcal{H}'$ into the adjoint Hilbert space $\overline{\mathcal{H}'}$, which in turn can be identified with the 
Hilbert space of linear functionals on $\mathcal{H}'$, by the Riesz representation theorem.

However we will interchangeably be using another, unitarily equivalent, realization of the annihilation 
and creation operators in the Fock space, which is more frequently used by mathematicians
(and fits well with that used e.g. in \cite{hida}, \cite{obataJFA}, \cite{obata-book}, \cite{HKPS}
for bosons, when adopting their results to the fermion case), 
because we will refer to the works \cite{hida}, \cite{obataJFA}, \cite{obata-book},
in the following part of our work. Let us call it the 
\emph{modified realization of annihilation-creation operators}
in the Fock space. This realization used by mathematicians is more natural 
for the interpretation of the creation and annihilation operators as derivations
(or graded derivations in case of fermi Fock space) on a nuclear (skew-commutative, or say Grassmann, 
in case of fermi Fock space) algebra of 
Hida test functions on an (infinite-dimensional) strong dual space to a nuclear space.  

In order to define it we first slightly modify the norm (\ref{NormGeneralPsiInGamma(H')})  of a 
general element  (\ref{GeneralPsiInGamma(H')}) and put for its square instead
\[
\|\Phi\|_{0}^2 = \sum \limits_{n \geq 0} \, n! \, \| \Phi_n \|^2.
\] 
Then we define the annihilation and creation operators through their action on general such elemet
$\Phi$ given by the following formulae 
\[
\begin{split}
1) \,\,\,\,\, a(u) \big(\Phi = \Phi_0 \big) = 0, \\
2) \,\,\,\,\, a(u) \Phi =  \sum \limits_{n \geq 0} \, n \, \overline{u} \, \widehat{\otimes}_1 \, \Phi_n, \\
3) \,\,\,\,\, a(u)^+ \Phi =  \sum \limits_{n \geq 0} \, u \, \widehat{\otimes} \, \Phi_n.  
\end{split}
\]

The unitary operator:
\[
U \Big( \sum \limits_{n \geq 0} \Phi_n \Big) = 
\sum \limits_{n \geq 0} (n!)^{-1/2} \, \Phi_n, \,\,\,
U^{-1} \Big( \sum \limits_{n \geq 0} \Phi_n \Big) = 
\sum \limits_{n \geq 0} (n!)^{1/2} \, \Phi_n, 
\]
with the convention that $0!=1$,
gives the unitary equivalence between the two realizations of the annihilation and creation operators
in the Fock spaces, as well as of the representations of $T_4 \circledS SL(2, \mathbb{C})$
in the corresponding Fock spaces. 
\end{rem}
\vspace*{1cm}

\subsection{Application of the Segal second quantization functor to the space $\mathcal{H}_{-m,0}^{\ominus c}$ of spinors conjugated to the spinors of the subspace $\mathcal{H}_{-m,0}^{\ominus}$}\label{positron}

In the next step we apply the functor $\Gamma$ of fermionic second quantization to the subspace
$\mathcal{H}_{-m,0}^{\ominus}$ and obtain the fermionic Fock space
\[
\Gamma(\mathcal{H}_{-m,0}^{\ominus}) = \mathbb{C} \oplus \mathcal{H}_{-m,0}^{\ominus} \oplus
\big( \mathcal{H}_{-m,0}^{\ominus} \big)^{\widehat{\otimes} 2} \oplus 
\big( \mathcal{H}_{-m,0}^{\ominus} \big)^{\widehat{\otimes} 3} \oplus \ldots;
\]
but the above mentioned interchange of the emission and absorption operators in 
$\Gamma\big(\mathcal{H}_{-m,0}^{\ominus} \big)$ results in 
replacing the single particle Hilbert space $\mathcal{H}_{-m,0}^{\ominus} = E_-\mathcal{H}$ 
with a conjugated one 
$\mathcal{H}_{-m,0}^{\ominus c}$ and in replacing of the representation (\ref{rep-on-H-}) acting in $\mathcal{H}_{-m,0}^{\ominus}$ with another conjugated representation acting in the Hilbert space $\mathcal{H}_{-m,0}^{\ominus c}$. 

This procedure is the well known basis for the solution of the ``negative energy states problem'' in relativistic
quantum field theory, therefore we only sketch briefly the general lines, presenting only the final results
in case of the free quantum Dirac field respecting the Dirac equation. 
Namely the solution is based on the observation that the negative energy solutions
lying in $\mathcal{H}_{-m,0}^{\ominus} = E_-\mathcal{H}$ (classicaly the negative energy solutions of the 
equation which is to be fulfiled by the quantized field, here of the Dirac equation $D \phi = m\phi$ 
(30) of Subsection 2.1 of \cite{wawrzycki2018}), should not be interpreted
as negative energy solutions of the original equation (here Dirac equation), but rather as a
kind of conugation of \emph{positive} energy solutions of a conjugation of the original (here Dirac) equation, 
with the conjugation depending on the actual kind of field. In particular for the scalar (complex) field fulfilling  the Klein-Gordon equation the conjugation coincides with the ordinary complex conjugation (but only accidentally).

For (free) Dirac field respecting  Dirac equation the conjugation is slightly more 
complicated and the conjugated equation
does not coincide with the original Dirac equation. In the more general higher spin local fields
the conjugation is similar as for the Dirac equation, and is easy to guess  with its general definition beig naturaly determined by the general construction of the single particle Hilbert space of the field (with local transformation law). 

Namely in general case of globally hyperbolic space-time
and a free field, say $\phi$, on it we can extract the essential points of the construction of the free field
on the flat Minkowski manifold, although the particular computations would be much less easy to handle.
In any case the space-time manifold with its globally hyperbolic causal structure (given by a Lorenzian metric)
is crucial, together with the type of field $\phi$ with its local transformation rule
fixing the associated type of bundle with $\phi$ ranging over its sections, and respecting a hyperbolic differential equation $D\phi = m\phi$. 
If a preferable and natural assumptions of analytic type are put
on the pseudo-riemannian space-time manifold (compare e.g. \cite{Stro}, \cite{HBaum}) then the 
Lorezian metric induces a Krein structure in the space of sections $\phi$ (compare
the formulas (37), (38) of Subsect. 2.3 of \cite{wawrzycki2018} in the special case of 
flat Minkowski space-time and the Dirac bispinors $\phi$ on it with the transformation law (39) of 
Subsect. 2.1 of \cite{wawrzycki2018}).
We expect the corresponding differental operator $D$ to be not merely Krein-self-adjoint, but moreover that it
allows a Krein-orthogonal spectral decomposition similar to that obtained in 
Subsect. 2.3 of \cite{wawrzycki2018} for the ordinary Dirac operator $D$ (in particular it is of spectral-type). This assumption is nontrivial, as in the Krein space
Krein-self-adjoint operator in general does not allow any spectral decomposition of the type obtained
in Subsect. 2.3 of \cite{wawrzycki2018} for $D$ (compare e.g. the classic Dunford-Schwartz analysis of the type of generalized spectral decompositions
of non-normal operators). In particular the method of extension of the construction of a free field on 
more general space-times proposed here have a rather restricted domain of validity, and is confined to situations with rather very special kind of corresponding hyperbolic differential operators $D$
allowing ``regular'' Krein-orthogonal spectral decompositions.  
Of course in general the spectral Krein-orthogonl decomposition of $D$ may contain a discrete component, 
or even consist of purely discrete part, depending on topology of the space-time manifold.

Next we consider the generalized eigenspace, which we agreed to denote by $\mathcal{H}$, of the Krein-self-adjoint operator $D$, corresponding to the eigenvalue $m$, and which consists of all
distributional solutions $\phi$ of the equation $D\phi = m\phi$. The closed subspaces of generalized eigenspaces corresponding to the generalized eigenvalues of $D$ inherit nondegenerate Krein-space structure
from the initial Krein space of sections $\phi$ in which $D$ acts. The restriction of the Krein-self-adjoint operator $D$
to this subspace $\mathcal{H}$ is not only Krein-self-adjoint but likewise self-adjoint with respect to the inherited  Krein space and Hilbert space structures on $\mathcal{H}$, with well defined direct sum stucure $\mathcal{H}=
E_+\mathcal{H} \oplus E_\mathcal{H}$ with closed subspaces $E_{\pm}\mathcal{H}$ which are orhogonal 
and with nondegenerate Krein space structure. Moreover the operator $D$ is of spectral-type and admits generalized spectral decomposition in the sense of Neumark-Lanze, explicitly computed in Subsections 2.1-2.3
of \cite{wawrzycki2018},
with each generalized eigenspace which inherits nondegenrate Hilbert space and Krein space structure.
This is far not the case for general Krein-selfadjoint operator, compare \cite{Bog}. 
In particular the space $\mathcal{H}$ of generalized 
eigenvectors of $D$ corresponding to the generalized real eigenvalue $m>0$ (say mass) inherits 
nondegenrate and natural Krein space structure, in particular Hilbert space structure. 
We expect that the space-time manifold, especially its causal structure, allows to pick up
the natural discrete operation of time-orientation-reversing in terms of an involutive unitary operator 
(say the $\textrm{sign} \, (H)$ = $H|H|^{-1}$ of the Hamiltonian 
operator $H$ in $\mathcal{H}$) with the property that the change of time orientation transformation acts 
through  $\textrm{sign} \, (H)$ as an involutive unitary which exchanges 
positive energy subspace $E_+\mathcal{H}$ with the negative 
energy subspace $E_-\mathcal{H}$ of $\mathcal{H}$. In case of globally hyperbolic and highly 
symmetric spacetimes with time symmetry (e.g. Einstein Static Universe) this plan
is within our grasp. In particular the harmonic analysis of \cite{PaneitzSegalI}- \cite{PaneitzSegalIII} is sufficiently effective on the Einstein Universe
to allow e.g. construction of QED on it together with the proof of its convergence, compare \cite{SegalZhouQED}.
In general the conjugation corresponding to the division of ``positive'' and ``negative energy'' 
solution subspaces $E_+\mathcal{H}$ and $E_+\mathcal{H}$ of the space of distributional solutions
of $D\phi=m\phi$ is easy to guess and is strongly suggested by the geometric context. 
Construction of the involutive unitary which corresponds to the division into ``positive'' and ``negative
energy'' solution subspaces is more tricky when time symmetry is lacking at the space-time geometry level, and reflects the conformal (causal) structure of space-time
in the operator-spectral format. In fact construction of this division involves spectral decomposition of non-normal, Krein-self-adjoint operator $D$, and as we know there are no general therorems which would assure existence of such decompositions nor its sufficiently regular behaviour. This is the essential source of difficulty in achieving the honest 
division into ``positive'' and ``negative'' frequency modes. 
Once a generalized spectral Krein-orthogonal decomposition of $D$, similar to that presented in Subsections 
2.1-2.3 of \cite{wawrzycki2018} is successful, the involutive unitary and
the corresponding conjugation can be easily guessed. This is the case e.g. for the Einstein Universe, compare
\cite{PaneitzSegalI}- \cite{PaneitzSegalIII}. It can be achieved by explicit expansion 
of the general solution of the Dirac equation $D\phi = m\phi$ into ``Einstein spinor modes'' (as called by 
Segal and Zhou) and explicit division of the modes into positive and negative frequency parts. 
This is a good example to study the relationship of the conformal structure and the corresponding
involutive unitary operator. Still more interesting case we obtain for de Sitter spacetime
lacking time symmetry, but with the sufficiently reach harmonic analysis to study quantum fields on it.
At least one example (of scalar quantum field on the three dimensional de Sitter spacetime), which comes naturally, we will encounter when studying infrared fields in Section 7 of \cite{wawrzycki2018}. 
The generalized regular Krein-isometric decomposition of $D$ (with finite but arbitrary high dimenion 
of the fibre of the fibre bundle of sections of the corresponding Clifford module), providing the corresponding
Krein-orthogonal decomposition of the initial Krein space acted on by $D$, serves as the generalization
of the Fourier transform $V_\mathcal{F}$ of Subsections 2.1-2.8 of \cite{wawrzycki2018} in case 
of less symmetric globally hyperbolic spacetimes.
      
After this general remark concerning construction of free fields on more general space-time manifolds,
let us back to the construction of the free Dirac field on the flat Minkowski space-time, or more precisely,
to the conjugation, which accompany the division $\mathcal{H} = E_+\mathcal{H} \oplus E_+\mathcal{H}$
into positive and negative energy solutions of the ordinary Dirac equation $D\phi = m\phi$
constructed as above.

As remarked earlier, the negative energy solutions $\phi$ should be interpreted as conjugations of
positive energy solutions $\phi^c$ of the conjugated
\begin{equation}\label{DiracConjugatedEq}
-i \partial_\mu \phi^c \big(\gamma^\mu\big)^c = m\phi^c
\end{equation}  
Dirac equation\footnote{In the standard notation used by physicist the conjugated spinor $\phi^c$
is written as $\phi^+ = \overline{\phi}^T$, which we have already reserved for the 
operator conjugation of operators in the Fock space. The complex conjugation followed by transposition we agree to denote in this section by 
using the $+$ superscript interchangibly with the conjugation superscript $c$, 
which is customary in physical literature concerning Dirac bispinors and Dirac equation.}.
The representation space of the conjugated representation is defined as the Hilbert space $\mathcal{H}_{-m,0}^{\ominus c}$ of conjugated bispinors 
\begin{equation}\label{ConjugationForDiracField}
(\widetilde{\phi})^c (p) = {\widetilde{\phi}(-p)}^{+}
= \big(\overline{\widetilde{\phi}(-p)} \big)^T 
\end{equation}
with
$\widetilde{\phi} = V^{\ominus} {\widetilde{\psi}_{{}_{-m,0}}}$ 
ranging over the Hilbert space $\mathcal{H}_{-m,0}^{\ominus}$ of bispinors 
concentrated on the orbit $\mathscr{O}_{{}_{-m,0,0,0}}$
(i.e. with ${\widetilde{\psi}_{{}_{-m,0}}}$ ranging over the Hilbert space of the representation 
\[
U^{{}_{(-m,0,0,0)} L^{{}^{1/2}}}
\]
concentrated on $\mathscr{O}_{{}_{-m,0,0,0}}$, compare Example 1, Subsection 2.1 of \cite{wawrzycki2018}).
Here $(\cdot)^T$ stands for tansposition operation and 
\[
\big(\gamma^\mu\big)^c = \big(\overline{\gamma^\mu}\big)^T = \gamma^{\mu +}.
\]
In the space-time coordinates, i.e. after Fourier transformation, the formula for conjugation
is equivalent to
\[
\phi^c(x) = \phi(x)^+ = \big(\overline{\phi(x)} \big)^T.
\]
On the Hilbert space $\mathcal{H}_{-m,0}^{\ominus c}$ of conjugated bispinors there is defined
the (conjugated) inner product
\begin{multline*}
\big(\phi^c, {\phi'}^{c}\big)_c 
= \big((\widetilde{\phi})^c, (\widetilde{\phi}')^c)_c = 
(\phi', \phi)  \\ = 
\int \limits_{x^0 =  t = const.} \Big(\phi'(\vec{x}, t), \phi(\vec{x}, t) \Big)_{{}_{\mathbb{C}^4}}
\, \ud^3 x \\
= \int \limits_{\mathbb{R}^3} \Big(\widetilde{\phi}'(-\vec{p}, -|p_0(\vec{p})|), 
\widetilde{\phi}(-\vec{p}, -|p_0(\vec{p})|) \Big)_{{}_{\mathbb{C}^4}}
\, \frac{\ud^3 \vec{p}}{(2 p_0)^2} \\=
 -\int \limits_{\mathbb{R}^3} \Big(\widetilde{\phi}'(\vec{p}, p_0(\vec{p})), 
\widetilde{\phi}(\vec{p}, p_0(\vec{p})) \Big)_{{}_{\mathbb{C}^4}}
\, \frac{\ud^3 \vec{p}}{(2 p_0)^2} \\ =
-\int \limits_{\mathscr{O}_{-m,0,0,0}} \Big(\widetilde{\phi}'(p), 
\phi(p) \Big)_{{}_{\mathbb{C}^4}} 
\, \frac{\ud \mu_{{}_{m,0}}(p)}{2 |p_0|} =
\big(\widetilde{\phi}',\widetilde{\phi}\big)_{{}_{\mathcal{H}_{-m,0}^{\ominus}}},
 \,\,\,\, p_0(\vec{p}) = -\sqrt{\vec{p} \cdot \vec{p} + m^2}.
\end{multline*}
where $(\cdot, \cdot )$ is the inner product (\ref{Inn-Prod-Single-Dirac}) in the Hilbert space
$\mathcal{H}_{-m,0}^{\ominus} \subset \mathcal{H}$ of distributional solutions (whose Fourier transforms 
are concentrated on $\mathscr{O}_{-m,0,0,0}$) 
of Dirac equation defined above, which induces, through Fourier transform,
the inner product $\big( \cdot, \cdot \big)_{{}_{\mathcal{H}_{-m,0}^{\ominus}}}$ on their Fourier transforms. 
In the Hilbert space $\mathcal{H}_{-m,0}^{\ominus c}$ there are defined the operations
of multiplication by a number $\alpha \in \mathbb{C}$ and addition by the respective ordinary operations 
in $\mathcal{H}_{-m,0}^{\ominus}$, in the following manner
\[
\alpha \cdot (\widetilde{\phi})^c = (\overline{\alpha} \widetilde{\phi})^c = \alpha (\widetilde{\phi})^c, \,\,\,
(\widetilde{\phi})^c + (\widetilde{\phi}')^c = (\widetilde{\phi} + \widetilde{\phi})^c, \,\,\,
\widetilde{\phi}, \widetilde{\phi}' \in \mathcal{H}_{-m,0}^{\ominus}.
\]

From the formula (\ref{ConjugationForDiracField}) one easily see that the Fourier transforms of 
the conjugated bispinors are concentrated on the positive energy orbit $\mathscr{O}_{m,0,0,0}$
in the momentum space, and thus they are positive energy solutions of the conjugated Dirac
equation (\ref{DiracConjugatedEq}).   

Then on the conjugated Hilbert space $\mathcal{H}_{-m,0}^{\ominus c}$ (of conjugated bispinors 
concentrated on
the positive energy orbit $\mathscr{O}_{{}_{m,0,0,0}}$) there acts naturally the representation 
\begin{equation}\label{rep-c-on-Hc}
\big\{\widetilde{V^\ominus} U^{{}_{(-m,0,0,0)} L^{{}^{1/2}}} \, (\widetilde{V^\ominus})^{-1} \big\}^c
\end{equation}
conjugated to
\[
\widetilde{V^\ominus} U^{{}_{(-m,0,0,0)} L^{{}^{1/2}}} \, (\widetilde{V^\ominus})^{-1} 
\]
with the general definition of conjugation
\[
U^c(\widetilde{\phi})^c = (U\widetilde{\phi})^c.
\]
Because the spin corresponding to the conjugated representation (\ref{rep-c-on-Hc})
is likewise $1/2$ and the orbit is equal $\mathscr{O}_{m,0,0,0}$, then one can guess
that (\ref{rep-c-on-Hc}) is likewise equivalent to (\ref{m}), by Mackey's classification.
Indeed one can construct explicit equivalence similarly as $V^\oplus$ in Example 1 
(Subsection 2.1 of \cite{wawrzycki2018}) with additional transpositions and complex conjugations in this construction.

Thus to the space  $\mathcal{H}_{-m,0}^{\ominus c}$ we apply the Segal's functor $\Gamma$ of fermionic
quantization and obtain the fermionic Fock space
\[ 
\mathcal{H}^{\ominus}_{F} = \Gamma\big(\mathcal{H}_{-m,0}^{\ominus c}\big) 
= \mathbb{C} \oplus \mathcal{H}_{-m,0}^{\ominus c} \oplus
\big( \mathcal{H}_{-m,0}^{\ominus c} \big)^{\widehat{\otimes} 2} \oplus 
\big( \mathcal{H}_{-m,0}^{\ominus c} \big)^{\widehat{\otimes} 3}
\oplus \ldots;
\] 
with the unitary representation
\begin{multline*} 
\Gamma \Bigg(\big\{\widetilde{V^\ominus} U^{{}_{(-m,0,0,0)} L^{{}^{1/2}}} \, (\widetilde{V^\ominus})^{-1} \big\}^{c} \Bigg)
= \bigoplus \limits_{n = 0,1,2 \ldots}
\Bigg(\big\{\widetilde{V^\ominus} U^{{}_{(-m,0,0,0)} L^{{}^{1/2}}} \, (\widetilde{V^\ominus})^{-1} \big\}^{c} \Bigg)^{\widehat{\otimes} n}.
\end{multline*}

The conjugation $\big({\widetilde{\phi}} \big)^c$ of
the bispinor function concentrated on $\mathscr{O}_{-m,0,0,0}$
will be sometimes denoted by ${\widetilde{\phi}}^{c}$
in order to simplify notation.
We construct in the standard manner the map 
\[
\mathcal{H}_{-m,0}^{\ominus c} \ni 
{\widetilde{\phi}}^{c} \,\, \longrightarrow \,\,\,
a_{\ominus}\big({\widetilde{\phi}}^{c}\big), \,\,\, 
a_{\ominus}^+\big({\widetilde{\phi}}^{c}\big) 
= {a_{\ominus}\big({\widetilde{\phi}}^{c}\big)}^+
\] 
from $\mathcal{H}_{-m,0}^{\ominus c}$ to the families of (ordinary operators, not distributions) of 
annihilation and creation operators acting in the fermionic Fock space  
$\Gamma\big(\mathcal{H}_{-m,0}^{\ominus c}\big)$, fulfilling the canonical anticommutation relations:
\begin{multline*}
\Big\{a_{\ominus}\big({\widetilde{\phi}}^{c}\big), \,\,\, 
{a_{\ominus}\big(\widetilde{\phi}'^{c}\big)}^+ \Big\}
= \big({\widetilde{\phi}}^c, 
\widetilde{\phi}'^{c}\big)_{{}_{\mathcal{H}_{-m,0}^{\ominus c}}} \\
= \big({\widetilde{\phi}}^c, 
\widetilde{\phi}'^{c}\big)_c \\
= \Big(\widetilde{\phi}', 
\widetilde{\phi}\Big)_{{}_{\mathcal{H}_{-m,0}^{\ominus}}} \\
= -\int \limits_{\mathscr{O}_{-m,0,0,0}} \Big(\widetilde{\phi}'(p), 
\widetilde{\phi}(p) \Big)_{{}_{\mathbb{C}^4}}
\, \frac{\ud \mu_{{}_{-m,0}}(p)}{2 |p_0|}  \\
= \int \limits_{\mathbb{R}^3} 
\Big(\widetilde{\phi}'(-\vec{p}, -|p_0(\vec{p})|), \,\, 
\widetilde{\phi}(-\vec{p}, -|p_0(\vec{p})|) \Big)_{{}_{\mathbb{C}^4}}
\, \frac{\ud^3 \vec{p}}{(2 |p_0(\vec{p})|)^2}, \\
p_0(\vec{p}) = -\sqrt{\vec{p} \cdot \vec{p} + m^2}.
\end{multline*}

In particular the representation of the group $T_4 \circledS SL(2, \mathbb{C})$ which acts in the Fock 
space $\mathcal{H}^{\ominus}_{F}$ is equal 
\[
\Gamma \Bigg(\big\{\widetilde{V^\ominus} U^{{}_{(-m,0,0,0)} L^{{}^{1/2}}} \, (\widetilde{V^\ominus})^{-1} \big\}^c \Bigg). 
\]

Of course on the Fock space $\mathcal{H}^{\ominus}_{F}=\Gamma\big(\mathcal{H}_{-m,0}^{\ominus c}\big)$
we have the corresponding parity number (untary and involutive) operator  ${\In}_\ominus$  fulfilling 
\[
{\In}_\ominus^2 = \boldsymbol{1}, \,\,\,  {\In}_\ominus^* = {\In}_\ominus,
\] 
and such that ${\In}_\ominus$ anticommutes with the annihilation (and creation) operators:
\[
\Big\{ a_{\ominus}\big((\widetilde{\phi}|_{{}_{\mathscr{O}_{-m,0,0,0}}})^c\big), \, {\In}_\ominus \Big\} = 0.
\]

Of course the operator ${\In}_\ominus$ commutes:
\[
\Bigg[\,\,\,
\Gamma \Big(\big\{\widetilde{V^\ominus} U^{{}_{(-m,0,0,0)} L^{{}^{1/2}}} \, (\widetilde{V^\ominus})^{-1} \big\}^{c} \Big) 
\,\,\,,
\,\,\,
{\In}_\ominus \,\,\, \Bigg] =0
\]
with the representation of $T_4 \circledS SL(2, \mathbb{C})$ acting in the Fock space
$\Gamma\big(\mathcal{H}_{-m,0}^{\oplus c}\big)$ and with any operator of the form
$\Gamma(A)$ (bounded or unbouded with linear $\Dom \, \Gamma(A)$
in $\Gamma\big(\mathcal{H}_{-m,0}^{\oplus c}\big)$).

\subsection{The Fock-Hilbert space $\mathcal{H}_F$ of the free Dirac 
field $\boldsymbol{\psi}$}\label{electron+positron}

The Hilbert space $\mathcal{H}_F$ of the free Dirac field is defined as the application
of the fermion second quantization functor $\Gamma$ to the ``single particle'' Hilbert space 
$\mathcal{H}' = \mathcal{H}_{m,0}^{\oplus} \oplus \mathcal{H}_{-m,0}^{\ominus c}$--orthogonal sum of the 
Hilbert spaces $\mathcal{H}_{m,0}^{\oplus}$ and $\mathcal{H}_{-m,0}^{\ominus c}$. Therefore,
by the known propery of the functor $\Gamma$, 
it is equal to the tensor product
\[
\mathcal{H}_F = \mathcal{H}^{\oplus}_{F} \otimes \mathcal{H}^{\ominus}_{F}
= \Gamma\big(\mathcal{H}_{m,0}^{\oplus}\big) \otimes \Gamma\big(\mathcal{H}_{-m,0}^{\ominus c}\big) 
= \Gamma\big(\mathcal{H}_{m,0}^{\oplus} \oplus \mathcal{H}_{-m,0}^{\ominus c}\big)
\]
of the fermion Fock
spaces $\mathcal{H}^{\oplus}_{F}= \Gamma\big(\mathcal{H}_{m,0}^{\oplus}\big)$ and 
$\mathcal{H}^{\ominus}_{F}= \Gamma\big(\mathcal{H}_{-m,0}^{\ominus c}\big)$ with the representation
\[
\bigg[ \bigoplus \limits_{n=0,1,2, \ldots}
\Big(\widetilde{V^\oplus} \, U^{{}_{(m,0,0,0)} L^{{}^{1/2}}} \, 
(\widetilde{V^\oplus})^{-1} \Big)^{\widehat{\otimes} n}\bigg] \otimes 
\bigg[  
\bigoplus \limits_{n=0,1,2, \ldots} \Big(\big\{\widetilde{V^\ominus} U^{{}_{(-m,0,0,0)} L^{{}^{1/2}}} \, 
(\widetilde{V^\ominus})^{-1} \big\}^c  \Big)^{\widehat{\otimes} n} \bigg]
\] 
of the group $T_4 \circledS SL(2, \mathbb{C})$ acting in the Hilbert space $\mathcal{H}_F$.

Now observe that 
\[
\big\{\widetilde{V^\ominus} \, U^{{}_{(-m,0,0,0)} L^{{}^{1/2}}} \, (\widetilde{V^\ominus})^{-1} \big\}^c
= (\widetilde{V^\ominus})^{+ \, -1} \, \{ U^{{}_{(-m,0,0,0)} L^{{}^{1/2}}} \}^c \, (\widetilde{V^\ominus})^+. 
\]

Because by Mackey's construction of induced representation it follows that
\[
\big\{ U^{{}_{(-m,0,0,0)} L^{{}^{1/2}}} \big\}^c
= {S}^{-1} \, U^{{}_{(m,0,0,0)} L^{{}^{1/2}}} \, S
\]
with some (involutive) unitary operator $S$, we have
\[
\big\{\widetilde{V^\ominus} U^{{}_{(-m,0,0,0)} L^{{}^{1/2}}} \, (\widetilde{V^\ominus})^{-1} \big\}^c  
= {U_0}^{-1} \,  U^{{}_{(m,0,0,0)} L^{{}^{1/2}}} \, U_0 , \,\,\, U_0 = S \, (\widetilde{V^\ominus})^+.
\]
Thus the joint spectrum of the translation generators of the representation acting
in the Hilbert space $\mathcal{H}_F$ of the free Dirac field thus constructed is concentrated on the
positive energy cone $C_+$, i.e. it is a positive energy field.

Into the Fock-Hilbert space $\mathcal{H}_F$ of the free Dirac field we again introduce in the standard manner the families 
\[
\mathcal{H}_{m,0}^{\oplus} \oplus \mathcal{H}_{-m,0}^{\ominus c} \ni 
\widetilde{\phi}_1 \oplus \widetilde{\phi}_2 \,\,\, \longrightarrow 
a'\big(\widetilde{\phi}_1 \oplus \widetilde{\phi}_2\big), \,\,\, 
a'^+\big(\widetilde{\phi}_1 \oplus \widetilde{\phi}_2\big) 
= {a'\big(\widetilde{\phi}_1 \oplus \widetilde{\phi}_2\big)}^+,
\] 
fulfilling canonical anticummutation relations
\begin{multline}\label{AntiComRelFor-a-InH_F}
\Big\{a'\big(\widetilde{\phi}_1 \oplus \widetilde{\phi}_2\big), 
{a'\big(\widetilde{\phi}'_{1} \oplus \widetilde{\phi}'_{2}\big)}^+ \Big\} = \\ 
\Big(\, \widetilde{\phi}_1 \oplus \widetilde{\phi}_2, \,\,\, 
\widetilde{\phi}'_{1} \oplus \widetilde{\phi}'_{2} \,\Big)_{{}_{\mathcal{H}_{m,0}^{\oplus} \oplus \mathcal{H}_{-m,0}^{\ominus c}}} 
=\Big( \widetilde{\phi}_1,  
\widetilde{\phi}'_{1}\Big)_{{}_{\mathcal{H}_{m,0}^{\oplus}}} 
+
\Big( \widetilde{\phi}_2,  
\widetilde{\phi}'_{2}\Big)_{{}_{\mathcal{H}_{-m,0}^{\ominus c}}},
\end{multline}
where $(\cdot, \cdot)_{{}_{\mathcal{H}}}$ stands for the inner product on the Hilbert 
space $\mathcal{H}$. Here $\widetilde{\phi}_{1}, \widetilde{\phi}'_{1} \in \mathcal{H}_{m,0}^{\oplus}$
and $\widetilde{\phi}_{2}, \widetilde{\phi}'_{2} \in \mathcal{H}_{-m,0}^{\ominus c}$.

It follows that\footnote{Note that the equality $\Gamma(\mathcal{H}_1 \oplus \mathcal{H}_2)
= \Gamma(\mathcal{H}_1) \otimes \Gamma(\mathcal{H}_2)$ expresses in fact existence of a 
\emph{canonical} unitary isomorhism respecting the relevant Fock structure 
with paricular importance of the canoninal nature of the indentification (a mere existence of a
unitary map, here in the context of separable Hilbert spaces, is trival and would tell us nothing as there  is plenty of such maps devoid of any relevance).
The point is that the identification makes the following equality to hold
\[
a(u\oplus v) = a_{{}_{1}}(u) \otimes {\In}_{{}_{2}} + \boldsymbol{1} \otimes a_{{}_{2}}(v),
\]
for the corresponding annihilation and creation operators: 
$a(u\oplus v), a(u\oplus v)^+$ acting in 
$\Gamma(\mathcal{H}_1 \oplus \mathcal{H}_2)$, $a_{{}_{1}}(u), a_{{}_{1}}(u)^+$ acting in 
$\Gamma(\mathcal{H}_1)$
and $a_{{}_{2}}(v), a_{{}_{2}}(v)^+$ in $\Gamma(\mathcal{H}_2)$. Recall that ${\In}_2$ is the 
involutive unitary (and self-adjoint) parity number operator in Fock space $\Gamma(\mathcal{H}_2)$.
In fact in case of the fermionic Fock spaces we have two canonical choices for the identification
of the spaces $\Gamma(\mathcal{H}_1 \oplus \mathcal{H}_2)$ and 
$\Gamma(\mathcal{H}_1) \otimes \Gamma(\mathcal{H}_2)$. The second identification makes the following equality
to hold
\[
a(u\oplus v) = a_{{}_{1}}(u) \otimes \boldsymbol{1} +  {\In}_{{}_{1}} \otimes a_{{}_{2}}(v)
\]
with the parity number involution ${\In}_{{}_{1}}$ of te Fock space $\Gamma(\mathcal{H}_1)$. 
Thus in paricular we can use the other canonical idetification, where instead of 
(\ref{a_plusInH_F}), (\ref{a_minusInH_F}), (\ref{aInH_F}) we had
\[
\begin{split}
a'\big(\widetilde{\phi}_1 \oplus 0 \big) =
a_\oplus\big(\widetilde{\phi}_1\big) \otimes \boldsymbol{1},
\,\,\
\widetilde{\phi}_{1} \in \mathcal{H}_{m,0}^{\oplus}, \\
a'\big(0 \oplus \widetilde{\phi}_2\big) =
{\In}_\oplus \otimes a_\ominus\big(\widetilde{\phi}_2\big), 
\,\,\
\widetilde{\phi}_{2} \in \mathcal{H}_{-m,0}^{\ominus c}, \\
a'\big(\widetilde{\phi}_1 \oplus \widetilde{\phi}_2) = a_\oplus\big(\widetilde{\phi}_1\big) \otimes 
\boldsymbol{1}
+ {\In}_\oplus  \otimes a_\ominus\big(\widetilde{\phi}_2\big).
\end{split}
\]

In case of the boson Fock spaces we have essentially one canonical identification
of the Fock spaces $\Gamma(\mathcal{H}_1 \oplus \mathcal{H}_2)$ and 
$\Gamma(\mathcal{H}_1) \otimes \Gamma(\mathcal{H}_2)$ which makes the following equality 
to hold
\[
a(u\oplus v) = a_{{}_{1}}(u) \otimes \boldsymbol{1} +  \boldsymbol{1} \otimes a_{{}_{2}}(v).
\]
Therefore during the construction of a field with integer spin,
which is not essetially neutral (with antiparicles), when the the fermionic functor $\Gamma$ 
is replaced with bosinic
and the anticommutatuion relations are replaced with commutation relations, the 
involutive unitary and selfadjoint operators ${\In}_\oplus$ and ${\In}_\ominus$ are replaced here 
with the unital operator $\boldsymbol{1}$.} 
\begin{equation}\label{a_plusInH_F}
a'\big(\widetilde{\phi}_1 \oplus 0 \big) =
a_\oplus\big(\widetilde{\phi}_1\big) \otimes {\In}_\ominus,
\,\,\
\widetilde{\phi}_{1} \in \mathcal{H}_{m,0}^{\oplus},
\end{equation} 
\begin{equation}\label{a_minusInH_F}
a'\big(0 \oplus \widetilde{\phi}_2\big) =
\boldsymbol{1} \otimes a_\ominus\big(\widetilde{\phi}_2\big), 
\,\,\
\widetilde{\phi}_{2} \in \mathcal{H}_{-m,0}^{\ominus c}
\end{equation}
and
\begin{equation}\label{aInH_F}
a'\big(\widetilde{\phi}_1 \oplus \widetilde{\phi}_2) = a_\oplus\big(\widetilde{\phi}_1\big) \otimes {\In}_\ominus
+ \boldsymbol{1} \otimes a_\ominus\big(\widetilde{\phi}_2\big).
\end{equation}
Here  ${\In}_\ominus$ is the parity numer (involutive and self-adjoint unitary) 
opertor in the Fock space
$\Gamma\big(\mathcal{H}_{-m,0}^{\ominus c}\big)$
anticommuting with $a_\ominus\big(\widetilde{\phi}_2\big)$.
The operators $a_\oplus(\widetilde{\phi}_1)$ act on $\Gamma\big(\mathcal{H}_{m,0}^{\oplus}\big)$
and $a_\ominus(\widetilde{\phi}_2)$, ${\In}_\ominus$ act on $\Gamma\big(\mathcal{H}_{-m,0}^{\ominus c}\big)$.

In order to simplify notation the operators (\ref{a_plusInH_F}) and (\ref{a_minusInH_F})
undersood as operators in the total Fock space
\[
\mathcal{H}_F = \mathcal{H}^{\oplus}_{F} \otimes \mathcal{H}^{\ominus}_{F}
= \Gamma\big(\mathcal{H}_{m,0}^{\oplus}\big) \otimes \Gamma\big(\mathcal{H}_{-m,0}^{\ominus c}\big) 
= \Gamma\big(\mathcal{H}_{m,0}^{\oplus} \oplus \mathcal{H}_{-m,0}^{\ominus c}\big)
\] 
of the free Dirac field will likewise be denoted by $a_\oplus(\widetilde{\phi}_1)$
and $a_\ominus(\widetilde{\phi}_2)$, where $\widetilde{\phi}_1$ and $\widetilde{\phi}_2$ 
are understood as elements $\widetilde{\phi}_1\oplus0$ and $0\oplus \widetilde{\phi}_2$
of the Hilbert space $\mathcal{H}_{m,0}^{\oplus} \oplus \mathcal{H}_{-m,0}^{\ominus c}$
respectively, especially when the context suggest with what Fock
space we are working.

Note in paricular that 
for the operators (\ref{a_plusInH_F}) and (\ref{a_minusInH_F}),
undersood as operators on $\mathcal{H}_F$ and denoted simply by
$a_\oplus(\widetilde{\phi}_1)$ and $a_\ominus(\widetilde{\phi}_2)$, we have the following
canonical aticommutation relations (which follow from (\ref{AntiComRelFor-a-InH_F}))
\begin{equation}\label{AnticommutationRelationsInDiracFockSpace}
\begin{split}
\big\{a_\oplus(\widetilde{\phi}_{1}), a_\oplus(\widetilde{\phi}'_{1})^+\big\} =
\big(\widetilde{\phi}_{1},\widetilde{\phi}'_{1}\big)_{{}_{\mathcal{H}_{m,0}^{\oplus}}}, \\
\big\{a_\ominus(\widetilde{\phi}_{2}), a_\ominus(\widetilde{\phi}'_{2})^+\big\} =
\big(\widetilde{\phi}_{2},\widetilde{\phi}'_{2}\big)_{{}_{\mathcal{H}_{-m,0}^{\ominus c}}}, \\
\big\{a_\oplus(\widetilde{\phi}_{1}), a_\oplus(\widetilde{\phi}'_{1})\big\} =
\big\{a_\ominus(\widetilde{\phi}_{2}), a_\ominus(\widetilde{\phi}'_{2})\big\} = 0, \\
\big\{a_\oplus(\widetilde{\phi}_{1}), a_\ominus(\widetilde{\phi}'_{2})^+\big\} = 
\big\{a_\oplus(\widetilde{\phi}_{1}), a_\ominus(\widetilde{\phi}'_{2}) \big\} = 0,
\end{split}
\end{equation}
where again $\widetilde{\phi}_1, \widetilde{\phi}_1$ and $\widetilde{\phi}_2,\widetilde{\phi}'_2$ 
are understood respectively as elements $\widetilde{\phi}_1\oplus0, \widetilde{\phi}'_1\oplus0$ and 
$0\oplus \widetilde{\phi}_2, 0\oplus \widetilde{\phi}'_2$
of the Hilbert space $\mathcal{H}_{m,0}^{\oplus} \oplus \mathcal{H}_{-m,0}^{\ominus c}$.

The functor $\Gamma$ allows us to
have a clear insight into the structure of the represntation of $T_4 \circledS SL(2, \mathbb{C})$ 
acting in $\mathcal{H}_F$, as by construction it behaves functorially under the application
of $\Gamma$, applied separately to $\mathcal{H}_{m,0}^{\oplus}$ and $\mathcal{H}_{-m,0}^{\ominus c}$, and preserves
the structure $\mathcal{H}_F  
= \Gamma\big(\mathcal{H}_{m,0}^{\oplus} \big) \otimes \Gamma\big(\mathcal{H}_{-m,0}^{\ominus c} \big)$
because both $\mathcal{H}_{m,0}^{\oplus}$ and $\mathcal{H}_{-m,0}^{\ominus}$ are invariant for the representation 
of $T_4 \circledS SL(2, \mathbb{C})$ in the single particle Hilbert space 
$\mathcal{H}' = \mathcal{H}_{m,0}^{\oplus} \oplus \mathcal{H}_{-m,0}^{\ominus c}$.
In particular by the general properties of $\Gamma$ the representation of $T_4 \circledS SL(2, \mathbb{C})$
acting in $\mathcal{H}_F$ is naturally equivalent to the representation 
\begin{multline*}
\Gamma\Big( U^{{}_{(m,0,0,0)} L^{{}^{1/2}}} \Big) \otimes \Gamma\Big( U^{{}_{(m,0,0,0)} L^{{}^{1/2}}} \Big) \\
= \bigg[ \bigoplus \limits_{n=0,1,2, \ldots}
\Big( U^{{}_{(m,0,0,0)} L^{{}^{1/2}}} \,  \Big)^{\widehat{\otimes} n}\bigg] \otimes 
\bigg[ 
 \bigoplus \limits_{n=0,1,2, \ldots} \Big( U^{{}_{(m,0,0,0)} L^{{}^{1/2}}} \Big)^{\widehat{\otimes} n} \bigg],
\end{multline*} 
with the equivalence given by the unitary operator $\Gamma(V^\oplus) \otimes 
\Gamma\big(S \, (\widetilde{V^\ominus})^+\big)$.

Recall also the simple functorial property of $\Gamma$: for any group representations $U_1$ and $U_2$,
$\Gamma(U_1 \oplus U_2)$ is naturally equivalent to $\Gamma(U_1) \otimes \Gamma(U_2)$. Thus the Hilbert space
$\mathcal{H}_{F}$ is naturally equivalent to the ordinary (in the mathematical sense) Fock space
with the representation of $T_4 \circledS SL(2, \mathbb{C})$ in the single particle Hilbert space
$\mathcal{H}' = \mathcal{H}_{m,0}^{\oplus} \oplus \mathcal{H}_{-m,0}^{\ominus c}$ equivalent to 
$U^{{}_{(m,0,0,0)} L^{{}^{1/2}}} \oplus U^{{}_{(m,0,0,0)} L^{{}^{1/2}}}$.

\subsection{Quantum Dirac free field $\boldsymbol{\psi}$ as a Wightman 
operator-valued distribution}\label{psiWightman}

In order to construct quantum Dirac field, $\boldsymbol{\psi}$, we need a more subtle structure than just the Fock space,
as the quantum field is something which could be called suggestively ``operator-valued distribution'',
and which in turn is motivated by the classic analysis of measurement of quantum fields due to Bohr and Rosenfeld. In fact the precise mathematical interpretation is in fact still on the way.  
Intentionally (direction initiated by Wightman) quantum field, say $\boldsymbol{\psi}$, 
is regarded as a map  
$f \mapsto \boldsymbol{\psi}(f)$ with $\boldsymbol{\psi}(f)$, intentionally equal 
\begin{equation}\label{psi(f)-symbollical}
\int \boldsymbol{\psi}(x) f(x) \, \ud^4 x = \sum_{a} \int \boldsymbol{\psi}^a (x) f^a(x) \, \ud^4 x,
\end{equation}
which maps continously
a specified test space (here the Schwartz's space $\mathcal{S}(\mathbb{R}^4; \mathbb{C}^4)$ 
of bispinors $f$ on the space-time) 
into a specified class of (in general unbounded) operators 
$L(\mathcal{D})$ on a dense domain $\mathcal{D}$ of the Hilbert space, i.e. of the Fock space 
$\mathcal{H}^{\oplus}_{F}= \Gamma\big(\mathcal{H}_{m,0}^{\oplus}\big) \otimes \Gamma\big(\mathcal{H}_{-m,0}^{\ominus c}\big) = \Gamma\big(\mathcal{H}_{m,0}^{\oplus} \oplus \mathcal{H}_{-m,0}^{\ominus c}\big)$
in case of the field $\boldsymbol{\psi}$ in question,
with a specified sequentially complete topology on $L(\mathcal{D})$ respecting the nuclear theorem
and a nuclear topology on the test space, compare \cite{wig} and \cite{Woronowicz}
for a more detailed treatment. This should be regarded as the first step toward the precise mathematical interpretation of the notion of quantum field
introduced by the founders of QED, and in fact this is one possible approach, most popular among mathematical 
physicists working within the  ``axiomatic approach to QFT''. There is also another possible approach, 
initiated by Berezin \cite{Berezin} and developed by mathematicians \cite{hida}, \cite{obataJFA}, 
\cite{obata-book}. Although Wightman's definition of the quantum
(free) field does not fit well with the causal approach to QFT, we give a general remark on it before passing
to the Berezin-Hida white noise construction -- more adequate here.

In the Wighman's construction of (free) quantum field the integral expression 
(\ref{psi(f)-symbollical}), and especially the quantum 
field $\boldsymbol{\psi}(x)$ at a specified space-time point, has only symbolic character, lacking any immediate meaninig  even when  
considering free field(s), such as $\boldsymbol{\psi}$. This is just like the symbol $\psi(x)$ for a symbolic evaluation at $x$ of a ``function'' which symbolizes (when -- again symbolically -- integrated with a test function $f$) the value at $f$ of a proper distribution -- singular generalized function. 
In particular when considering a free field 
$\boldsymbol{\psi}$, the value $\boldsymbol{\psi}(f)$ for a space-time test (say bispinor
function $f \in \mathcal{S}(\mathbb{R}^4; \mathbb{C}^4)$) is obtained through the creation and annihilation operators evaluated at
the Fourier transform $\widetilde{f}$ restricted  to the orbit 
$\mathscr{O}$ pertinent to the representation defining the 
field(s) $\boldsymbol{\psi}$ 
(in case of presence of antiparticles the representation is not irreducible and
evaluation of the creation operator, acting over the Fock space over the single particle Hilbert space of conjugated solutions is involved, and even in general one has to consider many orbits in presence of more complicated fields or several fields\footnote{One can consider even spectral measure of traslation generators conentrated on the set of orbits with a finite range of possible mass parameters and the corresponding field which is called in this case a \emph{generalized free field}. We describe the case of the quantum Dirac field in details below.}). 
The experession (\ref{psi(f)-symbollical}) is given a meaning 
whenever applied to the vectors of the allowed domain $\mathcal{D}$,
only very indirectly, utilizing the quantity $\boldsymbol{\psi}(f)$, $f \in \mathcal{S}(\mathbb{R}^4; \mathbb{C}^4)$, which must be defined as the primary datum,
together with the appropriate domain $\mathcal{D}$, compare \cite{wig}, \S 3-3. For the free Dirac field
$\boldsymbol{\psi}$, the expression
$\boldsymbol{\psi}(f)$, $f \in \mathcal{S}(\mathbb{R}^4; \mathbb{C}^4)$, is defined through the creation 
$a_{\ominus}\big((P^\ominus\widetilde{f}|_{{}_{\mathscr{O}}})^c\big)^+$ and annihilation 
$a_{\oplus}\big(P^\oplus\widetilde{f}|_{{}_{\mathscr{O}}}\big)$ operators:
\begin{equation}\label{psi(f)=a_+(f)+a_-(f^c)^+}
\boldsymbol{\psi}(f) = a_{\oplus}\big(P^\oplus\widetilde{f}|_{{}_{\mathscr{O}_{m,0,0,0}}}\big) + 
a_{\ominus}\Big(\big(P^\ominus\widetilde{f}|_{{}_{\mathscr{O}_{-m,0,0,0}}}\big)^c\Big)^+,
\end{equation}
evaluated respectively at $P^\oplus\widetilde{f}|_{{}_{\mathscr{O}}}$ and 
$\big(P^\ominus\widetilde{f}|_{{}_{\mathscr{O}_{-m,0,0,0}}}\big)^c$. Here $\widetilde{f}$ is the ordinary 
Fourier transform of spacetime bispinor $f$, and $\widetilde{f}|_{{}_{\mathscr{O}_{m,0,0,0}}}$, 
$\widetilde{f}|_{{}_{\mathscr{O}_{-m,0,0,0}}}$ the respetive restrictions of $\widetilde{f}$ to
the orbits $\mathscr{O}_{m,0,0,0}$, $\mathscr{O}_{-m,0,0,0}$:
\[
\widetilde{f}|_{{}_{\mathscr{O}_{m,0,0,0}}}(p_0, \boldsymbol{\p}) = \widetilde{f}(\sqrt{|\boldsymbol{\p}|^2 +m^2},
\boldsymbol{\p}), \,\,\,
\widetilde{f}|_{{}_{\mathscr{O}_{-m,0,0,0}}}(p_0, \boldsymbol{\p}) = \widetilde{f}(-\sqrt{|\boldsymbol{\p}|^2 +m^2},
\boldsymbol{\p}).
\] 
Here $P^\oplus$ is the projection operator acting on bispinors 
$\widetilde{f}|_{{}_{\mathscr{O}_{m,0,0,0}}}$ concentrated on $\mathscr{O}_{m,0,0,0}$
and projecting on the Hilbert space $\mathcal{H}_{m,0}^{\oplus}$, defined in Subsection 2.1 of \cite{wawrzycki2018}.
$P^\ominus$ is the projection operator which projects bispinors 
$\widetilde{f}|_{{}_{\mathscr{O}_{-m,0,0,0}}}$ concentrated on $\mathscr{O}_{-m,0,0,0}$ on
 the Hilbert space $\mathcal{H}_{-m,0}^{\ominus}$, and defined in Subsection
2.1 of \cite{wawrzycki2018}, so that 
\[
\begin{split}
P^\oplus\widetilde{f}|_{{}_{\mathscr{O}_{m,0,0,0}}}(p) \overset{\textrm{df}}{=}
P^\oplus(p)\widetilde{f}(p), \,\,\,\, p = (\sqrt{|\boldsymbol{\p}|^2 +m^2},
\boldsymbol{\p}) \in \mathscr{O}_{m,0,0,0}, \\
P^\ominus\widetilde{f}|_{{}_{\mathscr{O}_{-m,0,0,0}}}(p) \overset{\textrm{df}}{=}
P^\ominus(p)\widetilde{f}(p), \,\,\,\, p = (-\sqrt{|\boldsymbol{\p}|^2 +m^2},
\boldsymbol{\p}) \in \mathscr{O}_{-m,0,0,0}.
\end{split}
\]
 Finally $(\cdot)^c$ stands for the conjugation defined in Subsection \ref{positron}.
By construction $P^\oplus\widetilde{f}|_{{}_{\mathscr{O}}}$ and 
$\big(P^\ominus\widetilde{f}|_{{}_{\mathscr{O}_{-m,0,0,0}}}\big)^c$
belong respectively to $\mathcal{H}_{m,0}^{\oplus}$ and $\mathcal{H}_{-m,0}^{\ominus c}$ whenever
$f \in \mathcal{S}(\mathbb{R}^4; \mathbb{C}^4)$,
and thus belong to the single particle Hilbert space 
$\mathcal{H}_{m,0}^{\oplus} \oplus \mathcal{H}_{-m,0}^{\ominus c}$, 
so that the expressions $a_\ominus\big((P^\ominus\widetilde{f}|_{{}_{\mathscr{O}_{-m,0,0,0}}})^c\big)^+$ and 
$a_\oplus(P^\oplus\widetilde{f}|_{{}_{\mathscr{O}}})$ make sense. 
Moreover both operators $P^\oplus, P^\ominus$ of multiplication by the projectors 
$P^\oplus(p)$, $p \in \mathscr{O}_{m,0,0,0}$ and respectively $P^\ominus(p)$, $p \in \mathscr{O}_{-m,0,0,0}$, 
commute by construction with the Fourier transformed 
Dirac operator of point-wise multiplication by the matrix $p_0\gamma^0 - p_k\gamma^k$ (summation with respect to 
$k=1,2,3$) on the Hilbert spaces $\mathcal{H}_{m,0}^{\oplus}$ and $\mathcal{H}_{-m,0}^{\ominus}$ of bispinors $\widetilde{f}|_{{}_{\mathscr{O}_{m,0,0,0}}}$ and respectively $\widetilde{f}|_{{}_{\mathscr{O}_{-m,0,0,0}}}$
concentrated respectively on $\mathscr{O}_{m,0,0,0}$ and $\mathscr{O}_{-m,0,0,0}$, so that
\[
\boldsymbol{\psi}\big((i\gamma^\mu \partial_\mu -m \boldsymbol{1})f\big) = 0, \,\,\,\,\, 
f \in \mathcal{S}(\mathbb{R}^4; \mathbb{C}^4),
\]
and the field $\boldsymbol{\psi}$ fulfills the free Dirac equation as expected, 
because the algebraic relation
\begin{equation}\label{AlgRelDiracEq}
\begin{split}
\big[p_0\gamma^0 - p_k\gamma^k - m\boldsymbol{1}\big]P^\oplus\widetilde{f}|_{{}_{\mathscr{O}_{m,0,0,0}}}(p) =0, 
\,\,\,\,\,\,\,\,\,\, 
p = (p_0, \boldsymbol{\p}) \in \mathscr{O}_{m,0,0,0} 
\\
\big[p_0\gamma^0 - p_k\gamma^k - m\boldsymbol{1}\big]P^\ominus\widetilde{f}|_{{}_{\mathscr{O}_{-m,0,0,0}}}(p) =0,
\,\,\,\,\,\,\,\,\,\, 
p = (p_0, \boldsymbol{\p}) \in \mathscr{O}_{-m,0,0,0},
\end{split}
\end{equation}
holds on the Hilbert spaces $\mathcal{H}_{m,0}^{\oplus}$ and $\mathcal{H}_{-m,0}^{\ominus}$ of bispinors
$\widetilde{f}|_{{}_{\mathscr{O}_{m,0,0,0}}}$ and respectively $\widetilde{f}|_{{}_{\mathscr{O}_{-m,0,0,0}}}$,
concentrated on $\mathscr{O}_{m,0,0,0}$ and respectively on $\mathscr{O}_{-m,0,0,0}$, compare Subsection 2.1 
of \cite{wawrzycki2018}. 
Indeed that $\boldsymbol{\psi}$ fulfills the homogeneous Dirac equation, can also be immediately seen by noting 
that the Fourier transformed
operator defining homogeneous Dirac equation is equal to point-wise multiplication by the matrix
\[
\big[p_0\gamma^0 - p_k\gamma^k - m\boldsymbol{1}_{{}_{4}}\big] = \big[ \slashed{p} - m \big]
\]
and that the projection operators $P^{\oplus}, P^\ominus$, commuting with it, are equal to operators 
of mutliplication by the projection matrices 
\[
\begin{split}
P^{\oplus}(p) = \frac{1}{2m} \big[ \slashed{p} + m \big], \,\,\, p \in \mathscr{O}_{m,0,0,0}, \\
P^{\ominus}(p) = \frac{1}{2m} \big[ \slashed{p} + m \big], \,\,\, p \in \mathscr{O}_{-m,0,0,0},
\end{split}
\]
compare Appendix \ref{fundamental,u,v}, formula (\ref{covariantPplusPminus}). From this 
and from the fact that
\[
\begin{split}
\big[ \slashed{p} + m \big]\big[ \slashed{p} - m \big] =
\big[ \slashed{p} - m \big]\big[ \slashed{p} + m \big] = [p \cdot p - m^2] \, \boldsymbol{1}_{{}_{4}} = 0, \,\,\,
p \in \mathscr{O}_{m,0,0,0}, \\
\big[ \slashed{p} + m \big]\big[ \slashed{p} - m \big] =
\big[ \slashed{p} - m \big]\big[ \slashed{p} + m \big] = [p \cdot p - m^2] \, \boldsymbol{1}_{{}_{4}} = 0, \,\,\,
p \in \mathscr{O}_{-m,0,0,0},
\end{split}
\]
the commutativity of  $\big[p_0\gamma^0 - p_k\gamma^k - m\boldsymbol{1}_{{}_{4}}\big]$
with $P^\oplus(p)$ on $\mathscr{O}_{m,0,0,0}$ and with $P^\ominus(p)$ on $\mathscr{O}_{-m,0,0,0}$,
as well as the relations  (\ref{AlgRelDiracEq}) are easily seen to hold, so that
our assertion follows. 

Note that in the formula (\ref{psi(f)=a_+(f)+a_-(f^c)^+}) we have used the simplified notation
for the operator (\ref{a_plusInH_F}) and for the operator adjoint to (\ref{a_minusInH_F}).
For the operator $a_{\oplus}\big(P^\oplus\widetilde{f}|_{{}_{\mathscr{O}_{m,0,0,0}}}\big)$
in the formula (\ref{psi(f)=a_+(f)+a_-(f^c)^+}) the reader should read
\begin{equation}\label{a'Inpsi}
a'\big(P^\oplus\widetilde{f}|_{{}_{\mathscr{O}_{m,0,0,0}}} \oplus 0\big) =
a_{\oplus}\big(P^\oplus\widetilde{f}|_{{}_{\mathscr{O}_{m,0,0,0}}}\big) \otimes {\In}_\ominus
\end{equation}
and for the operator $a_{\ominus}\Big(\big(P^\ominus\widetilde{f}|_{{}_{\mathscr{O}_{-m,0,0,0}}}\big)^c\Big)^+$
in (\ref{psi(f)=a_+(f)+a_-(f^c)^+}) the reader should read
\begin{equation}\label{a'^+Inpsi}
a'\Big( 0 \oplus \big(P^\ominus\widetilde{f}|_{{}_{\mathscr{O}_{-m,0,0,0}}}\big)^c \Big)^+ =
\boldsymbol{1} \otimes a_{\ominus}\Big(\big(P^\ominus\widetilde{f}|_{{}_{\mathscr{O}_{-m,0,0,0}}}\big)^c\Big)^+.
\end{equation}
On the left hand sides of the last two formulas we have the standard annihilation and creation
operators $a'(u\oplus v), a'(u\oplus v)^+$ acting on the Fock space 
\[
\mathcal{H}_F = \Gamma\big( \mathcal{H}_{m,0}^{\oplus} \oplus \mathcal{H}_{-m,0}^{\ominus c} \big)
= \Gamma\big( \mathcal{H}_{m,0}^{\oplus} \big) \otimes \Gamma \big(\mathcal{H}_{-m,0}^{\ominus c} \big)
\]
of the free Dirac field introduced in Subsection \ref{electron+positron}.
On the right hand sides of the last two formulas we have the annihilation and creation operators  
$a_{\oplus}\big(P^\oplus\widetilde{f}|_{{}_{\mathscr{O}_{m,0,0,0}}}\big)$ and 
$a_{\ominus}\Big(\big(P^\ominus\widetilde{f}|_{{}_{\mathscr{O}_{-m,0,0,0}}}\big)^c\Big)^+$
acting respectively in the Fock spaces $\Gamma\big(\mathcal{H}_{m,0}^{\oplus}\big)$
and $\Gamma\big(\mathcal{H}_{-m,0}^{\ominus c}\big)$, and defined respecively in Subsections \ref{electron}
and \ref{positron}. For definition of the unitary involutive (and thus self-adjoint) 
operator\footnote{The operator ${\In}_\ominus$ is replaced with the unital operator in case of integer spin (non-neutral) field.} ${\In}_\ominus$ we refer to Subsections \ref{electron} and \ref{positron}. 

Thus the formula (\ref{psi(f)=a_+(f)+a_-(f^c)^+}) should properly be written as 
\begin{equation}\label{psi(f)=a_+(f)+a_-(f^c)^+proper}
\boldsymbol{\psi}(f) = a'\big(P^\oplus\widetilde{f}|_{{}_{\mathscr{O}_{m,0,0,0}}} \oplus 0\big) + 
a'\Big( 0 \oplus \big(P^\ominus\widetilde{f}|_{{}_{\mathscr{O}_{-m,0,0,0}}}\big)^c \Big)^+.
\end{equation}

In fact $\boldsymbol{\psi}(f)$ is antilinear in $f$, but the additional complex 
conjugation will make it linear operator-valued distribution. We have not placed this conjugation
explicitly in order to simplify notation.

It should be stressed however that the structure 
$\mathcal{H}_F = \mathcal{H}^{\oplus}_{F} \otimes \mathcal{H}^{\ominus}_{F}
= \Gamma\big(\mathcal{H}_{m,0}^{\oplus} \big) \otimes \Gamma\big(\mathcal{H}_{-m,0}^{\ominus c} \big)$ of the 
Hilbert space of the free 
quantum Dirac field $\boldsymbol{\psi}$, as well as the tensor product form of the operators
(\ref{a'Inpsi}) and (\ref{a'^+Inpsi}) in (\ref{psi(f)=a_+(f)+a_-(f^c)^+proper})
 does not mean that the quantum Dirac field may be treated as sum of two independent fields of electrons and positrons. 
Indeed the quantized Dirac field, equal to the linear combination (\ref{psi(f)=a_+(f)+a_-(f^c)^+proper}) of operators
\footnote{Both treated as tensor product operators on $\Gamma\big(\mathcal{H}_{m,0}^{\oplus} \big) \otimes \Gamma\big(\mathcal{H}_{-m,0}^{\ominus c} \big)$, the first having the second factor trivial and 
equal to the fundamental unitary involution ${\In}_\ominus$ and vice versa for the second, with the first factor
trivial and equal to the unit operator.}, 
cannot be treated as sum of field operators respectively in $\Gamma\big(\mathcal{H}_{m,0}^{\oplus} \big)$
and $\Gamma\big(\mathcal{H}_{-m,0}^{\ominus c} \big)$ simply because the arguments 
\[
P^\oplus\widetilde{f}|_{{}_{\mathscr{O}_{m,0,0,0}}}
 \,\,\, \textrm{and} \,\,\
\big(P^\ominus\widetilde{f}|_{{}_{\mathscr{O}_{-m,0,0,0}}}\big)^c
\]
in the operators  (\ref{a'Inpsi}) and (\ref{a'^+Inpsi}) entering the formula
 (\ref{psi(f)=a_+(f)+a_-(f^c)^+proper}) for $\boldsymbol{\psi}(f)$ are not independent.
Indeed by choosing a function $f$ from the test space $\mathcal{S}(\mathbb{R}^4; \mathbb{C}^4)$
we predeterminate the restrictions
\[
\widetilde{f}|_{{}_{\mathscr{O}_{m,0,0,0}}}
 \,\,\, \textrm{and} \,\,\
\widetilde{f}|_{{}_{\mathscr{O}_{-m,0,0,0}}}
\] 
of its Fourier transform to the orbits $\mathscr{O}_{m,0,0,0}$ and $\mathscr{O}_{-m,0,0,0}$, which cannot be 
varied independetly one from another. This dependence, imposed on 
\[
\widetilde{f}_1 = \widetilde{f}|_{{}_{\mathscr{O}_{m,0,0,0}}}
 \,\,\, \textrm{and} \,\,\
\widetilde{f}_2 = \widetilde{f}|_{{}_{\mathscr{O}_{-m,0,0,0}}}
\] 
by the fact that they come from restrictions to the orbits of the Fourier transform of one and the same 
$f$, cannot be realized by any natural relation put on the two \emph{a priori} independent fields of 
electrons and positrons, and realized through (\ref{a'Inpsi}) and (\ref{a'^+Inpsi}) with two independent
arguments $f$, respectively, in (\ref{a'Inpsi}) and (\ref{a'^+Inpsi}).  

The domain $\mathcal{D}$ of the field $\boldsymbol{\psi}$, due to the interpretation initiated by Wightman,  is not determined uniquely
but in any case contains at least the
domain $\mathcal{D}_0$ which arises by the action of
polynomilal expressions in 
\[
\boldsymbol{\psi}(f_1), \boldsymbol{\psi}(f_2), \ldots, \,\,\,\,\, 
f_i \in \mathcal{S}(\mathbb{R}^4; \mathbb{C}^4)
\] 
on the vacuum $|0\rangle = \Psi_0$. 
However we know that the domain must be considerably larger if $L(\mathcal{D})$
is supposed to satisfy kernel theorem
in accordance to the result of \cite{Woronowicz}. 
In particular it must contain the domain called $\mathcal{D}_1$ in \cite{wig}, p. 107,
but it is even not clear for the free field determined by an irreducible representation corresponding to a single orbit that $L(\mathcal{D}_1)$
satisfies the theorem on kernel as stated in \cite{Woronowicz}. We only know, 
by the result of \cite{Woronowicz}, that such domain 
$\mathcal{D}$ exists on which $L(\mathcal{D})$ satisfies the theorem on kernel (with the ``strong topology''
on $L(\mathcal{D})$), and contains the domain called $\mathcal{D}_1$ in \cite{wig}, p. 107. 

More generally for any $f \in \mathcal{S}(\mathbb{R}^{4k}) = 
\mathcal{S}(\mathbb{R}^{4})^{\otimes k}$ and for any system of free fields
$\boldsymbol{\psi}_1, \ldots, \boldsymbol{\psi}_k$ one can give a meaning of a well defined vector in the 
dense domain $\mathcal{D}$ of the Fock space of the total system
to the expression of the form
\begin{equation}\label{PsiInD}
\Psi = \int \ud^4 x_1 \ldots \ud^4x_k \, f(x_1, \ldots, x_k ) \, \boldsymbol{\psi}_1(x_1) \ldots 
\boldsymbol{\psi}_k(x_k) \, \Psi_0,
\end{equation}
and then for any field $\boldsymbol{\psi}$ of the considered system of free fields and for any 
$\Psi$ of the form (\ref{PsiInD}) one can give a meaning by a limit process to the expression
\begin{equation}\label{psi(f)Psi}
\boldsymbol{\psi}(f)\Psi
\end{equation}
thus giving a meaning to $\boldsymbol{\psi}_1(x_1) \ldots 
\boldsymbol{\psi}_k(x_k)$ of an operator-valued distribution over the test space  
$\mathcal{S}(\mathbb{R}^{4})^{\otimes k}$ on the domain containing all vectors
of the form (\ref{PsiInD}), compare \cite{wig}, \S 3-3. This is achieved by noting first
that
\[
(\Psi_0, \boldsymbol{\psi}_1(f_1) \ldots 
\boldsymbol{\psi}_k(f_k) \, \Psi_0)
\]
is a well defined and separately continuous multilinear functional of the arguments
$f_i$ in the nuclear topology on the Schwartz space $\mathcal{S}(\mathbb{R}^4)$. 
Thus by the ordinary Schwartz kernel theorem it follows that there exists a unique distribution
$\mathscr{W}(x_1, \ldots, x_k)$ such that
\[
\int \mathscr{W}(x_1, \ldots, x_k) f_1(x_1)f_2(x_2) \ldots f_k(x_k) \, \ud^4x_1 \ldots \ud^4x_k
= (\Psi_0, \boldsymbol{\psi}_1(f_1) \ldots 
\boldsymbol{\psi}_k(f_k) \, \Psi_0)
\]
for any $f_i \in \mathcal{S}(\mathbb{R}^4)$. Using this fact (as in \cite{wig}, p. 107)
we next show that the states
\[
\Psi_J = \sum_{j=1}^{J} \boldsymbol{\psi}_1(f_{1j}) \ldots 
\boldsymbol{\psi}_k(f_{kj}) \, \Psi_0
\]
converge in norm of the Fock space whenever the functions
\[
f_J(x_1, \ldots, x_k) = \sum_{j=1}^{J} f_{1j}(x_1)f_2(x_2) \ldots f_{kj}(x_k) 
\]
converge to $f$ in $\mathcal{S}(\mathbb{R}^4)^k = \mathcal{S}(\mathbb{R}^{4k})$. The limit
of $\Psi_J$ is defined as the vector $\Psi$ giving the meaning to the expression
(\ref{PsiInD}). The value (\ref{psi(f)Psi}) is defined as the limit of $\boldsymbol{\psi}(f)\Psi_J$,
and gives a well defined ``operator-valued'' distribution by the pre-closed character
of the operators $\boldsymbol{\psi}(f)$ on the domains $\mathcal{D}_0 \subset \mathcal{D}_1$, compare
\cite{Woronowicz}. 

In Wightman approach it is the formula (\ref{psi(f)=a_+(f)+a_-(f^c)^+}) which gives
the meaning to the symbolic expression (\ref{psi(f)-symbollical}) when applied to the
elements of the domain $\mathcal{D}$.

For a given free field (or a system of free fields
$\boldsymbol{\psi}_1, \boldsymbol{\psi}_2, \ldots, \boldsymbol{\psi}_k$) one can give, within the mentioned Wightman approach, a meaning to the expression  
\begin{equation}\label{:psi(f):-symbollical}
:\partial^{\alpha_1}\boldsymbol{\psi}_1 \ldots \partial^{\alpha_k}\boldsymbol{\psi}_k:(f)
 = \int :\partial^{\alpha_1}\boldsymbol{\psi}_1(x) \ldots \partial^{\alpha_k}\boldsymbol{\psi}_k(x): \, f(x) \, \ud^4 x 
\end{equation}
as a limit, giving an operator-valued distribution \cite{WightmanGarding}. However here for definition of the ``Wick product'' due to \cite{WightmanGarding} and using Wightman's definition of the field the limit process involved here is devoid of any natural choice, as the ``Wick product field'' of Wightman and G{\aa}rding is obtained from an operator-valued distribution 
in several spacetime variables, and then as a limit we obtain operator valued distribution in just one space-time
variable. Such definition involves a considerable amount of unnatural and rather arbitrary choices
in selecting a (class of) limit(s) of passing from test function spaces in just one space-time variable
to the test space in several space-time variables, compare \cite{WightmanGarding} for one possible 
choice\footnote{For the opposite direction, i.e. for passing from distribution of one variable to distribution of several variables, we would have the natural choice given by the map defined by the restriction to the diagonal, which is continuous between the test spaces. Reverse direction is is by no means natural nor unique. The reader should also note that the ``definition'' of the 
Wick product in \cite{wig}, \S 3-2, p. 104, which merely says:
\[
:\partial^{\alpha_1}\boldsymbol{\psi}(x) \partial^{\alpha_k}\boldsymbol{\psi}(x):
=\lim \limits_{x_1, x_2 \rightarrow x} \Big[\partial^{\alpha}\boldsymbol{\psi}(x_1) \partial^{\beta}\boldsymbol{\psi}(x_2) - \big(\Psi_0, \partial^{\alpha}\boldsymbol{\psi}(x_1) \partial^{\beta}\boldsymbol{\psi}(x_2) \Psi_0\big)\Big],
\]
and 
\begin{multline*}
:\partial^{\alpha}\boldsymbol{\psi}(x) \partial^{\beta}\boldsymbol{\psi}(x)  
\partial^{\gamma}\boldsymbol{\psi}(x):
=\lim \limits_{x_1, x_2 \rightarrow x} \Big[\partial^{\alpha}\boldsymbol{\psi}(x_1) \partial^{\beta}\boldsymbol{\psi}(x_2)   \\
-\big(\Psi_0, \partial^{\alpha}\boldsymbol{\psi}(x_1) \partial^{\beta}\boldsymbol{\psi}(x_2)
\partial^{\gamma}\boldsymbol{\psi}(x_3) \Psi_0\big) \partial^{\gamma}\boldsymbol{\psi}(x_3) \\
-\big(\Psi_0, \partial^{\alpha}\boldsymbol{\psi}(x_1) \partial^{\gamma}\boldsymbol{\psi}(x_3)
\partial^{\gamma}\boldsymbol{\psi}(x_3) \Psi_0\big) \partial^{\beta}\boldsymbol{\psi}(x_2) \\
-\big(\Psi_0, \partial^{\beta}\boldsymbol{\psi}(x_2) \partial^{\gamma}\boldsymbol{\psi}(x_3)
\partial^{\gamma}\boldsymbol{\psi}(x_3) \Psi_0\big) \partial^{\alpha}\boldsymbol{\psi}(x_1)
\Big], \\
\,\,\,\, \textrm{and so on} \, \ldots
\end{multline*}
 is again only heuristic, and strictly speaking is meaningless
as a definition of operator-valued distribution,
as it involves limit process of passing from test space of one space-time variable to test space of several space-time variables, which is not specified there. The reader which would like to know the concrete choice 
of the possible limit process involved there which is meant by the authors will have to consult 
the paper \cite{WightmanGarding}.} of the limit process.  

Unfortunately the method of  \cite{WightmanGarding} is not efficient (for boson, and particularily 
for mass less fields) in the investigation of the closability
of the operator (\ref{:psi(f):-symbollical}) or its eventual self-adjointness nor for the proof of the 
``Wick theorem'' \cite{Bogoliubov_Shirkov}, Chap. III, useful in the causal perturbative approach to QED.
Similarly the space-time averaging as presented in \cite{WightmanGarding} is not applicable to the averaging
over space-like Cauchy hypersurfaces  of their ``Wick product fields'', necessary in construction of the conserved currents appearing in the Noether theorem for free fields. 
In particular the Quantization Postulate for free fields 
as formulated in \cite{Bogoliubov_Shirkov}, Chap. 2, \S 9.4, cannot be simply treated with Wightman-G{\aa}rding method, and for zero mass fields this Postulate seems to be intractable with 
Wightman-G{\aa}rding method\footnote{The mentioned weaknesses of Wightman-G{\aa}rding definition of the
``Wick product'' have also  been noted by I. E. Segal, compare e.g. \cite{Segal-NFWP.I}, \cite{Segal-ProcStone}.}.

\subsection{Motivation for introduction of Hida operators into causal perturbative QFT}\label{MotivationForHida}

This is somewhat unsatisfactory because the causal method, which is successful in 
avoiding ultraviolet infinities (also avoidning infrared infinities for the adiabatically 
switched off interaction at infinity), expresses the scattering operator $S(g)$ in terms
of time ordered products of Wick polynomials of free fields, and is substantially based
on the ``Wick theorem'' for free fields as stated in \cite{Bogoliubov_Shirkov}, Chap. III. 
These Wick plynomial fields are constructed with the help of the ``Wick theorem'' applied to 
the expressions including products of normally ordered factors in free fields, representing interaction terms, and some of the annihilation-creation operators as additional factors, with the annihilation-creation operators not necesary ``normally ordered''. Wick theorem allows us to equate such expressions with the corresponding expressions containing solely ``normally ordered products'' of creation-annihilation operators multiplied by the respective pairings. The point is that we explicitly utilize the commutation generalized functions and the Wick theorem and only at the end prove e. g. that the product of several factors with the space-time variable in each factor treated as independent of the variables of other factors and with each factor being normally oredered product of free fields representing interactin terms, is a well defined operator valued distribution on the domain $\mathcal{D}_0$ in the Wightman sense. 
Essentially this ``theorem'' allows to treat the (generalized) operators
of the type (compare Theorem 0 in \cite{Epstein-Glaser})
\begin{equation}\label{IntKerOpFreeWickField}
\kappa(x_1, \ldots, x_k) :\partial^{\alpha_1}\boldsymbol{\psi}_1(x_1) \ldots \partial^{\alpha_k}\boldsymbol{\psi}_k(x_k): 
\end{equation} 
with numerical,``translationally invariant'' ($\kappa(x_1+a, \ldots, x_k+a)=\kappa(x_1, \ldots, x_k)$), distributions\footnote{In fact we are interested here in distributions $\kappa$ which arise
as tensor products of the pairings of the corresponding free fields $\partial^{\alpha_i}\boldsymbol{\psi}_i$
and when the interaction does not contain derivatives we may confine attention in 
(\ref{IntKerOpFreeWickField}) to the case where derivatives are absent, i.e. with all the multiidices 
$\alpha_i = 0$. In paricular all such distributions have the mentioned invariance property.} 
$\kappa \in \mathcal{S}(\mathbb{R}^{4k})^* = 
\big(\mathcal{S}(\mathbb{R}^4)^*\big)^{\otimes k}$
which, when integrated with test functions $f \in \mathcal{S}(\mathbb{R}^{4k}) = 
\mathcal{S}(\mathbb{R}^4)^{\otimes k}$, define an operator valued distribution
\begin{equation}\label{IntKerOpFreeWickField(f)}
f \rightarrow \int \, f(x_1, \ldots, x_k) \, \kappa(x_1, \ldots, x_k) \, :\partial^{\alpha_1}\boldsymbol{\psi}_1(x_1) \ldots \partial^{\alpha_k}\boldsymbol{\psi}_k(x_k): \, \ud^4x_1 \ldots \ud^4x_k.
\end{equation}
Thus in practical computations we proceed from the ``kernel'' of the ``operator distribtion'' to the distribution, and not in the reverse direction which is pertinent to the Wightman approach in which the ``kernel''is only symbolic and difficult to handle.
It is therefore not satisfactory
that already at the free field level the ``Wick theorem'' in the form needed for the causal perturbative approach is not clearly related to the free field defined according to Wightman \cite{wig}. 

In spite of this inconvenience, ``Wick theorem'' of \cite{Bogoliubov_Shirkov}, Chap III, provides  
partially heuristic (but honest) basis for construction of ``operator-valued distributions'' of the type 
(\ref{IntKerOpFreeWickField(f)}), compare Theorem 0 of \cite{Epstein-Glaser}.  
This turned up to be effective in the realization of the causal
approach program of St\"uckelberg-Bogoliubov. As realized later by Epstein and Glaser \cite{Epstein-Glaser} 
the causal approach of St\"uckelberg-Bogoliubov provides a perturbative method which avoids ultraviolet infinities (and also infrared but with the unphysical adiabatically switched off interaction at infinity which, especially in case of QED, needs a further analysis of the behaviour of the theory when the physical interaction is restored, say by adiabatical switching on the interaction at infinity). The essential improvement of the causal method of St\"uckelberg-Bogoliubov added by Epstein and Glaser
is the carefull splitting of the operator-valued distributions of the type (\ref{IntKerOpFreeWickField(f)})
with causally supported distribution kernels $\kappa$ into the retarded and advanced parts -- a task which we
encounter in the causal construction of the perturbative series. Epstein and Glaser 
\cite{Epstein-Glaser} reduce this task
to the splitting of the numerical causally supported distribution kernels $\kappa$
into the retarded and advanced part.

Let us shortly summarize the causal perturbative approach due to 
St\"uckelberg-Bogoliubov-Epstein-Glaser on the example of QED. This approach, contrary to that based on the Hamiltonian, is not based on the Schr\"odinger-Tomonaga equaton in the interaction picture, with the main motivation lying in avoiding the problem with the singular character of the interaction Hamiltonian. In causal approach this is the scattering operator which plays the fundamental role, and the remaining quantities, \emph{i. e.} effective cross sections, and the local interacting fields as well as the definition of the product of interacting fields are obtained from the scattering opertor. The time evolution is encoded in the general principles put on the scattering matrix $S$, with the causality condition implemented by the ``switching off space-time function $g$'' multiplying the Lagrange ineraction  density $\mathcal{L}$, expressed as Wick polynomial of free fields. The scattering operator is treated as an (say operator-valued) functional of $g \mapsto S(g)$. The functional $S(g)$ is constructed perturbatively, and written as a formal functional power series in $g$
\[
S(g) = \boldsymbol{1} + \sum \limits_{n=1}^{\infty} {\textstyle\frac{1}{n!}} \int \ud^4 x_1 \cdots \ud^4 x_n
S_n(x_1, \ldots, x_n) \, g(x_1) \ldots g(x_n)
\]
\[
S(g) = \boldsymbol{1} + \sum \limits_{n=1}^{\infty} {\textstyle\frac{1}{n!}} \int \ud^4 x_1 \cdots \ud^4 x_n
\overline{S_n}(x_1, \ldots, x_n) \, g(x_1) \ldots g(x_n)
\]
The conditions put on $S(g)$ are the following. 
\emph{Causality}:
\[
(\textrm{I}) \,\,\,\,\,\,\,\,\,\,\,\,
S(g_1 + g_2) = S(g_1)S(g_2), \,\,\,\textrm{whenever}
\,\,\,
\textrm{supp} \, g_1 \preceq \textrm{supp} \, g_2
\]
where the spacetime region $G_1$ is said to causally precede $G_2$, in short $G_1 \preceq G_2$,
iff $\big(G_1 + \overline{V_{-}}\big) \cap G_2 =\emptyset$ or equivalently iff $G_1 \cap \big(G_2 + \overline{V_{+}}\big) = \emptyset$. Here $\overline{V_{-}}, \overline{V_{+}}$ are the closures of the interiors of the backward or foreward light cones. 
\emph{Covariance}
\[
(\textrm{II}) \,\,\,\,\,\,\,\,\,\,\,\,
U_{a,\lambda} S(g) U_{a,\lambda}^{-1} = S(g_{a,\lambda}), \,\,\,\,
g_{a, \Lambda}(x) = g(\Lambda^{-1}x - a)
\]
where $U$ is the representation of $T_4 \circledS SL(2, \mathbb{C})$ acting in the Fock space of all free fields of  the theory. Unitarity (Krein-isometricity in case when gauge fields are present, as is the case e. g. for QED)
\[
(\textrm{III}) \,\,\,\,\,\,\,\,\,\,\,\,
\mathfrak{J} S(g)^{+} \mathfrak{J} = S(g)^{-1}
\]
where $\mathfrak{J}= \mathfrak{J}^* = \mathfrak{J}^{-1}$ is the Krein-fundamental symmetry (or the Gupta-Bleuler operator in the particular case of QED). Finally we are using the 
\emph{Bohr's correspondence principle} (quasi-classical limit, compare \cite{Bogoliubov_Shirkov}),
which allows us to identify the first order contribution $S_1(x)$ to the scattering operator 
functional $S$:
\[
(\textrm{IV}) \,\,\,\,\,\,\,\,\,\,\,\,
S_1(x) = i \mathcal{L}(x)
\]
where $\mathcal{L}(x)$ is the Lagrange interaction density of the theory in question, 
expressed as a Wick polynomial of free fields underlying the theory. In case of spinor QED, 
$\mathcal{L}(x) = \boldsymbol{:}\overline{\boldsymbol{\psi}}(x) \gamma^\mu \boldsymbol{\psi}(x) A_\mu(x)\boldsymbol{:}$. The conditions (I) -- (IV), expressed in terms of the ``generalized operator kernels''
$S_n(x_1, \ldots, x_n)$ reads  
\begin{align*}
(\textrm{I}) & \,\,\,\,\,\,\,\,\,\,\,\, &
S_{n}(x_1, \ldots, x_n) = S_{k}(x_1, \ldots, x_k)S_{n-k}(x_{k+1}, \ldots, x_n), \\ 
& & \,\,\,\,\,\,\, \textrm{if} \, x_j \notin x_i + \overline{V_{+}}, \,\, i \in\{1, \ldots, k\}, j \in \{k+1, \ldots, n\}
\\
(\textrm{II} )& \,\,\,\,\,\,\,\,\,\,\,\, &
U_{a,\Lambda} S_n(x_1, \ldots, x_n)U_{a,\Lambda}^{-1} = S_n(\Lambda^{-1}x_1 + a, \ldots, \Lambda^{-1}x_n + a),
\\ 
(\textrm{III} )& \,\,\,\,\,\,\,\,\,\,\,\, &
\mathfrak{J} S_n(x_1, \ldots, x_n)^{+} \mathfrak{J} = \overline{S_n}(x_1, \ldots, x_n), \\
(\textrm{IV}) & \,\,\,\,\,\,\,\,\,\,\,\, &
S_1(x) = i \mathcal{L}(x)
\end{align*}

These conditions (I) -- (IV) should be understood as candidates for axioms of a theory which is to be formulated in well defined mathematical terms. Also the mathematical character of the operator
$S(g)$, as well as as these axioms should be understood properly. In fact the particular mathematical choices still remain to lie in very front of us, but have not been done yet. So we need to be very careful now. At the  present stage of the theory
we should proceed at the most general level, just taking care only not to fall into evident conflict with the computational practice we perform whenever we compute the effective cross section. For this purpose the operator $S(g)$ \emph{need not be an ordinary operator} acting on normalzable states, and even the the separate contributions 
\[
\int \ud^4 x_1 \cdots \ud^4 x_n
S_n(x_1, \ldots, x_n) \, g(x_1) \ldots g(x_n)
\]
of fixed order need not be ordinary operators. Before making further mathematical specifications and particular choices, let us look at the concrete example of QED. In that case the
``operator kernels'' $S_n(x_1, \ldots, x_n)$ have the general form  
\begin{multline}\label{QEDgeneralS_n(X)andA'_n(X)}
S_n(x_1, \ldots, x_n) = \\ =  \sum \limits_{k, a_j, b_i, \mu_m} t^{k, a_j b_i \mu_m}_{n}(x_1, \ldots, x_n) \,\, \boldsymbol{:} \sqcap_{j}
 \overline{\boldsymbol{\psi}}^{a_j}(x_j) \sqcap_{i} \boldsymbol{\psi}^{b_i}(x_i) \boldsymbol{:} \boldsymbol{:} \sqcap_{m} A_{\mu_m}(x_m)\boldsymbol{:}
\end{multline}
where the sum over $k$ is finite with the number of terms depending  on the order $n$, and with 
total number of factors under Wick products less than or equal $n$ and depending on the particular $k$.
The distribution kernels $t^{k}_{n}$ have causal supports and are determined by the axioms (I)--(IV).
Thus the $n$-th order contribution $S_n(g)$ to $S(g)$ written in the momentum picture has the general form 
\begin{multline}\label{generalS_n(g)}
 {\textstyle\frac{1}{n!}}  \int \ud^4 x_1 \cdots \ud^4 x_n
S_n(x_1, \ldots, x_n) \, g(x_1) \ldots g(x_n) = \\ =
\sum \limits_{0\leq l+m \leq n} \int 
\kappa_{l,m}^{(n)}(\boldsymbol{\p}_1, \ldots, \boldsymbol{\p}_{l+m}; g^{\otimes n}) \,
a_{s_1}(\boldsymbol{\p}_1)^{+} \cdots a_{s_{l+m}}(\boldsymbol{\p}_{l+m}), 
\end{multline}
where $a_{s_i}(\boldsymbol{\p}_i)^{+}, \ldots a_{s_j}(\boldsymbol{\p}_j)$ are the creation-annihilation operators of the free fields underlying the theory in the momentum representation, which respect canonical commutation/anticommutation relations.
Here the distributional kernels $\kappa_{l,m}^{(n)}(\boldsymbol{\p}_1, \ldots, \boldsymbol{\p}_{l+m}; g^{\otimes n}) $ are equal
\begin{multline}\label{kappa_l,m(p1,...p_l+m;g)}
\kappa_{l,m}^{(n)}(\boldsymbol{\p}_1, \ldots, \boldsymbol{\p}_{l+m}; g^{\otimes n}) = \\
= \int \ud^4x_1 \cdots \ud^4 x_n \, 
\kappa_{l,m}^{(n)}(\boldsymbol{\p}_1, \ldots, \boldsymbol{\p}_{l+m}; x_1, \ldots, x_n)
g^{\otimes n}(x_1, \ldots, x_n) \\ = 
 \int \ud^4x_1 \cdots \ud^4 x_n \, 
\kappa_{l,m}^{(n)}(\boldsymbol{\p}_1, \ldots, \boldsymbol{\p}_{l+m}; x_1, \ldots, x_n)
g(x_1) \cdots g(x_n),
\end{multline}
with $\kappa_{l,m}^{(n)}(\boldsymbol{\p}_1, \ldots, \boldsymbol{\p}_{l+m}; x_1, \ldots, x_n)$ 
representing kernels of distributions $\kappa_{l,m}$ belonging to 
\[
E^{* \otimes(l+m)} \otimes \mathscr{E}^{* \otimes n} \cong 
\mathscr{L}(E^{\otimes(l+m)}; \mathscr{E}^{*\otimes n}) \cong
\mathscr{L}(\mathscr{E}^{\otimes n}; E^{* \otimes (l+m)})
\]
which may be regarded as vecor-valued distributions $\kappa_{l,m}$ over the $n$-fold tensor product
\[
\mathscr{E}^{\otimes n}
\] 
of space-time test function space with values in distribution space 
\[
\big(E^{\otimes(l+m)} \big)^* = E^{* \otimes(l+m)}
\]
over the $l+m$-fold tensor product of the restrictions of Fourier transforms of space-time test functions to the respective orbits in momentum space defining the single particle Hilbert spaces of the respectve free fields of the theory. Of course $E$ will thus depend on the particual orbit and the corresponding species of the free field, but we disregard this dependence in notation here in order to simplify the notation. Also there will appear non-trivial analysis in case of light cone orbits corresponding to mass less free fields but we do not enter this point here, because it is not essential here in explaing the general line of our strategy. The importance of these subtleties will however appear at the further stage. 

Using the Wick theorem as in\footnote{Compare also Theorem 0 in \cite{Epstein-Glaser}.} \cite{Bogoliubov_Shirkov}, Chap. III \S 18, and the explicit form of the commutation generalized functions as kernels we can show that (\ref{QEDgeneralS_n(X)andA'_n(X)}) is a kernel of an operator valued distribution in the Wightman sense on the domain $\mathcal{D}_0$, or that the map $g \mapsto S_n(g)$ with $S_n(g)$ equal (\ref{generalS_n(g)}) is an operator valued distribution on the domain $\mathcal{D}_0$
in the Wightman sense. Although 
we shoud emphasize once again that the Wick product is not understood in this ``proof'' in the 
G{\aa}rding-Wightman sense, but rather that the Wick products are reconstructed from their kernels using the explicit form of the commutation generalized functions and the Wick theorem as stated in \cite{Bogoliubov_Shirkov}, Chap. III. Note, please, that the Wick theorem and its proof in 
\cite{Bogoliubov_Shirkov}, Chap. III utilizes the kernels of generalized operators, \emph{i. e.}
free field operators at particular space-time points and the commutation rules for the annihiliation-creation operators at particular points in the momentum representation. This kind of computation is rather intractable within the G{\aa}rding-Wightman approach, in which the ``kernels'' or fields at specified space-time point, so hardly used in \cite{Bogoliubov_Shirkov}, are only symbolic quantities. 
Nonetheless ``proof'' that (\ref{QEDgeneralS_n(X)andA'_n(X)})
is a kernel of a well defined operator-valued distribution on $\mathcal{D}_0$ in the Wightman sense and based on the Wick theorem of \cite{Bogoliubov_Shirkov} seems
to be honest, and concerning only this respect our remarks are rather only pedantic. We only like to emphasize that the computational practice in \cite{Bogoliubov_Shirkov} involves the free fields at particular space-time points explicitly
and the creation-annihilation operators at secific momenta, and ony afterwards with the explicit formulas for operator kernels, the statemets conscerning the analytic properties of Wick products are rather rigorously proved. 

But we should emphasize that we cannot \emph{a priori} insist that the free fields we should understand mathematically precisely in the sense of Wightman and similarily we cannot \emph{a priori} insist that the operator valued functional $g \mapsto S(g)$ we should understand as the operator valued distribution in the sense of Wightman. In fact looking at the effective cross section computation within the causal approach for the physical interaction with $g=1$, as presented e.g. in \cite{Bogoliubov_Shirkov}, Chap IV, \S \S 24, 25, we can even see that $g \mapsto S(g)$ cannot be understood as Wightman operator-valued distribution with $g$ in the Schwartz test space, if we are about to give precise mathematical basis for the computation of the effective cross section. Indeed because $g=1$ does not belong to the Schwartz space, we expect rather that $S(g=1)$ is in general meaningless if $g \mapsto S(g)$ is understood as Wightmann operator valued distribution.  In spite of some partial results, which try to find some sense of the limit $g=1$ of the expression $S(g)$
still understood as distribution in the Wightman sense, we know that for realistic interactions of fields including mass less fields (as QED) the limit $g=1$ cannot be given any sesnible mathematical meaning, although compare some partial results in \cite{epstein-glaser-al}, \cite{BlaSen}, \cite{duch}.   

Before giving the mathematical interpretation of free quantum fields and of the functionals 
$g \mapsto S_n(g)$, $g \mapsto S(g)$, which in our opinion is more adequate, and allows to convert the axioms (I) -- (IV) into 
precise mathematical statements as well the formal Wick theorem of \cite{Bogoliubov_Shirkov}
into a mathematical theorem, which give rigorous mathematical basis for the computation of effecive
cross sections, we stay for a while with the mathematical interpretation of $g \mapsto S(g)$
as the Wightman operator distribution, after Epstein and Glaser \cite{Epstein-Glaser}, with $g$ which cannot be put equal $g=1$ in the formulas for $S_n(g)$ and $S(g)$. We do this in order to remind shortly the ingenious contribution of Epstein and Glaser \cite{Epstein-Glaser}, who were the first which proved that indeed the conditions (I) -- (IV) are sufficient to construct $S_n(x_1, \ldots, x_n)$, provided we do have all $S_k(x_1, \ldots, x_k)$ for $k=1, \ldots, n-1$. In fact they have used (I) -- (IV) in a weaker form with the covariance condition (II) restricted only to the covariance under space-time translations.
We will simplify notation after  Epstein and Glaser \cite{Epstein-Glaser}, and for sets of 
space-time variables $\{x_1, \ldots, x_n\}$, with each $x_k \in \mathcal{M}$ representing a space-time point, we will use capital characters $X$, and respectively
$S(X)$ and $\overline{S}(X)$ for $S_n(x_1, \ldots, x_n)$ and $\overline{S_n}(x_1, \ldots, x_n)$,
with the indices $n$ ommited in  $S(X)$ and $\overline{S}(X)$, as they are understood to be always equal to the number of elemets in $X$. Order in $X$ understood as argument of $S_n$ can be ignored in view of the symmetricity of each $S_n$. So let us assume we have all $S_k(x_1, \ldots, x_k)$ for $k=1, \ldots, n-1$. Epstein and Glaser introduce the following operator-valued distributions 
\[
\begin{split}
A'_{(n)}(x_1, \ldots, x_{n-1}, x_n) = \sum \limits_{P_2} \overline{S}(X)S(Y,x_n), \\
R'_{(n)}(x_1, \ldots, x_{n-1}, x_n) = \sum \limits_{P_2} S(Y,x_n)\overline{S}(X), 
\end{split}
\]
where the sums run over all divisions $P_2$ of the set $\{x_1, \ldots, x_{n-1} \}$
into two disjoint subsets $X$ and $Y$:
\[
\{x_1, \ldots, x_{n-1} \} = X \sqcup Y,
\,\,\, \textrm{with} \,\,\,
X \neq \emptyset.
\] 
Thus by assumption $A'_{(n)}$ and $R'_{(n)}$ are known. Next Epstein and Glaser 
introduce the following operator-valued distributions
\[
\begin{split}
A_{(n)}(x_1, \ldots, x_{n-1}, x_n) = \sum \limits_{P_{2}^{0}} \overline{S}(X)S(Y,x_n)
= \sum \limits_{P_2} \overline{S}(X)S(Y,x_n) + S(x_1, \ldots, x_n), \\
R_{(n)}(x_1, \ldots, x_{n-1}, x_n) = \sum \limits_{P_{2}^{0}} S(Y,x_n)\overline{S}(X)
= \sum \limits_{P_2} S(Y,x_n)\overline{S}(X) + S(x_1, \ldots, x_n), 
\end{split}
\]
where now summation is extended over all divisions $P_{2}^{0}$ of the set 
 $\{x_1, \ldots, x_{n-1} \}$
into two disjoint subsets $X$ and $Y$, which include the empty set $X= \emptyset$. Note that
\[
D_{(n)} = R'_{(n)} - A'_{(n)} = R_{(n)} - A_{(n)}.
\]
In order to finish presentation of the essential point of Epstein-Glaser contribution, let us introduce after \cite{Epstein-Glaser} higher dimensional generalization of the backward and foreward cones:
\[
\Gamma_{\pm}^{(n)}(y) = \big\{X \in \mathcal{M}^n: x_j - y \in \overline{V_{\pm}} \big\}, \,\,\,\,\,
X = \{x_1, \ldots, x_n \}. 
\]
Then it is shown in \cite{Epstein-Glaser} that 
\[
\begin{split}
\textrm{supp} \, R_{(n)}(x_1, \ldots, x_{n-1}, x_n) \subseteq \Gamma_{+}^{(n-1)}(x_n),
\\
\textrm{supp} \, A_{(n)}(x_1, \ldots, x_{n-1}, x_n) \subseteq \Gamma_{-}^{(n-1)}(x_n),
\\
\textrm{supp} \, D_{(n)}(x_1, \ldots, x_{n-1}, x_n) \subseteq \Gamma_{+}^{(n-1)}(x_n)
\sqcup \Gamma_{-}^{(n-1)}(x_n),
\\
\end{split}
\]
and moreover that each $D_{(n)}$ can be (almost) uniquely 
splitted into sum of operator distributions each having the support, respectively, in
$\Gamma_{+}^{(n-1)}(x_n)$ or in $\Gamma_{-}^{(n-1)}(x_n)$ and that this splitting can be made explicitly
and independetly of the conditions (I) -- (IV). The essential point is that $R_{(n)}$ and 
$A_{(n)}$ can be separately computed as the spitting of $D_{(n)}$ into the advaced $A_{(n)}$ and retarded
$R_{(n)}$ parts, so that 
\[
\begin{split}
S_n(x_1, \ldots, x_{n-1}, x_n) = A_{(n)}(x_1, \ldots, x_{n-1}, x_n) - A'_{(n)}(x_1, \ldots, x_{n-1}, x_n)
\\
\,\,\,
\textrm{or equivalently}
\,\,\,
\\
S_n(x_1, \ldots, x_{n-1}, x_n) = R_{(n)}(x_1, \ldots, x_{n-1}, x_n) - R'_{(n)}(x_1, \ldots, x_{n-1}, x_n)
\end{split}
\]
and the inductive step from $n-1$ to $n$ can be computed without encountering any infinities and without any need for renormalization. In practical computations, in case of QED, the distributions $S_n(x_1, \ldots, x_n)$, $R_{(n)}(x_1, \ldots, x_n)$, $A_{(n)}(x_1, \ldots, x_n)$, $R'_{(n)}(x_1, \ldots, x_n)$ and $A'_{(n)}(x_1, \ldots, x_n)$, all have the general form (\ref{QEDgeneralS_n(X)andA'_n(X)}) with the respective translationally invariant and causally supported scalar distribution kernels $t^{k}_{n}$. These distributions, when evaluated at the test function $g^{\otimes n}$, all have the general form 
(\ref{generalS_n(g)}) with the corresponding distributions $\kappa_{l,m}^{(n)}$. 
Therefore the splitting problem is reduced to the splitting of scalar distributions $t^{k}_{n}$, or respectively $\kappa_{l,m}^{(n)}$, closely related to the tensor product of pairing functions of the respective free fields underlying the theory in question.
 That $S_n(x_1, \ldots, x_n)$, $R_{(n)}(x_1, \ldots, x_n)$, $A_{(n)}(x_1, \ldots, x_n)$, $R'_{(n)}(x_1, \ldots, x_n)$ and $A'_{(n)}(x_1, \ldots, x_n)$ are e. g. well defined operator distributions in Wightman sense on $\mathcal{D}_0$, we convince ourselves by utilizing the formal Wick theorem of \cite{Bogoliubov_Shirkov} at the level of operator kernels, e. g. free fields at specified space-time points and the commutation functions (``pairings''), as we have already said. 

Now we are ready to give the basis which allows us to solve the Adiabatic Limit Problem. This is in fact the choice of a concrete mathematical interpretation for the axioms (I) -- (IV). Namely we propose
to use the Hida white noise operators, \cite{hida}, \cite{HKPS}, \cite{obata-book}, as the creation-annihilation operators $a_{s_i}(\boldsymbol{\p}_i)^{+}, \ldots a_{s_j}(\boldsymbol{\p}_j)$ of free fields
in the momentum picture. All the rest of the causal principles (I) -- (IV) we keep totally unchanged. 
This in fact means that we have to construct the free fields of the theory using the creation-annihilation operators which are the Hida operators. Because the Hida operators indeed respect the canonical commutation relations, this mathematical realization fits in naturally into the laws of QFT, and in particular into the axioms (I) -- (IV). This allows us to treat the operators 
$a_{s}(\boldsymbol{\p})^{+}, a_{s}(\boldsymbol{\p})$ at each fixed point $\boldsymbol{\p}$ as a well defined operator transforming continously the test Hida space into its strong dual. Similarily this will allow us to treat the free fields at fixed space-time pont as well defined operators transforming continously the test Hida space into its strong dual. Moreover using Hida operators as creation-annihilation operators will allow us to interprete the free fields of the theory, in case of QED 
the Dirac $\boldsymbol{\psi}$ and electromagnetic potential free field $A$,
\[
\boldsymbol{\psi} = \Xi_{0,1}({}^{1}\kappa_{0,1}) + \Xi_{1,0}({}^{1}\kappa_{1,0}), \,\,\,
A = \Xi_{0,1}({}^{2}\kappa_{0,1}) + \Xi_{1,0}({}^{2}\kappa_{1,0}),
\] 
as integral kernel generalized operators with vector valued kernels (in the sense of Obata \cite{obataJFA}, \cite{obata-book}) 
\[
\begin{split}
{}^{1}\kappa_{0,1}, {}^{1}\kappa_{1,0} \in \mathscr{L}(E, \mathscr{E}^{*}) 
\cong  E^{*} \otimes \mathscr{E}^{*}, \\
{}^{2}\kappa_{0,1}, {}^{2}\kappa_{1,0} \in \mathscr{L}(E, \mathscr{E}^{*}) 
\cong  E^{*} \otimes  \mathscr{E}^{*}, 
\end{split}
\]
with the integral kernels of these distributions, which are equal to ordinary functions:

\[
\begin{split}
{}^{2}\kappa_{0,1}(\nu, \boldsymbol{\p}; \mu, x) =
\frac{\delta_{\nu \mu}}{\sqrt{2 p^0(\boldsymbol{\p})}}
e^{-ip\cdot x}, \,\,\,\,\,\,
p = (|p_0(\boldsymbol{\p})|, \boldsymbol{\p}), \, p\cdot p=0, \\
{}^{2}\kappa_{1,0}(\nu, \boldsymbol{\p}; \mu, x) = 
\frac{\delta_{\nu \mu}}{\sqrt{2 p^0(\boldsymbol{\p})}}
e^{ip\cdot x},
\,\,\,\,\,\,
p \cdot p = 0,
\end{split}
\]
\[ 
{}^{1}\kappa_{0,1}(s, \boldsymbol{\p}; a,x) = \left\{ \begin{array}{ll}
u_{s}^{a}(\boldsymbol{\p})e^{-ip\cdot x} \,\,\, \textrm{with $p = (|p_0(\boldsymbol{\p})|, \boldsymbol{\p}), \, p \cdot p = m^2$} & \textrm{if $s=1,2$}
\\
0 & \textrm{if $s=3,4$}
\end{array} \right.,
\]
\[
\kappa_{1,0}(s, \boldsymbol{\p}; a,x) = \left\{ \begin{array}{ll}
0 & \textrm{if $s=1,2$}
\\
v_{s-2}^{a}(\boldsymbol{\p})e^{ip\cdot x} \,\,\, \textrm{with $p \cdot p = m^2$} & \textrm{if $s=3,4$}
\end{array} \right. 
\]
which are in fact the respective plane wave solutions of d'Alembert and of Dirac equation, which span the corresponding generalized eigen-solution subspaces. All the important operations, which can be performed upon the free field operators, now understood as integral kernel operators, include: 1) differentation of the free field operator, 2) Wick product 
(with spaceitime variables in each factor treated as independent), 3) Wick product of free fields and their differentials at fixed space-time point, 4) splitting of integral kernel operators with causal support into retarded and advanced parts. These operations give another integral kernels operations, which can be realized by the corresponding operations performed upon the corresponding kernels $\kappa_{l,m}$: 1) differentiantion of the kernel, well defined for $\kappa_{l,m}$ being a distribution, 2) projective tensor product of the corresponding kernels, realized by ordinary product of functions representing the kernels with independend variables in each factor, 3) operation of pointwise multiplication (in space-time variables) performed on the functions corresponding to the kernels. 4) Splittinig of the corresponding
kernel distribution with causal support. In particular the formal Wick theorem
of \cite{Bogoliubov_Shirkov}, Chap. III, becomes a mathematical theorem (when using Hida operators), which becomes a particular case of the Theorem 4.5.1 \cite{obata-book}, which in case of Wick theorem simplifies to the case of finite decomposition of a generalized operator into integral kernel operators (with normally ordered creation-annihilation Hida operators). The higher order contributions $S_n(g)$ (\ref{generalS_n(g)}) evaluated at space-time test function $g$
belonging to a natural standard nuclear space $\mathscr{E}$, become finite sums of integral kernel operators, and $S(g)$ becomes to be equal to a Fock expansion into integral kernel operators transforming continously the Hida test space into its strong dual in the sense \cite{obata-book} or \cite{obataJFA}. Moreover the particular higher order contributions to the expansions
of more general scattering operators with additional interaction therms and the higher order contributions to the Bogoliubov-Shirkow functional derivatives of these scattering operators defining interacting fields gain the meaning of a (finite sum of) integral kernel operators with vector valued kernels, and the intearacting fields gain the meaning of generalized operators given by Fock expansions into integral kernel operators. 
The point is that now, when using Hida operators as creation-annihilation operators, each higher order constribution $S_n(g)\in \mathscr{L}((\boldsymbol{E}), (\boldsymbol{E})^*)$, \emph{i.e.} becomes a continous map $(\boldsymbol{E}) \longmapsto (\boldsymbol{E})^*$, for each fixed element $g$ of a natural
nucler test space $\mathscr{E}$ of functions on the space-time, which moreover defines a continous map
$\mathscr{E} \ni g \longmapsto \mathscr{L}((\boldsymbol{E}), (\boldsymbol{E})^*)$, which is sufficient for the computation of the effective coross section
in the adiabatic limit $g=1$, as explained in Introduction.

Moreover the function $g$ in the distributional kernel (\ref{kappa_l,m(p1,...p_l+m;g)}) in the causal pertutbative formula for the interacting fields  
does not have to belong to the Schwartz space or to $\mathscr{E}$, and in particular
the intearcitg field operators gain interpretation of Fock expansions into integral kernel operators with vector-valued kernels 
in the sense of  \cite{obataJFA} even with $g=1$. 

We obtain in this manner theory with the scattering operator function $\mathscr{E} \ni g \longmapsto S(g) \in 
\mathscr{L}((\boldsymbol{E}), (\boldsymbol{E})^*)$ and the particular contributions
$S_n(g)$, $g \in \mathscr{E}$, which are generalized operators transforming continously the Hida space into its strong dual,
and with the interacting quantum field operators with $g=1$ which are generalized operators transforming the test Hida space into its strong dual. This allows us to apply the theory only to the generalized states, e. g. the many particle plane wave states, and to the computation of the effective cross sections, provided we are interested in the high energy processes involving the plane wave states of the elementary free fields of the theory in question. This is sufficient for the computation of the effective cross sections in high energy scattering processes. Problems involving bound, and thus normalizable, states are beyond the scope of this theory, as we have already said in Introduction. 

In the following Sections we construct the free quantum Dirac field and the free quantum electromagnetic potential field with the help of Hida operators as the creation-annihilation operators. Next we prove the statements mentioned above, together with the proof that the interacting Dirac and electromagnetic potential fields are well defined generalized operators transforming continously the Hida space into its strong dual, with the higher order contributions equal to well defined integral kernel operators with vector valued kernels in the sense of Obata \cite{obata-book}. 

Here we only emphasize a novelity in comparison to the free fields in Wightman sense, that 
when constructing free fields with the help of Hida operators the standard nuclear spaces
$E$ which compose the standard Gelfand triples
\[
E \subset \mathcal{H}' \subset E^*
\]
in the single particle Hilbert spaces $\mathcal{H}'$ of the corresponding free fields (Dirac field $\boldsymbol{\psi}$ and $A$ in case of spinor QED) depend on the type of field and for mass less fields are different in comparison to massive fields. $E$ is unitarily equivalent (with the equivalence preserving the nuclear topology) to the space of restrictions $\widetilde{\phi}\big|_{{}_{\mathscr{O}}}$ to the corresponding orbits
\[
\mathscr{O} = \{p: p_0 \geq 0, p \cdot p = 0 \,\, \textrm{or} \,\, = m^2  \},
\]
of the Fourier transforms $\widetilde{\phi}$ of space-time test functions $\phi \in \mathscr{E}$, composing likewise a standard countably Hilbert nuclear space $\mathscr{E}$. It turns out that when constructing mass less fields with the help of Hida operators, the spacetime test space cannot be equal to the Schwartz space of scalar, four-vector, spinor, 
\emph{e. t. c.} functions but need to be chosen differently. When the free field is understood in the sense of Wightman, $\mathscr{E}$ can be put equal to the Schwartz space irrespectively if the field is massive or mass less.

\subsection{Quantum Dirac free field $\boldsymbol{\psi}$ 
as an integral kernel operator with vector-valued distributional kernel 
within the white noise 
construction of Berezin-Hida-Obata}\label{psiBerezin-Hida}

In constructing the quantum free Dirac field $\boldsymbol{\psi}$ according to Berezin-Hida,
we proceed, in a sense, in a totally opposite direction in comparison to Wightman.
Namely Wightman restricts the arguments 
$u\oplus v \in \mathcal{H}'= \mathcal{H}_{m,0}^{\oplus} \oplus \mathcal{H}_{-m,0}^{\ominus c}$ 
of the operators $a'(u\oplus v), a'(u\oplus v)^+$
in (\ref{psi(f)=a_+(f)+a_-(f^c)^+proper}) 
to the nuclear subspace $E \cong \mathcal{S}(\mathbb{R}^3; \mathbb{C}^4)$ 
of all those $u\oplus v$ for which $u$ are equal to 
\[
u = P^\oplus\widetilde{f}|_{{}_{\mathscr{O}_{m,0,0,0}}}, \,\,\
f \in \mathcal{S}(\mathbb{R}^4; \mathbb{C}^4)
\]
and 
\[
v = \big(P^\ominus\widetilde{f}|_{{}_{\mathscr{O}_{-m,0,0,0}}}\big)^c, \,\,\
f \in \mathcal{S}(\mathbb{R}^4; \mathbb{C}^4).
\]
In the following steps he keeps the arguments $u\oplus v$ of the annihilation
and creation operators $a'(u\oplus v), a'(u\oplus v)^+$ within the nuclear space $E$, and with the domain
$\mathcal{D}$ of the oparators $a'(u\oplus v), a'(u\oplus v)^+$ which is not uniquely nor naturally determined.

According to Berezin-Hida we choose quite an opposite direction: we extend the domain of the arguments
$u\oplus v$ of the creation
and annihilation operators $a'(u\oplus v), a'(u\oplus v)^+$ to include also generalized states 
(elements of the strong dual $E^* \cong \mathcal{S}(\mathbb{R}^3; \mathbb{C}^4)^*$ -- tempered distributions)
$u\oplus v$, like the plane wave solutions. This is exactly what is needed (and used but at the formal level) 
in the (formal) proof of the so called 
``Wick theorem'' for free fields, presented in
\cite{Bogoliubov_Shirkov}, Chap. III. By utilizing the rigorous construction of the Hida operators
 $a'(u\oplus v), a'(u\oplus v)^+$ we convert this formal proof into a rigorous one. 

This is achieved in the following manner. First we introduce the nuclear space $E$ as above, which composes
with the single particle Hilbert space 
$\mathcal{H}'= \mathcal{H}_{m,0}^{\oplus} \oplus \mathcal{H}_{-m,0}^{\ominus c}$,
a Gelfand triple
\[
\left. \begin{array}{ccccc}              E         & \subset &  \mathcal{H'} & \subset & E^*        \\
                                                   &         & \parallel      &         &  \\
                        &   & \mathcal{H}_{m,0}^{\oplus} \oplus \mathcal{H}_{-m,0}^{\ominus c} &  &  \end{array}\right..
\]  
We should do it in such a manner which allows lifting of this construction to the second quantized level
with the corresponding Gelfand triple
\[
\left. \begin{array}{ccccc}              (E)         & \subset &  \Gamma(\mathcal{H'}) & \subset & (E)^*        \\
                                                   &         & \parallel      &         &  \\
                        &   & \Gamma\big(\mathcal{H}_{m,0}^{\oplus} \oplus \mathcal{H}_{-m,0}^{\ominus c}\big) &  &  \end{array}\right.,
\]  
with a nuclear (Hida) dense subspace $(E)$ in the Fock space 
$\Gamma(\mathcal{H}')= \Gamma\big(\mathcal{H}_{m,0}^{\oplus} \oplus \mathcal{H}_{-m,0}^{\ominus c}\big)$.
For each $u\oplus v \in E^*$ the annihilation operators $a'(u\oplus v)$ become operators 
continously transforming the nuclear dense space $(E)$ into itself. Because the inclusion
of $(E)$ into the strong dual $(E)^*$ is continuous, the operators $a'(u\oplus v)$ 
can be naturally regarded as continous operators $(E) \rightarrow (E)^*$.
By construction the creation operators $a'(u\oplus v)^+$, $u\oplus v \in E^*$, are equal $\overline{(\cdot)}
\circ a'(u\oplus v)^*  \circ \overline{(\cdot)}$, \emph{i.e.} to the linear duals $a'(u\oplus v)^*$ of the
annihilation operators  $a'(u\oplus v)$ composed with complex conjugation, 
and thus transform continously the strong dual space $(E)^*$ into itself, and can be naturally regarded as continous operators $(E) = (E)^{**} \rightarrow (E)^*$ (because $(E)$ is reflexive).
For  $u\oplus v \in E$ the operators $a'(u\oplus v), a'(u\oplus v)^+$ become operators 
transforming continously the nuclear dense space $(E)$ into itself and thus belong to
$\mathscr{L}\big((E), (E)\big)$. Moreover
the maps
\[
\begin{split}
E \ni u\oplus v \, \longmapsto a'(u\oplus v) \in \mathscr{L}\big((E),(E)\big), \\
E \ni u\oplus v \, \longmapsto a'(u\oplus v)^+ \in \mathscr{L}\big((E),(E)\big),
\end{split}
\]
are continuous when $\mathscr{L}\big((E),(E)\big)$ -- the linear space of linear continuous operators
from $(E)$ into $(E)$ -- is given the natural nuclear topology of 
uniform convergence on bounded sets. 

Therefore it is important to have the Gelfand triple $E \subset \mathcal{H}' \subset E^*$ in the form which 
allows its lifting to the Fock space and the construction of
the Hida test space $(E)$ composing the Gelfand triple $(E) \subset \Gamma(\mathcal{H}') \subset (E)^*$.
This is in particular the case when we have the nuclear space $E \subset \mathcal{H}'$ in the standard form,
\cite{obata-book}.
Namely let $(\mathscr{O}, \ud\mu_{{}_{\mathscr{O}}})$ be a topological space $\mathscr{O}$ with a 
Baire (or Borel) measure $\ud\mu_{{}_{\mathscr{O}}}$. Then we assume that $\mathcal{H'}$ is naturally unitarily $U$
equivalent to the Hilbert 
space of $\mathbb{C}$-valued measurable (equivalence classes
modulo equality almost evereywhere) and square summable functios $L^2(\mathscr{O}, \ud\mu_{{}_{\mathscr{O}}})$.
Next we assume  that $E \subset \mathcal{H}'$ is naturally unitarily equivalent, with the same unitary equivalence $U$ which also defines an isomorphisim of $E$ with the standard countably Hilbert nuclear space
$\mathcal{S}_{A}(\mathscr{O}; \mathbb{C}) \subset L^2(\mathscr{O}, \ud\mu_{{}_{\mathscr{O}}};\mathbb{C})$,
composing a Gelfand triple
\[
\left. \begin{array}{ccccc}     \mathcal{S}_{A}(\mathscr{O}; \mathbb{C})    & \subset &  L^2(\mathscr{O}, \ud\mu_{{}_{\mathscr{O}}}; \mathbb{C}) & \subset & \mathcal{S}_{A}(\mathscr{O}; \mathbb{C})^*        
\end{array}\right.,
\]  
and fulfilling the Kubo-Takenaka conditions.  For standard construction of a nuclear space
$\mathcal{S}_{A}(\mathscr{O}; \mathbb{C}) \subset L^2(\mathscr{O}, \ud\mu_{{}_{\mathscr{O}}};\mathbb{C})$
as arising from a standard (self-adjoint with nuclear or Hilbert Schmidt  $A^{-1}$) operator $A$ on 
$L^2(\mathscr{O}, \ud\mu_{{}_{\mathscr{O}}};\mathbb{C})$, fulfilling Kubo-Takenaka conditions, compare 
\cite{obata-book}, or Subsection 5.1 of \cite{wawrzycki2018}. 

In this situation we have the natural lifting of the Gelfand triple over to the Fock space:
\[
\left. \begin{array}{ccccc}     \big( \mathcal{S}_{A}(\mathscr{O}; \mathbb{C})\big)   & \subset &  \Gamma\big(L^2(\mathscr{O}, \ud\mu_{{}_{\mathscr{O}}}; \mathbb{C})\big) &\subset & \big( \mathcal{S}_{A}(\mathscr{O}; \mathbb{C}) \big)^*        
\end{array}\right.,
\]  
constructed from the standard operator $\Gamma(A)$ in 
$\Gamma\big(L^2(\mathscr{O}, \ud\mu_{{}_{\mathscr{O}}}; \mathbb{C})\big)$. That the operator
$\Gamma(A)$ will be standard whenever $A$ is, also for the fermionic functor $\Gamma$ and under the same
assumptions for $A$ as in the boson case, can be proved in exactly the same way as in \cite{obata-book}, Lemma 3.1.2,
for the bosonic case (the proof is even simpler in fermi case because the occupation numbers assume only the values $0$
or $1$ in this case).  

Eventually we have the initial standard Gelfand triple in the single particle
Hilbert space $\mathcal{H}'$ given in the standard form only up to a unitary isomorphism:
\[
\left. \begin{array}{ccccc}              \mathcal{S}_{A}(\mathscr{O};\mathbb{C})         & \subset &  L^2(\mathscr{O};\mathbb{C}) & \subset & \mathcal{S}_{A}(\mathscr{O}; \mathbb{C})^*        \\
                               \downarrow \uparrow &         & \downarrow \uparrow      &         & \downarrow \uparrow  \\
                                         E         & \subset &  \mathcal{H'} & \subset & E^*        \\
                                                    &         & \parallel     &         &  \\
                         &  & \mathcal{H}_{m,0}^{\oplus} \oplus \mathcal{H}_{-m,0}^{\ominus c} &  & 
\end{array}\right.,
\]  
with the vertical arrows indicating the unitary operator (and its inverse) $U: \mathcal{H}' \rightarrow  L^2(\mathscr{O};\mathbb{C})$ whose restriction to $E$ defines an isomorphism 
$U: E \rightarrow \mathcal{S}_{A}(\mathscr{O}; \mathbb{C})$ of nuclear spaces
and whose linear transposition $U^*$ defines isomorphism 
$\mathcal{S}_{A}(\mathscr{O}; \mathbb{C})^* \rightarrow E^*$. 
The nuclear space $E\subset \mathcal{H}'$ then corresponds to the standard operator
$U^{-1}AU$ on $\mathcal{H}'$, and can be be constructed from it (compare  \cite{obata-book} or 
Subsection 5.1 of \cite{wawrzycki2018}). 

The last Gelfand triples can be lifted to the corresponding Fock spaces together with the corresponding
isomorphisms determined by the unitary operator $\Gamma(U)$: its restriction to $(E) \subset 
\Gamma\big(L^2(\mathscr{O}; \mathbb{C})\big)$ transforming continously 
$(E) \rightarrow \big(\mathcal{S}_{A}(\mathscr{O}; \mathbb{C})\big)$, 
or linear transposition of this restriction, defining the isomorphism 
$(E)^* \rightarrow \big(\mathcal{S}_{A}(\mathscr{O}; \mathbb{C}))^*$: 
\[
\left. \begin{array}{ccccc}              \big(\mathcal{S}_{A}(\mathscr{O}; \mathbb{C})\big)       & \subset &  \Gamma\big(L^2(\mathscr{O};\mathbb{C})\big) & \subset & \big(\mathcal{S}_{A}(\mathscr{O}; \mathbb{C}))^*        \\
                               \downarrow \uparrow &         & \downarrow \uparrow      &         & \downarrow \uparrow  \\
                                         (E)         & \subset &  \Gamma(\mathcal{H'}) & \subset & (E)^*        \\
                                                    &         & \parallel     &         &  \\
                         &  & \Gamma\big(\mathcal{H}_{m,0}^{\oplus} \oplus \mathcal{H}_{-m,0}^{\ominus c}\big) &  & 
\end{array}\right..
\]  
In this case we have the following relations for the annihilation (and correspondingly creation)
operators
\begin{multline}\label{G(U)^+a(U(u+v))G(U)=a'(u+v)}
\Gamma(U)^{+} \, a\big(U^{+-1}(u\oplus v)\big) \, \Gamma(U)=  a'(u\oplus v), \\
\Gamma(U)^{+} \, a\big(U^{+-1}(u\oplus v)\big)^+ \, \Gamma(U)=  a'(u\oplus v)^+, \\
u\oplus v \in E^*.
\end{multline}
Here the Hida operators $a'(u\oplus v), a'(u\oplus v)^+$ coincide with the ordinary annihilation and 
creation operators 
$a'(u\oplus v), a'(u\oplus v)^+$ (defined in Subsection \ref{electron+positron}) on the Hida 
subspace $(E) \subset \Gamma(\mathcal{H'})
\subset (E)^*$ of the Fock space 
$\Gamma(\mathcal{H'}) = \Gamma\big(\mathcal{H}_{m,0}^{\oplus} \oplus \mathcal{H}_{-m,0}^{\ominus c}\big)$, 
whenever $u\oplus v \in E \subset \mathcal{H'}= \mathcal{H}_{m,0}^{\oplus} \oplus \mathcal{H}_{-m,0}^{\ominus c}
\subset E^*$. Similarly $a(w),a(w)^+$ coincide with the standard annihilation and creation operators
on the Hida subspace $\big(\mathcal{S}_{A}(\mathscr{O}; \mathbb{C})\big)$
of the Fock space $\Gamma\big(L^2(\mathscr{O};\mathbb{C})\big)$, whenever 
$w \in \mathcal{S}_{A}(\mathscr{O}; \mathbb{C}) \subset  L^2(\mathscr{O};\mathbb{C}) \subset 
\mathcal{S}_{A}(\mathscr{O}; \mathbb{C})^*$. 
In this case we can restrict the creation and annihilation operators 

$a'(u\oplus v), a'(u\oplus v)^+$ to the Hida subspace $(E)$
and regard them as elements of $\mathscr{L}\big((E),(E)\big)$
(and respectively 
$a(w),a(w)^+ \in \mathscr{L}\big((\mathcal{S}_{A}(\mathscr{O}; \mathbb{C})), (\mathcal{S}_{A}(\mathscr{O}; \mathbb{C}))\big)$) and similarily restrict the linear 
dual composed with complex conjugation $\Gamma(U)^{+} = \overline{(\cdot)} \circ \Gamma(U)^{*} \circ \overline{(\cdot)}
: \big(\mathcal{S}_{A}(\mathscr{O}; \mathbb{C})\big)^* \rightarrow (E)^*$ 
to the subspace $(E)$, where it coincides with the ordinary inverse 
$\Gamma(U)^{-1}$ of the unitary operator $\Gamma(U)$, and with the inverse $U^{+-1} = \overline{(\cdot)} \circ
U^{*-1} \circ\overline{(\cdot)}$ of the linear dual 
$U^*: \mathcal{S}_{A}(\mathscr{O}; \mathbb{C})^* \rightarrow E^*$ to $U$ composed with conjugations
degenerating to $U^{+-1} = U$ on the subspace $E \subset E^*$. 
In this particual case the general formula (\ref{G(U)^+a(U(u+v))G(U)=a'(u+v)}) degenerates to
\begin{multline}\label{G(U)^+a(U(u+v))G(U)=a'(u+v)degenerated}
\Gamma(U)^{-1} \, a\big(U(u\oplus v)\big) \, \Gamma(U)=  a'(u\oplus v), \\
\Gamma(U)^{-1} \, a\big(U(u\oplus v)\big)^+ \, \Gamma(U)=  a'(u\oplus v)^+, \\
u\oplus v \in E \subset E^*.
\end{multline}
But the formula (\ref{G(U)^+a(U(u+v))G(U)=a'(u+v)}) is valid generally for the 
operators $a'(u\oplus v), a'(u\oplus v)^+ \in \mathscr{L}\big((E), (E)^*\big)$, 
\[
a(w), a(w)^+ \in \mathscr{L}\Big(\big(\mathcal{S}_{A}(\mathscr{O}; \mathbb{C})\big), \big(\mathcal{S}_{A}(\mathscr{O}; \mathbb{C})\big)^*\Big),
\]
uderstood in the sense of Hida with $u\oplus v \in E^*$, or
respectively $w \in \mathcal{S}_{A}(\mathscr{O}; \mathbb{C})^*$, and with $\Gamma(U)$
undestood as a continous isomorhism 
\[
(E) \, \longrightarrow \, \big(\mathcal{S}_{A}(\mathscr{O}; \mathbb{C})\big)
\]
of nuclear spaces in the first formula of (\ref{G(U)^+a(U(u+v))G(U)=a'(u+v)})
and with $\Gamma(U)^+ = \overline{(\cdot)} \circ \Gamma(U)^* \circ \overline{(\cdot)}$ 
as its continous dual isomorhism 
\[
\big(\mathcal{S}_{A}(\mathscr{O}; \mathbb{C})\big)^* \, \longrightarrow \, (E)^*
\]
composed with complex conjugation in 
(\ref{G(U)^+a(U(u+v))G(U)=a'(u+v)}). Below we give generalized operators
$a'(u\oplus v), a'(u\oplus v)^+$ (and respectively $a(w),a(w)^+$), due to Hida, which make sense
also for $u\oplus v$ (respectively $w$), lying in the space dual to $E$, respectively
dual to $\mathcal{S}_{A}(\mathscr{O}; \mathbb{C})$.

In order to simplify notation we agree to write the last isomorphisms 
(\ref{G(U)^+a(U(u+v))G(U)=a'(u+v)}) (and their particular case 
(\ref{G(U)^+a(U(u+v))G(U)=a'(u+v)degenerated})) induced by $U$
simply idetifying the corresponding operators, namely
\begin{multline}\label{a(U(u+v))=a'(u+v)}
\begin{split}
a\big(U^{+-1}(u\oplus v)\big) = a'(u\oplus v), \,\,\,
a\big(U^{+-1}(u\oplus v)\big)^+ = a'(u\oplus v)^+, \,\,\,
u\oplus v \in E^*, \\ 
a\big(U(u\oplus v)\big) = a'(u\oplus v), \,\,\,
a\big(U(u\oplus v)\big)^+ = a'(u\oplus v)^+, \,\,\,
u\oplus v \in E \subset E^*,
\end{split}
\end{multline}
as operators transforming continously Hida spaces into their strong duals
(in the first case) or as operators transforming continously Hida spaces
into Hida spaces (in the second case). 

Note that in our case the initial Gelfand triple
$E  \subset \mathcal{H}_{m,0}^{\oplus} \oplus \mathcal{H}_{-m,0}^{\ominus c}
\subset E^*$ over the single particle Hilbert space  $\mathcal{H'} = \mathcal{H}_{m,0}^{\oplus} \oplus \mathcal{H}_{-m,0}^{\ominus c}$ does not have the standard form, because the single particle Hilbert space $\mathcal{H}'$
does not have the form $L^2(\mathscr{O}, \ud\mu_{{}_{\mathscr{O}}};\mathbb{C})$.
Indeed note that the Hilbert space 
\begin{multline*}
L^2(\mathbb{R}^3, \ud^3 \boldsymbol{\p}/(2p_0(\boldsymbol{\p}))^2; \mathbb{C}^4)
= \oplus_{1}^{4} L^2(\mathbb{R}^3, \ud^3 \boldsymbol{\p}/(2p_0(\boldsymbol{\p}))^2; \mathbb{C}) \\
= L^2(\mathbb{R}^3 \sqcup \mathbb{R}^3 \sqcup \mathbb{R}^3 \sqcup \mathbb{R}^3, \ud^3 \boldsymbol{\p}/(2p_0(\boldsymbol{\p}))^2; \mathbb{C})
\end{multline*}
does have the required form $L^2(\mathscr{O}, \ud\mu_{{}_{\mathscr{O}}};\mathbb{C})$,
with 
\[
\mathscr{O} = \mathbb{R}^3 \sqcup \mathbb{R}^3 \sqcup \mathbb{R}^3 \sqcup \mathbb{R}^3
\]
equal to the disjoint sum of four copies of $\mathbb{R}^3$ and the direct sum measure 
$\ud\mu_{{}_{\mathscr{O}}}$ coinciding with $\frac{\ud^3 \boldsymbol{\p}}{(2p_0(\boldsymbol{\p}))^2}$
on each copy $\mathbb{R}^3$. But recall that although in our case the values $\widetilde{\phi}(p)$ of
the bispinors $\widetilde{\phi} \in \mathcal{H}_{m,0}^{\oplus}$ concentrated on the positive energy orbit
$\mathscr{O}_{m,0,0,0}$ range over $\mathbb{C}^4$, nonetheless $\mathcal{H}_{m,0}^{\oplus}$ does not have the standard form
\[
L^2(\mathbb{R}^3, \ud^3 \boldsymbol{\p}/(2p_0(\boldsymbol{\p}))^2; \mathbb{C}^4),
\]
because for each fixed $\boldsymbol{p}$ the vectors $\widetilde{\phi}(\boldsymbol{\p}, p_0(\boldsymbol{\p}))$,
with $\widetilde{\phi}$ ranging over $\mathcal{H}_{m,0}^{\oplus}$, 
do not span $\mathbb{C}^4$, but are equal to the image 
$\textrm{Im} \, P^\oplus(\boldsymbol{\p}, p_0(\boldsymbol{\p})) \neq \mathbb{C}^4$, 
for $p = (\boldsymbol{\p}, p_0(\boldsymbol{\p})) \in \mathscr{O}_{m,0,0,0}$,
because $\textrm{rank} \, P^\oplus(\boldsymbol{\p}, p_0(\boldsymbol{\p})) = 2 \neq 4$ 
(compare Subsection 2.1 of \cite{wawrzycki2018}, where the projection operator $P^\oplus$ of point-vise multiplication by
$P^\oplus(p)$, $p \in \mathscr{O}_{m,0,0,0}$, acting on bispinors concentrated on the orbit 
$\mathscr{O}_{m,0,0,0}$ is defined). 

Similarily $\mathcal{H}_{-m,0}^{\ominus c}$ does not have the standard form
\[
L^2(\mathbb{R}^3, \ud^3 \boldsymbol{\p}/(2p_0(\boldsymbol{\p}))^2; \mathbb{C}^4)
\]
in spite of the fact that the conjugations $\widetilde{\phi}^c \in \mathcal{H}_{-m,0}^{\ominus c}$ of the 
bispinors $\widetilde{\phi} \in \mathcal{H}_{-m,0}^{\ominus}$ concetrated on the negative energy orbit 
$\mathscr{O}_{-m,0,0,0}$ take their values in $\mathbb{C}^4$, 
because $\{\widetilde{\phi}(\boldsymbol{\p}, p_0(\boldsymbol{\p})), \widetilde{\phi} \in \mathcal{H}_{-m,0}^{\ominus}\} = \textrm{Im} \, P^\ominus(\boldsymbol{\p}, p_0(\boldsymbol{\p})) \neq \mathbb{C}^4$ with 
$\textrm{rank} \, P^\ominus(\boldsymbol{\p}, p_0(\boldsymbol{\p})) = 2 \neq 4$, for 
$p = (\boldsymbol{\p}, p_0(\boldsymbol{\p})) \in \mathscr{O}_{-m,0,0,0}$. 

But there exists a natural unitary isomorphism $U$ (in fact a class of such natural $U$)
\[
U: \,\, \mathcal{H'} = \mathcal{H}_{m,0}^{\oplus} \oplus \mathcal{H}_{-m,0}^{\ominus c} \,\,\,
\longrightarrow \,\,\,
L^2(\mathbb{R}^3, \ud^3 \boldsymbol{\p}; \mathbb{C}^4) 
\]
between the single particle Hilbert space $\mathcal{H}'$ and the Hilbert space 
\[
L^2(\mathbb{R}^3, \ud^3 \boldsymbol{\p}; \mathbb{C}^4)  = \oplus L^2(\mathbb{R}^3, \ud^3 \boldsymbol{\p}; \mathbb{C})
=L^2\big(\mathbb{R}^3 \sqcup \mathbb{R}^3 \sqcup \mathbb{R}^3 \sqcup \mathbb{R}^3, \ud^3 \boldsymbol{\p};
\mathbb{C}\big),
\]
which moreover restricts to an isomorphism between the nuclear spaces 
of Schwartz bispinors in $E  \subset \mathcal{H}'$
and Schwartz functions in $\mathcal{S}(\mathbb{R}^3; \mathbb{C}^4)
=\mathcal{S}_{A}(\mathbb{R}^3; \mathbb{C}^4) \subset L^2(\mathbb{R}^3, \ud^3 \boldsymbol{\p}; \mathbb{C}^4)$. 

Indeed for $\widetilde{\phi} \in \mathcal{H}_{m,0}^{\oplus}$, 
$\widetilde{\phi}' \in \mathcal{H}_{-m,0}^{\ominus}$ we put
\begin{multline}\label{isomorphismU}
U\Big( \widetilde{\phi} \oplus (\widetilde{\phi}')^c\Big) \overset{\textrm{df}}{=}
(\widetilde{\phi})_{1+} \oplus (\widetilde{\phi})_{2+} \oplus (\widetilde{\phi}')_{1-}
\oplus (\widetilde{\phi}')_{2-} \\ =
(\widetilde{\phi})_{1} \oplus (\widetilde{\phi})_{2} \oplus (\widetilde{\phi}')_{3}
\oplus (\widetilde{\phi}')_{4} \,\,\, \in \oplus_{1}^{4} L^2(\mathbb{R}^3; \mathbb{C}) = 
L^2(\mathbb{R}^3; \mathbb{C}^4),
\end{multline}
where
\[
\begin{split}
(\widetilde{\phi})_{1}(\boldsymbol{\p}) = (\widetilde{\phi})_{1+}(\boldsymbol{\p}) \overset{\textrm{df}}{=}
\frac{1}{2p_0(\boldsymbol{\p})} u_1(\boldsymbol{\p})^+ \widetilde{\phi}(p_0(\boldsymbol{\p}), \boldsymbol{\p}), 
 \,\,\,\,\, p_0(\boldsymbol{\p}) = \sqrt{|\boldsymbol{\p}|^2 + m^2}, \\
(\widetilde{\phi})_{2}(\boldsymbol{\p}) = (\widetilde{\phi})_{2+}(\boldsymbol{\p}) \overset{\textrm{df}}{=}
\frac{1}{2p_0(\boldsymbol{\p})} u_2(\boldsymbol{\p})^+ \widetilde{\phi}(p_0(\boldsymbol{\p}), \boldsymbol{\p}), 
 \,\,\,\,\, p_0(\boldsymbol{\p}) = \sqrt{|\boldsymbol{\p}|^2 + m^2},
\end{split}
\]
and
\begin{multline*}
(\widetilde{\phi}')_{3}(\boldsymbol{\p}) = (\widetilde{\phi}')_{1-}(\boldsymbol{\p}) \overset{\textrm{df}}{=}
\overline{\frac{1}{2|p_0(\boldsymbol{\p})|} v_1(\boldsymbol{\p})^+ 
\widetilde{\phi}'(-|p_0(\boldsymbol{\p})|, -\boldsymbol{\p})} \\ =
\overline{\frac{1}{2|p_0(\boldsymbol{\p})|} v_1(\boldsymbol{\p})^+ 
\big(\overline{(\widetilde{\phi}')^c(|p_0(\boldsymbol{\p})|, \boldsymbol{\p})}\big)^T}, \\
\,\,\,\,\, p_0(\boldsymbol{\p}) = - \sqrt{|\boldsymbol{\p}|^2 + m^2}, 
\end{multline*}
\begin{multline*}
(\widetilde{\phi}')_{4}(\boldsymbol{\p}) =
(\widetilde{\phi}')_{2-}(\boldsymbol{\p}) \overset{\textrm{df}}{=}
\overline{\frac{1}{2|p_0(\boldsymbol{\p})|} v_2(\boldsymbol{\p})^+ 
\widetilde{\phi}'(-|p_0(\boldsymbol{\p})|, -\boldsymbol{\p})} \\ =
\overline{\frac{1}{2|p_0(\boldsymbol{\p})|} v_2(\boldsymbol{\p})^+ 
\big(\overline{(\widetilde{\phi}')^c(|p_0(\boldsymbol{\p})|, \boldsymbol{\p})}\big)^T}, \\
\,\,\,\,\, p_0(\boldsymbol{\p}) = - \sqrt{|\boldsymbol{\p}|^2 + m^2}
\end{multline*}
Here $u_s(\boldsymbol{\p}), v_s(-\boldsymbol{\p})$, $s=1,2$, are the Fourier transforms
of the complete system of solutions of the Dirac equation, given by the formula (\ref{chiral,u,v})
of Appendix \ref{fundamental,u,v} in the so-called chiral representation of Dirac gamma
matrices (which we have used in Subsection 2.1 of \cite{wawrzycki2018}); or by the formula (\ref{standard,u,v})
of Appendix \ref{fundamental,u,v} in the so-called standard representation of the Dirac gamma
matrices. It follows that for any $(\widetilde{\phi})_{1}=(\widetilde{\phi})_{1+}, 
(\widetilde{\phi})_{2} = (\widetilde{\phi})_{2+}, (\widetilde{\phi}')_{3} = (\widetilde{\phi}')_{1-},
(\widetilde{\phi}')_{4} = (\widetilde{\phi}')_{2-} \in L^2(\mathbb{R}^3; \mathbb{C})$ we have
\begin{equation}\label{isomorphismU^-1}
U^{-1} \Big( (\widetilde{\phi})_{1+} \oplus (\widetilde{\phi})_{2+} \oplus (\widetilde{\phi}')_{1-}
\oplus (\widetilde{\phi}')_{2-}\Big) \overset{\textrm{df}}{=}
 \widetilde{\phi} \oplus (\widetilde{\phi}')^c \,\,\, \in \mathcal{H}_{m,0}^{\oplus} \oplus \mathcal{H}_{-m,0}^{\ominus c},
\end{equation}
where 
\[
\widetilde{\phi}(p_0(\boldsymbol{\p}), \boldsymbol{\p}) \overset{\textrm{df}}{=}
\sum_{s=1,2} 2p_0(\boldsymbol{\p}) \, (\widetilde{\phi})_{s+} (\boldsymbol{\p}) \, u_s(\boldsymbol{\p}),
\,\,\,\,\, p_0(\boldsymbol{\p}) = \sqrt{|\boldsymbol{\p}|^2 + m^2} \\
\]
and
\begin{multline*}
\big((\widetilde{\phi}')^c(|p_0(\boldsymbol{\p})|, \boldsymbol{\p})\big)^T =
\overline{\widetilde{\phi}'(-|p_0(\boldsymbol{\p})|, -\boldsymbol{\p})} 
\overset{\textrm{df}}{=}
\sum_{s=1,2} 2|p_0(\boldsymbol{\p})| \, (\widetilde{\phi}')_{s-} (\boldsymbol{\p}) \, 
\overline{v_s(\boldsymbol{\p})}, \\
p_0(\boldsymbol{\p}) = - \sqrt{|\boldsymbol{\p}|^2 + m^2}.
\end{multline*}
That $U^{-1}$ is indeed equal to the inverse of the operator $U$ follows immediately from the 
relations (\ref{E_+Phi=Phi}) for $\widetilde{\phi} \in \mathcal{H}_{m,0}^{\oplus}$
and from the relations (\ref{E_-Phi=Phi}) for $\widetilde{\phi}' \in \mathcal{H}_{-m,0}^{\ominus}$
of Appendix \ref{fundamental,u,v}. That $U^{-1}$ is isometric follows immediatelly from the orthonormality
relations (\ref{u^+u=delta}) for $u_s(\boldsymbol{\p}), v_s(\boldsymbol{\p})$, $s=1,2$.
That $U$ is isometric follows immediately from the 
relations (\ref{E_+Phi=Phi}) for $\widetilde{\phi} \in \mathcal{H}_{m,0}^{\oplus}$
and from the relations (\ref{E_-Phi=Phi}) for $\widetilde{\phi}' \in \mathcal{H}_{-m,0}^{\ominus}$
of Appendix \ref{fundamental,u,v}.
That $U$ transforms isomorphically the indicated nuclear spaces follows from the fact
that the components of  $u_s(\boldsymbol{\p}), v_s(\boldsymbol{\p})$, $s=1,2$, are all multilpliers
of the Schwartz algebra $\mathcal{S}(\mathbb{R}^3; \mathbb{C})$. 

Note here that there are more than just one canonical choice of the solutions
$u_s(\boldsymbol{\p}), v_s(\boldsymbol{-\p})$, $s=1,2$, with smooth components
belonging to the algebra of multipliers or even convolutors of $\mathcal{S}(\mathbb{R}^3; \mathbb{C})$.
Indeed having given one choice $u_s(\boldsymbol{\p}), v_s(\boldsymbol{-\p})$, $s=1,2$,
we can apply the unitary operator to $u_s(\boldsymbol{\p}), v_s(\boldsymbol{-\p})$, $s=1,2$, 
 of multiplication by a unitary matrix with components smoothly
depending on $\boldsymbol{\p}$ and belonging to the algebra of multipliers of  $\mathcal{S}(\mathbb{R}^3; \mathbb{C})$,
and which rotates the initial $u_s(\boldsymbol{\p}), v_s(\boldsymbol{-\p})$, $s=1,2$,
within the $2$-dimentional images respectively of $P^{\oplus}(p_0(\boldsymbol{\p}), \boldsymbol{\p})$
or $P^{\ominus}(-|p_0(\boldsymbol{\p})|, \boldsymbol{\p})$. We obtain in this way
various isomorphisms $U$ and the corresponding unitary equivalent realizations of the Dirac field.  

Recall, please, that the nuclear Schwartz space $\mathcal{S}(\mathbb{R}^3; \mathbb{C}^4)$
can be obtained as a standard countably Hilbert nuclear space 
\[
\mathcal{S}(\mathbb{R}^3; \mathbb{C}^4)
=\mathcal{S}_{A}(\mathbb{R}^3; \mathbb{C}^4) \subset L^2(\mathbb{R}^3, \ud^3 \boldsymbol{\p}; \mathbb{C}^4)
= \oplus_{1}^{4}  L^2(\mathbb{R}^3, \ud^3 \boldsymbol{\p}; \mathbb{C})
\]
with the standard operator $A$ on
\[
 L^2(\mathbb{R}^3, \ud^3 \boldsymbol{\p}; \mathbb{C}^4) 
= \oplus_{1}^{4}  L^2(\mathbb{R}^3, \ud^3 \boldsymbol{\p}; \mathbb{C})
\]
equal to the direct sum 
\begin{equation}\label{AinL^2(R^3;C^4)}
A = \oplus H_{(3)}
\end{equation}
of four copies of the three dimensional oscillator hamiltonian operator 
\[
H_{(3)} = - \Delta_{{}_{\boldsymbol{\p}}} +  \boldsymbol{\p} \cdot \boldsymbol{\p} + 1
\] 
on
\[
L^2(\mathbb{R}^3, \ud^3 \boldsymbol{\p}; \mathbb{C}),
\]
compare e.g. \cite{Hida1}, Appendix 9 of \cite{wawrzycki2018}, or \cite{Simon}.

Summing up we will construct the Gelfand triples
\begin{equation}\label{SinglePartGelfandTriplesForPsi}
\left. \begin{array}{ccccc}   & & L^2(\mathbb{R}^3 \sqcup \mathbb{R}^3 \sqcup \mathbb{R}^3 \sqcup \mathbb{R}^3, \ud^3 \boldsymbol{\p}; \mathbb{C}) & & \\
 & & \parallel & & \\
           \mathcal{S}_{A}(\mathbb{R}^3; \mathbb{C}^4)         & \subset & \oplus L^2(\mathbb{R}^3; \mathbb{C}) & \subset & \mathcal{S}_{A}(\mathbb{R}^3; \mathbb{C}^4)^*        \\
                               \downarrow \uparrow &         & \downarrow \uparrow      &         & \downarrow \uparrow  \\
                                         E         & \subset &  \mathcal{H'} & \subset & E^*        \\
                                                    &         & \parallel     &         &  \\
                         &  & \mathcal{H}_{m,0}^{\oplus} \oplus \mathcal{H}_{-m,0}^{\ominus c} &  & 
\end{array}\right.,
\end{equation}
related by vertical isomorhisms induced by the unitary operator (\ref{isomorphismU})
\[
U: \mathcal{H'} = \mathcal{H}_{m,0}^{\oplus} \oplus \mathcal{H}_{-m,0}^{\ominus c} 
\longrightarrow  \oplus L^2(\mathbb{R}^3; \mathbb{C})
\] 
with restriction to the nuclear space $E$ mapping isomorphically
\[
E \longrightarrow \mathcal{S}_{A}(\mathbb{R}^3; \mathbb{C}^4) = \mathcal{S}(\mathbb{R}^3; \mathbb{C}^4)
\]
with $A$ defined by (\ref{AinL^2(R^3;C^4)}). The first triple has the standard form, 
and can be lifted with the help of $\Gamma(A)$.  Thus we may define in the standard form the
Hida operators $a(w), a(w)^+$ in the Fock space $\Gamma\big(\oplus L^2(\mathbb{R}^3; \mathbb{C})\big)$.
The corresponding Hida operators $a'(u\oplus v),a'(u\oplus v)^+$ in the Fock space
$\Gamma(\mathcal{H}')$ of the free Dirac field need not be separately constructed, and 
can be expressed with the help of the standard Hida operators $a(w), a(w)^+$ in the Fock space 
$\Gamma\big(\oplus L^2(\mathbb{R}^3; \mathbb{C})\big)$, by utilizing the isomorphism induced by $U$.
Namely Hida operators $a'(u\oplus v),a'(u\oplus v)^+$ can be expressed by the Hida operators
$a(w), a(w)^+$ as in the formula (\ref{a(U(u+v))=a'(u+v)}), namely:
\[
\begin{split}
a\big(U^{+-1}(u\oplus v)\big) = a'(u\oplus v), \,\,\,
a\big(U^{+-1}(u\oplus v)\big)^+ = a'(u\oplus v)^+, \,\,\,
u\oplus v \in E^*, \\ 
a\big(U(u\oplus v)\big) = a'(u\oplus v), \,\,\,
a\big(U(u\oplus v)\big)^+ = a'(u\oplus v)^+, \,\,\,
u\oplus v \in E \subset E^*.
\end{split}
\]

The plan of the rest part of this Subsection is the following.
First, we give the white noise constrution of the
Hida operators $a(w), a(w)^+$ obtained by lifting to the Fock space of the first (standard)
Gelfand triple in (\ref{SinglePartGelfandTriplesForPsi}). In the next step we utilize
the natural unitary isomorphism $U$ given by (\ref{isomorphismU}),
which induces the isomorphism of the Gelfand triples in (\ref{SinglePartGelfandTriplesForPsi}).
Namely, using the unitary isomorphism $U$ and the Hida operators 
$a(w), a(w)^+$ corresponding to the lifting of the first triple in 
(\ref{SinglePartGelfandTriplesForPsi}) we compute the 
Hida operators $a'(u\oplus v),a'(u\oplus v)^+$ in the Fock space
$\Gamma(\mathcal{H}')$ (which enter into the Dirac field (\ref{psi(f)=a_+(f)+a_-(f^c)^+proper})),
using the formula (\ref{a(U(u+v))=a'(u+v)}).

Let us concetrate now on the first (standard) of the Gelfand triples 
in (\ref{SinglePartGelfandTriplesForPsi}) and its lifting to the Fock space
$\Gamma\big(\oplus L^2(\mathbb{R}^3; \mathbb{C})\big)$, together with the Hida
definition of the Hida operators 
$a(w), a(w)^+$, $w \in \mathcal{S}_{A}(\mathbb{R}^3)^* = \mathcal{S}(\mathbb{R}^3)^*$.
We only recall definition and some basic facts, reffering e.g. to \cite{obata-book}, \cite{HKPS},
\cite{obataJFA}, \cite{Shimada}, for more information. 

We are using here the \emph{modified realization of annihilation-creation operators} in the Fock space,
defined in the Remark \ref{TwoRepOfaa^+InFermiFock} of Subsection \ref{electron}.
It fits well with that used by Hida, Obata, Sait\^o, \cite{hida}, \cite{obata-book}, \cite{obataJFA},
for boson case, when adopting the results of \cite{hida}, \cite{obata-book}, \cite{obataJFA}, concerning integral kernel 
operators, to fermion case.

\begin{rem}\label{TretmentFermiIntegKerOp=TretmentBoseIntegKerOp}
It should be emphasized here that the results of 
\cite{hida}, \cite{obata-book}, \cite{obataJFA}, concerning the so called integral kernel
operators and their Fock expansions, can be proved without any essental changes
also for the fermi case
after \cite{hida}, \cite{obata-book}, \cite{obataJFA}. Note that these
theorems (e.g. Lemma 2.2, Thm. 2.2, Thm. 2.6. of \cite{hida}, or Thm. 3.13 of \cite{obataJFA})
could have been formulated and proved as well for the so called general Fock space 
\[
\Gamma_\textrm{general}(\mathcal{H}) = \bigoplus_{n=0}^{\infty} \mathcal{H}^{\otimes n}
\] 
without symmetrizing or antisymmetrizing the tensor products. In particular symmertization
(antisymmetrization) plays no fundamental role in the proof of these theorems, which are based on
the norm estimations of the $m$-contractions $\otimes_m, \otimes^m$. Their eventual symmetrizations 
$\widehat{\otimes}_m, \widehat{\otimes}^m$
(or antisymmetrizations), which arise in the latter stage when restricting attention to the boson (or fermion) 
case, has nothing to do with these estimations and allows to state the analogous results for 
boson as well as for the fermion case.

Although differences between the fermi and bose case which arise have nothing to do with the analysis of 
integral kernel operators (in which we are mostly interested), 
we should mention here some of them. The fundamental difference is that the algebra structure of the 
nuclear Hida test space, determined by the tensor product, is not commutative but skew commutative, due to the atisymmetricity of the tensors in the fermi Fock space, and cannot be naturally realized as a nuclear function 
space on the strong dual $E^*$ with multiplication defined by point wise multiplication 
(because such multiplication is always commutative). In connection with this we have no natural isomorphism
of the Fermi Fock space to the space of square integrable functions on $E^*$ with the Gaussian measure
on $E^*$ (no Wiener-It\^o-Segal decomposition based on commutative infinite-dimensional measure space is possible).
Of course a mere existence of a unitary map between the fermi Fock space and an $L^2$ space over a Gaussian
measure space is trivial, but there are plenty of such maps devoid of any relevance. \emph{Naturality} of the 
Wiener-It\^o-Segal decomposition for the bose case is crucial. In order to keep a \emph{natural nature}, e.g. preserving
the algebra structure of the Hida test space (now skew commutative), in extending Wiener-It\^o-Segal decomposition
to the fermi case, a non-commutative extension
of abstract integration is needed, and has been provided by Segal (note however that Segal 
\cite{Segal-TnsorAlg-II} is not using 
a non-commutative extension of ordinary measure -- but of a weak distribution on a Hilbert space).
Because these questions concerning non commutative character of the multiplicative structure of the 
Hida test space in case of fermi
case are not immediately related to the calculus of Fock expansions of integral kernel operators, 
developed in \cite{hida}, \cite{obata-book}, \cite{obataJFA}, we do not enter these questions
in our work. In particular we do not exploit in any susbstantial manner the fact that Hida annihilation operators can be interpreted as graded derivations on the $\mathbb{Z}_2$ graded skew commutative nuclear algebra of Hida test functionals.
The only practical consequence of this fact we feel in computations concerning integral kernel operators is that we 
confine ourselves to skew-symmetric kernels (in variables corresponding to fermi 
Hida creation-annihilation operators) in order to keep one-to-one correspondence between the 
kernels and corresponding operators.

But there is a relevant tool for computations which must be treated in slightly different manner in the two cases -- bose and fermi case. Namely the \emph{symbol calculus}, initiated by Berezin \cite{Berezin} and developed mainly 
by Obata \cite{obata}, \cite{obataJFA}, must be realized in a slightly different manner for fermi case in 
comparison with the bose case.  It order to adopt the symbol calculus of Obata
to the fermi case it is convenient first to divide the fermi fock space $\Gamma(\mathcal{H}')$
into the subspaces $\Gamma_+(\mathcal{H}')$ of even elements 
\[
\Phi = \sum \limits_{n=0}^{\infty} \Phi_n, 
\]
(with even $n$ in this decomposition), and $\Gamma_-(\mathcal{H}')$ of odd
elements $\Phi$ (with $n$ odd in this decomposition).
Similarily we do for the nuclear spaces 
$(E) = (E)_+ \oplus (E)_-, (E)^* = (E)_{+}^{*} \oplus (E)_{-}^{*}$. Next we note that
for $\xi \in E^{\widehat{\otimes} \, 2}$ (and generally $\xi \in E^{\widehat{\otimes} m}$
with even $m$) the exponetial map 
\[
\xi \mapsto \Phi_\xi = \sum \limits_{n=0}^\infty \frac{1}{(2n)!} \xi^{\widehat{\otimes} \, n}
\in (E)_+
\] 
is well defined and continuous. Using this exponential map we utilize the Obata symbol 
for even operators, 
i.e. transforming $(E)_+ \rightarrow (E)_{+}^{*}$
and $(E)_- \rightarrow (E)_{-}^{*}$. The odd operators, i.e. transforming 
$(E)_+ \rightarrow (E)_{-}^{*}$ and $(E)_- \rightarrow (E)_{+}^{*}$ are reduced to even by 
muliplication by one Hida (creation, respectively annihilation) operator.
Finally we note that any continuous operator $(E) \rightarrow (E)^*$ is naturally a 
direct sum of an even and an odd operator; compare \cite{Shimada}.
\end{rem}

Let $|\cdot |_0$, $(\cdot, \cdot)_0$ denote the standard $L^2$ norm and inner product on 
\[
 L^2(\mathbb{R}^3, \ud^3 \boldsymbol{\p}; \mathbb{C}^4) 
= \oplus_{1}^{4}  L^2(\mathbb{R}^3, \ud^3 \boldsymbol{\p}; \mathbb{C})
\]
and by the same symbol $|\cdot|_0$, after \cite{hida} and \cite{obata-book},  
we denote the Hilbert space norm on the Hilbert space tensor
product
\[
L^2(\mathbb{R}^3, \ud^3 \boldsymbol{\p}; \mathbb{C}^4)^{\otimes  n},
\]
as well as its restriction to the antisymmetrized tensor product
\[
L^2(\mathbb{R}^3, \ud^3 \boldsymbol{\p}; \mathbb{C}^4)^{\widehat{\otimes} \,  n}.
\]
Recall that  
\[
|f|_k = |(A^{\otimes n})^k f|_0 \,\,\,\,\,
f \in \Dom \, (A^{\otimes n})^k \subset L^2(\mathbb{R}^3, \ud^3 \boldsymbol{\p}; \mathbb{C}^4)^{\otimes  n}
\]
(in particular well defined for $f \in \mathcal{S}_{A}(\mathbb{R}^3; \mathbb{C}^4)^{\widehat{\otimes} \, n}$).

Let $\|\cdot\|_0$, $((\cdot, \cdot))_0$ denote the Hilbert space norm and the corresponding inner product  
on Fock space defined by the formula (convetion used by \cite{hida}, \cite{obataJFA}, compare Remark \ref{TwoRepOfaa^+InFermiFock} of Subsection \ref{electron})
\[
\| \Phi\|_{0}^2 = \sum \limits_{n=0}^{\infty} \, n! \, |\Phi_n |_{0}^2 
\]
for $\Phi$ with decomposition
\[
\Phi = \sum \limits_{n=0}^{\infty} \Phi_n, \,\,\, \textrm{with} \,\,\,
\Phi_n \in L^2(\mathbb{R}^3, \ud^3 \boldsymbol{\p}; \mathbb{C}^4)^{\widehat{\otimes} \,  n}.
\]
Recall that by definition
\[
\|\Phi\|_k = \|\Gamma(A)^k \Phi \|_0 \,\,\, \textrm{and} \,\,\,
|\Phi_n|_k = |(A^{\otimes n})^k \Phi_n|_0
\]
for $\Phi \in \Gamma\big(L^2(\mathbb{R}^3, \ud^3 \boldsymbol{\p}; \mathbb{C}^4) \big)$ 
and $\Phi_n \in L^2(\mathbb{R}^3, \ud^3 \boldsymbol{\p}; \mathbb{C}^4)^{\widehat{\otimes} \,  n}$.

It follows in particular that the general element
\begin{equation}\label{HidaPhi}
\Phi = \sum \limits_{n=0}^{\infty} \Phi_n, \,\,\, \textrm{with} \,\,\,
\| \Phi\|_{0}^2 = \sum \limits_{n=0}^{\infty} \, n! \, |\Phi_n |_{0}^2  < \infty,
\end{equation}
of the Fock space 
$\Gamma\big(L^2(\mathbb{R}^3, \ud^3 \boldsymbol{\p}; \mathbb{C}^4) \big)$
belongs to the Hida test space $\big(\mathcal{S}_{A}(\mathbb{R}^3; \mathbb{C}^4)\big)
\subset \Gamma\big(L^2(\mathbb{R}^3, \ud^3 \boldsymbol{\p}; \mathbb{C}^4) \big)$
iff $\Phi_n \in \mathcal{S}_{A}(\mathbb{R}^3; \mathbb{C}^4)^{\widehat{\otimes} \, n}$ 
for all $n= 0, 1,2, \ldots$ and
\[
\sum \limits_{n=0}^{\infty} \, n! \, |\Phi_n|_k < \infty \,\,\,
\textrm{for all} \,\,\, k\geq 0.
\] 
In this case 
\begin{equation}\label{||Phi||_kPhiInHida}
\|\Phi\|_{k}^2 = \sum \limits_{n=0}^{\infty} \, n! \, |\Phi_n|_k < \infty \,\,\,
\textrm{for all} \,\,\, k\geq 0.
\end{equation}

Note that the norms 
\[
\|\Phi\|_{k} = \|\Gamma(A)^k \Phi \|_0 \,\,\, \textrm{with} \,\,\,
\Phi \in \big(\mathcal{S}_{A}(\mathbb{R}^3; \mathbb{C}^4)\big)
\]
are well defined on the Hida space $\big(\mathcal{S}_{A}(\mathbb{R}^3; \mathbb{C}^4)\big)
\subset \Gamma\big(L^2(\mathbb{R}^3, \ud^3 \boldsymbol{\p}; \mathbb{C}^4) \big)$
also for $k$ equal to any negative integer. Completion of $\big(\mathcal{S}_{A}(\mathbb{R}^3; \mathbb{C}^4)\big)$ 
with respect to the Hilbertian norm
\[
\|\cdot \|_{-k} = \|\Gamma(A)^{-k} \cdot \|_0 \,\,\, \textrm{with fixed} \,\,\,
k \in \mathbb{N}
\]
is equal to a Hilbert space, which we denote
\begin{equation}\label{Hida-k}
\Big(\mathcal{S}_{A}(\mathbb{R}^3; \mathbb{C}^4)\Big)_{-k},
\end{equation}
and which is also equal do the completion  of $\Dom \, \Gamma(A)^{-k}$ (equal to the whole Fock space
$\Dom \, \Gamma(A)^{-k} = \Gamma\big(L^2(\mathbb{R}^3, \ud^3 \boldsymbol{\p}; \mathbb{C}^4) \big)$ for
 $k=0,1,2, \ldots $) with respect to the norm $\|\cdot \|_{-k}$.
The Hilbert space (\ref{Hida-k}) is for each $k\geq0$ canonically isomorphic, including the case $k=0$, 
(Riesz isomorphism) to the Hilbert space dual
of the Hilbert space
\begin{equation}\label{Hida+k}
\Big(\mathcal{S}_{A}(\mathbb{R}^3; \mathbb{C}^4)\Big)_{k},
\end{equation}
compare \cite{obata-book}. Recall that the Hilbert space (\ref{Hida+k}) is equal to the completion of the domain
$\Dom \, \Gamma(A)^{k}$ with respect to the
norm $\| \cdot\|_k$. The Hilbert spaces 
\[
\Big(\mathcal{S}_{A}(\mathbb{R}^3; \mathbb{C}^4)\Big)_{-k}, \,\,\, k = 0, 1, 2, \ldots
\]
compose an inductive system, \cite{GelfandIV}, \cite{obata-book}, with natural continuous inclusions
\begingroup\makeatletter\def\f@size{5}\check@mathfonts
\def\maketag@@@#1{\hbox{\m@th\large\normalfont#1}}%
\begin{equation}\label{IndSystemHida}
\left. \begin{array}{cccccccc}                       \Big(\mathcal{S}_{A}(\mathbb{R}^3; \mathbb{C}^4)\Big)_{-0}& 
\subset  & \Big(\mathcal{S}_{A}(\mathbb{R}^3; \mathbb{C}^4)\Big)_{-1} & \subset & 
\Big(\mathcal{S}_{A}(\mathbb{R}^3; \mathbb{C}^4)\Big)_{-2} & \subset \ldots \subset & 
\Big(\mathcal{S}_{A}(\mathbb{R}^3; \mathbb{C}^4)\Big)^*&        \\
                                                   \parallel &&&&&&&  \\
\overline{\Gamma\big(L^2(\mathbb{R}^3, \ud^3 \boldsymbol{\p}; \mathbb{C}^4) \big)}&&&&&&&  \\
\parallel &&&&&&&  \\
\Gamma\big(L^2(\mathbb{R}^3, \ud^3 \boldsymbol{\p}; \mathbb{C}^4) \big)^*&&&&&&&  \end{array}\right..
\end{equation}
\endgroup
which is dual to the projective system 
\begingroup\makeatletter\def\f@size{5}\check@mathfonts
\def\maketag@@@#1{\hbox{\m@th\large\normalfont#1}}%
\begin{equation}\label{ProjSystemHida}
\left. \begin{array}{cccccccc}                       \big(\mathcal{S}_{A}(\mathbb{R}^3; \mathbb{C}^4)\big)& 
\subset \ldots   &  \ldots \subset \Big(\mathcal{S}_{A}(\mathbb{R}^3; \mathbb{C}^4)\Big)_{2} & \subset & 
\Big(\mathcal{S}_{A}(\mathbb{R}^3; \mathbb{C}^4)\Big)_{1}& \subset & 
&  \Big(\mathcal{S}_{A}(\mathbb{R}^3; \mathbb{C}^4)\Big)_0      \\
                                                    &&&&&&& \parallel \\
&&&&&&& \Gamma\big(L^2(\mathbb{R}^3, \ud^3 \boldsymbol{\p}; \mathbb{C}^4) \big) \end{array}\right..
\end{equation}
\endgroup
defining the Hida space $\big(\mathcal{S}_{A}(\mathbb{R}^3; \mathbb{C}^4)\big)$. 
The two systems (\ref{ProjSystemHida}) and (\ref{IndSystemHida}) can be joined into single system of Hilbert spaces
with comparable and compatible norms, by using the natural isomorphism of the dual to the adjoint space 
\[
\Gamma\big(L^2(\mathbb{R}^3, \ud^3 \boldsymbol{\p}; \mathbb{C}^4) \big)^* \cong
\overline{\Gamma\big(L^2(\mathbb{R}^3, \ud^3 \boldsymbol{\p}; \mathbb{C}^4) \big)} =
\Big(\mathcal{S}_{A}(\mathbb{R}^3; \mathbb{C}^4)\Big)_{-0}
\]
to the Hilbert space 
\[ 
\Gamma\big(L^2(\mathbb{R}^3, \ud^3 \boldsymbol{\p}; \mathbb{C}^4) \big)
= \Big(\mathcal{S}_{A}(\mathbb{R}^3; \mathbb{C}^4)\Big)_{0}
\]
(Riesz isomorphism, compare \cite{GelfandIV}, \cite{obata-book}), 
and noting that the elemets of the Hilbert space $H$  and its adjoint space $\overline{H}$ are the same: 
\begingroup\makeatletter\def\f@size{5}\check@mathfonts
\def\maketag@@@#1{\hbox{\m@th\large\normalfont#1}}%
\begin{multline*} 
\left. \begin{array}{ccccccccc}                       \big(\mathcal{S}_{A}(\mathbb{R}^3; \mathbb{C}^4)\big)& 
\subset \ldots   &  \ldots \subset \Big(\mathcal{S}_{A}(\mathbb{R}^3; \mathbb{C}^4)\Big)_{2} & \subset & 
\Big(\mathcal{S}_{A}(\mathbb{R}^3; \mathbb{C}^4)\Big)_{1}& \subset & 
&  \Big(\mathcal{S}_{A}(\mathbb{R}^3; \mathbb{C}^4)\Big)_0 & =   \\
                                                    &&&&&&& \parallel& \\
&&&&&&& \Gamma\big(L^2(\mathbb{R}^3, \ud^3 \boldsymbol{\p}; \mathbb{C}^4) \big)& \end{array}\right.
\,\,\,
\\
\,\,\,
\left. \begin{array}{ccccccccc}                  =     &\Big(\mathcal{S}_{A}(\mathbb{R}^3; \mathbb{C}^4)\Big)_{-0}& 
\subset  & \Big(\mathcal{S}_{A}(\mathbb{R}^3; \mathbb{C}^4)\Big)_{-1} & \subset & 
\Big(\mathcal{S}_{A}(\mathbb{R}^3; \mathbb{C}^4)\Big)_{-k} & \subset \ldots \subset & 
\Big(\mathcal{S}_{A}(\mathbb{R}^3; \mathbb{C}^4)\Big)^*&        \\
                                                   & \parallel &&&&&&&  \\
&\overline{\Gamma\big(L^2(\mathbb{R}^3, \ud^3 \boldsymbol{\p}; \mathbb{C}^4) \big)}&&&&&&&  \end{array}\right.
\end{multline*}
\endgroup

The strong dual $\big(\mathcal{S}_{A}(\mathbb{R}^3; \mathbb{C}^4)\big)^*$ 
of the Hida space $\big(\mathcal{S}_{A}(\mathbb{R}^3; \mathbb{C}^4)\big)$ is equal
to the inductive limit of the system (\ref{IndSystemHida}). 
Recall that the Hida space $\big(\mathcal{S}_{A}(\mathbb{R}^3; \mathbb{C}^4)\big)$
itself is equal to the projective limit 
of the system (\ref{ProjSystemHida}), compare \cite{obata-book}.

Similarily as for the elements of Hida (or Fock) space, likewise 
each element $\Phi \in \big(\mathcal{S}_{A}(\mathbb{R}^3; \mathbb{C}^4)\big)^*$ 
of the strong dual to the Hida space has a unique decomposition
\begin{equation}\label{Hida*Phi}
\Phi = \sum \limits_{n=0}^{\infty} \Phi_n, \,\,\, \textrm{with} \,\,\,
\Phi_n \in \big(\mathcal{S}_{A}(\mathbb{R}^3; \mathbb{C}^4)^{\widehat{\otimes} \, n} \big)^*.
\end{equation}
In this case there exists a natural $k$ such that 
\[
\|\Phi\|_{-k}^2 = \sum \limits_{n=0}^{\infty} \, n! \, |\Phi_n|_{-k}^{2} < \infty.
\]

Note that we have natural real and complex structure on the spaces we encounter here with well defined complex conjugation
$\overline{(\cdot)}$. In partiular, if we denote the dual pairings 
on $\mathcal{S}_{A}(\mathbb{R}^3; \mathbb{C}^4)^* \times \mathcal{S}_{A}(\mathbb{R}^3; \mathbb{C}^4)$
and on 
$\big(\mathcal{S}_{A}(\mathbb{R}^3; \mathbb{C}^4)\big)^* \times \big(\mathcal{S}_{A}(\mathbb{R}^3; \mathbb{C}^4)\big)$
by $\langle \cdot, \cdot \rangle$ and respectively by $\langle \langle \cdot, \cdot \rangle \rangle$
then we have
\[
\begin{split}
\langle \xi, \eta \rangle = (\overline{\xi}, \eta)_0, \,\,\,
\textrm{for} \,\,\,
\xi \in \mathcal{S}_{A}(\mathbb{R}^3; \mathbb{C}^4) \subset \mathcal{S}_{A}(\mathbb{R}^3; \mathbb{C}^4)^*,
\eta \in \mathcal{S}_{A}(\mathbb{R}^3; \mathbb{C}^4), \\
\langle\langle \Psi, \Phi \rangle \rangle = (( \, \overline{\Psi} \, , \Phi \, ))_0, \,\,\,
\textrm{for} \,\,\,
\Psi \in \big(\mathcal{S}_{A}(\mathbb{R}^3; \mathbb{C}^4)\big) 
\subset \big(\mathcal{S}_{A}(\mathbb{R}^3; \mathbb{C}^4)\big)^*,
\Phi \in \big(\mathcal{S}_{A}(\mathbb{R}^3; \mathbb{C}^4)\big).
\end{split}
\]

Now we are ready to define the Hida operators $a(w), a(w)^+$, $w \in \mathcal{S}_{A}(\mathbb{R}^3; \mathbb{C}^4)^*$
in the Fock space $\Gamma\big(L^2(\mathbb{R}^3, \ud^3 \boldsymbol{\p}; \mathbb{C}^4) \big)$
corresponding to the first (standard) Gelfand triple in (\ref{SinglePartGelfandTriplesForPsi}).

Namely for each $w \in \mathcal{S}_{A}(\mathbb{R}^3; \mathbb{C}^4)^*$,
and each general element (\ref{HidaPhi}) of the Hida space 
we define Hida annihilation operator $a(w)$ which by definition acts on the element
$\Phi$ given by (\ref{HidaPhi}) according to the following formula
\[
\begin{split}
1) \,\,\,\,\, a(w) \big(\Phi = \Phi_0 \big) = 0, \\
2) \,\,\,\,\, a(w) \Phi =  \sum \limits_{n \geq 0} \, n \, \overline{w} \, \widehat{\otimes}_1 \, \Phi_n.   
\end{split}
\]

Now we define the Hida creation operator $a(w)^+$, $w \in \mathcal{S}_{A}(\mathbb{R}^3; \mathbb{C}^4)^*$,
transforming the strong dual $\big(\mathcal{S}_{A}(\mathbb{R}^3; \mathbb{C}^4)\big)^*$ of the Hida space
into itself. Namely let $w \in \mathcal{S}_{A}(\mathbb{R}^3; \mathbb{C}^4)^*$ and let $\Phi$
be any general element (\ref{Hida*Phi}) of the strong dual 
$\big(\mathcal{S}_{A}(\mathbb{R}^3; \mathbb{C}^4)\big)^*$.
The action of the Hida creation
operator $a(w)^+$, $w \in \mathcal{S}_{A}(\mathbb{R}^3; \mathbb{C}^4)^*$, on such $\Phi$
is by definition equal 
\[
\begin{split}
a(w)^+ \Phi =  \sum \limits_{n \geq 0}  \, w \, \widehat{\otimes} \, \Phi_n.  
\end{split}
\]
Here as well as in the definition of the Hida annihilation operator
the tensor product $\otimes$ and its $1$-contraction $\otimes_1$
(antisymmetrized  $\widehat{\otimes}$, $\widehat{\otimes}_1$) is equal to the 
projective tensor product over the respective nuclear spaces:
\begin{multline*}
\mathcal{S}_{A}(\mathbb{R}^3; \mathbb{C}^4)^*, \mathcal{S}_{A}(\mathbb{R}^3; \mathbb{C}^4)^{\otimes n},
\mathcal{S}_{A}(\mathbb{R}^3; \mathbb{C}^4)^{\widehat{\otimes} \, n}, \\
\big(\mathcal{S}_{A}(\mathbb{R}^3; \mathbb{C}^4)^{\otimes n}\big)^*,
\big(\mathcal{S}_{A}(\mathbb{R}^3; \mathbb{C}^4)^{\widehat{\otimes} \, n} \big)^*, 
\end{multline*}
In this case (of nuclear spaces) tensor product is essentially unique with the projective tensor product coinciding
with the equicontinuous tensor product. 
Recall that 
\[
v_{{}_{1}} \, \widehat{\otimes} \, \cdots \widehat{\otimes} \, v_{{}_{n}} =
(n!)^{-1} \sum \limits_{\pi} \textrm{sign} \, (\pi) \, v_{{}_{\pi(1)}} \otimes
\cdots \otimes v_{{}_{\pi(n)}},
\]
with $v_{{}_{i}}$ in the respective space, and that the antisymmetrized $1$-contraction $\widehat{\otimes}_1$
is uniquely determined by the formula
\begin{multline*}
u \, \widehat{\otimes}_1 v_{{}_{1}} \, \widehat{\otimes} \, \cdots \widehat{\otimes} \, v_{{}_{n}}
= (n!)^{-1} \sum \limits_{\pi} \textrm{sign} \, (\pi) \, \langle u,v_{{}_{\pi(1)}} \rangle \, 
 v_{{}_{\pi(2)}} \otimes
\cdots \otimes v_{{}_{\pi(n)}}, \\
u \in \mathcal{S}_{A}(\mathbb{R}^3; \mathbb{C}^4)^{*}, v_{{}_{i}} \in \mathcal{S}_{A}(\mathbb{R}^3; \mathbb{C}^4),
\end{multline*}
with the sums ranging over all permutations $\pi$ of the natural numbers $1, \ldots, n$, and with
the evaluation $\langle u,v_{{}_{\pi(1)}} \rangle$ of $u$ 
on $v_{{}_{\pi(1)}}$, which restricts to
\[
\langle u,v_{{}_{\pi(1)}} \rangle = (\overline{u},v_{{}_{\pi(n)}})_0 \,\,\,
\textrm{whenever} \,\,\,
u \in \mathcal{S}_{A}(\mathbb{R}^3; \mathbb{C}^4) \subset \mathcal{S}_{A}(\mathbb{R}^3; \mathbb{C}^4)^{*}.
\] 

It follows that $a(w)$, $w \in \mathcal{S}_{A}(\mathbb{R}^3; \mathbb{C}^4)^*$,
transforms continously the Hida space into the Hida space
\[
a(w): \, \big(\mathcal{S}_{A}(\mathbb{R}^3; \mathbb{C}^4)\big) \, \longrightarrow \,
\big(\mathcal{S}_{A}(\mathbb{R}^3; \mathbb{C}^4)\big),
\]
for a proof compare e.g. \cite{obata-book}, \cite{Shimada}. By composig it with the natural continous
inclusion $\big(\mathcal{S}_{A}(\mathbb{R}^3; \mathbb{C}^4)\big) 
\subset \big(\mathcal{S}_{A}(\mathbb{R}^3; \mathbb{C}^4)\big)^*$, we can also regard the Hida
annihilation operator $a(w)$, $w \in \mathcal{S}_{A}(\mathbb{R}^3; \mathbb{C}^4)^*$, as  a continuous operator
\[
a(w): \, \big(\mathcal{S}_{A}(\mathbb{R}^3; \mathbb{C}^4)\big) \, \longrightarrow \,
\big(\mathcal{S}_{A}(\mathbb{R}^3; \mathbb{C}^4)\big)^*.
\]

It follows by general property of transposition, \cite{treves}, that 
$a(w)^*$, $w \in \mathcal{S}_{A}(\mathbb{R}^3; \mathbb{C}^4)^*$, 
maps continously the strong dual of the Hida space into itself
\[
a(w)^*: \, \big(\mathcal{S}_{A}(\mathbb{R}^3; \mathbb{C}^4)\big)^* \, \longrightarrow \,
\big(\mathcal{S}_{A}(\mathbb{R}^3; \mathbb{C}^4)\big)^*.
\] 
By composig it with the dual 
\[
\big(\mathcal{S}_{A}(\mathbb{R}^3; \mathbb{C}^4)\big) \cong \big(\mathcal{S}_{A}(\mathbb{R}^3; \mathbb{C}^4)\big)^{**} 
\subset \big(\mathcal{S}_{A}(\mathbb{R}^3; \mathbb{C}^4)\big)^*
\]
of the natural inclusion $\big(\mathcal{S}_{A}(\mathbb{R}^3; \mathbb{C}^4)\big) 
\subset \big(\mathcal{S}_{A}(\mathbb{R}^3; \mathbb{C}^4)\big)^*$, we can regard the Hida 
creation operator $a(w)^*$, $w \in \mathcal{S}_{A}(\mathbb{R}^3; \mathbb{C}^4)^*$,
as a continuous operator 
\[
a(w)^*: \, \big(\mathcal{S}_{A}(\mathbb{R}^3; \mathbb{C}^4)\big) \, \longrightarrow \,
\big(\mathcal{S}_{A}(\mathbb{R}^3; \mathbb{C}^4)\big)^*.
\]
It turns out that 
\[
a(w)^+  = \overline{(\cdot)} \circ a(w)^* \circ \overline{(\cdot)}, \,\,\,
w \in \mathcal{S}_{A}(\mathbb{R}^3; \mathbb{C}^4)^*,
\] 
for $a(w)^*, a(w)^+$
understood as maps of the strong dual of the Hida space into itself 
(or resp. as maps transforming the 
Hida space into its strong dual); compare \cite{obata-book}, \cite{Shimada}.  

\begin{rem*}
Note that in fact the definition of the Hida operator used by mathematicians
is slightly different  in comaprison to ours with the additional complex conjugation
\[
\textrm{mathematicians's} \,\,\, a(w) = \,\,\, \textrm{ours} \,\,\,a(\overline{w}).  
\] 
In particular ours $a(w)$ is anti-linear in $w$, which is the convetion accepted in physical 
literature.  This is the conjugation $A^+ = \overline{(\cdot)} \circ A^{^*} \circ \overline{(\cdot)}$
equal to the linear transpose composed with complex conjugations, which connects the Hida generalized
annihilation $a(w)$ and creation operators $a(w)^+$, due to the convention which we have accepted,
and which is used by physicists. In the convention accepted by mathematicians it is the ordinary linear 
transpose which connects the generalized Hida annihilation
$a(w)$ and creation operators $a(w)^*$.
\end{rem*}

In the mathematical literature the fact that the Hida annihilation operator $a(w)$ is a ($\mathbb{Z}_2$-graded
in fermi case)  derivation on the Hida nuclear algebra (with the multiplication defined by the antisymmetrized tensor
product $\widehat{\otimes}$) is reflected by the following notation introduced by Hida:
\[
D_{w} \overset{\textrm{df}}{=} a(w), w \in  \mathcal{S}_{A}(\mathbb{R}^3; \mathbb{C}^4)^* = \mathcal{S}(\mathbb{R}^3; \mathbb{C}^4)^*.
\]
(here the convention used by mathematicians is better because their 
\[
D_{w} \overset{\textrm{df}}{=} a(w)
\]
is linear in $w$, and in bose case when the Hida space is realized as commutative algebra of functions on 
$\mathcal{S}_{A}(\mathbb{R}^3; \mathbb{C}^4)^*$, the Hida annihilation operator $a(w)$ is indeed 
equal to the G{\aa}teaux derivation in the direction of $w$ and not in direction
$\overline{w}$).

Recall that $\mathcal{S}_{A}(\mathbb{R}^3; \mathbb{C}^4) = \mathcal{S}(\mathbb{R}^3; \mathbb{C}^4)
= \oplus_{1}^{4} \mathcal{S}(\mathbb{R}^3; \mathbb{C})$ 
we regard as the nuclear space of complex valued functions $f$ on four disjoint copies
of $\mathbb{R}^3$ whose restrictions $f_s$ to each $s$-th copy coincide with 
the Schwartz functions in 
$\mathcal{S}_{H_{(3)}}(\mathbb{R}^3; \mathbb{C}) = \mathcal{S}(\mathbb{R}^3; \mathbb{C})$. In particular for each
value of the discrete index $s \in \{1,2,3,4\}$, correspoduing to each copy, and for each point 
$\boldsymbol{\p} \in \mathbb{R}^3$, we have well defined Dirac delta-functional 
$\delta_{s,\boldsymbol{\p}} \in \mathcal{S}_{A}(\mathbb{R}^3; \mathbb{C}^4)^* 
= \mathcal{S}(\mathbb{R}^3; \mathbb{C}^4)^*$
defined by 
\[
\delta_{s,\boldsymbol{\p}} (f) = f_s(\boldsymbol{\p}),
\] 
i.e. the evaluation of the restriction of $f$ to the $s$-th copy of $\mathbb{R}^3$ at the point
$\boldsymbol{\p}$ of that copy. Simply speaking $\delta_{s,\boldsymbol{\p}}$ is the evaluation functional at 
fixed point $(s,\boldsymbol{\p})$ of the disjoint sum 
$\mathbb{R}^3 \sqcup \mathbb{R}^3 \sqcup \mathbb{R}^3 \sqcup \mathbb{R}^3$.

The generalized Hida annihilation and creation operators $a(w), a(w)^+$ evaluated
at $w=\delta_{s,\boldsymbol{\p}}$ equal to the Dirac delta functionals $\delta_{s,\boldsymbol{\p}}$ have special importance,
and have special notation in mathematical literature
\[
\partial_{s, \boldsymbol{\p}} \overset{\textrm{df}}{=} D_{{}_{\delta_{s,\boldsymbol{\p}}}} 
\overset{\textrm{df}}{=} a(\delta_{s,\boldsymbol{\p}}),  \,\,\,
\partial_{s, \boldsymbol{\p}}^+ =  D_{{}_{\delta_{s,\boldsymbol{\p}}}}^+ 
= a(\delta_{s,\boldsymbol{\p}})^+
\]
reflecting the derivation-like character of these generalized Hida operators, and are called
\emph{Hida's differential operators}. 
But we have also widely used notation for operators in physical literature, with whom the Hida differential
operators should be identified. Namely generalized Hida 
operators should be identified with the operators frequently written by physicists in the following manner
\[
\begin{split}
 a_s(\boldsymbol{\p}) \overset{\textrm{df}}{=} D_{{}_{\delta_{s,\boldsymbol{\p}}}} \overset{\textrm{df}}{=}
 \partial_{s, \boldsymbol{\p}} 
\overset{\textrm{df}}{=} a(\delta_{s,\boldsymbol{\p}}), \\
a_s(\boldsymbol{\p})^+ \overset{\textrm{df}}{=}  D_{{}_{\delta_{s,\boldsymbol{\p}}}}^+ \overset{\textrm{df}}{=}
\partial_{s, \boldsymbol{\p}}^+ \overset{\textrm{df}}{=} a(\delta_{s,\boldsymbol{\p}})^+.
\end{split}
\]
More precisely the operators 
$a_s(\boldsymbol{\p}), a_s(\boldsymbol{\p})^+$ for $s=1,2$
should be identified with the operators 
$b_s(\boldsymbol{\p}), b_s(\boldsymbol{\p})^+$ for $s=1,-1$
of the book \cite{Scharf}, p. 82 (or with the operators 
$\overset{*}{a}_{s}^{-}(\boldsymbol{\p}), {a}_{s}^{+}(\boldsymbol{\p})$, $s=1,2$, 
of the book \cite{Bogoliubov_Shirkov}, p. 123)). 
The operators $a_s(\boldsymbol{\p}), a_s(\boldsymbol{\p})^+$ for $s=3,4$
should respectively be identified with the operators $d_{s}(\boldsymbol{\p}), d_{s}(\boldsymbol{\p})^+$
for $s=1,-1$, of the book \cite{Scharf}, p. 82 (or respectively with the operators 
${a}_{s}^{-}(\boldsymbol{\p}), \overset{*}{a}_{s}^{+}(\boldsymbol{\p})$, $s= 1,2$, of the book 
\cite{Bogoliubov_Shirkov}, p. 123).

Note that because the Dirac delta fuctional $\delta_{s,\boldsymbol{\p}}$ is real 
$\overline{\delta_{s,\boldsymbol{\p}}} = \delta_{s,\boldsymbol{\p}}$ (i.e. commutes with complex conjugation),
then
\[
a(\delta_{s,\boldsymbol{\p}})^+ = \partial_{s, \boldsymbol{\p}}^+ =
a(\delta_{s,\boldsymbol{\p}})^* = \partial_{s, \boldsymbol{\p}}^*, 
\]
so that for Hida's differential operators the linear adjunction  $\partial_{s, \boldsymbol{\p}}^*$ coincides with the Hermitean adjunction $\partial_{s, \boldsymbol{\p}}^+$.

We may thus summarize the notation used here with that used by other authors
in the following table
\begin{center}
\begin{tabular}{c|c|c|c}
\hline
& {\small Hida-Obata \cite{obata-book}} & Scharf \cite{Scharf} & 
Bogoliubov-Shirkov \cite{Bogoliubov_Shirkov} \\
\hline
$a_{s=1}(\boldsymbol{\p}) \overset{\textrm{df}}{=}
a(\delta_{s=1,\boldsymbol{\p}})$ & $\partial_{s=1, \boldsymbol{\p}}$ & $b_{s=1}(\boldsymbol{\p})$ &
$\overset{*}{a}_{s=1}^{-}(\boldsymbol{\p})$ \\
\hline
$a_{s=2}(\boldsymbol{\p}) \overset{\textrm{df}}{=}
a(\delta_{s=2,\boldsymbol{\p}})$ & $\partial_{s=2, \boldsymbol{\p}}$ & $b_{s=-1}(\boldsymbol{\p})$ &
$\overset{*}{a}_{s=2}^{-}(\boldsymbol{\p})$ \\
\hline
$a_{s=3}(\boldsymbol{\p}) \overset{\textrm{df}}{=}
a(\delta_{s=3,\boldsymbol{\p}})$ & $\partial_{s=3, \boldsymbol{\p}}$ & $d_{s=1}(\boldsymbol{\p})$ &
${a}_{s=1}^{-}(\boldsymbol{\p})$ \\
\hline
$a_{s=4}(\boldsymbol{\p}) \overset{\textrm{df}}{=}
a(\delta_{s=4,\boldsymbol{\p}})$ & $\partial_{s=4, \boldsymbol{\p}}$ & $d_{s=-1}(\boldsymbol{\p})$ &
${a}_{s=2}^{-}(\boldsymbol{\p})$ \\
\hline
$a_{s=1}(\boldsymbol{\p})^+ \overset{\textrm{df}}{=}
a(\delta_{s=1,\boldsymbol{\p}})^+$ & $\partial_{s=1, \boldsymbol{\p}}^*$ & $b_{s=1}(\boldsymbol{\p})^+$ &
${a}_{s=1}^{+}(\boldsymbol{\p})$ \\
\hline
$a_{s=2}(\boldsymbol{\p})^+ \overset{\textrm{df}}{=}
a(\delta_{s=2,\boldsymbol{\p}})^+$ & $\partial_{s=2, \boldsymbol{\p}}^*$ & $b_{s=-1}(\boldsymbol{\p})^+$ &
${a}_{s=2}^{+}(\boldsymbol{\p})$ \\
\hline
$a_{s=3}(\boldsymbol{\p})^+ \overset{\textrm{df}}{=}
a(\delta_{s=3,\boldsymbol{\p}})^+$ & $\partial_{s=3, \boldsymbol{\p}}^*$ & $d_{s=1}(\boldsymbol{\p})^+$ &
$\overset{*}{a}_{s=1}^{+}(\boldsymbol{\p})$ \\
\hline
$a_{s=4}(\boldsymbol{\p})^+ \overset{\textrm{df}}{=}
a(\delta_{s=4,\boldsymbol{\p}})^+$ & $\partial_{s=4, \boldsymbol{\p}}^*$ & $d_{s=-1}(\boldsymbol{\p})^+$ &
$\overset{*}{a}_{s=2}^{+}(\boldsymbol{\p})$ \\
\hline
\end{tabular}
\end{center}

Now we remind some basic results of the calculus of integral kernel operators
constructed mainly by Hida, Obata, and Sait\^o, which we will use here and in the following
Sections (especially in Section \ref{A(1)psi(1)}). 

Before doing it we make a general remark concerning norm estimations of the left $\widehat{\otimes_l}$ and right
$\widehat{\otimes^l}$ antisymmetrized (or symmetrized) $l$-contractions
(compare \cite{obata-book}) 
\[
|\widehat{f} \widehat{\otimes^l} \widehat{g}|_k, |\widehat{F} \widehat{\otimes}^l \widehat{g}|_{-k},  
|\widehat{F} \widehat{\otimes_l} \widehat{g}|_{-k}, 
\,\,\,
\widehat{F} \in  
\Big(\mathcal{S}_{A}(\mathbb{R}^3; \mathbb{C}^4)^{\widehat{\otimes} \, (l+m)} \Big)^{*},
\widehat{f}, \widehat{g} \in \mathcal{S}_{A}
(\mathbb{R}^3; \mathbb{C}^4)^{\widehat{\otimes} \, (l+n)}.
\] 
Namely passing from estimations for the norms
\[
|f\otimes^l g|_k, |F \otimes^l g|_{-k},  |F \otimes_l g|_{-k}, \,\,\, \textrm{for}  \,\,\,
F \in  
\Big(\mathcal{S}_{A}(\mathbb{R}^3; \mathbb{C}^4)^{\otimes (l+m)} \Big)^{*},
f, g \in \mathcal{S}_{A}
\mathbb{R}^3; \mathbb{C}^4)^{\otimes  (l+n)},
\]
with non antisymmetrized (or non symmetrized $F$, $f$ and $g$),
summarized in Prop. 3.4.3, Lemma 3.4.4, 3.4.5 of \cite{obata-book}, to estimations with
symmetrized or antisymmetrized $\widehat{F}$, $\widehat{f}$ and $\widehat{g}$  we note that 
we have
\begin{multline*}
F \widehat{\otimes}^l g = F \otimes^l g = \pm  F\otimes_l g = \pm F \widehat{\otimes}_l g, \\
\textrm{for} \,\,\,
F \in  
\Big(\mathcal{S}_{A}(\mathbb{R}^3; \mathbb{C}^4)^{\widehat{\otimes} (l+m)} \Big)^{*},
g \in \mathcal{S}_{A}(\mathbb{R}^3; \mathbb{C}^4)^{\widehat{\otimes}  (l+n)},
\end{multline*}
and  
\[
|\widehat{f}|_{k} \leq |f|_{k}, \,\,\, f \in 
\mathcal{S}_{A}(\mathbb{R}^3; \mathbb{C}^4)^{\otimes  n}, k \in \mathbb{Z},
\] 
in each case: for symmetrization as well as for antisymmetrization $\widehat{(\cdot)}$.
This allows to restate the estimations for non symmetrized/antisymmetrized
$F$, $f$ and $g$ (summarized in Prop. 3.4.3, Lemma 3.4.4, 3.4.5 of \cite{obata-book}) 
in the form of propositions
analogous to Prop. 3.4.7, 3.4.8, 3.4.9 in \cite{obata-book} for the contractions
of antisymmetrized $\widehat{F},\widehat{G},\widehat{g}, \widehat{f}$ on exactly the same footing
as for symmetrized $\widehat{F},\widehat{G},\widehat{g}, \widehat{f}$ (as we have already mentioned
in Remark \ref{TretmentFermiIntegKerOp=TretmentBoseIntegKerOp}). In particular
theorems concernig integral kernel operators and Fock expansions, in both cases 
1) of scalar-valued kernels \cite{hida}, \cite{obata}, and 2) of vector-valued kernels \cite{obataJFA}, 
can be stated and proved exactly as in \cite{hida}, \cite{obata}, \cite{obataJFA} 
also for the fermi case. The only difference which arises in fermi case (compared to the bose case)
comes from additional factor $(-1)$ depending on the degree of the involved tensors.
In particular we should note that for nonsymmetrized 
$F \in \Big( \mathcal{S}_{A}(\mathbb{R}^3; \mathbb{C}^4)^{\otimes  k} \Big)^*$,
$G \in \Big( \mathcal{S}_{A}(\mathbb{R}^3; \mathbb{C}^4)^{\otimes  l} \Big)^*$,
and $h \in \mathcal{S}_{A}(\mathbb{R}^3; \mathbb{C}^4)^{\otimes  (k+l+m)}$,
we have
\[
F \otimes_k \big(G \otimes_l h \big) =
\big(G \otimes F \big) \otimes_{k+l} h \,\,\, \textrm{in this order!}
\]
and thus by  antisymmetrization $\widehat{(\cdot)}$ we get
\begin{multline*}
\widehat{F} \widehat{\otimes_k} \big(\widehat{G} \widehat{\otimes_l} \widehat{h} \big) =
\big(\widehat{G} \widehat{\otimes} \widehat{F} \big) \widehat{\otimes_{k+l}} \widehat{h} =
(-1)^{(\textrm{deg} \, \widehat{F})(\textrm{deg} \, \widehat{G})} \,
\big(\widehat{F} \widehat{\otimes} \widehat{G} \big) \widehat{\otimes_{k+l}} \widehat{h}, \\
\,\,\,
\textrm{deg} \, \widehat{F} \overset{\textrm{df}}{=} k,
\textrm{deg} \, \widehat{G} \overset{\textrm{df}}{=} l;
\end{multline*}
(instead of Proposition 3.4.8 of \cite{obata-book} with symmetrization $\widehat{(\cdot)}$ in bose case, where the factor
$(-1)^{(\textrm{deg} \, \widehat{F})(\textrm{deg} \, G)}$ degenerates to $1$). 

Similarily we have for $F \in \Big( \mathcal{S}_{A}(\mathbb{R}^3; \mathbb{C}^4)^{\otimes  l} \Big)^*$,
$G \in \Big( \mathcal{S}_{A}(\mathbb{R}^3; \mathbb{C}^4)^{\otimes  m} \Big)^*$,
and $f \in \mathcal{S}_{A}(\mathbb{R}^3; \mathbb{C}^4)^{\otimes  (l+n)}$
\[
\langle F \otimes_l f, G \otimes_m g \rangle = \langle F \otimes G, f \otimes^n g \rangle.
\]
Again passing to the subspaces of antisymmetrized tensors we obtain
\[
\langle \widehat{F} \widehat{\otimes_l} \widehat{f}, \widehat{G} 
\widehat{\otimes_m} \widehat{g} \rangle = \langle \, \widehat{F} \, \widehat{\otimes} \, \widehat{G}, 
\widehat{f} \widehat{\otimes^n} \widehat{g} \rangle = 
(-1)^{m (\textrm{deg} \, \widehat{f})} \, \langle \, \widehat{F} \, \widehat{\otimes} \, \widehat{G}, 
\, \widehat{f} \, \widehat{\otimes_n} \, \widehat{g} \rangle,
\]
(instead of Prop. 3.4.9 in \cite{obata-book} with symmetrization $\widehat{(\cdot)}$ for bose case).

The replacements of symmetrization $\widehat{(\cdot)}$ with antisymmetrization
$\widehat{(\cdot)}$ (with the appropriate factors $-1$) in the analysis of 
integral kernel operators in \cite{obata-book}, are rather obvious, thus we 
leave the detailed inspection to the reader as an exercise. 
We mention only some particular cases in explicit form.

In particular we have the following analogue of Thm 4.1.7 of \cite{obata-book}. 
\begin{twr}\label{Dy1...Dym}
Let $\Phi \in \big(\mathcal{S}_{A}(\mathbb{R}^3; \mathbb{C}^4)\big)$ be any element of the Hida space,
and let
\[
\Phi = \sum \limits_{n=0}^{\infty} \Phi_n, \,\,\,
\Phi_n \in \mathcal{S}_{A}(\mathbb{R}^3; \mathbb{C}^4)^{\widehat{\otimes} \, n}
\]
be its decomposition (thus fulfiling (\ref{||Phi||_kPhiInHida})). Then for 
\[
y_1, \dots y_m \in \mathcal{S}_{A}(\mathbb{R}^3; \mathbb{C}^4)^*
\]
we have
\[
D_{y_1} \cdots D_{y_m} \Phi =
\sum \limits_{n=0}^{\infty} \, (-1)^{m-1} \frac{(n+m)!}{n!} \, (\overline{y_1} \widehat{\otimes}
\cdots \widehat{\otimes} \overline{y_m}) \widehat{\otimes}_m \Phi_{m+n}. 
\]
Moreover,  for any $k\geq 0$, $q>0$ and 
$\Phi \in \big(\mathcal{S}_{A}(\mathbb{R}^3; \mathbb{C}^4)\big)$ we have
\[
\| D_{y_1} \cdots D_{y_m} \Phi \|_{k} \leq
\rho^{-q/2} m^{m/2} \Bigg( \frac{\rho^{-q}}{-2qe\textrm{ln} \rho}\Bigg)^{m/2}
|y_1|_{{}_{-(k+q)}} \cdots |y_m|_{{}_{-(k+q)}} \, \| \Phi \|_{k+q}.
\]
\end{twr}

Here 
\[
\rho \overset{\textrm{df}}{=} \|A^{-1}\|_{\textrm{op}} = \lambda_{0}^{-1}, \,\,\,
\lambda_{0} = \textrm{inf Spec} \, A >1,
\]
which we achieve by eventually  adding the unit operator to the ordinary $3$-dimensional oscillator hamiltonian operator and taking the sum as the direct summand $H_{(3)}$  in $A$ defined by (\ref{AinL^2(R^3;C^4)}). 

Using this theorem (analogue of Thm. 4.1.7 of \cite{obata-book}) as well as the mentioned above
analogue of Prop. 3.4.9 of \cite{obata-book} as does Obata in \cite{obata-book})
we prove in particular the following (analogue of Lemma 4.3.1 in \cite{obata-book} or 
Lemma 2.1 in \cite{hida}):

\begin{lem}\label{etaPhiPsi}
For any elements $\Phi, \Psi \in \big(\mathcal{S}_{A}(\mathbb{R}^3; \mathbb{C}^4)\big)$
of the Hida space we put ($s_i, t_i \in \{1, \ldots, 4\}$, $\boldsymbol{\q}_i, \boldsymbol{\p}_i \in \mathbb{R}^3$)
\[
\eta_{{}_{\Phi, \Psi}}(s_1,\boldsymbol{\q}_1, \ldots, s_l, \boldsymbol{\q}_l, t_1, \boldsymbol{\p}_1, 
\ldots, t_m, \boldsymbol{\p}_m) 
= \big\langle \big\langle \partial_{s_1, \boldsymbol{\q}_1}^* \cdots \partial_{s_l, \boldsymbol{\q}_l}^* 
\partial_{t_1, \boldsymbol{\p}_1} \cdots \partial_{t_m, \boldsymbol{\p}_m} \, \Phi, \, \Psi \big\rangle \big\rangle ,
\]
then for any $k>0$ we have
\[
|\eta_{{}_{\Phi, \Psi}}|_{{}_{k}} \leq \rho^{-k} \big(l^lm^m\big)^{1/2}
\Bigg( \frac{\rho^{-k}}{-2ke \textrm{ln} \rho} \Bigg)^{(l+m)/2} \, \| \Phi \|_{k} \| \Psi \|_k.
\]
In particular, $\eta_{{}_{\Phi, \Psi}} \in \mathcal{S}_{A}(\mathbb{R}^3; \mathbb{C}^4)^{\otimes (l+m)}$.
\end{lem}

This allows analysis of an important class of \emph{integral kernel operators} $\Xi_{l,m}(\kappa_{l,m}) \in 
\mathscr{L}\big( \, \big(\mathcal{S}_{A}(\mathbb{R}^3; \mathbb{C}^4)\big)  \, ,
 \, \big(\mathcal{S}_{A}(\mathbb{R}^3; \mathbb{C}^4)\big)^* \, \big)$,
corresponding to $\kappa_{l,m} \in \big(\mathcal{S}_{A}(\mathbb{R}^3; \mathbb{C}^4)^{\otimes (l+m)}\big)^*
= \mathcal{S}(\mathbb{R}^3; \mathbb{C}^4)^{* \, \otimes (l+m)}$, and written 
\begin{multline}\label{Xilm(kappalm)Fermi}
\Xi_{l,m}(\kappa_{l,m})  \\
= \sum \limits_{s_1, \ldots s_l, t_1, \ldots t_m =1}^{4} \int \limits_{(\mathbb{R}^3)^{l+m}} \,
\kappa_{l,m}(s_1,\boldsymbol{\q}_1, \ldots, s_l, \boldsymbol{\q}_l, t_1, \boldsymbol{\p}_1, 
\ldots, t_m, \boldsymbol{\p}_m) \,\times \\
\times
\partial_{s_1, \boldsymbol{\q}_1}^* \cdots \partial_{s_l, \boldsymbol{\q}_l}^* 
\partial_{t_1, \boldsymbol{\p}_1} \cdots \partial_{t_m, \boldsymbol{\p}_m} \,
\ud^3 \boldsymbol{\q}_1 \ldots \ud^3 \boldsymbol{\q}_l \ud^3 \boldsymbol{\p}_1 \ldots \ud^3 \boldsymbol{\p}_m.
\end{multline}

\begin{twr}\label{Xi_l,m}
Namely (compare Thm.4.3.2 in \cite{obata-book} or Thm. 2.2. of \cite{hida}) 
for any $\kappa_{l,m} \in \big(\mathcal{S}_{A}(\mathbb{R}^3; \mathbb{C}^4)^{\otimes (l+m)}\big)^*
= \mathcal{S}(\mathbb{R}^3; \mathbb{C}^4)^{* \, \otimes (l+m)}$ there exists (uniquely corresponding to
$\kappa_{l,m}$ if $\kappa_{l,m}$ is antisymmetric: 
$\kappa_{l,m} \in \big(\mathcal{S}(\mathbb{R}^3; \mathbb{C}^4)^{\widehat{\otimes} \, l}
\otimes \mathcal{S}(\mathbb{R}^3; \mathbb{C}^4)^{\widehat{\otimes} \, m}\big)^*$  in fermi case, or symmetric in bose case) continuous operator $\Xi_{l,m}(\kappa_{l,m}) \in 
\mathscr{L}\big( \, \big(\mathcal{S}_{A}(\mathbb{R}^3; \mathbb{C}^4)\big)  \, ,
 \, \big(\mathcal{S}_{A}(\mathbb{R}^3; \mathbb{C}^4)\big)^* \, \big)$, written as in (\ref{Xilm(kappalm)Fermi}),
such that
\[
\big\langle \big\langle \Xi_{l,m}(\kappa_{l,m}) \Phi, \, \Psi \big\rangle\big\rangle
= \langle \kappa_{l,m}, \eta_{{}_{\Phi, \Psi}} \rangle, \,\,\,\,\,\,
\Phi, \Psi \in \big(\mathcal{S}_{A}(\mathbb{R}^3; \mathbb{C}^4)\big),
\] 
where
\[
\eta_{{}_{\Phi, \Psi}}(s_1,\boldsymbol{\q}_1, \ldots, s_l, \boldsymbol{\q}_l, t_1, \boldsymbol{\p}_1, 
\ldots, t_m, \boldsymbol{\p}_m) 
= \big\langle \big\langle \partial_{s_1, \boldsymbol{\q}_1}^* \cdots \partial_{s_l, \boldsymbol{\q}_l}^* 
\partial_{t_1, \boldsymbol{\p}_1} \cdots \partial_{t_m, \boldsymbol{\p}_m} \, \Phi, \, \Psi \big\rangle \big\rangle. 
\]
Moreover, for any $k>0$ with $|\kappa_{l,m}|_{-k} <\infty$ it holds
\[
\| \Xi_{l,m}(\kappa_{l,m}) \Phi\|_{-k} \leq 
\rho^{-k} \big(l^lm^m\big)^{1/2}
\Bigg( \frac{\rho^{-k}}{-2ke \textrm{\emph{ln}} \rho} \Bigg)^{(l+m)/2} \, |\kappa_{l,m}|_{{}_{-k}} \, \| \Phi \|_{k} .
\]
\end{twr}

We have the following important theorem (Thm. 4.3.9 of  \cite{obata-book}, Thm. 2.6 of \cite{hida})
which provides neccessary and sufficient condition for the integral kernel operator
(\ref{Xilm(kappalm)Fermi}) to be continuous not merely as an operator on the Hida space into its strong dual,
but likewise as operator transforming continously the Hida space into itself (thus becoming ordinary densely defined
operator in the Fock space):
\begin{twr}\label{Xi_l,m:Hida->Hida}
Let $\kappa_{l,m} \in \big(\mathcal{S}_{A}(\mathbb{R}^3; \mathbb{C}^4)^{\otimes (l+m)}\big)^*$.
Then 
\[
\Xi_{l,m}(\kappa_{l,m}) \in 
\mathscr{L}\big( \, \big(\mathcal{S}_{A}(\mathbb{R}^3; \mathbb{C}^4)\big)  \, ,
 \, \big(\mathcal{S}_{A}(\mathbb{R}^3; \mathbb{C}^4)\big) \, \big)
\]
 if and only if 
 $\kappa_{l,m} \in \mathcal{S}_{A}(\mathbb{R}^3; \mathbb{C}^4)^{\otimes l} \otimes \big(\mathcal{S}_{A}(\mathbb{R}^3; \mathbb{C}^4)^{\otimes m}\big)^*$. In that case, for any $k \in \mathbb{Z}$, $q>0$ with 
$\alpha+\beta \leq 2q$, it holds
\begin{multline*}
\| \Xi_{l,m}(\kappa_{l,m}) \Phi\|_{k}  \\ \leq 
\rho^{-q/2} \big(l^lm^m\big)^{1/2}
\Bigg( \frac{\rho^{-\alpha/2}}{-\alpha e \textrm{\emph{ln}} \rho} \Bigg)^{l/2} \, 
\Bigg( \frac{\rho^{-\beta/2}}{-\beta e \textrm{\emph{ln}} \rho} \Bigg)^{m/2} \, 
|\kappa_{l,m}|_{{}_{l,m;k,-(k+q)}}
\| \Phi \|_{k+q}, 
\end{multline*}
for all $\Phi \in \big(\mathcal{S}_{A}(\mathbb{R}^3; \mathbb{C}^4)\big)$. 
\end{twr}

Here for $f \in \big(\mathcal{S}_{A}(\mathbb{R}^3; \mathbb{C}^4)^{\otimes (l+m)}\big)^*$
we have defined after \cite{obata-book}, Chap. 3.4
\[
|f|_{{}_{l,m;k,q}} \overset{\textrm{df}}{=} 
\Bigg( \sum \limits_{\boldsymbol{\textrm{i}}, \boldsymbol{\textrm{j}}} |\langle f, 
e(\boldsymbol{\textrm{i}}) \otimes e(\boldsymbol{\textrm{j}}) \rangle |^{2} |e(\boldsymbol{\textrm{i}})|_{{}_{k}}^{2} 
|e(\boldsymbol{\textrm{j}})|_{{}_{q}}^{2} \Bigg)^{1/2}, \,\,\, k,q \in \mathbb{R}.
\]
Recall that here we have used (after \cite{obata-book}) the multiindex notation
\[
\begin{split}
e(\boldsymbol{\textrm{i}}) = e_{{}_{i_1}} \otimes \cdots \otimes  e_{{}_{i_l}}, \,\,\,\,
\boldsymbol{\textrm{i}} = (i_1, \ldots, i_l), \\
e(\boldsymbol{\textrm{j}}) = e_{{}_{j_1}} \otimes \cdots \otimes  e_{{}_{j_m}}, \,\,\,\,
\boldsymbol{\textrm{j}} = (j_1, \ldots, j_m), \\
\end{split}
\]
with $\{e_{{}_{j}}\}_{j=0}^{\infty}$ being the complete orthonormal system in 
\[
L^2(\mathbb{R}^3; \mathbb{C}^4) 
= L^2(\mathbb{R}^3 \sqcup \mathbb{R}^3 \sqcup \mathbb{R}^3 \sqcup \mathbb{R}^3; \mathbb{C})
\]
of eigenvectors of the operator $A$ defined by (\ref{AinL^2(R^3;C^4)}): 
$Ae_{{}_{j}} = \lambda_{{}_{j}} e_{{}_{j}}$, which belong to the nuclear Schwartz space 
\[
e_{{}_{j}} \in \mathcal{S}_{A}(\mathbb{R}^3; \mathbb{C}^4) = 
\mathcal{S}_{A}(\mathbb{R}^3 \sqcup \mathbb{R}^3 \sqcup \mathbb{R}^3 \sqcup \mathbb{R}^3; \mathbb{C}). 
\]
In our case 
\[
\mathbb{R}^3 \sqcup \mathbb{R}^3 \sqcup \mathbb{R}^3 \sqcup \mathbb{R}^3 \ni 
(s, \boldsymbol{\p}) \longmapsto e_{{}_{j}}(s, \boldsymbol{\p}) = \varepsilon_{{}_{j}}(\boldsymbol{\p}), \,\,\, s \in \{1,2,3,4\},
\] 
where $\{\varepsilon_{{}_{j}}\}_{j=0}^{\infty}$ is the system of products $\varepsilon_{{}_{j}}
= h_{{}_{n_j}}h_{{}_{m_j}}h_{{}_{l_j}}$, $\lambda_j = \mu_{{}_{n_j}} + \mu_{{}_{m_j}} + \mu_{{}_{l_j}} +1$ 
of Hermite functions --
composing the complete orthonormal system of eigenfunctions of the hamiltonian operator
$H_{(3)}$ in $L^2(\mathbb{R}^3; \mathbb{C})$ of the three dimensional oscillator (here 
$\mu_{{}_{i}}$ is the eigenvalue corresponding to the Hermite fuction $h_{{}_{i}}$
of the one dimensional oscillator hamiltonian $H_{(1)}$).  When considering the white noise
construction of zero mass fields we will likewise encounter another family of nuclear spaces
$\mathcal{S}_{A}(\mathbb{R}^3, \mathbb{C}^4) = \mathcal{S}^{0}(\mathbb{R}^3, \mathbb{C}^4)$, or 
$\mathcal{S}_{A}(\mathbb{R}^3, \mathbb{C}^n) = \mathcal{S}^{0}(\mathbb{R}^3, \mathbb{C}^n)$
with another standard operator $A = \oplus A^{(3)}$ on 
$L^2(\mathbb{R}^3; \mathbb{C}^4) = \oplus L^2(\mathbb{R}^3; \mathbb{C})$, or on  $L^2(\mathbb{R}^3; \mathbb{C}^n) = \oplus L^2(\mathbb{R}^3; \mathbb{C})$, with $A^{(3)} \neq H_{(3)}$.

In particular we have the following Corollary (the fermi analogue of Prop. 4.3.10 of \cite{obata-book}) 
\begin{cor}\label{D_xi=int(xiPartial)}
For $y \in \mathcal{S}_{A}(\mathbb{R}^3, \mathbb{C}^4)^*$ it holds that
\[
D_{\overline{y}} = \Xi_{0,1}(y) = \sum \limits_{s = 1}^{4} \int \limits_{\mathbb{R}^3} y(s, \boldsymbol{\p}) 
\partial_{s, \boldsymbol{\p}} \ud^3 \boldsymbol{\p}, \,\,\,\,
D_{y}^+ = \Xi_{1,0}(y) = \sum \limits_{s = 1}^{4} \int \limits_{\mathbb{R}^3} y(s, \boldsymbol{\p}) 
\partial_{s, \boldsymbol{\p}}^{*} \ud^3 \boldsymbol{\p}.
\]
In particular,
\[
\partial_{s, \boldsymbol{\p}} = \Xi_{0,1}(\delta_{s, \boldsymbol{\p}}), \,\,\,\,\,
\partial_{s, \boldsymbol{\p}}^{*} = \Xi_{1,0}(\delta_{s, \boldsymbol{\p}}).
\] 
For $y \in \mathcal{S}_{A}(\mathbb{R}^3, \mathbb{C}^4) \subset \mathcal{S}_{A}(\mathbb{R}^3, \mathbb{C}^4)^*$
\[
\Xi_{0,1}(y), \Xi_{1,0}(y) \in \mathscr{L}\Big( \, \big( \mathcal{S}_{A}(\mathbb{R}^3, \mathbb{C}^4)\big) \, , \, \big( \mathcal{S}_{A}(\mathbb{R}^3, \mathbb{C}^4)\big)  \, \Big)
\]
and the linear maps
\[
\begin{split}
\mathcal{S}_{A}(\mathbb{R}^3, \mathbb{C}^4) \ni y \longmapsto 
\Xi_{0,1}(y) = D_{\overline{y}} \in 
\mathscr{L}\Big( \, \big( \mathcal{S}_{A}(\mathbb{R}^3, \mathbb{C}^4)\big),  \, 
\big( \mathcal{S}_{A}(\mathbb{R}^3, \mathbb{C}^4)\big)  \, \Big) \\
\mathcal{S}_{A}(\mathbb{R}^3, \mathbb{C}^4) \ni y \longmapsto 
\Xi_{1,0}(y) = D_{y}^+ \in \mathscr{L}\Big( \, \big( \mathcal{S}_{A}(\mathbb{R}^3, \mathbb{C}^4)\big),
\, \big( \mathcal{S}_{A}(\mathbb{R}^3, \mathbb{C}^4)\big)  \, \Big)
\end{split}
\]
are continuous. 

Moreover, for $y_1, \ldots, y_m \in \mathcal{S}_{A}(\mathbb{R}^3, \mathbb{C}^4)^*$ it holds
\begin{multline*}
D_{\overline{y_1}} \cdots D_{\overline{y_m}} = \Xi_{0,m}(y_1 \otimes \cdots \otimes y_m) \\ =
\Xi_{0,m}(y_1 \, \widehat{\otimes} \,  \cdots \,  \widehat{\otimes} \, y_m)  \\ =
\sum \limits_{s_1, \ldots, s_m = 1}^{4} \int \limits_{(\mathbb{R}^3)^m} \, 
y_1(s_1, \boldsymbol{\p}_1) \cdots y_1(s_m, \boldsymbol{\p}_m) \, 
\partial_{s_1, \boldsymbol{\p}_1} \cdots \partial_{s_m, \boldsymbol{\p}_m} \,
\ud^3 \boldsymbol{\p}_1 \cdots \ud^3 \boldsymbol{\p}_m \\ =
(m!)^{-1} \sum \limits_{\pi \in \mathfrak{S}_m} \, \textrm{\emph{sign}} \, \pi \,\,
\sum \limits_{s_{1}, \ldots, s_{m} = 1}^{4} \int \limits_{(\mathbb{R}^3)^m} \, 
y_1(s_{{}_{\pi(1)}}, \boldsymbol{\p}_{{}_{\pi(1)}}) \cdots y_m
(s_{{}_{\pi(m)}}, \boldsymbol{\p}_{{}_{\pi(m)}}) \,
\times \\
\times \,
\partial_{s_1, \boldsymbol{\p}_1} \cdots 
\partial_{s_m, \boldsymbol{\p}_m} \,
\ud^3 \boldsymbol{\p}_1 \cdots \ud^3 \boldsymbol{\p}_m,
\end{multline*}
where $\pi$ runs over the set $\mathfrak{S}_m$ of all permutations of the numbers $1,2,\ldots, m$.
\end{cor}

Note that because for $y,y' \in \mathcal{S}_{A}(\mathbb{R}^3, \mathbb{C}^4)^*$,
$\xi,\xi' \in \mathcal{S}_{A}(\mathbb{R}^3, \mathbb{C}^4)$ all the operators
\[
\begin{split}
D_{\overline{y}} = \Xi_{0,1}(y) = \sum \limits_{s = 1}^{4} \int \limits_{\mathbb{R}^3} y(s, \boldsymbol{\p}) 
\partial_{s, \boldsymbol{\p}} \ud^3 \boldsymbol{\p}, \,\, \textrm{and} \,\,
D_{\xi}^+ = \Xi_{1,0}(\xi) = \sum \limits_{s = 1}^{4} \int \limits_{\mathbb{R}^3} \xi(s, \boldsymbol{\p}) 
\partial_{s, \boldsymbol{\p}}^{*} \ud^3 \boldsymbol{\p}, \\
D_{\overline{y'}} = \Xi_{0,1}(y') = \sum \limits_{s = 1}^{4} \int \limits_{\mathbb{R}^3} y'(s, \boldsymbol{\p}) 
\partial_{s, \boldsymbol{\p}} \ud^3 \boldsymbol{\p}, \,\, \textrm{and} \,\,
D_{\xi'}^+ = \Xi_{1,0}(\xi') = \sum \limits_{s = 1}^{4} \int \limits_{\mathbb{R}^3} \xi'(s, \boldsymbol{\p}) 
\partial_{s, \boldsymbol{\p}}^{*} \ud^3 \boldsymbol{\p}, 
\end{split}
\]
belong to $\mathscr{L}\Big( \, \big( \mathcal{S}_{A}(\mathbb{R}^3, \mathbb{C}^4)\big) \, , \, \big( \mathcal{S}_{A}(\mathbb{R}^3, \mathbb{C}^4)\big)  \, \Big)$ then their products as operators transforming Hida space into Hida space
are meaningfull. We have in this case the canonical anticommutation rules
\begin{equation}\label{[Xi_01, Xi_1,0]}
\big\{ \Xi_{0,1}(y), \Xi_{1,0}(\xi) \big\} = \langle y, \xi \rangle \, \boldsymbol{1},
\,\,\,
\big\{ \Xi_{0,1}(y), \Xi_{0,1}(y') \big\} 
= \big\{ \Xi_{1,0}(\xi), \Xi_{1,0}(\xi') \big\} = 0,
\end{equation}
or
\[
\big\{ D_{\overline{y}}, D_{\xi}^+ \big\} = \langle y, \xi \rangle \, \boldsymbol{1},
\,\,\,
\big\{ D_{\overline{y}}, D_{\overline{y'}} \big\} 
= \big\{ D_{\xi}^+, D_{\xi'}^+ \big\} = 0.
\]
They are frequently written in the form (which should be understood properly
in a rigorous sense explained below)
\begin{equation}\label{[partial, partial^*]}
\big\{\partial_{s, \boldsymbol{\p}}, \partial_{s', \boldsymbol{\p}'}^* \big\} = 
\delta_{s, \boldsymbol{\p}}(s', \boldsymbol{\p}'), \,\,\, 
\big\{\partial_{s, \boldsymbol{\p}}, \partial_{s', \boldsymbol{\p}'} \big\} = 
\big\{\partial_{s, \boldsymbol{\p}}^*, \partial_{s', \boldsymbol{\p}'}^* \big\} = 0, 
\end{equation}
or using the notation of physicists
\begin{multline*}
\big\{ a_{s}(\boldsymbol{\p}), a_{s'}(\boldsymbol{\p}')^+ \big\} = 
\delta_{ss'} \delta(\boldsymbol{\p} -\boldsymbol{\p}'), \,\,\,
\big\{ a_{s}(\boldsymbol{\p}), a_{s'}(\boldsymbol{\p}') \big\} = 
\big\{ a_{s}(\boldsymbol{\p})^+, a_{s'}(\boldsymbol{\p}')^+ \big\} = 0, \\
s,s' \in \{1,2,3,4\}
\end{multline*}
or (like in \cite{Scharf}, p. 82)
\begin{equation}\label{[b,b^+],[d,d^+]}
\boxed{
\begin{split}
\big\{ b_{s}(\boldsymbol{\p}), b_{s'}(\boldsymbol{\p}')^+ \big\} = 
\delta_{ss'} \delta(\boldsymbol{\p} -\boldsymbol{\p}'), \,\,\,
\big\{ b_{s}(\boldsymbol{\p}), b_{s'}(\boldsymbol{\p}') \big\} = 
\big\{ b_{s}(\boldsymbol{\p})^+, b_{s'}(\boldsymbol{\p}')^+ \big\} =0, \\
\big\{ d_{s}(\boldsymbol{\p}), d_{s'}(\boldsymbol{\p}')^+ \big\} = 
\delta_{ss'} \delta(\boldsymbol{\p} -\boldsymbol{\p}'), \,\,\,
\big\{ d_{s}(\boldsymbol{\p}), d_{s'}(\boldsymbol{\p}') \big\} = 
\big\{ d_{s}(\boldsymbol{\p})^+, d_{s'}(\boldsymbol{\p}')^+ \big\} =0,\\
\big\{ b_{s}(\boldsymbol{\p}), d_{s'}(\boldsymbol{\p}')^+ \big\} = 0, \,\,\, s,s'=1,-1,
\end{split}
}
\end{equation}
with the obvious identifications 
\[
\begin{split}
D_{y}= a(y) = a(y|_{{}_{s=1}} \oplus y|_{{}_{s=2}} \oplus y|_{{}_{s=3}} \oplus y|_{{}_{s=4}}) \\ =
b(y|_{{}_{s=1}} \oplus y|_{{}_{s=2}} \oplus 0 \oplus 0) + 
d(0 \oplus 0 \oplus y|_{{}_{s=3}} \oplus y|_{{}_{s=4}}) \\
a(y) = \sum \limits_{s = 1}^{4} \int \limits_{\mathbb{R}^3} \overline{y(s, \boldsymbol{\p})} 
a_{s} (\boldsymbol{\p}) \ud^3 \boldsymbol{\p}, \\
b(y|_{{}_{s=1}} \oplus y|_{{}_{s=2}} \oplus 0 \oplus 0) = 
\sum \limits_{s = 1}^{2} \int \limits_{\mathbb{R}^3} \overline{y(s, \boldsymbol{\p})} 
a_{s} (\boldsymbol{\p}) \ud^3 \boldsymbol{\p} =
\sum \limits_{s = 1}^{2} \int \limits_{\mathbb{R}^3} \overline{y(s, \boldsymbol{\p})} 
b_{-2s+3} (\boldsymbol{\p}) \ud^3 \boldsymbol{\p}, \\
d(0 \oplus 0 \oplus y|_{{}_{s=3}} \oplus y|_{{}_{s=4}}) = 
\sum \limits_{s = 3}^{4} \int \limits_{\mathbb{R}^3} \overline{y(s, \boldsymbol{\p})} 
a_{s} (\boldsymbol{\p}) \ud^3 \boldsymbol{\p} =
\sum \limits_{s = 3}^{4} \int \limits_{\mathbb{R}^3} \overline{y(s, \boldsymbol{\p})} 
d_{-2s+7} (\boldsymbol{\p}) \ud^3 \boldsymbol{\p}
\end{split}
\]
for 
\[
y \in \mathcal{S}_{A}(\mathbb{R}^3, \mathbb{C}^4)^*.
\]

The relations (\ref{[partial, partial^*]}) or equivalently (\ref{[b,b^+],[d,d^+]})
should be interpreted properly. Namely the first set of relations (\ref{[Xi_01, Xi_1,0]})
in the particular case $y,\xi \in \mathcal{S}_{A}(\mathbb{R}^3, \mathbb{C}^4)$
reduces to 
\[
\big\{ \Xi_{0,1}(y), \Xi_{1,0}(\xi) \big\} = ( \overline{y}, \xi)_0 \, \boldsymbol{1}
\]
with the inner product $(\cdot, \cdot)_0$ on $L^2(\mathbb{R}^3; \mathbb{C}^4)$.
Using the continuity of the inner product $(\cdot, \cdot)_0$
in the nuclear topology of $\mathcal{S}_{A}(\mathbb{R}^3, \mathbb{C}^4)
\subset L^2(\mathbb{R}^3; \mathbb{C}^4)$ (compare \cite{GelfandIV}, Ch. I.4.2) and  the fact that 
$\mathcal{S}_{A}(\mathbb{R}^3, \mathbb{C}^4)$ is a Frech\'et space,
it follows that the bilinear map $y\times \xi \mapsto 
(\overline{y}, \xi)_{0} \bold{1}$
defines an operator-valued distribution (compare e. g. Proposition 1.3.11 of \cite{obata-book}):
\begin{multline*}
\mathcal{S}_{A}(\mathbb{R}^3, \mathbb{C}^4) \otimes \mathcal{S}_{A}(\mathbb{R}^3, \mathbb{C}^4) \ni \zeta \mapsto \Xi_{0,0}(\zeta)  \\
= \int \limits_{\mathbb{R}^{3} \times \mathbb{R}^{3}}  
\zeta (s,\boldsymbol{\p}, s' ,\boldsymbol{\p}')
\tau(s,\boldsymbol{\p}, s', \boldsymbol{\p}')\, \bold{1} \, \ud^3 p \ud^3 p'  = 
\tau(\zeta) \bold{1}
\end{multline*}
where $\tau \in (\mathcal{S}_{A}(\mathbb{R}^3, \mathbb{C}^4) \otimes \mathcal{S}_{A}(\mathbb{R}^3, \mathbb{C}^4))^*$ 
is defined by
\[
\langle \tau, y \otimes \xi \rangle = (\overline{y}, \xi)_0
= \langle y, \xi \rangle, \,\,\, y, \xi \in \mathcal{S}_{A}(\mathbb{R}^3, \mathbb{C}^4), 
\]
therefore we have
\begin{multline*}
\Xi_{0,0}(y \otimes \xi) 
= \big\{ \Xi_{0,1}(y), \Xi_{1,0}(\xi) \big\} \\
= \sum \limits_{s,s'} \, \int \limits_{\mathbb{R}^{3} \times \mathbb{R}^{3}}  
y \otimes \xi (s',\boldsymbol{\p}', s,\boldsymbol{\p}) \,\,
\delta_{s s'} \, \delta(\boldsymbol{\p}- \boldsymbol{\p}') \, \bold{1} \, \ud^3 p \ud^3 p' \\
= \sum \limits_{s,s'} \, \int \limits_{\mathbb{R}^{3} \times \mathbb{R}^{3}}  
y(s',\boldsymbol{\p}') \, \xi(s,\boldsymbol{\p}) \,\,
\delta_{s s'} \, \delta(\boldsymbol{\p}- \boldsymbol{\p}') \, \bold{1} \, \ud^3 p \ud^3 p',
\end{multline*}
and
\[
\big\{\partial_{s, \boldsymbol{\p}}, \partial_{s', \boldsymbol{\p}'}^* \big\} 
= \delta_{s s'} \, \delta(\boldsymbol{\p} - \boldsymbol{\p}') \bold{1}.
\]

Note here that within the white noise construction of Hida
the operators $\partial_{s,\boldsymbol{\p}}, \partial_{s, \boldsymbol{\p}}^*$
are well defined at each point $(s,\boldsymbol{\p}) \in \sqcup \, \mathbb{R}^3
= \mathbb{R}^3 \sqcup \mathbb{R}^3 \sqcup \mathbb{R}^3 \sqcup \mathbb{R}^3$, and 
there is no need for treating them as operator-valued distributions
when using the calculus for integral kernel operators.

The exceptional situations, which involve more factors 
$\partial_{s,\boldsymbol{\p}}, \partial_{s, \boldsymbol{\p}}^*$ in non ``normal'' order, 
in which we are forced to treat them as distributions
are however easily and naturally grashped within the white noise calculus. 
The first such situation where we need to use distributional interpretation we encounter when
trying to give proper meaning to (\ref{[partial, partial^*]}) or equivalently
(\ref{[b,b^+],[d,d^+]}) which formally involve both
\begin{equation}\label{HiaPartial*Partial,PartialPartial*}
\partial_{s', \boldsymbol{\p}'}^* \partial_{s, \boldsymbol{\p}}  \,\,\,
\textrm{and} \,\,\,
\partial_{s, \boldsymbol{\p}} \partial_{s', \boldsymbol{\p}'}^*,
\end{equation}  
with more than just one factor of the type $\partial_{s,\boldsymbol{\p}}, \partial_{s, \boldsymbol{\p}}^*$
containing both $\partial_{s,\boldsymbol{\p}}$ and the adjoint operator
$\partial_{s, \boldsymbol{\p}}^*$. Note that the first of the expressions (that in the ``normal'' order)
in (\ref{HiaPartial*Partial,PartialPartial*}) is meaningfull as a continuous operator transforming the Hida space into its dual. But the second expression in (\ref{HiaPartial*Partial,PartialPartial*}) is meaningless
as a generalized operator on the Hida space (or its dual). Nonetheless both expressions
in (\ref{HiaPartial*Partial,PartialPartial*}) are well defined as operator-valued distributions.
Indeed the coresponding maps
\[
\chi \times \xi \longmapsto \Xi_{1,0}(\xi) \circ \Xi_{0,1}(\chi), \,\,\,\,
\chi \times \xi \longmapsto  \Xi_{0,1}(\chi) \circ \Xi_{1,0}(\xi)
\]
are bilinear and separately continuous as maps 
\[
\mathcal{S}_{A}(\mathbb{R}^3, \mathbb{C}^4) \times 
\mathcal{S}_{A}(\mathbb{R}^3, \mathbb{C}^4) \longrightarrow 
\mathscr{L}\Big( \, \big( \mathcal{S}_{A}(\mathbb{R}^3, \mathbb{C}^4)\big),
\, \big( \mathcal{S}_{A}(\mathbb{R}^3, \mathbb{C}^4)\big)  \, \Big). 
\]
Therefore because $\mathcal{S}_{A}(\mathbb{R}^3, \mathbb{C}^4)$ is a Fr\'echet space then by Proposition 1.3.11 of \cite{obata-book}
there exist the corresponding operator-valued distributions, written 
\begin{multline}\label{Distribution:Partial^*Partial}
\chi \otimes \xi \longmapsto  \\
\sum \limits_{s,s' = 1}^{4} \int \limits_{\mathbb{R}^3} \chi \otimes \xi(s', \boldsymbol{\p}', s, \boldsymbol{\p}) \,
\partial_{s', \boldsymbol{\p}'}^* \partial_{s, \boldsymbol{\p}} \, \ud^3 \boldsymbol{\p}' \ud^3 \boldsymbol{\p}
= \Xi_{1,1}(\chi \otimes \xi) = \Xi_{1,0}(\xi) \circ \Xi_{0,1}(\chi),
\end{multline} 
and
\begin{equation}\label{Distribution:PartialPartial^*}
\chi \otimes \xi \longmapsto 
\sum \limits_{s,s' = 1}^{4} \int \limits_{\mathbb{R}^3} \chi \otimes \xi(s', \boldsymbol{\p}', s, \boldsymbol{\p}) \,
\partial_{s, \boldsymbol{\p}} \partial_{s', \boldsymbol{\p}'}^*  \, \ud^3 \boldsymbol{\p}' \ud^3 \boldsymbol{\p}
= \Xi_{0,1}(\chi) \circ \Xi_{1,0}(\xi),
\end{equation}
continuous as maps
\[
\mathcal{S}_{A}(\mathbb{R}^3, \mathbb{C}^4)^{\otimes 2} \longrightarrow 
\mathscr{L}\Big( \, \big( \mathcal{S}_{A}(\mathbb{R}^3, \mathbb{C}^4)\big),
\, \big( \mathcal{S}_{A}(\mathbb{R}^3, \mathbb{C}^4)\big)  \, \Big). 
\]
Here in the formula  (\ref{Distribution:PartialPartial^*}) the 
``distributional integral kernel'', say operator-valued distribution
$\partial_{s, \boldsymbol{\p}} \partial_{s', \boldsymbol{\p}'}^*$, has only formal meaning, 
and cannot be interpreted as any actual generalized operator on the Hida space.
But the integral in the formula (\ref{Distribution:Partial^*Partial}) 
represents an integral kernel operator so that the equalities in the formula 
(\ref{Distribution:Partial^*Partial}) is actally a theorem which can immediatelly be 
checked by application of definition of Hida operators. But likewise the operator 
$\Xi_{0,1}(\chi) \circ \Xi_{1,0}(\xi)$ in the formula 
(\ref{Distribution:PartialPartial^*}), transforming continously the Hida space into itself,
 can be expressed as a (here finite) sum of integral kernel operators. This follows from the general theorem,
\cite{obata} Thm. 6.1 or \cite{obata-book}, Thm 4.5.1 (which can as well be proved for fermi case without any essential
changes in the proof of \cite{obata}, \cite{obata-book}). However our case is so simple that the corresponding
decomposition of the operator $\Xi_{0,1}(\chi) \circ \Xi_{1,0}(\xi)$ into the sum of integral kernel 
operators can be proven to be equal   
\begin{multline}\label{:PartialPartial*:}
\chi \otimes \xi \longmapsto \Xi_{0,1}(\chi) \circ \Xi_{1,0}(\xi)  \\ = 
- \Xi_{1,1}(\chi \otimes \xi)
\,\,\,\,
+
\,\,\,\,
\Xi_{0,0}(\chi \otimes \xi)  \\
- \sum \limits_{s,s' = 1}^{4} \int \limits_{\mathbb{R}^3} \chi \otimes \xi(s', \boldsymbol{\p}', s, \boldsymbol{\p}) \,
\partial_{s', \boldsymbol{\p}'}^* \partial_{s, \boldsymbol{\p}} \, \ud^3 \boldsymbol{\p}' \ud^3 \boldsymbol{\p}
\,\,\,\,
+
\,\,\,\,
 ( \overline{\chi}, \xi)_0 \, \boldsymbol{1} \\
= - \sum \limits_{s,s' = 1}^{4} \int \limits_{\mathbb{R}^3} \chi \otimes \xi(s', \boldsymbol{\p}', s, \boldsymbol{\p}) \,
\partial_{s', \boldsymbol{\p}'}^* \partial_{s, \boldsymbol{\p}} \, \ud^3 \boldsymbol{\p}' \ud^3 \boldsymbol{\p}
\\
+
\sum \limits_{s,s' = 1}^{4} \, \int \chi \otimes \xi(s', \boldsymbol{\p}', s, \boldsymbol{\p}) \,
 \big\{\partial_{s, \boldsymbol{\p}}, \partial_{s', \boldsymbol{\p}'}^* \big\} \,
\ud^3 \boldsymbol{\p}' \ud^3 \boldsymbol{\p},
\end{multline}
using the definition of Hida operators and the relations  (\ref{[Xi_01, Xi_1,0]}).

The operator-valued distribution (\ref{:PartialPartial*:}) is called the normal order form distribution
$\boldsymbol{:} \partial_{s, \boldsymbol{\p}} \partial_{s', \boldsymbol{\p}'}^* \boldsymbol{:} + pairing$ 
of the operator-valued distribution (\ref{Distribution:PartialPartial^*}) symbolized by 
$\partial_{s, \boldsymbol{\p}} \partial_{s', \boldsymbol{\p}'}^*$, which is written symbolically
\[
\partial_{s, \boldsymbol{\p}} \partial_{s', \boldsymbol{\p}'}^* = \,\,\,
\boldsymbol{:} \partial_{s, \boldsymbol{\p}} \partial_{s', \boldsymbol{\p}'}^* \boldsymbol{:} 
\,\,\, + pairing = - \partial_{s', \boldsymbol{\p}'}^* \partial_{s, \boldsymbol{\p}} 
+ 
\big\{\partial_{s, \boldsymbol{\p}}, \partial_{s', \boldsymbol{\p}'}^* \big\}
\]

Similarily we have for decomposition of the operator-valued distributions
involving more factors 
\begin{equation}\label{...Pariali...Partial*j...}
\cdots \partial_{s_i,\boldsymbol{\p}_i}  \cdots \cdots  \partial_{s_j, \boldsymbol{\p}_j}^*  \cdots
\end{equation} 
of the type $\partial_{s,\boldsymbol{\p}}, \partial_{s, \boldsymbol{\p}}^*$, not necessary normally ordered, 
into sum of components  with ``normally'' ordered Hida's differential operators, and similarily
as in the ``Wick theorem'' in \cite{Bogoliubov_Shirkov}, Chap. III.
Note that although reduction of such distributions into ``normal form'' follows from the general theorem
for decompostions of the corresponding operators
\begin{equation}\label{...Xi01(chii)...Xi10(xij)...}
\cdots \circ \Xi_{0,1}(\chi_i) \circ \cdots \cdots \circ \Xi_{1,0}(\xi_j) \circ \cdots
\end{equation} 
transforming continously the Hida space into itself into sums of integral kernel operators
(\cite{obata} Thm. 6.1 or \cite{obata-book}, Thm 4.5.1 ), 
the simple operator (\ref{...Xi01(chii)...Xi10(xij)...}) can be decomposed
by induction, using the definition of Hida operators and the relations (\ref{[Xi_01, Xi_1,0]}).
We may also compute decompositions of more involved  distributions then (\ref{...Pariali...Partial*j...}) 
which contain ``normally orderred'' factors $\partial_{s, \boldsymbol{\p}}^* \partial_{s, \boldsymbol{\p}}$
with both $\partial_{s, \boldsymbol{\p}}^*$ and $ \partial_{s, \boldsymbol{\p}}$ evaluated 
at the same point $(s,\boldsymbol{\p})$, as well defined distributions:
\begin{equation}\label{...Pariali...Partial*pjPartialpj...}
\cdots \partial_{s_i,\boldsymbol{\p}_i}  \cdots \cdots  
\partial_{s_j, \boldsymbol{\p}_j}^*\partial_{s_j, \boldsymbol{\p}_j}  \cdots
\end{equation} 
with the correspoding operators
\begin{equation}\label{...Xi01(chii)...Xi11(xij(tau))...}
\cdots \circ \Xi_{0,1}(\chi_i) \circ \cdots \cdots \circ \Xi_{1,1}\big((\xi_j\otimes 1)\tau\big) \circ \cdots
\end{equation} 
transforming continously the Hida space into itself. Here $\tau \in \mathcal{S}_{A}(\mathbb{R}^3, \mathbb{C}^4) \otimes \mathcal{S}_{A}(\mathbb{R}^3, \mathbb{C}^4)^*$ 
is uniquelly determined by the formula
\[
\langle \tau, y \otimes \xi \rangle  
= \langle y, \xi \rangle = (\overline{y}, \xi )_0, \,\,\, y, \xi \in \mathcal{S}_{A}(\mathbb{R}^3, \mathbb{C}^4). 
\]
By Theorem \ref{Xi_l,m:Hida->Hida} the operator $\Xi_{1,1}\big((\xi_j \otimes 1)\tau \big)$, with
$\xi_j\in \mathcal{S}_{A}(\mathbb{R}^3, \mathbb{C}^4)$, belongs to 
\[
\mathscr{L}\Big( \, \big( \mathcal{S}_{A}(\mathbb{R}^3, \mathbb{C}^4)\big),
\, \big( \mathcal{S}_{A}(\mathbb{R}^3, \mathbb{C}^4)\big)  \, \Big),
\] 
and the map 
\[
\mathcal{S}_{A}(\mathbb{R}^3, \mathbb{C}^4) \ni \xi_j \longmapsto 
\Xi_{1,1}\big((\xi_j \otimes 1)\tau\big) \in \mathscr{L}\Big( \, \big( \mathcal{S}_{A}(\mathbb{R}^3, \mathbb{C}^4)\big),
\, \big( \mathcal{S}_{A}(\mathbb{R}^3, \mathbb{C}^4)\big)  \, \Big)
\]
is continuous, similarly as for the remaining integral kernel operators $\Xi_{0,1}(\chi_i), \ldots$ 
in (\ref{...Xi01(chii)...Xi11(xij(tau))...}), so that indeed 
(\ref{...Xi01(chii)...Xi11(xij(tau))...})
determines a well defined distribution transforming continously 
\[
\mathcal{S}_{A}(\mathbb{R}^3, \mathbb{C}^4)^{\otimes n} \longrightarrow 
\mathscr{L}\Big( \, \big( \mathcal{S}_{A}(\mathbb{R}^3, \mathbb{C}^4)\big),
\, \big( \mathcal{S}_{A}(\mathbb{R}^3, \mathbb{C}^4)\big)  \, \Big).
\]
By the general theorem (\cite{obata} Thm. 6.1 or \cite{obata-book}, Thm 4.5.1 ) the operator 
(\ref{...Xi01(chii)...Xi11(xij(tau))...}) can be uniquely decomposed into (here finite)
sum of integral kernel operators, thus providing the decomposition of the distribution
(\ref{...Pariali...Partial*pjPartialpj...}) into sum of components, each in the ``normal order''.
We do not enter here into the investigation of the ``Wick theorem'' for distributions expressed as 
simple monomials in the Hida differential operators. 
In fact the ``Wick theorem'' of \cite{Bogoliubov_Shirkov}, Chap III, involves the free field operators
and not merely the (simpler) operators 
$a(\delta_{s,\boldsymbol{\p}}) = \partial_{s, \boldsymbol{\p}} = a_s(\boldsymbol{\p}), a(\delta_{s,\boldsymbol{\p}})^+ = \partial_{s', \boldsymbol{\p}'}^* 
= a_{s'}(\boldsymbol{\p}')^+$. It is true that Wick theorem for free field operators may be immediately reduced to the 
Wick theorem for the corresponding $\partial_{s, \boldsymbol{\p}} = a_s(\boldsymbol{\p}), \partial_{s', \boldsymbol{\p}'}^* 
= a_{s'}(\boldsymbol{\p}')^+$ by utilizing the corresponding unitary isomorphisms $U$ (relating the standard Gelfand triples over the corresponding $L^2(\mathbb{R}^3; \mathbb{C}^n)$ with that over the single particle Hilbert spaces), in our case of Dirac field the isomorphism $U$ relating the Gelfand triples 
(\ref{SinglePartGelfandTriplesForPsi}), which serves to construct the field out of the standard Hida operators through the formula (\ref{a(U(u+v))=a'(u+v)}). However starting with ``Wick theorem'' for the standard Hida differential operators 
woud not be the correct succession for doing things, 
because we are interested in very special kind
of distributions to be decomposed, which arise as polynomials of free fields containing concrete form of (Wick ordered)
interacting term (or terms).  
Therefore we should first construct explicitly the free fields in therms of Hida differential operators 
(as special kinds of integral kernel operators, with vector-valued kernels), and then prove ``Wick theorem''
for polynomilas of free fields containing the Wick ordered polynomials as interaction terms. 

Here we have only taken the opportunity to emphasize the proper mathematical basis for the 
``Wick theorem for free fields'' as stated in \cite{Bogoliubov_Shirkov}, Chap. III, which 
becomes a particular case of general theorem, \cite{obata} Thm. 6.1 or \cite{obata-book}, Thm 4.5.1  (extended on genealized operators in the tensor product of several Fock -- bose and fermi -- spaces) on decomposition
of operators transforming continously the Hida space into itself into a series of integral kernel
operators. 

Summing up the discussion of the relations  (\ref{[partial, partial^*]}) or equivalently (\ref{[b,b^+],[d,d^+]})
and of the ``Wick theorem for Hida differential operators'', 
we should emphasize that (\ref{[partial, partial^*]}) or (\ref{[b,b^+],[d,d^+]}) 
should be understood as equalities of operator valued distributions, transforming continously 
\[
\mathcal{S}_{A}(\mathbb{R}^3, \mathbb{C}^4)^{\otimes 2} \longrightarrow 
\mathscr{L}\Big( \, \big( \mathcal{S}_{A}(\mathbb{R}^3, \mathbb{C}^4)\big),
\, \big( \mathcal{S}_{A}(\mathbb{R}^3, \mathbb{C}^4)\big)  \, \Big). 
\]

Now having given the Hida operators $a(\delta_{s,\boldsymbol{\p}}) = \partial_{s, \boldsymbol{\p}} = a_s(\boldsymbol{\p}), \partial_{s', \boldsymbol{\p}'}^* = a_{s'}(\boldsymbol{\p}')^+$, $a(w), a(w)^*$, $w \in 
\mathcal{S}_{A}(\mathbb{R}^3, \mathbb{C}^4)\big)^*$ corresponding to the Fock  lifting $\Gamma$
of the first standard Gelfand triple in (\ref{SinglePartGelfandTriplesForPsi}), we can now utilize the unitary isomorphism $U$, given by (\ref{isomorphismU}), 
relating the triples in (\ref{SinglePartGelfandTriplesForPsi}), and then construct the free Dirac field
as Hida generalized operator, using $a(\delta_{s,\boldsymbol{\p}}) = \partial_{s, \boldsymbol{\p}} = a_s(\boldsymbol{\p}), 
a(\delta_{s,\boldsymbol{\p}})^+ = \partial_{s', \boldsymbol{\p}'}^* = a_{s'}(\boldsymbol{\p}')^+$,
$a(w), a(w)^*$, $w \in \mathcal{S}_{A}(\mathbb{R}^3, \mathbb{C}^4)\big)^*$ and the formula 
(\ref{a(U(u+v))=a'(u+v)}):
\begin{multline*}
\boldsymbol{\psi}(\phi) = a'\big(P^\oplus\widetilde{\phi}|_{{}_{\mathscr{O}_{m,0,0,0}}} \oplus 0\big) + 
a'\Big( 0 \oplus \big(P^\ominus\widetilde{\phi}|_{{}_{\mathscr{O}_{-m,0,0,0}}}\big)^c \Big)^+ \\
 = a\Big(U\big(P^\oplus\widetilde{\phi}|_{{}_{\mathscr{O}_{m,0,0,0}}} \oplus 0\big)\Big) + 
a\Bigg(U\Big( 0 \oplus \big(P^\ominus\widetilde{\phi}|_{{}_{\mathscr{O}_{-m,0,0,0}}}\big)^c \Big)\Bigg)^+, 
\end{multline*}
for 
\begin{multline*}
0 \oplus \big(P^\ominus\widetilde{\phi}|_{{}_{\mathscr{O}_{-m,0,0,0}}}\big)^c, \\
\,\,\, \textrm{and} \,\,\,
P^\oplus\widetilde{\phi}|_{{}_{\mathscr{O}_{m,0,0,0}}} \oplus 0 \in
E, \phi \in \mathscr{E} = \mathcal{S}(\mathbb{R}^4; \mathbb{C}^4) = 
\mathcal{S}_{\oplus H_{(4)}}(\mathbb{R}^4, \mathbb{C}^4)\big).
\end{multline*}

But the (free) Dirac field $\boldsymbol{\psi}$ (and in general quantum free field) is naturally an integral 
kernel operator with well defined kernel equal to integral kernel operator
\begin{multline*}
\boldsymbol{\psi}^a(x) = \sum_{s=1}^{4} \, \int \limits_{\mathbb{R}^3} 
\kappa_{0,1}(s, \boldsymbol{p}; a, x) \,\, \partial_{s, \boldsymbol{\p}} \, \ud^3 \boldsymbol{\p}
+
\sum_{s=1}^{4} \, \int \limits_{\mathbb{R}^3} 
\kappa_{1,0}(s, \boldsymbol{p}; a, x) \,\, \partial_{s, \boldsymbol{\p}}^* \, \ud^3 \boldsymbol{\p} \\
= \Xi_{0,1}\big(\kappa_{0,1}(a,x)\big) + \Xi_{1,0}\big(\kappa_{1,0}(a,x)\big),
\end{multline*}
with vector-valued distributional kernels $\kappa_{lm}(a,x)$ representing distributions 
\begin{multline*}
\kappa_{lm} \in \mathscr{L}\big( \mathcal{S}_{A}(\mathbb{R}^3, \mathbb{C}^4)^{\otimes(l+m)}, \,\,
\mathscr{L}(\mathscr{E}, \mathbb{C})  \big) \cong 
\mathscr{L}\big( \mathcal{S}_{A}(\mathbb{R}^3, \mathbb{C}^4)^{\otimes(l+m)}, \,\,
\mathscr{E}^* \big) \\ 
\cong \big( \mathcal{S}_{A}(\mathbb{R}^3, \mathbb{C}^4)^{\otimes(l+m)} \big)^* \otimes
\mathscr{E}^* \cong
\mathscr{L}\Big( \mathscr{E}, \,\, \big(\mathcal{S}_{A}(\mathbb{R}^3, \mathbb{C}^4)^{\otimes(l+m)}\big)^* 
\Big),
\end{multline*}
in the sense of Obata \cite{obataJFA}. In fact we have used the standard
nuclear space $\mathcal{S}_{A}(\mathbb{R}^3, \mathbb{C}^4)$ instead of the isomorphic
nuclear space $E$, because we have discarded the isomorphism $\Gamma(U)$ in 
(\ref{G(U)^+a(U(u+v))G(U)=a'(u+v)}) or in
(\ref{G(U)^+a(U(u+v))G(U)=a'(u+v)degenerated})), and realize the Hida operators $a'$
in the Fock lifting of the standard Gelfand triple in  
(\ref{SinglePartGelfandTriplesForPsi}). We will find such 
$\mathscr{L}\big( \mathscr{E}, \,\, \mathbb{C} \big) \cong
\mathscr{E}^*$-valued 
distribution kernels $\kappa_{0,1}, \kappa_{1,0} \in 
\mathscr{L}\big( \mathscr{E}, \,\, \mathcal{S}_{A}(\mathbb{R}^3, \mathbb{C}^4)^* 
\big) \cong \mathscr{L}\big( \mathcal{S}_{A}(\mathbb{R}^3, \mathbb{C}^4), \,\,
\mathscr{L}(\mathscr{E}, \mathbb{C})  \big)$ that
\begin{multline}\label{psi=IntKerOpVectValKer}
\boldsymbol{\psi}(\phi) = a'\big(P^\oplus\widetilde{\phi}|_{{}_{\mathscr{O}_{m,0,0,0}}} \oplus 0\big) + 
a'\Big( 0 \oplus \big(P^\ominus\widetilde{\phi}|_{{}_{\mathscr{O}_{-m,0,0,0}}}\big)^c \Big)^+ \\
=  a\Big(U\big(P^\oplus\widetilde{\phi}|_{{}_{\mathscr{O}_{m,0,0,0}}} \oplus 0\big)\Big) + 
a\Bigg(U\Big( 0 \oplus \big(P^\ominus\widetilde{\phi}|_{{}_{\mathscr{O}_{-m,0,0,0}}}\big)^c \Big)\Bigg)^+ \\
 = \sum_{s=1}^{4} \, \int \limits_{\mathbb{R}^3} 
\kappa_{0,1}(\overline{\phi})(s, \boldsymbol{p}) \,\, \partial_{s, \boldsymbol{\p}} \, \ud^3 \boldsymbol{\p}
+
\sum_{s=1}^{4} \, \int \limits_{\mathbb{R}^3} 
\kappa_{1,0}(\overline{\phi})(s, \boldsymbol{p}) \,\, \partial_{s, \boldsymbol{\p}}^* \, \ud^3 \boldsymbol{\p} \\
= \Xi_{0,1}\big(\kappa_{0,1}(\overline{\phi})\big) + \Xi_{1,0}\big(\kappa_{1,0}(\overline{\phi})\big), \,\,\,\,\,\,
\phi \in \mathscr{E} = \mathcal{S}(\mathbb{R}^4; \mathbb{C}^4).
\end{multline}
Here $\kappa_{0,1}, \kappa_{1,0} \in
\mathscr{L}\big( \mathscr{E}, \,\, \mathcal{S}_{A}(\mathbb{R}^3, \mathbb{C}^4)^* 
\big) \cong \mathscr{L}\big( \mathcal{S}_{A}(\mathbb{R}^3, \mathbb{C}^4), \,\,
\mathscr{L}(\mathscr{E}, \mathbb{C})  \big)$ are vector valued distributions represented with the 
following distribution kernels
\begin{equation}\label{kappa_0,1}
\boxed{
\kappa_{0,1}(s, \boldsymbol{\p}; a,x) = \left\{ \begin{array}{ll}
\frac{1}{2|p_0(\boldsymbol{\p})|}u_{s}^{a}(\boldsymbol{\p})e^{-ip\cdot x} \,\,\, \textrm{with $p = (|p_0(\boldsymbol{\p})|, \boldsymbol{\p}) \in \mathscr{O}_{m,0,0,0}$} & \textrm{if $s=1,2$}
\\
0 & \textrm{if $s=3,4$}
\end{array} \right.,
}
\end{equation}
\begin{equation}\label{kappa_1,0}
\boxed{
\kappa_{1,0}(s, \boldsymbol{\p}; a,x) = \left\{ \begin{array}{ll}
0 & \textrm{if $s=1,2$}
\\
\frac{1}{2|p_0(\boldsymbol{\p})|}v_{s-2}^{a}(\boldsymbol{\p})e^{ip\cdot x} \,\,\, \textrm{with $p = (|p_0(\boldsymbol{\p})|, \boldsymbol{\p}) \in \mathscr{O}_{m,0,0,0}$} & \textrm{if $s=3,4$}
\end{array} \right.
}
\end{equation}
Here $\kappa_{0,1}(\phi), \kappa_{1,0}(\phi)$ denote the kernels representing distributions
in $\mathcal{S}_{A}(\mathbb{R}^3, \,\, \mathbb{C}^4)^*$ which are defined in the standard manner
\[
\kappa_{0,1}(\phi)(s, \boldsymbol{\p})
=  \sum_{a=1}^{4} \int \limits_{\mathbb{R}^3}
\kappa_{0,1}(s, \boldsymbol{\p}; a,x) \phi^{a}(x) \, \ud^4 x
\]
and analogously for $\kappa_{1,0}(\phi)$, and such that
\[
\begin{split}
\kappa_{0,1}: \mathscr{E} \ni \phi \longmapsto \kappa_{0,1}(\phi) 
\in \mathcal{S}_{A}(\mathbb{R}^3, \,\, \mathbb{C}^4)^*, \\
\kappa_{1,0}: \mathscr{E} \ni \phi \longmapsto \kappa_{1,0}(\phi) 
\in \mathcal{S}_{A}(\mathbb{R}^3, \,\, \mathbb{C}^4)^*
\end{split}
\]  
belong to $\mathscr{L}\big( \mathscr{E}, \,\, \big(\mathcal{S}_{A}(\mathbb{R}^3, \mathbb{C}^4)^* 
\big) \cong \mathscr{L}\big( \mathcal{S}_{A}(\mathbb{R}^3, \mathbb{C}^4), \,\,
\mathscr{L}(\mathscr{E}, \mathbb{C})  \big)$. We should emphasize here that in case of free fields the
the vector-valued distributions $\kappa_{0,1}, \kappa_{1,0}$ are regular function like distributions with
distribution kernels $\kappa_{0,1}(s, \boldsymbol{\p}; a,x), \kappa_{0,1}(s, \boldsymbol{\p}; a,x)$
equal to ordinary functions, determining functions 
\begin{equation}\label{multiplicative-kappa_{0,1},kappa_{1,0}}
\begin{split}
\Bigg( \, (a,x) \mapsto \kappa_{0,1; s, \boldsymbol{\p}}(a,x) 
 \overset{\textrm{df}}{=} 
\kappa_{0,1}(s, \boldsymbol{\p}; a,x) \, \Bigg) \in \mathcal{O}_{M} \subset \mathscr{E}^*, 
\,\,\,(s,\boldsymbol{\p}) \in \sqcup \, \mathbb{R}^3, \\
\Bigg( \, (a,x) \mapsto \kappa_{1,0; s, \boldsymbol{\p}}(a,x) 
 \overset{\textrm{df}}{=} 
\kappa_{1,0}(s, \boldsymbol{\p}; a,x) \, \Bigg) \in \mathcal{O}_{M} \subset \mathscr{E}^*, 
\,\,\,(s,\boldsymbol{\p}) \in \sqcup \, \mathbb{R}^3, \\
\Bigg( \, (s,\boldsymbol{\p}) \mapsto \kappa_{0,1; a,x }(s, \boldsymbol{\p}) 
\overset{\textrm{df}}{=} 
\kappa_{0,1}(s, \boldsymbol{\p}; a,x) \, \Bigg) \in \mathcal{O}_{M, A} 
\subset \mathcal{S}_{A}(\mathbb{R}^3, \mathbb{C}^4)^*, \\
\Bigg( \, (s,\boldsymbol{\p}) \mapsto \kappa_{1,0; a,x }(s, \boldsymbol{\p}) 
\overset{\textrm{df}}{=} 
\kappa_{1,0}(s, \boldsymbol{\p}; a,x) \, \Bigg) \in \mathcal{O}_{M, A}
\subset \mathcal{S}_{A}(\mathbb{R}^3, \mathbb{C}^4)^*,
\end{split}
\end{equation}
which belong respectively to the function algebra of multipliers $\mathcal{O}_{M}$ of the nuclear algebra 
$\mathscr{E} = \mathcal{S}(\mathbb{R}^4; \mathbb{C}^4) = 
\mathcal{S}_{\oplus H_{(4)}}(\mathbb{R}^3, \mathbb{C}^4)$ (in the first two cases),
and respectively to the algebra of multipliers $\mathcal{O}_{M, A}$ of the nuclear algebra
$\mathcal{S}_{A}(\mathbb{R}^3, \mathbb{C}^4) = \mathcal{S}(\mathbb{R}^3, \mathbb{C}^4)$
(in the last two cases). These statements can be understood in the sense that for each fixed value of the respective
discrete index, $a$ or $s$, the functions $x \mapsto \kappa_{l,m}(s, \boldsymbol{\p}; a,x)$
or $\boldsymbol{\p} \mapsto \kappa_{0,1}(s, \boldsymbol{\p}; a,x)$, belong respectively to the algebra of multipliers of
$\mathcal{S}(\mathbb{R}^4; \mathbb{C}) = 
\mathcal{S}_{H_{(4)}}(\mathbb{R}^3, \mathbb{C})$ or convolutors of $\mathcal{S}_{H_{(3)}}(\mathbb{R}^3, \mathbb{C}) = \mathcal{S}(\mathbb{R}^3, \mathbb{C})$. But according to our general prescription, we should also note that
$\mathscr{E} = \mathcal{S}(\mathbb{R}^4; \mathbb{C}^4) = 
\mathcal{S}_{\oplus H_{(4)}}(\mathbb{R}^3, \mathbb{C}^4) = \mathcal{S}_{\oplus H_{(4)}}(\sqcup \mathbb{R}^4; \mathbb{C})$
can be treated as nuclear algebra of $\mathbb{C}$-valued functions on the disjoint sum $\sqcup \mathbb{R}^4$
of four disjoint copies of $\mathbb{R}^4$, with the natural point-wise multiplication rule of any two such functions. 
So that the algebra $\mathcal{O}_{M}$ of multipliers is well defined and coincides with all those functons whose restrictions to each copy $\mathbb{R}^4$ belongs to the algebra of multipliers of 
$\mathcal{S}(\mathbb{R}^4; \mathbb{C}) = 
\mathcal{S}_{H_{(4)}}(\mathbb{R}^3, \mathbb{C})$. 
The algebra of convolutors $\mathcal{O}_{C}$ of $\mathscr{E}$, is also well defined with the ordinary Fourier transform exchanging the convolution and point-wise multiplication if we define action of translation $T_{b}$, $b \in \mathbb{R}^4$
on $(a, x) \in \sqcup \mathbb{R}^4$ as equal $T_{b}(a,x) = (a, x + b)$. Similarily the algebras 
$\mathcal{O}_{M, A}(\mathbb{R}^3; \mathbb{C}^4)$,
$\mathcal{O}_{M, A}(\mathbb{R}^3; \mathbb{C}^4)$, of multipliers and convolutors of 
$\mathcal{S}_{A}(\mathbb{R}^3, \mathbb{C}^4) = \mathcal{S}(\mathbb{R}^3, \mathbb{C}^4)
= \mathcal{S}(\sqcup \mathbb{R}^3, \mathbb{C})$ are well defined, where the last is the algebra of all such functions
on $\sqcup \mathbb{R}^4$ with restrictions to each copy $\mathbb{R}^3$ belonging to $\mathcal{S}(\mathbb{R}^3; \mathbb{C})
= \mathcal{S}_{H_{(3)}}(\mathbb{R}^3; \mathbb{C})$.

Note in particular that the integrals in the pairings
\begin{multline*}
\langle \kappa_{0,1}(\phi), \xi \rangle 
= \sum_{s=1}^{4} \, \int \limits_{ \mathbb{R}^4 \times \mathbb{R}^3} 
\kappa_{0,1}(\phi)(s, \boldsymbol{p}) \,\, \xi(s, \boldsymbol{\p}) \, \ud^3 \boldsymbol{\p} \\ 
= \sum_{s=1}^{4} \, \sum_{a=1}^{4} \, \int \limits_{\mathbb{R}^3} 
\kappa_{0,1}(s, \boldsymbol{p}; a, x) \, \phi^{a}(x) \,\, \xi(s, \boldsymbol{\p}) \, \ud^4 x \, \ud^3 \boldsymbol{\p},
\,\,\, \xi \in \mathcal{S}_{A}(\mathbb{R}^3, \mathbb{C}^4), \phi \in  \mathscr{E},
\end{multline*}
are not merely symbolic but actual well defined Lebesgue integrals.\footnote{Here for the case of the Dirac field. 
But we have analogous situation for other fields with the standard Hilbert space 
$L^2(\mathbb{R}^3; \mathbb{C}^4)$ and the standard operator $A$
in (\ref{SinglePartGelfandTriplesForPsi}) possibly replaced with corresponding standard
$L^2(\mathbb{R}^3; \mathbb{C}^n)$ and $A= \oplus H_{(3)}$ or $=\oplus A^{(3)}$. In this case 
$\mathcal{S}_{A= \oplus H_{(3)}}(\mathbb{R}^3; \mathbb{C}^n) = \mathcal{S}(\mathbb{R}^3; \mathbb{C}^n)$
or $\mathcal{S}_{A = \oplus A^{(3)}}(\mathbb{R}^3; \mathbb{C}^n) = \mathcal{S}^{0}(\mathbb{R}^3; \mathbb{C}^n)$,  
$\mathscr{E} = \mathcal{S}_{\oplus H_{(4)}}(\mathbb{R}^4;\mathbb{C}^n) = \mathcal{S}(\mathbb{R}^4;\mathbb{C}^n)$ or 
$\mathscr{E} = \widetilde{\mathcal{S}_{\oplus A_{(4)}}(\mathbb{R}^4;\mathbb{C}^n)} 
= \widetilde{\mathcal{S}^{0}(\mathbb{R}^4;\mathbb{C}^n)} = \mathcal{S}^{00}(\mathbb{R}^4;\mathbb{C}^n)$ 
(compare Section 5 of \cite{wawrzycki2018}) 
and with the corresponding unitary isomorphism $U$ joining the corresponding spectral triples analugous to
(\ref{SinglePartGelfandTriplesForPsi}). In this case the summation with respect to the indices $s,a$
runs over $\{1, 2, \ldots, n\}$.} 

We have the following
\begin{lem}\label{kappa_0,1(barphi),kappa_1,0(barphi)}
Let $\phi \in \mathscr{E} = \mathcal{S}(\mathbb{R}^4; \mathbb{C}^4)$
and $\kappa_{0,1}, \kappa_{1,0}$ be the vector-valued ditributions (\ref{kappa_0,1})
and respectively (\ref{kappa_1,0}). Then 
\[
\begin{split}
\kappa_{0,1}(\overline{\phi})(s, \boldsymbol{\p}) = 
\overline{\big(P^\oplus\widetilde{\phi}|_{{}_{\mathscr{O}_{m,0,0,0}}}\big)_{s+}(\boldsymbol{\p})}=
\overline{\big(P^\oplus\widetilde{\phi}|_{{}_{\mathscr{O}_{m,0,0,0}}}\big)_{s}(\boldsymbol{\p})}, \,\,\,
s=1,2, \\
\kappa_{0,1}(\overline{\phi})(s, \boldsymbol{\p}) = 0, \,\,\, s=3,4, \\
\kappa_{1,0}(\overline{\phi})(s, \boldsymbol{\p}) = 0, \,\,\, s=1,2, \\
\kappa_{1,0}(\overline{\phi})(s, \boldsymbol{\p}) = 
\big(P^\ominus\widetilde{\phi}|_{{}_{\mathscr{O}_{-m,0,0,0}}}\big)_{s}(\boldsymbol{\p}), \,\,\,
s=3,4,
\end{split}
\]  
where $\big(P^\oplus\widetilde{\phi}|_{{}_{\mathscr{O}_{m,0,0,0}}}\big)_{s}$
stands for the $s$-th component of 
\[
U\Big( P^\oplus\widetilde{\phi}|_{{}_{\mathscr{O}_{m,0,0,0}}} \oplus 0 \Big), \,\,\, \textrm{for} \,\, s= 1,2
\]
or respectively $\big(P^\ominus\widetilde{\phi}|_{{}_{\mathscr{O}_{-m,0,0,0}}}\big)_{s}$ stands for the $s$-th component of 
\[
U\Big( 0 \oplus \big(P^\ominus\widetilde{\phi}|_{{}_{\mathscr{O}_{-m,0,0,0}}}\big)^c \Big), \,\,\, \textrm{for} \,\, s= 3,4
\]
in the image of the unitary isomorphism (\ref{isomorphismU}).
\end{lem}
\qedsymbol \,
We have by definition for $s=1,2$
\begin{multline*}
\kappa_{0,1}(\overline{\phi})(s, \boldsymbol{\p})
= \sum_{a=1}^{4} \frac{u_{s}^{a}(\boldsymbol{\p})}{2p_0(\boldsymbol{\p})}
\overline{\int \limits_{\mathbb{R}^4} \phi^a(x)e^{ip\cdot x} \, \ud^4 x}
=  \sum_{a=1}^{4} \frac{u_{s}^{a}(\boldsymbol{\p})}{2p_0(\boldsymbol{\p})}
\overline{\widetilde{\phi}^a(p_0(\boldsymbol{\p}), \boldsymbol{\p})} \\ 
= \sum_{a=1}^{4}
\overline{\frac{\overline{u_{s}^{a}(\boldsymbol{\p})}}{2p_0(\boldsymbol{\p})} 
\widetilde{\phi}^a(p_0(\boldsymbol{\p}), \boldsymbol{\p}) } = 
\overline{\frac{1}{p_0(\boldsymbol{\p})} u_{s}(\boldsymbol{\p})^+ 
\widetilde{\phi}(p_0(\boldsymbol{\p}), \boldsymbol{\p})} \\ =
\overline{\frac{1}{2p_0(\boldsymbol{\p})} u_{s}(\boldsymbol{\p})^+
\big(P^\oplus\widetilde{\phi}\big)(p_0(\boldsymbol{\p}), \boldsymbol{\p}) } =
\overline{\big(P^\oplus\widetilde{\phi}|_{{}_{\mathscr{O}_{m,0,0,0}}}\big)_{s}(\boldsymbol{\p})}, \,\,\,
\textrm{for $s=1,2$}.
\end{multline*}
Here the first four equalities follow by definition, the fifth equality follows from the property
(\ref{u^+P^plusPhi=u^+Phi}) (compare Appendix \ref{fundamental,u,v}) of $u_s(\boldsymbol{\p})$, and recall that
the last term $\overline{\big(P^\oplus\widetilde{\phi}|_{{}_{\mathscr{O}_{m,0,0,0}}}\big)_{s}}$ 
is equal to the complex conjugation of 
the $s$-th direct summand in 
\[
U\Big( P^\oplus\widetilde{\phi}|_{{}_{\mathscr{O}_{m,0,0,0}}} \oplus 0 \Big), \,\,\, \textrm{for} \,\, s= 1,2
\]
by definition (\ref{isomorphismU}) of the unitary isomorphism $U$. 

Similarily we have  by definition for $s=3,4$
\begin{multline*}
\kappa_{1,0}(\overline{\phi})(s, \boldsymbol{\p})
= \sum_{a=1}^{4} \frac{v_{s-2}^{a}(\boldsymbol{\p})}{2|p_0(\boldsymbol{\p})|}
\overline{\int \limits_{\mathbb{R}^4} \phi^a(x)e^{-ip\cdot x} \, \ud^4 x}
=  \sum_{a=1}^{4} \frac{v_{s-2}^{a}(\boldsymbol{\p})}{2|p_0(\boldsymbol{\p})|}
\overline{\widetilde{\phi}^a(-|p_0(\boldsymbol{\p})|, -\boldsymbol{\p})} \\ 
= \sum_{a=1}^{4}
\overline{\frac{\overline{v_{s-2}^{a}(\boldsymbol{\p})}}{2|p_0(\boldsymbol{\p})|} 
\widetilde{\phi}^a(-|p_0(\boldsymbol{\p})|, -\boldsymbol{\p}) } = 
\overline{\frac{1}{2|p_0(\boldsymbol{\p})|} v_{s-2}(\boldsymbol{\p})^+ 
\widetilde{\phi}(-|p_0(\boldsymbol{\p})|, -\boldsymbol{\p})} \\ =
\overline{\frac{1}{2|p_0(\boldsymbol{\p})|} v_{s-2}(\boldsymbol{\p})^+
\big(P^\ominus\widetilde{\phi}\big)(-|p_0(\boldsymbol{\p})|, -\boldsymbol{\p}) } =
\big(P^\ominus\widetilde{\phi}|_{{}_{\mathscr{O}_{-m,0,0,0}}}\big)_{s}(\boldsymbol{\p}), \,\,\,
\textrm{for $s=3,4$}.
\end{multline*}
Here the equalities follow by definition, except the fifth equality, which follows from the property
(\ref{v^+P^minusPhi=v^+Phi}) (compare Appendix \ref{fundamental,u,v}) of $v_s(\boldsymbol{\p})$, and recall that
the last term $\big(P^\ominus\widetilde{\phi}|_{{}_{\mathscr{O}_{-m,0,0,0}}}\big)_{s}$ is equal to the  
$s$-th direct summand in 
\[
U\Big( 0 \oplus \big(P^\ominus\widetilde{\phi}|_{{}_{\mathscr{O}_{-m,0,0,0}}}\big)^c \Big), \,\,\, 
\textrm{for} \,\, s= 3,4,
\]
by definition (\ref{isomorphismU}) of the unitary isomorphism $U$. 

The rest part:
\[
\begin{split}
\kappa_{0,1}(\overline{\phi})(s, \boldsymbol{\p}) = 0, \,\,\, s=3,4, \\
\kappa_{1,0}(\overline{\phi})(s, \boldsymbol{\p}) = 0, \,\,\, s=1,2, \\
\end{split}
\] 
of our Lemma follows immediately from definition (\ref{kappa_0,1})
and respectively (\ref{kappa_1,0}) of the distributions
$\kappa_{0,1}, \kappa_{1,0}$.
\qed

From Lemma \ref{kappa_0,1(barphi),kappa_1,0(barphi)} and from (\ref{a(U(u+v))=a'(u+v)})
it follows
\begin{lem}\label{psi=integKerOpVecValProof} 
Let $\kappa_{0,1}$ and $\kappa_{1,0}$ be the vector-valued distributions
(\ref{kappa_0,1}) and respectively (\ref{kappa_1,0}). Then
the equality (\ref{psi=IntKerOpVectValKer}) holds true:
\begin{multline*}
\boldsymbol{\psi}(\phi) = a'\big(P^\oplus\widetilde{\phi}|_{{}_{\mathscr{O}_{m,0,0,0}}} \oplus 0\big) + 
a'\Big( 0 \oplus \big(P^\ominus\widetilde{\phi}|_{{}_{\mathscr{O}_{-m,0,0,0}}}\big)^c \Big)^+ \\
=  a\Big(U\big(P^\oplus\widetilde{\phi}|_{{}_{\mathscr{O}_{m,0,0,0}}} \oplus 0\big)\Big) + 
a\Bigg(U\Big( 0 \oplus \big(P^\ominus\widetilde{\phi}|_{{}_{\mathscr{O}_{-m,0,0,0}}}\big)^c \Big)\Bigg)^+ \\
 = \sum_{s=1}^{4} \, \int \limits_{\mathbb{R}^3} 
\kappa_{0,1}(\overline{\phi})(s, \boldsymbol{p}) \,\, \partial_{s, \boldsymbol{\p}} \, \ud^3 \boldsymbol{\p}
+
\sum_{s=1}^{4} \, \int \limits_{\mathbb{R}^3} 
\kappa_{1,0}(\overline{\phi})(s, \boldsymbol{p}) \,\, \partial_{s, \boldsymbol{\p}}^* \, \ud^3 \boldsymbol{\p} \\
= \Xi_{0,1}\big(\kappa_{0,1}(\overline{\phi})\big) + \Xi_{1,0}\big(\kappa_{1,0}(\overline{\phi})\big), \,\,\,\,\,\,
\phi \in \mathscr{E} = \mathcal{S}(\mathbb{R}^4; \mathbb{C}^4).
\end{multline*}
\end{lem}
\qedsymbol \, 
Indeed, we have
\begin{multline*}
\sum_{s=1}^{4} \int \limits_{\mathbb{R}^4} \kappa_{0,1}(\overline{\phi})(s, \boldsymbol{\p}) \, 
\partial_{s, \boldsymbol{\p}} \, \ud^3 \boldsymbol{p} =
\sum_{s=1}^{2} \int \limits_{\mathbb{R}^3} 
\overline{\big( P^\oplus \widetilde{\phi}|_{{}_{\mathscr{O}_{m,0,0,0}}} \big)_{s}(\boldsymbol{\p})} \, \partial_{s, \boldsymbol{\p}}
\, \ud^3 \boldsymbol{\p} \\
= a\Big( \,  \big( P^\oplus \widetilde{\phi}|_{{}_{\mathscr{O}_{m,0,0,0}}} \big)_{1} \oplus
\big( P^\oplus \widetilde{\phi}|_{{}_{\mathscr{O}_{m,0,0,0}}} \big)_{2} \oplus 0 \oplus 0 \Big)
= a\Big( \,  U \big( P^\oplus \widetilde{\phi}|_{{}_{\mathscr{O}_{m,0,0,0}}} \oplus 0 \big) \, \Big) \\ =
a'\Big( P^\oplus \widetilde{\phi}|_{{}_{\mathscr{O}_{m,0,0,0}}} \oplus 0 \Big).
\end{multline*}
Here the first three equalites follow from Lemma \ref{kappa_0,1(barphi),kappa_1,0(barphi)},
and Corollary \ref{D_xi=int(xiPartial)}, the last equality follows from 
(\ref{a(U(u+v))=a'(u+v)}).

Similarily we have
\begin{multline*}
\sum_{s=1}^{4} \int \limits_{\mathbb{R}^4} \kappa_{1,0}(\overline{\phi})(s, \boldsymbol{\p}) \, 
\partial_{s, \boldsymbol{\p}}^{*} \, \ud^3 \boldsymbol{p} =
\sum_{s=3}^{4} \int \limits_{\mathbb{R}^3} 
\big( P^\ominus \widetilde{\phi}|_{{}_{\mathscr{O}_{-m,0,0,0}}} \big)_{s}(\boldsymbol{\p}) \, 
\partial_{s, \boldsymbol{\p}}^{*}
\, \ud^3 \boldsymbol{\p} \\
= a\Big( \,  0 \oplus 0 \oplus  \big( P^\ominus \widetilde{\phi}|_{{}_{\mathscr{O}_{m,0,0,0}}} \big)_{3} \oplus
\big( P^\ominus \widetilde{\phi}|_{{}_{\mathscr{O}_{m,0,0,0}}} \big)_{4} \, \Big)
= a\Big( \,  U \big( 0 \oplus (P^\ominus \widetilde{\phi}|_{{}_{\mathscr{O}_{-m,0,0,0}}})^c  \big) \, \Big) \\ =
a'\Big( 0 \oplus (P^\ominus \widetilde{\phi}|_{{}_{\mathscr{O}_{-m,0,0,0}}})^c \Big).
\end{multline*}
Here the first three equalites follow from Lemma \ref{kappa_0,1(barphi),kappa_1,0(barphi)},
and Corollary \ref{D_xi=int(xiPartial)}, the last equality follows from 
(\ref{a(U(u+v))=a'(u+v)}).
\qed

Let $\mathcal{O}_C = \mathcal{O}_C(\mathbb{R}^4; \mathbb{C}^4)$ be the predual of 
of the Schwartz algebra of convolutors $\mathcal{O}'_{C} = \mathcal{O}'_{C}(\mathbb{R}^4; \mathbb{C}^4)$,
which means that each component of each elemet of $\mathcal{O}_C$ belongs to the Horv\'ath
predual $\mathcal{O}_{C}(\mathbb{R}^4; \mathbb{C})$ of the ordinary Schwartz convolution
algebra $\mathcal{O}'_{C}(\mathbb{R}^4; \mathbb{C})$. For detailed construction and definition of  
$\mathcal{O}'_{C}(\mathbb{R}^4; \mathbb{C})$ and $\mathcal{O}_{C}(\mathbb{R}^4; \mathbb{C})$,
compare \cite{Schwartz}, \cite{Horvath} or \cite{Kisynski}, or finally compare the summary of their
properties presented in Appendix \ref{convolutorsO'_C}.

The following Lemma holds true (and we have in general analogous Lemma for a local field
understood as a sum of integral kernel operators with vector-valued kernels) 
\begin{lem}\label{kappa0,1,kappa1,0psi}
For the $\mathscr{L}(\mathscr{E},\mathbb{C})$-valued (or $\mathscr{E}^*$ -valued) distributions 
$\kappa_{0,1}, \kappa_{1,0}$, given by (\ref{kappa_0,1}) and (\ref{kappa_1,0}),
in the equality (\ref{psi=IntKerOpVectValKer}) defining the Dirac $\psi$ field we have
\begin{multline*}
\Bigg( \, (a,x) \mapsto \sum_{s} \, \int \limits_{\mathbb{R}^3}
\kappa_{0,1}(s, \boldsymbol{\p}; a,x)\, \xi(s,\boldsymbol{\p}) \,\ud^3\boldsymbol{\p} \,\, \Bigg) 
\in \mathcal{O}_C \subset \mathcal{O}_M \subset \mathscr{E}^*, \,\, 
\xi \in \mathcal{S}_{A}(\mathbb{R}^3, \mathbb{C}^4), \\
\Bigg( \, (a,x) \mapsto \sum_{s} \, \int \limits_{\mathbb{R}^3}
\kappa_{1,0}(s, \boldsymbol{\p}; a,x) \, \xi(s, \boldsymbol{\p}) \,\ud^3\boldsymbol{\p} \,\, \Bigg)  
\in \mathcal{O}_C \subset \mathcal{O}_M \subset \mathscr{E}^*, 
 \,\, \xi \in \mathcal{S}_{A}(\mathbb{R}^3, \mathbb{C}^4), \\
\Bigg( \, (s,\boldsymbol{\p}) \mapsto \sum_{a} \, \int \limits_{\mathbb{R}^4}
\kappa_{0,1}(s, \boldsymbol{\p}; a,x) \, \phi^a(x) \,\ud^4x \,\, \Bigg) \in 
\mathcal{S}_{A}(\mathbb{R}^3, \mathbb{C}^4), \,\, \phi \in \mathscr{E}, \\
\Bigg( \, (s,\boldsymbol{\p}) \mapsto \sum_{a} \, \int \limits_{\mathbb{R}^4}
\kappa_{1,0}(s, \boldsymbol{\p}; a,x) \, \phi^a(x) \,\ud^4x \,\, \Bigg) \in 
\mathcal{S}_{A}(\mathbb{R}^3, \mathbb{C}^4), 
 \,\, \phi \in \mathscr{E}. 
\end{multline*}
Moreover the maps 
\[
\begin{split}
\kappa_{0,1}: \mathscr{E} \ni \phi \longmapsto \kappa_{0,1}(\phi) 
\in \mathcal{S}_{A}(\mathbb{R}^3, \,\, \mathbb{C}^4), \\
\kappa_{1,0}: \mathscr{E} \ni \phi \longmapsto \kappa_{1,0}(\phi) 
\in \mathcal{S}_{A}(\mathbb{R}^3, \,\, \mathbb{C}^4)
\end{split}
\]
are continuous (for $\kappa_{0,1}, \kappa_{1,0}$ uderstood as maps in 
\[
\mathscr{L}\big( \mathscr{E}, \,\, \big(\mathcal{S}_{A}(\mathbb{R}^3, \mathbb{C}^4)^* 
\big) \cong \mathscr{L}\big( \mathcal{S}_{A}(\mathbb{R}^3, \mathbb{C}^4), \,\,
\mathscr{L}(\mathscr{E}, \mathbb{C})  \big)) 
\]
and, equivalently,
the maps $\xi \longmapsto \kappa_{0,1}(\xi)$, $\xi \longmapsto \kappa_{1,0}(\xi)$ can be extended to
continuous maps
\[
\begin{split}
\kappa_{0,1}: \mathcal{S}_{A}(\mathbb{R}^3, \mathbb{C}^4)^* \ni \xi \longmapsto \kappa_{0,1}(\xi) 
\in \mathscr{E}^*, \\
\kappa_{1,0}: \mathcal{S}_{A}(\mathbb{R}^3, \mathbb{C}^4)^* \ni \xi \longmapsto \kappa_{1,0}(\xi) 
\in \mathscr{E}^*,
\end{split}
\]
(for $\kappa_{0,1}, \kappa_{1,0}$ uderstood as maps 
$\mathscr{L}\big( \mathcal{S}_{A}(\mathbb{R}^3, \mathbb{C}^4), \,\,
\mathscr{L}(\mathscr{E}, \mathbb{C})  \big) \cong 
\mathscr{L}\big( \mathcal{S}_{A}(\mathbb{R}^3, \mathbb{C}^4), \,\,
\mathscr{E}^*  \big)$). Therefore 
not only $\kappa_{0,1}, \kappa_{1,0}
\in \mathscr{L}\big( \mathcal{S}_{A}(\mathbb{R}^3, \mathbb{C}^4), \,\,
\mathscr{L}(\mathscr{E}, \mathbb{C})  \big)$, but both $\kappa_{0,1}, \kappa_{1,0}$
can be (uniquely) extended to elements of 
\[
\mathscr{L}\big( \mathcal{S}_{A}(\mathbb{R}^3, \mathbb{C}^4)^*, \,\,
\mathscr{L}(\mathscr{E}, \mathbb{C})  \big) \cong 
\mathscr{L}\big( \mathcal{S}_{A}(\mathbb{R}^3, \mathbb{C}^4)^*, \,\,
\mathscr{E}^*  \big)  \cong
\mathscr{L}\big( \mathscr{E}, \,\, 
\mathcal{S}_{A}(\mathbb{R}^3, \mathbb{C}^4)  \big).
\]
\end{lem}
\qedsymbol \,
That for each $\xi \in \mathcal{S}_{A}(\mathbb{R}^3, \mathbb{C}^4)$ the functions
$\kappa_{0,1}(\xi), \kappa_{1,0}(\xi)$ given by (here $x = (x_0, \boldsymbol{\x})$)
\[
\begin{split}
(a,x) \mapsto \sum_{s=1}^{4} \, \int \limits_{\mathbb{R}^3}
\kappa_{0,1}(s, \boldsymbol{\p}; a,x)\, \xi(s,\boldsymbol{\p}) \,\ud^3\boldsymbol{\p} =
\sum_{s =1}^{2} \, \int \limits_{\mathbb{R}^3}
\frac{u_{s}^{a}(\boldsymbol{\p})}{2p_0(\boldsymbol{\p})} \, \xi(s,\boldsymbol{\p}) e^{-ip_0(\boldsymbol{\p})x_0 + i\boldsymbol{\p} \cdot \boldsymbol{\x}} \, \ud^3 \boldsymbol{\p}, \\
(a,x) \mapsto \sum_{s=1}^{4} \, \int \limits_{\mathbb{R}^3}
\kappa_{1,0}(s, \boldsymbol{\p}; a,x)\, \xi(s,\boldsymbol{\p}) \,\ud^3\boldsymbol{\p} =
\sum_{s=3}^{4} \, \int \limits_{\mathbb{R}^3}
\frac{v_{s-2}^{a}(\boldsymbol{\p})}{2p_0(\boldsymbol{\p})} \, \xi(s,\boldsymbol{\p}) e^{i|p_0(\boldsymbol{\p})|x_0 - i\boldsymbol{\p} \cdot \boldsymbol{\x}} \, \ud^3 \boldsymbol{\p},
\end{split}
\]
belong to $\mathcal{O}_C \subset \mathcal{O}_M \subset \mathscr{E}^*$ is immediate. Indeed, that they are smooth is obvious, similarily as it is obvious the existence of such a natural $N$ (it is sufficient to take here $N=0$)
that for each multiindex $\alpha \in \mathbb{N}^4$ the functions
\[
(a,x) \mapsto (1 + |x|^2)^{-N} |D_{x^{\alpha}}^{\alpha}\kappa_{0,1}(\xi)(a,x)|, \,\,\,
(a,x) \mapsto (1 + |x|^2)^{-N} |D_{x^{\alpha}}^{\alpha}\kappa_{1,0}(\xi)(a,x)|
\]
are bounded. Here $D_{x^{\alpha}}^{\alpha}\kappa_{l,m}(\xi)$ denotes the ordinary derivative of 
the function $\kappa_{l,m}(\xi)$
of $|\alpha| = \alpha_0 + \alpha_1+\alpha_2+ \alpha_3$ order with respect to space-time 
variables $x= (x_0, x_1, x_2, x_3)$; and here 
$|x|^2= (x_{0})^2 + (x_{1})^2 + (x_{2})^2+ (x_{3})^2$.
The first statement of the Lemma equivalently means that if we fix the value of the discrete index $a$ in the above functions 
\[
(a,x) \mapsto \kappa_{0,1}(\xi)(a,x), \,\,\,\,\, (a,x) \mapsto \kappa_{1,0}(\xi)(a,x), 
\]
then we obtain functions which belong to the algebra of convolutors
of the algebra 
\[\mathcal{S}(\mathbb{R}^4; \mathbb{C}) = \mathcal{S}_{H_{(4)}}(\mathbb{R}^4; \mathbb{C}).
\]
of $\mathbb{C}$-valued functions.

Consider now the functions (in both formulas below the variable $p = (|p_0(\boldsymbol{\p})|, \boldsymbol{\p})$ 
is restricted to the \emph{positive} energy orbit $\mathscr{O}_{m,0,0,0}$)
\[
\begin{split}
(s, \boldsymbol{\p}) \mapsto \kappa_{0,1}(\phi)(s, \boldsymbol{\p})
= \sum_{a=1}^{4} \frac{u_{s}^{a}(\boldsymbol{\p})}{2|p_0(\boldsymbol{\p})|} \int \limits_{\mathbb{R}^3}
\phi^{a}(x) e^{-ip \cdot x} \, \ud^4 x \\
=  \sum_{a=1}^{4} \frac{u_{s}^{a}(\boldsymbol{\p})}{2|p_0(\boldsymbol{\p})|} 
\widetilde{\phi}^{a}|_{{}_{\mathscr{O}_{-m,0,0,0}}}(-p), \\
(s, \boldsymbol{\p}) \mapsto \kappa_{1,0}(\phi)(s, \boldsymbol{\p})
= \sum_{a=1}^{4} \frac{v_{s}^{a}(\boldsymbol{\p})}{2|p_0(\boldsymbol{\p})|} \int \limits_{\mathbb{R}^3}
\phi^{a}(x) e^{ip \cdot x} \, \ud^4 x \\
=  \sum_{a=1}^{4} \frac{v_{s}^{a}(\boldsymbol{\p})}{2|p_0(\boldsymbol{\p})|} 
\widetilde{\phi}^{a}|_{{}_{\mathscr{O}_{m,0,0,0}}}(p),
\end{split}
\] 
with $\phi \in \mathcal{S}(\mathbb{R}^4; \mathbb{C}^4)$. That both functions
$\kappa_{0,1}(\phi), \kappa_{1,0}(\phi)$ depend continously on $\phi$ as maps 
\[
\mathscr{E} = \mathcal{S}(\mathbb{R}^4; \mathbb{C}^4)
\longrightarrow \mathcal{S}_{A}(\mathbb{R}^3, \,\, \mathbb{C}^4) = \mathcal{S}(\mathbb{R}^3, \,\, \mathbb{C}^4)
\]
follows from: 1) continuity of the Fourier transform as a map on the Schwartz space, as well as 2)
from the continuity of the restriction to the orbits $\mathscr{O}_{m,0,0,0}$ and $\mathscr{O}_{-m,0,0,0}$
(with $m \neq 0$) regarded as a map from $\mathcal{S}(\mathbb{R}^4;\mathbb{C})$ into 
$\mathcal{S}(\mathbb{R}^3;\mathbb{C})$, and finally 3) from the fact that the 
functions $\boldsymbol{\p} \mapsto \frac{u_{s}^{a}(\boldsymbol{\p})}{2|p_0(\boldsymbol{\p})|}$ and 
$\boldsymbol{\p} \mapsto \frac{v_{s}^{a}(\boldsymbol{\p})}{2|p_0(\boldsymbol{\p})|}$
are multipliers of the Schwartz algebra $\mathcal{S}(\mathbb{R}^3;\mathbb{C})$,
compare Appendix \ref{fundamental,u,v} and Appendix \ref{convolutorsO'_C}. 
\qed

\begin{rem*}
Note here that the continuity of the maps
\[
\begin{split}
\kappa_{0,1}: \mathscr{E} \ni \phi \longmapsto \kappa_{0,1}(\phi) 
\in \mathcal{S}_{A}(\mathbb{R}^3, \,\, \mathbb{C}^4), \\
\kappa_{1,0}: \mathscr{E} \ni \phi \longmapsto \kappa_{1,0}(\phi) 
\in \mathcal{S}_{A}(\mathbb{R}^3, \,\, \mathbb{C}^4)
\end{split}
\]
is based on the continuity of the restriction to the orbits 
$\mathscr{O}_{m,0,0,0}$ and $\mathscr{O}_{-m,0,0,0}$, regarded as a map
$\widetilde{\mathscr{E}} = \mathcal{S}(\mathbb{R}^4;\mathbb{C}) \rightarrow \mathcal{S}(\mathbb{R}^3;\mathbb{C})$
between the ordinary Schwartz spaces. This continuity breaks down for the orbit 
equal to the light cone $\mathscr{O}_{1,0,0,1}$, because of the singularity at the apex.
Therefore the space-time test space 
\[
\mathscr{E} = \widetilde{\mathcal{S}_{\oplus A^{(4)}}(\mathbb{R}^4); \mathbb{C}^n)} =
\mathcal{S}^{00}(\mathbb{R}^4; \mathbb{C}^n) \neq \mathcal{S}(\mathbb{R}^4;\mathbb{C}^n)
\]
cannot be equal $\mathcal{S}(\mathbb{R}^4;\mathbb{C}^n)$
and the standard operator $A\neq \oplus H_{(3)}$ with
\[
\mathcal{S}_{A}(\mathbb{R}^3, \,\, \mathbb{C}^n) = 
\mathcal{S}_{\oplus A^{(3)}}(\mathbb{R}^3); \mathbb{C}^n) 
= \mathcal{S}^{0}(\mathbb{R}^3, \,\, \mathbb{C}^n)
\neq \mathcal{S}(\mathbb{R}^3;\mathbb{C}^n),
\]
for fields based on representations pertinent to the light cone orbit $\mathscr{O}_{1,0,0,1}$,
if the continuity of the said maps
$\phi \rightarrow \kappa_{0,1}(\phi)$, $\phi \rightarrow \kappa_{1,0}(\phi)$ is to be preserved. 
But the said continuity of the map $\phi \rightarrow \kappa_{1,0}(\phi)$ is necessary and sufficient
(as we will soon see, compare Corollary \ref{continuity-kappa_0,1,kappa_1,0=psiOpValDistr}) 
for the field 
$\boldsymbol{\psi} = \Xi_{0,1}(\kappa_{1,0}) + \Xi_{1,0}(\kappa_{1,0})$  to be continuous
\[
\phi \longmapsto \Xi_{0,1}(\kappa_{1,0}(\phi) + \Xi_{1,0}(\kappa_{1,0}(\phi))
\] 
as a map in
\[
\mathscr{L}\Big( \mathscr{E}, \, \mathscr{L}\big((E), \, (E) \big) \Big),
\]
i.e. necessary and syfficient condition for $\boldsymbol{\psi} = \Xi_{0,1}(\kappa_{1,0}) + \Xi_{1,0}(\kappa_{1,0})$ 
to be a well defined operator valued 
distribution. Therefore the space-time test function space $\mathscr{E}$
for zero mass fields must be modified and cannot coincide with the ordinary Schwartz space.
This is at least the case for zero mass fields constructed as above as 
integral kernel operators with vector-valued kernels in the sense of Obata
\cite{obataJFA}, within the white noise formalism, compare Thm. \ref{ZeromassTestspace} 
of Subsection \ref{A=Xi0,1+Xi1,0}. There exist even more profound reasons for the modification of the 
space-time test space when constructing mass less field with the help of Hida creation-annihilation operators, 
compare Section 5 of \cite{wawrzycki2018}.
When using Wightman's definition of quantum field no such modification of the test function space
is necessary in passing to zero mass fields. But Wightman's defintion is not very much useful
for the traditional perturbative approach to QED and other realistic perturbative QFT. 
For definition of the standard operators $A^{(m)}$ and the nuclear spaces 
$\mathcal{S}_{\oplus_{1}^{n} A^{(m)}}(\mathbb{R}^m; \mathbb{C}^n) = \mathcal{S}^{0}(\mathbb{R}^m; \mathbb{C}^n)$ 
and their Fourier transform images $\mathcal{S}^{00}(\mathbb{R}^m; \mathbb{C}^n)$
we refer to Section 5 of \cite{wawrzycki2018}.

\end{rem*}

Therefore, before
giving the construction of the Dirac field $\boldsymbol{\psi}$ as an integral kernel operator with vector-valued kernel we should give here  general theorems on integral kernel operators (\ref{Xilm(kappalm)Fermi})
\begin{multline*}
\Xi_{l,m}(\kappa_{l,m}(a,x))  \\
= \sum \limits_{s_1, \ldots s_l, t_1, \ldots t_m =1}^{4} \int \limits_{(\mathbb{R}^3)^{l+m}} \,
\kappa_{l,m}(s_1,\boldsymbol{\q}_1, \ldots, s_l, \boldsymbol{\q}_l, t_1, \boldsymbol{\p}_1, 
\ldots, t_m, \boldsymbol{\p}_m; a,x) \,\times \\
\times
\partial_{s_1, \boldsymbol{\q}_1}^* \cdots \partial_{s_l, \boldsymbol{\q}_l}^* 
\partial_{t_1, \boldsymbol{\p}_1} \cdots \partial_{t_m, \boldsymbol{\p}_m} \,
\ud^3 \boldsymbol{\q}_1 \ldots \ud^3 \boldsymbol{\q}_l \ud^3 \boldsymbol{\p}_1 \ldots \ud^3 \boldsymbol{\p}_m,
\end{multline*}
for which 
\begin{multline*}
\Xi_{l,m}(\kappa_{l,m}(\phi))  \\
= \sum \limits_{s_1, \ldots s_l, t_1, \ldots t_m =1}^{4} \int \limits_{(\mathbb{R}^3)^{l+m}} \,
\kappa_{l,m}(\phi)(s_1,\boldsymbol{\q}_1, \ldots, s_l, \boldsymbol{\q}_l, t_1, \boldsymbol{\p}_1, 
\ldots, t_m, \boldsymbol{\p}_m) \,\times \\
\times
\partial_{s_1, \boldsymbol{\q}_1}^* \cdots \partial_{s_l, \boldsymbol{\q}_l}^* 
\partial_{t_1, \boldsymbol{\p}_1} \cdots \partial_{t_m, \boldsymbol{\p}_m} \,
\ud^3 \boldsymbol{\q}_1 \ldots \ud^3 \boldsymbol{\q}_l \ud^3 \boldsymbol{\p}_1 \ldots \ud^3 \boldsymbol{\p}_m,
\end{multline*}
are equal to integral kernel operators (\ref{Xilm(kappalm)Fermi}) with scalar valued kernels 
$\kappa_{l,m}(\phi) \in \big(\mathcal{S}_{A}(\mathbb{R}^3, \,\, \mathbb{C}^4)^{\otimes(l+m)}\big)^*$,
and with 
\begin{multline*}
\kappa_{l,m} \in 
\mathscr{L}\big( \mathscr{E}, \,\, \big(\mathcal{S}_{A}(\mathbb{R}^3, \mathbb{C}^4)^{\otimes(l+m)}\big)^* 
\big) \cong \mathscr{L}\big( \mathcal{S}_{A}(\mathbb{R}^3, \mathbb{C}^4)^{\otimes(l+m)}, \,\,
\mathscr{L}(\mathscr{E}, \mathbb{C})  \big) \\
= \mathscr{L}\big( \mathcal{S}_{A}(\mathbb{R}^3, \mathbb{C}^4)^{\otimes(l+m)}, \,\,
\mathscr{E}^* \big),
\end{multline*}
worked out by Obata \cite{obataJFA}, \cite{obata-book}, Chap. 6.3.  Obata provided detailed analysis of the bose case,
but in a manner easily adopted to the fermi case, and moreover he analyzed slightly more general case
of integral kernel operators with 
$\mathscr{L}\big( \mathscr{E}, \,\, \mathscr{E}^* \big)$-valued distributions
\[
\kappa_{l,m} \in 
\mathscr{L}\big( \mathcal{S}_{A}(\mathbb{R}^3, \mathbb{C}^4)^{\otimes(l+m)}, \,\,
\mathscr{L}(\mathscr{E}, \mathscr{E}^*)  \big).
\]
We only need to analyse the 
special case
of $\mathscr{L}\big( \mathscr{E}, \,\, \mathbb{C} \big) \cong \mathscr{E}^*$-valued distribution kernels
\[
\kappa_{l,m} \in 
\mathscr{L}\big( \mathcal{S}_{A}(\mathbb{R}^3, \mathbb{C}^4)^{\otimes(l+m)}, \,\,
\mathscr{L}(\mathscr{E}, \mathbb{C})  \big) \cong 
\mathscr{L}\big( \mathcal{S}_{A}(\mathbb{R}^3, \mathbb{C}^4)^{\otimes(l+m)}, \,\,
\mathscr{E}^* \big).
\]

In fact in realistic QFT, such as QED, we have several free fields, coupled with lagrangian equal to a
Wick polynomial of free fields (we have in view the causal perturbative approach). Therefore
we need to consider a generalization of \cite{obataJFA} to the case of integral kernel operators in tensor product 
of, say $N$, (fermi and/or bose) Fock spaces $\Gamma(\mathcal{H}'_{i})$ over the corresponding single particle Hilbert spaces $\mathcal{H}'_{i}$, the corresponding standard Gelfand triples
\[
\left. \begin{array}{ccccc}   & & L^2( \sqcup \mathbb{R}^3, \ud^3 \boldsymbol{\p}; \mathbb{C}) & & \\
 & & \parallel & & \\
           \mathcal{S}_{A_i}(\mathbb{R}^3; \mathbb{C}^{{}^{r_i}})         & \subset & \oplus_{{}_{1}}^{{}^{r_i}} L^2(\mathbb{R}^3; \mathbb{C}) & \subset & \mathcal{S}_{A_{i}}(\mathbb{R}^3; \mathbb{C}^{{}^{r_i}})^*        \\
                               \downarrow \uparrow &         & \downarrow \uparrow      &         & \downarrow \uparrow  \\
                                         E_i         & \subset &  \mathcal{H}'_{i} & \subset & E_{i}^*        \\
                                                   
\end{array}\right., \,\,\, i=1,2, \dots, N,
\]
(the analogues of (\ref{SinglePartGelfandTriplesForPsi})) with the correspoding unitary isomorphisms $U_i$
(analogues of the isomorphism $U$ joining the Gelfand triples (\ref{SinglePartGelfandTriplesForPsi})). 
We only need to analyse the 
special case
of $\mathscr{L}\big( \mathscr{E}, \,\, \mathbb{C} \big) \cong \mathscr{E}^*$-valued distribution kernels
\begin{equation}\label{general-vect-valued--kappa_(lm)}
\kappa_{l,m} \in 
\mathscr{L}\big(\mathcal{S}_{A_{{}_{n_1}}}(\mathbb{R}^3, \mathbb{C}^{r_{{}_{1}}}) \otimes \cdots \otimes 
\mathcal{S}_{A_{{}_{n_i}}}(\mathbb{R}^3, \mathbb{C}^{r_{{}^{i}}}) \otimes \cdots 
\otimes \mathcal{S}_{A_{{}_{n_{l+m}}}}(\mathbb{R}^3, \mathbb{C}^{r_{{}^{l+m}}}), \,\,
\mathscr{L}(\mathscr{E}, \mathbb{C})  \big).
\end{equation}
Here 
\begin{multline}\label{general-mathscr(E)}
\mathscr{E} = \mathcal{S}_{B}\big(\sqcup \mathbb{R}^W; \mathbb{C}\big) 
= \mathcal{S}_{B_{{}_{p_1}}}(\mathbb{R}^4; \mathbb{C}^{q_{{}_{1}}}) \otimes \cdots \otimes
\mathcal{S}_{B_{{}_{p_M}}}(\mathbb{R}^4; \mathbb{C}^{q_{{}_{M}}}) \\
\subset L^2\big(\sqcup \mathbb{R}^W; \mathbb{C}\big)
= L^2(\mathbb{R}^4; \mathbb{C}^{q_{{}_{1}}})\otimes \dots \otimes 
L^2(\mathbb{R}^4; \mathbb{C}^{q_{{}_{M}}}),
\end{multline}
with
\begin{multline*}
B = B_{{}_{p_1}} \otimes \cdots \otimes B_{{}_{p_M}}, \,\,\,
p_k \in \{1,2\},
\\
\textrm{on} \,\,\  
L^2\big(\sqcup \mathbb{R}^W; \mathbb{C}\big) = L^2\big(\mathbb{R}^4; \mathbb{C}^{q_1}) \otimes \cdots 
\otimes L^2\big(\mathbb{R}^4; \mathbb{C}^{q_M}\big), \\ 
\,\,\, W= 4M,  \,\,\,\,\,\,\,\, q_k, M =1, 2, \ldots, \\
\,\,\,\,\ \sqcup \mathbb{R}^W = \textrm{$q_1 q_2 \cdots q_M$ disjoint copies of $\mathbb{R}^W$}
\end{multline*}
Moreover we have only two possibilities for $A_i,B_i$, $i=1,2$, on each respective 
$L^2(\mathbb{R}^3, \mathbb{C}^{r_i}), L^2(\mathbb{R}^4, \mathbb{C}^{q_i})$:
\[
\begin{split}
\mathcal{S}_{A_{n_{{}_{i}}}}(\mathbb{R}^3; \mathbb{C}^{r_{{}_{i}}}) = \mathcal{S}_{\oplus H_{(3)}}(\mathbb{R}^3; \mathbb{C}^{r_{{}_{i}}})
= \mathcal{S}(\mathbb{R}^3; \mathbb{C}^{r_{{}_{i}}}), \,\,\, \textrm{or} \\
\mathcal{S}_{A_{n_{{}_{i}}}}(\mathbb{R}^3; \mathbb{C}^{r_{{}_{i}}}) = 
\mathcal{S}_{\oplus A^{(3)}}(\mathbb{R}^3; \mathbb{C}^{r_{{}_{i}}})
= \mathcal{S}^{0}(\mathbb{R}^3; \mathbb{C}^{r_{{}_{i}}}), \,\,\, \\
 \mathcal{S}_{B_{p_{{}_{i}}}}(\mathbb{R}^4; \mathbb{C}^{q_{{}_{i}}}) = \mathcal{S}_{\oplus H_{(4)}}(\mathbb{R}^4; \mathbb{C}^{q_{{}_{i}}})
= \mathcal{S}(\mathbb{R}^4; \mathbb{C}^{q_{{}_{i}}}), \,\,\, \textrm{or} \\
\mathcal{S}_{B_{p_{{}_{i}}}}(\mathbb{R}^4; \mathbb{C}^{q_{{}_{i}}}) = \widetilde{\mathcal{S}_{\oplus A^{(4)}}(\mathbb{R}^4; \mathbb{C}^{q_{{}_{i}}})}
= \mathcal{S}^{00}(\mathbb{R}^4; \mathbb{C}^{q_{{}_{i}}}).
\end{split} 
\]
Here we have the nuclear spaces $\mathcal{S}^{00}(\mathbb{R}^4; \mathbb{C}^n),
\mathcal{S}^{0}(\mathbb{R}^3; \mathbb{C}^n)$, and 
the standard operators $A^{(n)}$ in $L^2(\mathbb{R}^n, \mathbb{C})$, constructed in 
Subsections 5.2-5.5 and 5.8 of \cite{wawrzycki2018}). 
$H_{(4)}$ is the hamiltonian operator  on $L^2(\mathbb{R}^4; \mathbb{C})$ of the  
$4$-dimensional oscillator, compare Appendix 9 of \cite{wawrzycki2018}. 
Here $\widetilde{(\cdot)} = \mathscr{F}(\cdot)$ 
stands for the Fourier transform image. Note that 
\[
\mathcal{S}_{\oplus A^{(4)}}(\mathbb{R}^4; \mathbb{C}^{q}) = \mathcal{S}^{0}(\mathbb{R}^4; \mathbb{C}^{q})
\]
is the nuclear subspace of all those functions in $\mathcal{S}(\mathbb{R}^4; \mathbb{C}^{q})$ which  together with all their derivatives vanish at zero,  so that $\mathcal{S}^{00}(\mathbb{R}^4; \mathbb{C}^{q})$ is the nuclear space of Fourier transforms of all such functions, compare Subsections 5.2-5.5 of \cite{wawrzycki2018}.

For QED it is sufficient to confine attention to just one case of all $r_i =4$ in
(\ref{general-vect-valued--kappa_(lm)}) and the case of integral kernel operators in the tensor product of two Fock liftings of the standard Gelfand triples 
$\mathcal{S}_{A_i}(\mathbb{R}^3; \mathbb{C}^4) \subset L^2(\mathbb{R}^3;\mathbb{C}^4)
\subset \mathcal{S}_{A_i}(\mathbb{R}^3; \mathbb{C}^4)^*$, $i=1,2$, both over $L^2(\mathbb{R}^3;\mathbb{C}^4)$.
Namely: one fermi Fock lifting of the standard triple in (\ref{SinglePartGelfandTriplesForPsi}), correspoding to the Dirac field, with the standard operators $A_1 = \oplus H_{(3)},
B_1 = \oplus H_{(4)}$ defined above,
and one boson Fock lifting of the standard triple in (272) of Subsect. 5.8 of \cite{wawrzycki2018},
correponding to the electromagnetic potential field with the standard operators 
$A_2 =\oplus A^{(3)}, B_2 = \mathscr{F}^{-1}\oplus A^{(4)}\mathscr{F}$ constructed in Subsection 5.8
of \cite{wawrzycki2018}.
Then we consider the standard Hida space $(\boldsymbol{E})= (E_1) \otimes (E_2)$ as arising from the standard
(with nuclear inverse) operator $\Gamma_{\textrm{Fermi}}(A_1) \otimes \Gamma_{\textrm{Bose}}(A_2)$ in the 
tensor product Fock space 
$\Gamma_{\textrm{Fermi}}\big( L^2(\mathbb{R}^3;\mathbb{C}^4)\big) 
\otimes \Gamma_{\textrm{Bose}}\big( L^2(\mathbb{R}^3;\mathbb{C}^4)\big)$ and equal to the tensor product of the 
Hida spaces
\[
(E_i) = \big(  \mathcal{S}_{A_i}(\mathbb{R}^3; \mathbb{C}^4) \big).
\]
The corresponding bose Hida differential operators acting on
$(E_2) \subset \Gamma_{\textrm{Bose}}\big( L^2(\mathbb{R}^3;\mathbb{C}^4)\big)$ (constructed in Section
5 of \cite{wawrzycki2018})
we denote here by $\partial_{\mu, \boldsymbol{\p}}$, $\mu \in \{0,1,2,3\}$, $\boldsymbol{\p} \in \mathbb{R}^3$.
We use the greek indices notation for the discrete parameter $\mu$ in order to distinguish them
from the fermi Hida differential operators $\partial_{s, \boldsymbol{\p}}$ acting on 
$(E_1) \subset \Gamma_{\textrm{Fermi}}\big( L^2(\mathbb{R}^3;\mathbb{C}^4)\big)$. In fact the Hida differential 
operators as acting on 
$(\boldsymbol{E}) = (E_1) \otimes (E_2) \subset \Gamma_{\textrm{Fermi}}\big( L^2(\mathbb{R}^3;\mathbb{C}^4)\big)
\otimes \Gamma_{\textrm{Bose}}\big( L^2(\mathbb{R}^3;\mathbb{C}^4)\big)$ should be uderstood respectively as
equal $\partial_{s, \boldsymbol{\p}} \otimes \boldsymbol{1}$ and 
$\boldsymbol{1} \otimes \partial_{\mu, \boldsymbol{\p}}$. However in order to simplify notation we will 
likewise write for them simply $\partial_{s, \boldsymbol{\p}}$ and $\partial_{\mu, \boldsymbol{\p}}$.
Of course in this notation $E_1, \mathcal{H}'_{1}$ is the standard nuclear space 
$E_1= \mathcal{S}_{\oplus H_{(3)}}(\mathbb{R}^3; \mathbb{C}^4)$
and the single particle Hilbert space $\mathcal{H}'$ in (\ref{SinglePartGelfandTriplesForPsi}); and
$E_2, \mathcal{H}'_{2}$ is the nuclear space $E_2 = E =  \mathcal{S}_{\oplus A^{(3)}}(\mathbb{R}^3; \mathbb{C}^4)$
and the single particle Hilbert space $\mathcal{H}'$ in
(272) of Subsection 5.8 of \cite{wawrzycki2018}.

Of course one can consider the generalization of \cite{obataJFA} for vector-valued kernels for integral kernel operators
on tensor product of any finite number of standard fermi and/or bose Fock spaces with the respective tensor product 
of the corresponding standard Gelfand triples. Having in view only the QED case we confine attention to the tensor product 
of just two mentioned above Fock spaces and the tensor produnct of the correspoding standard Gelfand triples
(\ref{SinglePartGelfandTriplesForPsi})(of this Subsection) and (272) (of Subsect. 5.8 of \cite{wawrzycki2018}). 
We consider integral kernel operators $\Xi_{l,m}(\kappa_{l,m})$ for general $\mathscr{L}\big( \mathscr{E}, \,\, \mathbb{C} \big) \cong \mathscr{E}^*$-valued kernel
\[
\kappa_{l,m} \in 
\mathscr{L}\big(\underbrace{\mathcal{S}_{A_{i_{{}_{1}}}}(\mathbb{R}^3, \mathbb{C}^{4}) \otimes \cdots \otimes 
\mathcal{S}_{A_{i_{{}_{l+m}}}}(\mathbb{R}^3, \mathbb{C}^{4})}_{\textrm{$(l+m)$-fold tesor product}}, \,\,
\mathscr{L}(\mathscr{E}, \mathbb{C})  \big), 
\]
with
\[
A_{i_{{}_{k}}} = A_1 = \oplus_{1}^{4} H_{(3)} \,\,\, \textrm{or} \,\,\, A_{i_{{}_{k}}} = A_2 = \oplus_{0}^{3} 
A^{(3)} \,\,\,
\textrm{on} \,\,\, L^2(\mathbb{R}^3; \mathbb{C}^4) = \oplus L^2(\mathbb{R}^3; \mathbb{C}).
\]
In this case $\Xi_{l,m}(\kappa_{l,m})$, if expressed as integral kernel operator
\begin{multline*}
\Xi_{l,m}(\kappa_{l,m})  \\
= \sum \limits_{ s_{i_{{}_{k}}}, \mu_{i_{{}_{k}}} } 
\int \limits_{(\mathbb{R}^3)^{l+m}}  
\kappa_{l,m}(\overbrace{s_{i_{{}_{1}}},\boldsymbol{\p}_{i_{{}_{1}}}, \ldots, 
\mu_{l}, \boldsymbol{\p}_l}^{\textrm{jointly $l$ terms 
$s_{i_{{}_{k}}},\boldsymbol{\p}_{i_{{}_{k}}}$ or $\mu_{i_{{}_{k}}}, \boldsymbol{\p}_{i_{{}_{k}}}$}}, 
\underbrace{s_{i_{{}_{l+1}}}, \boldsymbol{\p}_{i_{{}_{l+1}}}, 
\ldots, \mu_{i_{{}_{l+m}}}, \boldsymbol{\p}_{i_{{}_{l+m}}}}_{\textrm{jointly $m$ terms 
$s_{i_{{}_{k}}},\boldsymbol{\p}_{i_{{}_{k}}}$ or $\mu_{i_{{}_{k}}}, \boldsymbol{\p}_{i_{{}_{1}}}$}}) 
\,\times \\
\times
\overbrace{\partial_{s_{i_{{}_{1}}}, \boldsymbol{\p}_{i_{{}_{1}}}}^* \cdots 
\partial_{\mu_{i_{{}_{l}}}, \boldsymbol{\p}_{i_{{}_{l}}}}^*}^{\textrm{jointly $l$ terms 
$\partial_{s_{i_{{}_{k}}},\boldsymbol{\p}_{i_{{}_{k}}}}^*$ or 
$\partial_{\mu_{i_{{}_{k}}}, \boldsymbol{\p}_{i_{{}_{k}}}}^*$}}
\underbrace{\partial_{s_{i_{{}_{l+1}}}, \boldsymbol{\p}_{i_{{}_{l+1}}}} \cdots 
\partial_{\mu_{i_{{}_{l+m}}}, \boldsymbol{\p}_{i_{{}_{l+m}}}}}_{\textrm{jointly $m$ terms 
$\partial_{s_{i_{{}_{k}}},\boldsymbol{\p}_{i_{{}_{k}}}}$ or 
$\partial_{\mu_{i_{{}_{k}}}, \boldsymbol{\p}_{i_{{}_{k}}}}$}} \,
\ud^3 \boldsymbol{\p}_{i_{{}_{1}}} \ldots \ud^3 \boldsymbol{\p}_{i_{{}_{l}}} 
\ud^3 \boldsymbol{\p}_{i_{{}_{l+1}}} \ldots \ud^3 \boldsymbol{\p}_{i_{{}_{l+m}}} 
\end{multline*}
\begin{multline*}
= \sum \limits_{ s_{i_{{}_{k}}}, \mu_{i_{{}_{k}}}, t_{j_{{}_{k}}}, \nu_{j_{{}_{k}}}} 
\int \limits_{(\mathbb{R}^3)^{l+m}}  
\kappa_{l,m}(s_{i_{{}_{1}}},\boldsymbol{\q}_{i_{{}_{1}}}, \ldots, 
\mu_{i_{{}_{l}}}, \boldsymbol{\q}_{i_{{}_{l}}}, 
t_{j_{{}_{1}}}, \boldsymbol{\p}_{j_{{}_{1}}}, 
\ldots, \nu_{j_{{}_{m}}}, \boldsymbol{\p}_{j_{{}_{m}}}) 
\,\times \\
\times
\partial_{s_{i_{{}_{1}}}, \boldsymbol{\q}_{i_{{}_{1}}}}^* \cdots 
\partial_{\mu_{i_{{}_{l}}}, \boldsymbol{\q}_{i_{{}_{l}}}}^*
\partial_{t_{i_{{}_{1}}}, \boldsymbol{\p}_{i_{{}_{1}}}} \cdots 
\partial_{\nu_{j_{{}_{m}}}, \boldsymbol{\p}_{j_{{}_{m}}}} \,
\ud^3 \boldsymbol{\q}_{i_{{}_{1}}} \ldots \ud^3 \boldsymbol{\q}_{i_{{}_{l}}} \ud^3 
\boldsymbol{\p}_{j_{{}_{1}}} \ldots \ud^3 \boldsymbol{\p}_{j_{{}_{m}}},
\end{multline*}
transforming $(\boldsymbol{E}) \otimes \mathscr{E}$ into $(\boldsymbol{E})$, is understood as follows 
(compare \cite{obataJFA}): the operators 
$\partial_{s,\boldsymbol{\p}}^*, \partial_{\mu, \boldsymbol{\p}}^*$ 
and $\partial_{s,\boldsymbol{\p}}, \partial_{\mu, \boldsymbol{\p}}$  as operators
on $(\boldsymbol{E}) \otimes \mathscr{E} = (E_1) \otimes(E_2) \otimes \mathscr{E}$ are, respectively,
shortened notation for $\big((\partial_{s,\boldsymbol{\p}} \otimes \boldsymbol{1}) \otimes 
\boldsymbol{1}_{{}_{\mathscr{E}}}\big)^*, \big((\boldsymbol{1} \otimes \partial_{\mu, \boldsymbol{\p}}) \otimes 
\boldsymbol{1}_{{}_{\mathscr{E}}}\big)^*$ 
and $(\partial_{s,\boldsymbol{\p}} \otimes \boldsymbol{1}) \otimes 
\boldsymbol{1}_{{}_{\mathscr{E}}}, (\boldsymbol{1} \otimes \partial_{\mu, \boldsymbol{\p}}) \otimes 
\boldsymbol{1}_{{}_{\mathscr{E}}}$, and $\kappa_{l,m}$ is an 
$\mathscr{L}\big( \mathscr{E}, \,\, \mathbb{C} \big) \cong \mathscr{E}^*$-valued distribution
on $(\mathbb{R}^3)^{(l+m)}$, i.e. on the test space $E_{i_1} \otimes \cdots \otimes E_{i_{l+m}}$ ($(l+m)$-fold tensor product) and this distribution $\kappa_{l,m}$ in the above formula for the integral kernel operator should be
identified with $\boldsymbol{1}_{{}_{(\boldsymbol{E})}} \otimes \kappa_{l,m}$. 

Now any element $\Phi \in (\boldsymbol{E}) = (E_1) \otimes (E_2)$ has the unique absolutely convergent
decomposition (compare \cite{obataJFA}, Prop. 2.3)
\begin{equation}\label{PhiIn(E_1)otimes(E_2)}
\Phi = \sum_{n=0}^{\infty} \Phi_n, \,\,\,
\Phi_n \in \bigoplus_{n_1+n_2 =n} E_{1}^{\widehat{\otimes} \, n_1} \otimes E_{2}^{\widehat{\otimes} \, n_2}, 
\end{equation}
(here the tensor product $E_{1}^{\widehat{\otimes} \, n_1}$ is antisymmetrized $\widehat{\otimes}$
and symmetrized $\widehat{\otimes}$ in $E_{2}^{\widehat{\otimes} \, n_2}$). 
For any element 
\[
\Phi \otimes \phi \in (\boldsymbol{E}) \otimes \mathscr{E} = (E_1) \otimes (E_2) \otimes \mathscr{E}
\]
and any $\mathscr{L}(\mathscr{E}, \mathbb{C})$-valued distribution
\[
\kappa_{l,m} \in \mathscr{L} \big(\overbrace{E_{i_1} \otimes \cdots 
\otimes E_{i_{l+m}}}^{\textrm{$(l+m)$ terms $E_{i_j}$, $i_j\in \{1,2\}$}}, 
\,\, \mathscr{L}(\mathscr{E}, \mathbb{C}) \big)
\cong \mathscr{L} \big(E_{i_1} \otimes \cdots \otimes E_{j_{l+m}}, \,\, \mathscr{E}^* \big).
\]
we put after \cite{obataJFA}
\[
\Xi_{l,m}(\kappa_{l,m}) (\Phi \otimes \phi)
= \sum_{n=0}^{\infty}  \kappa_{l,m} \otimes_m (\Phi_{n+m} \otimes \phi).
\]
Note that here $\otimes_m$ denotes the $m$-contraction of $\Phi_{n+m} \otimes \phi$
with the $\mathscr{L}(\mathscr{E}, \mathbb{C})$-valued distribution uniquely determined 
(after \cite{obataJFA}) by the formula
\[
\begin{split}
\langle \kappa_{l,m} \otimes_m (f_0 \otimes \phi), g_0 \rangle =
\langle \kappa_{l,m}(g_0 \otimes_n f_0), \phi \rangle, \\
 f_0 \in E_{{}_{j_{{}_{1}}}} \otimes \cdots \otimes E_{{}_{j_{{}_{m}}}}
\otimes E_{{}_{i_{{}_{1}}}}\otimes E_{i_{n}}, \\
g_0 \in E_{{}_{j_{{}_{1}}}} \otimes \cdots \otimes E_{{}_{j_{{}_{m}}}}
\otimes E_{{}_{i_{{}_{1}}}} \otimes \ldots \otimes E_{{}_{i_{{}_{n}}}}, \,\,\,\,
\phi \in \mathscr{E}.
\end{split}
\] 
It follows that for any 
\begin{multline*}
\kappa_{l,m} \in \mathscr{L} \big(\overbrace{E_{i_1} \otimes \cdots 
\otimes E_{i_{l+m}}}^{\textrm{$(l+m)$ terms $E_{i_k}$, $i_k \in \{1,2\}$}}, 
\,\, \mathscr{L}(\mathscr{E}, \mathbb{C}) \big) \\
\cong \mathscr{L} \big(E_{i_1} \otimes \cdots \otimes E_{i_{l+m}}, \,\, \mathscr{E}^* \big) \\ \cong
\mathscr{L} \big(\mathscr{E}, \,\, \big( E_{i_1} \otimes \cdots \otimes E_{i_{l+m}} \big)^* \big),
\end{multline*}
the operator $\Xi_{l,m}(\kappa_{l,m})$, defined by contraction $\otimes_m$ with  $\kappa_{l,m}$,
belongs to 
\[
\mathscr{L}\big((\boldsymbol{E}) \otimes \mathscr{E}, (\boldsymbol{E})^* \big) \cong
\mathscr{L}\big( \mathscr{E}, \, \mathscr{L}((\boldsymbol{E}), \, (\boldsymbol{E})^*) \big)
\]
with a precise norm estimation (compare Thms. 3.6 and 3.9 of \cite{obataJFA}). 
Moreover $\Xi_{l,m}(\kappa_{l,m})$ is uniquely determined by the formula
\begin{equation}\label{VectValotimesXi=intKerOp}
\big\langle \big\langle \Xi_{l,m}(\kappa_{l,m})(\Phi \otimes \phi), \Psi  \big \rangle \big \rangle
= \langle \kappa_{l,m}(\eta_{\Phi, \Psi}), \phi \rangle,
\,\,\,
\Phi, \Psi \in (\boldsymbol{E}), \phi \in \mathscr{E}, 
\end{equation}
or equivalently
\begin{equation}\label{VectValotimesXi=intKerOp'}
\big\langle \big\langle \Xi_{l,m}(\kappa_{l,m})(\Phi \otimes \phi), \Psi  \big \rangle \big \rangle
= \langle \kappa_{l,m}(\phi),  \eta_{\Phi, \Psi} \rangle
= \langle \kappa_{l,m}(\eta_{\Phi, \Psi}), \phi \rangle,
\,\,\,
\Phi, \Psi \in (\boldsymbol{E}), \phi \in \mathscr{E},
\end{equation}
 for $\kappa_{l,m}$ understood as an element of 
\[
\mathscr{L} \big(E_{i_1} \otimes \cdots \otimes E_{i_{l+m}}, \,\, \mathscr{E}^* \big)
\,\,\, \textrm{or} \,\,\,
\mathscr{L} \big(\mathscr{E} , \,\, \big(E_{i_1} \otimes \cdots \otimes E_{i_{l+m}} \big)^* \, \big)
\cong \mathscr{L} \big(E_{i_1} \otimes \cdots \otimes E_{i_{l+m}}, \,\, \mathscr{E}^* \big)
\]
respectively in the first case (\ref{VectValotimesXi=intKerOp}) and in the second case
(\ref{VectValotimesXi=intKerOp'}).
Here 
\[
\eta_{\Phi, \Psi}(w_{i_1}, \ldots w_{i_l}, w_{i_{l+1}}, \ldots w_{i_{l+m}}) 
= \big\langle \big\langle \partial_{w_{i_1}}^* \cdots \partial_{w_{i_l}}^* \partial_{w_{i_{l+1}}} \cdots 
\partial_{w_{i_{l+m}}} \Phi, \Psi  
\big \rangle \big \rangle,
\]
and $w_{i_k} = (s_{i_k},\boldsymbol{\q}_{i_k})$ if $E_{i_k} = E_1$
or $w_{i_k} = (\mu_{i_k},\boldsymbol{\q}_{i_k})$ if $E_{i_k} = E_2$.

Note that 
\[
\eta_{\Phi, \Psi} \in E_{i_1} \otimes \cdots \otimes E_{i_{l+m}}.
\]
The formula (\ref{VectValotimesXi=intKerOp}), or equivalently (\ref{VectValotimesXi=intKerOp'}), 
justifies the identification of $\Xi_{l,m}(\kappa_{l,m})$, 
defined through the $m$-contraction $\otimes_m$ with vector valued distribution $\kappa_{l,m}$, 
with the integral kernel operator 
\begin{multline}\label{electron-positron-photon-Xi}
\Xi_{l,m}(\kappa_{l,m}) = 
\int \limits_{(\sqcup \mathbb{R}^3)^{(l+m)}}
\kappa_{l,m}(w_{i_1}, \ldots w_{i_l}, w_{i_{l+1}}, \ldots w_{i_{l+m}}) 
\, \\ \times
\partial_{w_{i_1}}^* \cdots \partial_{w_{i_l}}^* \partial_{w_{i_{l+1}}} \cdots \partial_{w_{i_{l+m}}}
\ud w_{i_1} \cdots \ud w_{i_l} \ud w_{i_{l+1}} \cdots \ud w_{i_{l+m}} = \\
\int \limits_{(\sqcup \mathbb{R}^3)^{(l+m)}}
\kappa_{l,m}(w_{i_1}, \ldots w_{i_l}, u_{j_{1}}, \ldots u_{j_{m}}) 
\,
\partial_{w_{i_1}}^* \cdots \partial_{w_{i_l}}^* \partial_{u_{j_{1}}} \cdots \partial_{u_{j_{m}}}
\ud w_{i_1} \cdots \ud w_{i_l} \ud u_{j_{1}} \cdots \ud u_{j_{m}} 
\end{multline} 
defined by $\mathscr{L}(\mathscr{E}, \mathbb{C})$-valued distribution
kernel $\kappa_{l,m}$.
Here of course
\[
\begin{split}
\int \limits_{\sqcup \mathbb{R}^3} f(w) \ud w \overset{\textrm{df}}{=}  \sum_{s=1}^{4} \, \int \limits_{\mathbb{R}^3} 
f(s, \boldsymbol{\p}) \ud^3 \boldsymbol{\p} \,\,\, \textrm{for} \,\,\, w = (s, \boldsymbol{\p}), \\
\int \limits_{\sqcup \mathbb{R}^3} f(w) \ud w \overset{\textrm{df}}{=}  \sum_{\mu=0}^{3} \, \int \limits_{\mathbb{R}^3} 
f(\mu, \boldsymbol{\p}) \ud^3 \boldsymbol{\p} \,\,\, \textrm{for} \,\,\, w = (\mu, \boldsymbol{\p}),
\end{split}
\]
and we have put $u_{j_k} = w_{i_{l+k}}$, $k=1,2, \ldots, m$.

In our work we are especially interested in (the generalization of) Thm. 3.13 of  \cite{obataJFA},
which gives necessary and sufficient condition for the $\mathscr{L}\big( \mathscr{E}, \,\, \mathbb{C} \big)
\cong \mathscr{E}^*$-valued
distribution $\kappa_{l,m}$ in order that the corresponding $\Xi_{l,m}(\kappa_{l,m})$ be a continuous
operator from $(\boldsymbol{E}) \otimes \mathscr{E}$ into $(\boldsymbol{E})$, thus belonging
to 
\[
\mathscr{L}\big((\boldsymbol{E}) \otimes \mathscr{E}, (\boldsymbol{E})\big) \cong 
\mathscr{L}\Big(\mathscr{E}, \,\, \mathscr{L}\big((\boldsymbol{E}), \, (\boldsymbol{E})\big) \Big)
\]
and thus determining a well defined operator-valued distribution on the test space $\mathscr{E}$. 

We formulate the generalization of Thm. 3.13 over to our tensor product of Fock spaces 
and the correponding tensor product of Gelfand triples (\ref{SinglePartGelfandTriplesForPsi})
(of this Subsect.)
and (272) (of Subsection 5.8 of \cite{wawrzycki2018}).  
We will use the (generalization of) Theorem 3.13 and Proposition 3.12 of \cite{obataJFA} 
for the construction of free fields and in Subsection \ref{psiBerezin-Hida} and Section \ref{A(1)psi(1)} 
when analysing the perturbative 
corrections (within the causal method of St\"uckelberg-Bogoliubov) to interacting fields, as integral kernel operators with 
$\mathscr{E}^*$-valued kernels, in QED. 

Exactly as for the analysis of
integral kernel operators with scalar valued kernels, also the results and proofs of \cite{obataJFA} for 
integral kernel operators with vector-valued kernels can be easily adopted to the fermi case, as well as for 
the more general case of several bose and fermi fields on the tensor product of the corresponding Fock spaces.

We have the following generalization of Thm. 3.13 of \cite{obataJFA}:
\begin{twr}\label{obataJFA.Thm.3.13}
Let 
\[
\kappa_{l,m} \in \mathscr{L} \big(\overbrace{E_{i_1} \otimes \cdots 
\otimes E_{i_{l+m}}}^{\textrm{$(l+m)$ terms $E_{i_j}$, $i_j\in \{1,2\}$}}, 
\,\, \mathscr{L}(\mathscr{E}, \mathbb{C}) \big)
\cong \mathscr{L} \big(E_{i_1} \otimes \cdots \otimes E_{i_{l+m}}, \,\, \mathscr{E}^* \big).
\]
Then 
\[
\Xi_{l,m}(\kappa_{l,m}) \in 
\mathscr{L}\big((\boldsymbol{E}) \otimes \mathscr{E}, (\boldsymbol{E})\big) \cong 
\mathscr{L}\Big(\mathscr{E}, \,\, \mathscr{L}\big((\boldsymbol{E}), \, (\boldsymbol{E})\big) \Big)
\]
if and only if the bilinear map
\begin{multline*}
\xi \times \eta \mapsto \kappa_{l,m}(\xi \otimes \eta), 
\\
\xi \in \overbrace{E_{i_1} \otimes \cdots 
\otimes E_{i_l}}^{\textrm{first $l$ terms $E_{i_j}$, $i_j \in \{1,2\}$}}, \\
\eta \in \overbrace{E_{i_{l+1}} \otimes \cdots 
\otimes E_{i_{l+m}}}^{\textrm{last $m$ terms $E_{i_j}$, $i_j\in \{1,2\}$}},
\end{multline*}
can be extended to a separately continuous bilinear map from
\[
\Big( \overbrace{E_{i_1} \otimes \cdots 
\otimes E_{i_l}}^{\textrm{first $l$ terms $E_{i_j}$}} \Big)^*
\times
\Big( \overbrace{E_{i_{l+1}} \otimes \cdots 
\otimes E_{i_{l+m}}}^{\textrm{last $m$ terms $E_{i_j}$}} \Big)
\,\,\, \textrm{into} \,\,\,\mathscr{L}(\mathscr{E}, \mathbb{C}) = \mathscr{E}^*.
\]
This is the case if and only if for any $k\geq 0$ there exist 
$r \in \mathbb{R}$ such that $|\kappa_{l,m}|_{{}_{l,m;k,r;k}} < \infty$;
and moreover in this case for any $k \in \mathbb{R}$ and $q_0 < q_1 < q$ we have
\begin{multline*}
\|\Xi_{l,m}(\kappa_{l,m}) (\Phi \otimes \phi)\|_{{}_{k}} \leq \rho^{-q/2} \delta^{-1} \sigma^{2} \sqrt{l^l m^m} \Delta_{q_1}^{(l+m)/2} \\
\times 
|\kappa_{l,m}|_{{}_{l,m;k+1, -(k+q+1);k+1}} \|\Phi\|_{{}_{k+q+2}}, \,\,\,
\Phi \in (\boldsymbol{E}), \phi \in \mathscr{E}.
\end{multline*}
\end{twr}

Here for any linear map
\[
\kappa_{l,m}:  \overbrace{E_{i_1} \otimes \cdots 
\otimes E_{i_{l+m}}}^{\textrm{$(l+m)$ terms $E_{i_j}$, $i_j\in \{1,2\}$}}
\longrightarrow \mathscr{L}(\mathscr{E}, \mathbb{C}) = \mathscr{E}^* 
\]
and $k,q,r \in \mathbb{R}$ we put (after \cite{obataJFA}):
\begin{multline*}
|\kappa_{l,m}|_{{}_{l,m;kq;r}} = \textrm{sup} \Bigg\{\sum_{\textrm{i}, \textrm{j}} 
|\langle \kappa_{l,m}(e(\textrm{i}) \otimes e(\textrm{j})), \phi\rangle|^2 |e(\textrm{i})|_{{}_{k}}^2
|e(\textrm{j})|_{{}_{q}}^2, \\ 
\phi \in \mathscr{E}, |\phi|_{{}_{-r}} \leq 1 \Bigg\}^{1/2}.
\end{multline*}
Note that we are using the multiindex notation
\[
e(\textrm{i}) = e_{{}_{i_{{}_{1}}}} \otimes \cdots \otimes e_{{}_{i_{{}_{l}}}} \in 
E_{{}_{i_{{}_{1}}}} \otimes \cdots \otimes E_{{}_{i_{{}_{l}}}}, \,\,\,
\textrm{i} = (i_{{}_{1}}, \ldots, i_{{}_{l}}) 
\]
\begin{multline*}
e(\textrm{j}) = e_{{}_{j_{{}_{1}}}} \otimes \cdots \otimes e_{{}_{j_{{}_{m}}}} =
e_{{}_{i_{{}_{l+1}}}} \otimes \cdots \otimes e_{{}_{i_{{}_{l+m}}}} \in 
E_{{}_{i_{{}_{l+1}}}} \otimes \cdots \otimes E_{{}_{i_{{}_{l+m}}}}, \\
\,\,\,\,\,\, \textrm{j} = (j_{{}_{1}}, \ldots, j_{{}_{m}}) = (i_{{}_{l+1}}, \ldots, i_{{}_{l+m}}),
\end{multline*}  
but now $e_{{}_{i_{{}_{k}}}}$ is the element of the complete orthonormal system of eigenvectors
of the standard operator $A_1$ whenever $e_{{}_{i_{{}_{k}}}} \in E_{{}_{i_{{}_{k}}}} = E_1$ or of the standard  
operator $A_2$ whenever $e_{{}_{i_{{}_{k}}}} \in E_{{}_{i_{{}_{k}}}} = E_2$. 
Note also that with the system of eigenvalues (counted with multiplicity) 
\[
\lambda_{i0}, \lambda_{i1}, \lambda_{i2}, \ldots \,\,\,\, \textrm{of $A_i$},
\]
we have put here 
\[
\delta_i = \Bigg(\sum_{j=0}^{\infty} \lambda_{ij} \Bigg)^{1/2}
= \|A_{i}^{-1} \|_{\textrm{HS}} < \infty, \,\,\,\,
\delta^{-1} \overset{\textrm{df}}{=} \underset{i=1,2}{\textrm{max}} \,\,\delta_{i}^{-1}
\]
for the maximum of the inverses of the Hilbert-Schmidt norms of the  nuclear operators $A_{i}^{-1}$, $i=1,2$. 
Similarily here
\[
\rho = \underset{i=1,2}{\textrm{max}} \,\,  \|A_{i}^{-1} \|_{\textrm{op}}
\]
for the operator norm $\| \cdot \|_{\textrm{op}}$. Here
\[
\Delta_q = \underset{i=1,2}{\textrm{max}} \,\, \Delta_{q_1,i}, \,\,\, q> \underset{i=1,2}{\textrm{max}} \,\, q_{0i} = q_0
\]
where for $i=1,2$
\[
\Delta_{q,i} = \frac{\delta_i}{-e \rho_{i}^{q/2} \textrm{ln}(\delta_{i}^2 \rho_{i}^q)}, \,\,\, 
q> q_{0i} = \textrm{inf} \,  \{q> 0, \delta_{i}^2 \rho_{i}^{q} \leq 1 \}
\]
is a finite constant uniquely determined by the standard 
operator $A_i$, $i=1,2$, if  $q>q_{0,i}$ for the positive constant $q_{0i}$ again depending on $A_i$,
compare \cite{obataJFA}, p. 210.  
Recall that 
\[
\rho_i = \|A_{i}^{-1} \|_{\textrm{op}}.
\]
Finally
\[
\sigma = (\textrm{inf \, Spec} B)^{-1} = \|B^{-1}\|_{\textrm{op}}
\]
for the standard operator $B = B_{{}_{p_1}} \otimes \cdots \otimes B_{{}_{p_M}}$,
$p_k \in \{1,2\}$ on $\otimes_{k=1}^{M}L^2(\mathbb{R}^4; \mathbb{C}^{q_k})$, defining the nuclear test space
\begin{multline*}
\mathscr{E} = \mathcal{S}_{B}(\sqcup \mathbb{R}^{4M}; \mathbb{C})  \\
= \mathcal{S}_{B_{{}_{p_1}}}(\mathbb{R}^4; \mathbb{C}^{q_1}) \otimes \cdots \otimes
\mathcal{S}_{B_{{}_{p_M}}}(\mathbb{R}^4; \mathbb{C}^{q_M}) \subset L^2(\sqcup \mathbb{R}^{4M}; \mathbb{C})
= \otimes_{k=1}^{M}L^2(\mathbb{R}^4; \mathbb{C}^{q_k})
\end{multline*}
(we need the general case with $M>1$ for the analysis of Wick products of $M$ free fields
or of their space-time derivatives or of their seperatate components).
Recall once more that here
\[
\begin{split}
B_{p_k} = \oplus H_{(4)} \,\,\,\, \textrm{on} \,\,\,\,
\oplus_{k=1}^{q_k} L^2(\mathbb{R}^4; \mathbb{C}) = L^2(\mathbb{R}^4; \mathbb{C}^{q_k}), \,\,\, \textrm{for} \,\,\,
p_k = 1 \\
B_{p_k} = \mathscr{F}^{-1} \oplus A^{(4)} \mathscr{F} \,\,\,\, \textrm{on} \,\,\,\,
\oplus_{k=1}^{q_k} L^2(\mathbb{R}^4; \mathbb{C}) = L^2(\mathbb{R}^4; \mathbb{C}^{q_k}),\,\,\, \textrm{for} \,\,\,
p_k = 2
\end{split}
\]
with the hamiltonian operator $H_{(4)}$ on $L^2(\mathbb{R}^4; \mathbb{C})$ of the  
$4$-dimensional oscillator, compare Appendix 9 of \cite{wawrzycki2018}. The standard operator
$A^{(4)}$ on $L^2(\mathbb{R}^4; \mathbb{C})$ is defined in Subsection 5.3 of \cite{wawrzycki2018}. 
\begin{equation}\label{mathscrE_1,mathscrE_2}
\begin{split}
\mathscr{E}_{p_k} = \mathcal{S}_{B_{p_k}}(\mathbb{R}^4; \mathbb{C}^{q_k}) = 
\mathcal{S}_{\oplus H_{(4)}}(\mathbb{R}^4; \mathbb{C}^{q_k}) = \mathcal{S}(\mathbb{R}^4; \mathbb{C}^{q_k}),
\,\,\, p_k = 1 \\
\mathscr{E}_{p_k} = \mathcal{S}_{B_{p_k}}(\mathbb{R}^4; \mathbb{C}^{q_k}) = 
\mathcal{S}_{\mathscr{F}\oplus A^{(4)}\mathscr{F}^{-1}}(\mathbb{R}^4; \mathbb{C}^{q_k}) 
= \mathcal{S}^{00}(\mathbb{R}^4; \mathbb{C}^{q_k}), \,\,\, p_k = 2.
\end{split}
\end{equation}
Recall that 
\begin{multline*}
|\phi|_{{}_{-r}} \overset{\textrm{df}}{=} \big|B^{-r} \phi \big|_0 =
\big| (B_{{}_{p_1}} \otimes \cdots \otimes B_{{}_{p_M}})^{-r} \phi\big|_0 
\\
= \big|(B_{{}_{p_1}} \otimes \cdots \otimes B_{{}_{p_M}})^{-r} \phi \big|_{{}{\otimes_{k=1}^{M} 
L^2(\mathbb{R}^4; \mathbb{C}^{q_k})}},
\,\,\,\,\,\,
\phi \in \mathscr{E}, r \in \mathbb{R}.
\end{multline*}
Recall that in computation of the operator or Hilbert-Schmidt norm the unitary Fourier transform $\mathscr{F}$
in definition of $B_2$ can be ignored and the respective norms can be simply computed for 
$\oplus A^{(4)}$. 

From Thm. \ref{obataJFA.Thm.3.13} we obtain the following
\begin{cor}\label{continuity-kappa_0,1,kappa_1,0=psiOpValDistr}
The Dirac free field 
\[
\boldsymbol{\psi} = \Xi_{0,1}(\kappa_{0,1}) + \Xi_{1,0}(\kappa_{1,0}) \in
\mathscr{L}\big( (E) \otimes \mathscr{E}, \, (E)^* \big) \cong
\mathscr{L}\big( \mathscr{E}, \,\, \mathscr{L}( (E), (E)^*) \big) 
\]
uderstood as integral kernel operator 
with vector-valued distributions 
\[
\kappa_{0,1}, \kappa_{1,0} \in \mathscr{L}\big( \mathcal{S}_{A}(\mathbb{R}^3, \mathbb{C}^4), \,\,
\mathscr{E}^*  \big) \cong \mathcal{S}_{A}(\mathbb{R}^3, \mathbb{C}^4)^* \otimes \mathscr{E}^*
\]
belongs to $\mathscr{L}\big( (E) \otimes \mathscr{E}, \, (E) \big) \cong
\mathscr{L}\big( \mathscr{E}, \,\, \mathscr{L}( (E), (E)) \big)$, i.e.
\[
\boldsymbol{\psi} = \Xi_{0,1}(\kappa_{0,1}) + \Xi_{1,0}(\kappa_{1,0}) \in
\mathscr{L}\big( (E) \otimes \mathscr{E}, \, (E) \big) \cong
\mathscr{L}\big( \mathscr{E}, \,\, \mathscr{L}( (E), (E)) \big),
\]
if and only if the map $\phi \mapsto \kappa_{1,0}(\phi)$
belongs to 
\[
\mathscr{L}\big( \mathscr{E}, \,\, \mathcal{S}_{A}(\mathbb{R}^3, \mathbb{C}^4) \, \big),
\] 
i.e. if and only if $\kappa_{1,0}$ can be extended to a map belonging to
\begin{multline*}
\mathscr{L}\big( \mathcal{S}_{A}(\mathbb{R}^3, \mathbb{C}^4)^*, \,\,
\mathscr{E}^*  \big) \cong \mathcal{S}_{A}(\mathbb{R}^3, \mathbb{C}^4) \otimes \mathscr{E}^* \\
\cong \mathscr{E}^* \otimes \mathcal{S}_{A}(\mathbb{R}^3, \mathbb{C}^4)
\cong \mathscr{L}\big( \mathscr{E}, \,\, \mathcal{S}_{A}(\mathbb{R}^3, \mathbb{C}^4) \, \big).
\end{multline*}
\end{cor}
Here of course we have the special case of Thm \ref{obataJFA.Thm.3.13} with the tensor product
of the two Fock spaces (corresponding to the Dirac field and the electromagnetic potential field)
degenerated to just one Fock space -- that corresponding to the Dirac field, and with
the Hida space $(\boldsymbol{E}) = (E_1) \otimes (E_2)$ degenerated to just the Hida space
$(E_1) \overset{df}{=} (E) \overset{df}{=} \big(\mathcal{S}_{A}(\mathbb{R}^3;\mathbb{C}^4)\big) = 
\big(\mathcal{S}_{\oplus H_{(3)}}(\mathbb{R}^3;\mathbb{C}^4)\big)$ corresponding to the Dirac field, with
the standard operator $A =A_1= \oplus H_{(3)}$ given by (\ref{AinL^2(R^3;C^4)}); and finally with $M=1$
and $B$ degenerated to $B_1$ with the nuclear test space $\mathscr{E}$ degenerated to
\[
\mathscr{E} = \mathcal{S}_{B}(\sqcup \mathbb{R}^4; \mathbb{C})
= \mathcal{S}_{B_1}(\mathbb{R}^4; \mathbb{C}^4) = \mathcal{S}_{\oplus H_{(4)}}(\mathbb{R}^4; \mathbb{C}^4) 
= \mathcal{S}(\mathbb{R}^4; \mathbb{C}^4) = \mathscr{E}_1 
\] 
of (\ref{mathscrE_1,mathscrE_2}).

Equivalently we may consider here the integral kernel operator $\boldsymbol{\psi} = \Xi_{0,1}(\kappa_{0,1}) + \Xi_{1,0}(\kappa_{1,0})$ as acting in the said tensor product of two Fock spaces, having the form of sum of tensor
product opertors on $(\boldsymbol{E}) = (E_1) \otimes (E_2)$  with the second factor operators acting on the
second factor $(E_2)$ trivially as the unit operator, in accordance with the identification
of the operator 
\[
\partial_w  = \left\{ \begin{array}{ll}
\partial_{s, \boldsymbol{\p}} \otimes \boldsymbol{1}, & \textrm{if $w = (s, \boldsymbol{\p})$ refers to fermi variables}, 
\\
\boldsymbol{1} \otimes \partial_{\mu, \boldsymbol{\p}},  & 
\textrm{if $w = (\mu, \boldsymbol{\p})$ refers to bose variables},
\end{array} \right.
\] 
in the general formula (\ref{electron-positron-photon-Xi}). But now we
have to replace the general formula (\ref{electron-positron-photon-Xi}) defining 
the operators $\Xi_{0,1}(\kappa_{0,1}),
\Xi_{1,0}(\kappa_{1,0})$ giving the Dirac field, with another one 
in which the integration variables are 
restricted only to the fermi variables. This is not the special case
of (\ref{electron-positron-photon-Xi}) for $l=0, m=1$ (or $l=1, m=0$)
of an integral operator in the tensor product of Fock spaces, because this is not true 
that the krnels $\kappa_{0,1}, \kappa_{1,0}$ inserted into the general formula  
(\ref{electron-positron-photon-Xi}) cancel out the unwanted boson
variables.  Thus $\boldsymbol{\psi} = \Xi_{0,1}(\kappa_{0,1}) + \Xi_{1,0}(\kappa_{1,0})$ considered 
as acting in the said tensor product of two Fock spaces is a special integral kernel operator with 
integration variables restricted to fermion variables. Similarily we have for the electromagnetic 
potential field, if considered as integral kernel operator in the said tensor product 
of Fock spaces: it is an exceptional integral kernel operator with the integration variables
in the general formula  (\ref{electron-positron-photon-Xi}) restricted only to boson variables.

From the Corrollary \ref{continuity-kappa_0,1,kappa_1,0=psiOpValDistr}
and Lemma \ref{kappa0,1,kappa1,0psi} it follows
\begin{cor}\label{psi=intKerOpVectVal=OpValDistr}
Let
\[
\boldsymbol{\psi} = \Xi_{0,1}(\kappa_{0,1}) + \Xi_{1,0}(\kappa_{1,0}) \in
\mathscr{L}\big( (E) \otimes \mathscr{E}, \, (E)^* \big) \cong
\mathscr{L}\big( \mathscr{E}, \,\, \mathscr{L}( (E), (E)^*) \big)
\]
be the Dirac field uderstood as an integral kernel operator with vector-valued kernels
\[
\kappa_{0,1}, \kappa_{1,0} \in \mathscr{L}\big( \mathcal{S}_{A}(\mathbb{R}^3, \mathbb{C}^4), \,\,
\mathscr{E}^*  \big) \cong \mathcal{S}_{A}(\mathbb{R}^3, \mathbb{C}^4)^* \otimes \mathscr{E}^*,
\]
defined by (\ref{kappa_0,1}) and (\ref{kappa_1,0}). Then the Dirac field operator
\[
\boldsymbol{\psi}  = \boldsymbol{\psi}^{(-)} + \boldsymbol{\psi}^{(+)} = \Xi_{0,1}(\kappa_{0,1}) + \Xi_{1,0}(\kappa_{1,0}),
\]
belongs to $\mathscr{L}\big( (E) \otimes \mathscr{E}, \, (E) \big) \cong
\mathscr{L}\Big( \mathscr{E}, \,\, \mathscr{L}\big( (E), (E)\big) \, \Big)$, i.e.
\[
\boldsymbol{\psi} = \Xi_{0,1}(\kappa_{0,1}) + \Xi_{1,0}(\kappa_{1,0}) \in
\mathscr{L}\big( (E) \otimes \mathscr{E}, \, (E) \big) \cong
\mathscr{L}\Big( \mathscr{E}, \,\, \mathscr{L}\big( (E), (E)\big) \, \Big),
\]
which means in particular that the Dirac field $\boldsymbol{\psi}$, understood as a sum 
$\boldsymbol{\psi} = \Xi_{0,1}(\kappa_{0,1}) + \Xi_{1,0}(\kappa_{1,0})$ of 
two integral kernel operators with vector-valued kernels, 
defines an operator valued distribution through the continuous map
\[
\mathscr{E} \ni \varphi \longmapsto
\Xi_{0,1}\big(\kappa_{0,1}(\varphi)\big) + \Xi_{1,0}\big(\kappa_{1,0}(\varphi)\big)
\in \mathscr{L}\big( (E), (E)\big).
\]
\end{cor}
Note here that the last Corollary \ref{psi=intKerOpVectVal=OpValDistr} follows immediately 
from the proved equality (\ref{psi=IntKerOpVectValKer}), i.e. Lemma
\ref{psi=integKerOpVecValProof}, Corollary \ref{D_xi=int(xiPartial)}, and continuity of the restriction
to the orbit $\mathscr{O}_{m,0,0,0}$ regarded as a map $\mathcal{S}(\mathbb{R}^4; \mathbb{C})
\rightarrow \mathcal{S}(\mathbb{R}^4; \mathbb{C})$.

We have introduced the decomposition of the Dirac field operator $\boldsymbol{\psi}$ into the positive and negative
frequency parts after the classic physical tradition
\[
\boldsymbol{\psi}^{(-)} \overset{df}{=} \Xi_{0,1}(\kappa_{0,1}), \,\,\,\,
\boldsymbol{\psi}^{(+)} \overset{df}{=} \Xi_{1,0}(\kappa_{1,0}).
\]  

Thus as a Corollary to Thm. \ref{obataJFA.Thm.3.13} we have obtained the Dirac field
$\boldsymbol{\psi}$ as a sum of two integral kernel operators with vector valued 
kernels $\kappa_{0,1}, \kappa_{1,0}$ (\ref{kappa_0,1}) and (\ref{kappa_1,0}).
But as we have seen the (free) Dirac field $\boldsymbol{\psi}$ (and in general a quantum free field
uderstood as sum of integral kernel operators with vector-valued kernels) is naturally an integral 
kernel operator with well defined kernel equal to (scalar) integral kernel operator
\begin{equation}\label{psi(x)}
\boxed{
\begin{split}
\boldsymbol{\psi}^a(x) = \sum_{s=1}^{4} \, \int \limits_{\mathbb{R}^3} 
\kappa_{0,1}(s, \boldsymbol{p}; a, x) \,\, \partial_{s, \boldsymbol{\p}} \, \ud^3 \boldsymbol{\p}
+
\sum_{s=1}^{4} \, \int \limits_{\mathbb{R}^3} 
\kappa_{1,0}(s, \boldsymbol{p}; a, x) \,\, \partial_{s, \boldsymbol{\p}}^* \, \ud^3 \boldsymbol{\p} \\
= \boldsymbol{\psi}^{(-) \, a}(x) + \boldsymbol{\psi}^{(+) \, a}(x)
= \Xi_{0,1}\big(\kappa_{0,1}(a,x)\big) + \Xi_{1,0}\big(\kappa_{1,0}(a,x)\big)  \\ =
\sum_{s=1}^{2} \, \int \limits_{\mathbb{R}^3} 
\frac{1}{2|p_0(\boldsymbol{\p})|}u_{s}^{a}(\boldsymbol{\p})e^{-ip\cdot x} \,\, \partial_{s, \boldsymbol{\p}} \, \ud^3 \boldsymbol{\p}
+
\sum_{s=1}^{2} \, \int \limits_{\mathbb{R}^3} 
\frac{1}{2|p_0(\boldsymbol{\p})|}v_{s}^{a}(\boldsymbol{\p})e^{ip\cdot x} \,\, \partial_{s+2, \boldsymbol{\p}}^* \, \ud^3 \boldsymbol{\p} \\ =
\sum_{s=1}^{2} \, \int \limits_{\mathbb{R}^3} 
\frac{1}{2|p_0(\boldsymbol{\p})|}u_{s}^{a}(\boldsymbol{\p})e^{-ip\cdot x} \,\, a_{s}(\boldsymbol{\p}) \, \ud^3 \boldsymbol{\p}
+
\sum_{s=1}^{2} \, \int \limits_{\mathbb{R}^3} 
\frac{1}{2|p_0(\boldsymbol{\p})|}v_{s}^{a}(\boldsymbol{\p})e^{ip\cdot x} \,\, a_{s+2}(\boldsymbol{\p})^+ \, \ud^3 \boldsymbol{\p}
\\ =
\sum_{s=1}^{2} \, \int \limits_{\mathbb{R}^3} 
\frac{1}{2|p_0(\boldsymbol{\p})|}u_{s}^{a}(\boldsymbol{\p})e^{-ip\cdot x} \,\, b_{s}(\boldsymbol{\p}) \, \ud^3 \boldsymbol{\p}
+
\sum_{s=1}^{2} \, \int \limits_{\mathbb{R}^3} 
\frac{1}{2|p_0(\boldsymbol{\p})|}v_{s}^{a}(\boldsymbol{\p})e^{ip\cdot x} \,\, d_{s}(\boldsymbol{\p})^+ \, \ud^3 \boldsymbol{\p}.
\end{split}
}
\end{equation}
\[
\,\,\,\,\,\,\,\,\,\,\,\, \textrm{with $p = (|p_0(\boldsymbol{\p})|, \boldsymbol{\p}) \in \mathscr{O}_{m,0,0,0}$},
\]
and where we have put $b_{s=1} (\boldsymbol{\p}), b_{s=2} (\boldsymbol{\p}), 
d_{s=1} (\boldsymbol{\p}), d_{s=2} (\boldsymbol{\p})$, respectively, for the operators 
$b_{s=1} (\boldsymbol{\p}), b_{s=-1} (\boldsymbol{\p}), d_{s=1} (\boldsymbol{\p}), d_{s=-1} (\boldsymbol{\p})$ 
used in \cite{Scharf}, p. 82, just changing the names of the  
summation index from $\{1,-1 \}$ into $\{1, 2\}$.
Here the expressions in (\ref{psi(x)}), for each fixed space-time point $x$, are not merely symbolic, 
but they are meaningfull integral kernel 
operators transforming continously the Hida space $(E)$ into its strong dual $(E)^*$, and moreover even the integral
signs in these experessions are not merely symbolic, but are meaningfull (point-wise) Pettis integrals 
(compare \cite{HKPS}, or Subsection 5.8 of \cite{wawrzycki2018}). 

We see that there is an addditional weight $|p_0(\boldsymbol{\p})|^{-1}$ factor under the integration sign 
in our formula for the local free Dirac field
$\boldsymbol{\psi}(x)$ in our formula (\ref{psi(x)}) in comparison to the standard formula for the free quantum 
Dirac field used in other books, compare \cite{Scharf} formula\footnote{In the formula (2.2.33)
of \cite{Scharf} the summation sign over $s$ has been lost (of course by a trivial misprint),
and the additional irrelevant constant factors equal to the respective powers of $2\pi$ appear in the litarature which are lost in our formula because we have not normalized the measures when using Fourier transformations.} (2.2.33) or 
the formula (7.32) of \cite{Bogoliubov_Shirkov} (with the respecive amplitudes $a_{\nu}^{\pm}$ replaced with 
the creation-annihilation operators). Our field $\boldsymbol{\psi}$ (\ref{psi(x)}) and the standard
Dirac field, given by the formula (\ref{standardpsi(x)}) of Subsection \ref{StandardDiracPsiField}, although
not equal, are mutually unitary isomorphic in a sense 
explained in Subsection \ref{StandardDiracPsiField}. Nonetheless there are
important differeneces between these two realizations of the field $\boldsymbol{\psi}$. 
We explain them in more details in Subsection
\ref{StandardDiracPsiField}.

\subsection{Fundamental rules for computations involving free fields understood 
as integral kernel operators with vector-valued kernels}\label{OperationsOnXi}

In this Subsection we give several useful computational rules, performed upon integral kernel operators
$\Xi_{l,m}(\kappa_{l,m})$ determined by $\mathscr{L}(\mathscr{E}, \mathbb{C})$-valued distributions,
$\kappa_{l,m}$, respecting the extendibility condition of Thm. \ref{obataJFA.Thm.3.13} of the preceding 
Subsection \ref{psiBerezin-Hida} (or resp. of Thm. 3.13 of \cite{obataJFA}). This property allows to treat 
such $\Xi_{l,m}(\kappa_{l,m})$ as well defined operator-valued distributions on the standard nuclear test space 
$\mathscr{E}$, which in our case will always be equal to the tensor product
\[
\mathscr{E} = \mathscr{E}_{{}_{n_1}} \otimes \cdots \otimes \mathscr{E}_{{}_{n_M}}, \,\,\, n_k \in \{1,2\},
\]
of $M$ space-time test spaces
$\mathscr{E}_1, \mathscr{E}_2$ given by (\ref{mathscrE_1,mathscrE_2}), Subsection  \ref{psiBerezin-Hida},
with $M =1$ and $p_k$ put equal $n_k$.
We encouner the cases with $M =1$ and (operator-valued distributions with one space-time variable)
or with $M>1$ space-time variables. 
In fact the integral kernel operators which are of importance for us are of still more special character, 
being obtainable from the integral kernel operators defined by the free fields underlying the considered
Quantum Field Theory, as a result of special operations: composition of Wick product, differetiation, integration and convolution with pairing functions. 

Having in view the causal perturbative QED we confine attention to integral kernel operators
$\Xi_{l,m}(\kappa_{l,m})$ in the tensor product of just two Fock spaces -- the first one fermionic and corresponding to the Dirac field and the second one bosonic and corresponding to the electromagnetic potential field, compare 
Subsection \ref{psiBerezin-Hida}. 
Thus considered here integral kernel operators $\Xi_{l,m}(\kappa_{l,m})$ act on the Hida space
$(\boldsymbol{E}) = (E_1) \otimes (E_2) \subset \Gamma_{\textrm{Fermi}}\big( L^2(\mathbb{R}^3;\mathbb{C}^4)\big)
\otimes \Gamma_{\textrm{Bose}}\big( L^2(\mathbb{R}^3;\mathbb{C}^4)\big)$, constructed as in the previous 
Subsection \ref{psiBerezin-Hida}. We have also formulated the Thm. \ref{obataJFA.Thm.3.13}, 
Subsection \ref{psiBerezin-Hida}, for the said tensor product of the two mentioned above Fock spaces. 
Of course analogous Theorem and corresponding rules of calculation
with integral kernel operators $\Xi_{l,m}(\kappa_{l,m})$ are valid on tensor product of 
more than just two indicated Fock spaces. 

The space $E_1 = \mathcal{S}_{A_1}(\mathbb{R}^3; \mathbb{C}^4)
= \mathcal{S}(\mathbb{R}^3; \mathbb{C}^4)$ 
with index $1$ and the standard operator $A_1 = A$ (\ref{AinL^2(R^3;C^4)}) refers to the standard 
nuclear space in (\ref{SinglePartGelfandTriplesForPsi})), corresponding to the Dirac field, with the space-time 
test space $\mathscr{E}_1 = \mathcal{S}_{\oplus H_{(4)}}(\mathbb{R}^4; \mathbb{C}^4) 
= \mathcal{S}(\mathbb{R}^4; \mathbb{C}^4)$.
The space  $E_2 = \mathcal{S}_{A_2}(\mathbb{R}^3; \mathbb{C}^4)
= \mathcal{S}^{0}(\mathbb{R}^3; \mathbb{C}^4)$ 
with index $2$ is the nuclear space $E$ determined by the standard operator 
$A_2 = \oplus_{0}^{3} A^{(3)} = A$, which enters the triple in (272)
of Subsect. 5.8 of \cite{wawrzycki2018}, and which serves to define 
the free quantum electromagnetic potential field, Subsection5.8 of \cite{wawrzycki2018}, with the space-time 
test space $\mathscr{E}_2 = \mathcal{S}_{\mathscr{F}^{-1}\oplus A^{(4)}\mathscr{F}}(\mathbb{R}^4; \mathbb{C}^4) 
= \mathcal{S}^{00}(\mathbb{R}^4; \mathbb{C}^4)$.

The vector-valued distributions $\kappa_{0,1}, \kappa_{1,0} \in \mathscr{L}(E_1, \mathscr{E}_{1}^{*})$ 
determined by the plane wave kernels
(\ref{kappa_0,1}) and (\ref{kappa_1,0}), defining the free Dirac field as the integral kernel operator
\[
\boldsymbol{\psi} = \Xi_{0,1}(\kappa_{0,1}) + \Xi_{1,0}(\kappa_{1,0}) 
= \boldsymbol{\psi}^{(-)} + \boldsymbol{\psi}^{(+)},
\]
and in general the vector-valued plane-wave distributions
$\kappa_{0,1}, \kappa_{1,0}, \ldots$ defining all free quantum fields of 
the theory play a fundametal role in the theory.
In QED we encouter besides the plane waves (\ref{kappa_0,1}) and (\ref{kappa_1,0}) 
the plane waves $\kappa_{0,1}, \kappa_{1,0} \in \mathscr{L}(E_2, \mathscr{E}_{2}^{*})$ 
(\ref{kappa_0,1kappa_1,0A'}), Subsection \ref{equivalentA-s}, defining the free quantum electromnagnetic
potential field:
\[
A = \Xi_{0,1}(\kappa_{0,1}) + \Xi_{1,0}(\kappa_{1,0}) = A^{(-)} + A^{(+)},
\]
if we change slightly the convention (used by mathematicians) of Subsection
\ref{psiBerezin-Hida} and use for $\partial_{w}^*$ in the general integral kernel operator 
(\ref{electron-positron-photon-Xi}),
on the tensor product of Fock spaces of the Dirac field $\boldsymbol{\psi}$ and the electromnagnetic potential field $A$,
the operators  $\eta\partial_{\mu, \boldsymbol{\p}}^{*} \eta$ whenever 
$w = (\mu, \boldsymbol{\p})$ corresponds to the photon variables $\mu, \boldsymbol{\p}$ 
in (\ref{electron-positron-photon-Xi}), 
insted of the ordinary transposed operators $\partial_{\mu, \boldsymbol{\p}}^{*}$. 
Here $\eta$ is the Gupta-Bleuler operator. This convention fits well with notation used by physicists, 
as they are using the Krein-adjoined annihilation operators of the photon variables in Fock 
normal expansions.

Indeed in terms of these kernels $\kappa_{0,1}, \kappa_{0,1}, \ldots$
all important quantities of the theory are expressed: 
\begin{enumerate}
\item[1)]
The Wick polynomials of free fields are expressed through 
(symmetrized in bose variables or respectively antisymmetrized in fermi variables) tensor product operation 
performed upon the plane wave kernels $\kappa_{0,1}, \kappa_{1,0}, \ldots$
defining the free fields of the theory,
\item[2)]
Wick polynomial of free fields at the same space-time point are expressed through the symmetrized or antisymmetrized
in $\xi_{1}, \ldots , \xi_M$ 
operation of pointwise product $\kappa_{l_1,m_1}(\xi_1) \cdot \kappa'_{l_1,m_1}(\xi_1) \cdot \ldots \cdot \kappa^{(M)}_{l_M,m_M}(\xi_M)$ utilizing the fact that 
$\kappa_{0,1}(\xi), \kappa_{1,0}(\xi), \kappa'_{0,1}(\xi), \kappa'_{1,0}(\xi), \ldots$,
with $\xi_i \in \mathcal{S}_{A_i}(\mathbb{R}^3, \mathbb{C}^4)$ belong to the algebra of multipliers 
of the respective nuclear algebra $\mathscr{E}_i= \mathcal{S}_{B_i}(\mathbb{R}^4; \mathbb{C}^4)$
(equal $\mathcal{S}(\mathbb{R}^4; \mathbb{C}^4)$ or respectively $\mathcal{S}^{00}(\mathbb{R}^4; \mathbb{C}^4)$)
of spaces of space-time test functions, and the fact that the maps
\begin{multline*}
 E_{i} \times E_{j} \ni \xi \times \zeta 
\mapsto \kappa_{1,0}(\xi)\cdot \kappa'_{1,0}(\zeta) \in \mathscr{E}_{k}^*, \\
 i,j,k \in \{1, 2\},
\end{multline*}
are jointly continuous in the ordinary nuclear topology on $E_{i}$ and strong dual topology on
$\mathscr{E}_{k}^*$ which secures the Wick product to be a well defined integral kernel operator belonging to 
\[
\mathscr{L}((\boldsymbol{E}) \otimes \mathscr{E}, \, (\boldsymbol{E})^*)
\]
for $\mathscr{E}$ equal to the test function space $\mathscr{E}_{1}= \mathcal{S}(\mathbb{R}^4)$ as well as for 
$\mathscr{E}_{2}= \mathcal{S}^{00}(\mathbb{R}^4)$.
Moreover if among the integral kernel operators defined by the plane waves defining free fields there are no 
factors corresponding to zero mass free fields, then
\begin{multline*}
E_{i}^{*} \times E_{j}^* \subset E_{i} \times E_{j} \ni \xi \times \zeta 
\mapsto \kappa_{1,0}(\xi)\cdot \kappa'_{1,0}(\zeta) \in \mathscr{E}_{k}^*, \\
 i,j,k \in \{1, 2\},
\end{multline*}
defined through ordinary point-wise product $\cdot$, are hypocontinuous in the topology inherited from 
the strong dual topology on $E_{i}^{*}$, and strong dual topology on $\mathscr{E}_{j}^{*}$,
which secures in this case the Wick product to be an integrl kernel operator which belongs even to
\[
\mathscr{L}((\boldsymbol{E}) \otimes \mathscr{E}, \, (\boldsymbol{E})) \cong
\mathscr{L}(\mathscr{E}, \, \mathscr{L}((\boldsymbol{E}), (\boldsymbol{E}))
\]
for $\mathscr{E}$ equal to the test function space $\mathscr{E}_{1}= \mathcal{S}(\mathbb{R}^4)$ as well as for 
$\mathscr{E}_{2}= \mathcal{S}^{00}(\mathbb{R}^4)$.
\item[3)]
The perturbative contributions to interacting fields are expressed through convolutions of the kernels corresponding to 
Wick polynomials of free fields with the respective pairing ``generalized functions'', and utilizing the fact that
$\kappa_{0,1}(\xi_{n_1}), \kappa_{0,1}(\xi_{n_2}), \kappa'_{0,1}(\xi_{n_3}), \ldots$, and their pointwise products
with $\xi_{n_{k}} \in \mathcal{S}_{A_{n_k}}(\mathbb{R}^3, \mathbb{C}^4)$ belong to the algebra of convolutors
of the respective nuclear algebra $\mathscr{E}_{n_k}$ ($n_k \in \{1,2\}$). 
\end{enumerate}

In all these constructions we apply the Theorem \ref{obataJFA.Thm.3.13}, and check validity of the condition stated in this Theorem, asserting that the constructed integral kernel operator belongs to
\[
\mathscr{L}((\boldsymbol{E}) \otimes \mathscr{E}, \, (\boldsymbol{E})) \cong
\mathscr{L}(\mathscr{E}, \, \mathscr{L}((\boldsymbol{E}), (\boldsymbol{E}))
\]
and defines an operator-valued distribution on the corresponding test space $\mathscr{E}$. 
Alternatively we check that the constructed operator $\Xi(\kappa)$ has the kernel which respect
weaker condition (\ref{general-vect-valued--kappa_(lm)}) 
\[
\kappa \in \mathscr{L}(E_{n_1}, \ldots, E_{n_{l+m}}, \, \mathscr{E}^*),
\]
which means by the generalization to tensor product of Fock spaces of Thm. 3.9  (compare Subsection 
\ref{psiBerezin-Hida}) that the integral kernel operator belongs to 
\[
\mathscr{L}((\boldsymbol{E}) \otimes \mathscr{E}, \, (\boldsymbol{E})^*) \cong
\mathscr{L}(\mathscr{E}, \, \mathscr{L}((\boldsymbol{E}), (\boldsymbol{E})^*).
\]

In general it cannot be asserted\footnote{As we will soon see the Wick product of integral kernel operators corresponding to zero mass fields  or their derivatives does not belong to 
\[
\mathscr{L}((\boldsymbol{E}) \otimes \mathscr{E}, \, (\boldsymbol{E})), \,\,\, \textrm{but belongs to} \,\, \mathscr{L}((\boldsymbol{E}) \otimes \mathscr{E}, \, (\boldsymbol{E})^*).
\]
} that the integral kernel operator $\Xi$ represented by the Wick product $\Xi$ of integral kernel operators defined 
by free fields belonging to 
\[
\mathscr{L}((\boldsymbol{E}) \otimes \mathscr{E}, \, (\boldsymbol{E})), 
\]
belongs to 
\[
\mathscr{L}((\boldsymbol{E}) \otimes \mathscr{E}, \, (\boldsymbol{E})). 
\]
This would be true only for the Wick product (at the fixed space-time point) $\Xi$  of integral kernel operators 
correponding to massive free fields 
(such as Dirac field) or their derivatives. But if among the factors in the Wick product there are present
integral kernel operators corresponding to zero mass fields (or their derivatives), then their Wick product (at the fixed space-time point) $\Xi$ 
represents a general integral kernel operator (with vector valued kernel) $\Xi(\kappa)$ which belongs to 
\[
\mathscr{L}((\boldsymbol{E}) \otimes \mathscr{E}, \, (\boldsymbol{E})^*). 
\] 
Therefore for any test function $\phi \in \mathscr{E}$ this Wick product operator $\Xi(\kappa)$ can be 
evaluated $\langle \langle \Xi(\kappa)(\Phi \otimes \phi), \, \Psi \rangle \rangle
= \langle \langle \Xi(\kappa(\phi))\Phi, \, \Psi \rangle \rangle$ at $\Phi \otimes \phi$ and  
$\Phi, \Psi \in (\boldsymbol{E})$, and for fixed $\Phi, \Psi \in (\boldsymbol{E})$ represents a 
scalar distribution (as a function of $\phi \in \mathscr{E}$ compare (\ref{VectValotimesXi=intKerOp}) or 
(\ref{VectValotimesXi=intKerOp})). Otherwise: for any test function $\phi \in \mathscr{E}$ the Wick 
product operator $\Xi(\kappa(\phi))$  can be evaluated at $\Phi, \Psi \in (\boldsymbol{E})$, and gives the value
$\langle \langle \Xi(\kappa(\phi))\Phi, \, \Psi \rangle \rangle$, which is equal to a distribution (as a functional of the space-time test function $\phi$).
This is what might have been expected since the very work of Wick himself or from the 
analysis of Bogoliuov and Shirkov \cite{Bogoliubov_Shirkov}, which alredy suggested that the general Wick product 
of free fields determines, at each fixed space-time point, is a well defined sesquilinear form for states ranging over a suitable dense domain. 

But what is most important each order contribution  to interatig Dirac and electromagnetic potential field, 
has the form of a finite sum of integral kernel operators 
\[
\Xi_{l,m}(\kappa_{l,m}) \in \mathscr{L}((\boldsymbol{E}) \otimes \mathscr{E}, \, (\boldsymbol{E})^*), 
\]
respectively with 
$\mathscr{E}_{i}^{*}$ -valued kernels $\kappa_{l,m}$, $i=1,2$, exactly as for the Wick polynomials of free fields
(at fixed space-time point), ant thus represent objects of the same class as the Wick polynomials of free fields,
i.e. finite sums of well defined  integral kernel operators with vector-valued kernels.  
Moreover the full interactig Dirac field and the interacting electromagnetic field (in all orders) have the form of Fock expansions (in the sense of \cite{obataJFA})
\[
\sum \limits_{l,m =0}^{\infty} \Xi_{l,m}(\kappa_{l,m}),
\]
which can be subject to computationally effective convergence ctriteria of \cite{obataJFA}, utilizing
symbol calculus of Obata. 

Thus all operators considered by the theory: free fields, Wick products of their derivatives, and contributions
to interacting fields are all finite sums of integral kernel operators in te sense of Obata \cite{obataJFA}
introduced in Subsection \ref{psiBerezin-Hida}. Among them the free field operators, their derivatives
and Wick polynomials of derivatives of massive fields behave most ``smoothly'' and  belong to
\[
\mathscr{L}((\boldsymbol{E}) \otimes \mathscr{E}, \, (\boldsymbol{E})) \cong
\mathscr{L}(\mathscr{E}, \, \mathscr{L}((\boldsymbol{E}), (\boldsymbol{E})).
\]
General Wick polynomials of derivatives of free fields (including zero mass fields) and
contributions to interacting fields, of which we can say that belong to the general class
of integral kernel oerators, belong to
\[
\mathscr{L}((\boldsymbol{E}) \otimes \mathscr{E}, \, (\boldsymbol{E})^*) \cong
\mathscr{L}(\mathscr{E}, \, \mathscr{L}((\boldsymbol{E}), (\boldsymbol{E})^*),
\]
and are in this sense slightly more singular integral kernel operators
than the free fields themselves. In particular we cannot say that they are operator-valued 
distributions in the white noise sense but nonetheless, when evaluated at fixed elements of Hida
subspace of the Fock space, they represent scalar-valued distributions on the space-time test function space $\mathscr{E}_2$ or $\mathscr{E}_1$.

Thus we start with the fundamental integral kernel operators $\Xi_{0,1}(\kappa_{0,1})$, $\Xi_{1,0}(\kappa_{1,0})$ 
defined by the free fields of the theory. 
But we should distinguish the free field integral kernel operators 
\[
\boldsymbol{\psi} = \Xi_{0,1}(\kappa_{0,1}) + \Xi_{1,0}(\kappa_{1,0}), \,\,\,
A = \Xi_{0,1}(\kappa_{0,1}) + \Xi_{1,0}(\kappa_{1,0}),
\]
acting in their own (resp. fermionic or bosonic) Fock spaces from the corresponding free field integral kernel 
operators 
\[
\boldsymbol{\psi} = \Xi_{0,1}({}^{1}\kappa_{0,1}) + \Xi_{1,0}({}^{1}\kappa_{1,0}), \,\,\,
A = \Xi_{0,1}({}^{2}\kappa_{0,1}) + \Xi_{1,0}({}^{2}\kappa_{1,0}),
\]
both acting in the tensor product Fock space. In the last case the integral kernel operators 
$\Xi_{0,1}({}^{1}\kappa_{0,1}), \Xi_{1,0}({}^{1}\kappa_{1,0})$ are defined by the integral 
formula (\ref{electron-positron-photon-Xi}) in which the integration is restricted to fermi variables $w$
only, and the operators $\Xi_{0,1}({}^{1}\kappa_{0,1}), \Xi_{1,0}({}^{1}\kappa_{1,0})$ act trivially as unit
operators on the second factor. Here ${}^{1}\kappa_{0,1}, {}^{1}\kappa_{0,1}$ are exactly the kernels
(\ref{kappa_0,1}) and (\ref{kappa_1,0}) corresponding to the Dirac field, and denoted with the additional
left-handed-superstript $1$, in order to distiguish them from the kernels ${}^{2}\kappa_{0,1}, {}^{2}\kappa_{0,1}$
(\ref{kappa_0,1kappa_1,0A'}), Subsection \ref{equivalentA-s}, in 
$A = \Xi_{0,1}({}^{2}\kappa_{0,1}) + \Xi_{1,0}({}^{2}\kappa_{1,0})$ acting trivially on the first factor 
in the tensor product of Fock spaces, and defined by the formula (\ref{electron-positron-photon-Xi}) 
in which the integration is restricted to bose variables $w$ only.

And generally kernels $\kappa_{0,1}, \kappa_{1,0}$ respecting the condition of Lemma \ref{kappa0,1,kappa1,0psi}, 
Subsection \ref{psiBerezin-Hida}, corresponding to integral kernel operators 
which act trivially as unit operators on the second bosonic Fock space factor with integration in their definition restricted to fermi variables, will be denoted by 
${}^{1}\kappa_{0,1}, {}^{1}\kappa_{0,1}$ with the additional superscript $1$; and vice versa for kernels corresponding to integral kernel operators acting trivially on the first fermionic Fock space factor with integration in their definition restricted to boson variables, 
denoted by ${}^{2}\kappa_{0,1}, {}^{2}\kappa_{0,1}$ with the additional left-handed- superstript $2$.

Thus we start with the following fundamental integral kernel operators 
\[
\Xi_{0,1}({}^{1}\kappa_{0,1}), \Xi_{1,0}({}^{1}\kappa_{1,0}), 
\Xi_{0,1}({}^{2}\kappa_{0,1}), \Xi_{1,0}({}^{2}\kappa_{1,0}), 
\]
determined by the free fields of the theory and their derivatives, coresponding to 
vector-valued distributions
\[
\begin{split}
{}^{1}\kappa_{0,1}, {}^{1}\kappa_{1,0} \in \mathscr{L}(E_{1}, \mathscr{E}_{1}^{*}) 
\cong  E_{1}^{*} \otimes \mathscr{E}_{1}^{*}, \\
{}^{2}\kappa_{0,1}, {}^{2}\kappa_{1,0} \in \mathscr{L}(E_{2}, \mathscr{E}_{2}^{*}) 
\cong  E_{2}^{*} \otimes  \mathscr{E}_{2}^{*}, 
\end{split}
\]
which have the property that they can be (uniquely) extended to elements (denoted by the same symbols) 
\[
\begin{split}
{}^{1}\kappa_{0,1}, {}^{1}\kappa_{1,0} \in \mathscr{L}(E_{1}^{*}, \mathscr{E}_{1}^{*}) 
\cong  E_{1} \otimes \mathscr{E}_{1}^{*}, \\
{}^{2}\kappa_{0,1}, {}^{2}\kappa_{1,0} \in \mathscr{L}(E_{2}^{*}, \mathscr{E}_{2}^{*}) 
\cong  E_{2}  \otimes \mathscr{E}_{2}^{*}, \\
{}^{1}\kappa_{0,1}(\xi), {}^{1}\kappa_{1,0}(\xi) \in \mathcal{O}_C = \mathcal{O}_{CB_1} 
\subset \mathcal{O}'_{CB_1} \,\,\, \textrm{if $\xi \in E_1$}, \\
{}^{2}\kappa_{0,1}(\xi), {}^{2}\kappa_{1,0}(\xi) \in \mathcal{O}_C
\subset \mathcal{O}'_{CB_2} \,\,\, 
\textrm{if $\xi \in E_2$},
\end{split}
\] 
compare Lemma \ref{kappa0,1,kappa1,0psi}, Subsection \ref{psiBerezin-Hida}
(for the kernels defining Dirac field), and respectively Lemma \ref{kappa0,1,kappa1,0ForA},
Subsection \ref{A=Xi0,1+Xi1,0} for the kernels defining the electromagnetic potential field.
Here $\mathcal{O}'_{C}(\mathbb{R}^4), \mathcal{O}'_{CB_2}(\mathbb{R}^4)$ denote the algebras of convolutors, respectively, of $\mathcal{S}_{B_1}(\mathbb{R}^4) = \mathcal{S}(\mathbb{R}^4), \mathcal{S}_{B_2}(\mathbb{R}^4)
= \mathcal{S}^{00}(\mathbb{R}^4)$, and $\mathcal{O}_{C}(\mathbb{R}^4), \mathcal{O}_{CB_2}(\mathbb{R}^4)$
are their preduals, compare Appendix \ref{convolutorsO'_C}. 
Because all the spaces $E_{i}, E_{i}^{*}, \mathscr{E}_{i}, \mathscr{E}_{i}^{*}$, $i=1,2$,
are nuclear then we have natural topological inclusions
\[
\mathscr{L}(E_{i}^{*}, \mathscr{E}_{i}^{*}) 
\cong  E_{i} \otimes \mathscr{E}_{i}^{*} \subset 
E_{i}^{*} \otimes \mathscr{E}_{i}^{*} \cong \mathscr{L}(E_{i}, \mathscr{E}_{i}^{*}), \,\,\,i=1,2 
\]
induced by the natural topological inclusions $E_{i} \subset E_{i}^{*}$ in both cases: if we endow $E_{i}$
with the topologies on $E_i$ inherited from $E_{i}^{*}$ and with their ordinary nuclear 
topologies, compare Prop. 43.7 and its Corollary  
in \cite{treves}. In the first case we obtain isomorphic inclusions by the cited Proposition,
as in case of nuclear spaces the projective tensor product coincides with the equicontinuous
and thus with the essentially unique tensor product in this category of linear topological 
spacs, compare \cite{treves}. Therefore we simply have

\[
\begin{split}
{}^{1}\kappa_{0,1}, {}^{1}\kappa_{1,0} \in \mathscr{L}(E^{*}_{1}, \mathscr{E}_{1}^{*}) 
\cong  E_{1} \otimes \mathscr{E}_{1}^{*}, \\
{}^{2}\kappa_{0,1}, {}^{2}\kappa_{1,0} \in \mathscr{L}(E^{*}_{2}, \mathscr{E}_{2}^{*}) 
\cong  E_{2}  \otimes \mathscr{E}_{2}^{*}, \\
{}^{1}\kappa_{0,1}(\xi), {}^{1}\kappa_{1,0}(\xi) \in \mathcal{O}_C  = \mathcal{O}_{CB_1}
\,\,\, \textrm{if $\xi \in E_1$}, \\
{}^{2}\kappa_{0,1}(\xi), {}^{2}\kappa_{1,0}(\xi) \in  \mathcal{O}_C \subset 
 \mathcal{O}'_{CB_2} \,\,\, 
\textrm{if $\xi \in E_2$}.
\end{split}
\] 
Recall that in case of kernels ${}^{1}\kappa_{0,1}, {}^{1}\kappa_{1,0}$,
respectively, ${}^{2}\kappa_{0,1}, {}^{2}\kappa_{1,0}$, defining the free fields 
$\boldsymbol{\psi}, A$ we have the spacetime
test spaces $\mathscr{E}_{1}$, respectively, $\mathscr{E}_{2}$, 
given by the formula (\ref{mathscrE_1,mathscrE_2}) with $p_k = n_k=1$, and respectively, 
$p_k= n_k = 2$ and with $q_k = 4$ and $M=1$ in (\ref{mathscrE_1,mathscrE_2}).

In fact we have two possible realizations of the free Dirac field $\boldsymbol{\psi}$,
having different commutation functions and pairings, which nonetheless are \emph{a priori}
equally good form the point of view of causal perturbative approach. This will be explained in 
Subsection \ref{StandardDiracPsiField}. Thus besides the plane wave distributions
${}^{1}\kappa_{0,1}, {}^{1}\kappa_{0,1}$ defined by (\ref{kappa_0,1}) and (\ref{kappa_1,0}),
Subsect. \ref{psiBerezin-Hida}, we can use (\ref{skappa_0,1}) and (\ref{skappa_1,0}) of
Subsection \ref{StandardDiracPsiField}. Similarily we have two possibilites 
for the realization of the free electromagnetic potential field $A$, both having the same commutation and pairing
functions, but with slightly different behaviour in the infrared regime. This will be explained in 
Subsection \ref{equivalentA-s}. Namely besides the formulas 
(\ref{kappa_0,1kappa_1,0A'}) for ${}^{2}\kappa_{0,1}, {}^{2}\kappa_{0,1}$ we can use 
(\ref{kappa_0,1kappa_1,0A}), Subsection \ref{A=Xi0,1+Xi1,0}. Correspondingly we have \emph{a priori}
four versions of perturbative QED, and although it seems that they all should be essentially equivalent,
they all should be subject to a systematic investigation. The formulas 
(\ref{kappa_0,1kappa_1,0A}) and (\ref{skappa_0,1}) and (\ref{skappa_1,0}) are the standard
(in the Gupta-Bleuler gauge of QED) but the remaining three possibilites should also be 
seriously considered.

Here we give definition and general rules in forming Wick product of integral kernel operators
\begin{equation}\label{FreeFieldOp}
\Xi_{l_1,m_1}\Big({}^{{}^{n_1}}_{{}_{1}}\kappa_{l_1,m_1}\Big), \ldots 
\Xi_{l_M,m_M}\Big({}^{{}^{n_M}}_{{}_{M}}\kappa_{l_M,m_M}\Big)
\end{equation}
with general (not necessary equal to plane wave distributions defining the free fields, as we have in view 
e.g. also their spatio-temporal-derivative fields) 
\[
{}^{{}^{n_k}}_{{}_{k}}\kappa_{l_k,m_k}
\in \mathscr{L}(E_{{}_{n_k}}, \mathscr{E}^{*}_{{}_{n_k}}) \cong 
E^{*}_{{}_{n_k}} \otimes \mathscr{E}^{*}_{{}_{n_k}},  \,\,\,\, k=1,2, \ldots M
\]
extendible to
\[
{}^{{}^{n_k}}_{{}_{k}}\kappa_{l_k,m_k}
\in \mathscr{L}(E^{*}_{{}_{n_k}}, \mathscr{E}^{*}_{{}_{n_k}}) \cong 
E_{{}_{p_k}} \otimes \mathscr{E}^{*}_{{}_{n_k}}
\]
and with the property that
\[
{}^{{}^{n_k}}_{{}_{k}}\kappa_{l_k,m_k}(\xi) \in \mathcal{O}_C, \,\,\, \xi \in E_{{}_{n_k}}.
\]
Here 
\[
n_k = \left\{ \begin{array}{l}
1 \\
\textrm{or} \\
2
\end{array} \right., \,\,\,\, \textrm{and} \,\,\,
(l_k, m_k) = \left\{ \begin{array}{l}
(0,1) \\
\textrm{or} \\
(1,0)
\end{array} \right.
\]
and the integral kernel operator 
\[
\Xi_{l_k,m_k}\Big({}^{{}^{n_k}}_{{}_{k}}\kappa_{l_k,m_k}\Big),
\]
regarded as the operator on the said tensor product of Fock spaces, has the exceptional form
(similarily as for the operators defined by the free fields $A$ and $\boldsymbol{\psi}$)
that the integraton in the general formula (\ref{electron-positron-photon-Xi})
for this operator is restricted to fermion variables, if $n_k = 1$, or to bose variables, if
$n_k = 2$. 

We then define the Wick product 
\[
\boldsymbol{}: \Xi_{l_1,m_1}\Big({}^{{}^{n_1}}_{{}_{1}}\kappa_{l_1,m_1}\Big) \cdots 
\Xi_{l_M,m_M}\Big({}^{{}^{n_M}}_{{}_{M}}\kappa_{l_M,m_M}\big) \boldsymbol{:}
\] 
of $M$ such operators as the ordinary product of these operators, but rearranged 
in such a manner that all operators 
\[
\Xi_{l_k,m_k}\Big({}^{{}^{n_k}}_{{}_{k}}\kappa_{l_k,m_k}\Big)
\]
with $(l_k, m_k) = (1,0)$ stand to the left of all operators
\[
\Xi_{l_k,m_k}\Big({}^{{}^{n_k}}_{{}_{k}}\kappa_{l_k,m_k}\Big)
\]
with $(l_k, m_k) = (0,1)$, multiplied in addition  by the factor $(-1)^p$
with $p$ equal to the parity of the 
permutation performed upon fermi operators, having $n_k = 1$ and corresponding to the fermi variables, 
required to bring the operators into the required ``normal'' order. 

\begin{center}
{\small RULE I}
\end{center}
\emph{We have the following computational rule}
\begin{multline*}
\boldsymbol{}: \Xi_{l_1,m_1}\Big({}^{{}^{n_1}}_{{}_{1}}\kappa_{l_1,m_1}\Big) \cdots 
\Xi_{l_M,m_M}\Big({}^{{}^{n_1}}_{{}_{M}}\kappa_{l_M,m_M}\big) \boldsymbol{:} \\
= \Xi_{l,m}(\kappa_{lm}), 
\\
 l = l_1 + \cdots l_M, \,\,\, m = m_1 + \cdots m_M
\end{multline*}
\emph{where} 
\[
\kappa_{l,m} = \Big({}^{{}^{n_1}}_{{}_{1}}\kappa_{l_1,m_1} \Big) \overline{\otimes} \cdots \overline{\otimes} \,\,
\Big( {}^{{}^{n_1}}_{{}_{M}}\kappa_{l_M,m_M} \Big)
\]
\emph{stands for the ordinary tensor product 
\begin{multline*}
\Big({}^{{}^{n_1}}_{{}_{1}}\kappa_{l_1,m_1}\Big) \otimes \cdots \otimes \,\,
\Big({}^{{}^{n_1}}_{{}_{M}}\kappa_{l_M,m_M}\Big) \in
E_{{}_{n_1}} \otimes \mathscr{E}^{*}_{{}_{n_1}} \otimes \cdots \otimes 
E_{{}_{n_M}} \otimes \mathscr{E}^{*}_{{}_{n_M}} \\
\cong E_{{}_{n_1}} \otimes \cdots \otimes  E_{{}_{n_M}} \otimes 
\mathscr{E}^{*}_{{}_{n_1}} \otimes  \cdots \otimes \mathscr{E}^{*}_{{}_{n_M}} \\
\cong \mathscr{L}(E^{*}_{{}_{n_1}} \otimes \cdots \otimes  E^{*}_{{}_{n_M}}, \,\,
\mathscr{E}^{*}_{{}_{n_1}} \otimes  \cdots \otimes \mathscr{E}^{*}_{{}_{n_M}})
\end{multline*}
1) separately symmetrized with respect to all bose variables, lying among the first $l$ variables,
2) separately symmetrized with respect to all bose variables, lying among the last $m$
variables, 3) separately antisymmetrized with respect to all fermi variables which 
lie among the first $l$ variables, 4) separately antisymmetrized with respect 
to all fermi variables lying among the 
last $m$ variables, finally 5) the result multiplied by the factor
$(-1)^p$, where $p$ is the parity of the  prmutation performed upon the fermi operators necessay to rearrange them into 
the order in which they stand in the general formula (\ref{electron-positron-photon-Xi})
for $\Xi_{l,m}(\kappa_{l,m})$.  Here by definition $n_k$ is counted among the first $l$ variables
iff the corresponding $(l_k, m_k) = (1,0)$, and $n_k$ is counted among last $m$
variables iff the corresponding $(l_k, m_k) = (0,1)$.}

This is effective computational rule because in practical situations, e.g. for the Wick product of 
integral kernel operators defined by free fields of the theory, the tensor product of the corresponding kernels
may be represented by ordinary products of the functions representing kernels:
\begin{multline*}
\Big({}^{{}^{n_1}}_{{}_{1}}\kappa_{l_1,m_1}\Big) \otimes \cdots \otimes \,\,
\Big({}^{{}^{n_1}}_{{}_{M}}\kappa_{l_M,m_M}\Big)(w_1, \ldots, w_M; X_1, \ldots, X_M) \\ =
\Big({}^{{}^{n_1}}_{{}_{1}}\kappa_{l_1,m_1}\Big)(w_1, X_1) \cdots 
\Big({}^{{}^{n_1}}_{{}_{M}}\kappa_{l_M,m_M}\Big)(w_M, X_M), \\
X_k = \left\{ \begin{array}{ll}
(a_k,x_k), & \textrm{for $X_k$ corresponding to fermi variables $w_k= (s_k, \boldsymbol{\p}_k)$} \\
\textrm{or} & \\
(\mu_k,x_k), & \textrm{for $X_k$ corresponding to bose variables $w_k= (\nu_k, \boldsymbol{\p}_k)$}
\end{array} \right., \\
w_k = \left\{ \begin{array}{ll}
(s_k,\boldsymbol{\p}_k), & \textrm{for fermi variables $w_k$} \\
\textrm{or} & \\
(\nu_k,\boldsymbol{\p}_k), & \textrm{for bose variables $w_k$}
\end{array} \right., 
\\
\textrm{$x_k$ denotes for each $k$ spacetime coordinates variable}, \\
s_k \in \{1,2,3,4\}, \,\,\, \mu_k, \nu_k \in \{0,1,2,3\}, a_k \in \{1,2,3,4\}.
\end{multline*}
In case of Wick product integral kernel operators corresponding to fixed components of the fields,
the respective values of $\mu_k$ and $a_k$ will be correspondingly fixed, and the test spaces $\mathscr{E}_{n_k}$
will be equal (\ref{mathscrE_1,mathscrE_2}) with $q_k = 1$, \emph{i.e.} scalar test spaces.
Thus the symmetrized/antisymmetrized
tensor product $\overline{\otimes}$ of the kernels corresponding to free fields can be easily and explicitly 
computed, by the indicated symmetrizations and antisymmetrizations applied to the kernel functions:
\[
\Big({}^{{}^{n_1}}_{{}_{1}}\kappa_{l_1,m_1}\Big) \otimes \cdots \otimes \,\,
\Big({}^{{}^{n_1}}_{{}_{M}}\kappa_{l_M,m_M}\Big)(w_1, \ldots, w_M; X_1, \ldots, X_M),
\]
remembering that the variable $(w_k, X_k)$ is counted among the first $l$ variables
iff $(l_k, m_k) = (1,0)$, and the variable $(w_k, X_k)$ is counted among the last $m$ variables
iff $(l_k, m_k) = (0,1)$. 

The Rule I can be justified by utilizing the fact that 
\[
\Xi_{l_k,m_k}\Big({}^{{}^{n_k}}_{{}_{k}}\kappa_{l_k,m_k}(X_k)\Big),
\]
exist point-wisely as Pettis integral for each fixed point $X_k$, with the scalar distribution 
\[
{}^{{}^{n_k}}_{{}_{k}}\kappa_{l_k,m_k}(X_k) 
\]
(with fixed $X_k$) represented by the scalar function 
\[
w_k \longmapsto {}^{{}^{n_k}}_{{}_{k}}\kappa_{l_k,m_k}(w_k,X_k) 
\]
kernel, as in the proof of Bogoliubov-Shirkov Hypothesis in Subsection 5.9 of \cite{wawrzycki2018}. 

From the Rule I it easily follows that the Wick product
of the class of integral kernel operators (\ref{FreeFieldOp}), subsuming free field operators, is a well defined 
(sum of) integral kernel operator(s) $\Xi(\kappa_{l,m})$ with the kernel(s)
\begin{equation}\label{StrongkappaProperty}
\kappa_{l,m} \in \mathscr{L}(E^{*}_{{}_{n_1}} \otimes \cdots \otimes  E^{*}_{{}_{n_M}}, \,\,
\mathscr{E}^{*}_{{}_{n_1}} \otimes  \cdots \otimes \mathscr{E}^{*}_{{}_{n_M}}), \,\,\, M= l+m
\end{equation}
and thus with
\[
\Xi_{l,m}(\kappa_{l,m}) \in 
\mathscr{L}\big((\boldsymbol{E}) \otimes \mathscr{E}, (\boldsymbol{E})\big) \cong 
\mathscr{L}\Big(\mathscr{E}, \,\, \mathscr{L}\big((\boldsymbol{E}), \, (\boldsymbol{E})\big) \Big)
\]
by Thm. \ref{obataJFA.Thm.3.13}, Subsection \ref{psiBerezin-Hida}, for
\begin{equation}\label{tensormathscrE}
\mathscr{E} = \mathscr{E}^{*}_{{}_{n_1}} \otimes  \cdots \otimes \mathscr{E}^{*}_{{}_{n_M}}.
\end{equation} 
In particular it defines an operator-valued distribution on the tensor product
(\ref{tensormathscrE}) of space-time test function spaces $\mathscr{E}_1, \mathscr{E}_2$ with
$\mathscr{E}_{{}_{n_k}} = \mathscr{E}_1$ iff $n_k = 1$ and $\mathscr{E}_{{}_{n_k}} = \mathscr{E}_2$ iff 
$n_k = 2$ (respectively for the fermi operator or bose operator in the Wick product). 

It is easily seen that we get in this way a Wick graded algebra which subsumes in particular all 
finite sums of integral kernel operators $\Xi_{l,m}(\kappa_{l,m})$ 
with kernels $\kappa_{l,m}$ having the property (\ref{StrongkappaProperty}). 
Let 
\[
\Xi(\kappa'_{l',m'}) \,\,\, \textrm{and} \,\,\,
\Xi_{l'',m''}(\kappa''_{l'',m''})
\]
be two such operators with 
\[
\begin{split}
\kappa'_{l',m'} \in \mathscr{L}(E^{*}_{{}_{n'_1}} \otimes \cdots \otimes  E^{*}_{{}_{n'_{M'}}}, \,\,
\mathscr{E}^{*}_{{}_{n'_1}} \otimes  \cdots \otimes \mathscr{E}^{*}_{{}_{n'_{M'}}}), \\
\kappa''_{l'',m''} \in \mathscr{L}(E^{*}_{{}_{n''_1}} \otimes \cdots \otimes  E^{*}_{{}_{n''_{M''}}}, \,\,
\mathscr{E}^{*}_{{}_{n''_1}} \otimes  \cdots \otimes \mathscr{E}^{*}_{{}_{n''_{M''}}})
\end{split}
\]
It is easily seen that 
we have the following rule for Wick product of such operators
\[
\boldsymbol{:} \Xi(\kappa'_{l',m'}) \, \Xi_{l'',m''}(\kappa''_{l'',m''}) \boldsymbol{:} =
\Xi_{l,m}(\kappa_{l,m}), \,\,\, l = l' + l'', \, m = m' + m'', 
\]
where 
\[
\kappa_{l,m} = \kappa'_{l',m'} \, \overline{\otimes} \,\, \kappa''_{l'',m''}
\]
is equal to the ordinary tensor product 
\begin{multline*}
\kappa'_{l',m'} \, \otimes  \,\, \kappa''_{l'',m''} \\
\in E_{{}_{n'_1}} \otimes \cdots \otimes  E_{{}_{n'_{M'}}} \otimes 
E_{{}_{n''_1}} \otimes \cdots \otimes  E_{{}_{n''_{M''}}} \otimes 
\mathscr{E}^{*}_{{}_{n'_1}} \otimes  \cdots \otimes \mathscr{E}^{*}_{{}_{n'_{M'}}} \otimes
\mathscr{E}^{*}_{{}_{n''_1}} \otimes  \cdots \otimes \mathscr{E}^{*}_{{}_{n''_{M''}}} \\ \cong
\mathscr{L}(E^{*}_{{}_{n'_1}} \otimes \cdots \otimes  E^{*}_{{}_{n'_{M'}}} \otimes 
E^{*}_{{}_{n''_1}} \otimes \cdots \otimes  E^{*}_{{}_{n''_{M''}}}, \,\,
\mathscr{E}^{*}_{{}_{n'_1}} \otimes  \cdots \otimes \mathscr{E}^{*}_{{}_{n'_{M'}}} \otimes
\mathscr{E}^{*}_{{}_{n''_1}} \otimes  \cdots \otimes \mathscr{E}^{*}_{{}_{n''_{M''}}}),
\end{multline*}
\begin{enumerate}
\item[1)] 
multiplied by $(-1)^p$ where $p$ is the parity of the permutation which has to be applied to the 
fermi operators lying among the Hida operators put in the order
\[
\partial_{w'_1}^* \cdots \partial_{w'_{M'}} \partial_{w''_{1}}^* \cdots \partial_{w''_{M''}}
\] 
in which the Hida operators are put formally together in the order in which they stand in the general formula 
(\ref{electron-positron-photon-Xi}) for $\Xi_{l',m'}(\kappa'_{l',m'})$ (first) and in the general formula
(\ref{electron-positron-photon-Xi}) for $\Xi_{l'',m''}(\kappa''_{l'',m''})$ (second), in order to rearrange them into 
the order in which they stand in the general formula  (\ref{electron-positron-photon-Xi}) 
for $\Xi_{l,m}(\kappa''_{l,m})$ 
\item[2)]
separately symmetrized with respect to all bose variables which lie within the
the first $l$ variables,
\item[3)]
separately symmetrized with respect to all bose variables which lie within the last $m$
variables,
\item[4)]
separately antisymmetrized with respect to all fermi variables which lie among the first $l$
variables,
\item[5)]
separately antisymmetrized with respect to all fermi variables which lie among the last
$m$ variables,
\item[6)]
the $n'_k$-th or respectively $n''_k$-th variable is counted as lying among the first $l$ variables if it lies
among the first $l'$ variables in $\kappa'_{l',m'}$ or among the first $l''$ variables 
of the kernel $\kappa''_{l'',m''}$. The remaining variables are counted as the last $m$ variables.  
\end{enumerate}

In fact Wick product is well defined on a much larger class of integral kernel operators
$\Xi_{l,m}(\kappa_{l,m})$, because for its validity it is sufficient that the kernels $\kappa_{l,m}$
respect the condition of Theorem \ref{obataJFA.Thm.3.13}, considerably weaker than the condition 
(\ref{StrongkappaProperty}). In this wider class of operators the last rule for computation
of the Wick product remains true.

A much more interesting case we encounter when among the integral kernel operators 
(\ref{FreeFieldOp}) there are present such, which are equal to Wick polynomials 
of free fields at one and the same space-time point. Now we give general definition of
such a Wick product of (fixed components of) free fields at one and the same space-time point, and show that the 
correponding integral kernel operator lies among the class which can be placed into the 
above Wick product. The resulting integral kernel operator $\Xi$ will be a finite sum of well defined
integral kernel operators $\Xi(\kappa_{l,m})$ with the kernel(s)
\begin{equation}\label{StrongkappaProperty}
\kappa_{l,m} \in \mathscr{L}(E_{{}_{n_1}} \otimes \cdots \otimes  E_{{}_{n_M}}, \,\,
\mathscr{E}^{*}_{{}_{n_1}} \otimes  \cdots \otimes \mathscr{E}^{*}_{{}_{n_M}}), \,\,\, M= l+m
\end{equation}
and thus with
\[
\Xi_{l,m}(\kappa_{l,m}) \in 
\mathscr{L}\big((\boldsymbol{E}) \otimes \mathscr{E}, (\boldsymbol{E})^*\big) \cong 
\mathscr{L}\Big(\mathscr{E}, \,\, \mathscr{L}\big((\boldsymbol{E}), \, (\boldsymbol{E})^*\big) \Big)
\]
by the generalization of Thm. 3.9 of \cite{obataJFA} to the tensor product of Fock spaces, 
compare Subsection \ref{psiBerezin-Hida}. Therefore the Wick product of free fields (or their derivatives) 
$\Xi$ at the fixed space-time point belongs to the general class of finite sums of integral kernel operators with vector-valued kernels,
which in general does not belong to 
\[
\mathscr{L}\big((\boldsymbol{E}) \otimes \mathscr{E}, (\boldsymbol{E})\big) \cong 
\mathscr{L}\Big(\mathscr{E}, \,\, \mathscr{L}\big((\boldsymbol{E}), \, (\boldsymbol{E})\big) \Big)
\]
if among the factors in the Wick product (at fixed point) there are zero mass fields or their derivatives.
But if among the fators there are no factors corresponding to zero mass fields (or their derivatives)
then  the resulting integral kernel operator $\Xi$ -- Wick product at fixet point -- will be a finite sum of 
well defined integral kernel operators $\Xi(\kappa_{l,m})$ with the kernels respecting the condition of Thm.  
\ref{obataJFA.Thm.3.13}, \emph{i. e.}  with 
\[
\Xi_{l,m}(\kappa_{l,m}) \in 
\mathscr{L}\big((\boldsymbol{E}) \otimes \mathscr{E}, (\boldsymbol{E})\big) \cong 
\mathscr{L}\Big(\mathscr{E}, \,\, \mathscr{L}\big((\boldsymbol{E}), \, (\boldsymbol{E})\big) \Big)
\]
by the generalization of Thm. 3.13 of \cite{obataJFA} to the tensor product of Fock spaces, 
compare Thm. \ref{obataJFA.Thm.3.13} of Subsection \ref{psiBerezin-Hida}, and
with $\mathscr{E}_{1}^{*}$-valued or respectively  
$\mathscr{E}_{2}^*$-valued distribution kernels, for both nuclear space-time test function spaces: 
$\mathscr{E}_{1}$ and for $\mathscr{E}_{2}$ given by the special case of (\ref{mathscrE_1,mathscrE_2}) 
with $M=1$ and $q_k=1$ in it, \emph{i.e.}
\[
\begin{split}
\mathscr{E}_{1} = \mathcal{S}_{H_{(4)}}(\mathbb{R}^4;\mathbb{C})=
\mathcal{S}(\mathbb{R}^4; \mathbb{C}) \,\,\, \textrm{or} \\
\mathscr{E}_{2} = \mathcal{S}_{\mathscr{F}A^{(4)}\mathscr{F}^{-1}}(\mathbb{R}^4;\mathbb{C})=
\mathcal{S}^{00}(\mathbb{R}^4; \mathbb{C}).
\end{split}
\]

For the need of causal perturbative construction of interacting fields it is sufficient to confine attention to integral kernel operators representing the respective components of free fields, of their spatio-temporal dervatives, their Wick products, their integrals with pairing functions (e.g. convolutions of Wick products of spatio-temporal 
derivatives of fixed components of free fields with pairing distributions, \emph{i. e.}`` pairing functions'').  
Therefore we confine ourselves to fixed components of the free fields and of their spatio-temporal derivatives 
and thus to scalar-valued space-time test 
function spaces $\mathscr{E}_{1}= \mathcal{S}(\mathbb{R}^4; \mathbb{C})$ or respectively 
$\mathscr{E}_{2} = \mathcal{S}^{00}(\mathbb{R}^4; \mathbb{C})$.
Correspondingly to this we consider integral kernel operators with the vector-valued kernels corresponding 
to fixed components of free fields which can be represented
by the functions
\begin{equation}\label{KernelsOfFreeFieldComponents}
\begin{split}
{}^{1}\kappa_{0,1}(w;X) = {}^{1}\kappa_{0,1}(s, \boldsymbol{\p}; a, x), \,\,\,
{}^{1}\kappa_{1,0}(w;X) = {}^{1}\kappa_{0,1}(s, \boldsymbol{\p}; a, x) \,\,\, \textrm{or}  \\
{}^{2}\kappa_{0,1}(w;X) = {}^{2}\kappa_{0,1}(\nu, \boldsymbol{\p}; \mu, x), \,\,\,
{}^{2}\kappa_{1,0}(w;X) = {}^{2}\kappa_{0,1}(\nu, \boldsymbol{\p}; \mu, x),
\end{split}
\end{equation}
with fixed values of the discrete  indices $a, \mu$. To this class (\ref{KernelsOfFreeFieldComponents}) of kernels 
we add their spatio-temporal derivatives 
\begin{equation}\label{KernelsOfDerivativesOfFreeFieldComponents}
\begin{split}
\partial^{\alpha} \,\, {}^{1}\kappa_{0,1}(w;X) = \partial^{\alpha} \,\, {}^{1}\kappa_{0,1}(s, \boldsymbol{\p}; a, x), \,\,\,
\partial^{\alpha} \,\, {}^{1}\kappa_{1,0}(w;X) = \partial^{\alpha} \,\, {}^{1}\kappa_{0,1}(s, \boldsymbol{\p}; a, x) \,\,\, \textrm{or}  \\
\partial^{\alpha} \,\, {}^{2}\kappa_{0,1}(w;X) = \partial^{\alpha} \,\, {}^{2}\kappa_{0,1}(\nu, \boldsymbol{\p}; \mu, x), \,\,\,
\partial^{\alpha} \,\, {}^{2}\kappa_{1,0}(w;X) = \partial^{\alpha} \,\, {}^{2}\kappa_{0,1}(\nu, \boldsymbol{\p}; \mu, x), \\
\textrm{where} \\
\alpha = (\alpha_0, \alpha_1, \alpha_2, \alpha_3) \in \mathbb{N}_{0}^{4} \,\,\, \textrm{and} \,\,\,
\partial^{\alpha} = \frac{\partial^{|\alpha_0|}}{(\partial x_0)^{\alpha_0}}
\frac{\partial^{|\alpha_1|}}{(\partial x_1)^{\alpha_1}}
\frac{\partial^{|\alpha_2|}}{(\partial x_2)^{\alpha_2}}
\frac{\partial^{|\alpha_3|}}{(\partial x_3)^{\alpha_3}}
\end{split}
\end{equation}

\begin{defin}\label{K_0}
The class $\mathfrak{K}_0$ of kernels we are considering in the sequel consists of the plane wave kernels
(\ref{KernelsOfFreeFieldComponents}) defining the free fields of the theory and of 
their spatio-temporal derivatives (\ref{KernelsOfDerivativesOfFreeFieldComponents}), with fixed values
of the indices $a, \mu, \alpha$. 
\end{defin}

Upon the integral kernel operators determined by the vector valued kernels
$\mathfrak{K}_0$ we perform the operations of Wick product (Rule I), Wick products at the same space-time point
(Rule II), spatio-temporal derivations (Rule III), integrations (IV and V) and finally convolutions with pairing functions
(Rule VI). Correspondingly to each of the said operations there exists the correponding Rule performed upon the kernels, corresponding to the operators. Of course the operations performed upon the kernels
in $\mathfrak{K}_0$ and determined by the Rules will extend the initial class $\mathfrak{K}_0$. 
We use a general notation 
\[
{}^{{}^{n}}_{{}_{k}}\kappa_{l,m}(s, \boldsymbol{\p};x), \,\, n = 1 
\]
for a kernel
\[
\partial^{\alpha} \,\, {}^{1}\kappa_{l,m}(s, \boldsymbol{\p}; a, x), \,\,\, (l,m) = (0,1) \,\, \textrm{or} \,\, =(1,0)
\]
with fixed indices $a, \alpha$ and 
with ${}^{1}\kappa_{0,1}(s, \boldsymbol{\p}; a, x)$ equal to the plane wave kernel defining the free Dirac field.
Similarily we will denote simply by
\[
{}^{{}^{n}}_{{}_{k}}\kappa_{l,m}(\nu, \boldsymbol{\p}; x), \,\, n = 2 
\]
the kernel
\[
\partial^{\alpha} \,\, {}^{2}\kappa_{l,m}(\nu, \boldsymbol{\p}; \mu, x),  \,\,\, (l,m) = (0,1) \,\, \textrm{or} \,\, =(1,0)
\]
with fixed indices $\mu, \alpha$ and 
with ${}^{2}\kappa_{l,m}(\nu, \boldsymbol{\p}; \mu, x)$ equal to the plane wave kernel defining 
the free electromagnetic potential field.

Assuming 
\[
{}^{{}^{n_k}}_{{}_{k}}\kappa_{l_k,m_k} \in \mathfrak{K}_0, \,\,\, k = 1, \ldots, M,
\]
we consider the following Wick monomials,
i.e. Wick products at the same space-time point, of the following operators 
\begin{equation}\label{FreeFieldOpAtx}
\Xi_{l_1,m_1}\Big({}^{{}^{n_1}}_{{}_{1}}\kappa_{l_1,m_1}\Big), \ldots 
\Xi_{l_M,m_M}\Big({}^{{}^{n_M}}_{{}_{M}}\kappa_{l_M,m_M}\Big)
\end{equation}
with general (not necessary equal to plane wave distributions defining the free fields, as we have in view 
also their spatio-temporal-derivative fields) kernels
\[
{}^{{}^{n_k}}_{{}_{k}}\kappa_{l_k,m_k}
\in \mathscr{L}(E_{{}_{n_k}}, \mathscr{E}^{*}_{{}_{n_k}}) \cong 
E^{*}_{{}_{n_k}} \otimes \mathscr{E}^{*}_{{}_{n_k}}, , \,\,\,\, k=1,2, \ldots M
\]
representable by ordinary functions, respecting the conditions expressed in Lemma \ref{kappa0,1,kappa1,0psi}, Subsection \ref{psiBerezin-Hida} or respectively Lemma \ref{kappa0,1,kappa1,0ForA},
Subsection \ref{A=Xi0,1+Xi1,0}, \emph{i.e.} extendible to elements
\begin{equation}\label{ExtedibilityCondition}
{}^{{}^{n_k}}_{{}_{k}}\kappa_{l_k,m_k}
\in \mathscr{L}(E^{*}_{{}_{n_k}}, \mathscr{E}^{*}_{{}_{n_k}}) \cong 
E_{{}_{n_k}} \otimes \mathscr{E}^{*}_{{}_{n_k}}
\end{equation}
with the property that
\begin{equation}\label{ConvolutabilityCondition}
{}^{{}^{n_k}}_{{}_{k}}\kappa_{l_k,m_k}(\xi) \in \mathcal{O}_C(\mathbb{R}^4; \mathbb{C}), 
\,\,\, \xi \in E_{{}_{n_k}}.
\end{equation}
Here 
\[
n_k = \left\{ \begin{array}{l}
1 \\
\textrm{or} \\
2
\end{array} \right., \,\,\,\, \textrm{and} \,\,\,
(l_k, m_k) = \left\{ \begin{array}{l}
(0,1) \\
\textrm{or} \\
(1,0)
\end{array} \right.
\]
and the integral kernel operator 
\[
\Xi_{l_k,m_k}\Big({}^{{}^{n_k}}_{{}_{k}}\kappa_{l_k,m_k}\Big),
\]
regarded as the operator on the said tensor product of Fock spaces, has the exceptional form
(similarily as for the operators defined by the free fields $A$ and $\boldsymbol{\psi}$)
that the integraton in the general formula (\ref{electron-positron-photon-Xi})
for this operator is restricted to fermion variables, if $n_k = 1$, or to bose variables, if
$n_k = 2$. 

Validity of (\ref{ExtedibilityCondition}) and (\ref{ConvolutabilityCondition}) for spatio-temporal
derivatives of the plane wave kernels (\ref{KernelsOfFreeFieldComponents}) can be proved exactly as for
kernels (\ref{KernelsOfFreeFieldComponents}) themselves by repeating the argumet of the proof of
Lemma \ref{kappa0,1,kappa1,0psi}, Subsection \ref{psiBerezin-Hida} or respectively 
Lemma \ref{kappa0,1,kappa1,0ForA},
Subsection \ref{A=Xi0,1+Xi1,0}. 
 
In fact in construction of interacting fields in the standard spinor QED it would be sufficient to consider only 
the kernels  (\ref{KernelsOfFreeFieldComponents}) and the kernels which arise by performing upon them the respective operations determined by the Rules I - VI, except the III-rd, given below. This is because no spatio-temporal derivatives of free fields enter the interaction lagrangian in spinor QED, but only free fields themselves. But in case of scalar QED the interaction lagrangian contains derivatives of 
free fields, so in that case spatio-temporal derivatives of the kernels determining the scalar free field has 
to be taken into consideration.  

So let 
\[
{}^{{}^{n_k}}_{{}_{k}}\kappa_{l_k,m_k} \in \mathfrak{K}_0, \,\,\, k = 1, \ldots, M.
\]
Then for each fixed space-time point $x$ the scalar integral kernel operators
\begin{equation}\label{FreeFieldOpAtx}
\Xi_{l_1,m_1}\Big({}^{{}^{n_1}}_{{}_{1}}\kappa_{l_1,m_1}(x)\Big), \ldots 
\Xi_{l_M,m_M}\Big({}^{{}^{n_M}}_{{}_{M}}\kappa_{l_M,m_M}(x)\big)
\end{equation}
determined by scalar kernel functions
\[
{}^{{}^{n_k}}_{{}_{k}}\kappa_{l_k,m_k}(x): w_{n_k} \longmapsto 
{}^{{}^{n_k}}_{{}_{k}}\kappa_{l_k,m_k}(w_{n_k}; x),
\]
are well defined generalized operators transforming continously the Hida space $(\boldsymbol{E})$
into its strong dual $(\boldsymbol{E})^*$, and exist point-wisey as Pettis integrals
(\ref{electron-positron-photon-Xi}) with integration in (\ref{electron-positron-photon-Xi})
restricted to fermi variables, iff $n_k = 1$, or to bose variables, iff $n_k = 2$, compare 
Subsection 5.9 of \cite{wawrzycki2018}.
Moreover for each fixed $x$ there exist a well defined Wick product of the operators 
(\ref{FreeFieldOpAtx}) 
\begin{equation}\label{WickFreeFieldOpAtx}
\boldsymbol{:}\Xi_{l_1,m_1}\Big({}^{{}^{n_1}}_{{}_{1}}\kappa_{l_1,m_1}(x)\Big), \ldots 
\Xi_{l_M,m_M}\Big({}^{{}^{n_1}}_{{}_{M}}\kappa_{l_M,m_M}(x)\big) \boldsymbol{:}
\end{equation}
defined as the ordinary product of these operators, but rearranged in the so called ``normal'' order,
in which all operators 
\begin{equation}\label{Xi0,1}
\Xi_{l_k,m_k}\Big({}^{{}^{n_k}}_{{}_{k}}\kappa_{l_k,m_k}(x)\Big)
\end{equation}
with $(l_k, m_k) = (1,0)$ stand to the left of all opertators
\begin{equation}\label{Xi1,0}
\Xi_{l_k,m_k}\Big({}^{{}^{n_k}}_{{}_{k}}\kappa_{l_k,m_k}(x)\Big)
\end{equation}
with $(l_k, m_k) = (0,1)$, multiplied in addition  by the factor $(-1)^p$
with $p$ equal to the parity of the 
permutation performed upon fermi operators, having $n_k = 1$ and corresponding to the fermi variables, 
required to bring the operators into the required ``normal'' order.

\begin{center}
{\small RULE II}
\end{center}
\emph{We have the following computational rule}
\begin{multline*}
\boldsymbol{}: \Xi_{l_1,m_1}\Big({}^{{}^{n_1}}_{{}_{1}}\kappa_{l_1,m_1}(x)\Big) \cdots 
\Xi_{l_M,m_M}\Big({}^{{}^{n_1}}_{{}_{M}}\kappa_{l_M,m_M}(x)\big) \boldsymbol{:} \\
= \Xi_{l,m}(\kappa_{lm}(x)), 
\\
 l = l_1 + \cdots l_M, \,\,\, m = m_1 + \cdots m_M
\end{multline*}
\emph{where the ordinary function representing the kernel $\kappa_{l,m}$} 
\[
\kappa_{l,m}(w_1, \ldots, w_M; x) = 
\Big({}^{{}^{n_1}}_{{}_{1}}\kappa_{l_1,m_1} \Big) \overline{\dot{\otimes}} \cdots \overline{\dot{\otimes}} \,\,
\Big( {}^{{}^{n_M}}_{{}_{M}}\kappa_{l_M,m_M} \Big)(w_1, \ldots, w_M; x)
\]
\emph{is equal to the ordinary product 
\begin{multline*}
\Big({}^{{}^{n_1}}_{{}_{1}}\kappa_{l_1,m_1}\Big) \dot{\otimes} \cdots \dot{\otimes} \,\,
\Big({}^{{}^{n_M}}_{{}_{M}}\kappa_{l_M,m_M}\Big)(w_1, \ldots, w_M; x) \\ =
\Big({}^{{}^{n_1}}_{{}_{1}}\kappa_{l_1,m_1}\Big)(w_1;x) 
\cdots \Big({}^{{}^{n_M}}_{{}_{M}}\kappa_{l_M,m_M}\Big)(w_M;x),
\end{multline*}
1) separately symmetrized with respect to all bose variables, lying among the first $l$ variables,
2) separately symmetrized with respect to all bose variables, lying among the last $m$
variables, 3) separately antisymmetrized with respect to all fermi variables which 
lie among the first $l$ variables, 4) separately antisymmetrized with respect 
to all fermi variables lying among the 
last $m$ variables, finally 5) the result multiplied by the factor
$(-1)^p$, where $p$ is the parity of the  permutation performed upon the fermi operators necessay to rearrange them into 
the order in which they stand in the general formula (\ref{electron-positron-photon-Xi})
for $\Xi_{l,m}(\kappa_{l,m})$.  Here by definition $n_k$ is counted among the first $l$ variables
iff the corresponding $(l_k, m_k) = (1,0)$, and $n_k$ is counted among last $m$
variables iff the corresponding $(l_k, m_k) = (0,1)$.}

Again the Rule II can be justified by using the fact that the operators (\ref{Xi0,1})
exist point-wisely as Pettis integrals, and represent operators mapping continously 
the strong dual $(\boldsymbol{E})^*$ of the Hida space into its strong dual $(\boldsymbol{E})^*$
(continuous as well as operators $(\boldsymbol{E}) \rightarrow (\boldsymbol{E})^*$), and similarly
we have for the operators (\ref{Xi1,0}), representing continous operators 
$(\boldsymbol{E}) \rightarrow (\boldsymbol{E})$ 
(as well  continuous as operators $(\boldsymbol{E}) \rightarrow (\boldsymbol{E})^*$). The proof, using essentially the same arguments as that used in the proof of Bogoliubov-Shirkov Hypothesis in Subsection 5.9 of \cite{wawrzycki2018}, 
can be omitted, compare Subsection 5.9 of \cite{wawrzycki2018}.

From the Rule II it easily follows that the Wick product (\ref{WickFreeFieldOpAtx}) determines 
integral kernel operator 
\[
\Xi_{l,m}(\kappa_{l,m}) = 
\Xi_{l,m} \Bigg( \Big({}^{{}^{n_1}}_{{}_{1}}\kappa_{l_1,m_1} \Big) \overline{\dot{\otimes}} \cdots \overline{\dot{\otimes}} \,\, \Big( {}^{{}^{n_M}}_{{}_{M}}\kappa_{l_M,m_M} \Big)  \Bigg)
\]
with vector valued kernel 
\begin{multline}\label{kappaInE^*xE^*x...xE^*xmathscrE^*}
\kappa_{l,m} = \Big({}^{{}^{n_1}}_{{}_{1}}\kappa_{l_1,m_1} \Big) \overline{\dot{\otimes}} \cdots \overline{\dot{\otimes}} \,\, \Big( {}^{{}^{n_M}}_{{}_{M}}\kappa_{l_M,m_M} \Big) \\
\in
E^{*}_{{}_{n_1}} \otimes \cdots \otimes 
E^{*}_{{}_{n_M}} \otimes \mathscr{E}^{*}_{{}_{i}} 
\cong \mathscr{L}(E_{{}_{n_1}} \otimes \cdots \otimes  E_{{}_{n_M}}, \,\,
\mathscr{E}^{*}_{{}_{i}}), \,\,\, i=1,2,
\end{multline}
and, when all $n_k =1$ (\emph{i.e.}  all ${}^{{}^{n_k}}_{{}_{k}}\kappa_{l_k,m_k}$ are the plane wave kernels correponding to derivatives of the Dirac field), defines the bilinear map
\begin{multline}\label{Bilin.kappa.kappaInExExE^*}
\xi \times \eta \mapsto \kappa_{l,m}(\xi \otimes \eta), 
\\
\xi \in \overbrace{E_{i_1} \otimes \cdots 
\otimes E_{i_l}}^{\textrm{first $l$ terms $E_{i_j}$, $i_j \in \{1,2\}$}}, \\
\eta \in \overbrace{E_{i_{l+1}} \otimes \cdots 
\otimes E_{i_{l+m}}}^{\textrm{last $m$ terms $E_{i_j}$, $i_j\in \{1,2\}$}},
\end{multline}
which can be extended to a separately continuous bilinear map from
\begin{equation}\label{kappa.kappaInExExE^*}
\Big( \overbrace{E_{i_1} \otimes \cdots 
\otimes E_{i_l}}^{\textrm{first $l$ terms $E_{i_j}$}} \Big)^*
\times
\Big( \overbrace{E_{i_{l+1}} \otimes \cdots 
\otimes E_{i_{l+m}}}^{\textrm{last $m$ terms $E_{i_j}$}} \Big)
\,\,\, \textrm{into} \,\,\,\mathscr{L}(\mathscr{E}, \mathbb{C}) = \mathscr{E}^*.
\end{equation}
Thus in each case 
\begin{multline*}
\Xi_{l,m}(\kappa_{l,m})  = 
\Xi_{l,m} \Bigg( \Big({}^{{}^{n_1}}_{{}_{1}}\kappa_{l_1,m_1} \Big) \overline{\dot{\otimes}} \cdots \overline{\dot{\otimes}} \,\, \Big( {}^{{}^{n_M}}_{{}_{M}}\kappa_{l_M,m_M} \Big)  \Bigg) \\
\in \mathscr{L}\big((\boldsymbol{E}) \otimes \mathscr{E}_i, (\boldsymbol{E})^{*} \big) \cong 
\mathscr{L}\Big(\mathscr{E}_i, \,\, \mathscr{L}\big((\boldsymbol{E}), \, (\boldsymbol{E})^{*}\big) \Big),
\,\,\, i = 1,2,
\end{multline*}
by Theorem 3.9 of \cite{obataJFA} (or its generaliztion to the case of tensor product of Fock spaces,
compare Subsection \ref{psiBerezin-Hida}). 

In case in which there are no factors 
\[
\Xi_{l_k, m_k}\Big({}^{{}^{n_k}}_{{}_{k}}\kappa_{l_1,m_1}\Big) \,\,\, \textrm{with} \,\,\, n_k = 2
\]
\emph{i.e.} no factors corresponding to the (derivatives) of the zero mass free fields of the theory, e.g. of the electromagnetic potential field in case of QED, we have
\begin{multline*}
\Xi_{l,m}(\kappa_{l,m})  = 
\Xi_{l,m} \Bigg( \Big({}^{{}^{n_1}}_{{}_{1}}\kappa_{l_1,m_1} \Big) \overline{\dot{\otimes}} \cdots \overline{\dot{\otimes}} \,\, \Big( {}^{{}^{n_M}}_{{}_{M}}\kappa_{l_M,m_M} \Big)  \Bigg) \\
\in \mathscr{L}\big((\boldsymbol{E}) \otimes \mathscr{E}_i, (\boldsymbol{E})\big) \cong 
\mathscr{L}\Big(\mathscr{E}_i, \,\, \mathscr{L}\big((\boldsymbol{E}), \, (\boldsymbol{E})\big) \Big),
\,\,\, i = 1,2,
\end{multline*}
by Theorem  \ref{obataJFA.Thm.3.13}, Subsection \ref{psiBerezin-Hida} (generalization of
Thm. 3.13 in \cite{obataJFA}).

Indeed we use several technical Lemmas which allow us to show (\ref{kappaInE^*xE^*x...xE^*xmathscrE^*})
as well as the extedibility
(\ref{kappa.kappaInExExE^*}) property of the bilinear map (\ref{Bilin.kappa.kappaInExExE^*})
in case in which the zero mass terms are absent. 
We need the following technical definition

\begin{defin}\label{mathfrakS_i}
Let $\mathfrak{S}_i$, $i=1,2$, denote the family of subsets of $E_{i} \subset E_{i}^{*}$
which are bounded in the topology on  $E_{i}$ 
induced by the strong dual topology on $E_{i}^{*}$. Otherwise: 
$\mathfrak{S}_i$ is the family of intersections of all sets bounded in
the strong dual space $E_{i}^{*}$ with the subset $E_{i}$ of $E_{i}^{*}$.
\end{defin}

\begin{lem}\label{Hypocont.Ofkappa.kappa}
Let
\[
{}^{{}^{1}}_{{}_{1}}\kappa_{1,0}, {}^{{}^{1}}_{{}_{2}}\kappa_{1,0} \in \mathfrak{K}_0, 
\]
i.e. let the above two kernels be equal to fixed components of plane wave kernels 
defininig the massive free fields of the theory (\emph{i. e.} the Dirac field in case of QED), 
or to their spatio-temporal derivatives $\partial^{\alpha}$
with fixed value of the multiindex $\alpha \in \mathbb{N}_{0}^{4}$.
Then the map 
\[
E_{1}^{*} \times E_{1}^{*} \supset E_{1} \times E_{1} \ni \xi_1 \times \xi_2
\longmapsto {}^{{}^{1}}_{{}_{1}}\kappa_{1,0}(\xi_1) 
\cdot {}^{{}^{1}}_{{}_{2}}\kappa_{1,0}(\xi_2) \in \mathscr{E}_{k}^{*}, 
\]
is $\big(\mathfrak{S}_{1}, \mathfrak{S}_{1}\big)$-hypocontinuous as a map
\[
E_{1} \times E_{1} \longrightarrow \mathscr{E}_{k}^{*}, \,\,\, k =1,2
\]
with the topology on $E_{1} \subset E_{1}^{*}$, induced by the strong dual topology 
on $E_{1}^{*}$, and with the strong dual topology on $\mathscr{E}_{k}^{*}$, $k=1,2$.
\end{lem}

\qedsymbol \, (An outline of the proof)
$\mathscr{E}_2 = \mathcal{S}^{00}(\mathbb{R}^4; \mathbb{C})$ is continously
inserted into $\mathcal{S}(\mathbb{R}^4; \mathbb{C})$, and thus 
the strong dual $\mathscr{E}_{1}^{*} = \mathcal{S}(\mathbb{R}^4; \mathbb{C})^*$ is continously inserted 
into the strong dual $\mathscr{E}_{2}^{*} = \mathcal{S}^{00}(\mathbb{R}^4; \mathbb{C})^*$,
for the proof compare Subsection 5.5 of \cite{wawrzycki2018}. It is therefore sufficient 
to prove the Lemma for the case
$\mathscr{E}_{1}^{*} = \mathcal{S}(\mathbb{R}^4; \mathbb{C})^{*}$ with  $k=1$.

Consider for example the case of the plane wave kernel $\kappa_{1,0}$ given by the formula
(\ref{kappa_1,0}),
Subsect. \ref{psiBerezin-Hida} or (\ref{skappa_1,0}) of
Subsection \ref{StandardDiracPsiField} which defines (one of the two \emph{a priori} possible)
Dirac free fields (the analysis of their fixed satio-temporal derivation components is identical).

Recall that for $\phi \in \mathscr{E}_1 = \mathcal{S}(\mathbb{R}^4; \mathbb{C})$,
$\xi_1, \xi_2 \in E_1 = \mathcal{S}(\mathbb{R}^3; \mathbb{C}^4)$ 
(here we fix once for all the spinor indices $a_1, a_2$ and in case of spatio-temporal derivatives
$\partial^{\alpha_1}\kappa_{1,0}$ and $\partial^{\alpha_2}\kappa_{1,0}$ the additional multiindices 
$\alpha_1, \alpha_2 \in \mathbb{N}_{0}^{4}$ would also be fixed) we have
\begin{multline*}
\langle \kappa_{1,0}(\xi_1) \cdot\kappa_{1,0}(\xi_2), \phi \rangle =\\
\sum \limits_{s_1, s_2} \int \limits_{\mathbb{R}^3 \times \mathbb{R}^3 \times \mathbb{R}^4}
\kappa_{1,0}(s_1, \boldsymbol{\p}_1; a_1, x) \cdot \kappa_{1,0}(s_2, \boldsymbol{\p}_2; a_2, x)
\, \xi_{1}(s_1, \boldsymbol{\p}_1) \xi_{2}(s_2, \boldsymbol{\p}_1) \phi(x) \,
\ud^3 \boldsymbol{\p}_1 \, \ud^3 \boldsymbol{\p}_2 \, \ud^4x. 
\end{multline*}
\[
\begin{split}
\kappa_{1,0}(\xi_1)(a_1,x) = 
\sum \limits_{s_1}\int \limits_{\mathbb{R}^3} 
\kappa_{1,0}(s_1, \boldsymbol{\p}_1; a_1, x) \, \xi_{1}(s_1, \boldsymbol{\p}_1)
\, \ud^3 \boldsymbol{\p}_1, \\
\kappa_{1,0}(\xi_2)(a_2,x) = 
\sum \limits_{s_2}\int \limits_{\mathbb{R}^3} 
\kappa_{1,0}(s_2, \boldsymbol{\p}_2; a_2, x) \, \xi_{2}(s_2, \boldsymbol{\p}_2)
\, \ud^3 \boldsymbol{\p}_1.
\end{split}
\]

Next we show that if $\xi_1 \in E_1 = \mathcal{S}(\mathbb{R}^3; \mathbb{C})$ ranges over a set 
$S \in \mathfrak{S}_1$, i.e. over $S \subset E_1 \subset E_{1}^{*}$ bounded in the strong 
dual topology on $E_{1}^{*}$, and if $\phi \in \mathscr{E}_{1} = \mathcal{S}(\mathbb{R}^4; \mathbb{C})$
ranges over a set $B \subset \mathscr{E}_{1} = \mathcal{S}(\mathbb{R}^4; \mathbb{C})$ 
bounded in $\mathscr{E}_{1} = \mathcal{S}(\mathbb{R}^4; \mathbb{C})$ (with respectto the ordinary 
nuclear Schwartz topolody on $\mathcal{S}(\mathbb{R}^4;\mathbb{C})$, then the set
$B^{+}(S, B)$ of functions (spinor indices $a_1, a_2$ are fixed)
\[
(s_2, \boldsymbol{\p}_2) \longmapsto
\sum \limits_{s_1}
\int \limits_{\mathbb{R}^3 \times \mathbb{R}^4} 
\kappa_{1,0}(s_1, \boldsymbol{\p}_1; a_1, x) \cdot \kappa_{1,0}(s_2, \boldsymbol{\p}_2; a_2, x)
\, \xi_1(s_1, \boldsymbol{\p}_1) \, \phi(x) \, \ud^3 \boldsymbol{\p}_1 \, \ud^4x
\]
and the set $B^{+}(B, S)$ of functions 
\[
(s_1, \boldsymbol{\p}_1) \longmapsto
\sum \limits_{s_1}
\int \limits_{\mathbb{R}^3 \times \mathbb{R}^4} 
\kappa_{1,0}(s_1, \boldsymbol{\p}_1; a_1, x) \cdot \kappa_{1,0}(s_2, \boldsymbol{\p}_2; a_2, x)
\, \xi_2(s_2, \boldsymbol{\p}_2) \, \phi(x) \, \ud^3 \boldsymbol{\p}_2 \, \ud^4x
\]
with $\xi_2$ ranging over $S \in \mathfrak{S}_1$
and $\phi \in B$
are bounded in $E_1= \mathcal{S}(\mathbb{R}^3; \mathbb{C}^4)$. The proof, being a simple
verification of definition of boundedness, can be omitted, but we encourage 
the reader to perform the computations explicitly.

Next we observe that for any $S \in \mathfrak{S}_1$ and any strong zero-neighborhood $W(B, \epsilon)$ in 
$\mathscr{E}_{1}^{*} = \mathcal{S}(\mathbb{R}^4; \mathbb{C})^{*}$, determined by a bounded
set $B$ in $\mathscr{E}_1 = \mathcal{S}(\mathbb{R}^4; \mathbb{C})$ and $\epsilon >0$, for 
the strong zero-neighborhoods $V\big(B^{+}(S,B), \, \epsilon\big)$ and $V\big(B^{+}(B,S), \, \epsilon\big)$
we have
\[
|\langle \kappa_{1,0}(\xi_1) \cdot\kappa_{1,0}(\xi_2), \phi \rangle| < \epsilon
\] 
whenever 
\[
\xi_1 \in S, \,\,\,\, \xi_2 \in  V\big(B^{+}(S,B), \, \epsilon\big)
\]
or whenever 
\[
\xi_1 \in V\big(B^{+}(B,S), \, \epsilon\big), \,\,\,\, \xi_2 \in S.
\]
Put otherwise
\[
\begin{split}
\kappa_{1,0}(S) \cdot \kappa_{1,0}\Big(V\big(B^{+}(S,B), \, \epsilon\big)\Big) \subset W(B, \epsilon), \\
\kappa_{1,0}\Big( V\big(B^{+}(B,S), \, \epsilon\big) \Big) \cdot \kappa_{1,0}(S) \subset W(B, \epsilon).
\end{split}
\]
\qed

\begin{lem}\label{Cont.Ofkappa.kappa}
\begin{enumerate}
\item[1)]
Let $\phi \in \mathscr{E}_1 = \mathcal{S}(\mathbb{R}^4; \mathbb{C})$ and let 
$\widetilde{\phi}$ be equal to its Fourier transform
\[
\widetilde{\phi}(p) = \int \limits_{\mathbb{R}^4} \phi(x) \, e^{i p \cdot x} \, \ud^4 x.
\]
Then if $\phi \in \mathcal{S}(\mathbb{R}^4; \mathbb{C})$ ranges over a bounded set $B$
in the Schwartz space $\mathcal{S}$, equivalently, if $\widetilde{\phi}$ ranges over a 
bounded set $\widetilde{B}$ in $\mathcal{S}(\mathbb{R}^4; \mathbb{C})$, then there exists
a constant $C_{B}$ depending on $B$ such that
\[
|\widetilde{\phi}(\boldsymbol{\p} \pm \boldsymbol{\p}', p_0(\boldsymbol{\p}) \pm p'_{0}(\boldsymbol{\p}'))|
\leq C_{B}, \,\,\,\,\,\, 
\boldsymbol{\p}, \boldsymbol{\p}' \in \mathbb{R}^3, \phi \in B
\]
in each case 
\[
\begin{split}
p_0(\boldsymbol{\p}) = \sqrt{|\boldsymbol{\p}|^2 + m}, \,\,\, \textrm{or} \,\,\,
p_0(\boldsymbol{\p}) = \sqrt{|\boldsymbol{\p}|^2} = |\boldsymbol{\p}| \\
p'_{0}(\boldsymbol{\p}') = \sqrt{|\boldsymbol{\p}'|^2 + m}, \,\,\, \textrm{or} \,\,\,
p'_{0}(\boldsymbol{\p}') = \sqrt{|\boldsymbol{\p}'|^2} = |\boldsymbol{\p}'|.
\end{split}
\]
\item[2)]
Let
\[
{}^{{}^{n_1}}_{{}_{1}}\kappa_{l_1,m_1}, {}^{{}^{n_2}}_{{}_{2}}\kappa_{l_2,m_2} \in \mathfrak{K}_0, 
\,\,\, (l_k, m_k) \in \{(0,1), (1,0) \}, n_k \in \{1, 2\}, k =1,2,
\]
i.e. let the above two kernels be equal to fixed components of plane wave kernels 
defininig free fields of the theory, or to their spatio-temporal derivatives $\partial^{\alpha}$
with fixed value of the multiindex $\alpha \in \mathbb{N}_{0}^{4}$.
Then the map 
\[
E_{n_1} \times E_{n_2} \ni \xi_1 \times \xi_2
\longmapsto {}^{{}^{n_1}}_{{}_{1}}\kappa_{l_1,m_1}(\xi_1) 
\cdot {}^{{}^{n_2}}_{{}_{2}}\kappa_{l_2,m_2}(\xi_2) \in \mathscr{E}_{k}^{*}, 
\]
is continuous as a map
\[
E_{n_1} \times E_{n_2} \longrightarrow \mathscr{E}_{k}^{*}, \,\,\, k =1,2
\]
with the ordinary nuclear topology on $E_{n_k}$, $k=1,2$, and with the strong 
dual topology on $\mathscr{E}_{k}^{*}$, $k=1,2$.
\end{enumerate}
\end{lem}
\qedsymbol \,
The first part 1) is obvious.

Concerning 2) we will use the the following two facts. 
\begin{enumerate}
\item[I)]
The functions 
\[
\boldsymbol{\p} \rightarrow \frac{P(\boldsymbol{\p})}{p_0(\boldsymbol{\p})} 
= \frac{P(\boldsymbol{\p})}{\sqrt{|\boldsymbol{\p}|^2 + m}}, \,\,\, m \neq 0
\] 
with $P(\boldsymbol{\p})$ being equal to polynomials in four real variables 
$(\boldsymbol{\p}, p_0(\boldsymbol{\p}))= (p_1, p_2, p_3, \sqrt{|\boldsymbol{\p}|^2 + m})$ 
are multipliers of the Schwartz algebra 
$\mathcal{S}(\mathbb{R}^3; \mathbb{C})$, compare \cite{Schwartz} or Appendix \ref{convolutorsO'_C}.
\item[II)]
The functions 
\[
\boldsymbol{\p} \rightarrow \frac{P(\boldsymbol{\p})}{p_0(\boldsymbol{\p})} 
= \frac{P(\boldsymbol{\p})}{|\boldsymbol{\p}|}, 
\]
with $P(\boldsymbol{\p})$ being equal to polynomials in four real variables 
$(\boldsymbol{\p}, p_0(\boldsymbol{\p}))= (p_1, p_2, p_3, |\boldsymbol{\p}|)$ 
are multipliers of the nuclear algebra $\mathcal{S}^{0}(\mathbb{R}^3; \mathbb{C})$, 
for a proof compare Subsections 5.2-5.5 of \cite{wawrzycki2018}. 
\end{enumerate}

Recall that in case of QED we have
\[
\begin{split}
E_1 = \mathcal{S}_{A_1}(\mathbb{R}^3; \mathbb{C}^4) 
= \mathcal{S}(\mathbb{R}^3; \mathbb{C}^4) = \oplus \mathcal{S}(\mathbb{R}^3; \mathbb{C}) 
\,\,\, \textrm{and} \\
E_2 = \mathcal{S}_{A_2}(\mathbb{R}^3; \mathbb{C}^4) 
= \mathcal{S}^{0}(\mathbb{R}^3; \mathbb{C}^4) = \oplus \mathcal{S}^{0}(\mathbb{R}^3; \mathbb{C}).
\end{split}
\]
with $A_2 = \oplus_{0}^{3} A^{(3)}$ and $A^{(3)}$ on $L^2(\mathbb{R}^3; \mathbb{C})$ constructed in Subsection
5.3 of \cite{wawrzycki2018}, and with $A_1 = \oplus_{1}^{4} H_{(3)}$ equal to the direct sum of four 
copies of the three dimensional
oscillator hamiltonian, \emph{i. e.} $A_1$ is equal to the operator 
$A$ given by (\ref{AinL^2(R^3;C^4)}). 

In particular let us consider the distribution defined by the kernel
\begin{equation}\label{kappa.kappa(x)}
\kappa_{1,0} \overset{\cdot}{\otimes} \kappa_{1,0}(\nu_1, \boldsymbol{p}_1, \nu_2, \boldsymbol{p}_2; x) = 
\kappa_{1,0}(\nu_1, \boldsymbol{p}_1; \mu, x) \cdot \kappa_{1,0}(\nu_2, \boldsymbol{\p}_2; \lambda, x), \,\,\, 
\textrm{with fixed $\mu, \lambda$}
\end{equation}
and with $\kappa_{1,0}$ equal to the plane wave kernel defininig the free electromagnetic potential field,
and given by the formula (\ref{kappa_0,1kappa_1,0A'}), Subsection \ref{equivalentA-s}.
For each $\xi_1, \xi_2 \in E_2 = \mathcal{S}^{0}(\mathbb{R}^4; \mathbb{C})$ the value
 of the distribution 
\begin{multline*}
\kappa_{0,1} \overset{\cdot}{\otimes} \kappa_{1,0}(\xi_1 \otimes \xi_2)(x) = 
\kappa_{1,0}(\xi_1)(\mu,x) \cdot \kappa_{1,0}(\xi_2)(\lambda, x) \\
= \int \limits_{\mathbb{R}^3 \times \mathbb{R}^3} 
\frac{\ud^3 \boldsymbol{\p}_1 \, \ud^3 \boldsymbol{\p}_2}{|\boldsymbol{\p}_1| |\boldsymbol{\p}_2|}
\xi_{1}^{\mu}(\boldsymbol{\p}_1) \xi_{2}^{\lambda}(\boldsymbol{\p}_2) \, e^{i(p_1 + p_2) \cdot x}, \\
 \,\,\, 
\xi_1 \otimes \xi_2 (\boldsymbol{p}_1 \times \boldsymbol{\p}_2) = \xi_1(\boldsymbol{\p}_1)\xi_2(\boldsymbol{\p}_2)
\end{multline*}
 on $\phi \in \mathcal{S}(\mathbb{R}^4; \mathbb{C})$ is equal
\[
\langle \kappa_{1,0}(\xi_1) \cdot \kappa_{1,0}(\xi_{2}), \phi \rangle
=  \int \limits_{\mathbb{R}^3 \times \mathbb{R}^3} 
\frac{\ud^3 \boldsymbol{\p}_1 \, \ud^3 \boldsymbol{\p}_2}{|\boldsymbol{\p}_1| |\boldsymbol{\p}_2|}
\xi_{1}^{\mu}(\boldsymbol{\p}_1) \xi_{2}^{\lambda}(\boldsymbol{\p}_2) \, 
\widetilde{\phi}(\boldsymbol{\p}_1 + \boldsymbol{\p}_2, |\boldsymbol{\p}_1| + |\boldsymbol{\p}_2|).
\]
Now let $\xi_1$, $\xi_2$ range respectively over the bounded sets $B_1$ and $B_2$ in $E_2 = 
\mathcal{S}^{0}(\mathbb{R}^3; \mathbb{C}^4)$. Let $\phi$ range over a bounded set $B$ in 
$\mathcal{S}(\mathbb{R}^4; \mathbb{C})$, equivalently, $\widetilde{\phi}$ range over a bounded
set $\widetilde{B}$ in $\mathcal{S}(\mathbb{R}^4; \mathbb{C})$. 
Because the function 
\[
\boldsymbol{\p} \mapsto \frac{1}{|\boldsymbol{\p}|}
\]
is a multiplier of the nuclear algebra $\mathcal{S}^{0}(\mathbb{R}^3; \mathbb{C})$
(Subsections 5.4 and 5.5 of \cite{wawrzycki2018}) then the sets of functions
\[
\begin{split}
B'_1 = \big\{\xi'_{1}, \xi_1 \in B_1 \big\}  \,\,\,\,\,\,\,  \textrm{where} \,\,\,
\xi'_1 (\boldsymbol{\p}_1) = \frac{\xi_1(\boldsymbol{\p}_1)}{|\boldsymbol{\p}_1|}, \\
B'_2 = \big\{\xi'_{2}, \xi_2 \in B_2 \big\}  \,\,\,\,\,\,\,  \textrm{where} \,\,\,
\xi'_2 (\boldsymbol{\p}_2) = \frac{\xi_2(\boldsymbol{\p}_2)}{|\boldsymbol{\p}_2|},
\end{split}
\]
are bounded in $E_2 = \mathcal{S}(\mathbb{R}^3; \mathbb{C}^4)$, and the set $B'_{1} \otimes B'_{2}$
is bounded in $E_2 \otimes E_2$. 
This in particular means that each of the norms (values  of the indeces $\mu, \nu \in \{0,1,2,3\}$
are fixed and $\zeta^{(q)}$ denotes derivative of $q$-th order $q \in \mathbb{N}_{0}^{6}$ of a function 
$\zeta$ on $\mathbb{R}^{6}$)
\[
\rceil \rceil \xi_{1}^{\mu} \otimes \xi_{2}^{\lambda} \lceil \lceil_{{}_{m}} \overset{\textrm{df}}{=}
\underset{|q| \leq m}{\textrm{sup}}(1 + |\boldsymbol{\p}_1 \times \boldsymbol{2}|^2)^m 
\Big| \big(\xi_{1}^{\mu} \otimes \xi_{2}^{\lambda} \big)^{(q)}\Big|
\]
is separately bounded on $B'_{1} \otimes B'_{2}$, \emph{i. e.} for each
$m= 0,1,2, \ldots$ there exists a finite constant $C'_{{}_{m}}$ such that
\[
\rceil \rceil \xi_{1}^{\mu} \otimes \xi_{2}^{\lambda} \lceil \lceil_{{}_{m}}
\leq C'_{{}_{m}}, \,\,\, \xi_1 \in B'_{1}, \xi_2 \in B'_{2},
\]
and moreover for each $m = 0,1,2, \ldots$ there exists $m'(m) \in \mathbb{N}_{0}$ and $C(m) < \infty $ such that
\begin{equation}\label{||.||<C|.|.|.|}
\Bigg\rceil \Bigg\rceil
\frac{(1+|\boldsymbol{\p}_1 \times \boldsymbol{\p}_2|^2)^4}{|\boldsymbol{\p}_1| \, |\boldsymbol{\p}_2|}
\xi_1 \otimes \xi_2
\Bigg\lceil \Bigg\lceil_{{}_{m}} \leq \,\, C(m) \,
\rceil \xi_1 \lceil_{{}_{m'}}  \rceil \xi_2 \lceil_{{}_{m'}}
\end{equation}
where $\{ \rceil \cdot \lceil_{{}_{m}} \}_{m \in \mathbb{N}_{0}}$ is one of the equivalent systems of norms defining 
$\mathcal{S}^{0}(\mathbb{R}^3; \mathbb{C})$ and given in Subsection 5.5 of \cite{wawrzycki2018}.

Now using the part 1) of the Lemma and the inequality (\ref{||.||<C|.|.|.|}) we obtain the following inequalities 
(with fixed values of the indices $\mu$ and $\lambda$ in each factor 
$\kappa_{1,0}(\xi_1)$ and $\kappa_{1,0}(\xi_1)$)
\begin{multline}\label{|<kappa(xi).kappa(xi),phi>|}
|\langle \kappa_{1,0}(\xi_1) \cdot \kappa_{1,0}(\xi_2), \phi \rangle|
= \Bigg|\int \limits_{\mathbb{R}^3 \times \mathbb{R}^3} 
\frac{\ud^3 \boldsymbol{\p}_1 \, \ud^3 \boldsymbol{\p}_2}{|\boldsymbol{\p}_1| |\boldsymbol{\p}_2|}
\xi_{1}^{\mu}(\boldsymbol{\p}_1) \xi_{2}^{\lambda}(\boldsymbol{\p}_2) \, 
\widetilde{\phi}(\boldsymbol{\p}_1 + \boldsymbol{\p}_2, |\boldsymbol{\p}_1| + |\boldsymbol{\p}_2|) \Bigg| \\
\leq 
\int \limits_{\mathbb{R}^3 \times \mathbb{R}^3} 
\frac{\ud^3 \boldsymbol{\p}_1 \, \ud^3 \boldsymbol{\p}_2}{|\boldsymbol{\p}_1| |\boldsymbol{\p}_2|}
|\xi_{1}^{\mu}(\boldsymbol{\p}_1) \xi_{2}^{\lambda}(\boldsymbol{\p}_2)| \, 
|\widetilde{\phi}(\boldsymbol{\p}_1 + \boldsymbol{\p}_2, |\boldsymbol{\p}_1| + |\boldsymbol{\p}_2|)| \\
\leq 
C_B \, \int \limits_{\mathbb{R}^3 \times \mathbb{R}^3} 
\frac{\ud^3 \boldsymbol{\p}_1 \, \ud^3 \boldsymbol{\p}_2}{|\boldsymbol{\p}_1| |\boldsymbol{\p}_2|}
|\xi_{1}^{\mu}(\boldsymbol{\p}_1) \xi_{2}^{\lambda}(\boldsymbol{\p}_2)| \\
\leq 
C_B \, \int \limits_{\mathbb{R}^3 \times \mathbb{R}^3} 
\ud^3 \boldsymbol{\p}_1 \, \ud^3 \boldsymbol{\p}_2 \frac{1}{(1+|\boldsymbol{\p}_1 \times \boldsymbol{\p}_2|^2)^{4}}
\frac{(1+|\boldsymbol{\p}_1 \times \boldsymbol{\p}_2|^2)^{4}|\xi_{1}^{\mu}(\boldsymbol{\p}_1) \xi_{2}^{\lambda}(\boldsymbol{\p}_2)|}{|\boldsymbol{\p}_1| |\boldsymbol{\p}_2|} \\
\leq 
C_B \, \Bigg| \frac{1}{(1+|\boldsymbol{\p}_1 \times \boldsymbol{\p}_2|^2)^{4}} \Bigg|_{{}_{L^2(\mathbb{R}^6)}} \,
\Bigg| \frac{(1+|\boldsymbol{\p}_1 \times \boldsymbol{\p}_2|^2)^{4}}{|\boldsymbol{\p}_1| |\boldsymbol{\p}_2|}
\xi_{1}^{\mu} \otimes \xi_{2}^{\lambda}
\Bigg|_{{}_{\infty}} \\
 \leq C' \Bigg\rceil \Bigg\rceil
\frac{(1+|\boldsymbol{\p}_1 \times \boldsymbol{\p}_2|^2)^4}{|\boldsymbol{\p}_1||\boldsymbol{\p}_2|}
\xi_{1}^{\mu} \otimes \xi_{2}^{\lambda}
\Bigg\lceil \Bigg\lceil_{{}_{4}} \\
\, \leq \,\, C' C(4) \,
\rceil \xi_{1}^{\mu} \lceil_{{}_{m'}}  \rceil \xi_{2}^{\lambda} \lceil_{{}_{m'}}
\end{multline}
for some finite $m' \in \mathbb{N}_0$. 

Therefore for any strong zero-neighborhood $V(B, \epsilon)$ in $\mathcal{S}(\mathbb{R}^4; \mathbb{C})^*$
determined by a bounded subset $B$ in $\mathcal{S}(\mathbb{R}^4; \mathbb{C})$ and $\epsilon >0$
there exist zero-neighboorhods $V_1$ and $V_2$ in $E_2 = \mathcal{S}^{0}(\mathbb{R}^3; \mathbb{C}^4)$
such that
\[
|\langle \kappa_{1,0}(\xi_1) \cdot \kappa_{1,0}(\xi_2), \phi \rangle| \leq \epsilon, 
\,\,\,\,
\xi_1 \in V_1, \xi_2 \in V_2, \phi \in B,
\]
or equivalently
\[
\kappa_{1,0}(\xi_1) \cdot \kappa_{1,0}(\xi_2) \in V(B, \epsilon), \,\,\,
\xi_1 \in V_1, \xi_2 \in V_2,
\]
if we define 
\[
V_1 = \Bigg\{\xi, \rceil \xi^{\mu} \lceil_{{}_{m'}} < \sqrt{\frac{\epsilon}{C' C(4)}}  \Bigg\}, \,\,\,
V_2 = \Bigg\{\xi, \rceil \xi^{\lambda} \lceil_{{}_{m'}} < \sqrt{\frac{\epsilon}{C' C(4)}} \Bigg\},
\]
which follows from the inequalities (\ref{|<kappa(xi).kappa(xi),phi>|}). 

The same proof holds if we replace one or both the kernels $\kappa_{1,0}$ 
by the kernel $\kappa_{0,1}$ defined by (\ref{kappa_0,1kappa_1,0A'}), 
Subsection \ref{equivalentA-s}, or by their derivatives because for any polynomial
$P(\boldsymbol{\p}_1, \boldsymbol{\p}_2)$ in eight real variables 
\[
(\boldsymbol{\p}_1, p_{10}(\boldsymbol{\p}_1), \boldsymbol{\p}_2, p_{20}(\boldsymbol{\p}_2))
= (\boldsymbol{\p}_1, |\boldsymbol{\p}_1|, \boldsymbol{\p}_2, |\boldsymbol{\p}_2|)
\]
and for each $m = 0,1,2, \ldots$ there exists $m'(m) \in \mathbb{N}_{0}$ and $C(m) < \infty $ such that
\begin{equation}\label{||P(p1p2).||<C|.|.|.|}
\Bigg\rceil \Bigg\rceil
\frac{(1+|\boldsymbol{\p}_1 \times \boldsymbol{\p}_2|^2)^4 P(\boldsymbol{\p}_1, \boldsymbol{\p}_2)}{|\boldsymbol{\p}_1|
\, |\boldsymbol{\p}_2|}
\xi_{1}^{\mu} \otimes \xi_{2}^{\lambda}
\Bigg\lceil \Bigg\lceil_{{}_{m}} \leq \,\, C(m) \,
\rceil \xi_1 \lceil_{{}_{m'}}  \rceil \xi_2 \lceil_{{}_{m'}}.
\end{equation}
Analogous proof can be repeated for all $\kappa_{1,0}, \kappa_{0,1}$ defined by 
(\ref{kappa_0,1kappa_1,0A}), Subsection \ref{A=Xi0,1+Xi1,0} (for plane wave kernels defining the free electromagnetic
potential field) and their derivatives; or for plane wave kernels
(\ref{kappa_0,1}) and (\ref{kappa_1,0}),
Subsect. \ref{psiBerezin-Hida} or (\ref{skappa_0,1}) and (\ref{skappa_1,0}) of
Subsection \ref{StandardDiracPsiField} (for kernels defining the Dirac field)
and their derivatives. We have to remember that if the kernel corresponds to the electromagnetic potential field
then the nuclear space on which it is defined is equal $E_2 = \mathcal{S}^{0}(\mathbb{R}^3; \mathbb{C}^4)$
and if the kernel corresponds to the Dirac field then it is defined on the nuclear space $E_1 = \mathcal{S}(\mathbb{R}^3; \mathbb{C}^4)$. In the last case we can use the standard system of norms defining the Schwartz topology
on $\mathcal{S}(\mathbb{R}^3; \mathbb{C})$. In particular if both factors\footnote{$(l_k,m_k) = (1,0)$ or $(l_k,m_k) = (0,1)$ for $k = 1, 2$.} 
$\kappa_{l_1,m_1}(\xi_1)$
and $\kappa_{l_2,m_2}(\xi_2)$ in  the pointwise product $\kappa_{l_1,m_1}(\xi_1) \cdot \kappa_{l_2,m_2}(\xi_2)$
corespond to kernels defining a fixed component of the Dirac field (or its fixed component derivative) then
we are using the inequality (\ref{||P(p1p2).||<C|.|.|.|}) with the the same system of norms 
$\{ \rceil \rceil \cdot \lceil \lceil_{{}_{m}} \}_{m \in \mathbb{N}_{0}}$ on the left hand side
but with the system of norms 
$\{ \rceil \cdot \lceil_{{}_{m}} \}_{m \in \mathbb{N}_{0}}$ replaced by the standard system of norms defining
the Schwartz topology on $\mathcal{S}(\mathbb{R}^3; \mathbb{C})$
and with 
\[
\frac{(1+|\boldsymbol{\p}_1 \times \boldsymbol{\p}_2|^2)^4 P(\boldsymbol{\p}_1, \boldsymbol{\p}_2)}{|\boldsymbol{\p}_1|
\, |\boldsymbol{\p}_2|}
\]
in (\ref{||P(p1p2).||<C|.|.|.|}) replaced by
\[
\frac{(1+|\boldsymbol{\p}_1 \times \boldsymbol{\p}_2|^2)^4 P(\boldsymbol{\p}_1, \boldsymbol{\p}_2)}
{\sqrt{|\boldsymbol{\p}_1|^2 +m} \, \sqrt{|\boldsymbol{\p}_2|^2+m}} \,\,\,
\textrm{or} \,\,\,
(1+|\boldsymbol{\p}_1 \times \boldsymbol{\p}_2|^2)^4 P(\boldsymbol{\p}_1, \boldsymbol{\p}_2)
\]
with
$P(\boldsymbol{\p}_1, \boldsymbol{\p}_2)$ equal to any polynomial 
in eight real variables 
\[
(\boldsymbol{\p}_1, p_{10}(\boldsymbol{\p}_1), \boldsymbol{\p}_2, p_{20}(\boldsymbol{\p}_2))
= (\boldsymbol{\p}_1, \sqrt{|\boldsymbol{\p}_1|^2 +m}, \boldsymbol{\p}_2, \sqrt{|\boldsymbol{\p}_2|^2+m}).
\]
If the first factor $\kappa_{l_1,m_1}(\xi_1)$ corresponds to a fixed component of the Dirac field 
(or its fixed component derivative)
and the second factor  $\kappa_{l_2,m_2}(\xi_2)$ to a fixed component of the electromagnetic potential field
(or its fixed component derivative) then we are using the inequality 
(\ref{||P(p1p2).||<C|.|.|.|}), with the the same system of norms 
$\{ \rceil \rceil \cdot \lceil \lceil_{{}_{m}} \}_{m \in \mathbb{N}_{0}}$ on the left hand side,
the same system of norms 
$\{ \rceil \xi_2 \lceil_{{}_{m}} \}_{m \in \mathbb{N}_{0}}$ defining the nuclear topology 
$S^{0}(\mathbb{R}^3; \mathbb{C})$ (inherited from $\mathcal{S}(\mathbb{R}^3; \mathbb{C})$, compare Subsections
5,2-5.5 of \cite{wawrzycki2018}), but with the system of norms $\{ \rceil \xi_1 \lceil_{{}_{m}} \}_{m \in \mathbb{N}_{0}}$
replaced by any standard which defines the Schwartz topology on $\mathcal{S}(\mathbb{R}^3; \mathbb{C})$,
and with 
\[
\frac{(1+|\boldsymbol{\p}_1 \times \boldsymbol{\p}_2|^2)^4 P(\boldsymbol{\p}_1, \boldsymbol{\p}_2)}{|\boldsymbol{\p}_1|
\, |\boldsymbol{\p}_2|}
\]
in (\ref{||P(p1p2).||<C|.|.|.|}) replaced by
\[
\frac{(1+|\boldsymbol{\p}_1 \times \boldsymbol{\p}_2|^2)^4 P(\boldsymbol{\p}_1, \boldsymbol{\p}_2)}
{\sqrt{|\boldsymbol{\p}_1|^2 +m} \, |\boldsymbol{\p}_2|} \,\,\,
\textrm{or} \,\,\,
\frac{(1+|\boldsymbol{\p}_1 \times \boldsymbol{\p}_2|^2)^4 P(\boldsymbol{\p}_1, \boldsymbol{\p}_2)}
{|\boldsymbol{\p}_2|}
\]
with $P(\boldsymbol{\p}_1, \boldsymbol{\p}_2)$ equal to any polynomial 
in eight real variables 
\[
(\boldsymbol{\p}_1, p_{10}(\boldsymbol{\p}_1), \boldsymbol{\p}_2, p_{20}(\boldsymbol{\p}_2))
= (\boldsymbol{\p}_1, \sqrt{|\boldsymbol{\p}_1|^2 +m}, \boldsymbol{\p}_2, |\boldsymbol{\p}_2|).
\]
\qed

\begin{lem}\label{kappaBarDotOtimeskappa}
Let
\[
{}^{{}^{n_k}}_{{}_{k}}\kappa_{l_k,m_k} \in \mathfrak{K}_0, \,\,\, k = 1, \ldots, M.
\]
i.e. we have the kernels belonging to the class\footnote{Recall that each element of $\mathfrak{K}_0$ is equal to a component of a plane wave kernel defining free field of the theory or to its spatio-temporal derivative
$\partial^{\alpha}$ with fixed $\alpha$, compare Definition \ref{K_0}.} 
$\mathfrak{K}_0$. 
\begin{enumerate}
\item[1)]
Then it follows in particular that
\[
{}^{{}^{n_k}}_{{}_{k}}\kappa_{l_k,m_k}
\in \mathscr{L}(E_{{}_{n_k}}, \mathscr{E}^{*}_{{}_{n_k}}) \cong 
E^{*}_{{}_{n_k}} \otimes \mathscr{E}^{*}_{{}_{n_k}}, \,\,\, k=1, \ldots, M,
\]
are regular vector-valued distributions defined by ordinary functions, which 
fulfil the condition (\ref{ExtedibilityCondition}), \emph{i.e.} are extendible to elements 
\[
{}^{{}^{n_k}}_{{}_{k}}\kappa_{l_k,m_k}
\in \mathscr{L}(E^{*}_{{}_{n_k}}, \mathscr{E}^{*}_{{}_{n_k}}) \cong 
E_{{}_{n_k}} \otimes \mathscr{E}^{*}_{{}_{n_k}}, \,\,\, k=1, \ldots, M,
\]
and have the property (\ref{ConvolutabilityCondition}) that
\[
{}^{{}^{n_k}}_{{}_{k}}\kappa_{l_k,m_k}(\xi) \in \mathcal{O}_C(\mathbb{R}^4; \mathbb{C}), 
\,\,\, \xi \in E_{{}_{n_k}}.
\]
\item[2)]
The  ``point-wise'' multiplicative tensor product $\dot{\otimes}$ of these distributions,
defined as in Rule II, gives a vector valued kernel 
\[
\kappa_{l,m} = \Big({}^{{}^{n_1}}_{{}_{1}}\kappa_{l_1,m_1} \Big) \overline{\dot{\otimes}} \cdots \overline{\dot{\otimes}} \,\, \Big( {}^{{}^{n_M}}_{{}_{M}}\kappa_{l_M,m_M} \Big) \\
\in \mathcal{O}_{C}(\mathbb{R}^4; \mathbb{C}).
\]
\item[3)]
The  ``point-wise'' multiplicative tensor product $\dot{\otimes}$ of these distributions,
defined as in Rule II, gives a vector valued kernel 
\begin{multline*}
\kappa_{l,m} = \Big({}^{{}^{n_1}}_{{}_{1}}\kappa_{l_1,m_1} \Big) \overline{\dot{\otimes}} 
\cdots \overline{\dot{\otimes}} \,\, \Big( {}^{{}^{n_M}}_{{}_{M}}\kappa_{l_M,m_M} \Big) \\
\in
E^{*}_{{}_{n_1}} \otimes \cdots \otimes 
E^{*}_{{}_{n_M}} \otimes \mathscr{E}^{*}_{{}_{i}} 
\cong \mathscr{L}(E_{{}_{n_1}} \otimes \cdots \otimes  E_{{}_{n_M}}, \,\,
\mathscr{E}^{*}_{{}_{i}}), \,\,\, i=1,2.
\end{multline*}
\item[4)] 
If all $n_1, \ldots, n_M$ are equal $1$, \emph{i. e.} if all factors
\[
\Xi_{l_k, m_k}\Big({}^{{}^{n_k}}_{{}_{k}}\kappa_{l_1,m_1}\Big) \,\,\, \textrm{with} \,\,\, n_k = 1
\]
correspond to (derivatives) of the free massive fields of the theory (\emph{i. e.} 
derivatives of the Dirac free field in case of spinor QED), then the bilinear map
\begin{multline*}
\xi \times \eta \mapsto \kappa_{l,m}(\xi \otimes \eta), 
\\
\xi \in \overbrace{E_{i_1} \otimes \cdots 
\otimes E_{i_l}}^{\textrm{first $l$ terms $E_{i_j}$, $i_j \in \{1,2\}$}}, \\
\eta \in \overbrace{E_{i_{l+1}} \otimes \cdots 
\otimes E_{i_{l+m}}}^{\textrm{last $m$ terms $E_{i_j}$, $i_j\in \{1,2\}$}},
\end{multline*}
can be extended to a separately continuous bilinear map from
\[
\Big( \overbrace{E_{i_1} \otimes \cdots 
\otimes E_{i_l}}^{\textrm{first $l$ terms $E_{i_j}$}} \Big)^*
\times
\Big( \overbrace{E_{i_{l+1}} \otimes \cdots 
\otimes E_{i_{l+m}}}^{\textrm{last $m$ terms $E_{i_j}$}} \Big)
\,\,\, \textrm{into} \,\,\,\mathscr{L}(\mathscr{E}, \mathbb{C}) = \mathscr{E}^*.
\]
\end{enumerate}
\end{lem}

\qedsymbol \, 
The first two parts 1) and 2) can be proved exactly as Lemma \ref{kappa0,1,kappa1,0psi}, Subsection \ref{psiBerezin-Hida} or respectively Lemma \ref{kappa0,1,kappa1,0ForA},
Subsection \ref{A=Xi0,1+Xi1,0}.

Concerning 3) it is sufficient to consider the case $M=2$. But the case $M=2$ follows immediately
from the part 2) of Lemma \ref{Cont.Ofkappa.kappa}.

Concerning 4) it is sufficient to consider the case $M=2$. 
Let us consider first the case in which the first factor has $(l_1,m_1) =(1,0)$ and the second $(l_2, m_2) = (0,1)$.
That the map
\[
\xi_1 \times \xi_2 \longmapsto 
{}^{{}^{1}}_{{}_{1}}\kappa_{1,0} \overset{\cdot}{\otimes} {}^{{}^{1}}_{{}_{2}}\kappa_{0,1} (\xi_1 \otimes \xi_2)
=
{}^{{}^{1}}_{{}_{1}}\kappa_{1,0}(\xi_1) \cdot {}^{{}^{1}}_{{}_{2}}\kappa_{0,1}(\xi_2) 
\]
can be extedned to a map which is separately continous as a map
\[
E_{1}^{*} \times E_{1} \mapsto \mathscr{E}_{k}^{*}, \,\,\, k= 1,2
\]
follows immediately from the extendibility property (\ref{ExtedibilityCondition})
asserted in the first part of our Lemma and from the property (\ref{ConvolutabilityCondition}) 
which assures that
\[
{}^{{}^{n_k}}_{{}_{k}}\kappa_{l_k,m_k}(\xi) \in \mathcal{O}_C(\mathbb{R}^4; \mathbb{C}), 
\,\,\, \xi \in E_{{}_{n_k}}.
\]
and in particular assures that
\[
{}^{{}^{n_k}}_{{}_{k}}\kappa_{l_k,m_k}(\xi), \,\,\, \xi \in E_{{}_{n_k}}
\]
is contained within the algebra of multipliers of $\mathscr{E}_{k}$, $k=1,2$ and 
of $\mathscr{E}_{k}^{*}$. This is because $\mathcal{O}_C(\mathbb{R}^4; \mathbb{C})$ is contained
in both the algebras of multipliers $\mathcal{O}_{MB_1} = \mathcal{O}_{M}, \mathcal{O}_{MB_2}$, 
respectively, of $\mathscr{E}_1, \mathscr{E}_2$,
compare Subsections 5.4, 5.5 of \cite{wawrzycki2018} and Appendix \ref{convolutorsO'_C}.  
In particular the operator of pointwise multiplication by a fixed 
\[
{}^{{}^{n_k}}_{{}_{k}}\kappa_{l_k,m_k}(\xi), \,\,\, \xi \in E_{{}_{n_k}}
\]
transforms continously $\mathscr{E}_{k}$, $k=1,2$ and 
$\mathscr{E}_{k}^{*}$, $k=1,2$ into themselves.

Let us consider now the case $M=2$ in which both factors have $(l_1,m_1) = (l_2,m_2) = (1,0)$:
\begin{equation}\label{kappaDirac.kappaDirac}
\xi_1 \times \xi_2 \longmapsto 
{}^{{}^{1}}_{{}_{1}}\kappa_{1,0} \overset{\cdot}{\otimes} {}^{{}^{1}}_{{}_{2}}\kappa_{1,0} (\xi_1 \otimes \xi_2)
=
{}^{{}^{1}}_{{}_{1}}\kappa_{1,0}(\xi_1) \cdot {}^{{}^{1}}_{{}_{2}}\kappa_{1,0}(\xi_2) 
\end{equation}
and the plane wave kernels 
\[
{}^{{}^{1}}_{{}_{1}}\kappa_{1,0}, {}^{{}^{1}}_{{}_{2}}\kappa_{1,0}
\]
correspond to some fixed components of the Dirac field or its fixed component derivative.
In this case the above map (\ref{kappaDirac.kappaDirac}) coincides with a particular case of the map
of Lemma \ref{Hypocont.Ofkappa.kappa}. 
From Lemma \ref{Hypocont.Ofkappa.kappa}
and the Proposition of Chap III \S 5.4, p. 90 of \cite{Schaefer}, it follows
that the $\big(\mathfrak{S}_{n_1}, \mathfrak{S}_{n_2}\big)$-hypocontinuous map
\[
E_{n_1} \times E_{n_2} \longrightarrow \mathscr{E}_{k}^{*}, \,\,\, k =1,2
\]
of Lemma \ref{Hypocont.Ofkappa.kappa}, can be uniquely extended to 
$\big(\mathfrak{S}_{n_1}^{*}, \mathfrak{S}_{n_2}^{*}\big)$-hypocontinuous map
\[
E_{n_1}^{*} \times E_{n_2}^{*} \longrightarrow \mathscr{E}_{k}^{*}, \,\,\, k =1,2
\]
with respect to the strong dual topology on each indicated space, where 
$\mathfrak{S}_{n_k}^{*}$, $k=1,2$, is the family of all bounded sets on 
strong dual space $E_{n_k}^{*}$, which simply means that the map of Lemma
\ref{Hypocont.Ofkappa.kappa} can be uniquely extended to a hypocontinuous map
\[
E_{n_1}^{*} \times E_{n_2}^{*} \longrightarrow \mathscr{E}_{k}^{*}, \,\,\, k =1,2
\]
or in particular to separately continuous map
\[
E_{n_1}^{*} \times E_{n_2}^{*} \longrightarrow \mathscr{E}_{k}^{*}, \,\,\, k =1,2
\]
with respect to the strong dual topology. Because $E_{n_k}^{*}$, $\mathscr{E}_{k}^{*}$, $k=1,2$
are all equal to strong dual spaces of reflexive Fr\'echet spaces 
$E_{n_k}$, $\mathscr{E}_{k}$, then by Thm. 41.1 the map of Lemma \ref{Hypocont.Ofkappa.kappa}
can be uniquely extended to (jointly) continuous map
\[
E_{n_1}^{*} \times E_{n_2}^{*} \longrightarrow \mathscr{E}_{k}^{*}, \,\,\, k =1,2
\]
with respect to the strong dual topology.
\qed

Before continuing we give a commentary concerning the proof of 4), case $M=2$ of the last Lemma.
Namey in this proof we can proceed as in the proof of the second part of
Lemma \ref{kappa0,1,kappa1,0psi}, Subsection \ref{psiBerezin-Hida} or respectively of 
Lemma \ref{kappa0,1,kappa1,0ForA},
Subsection \ref{A=Xi0,1+Xi1,0}. Namely 
\[
{}^{{}^{1}}_{{}_{1}}\kappa_{1,0} \overset{\cdot}{\otimes} {}^{{}^{1}}_{{}_{2}}\kappa_{0,1} 
\]
we can treat as an element of 
\[
\mathscr{L}(\mathscr{E}_{{}_{i}}, \, E_{{}_{n_1}}^{*} 
\otimes  E_{{}_{n_2}}^{*}) \cong
\mathscr{L}(E_{{}_{n_1}} \otimes  E_{{}_{n_2}}, \,\,
\mathscr{E}^{*}_{{}_{i}}). 
\]
Assertion 4), case $M=2$, will be proved if we show that 
\[
{}^{{}^{1}}_{{}_{1}}\kappa_{1,0} \overset{\cdot}{\otimes} {}^{{}^{1}}_{{}_{2}}\kappa_{0,1}
\in
\mathscr{L}(\mathscr{E}_{{}_{i}}, \, E_{{}_{n_1}}^{*} 
\otimes   E_{{}_{n_2}}^{*}) 
\]
actually belongs to
\[
\mathscr{L}(\mathscr{E}_{{}_{i}}, \, E_{{}_{n_1}} 
\otimes  E_{{}_{n_2}}). 
\]
Similarily 
\[
{}^{{}^{2}}_{{}_{1}}\kappa_{1,0} \overset{\cdot}{\otimes} {}^{{}^{2}}_{{}_{2}}\kappa_{0,1}
\in
\mathscr{L}(\mathscr{E}_{{}_{i}}, \, E_{{}_{n_1}}^{*} 
\otimes   E_{{}_{n_2}}^{*}) \cong
\mathscr{L}(E_{{}_{n_1}} \otimes  E_{{}_{n_2}}, \,\,
\mathscr{E}^{*}_{{}_{i}}). 
\]
wlould be extedible to an element of 
\[
\mathscr{L}(E_{{}_{n_1}}^{*} \otimes  E_{{}_{n_2}}^{*}, \,\,
\mathscr{E}^{*}_{{}_{i}}) \cong
\mathscr{L}(\mathscr{E}_{{}_{i}}, \, E_{{}_{n_1}} 
\otimes  E_{{}_{n_2}}) 
\]
if 
\[
{}^{{}^{2}}_{{}_{1}}\kappa_{1,0} \overset{\cdot}{\otimes} {}^{{}^{2}}_{{}_{2}}\kappa_{0,1}
\in \mathscr{L}(\mathscr{E}_{{}_{i}}, \, E_{{}_{n_1}}^{*} 
\otimes E_{{}_{n_2}}^{*})
\]
actually belongs to 
\[
\mathscr{L}(\mathscr{E}_{{}_{i}}, \, E_{{}_{n_1}} 
\otimes E_{{}_{n_2}}).
\]
This however is imposible because if both kernels ${}^{{}^{2}}_{{}_{1}}\kappa_{1,0}, 
{}^{{}^{2}}_{{}_{2}}\kappa_{0,1}$ are acssociated to a fixed component of the free zero mass electromagnetic 
potential field (or its derivative ), then easy computation shows that 
${}^{{}^{2}}_{{}_{1}}\kappa_{1,0} \overset{\cdot}{\otimes} {}^{{}^{2}}_{{}_{2}}\kappa_{0,1}(\phi)$,
$\phi \in \mathscr{E}_2$, has the following general form
\[
{}^{{}^{2}}_{{}_{1}}\kappa_{1,0} \overset{\cdot}{\otimes} {}^{{}^{2}}_{{}_{2}}\kappa_{0,1}(\phi)
(\boldsymbol{\p}_1, \boldsymbol{\p}_2)  =
M_{1}^{\nu_1}(\boldsymbol{\p}_1) M_{2}^{\mu_2}(\boldsymbol{\p}_2) \widetilde{\phi}(-\boldsymbol{\p}_1 + \boldsymbol{\p}_2,
p_{10}(-\boldsymbol{\p}_1) + p_{20}(\boldsymbol{\p}_2)),
\]
where $M_{i}^{\nu_i}$ is a multiplier of $E_{{}_{n_i}}$, $i=1,2$, and 
\[
p_{10}(\boldsymbol{\p}_1) = |\boldsymbol{\p}_1|,  \,\,\,\,\, p_{20}(\boldsymbol{\p}_2) = |\boldsymbol{\p}_2|.
\]
We can now easily see that
\[
{}^{{}^{2}}_{{}_{1}}\kappa_{1,0} \overset{\cdot}{\otimes} {}^{{}^{2}}_{{}_{2}}\kappa_{0,1}(\phi)
\]
cannot even belong to $\mathscr{C}^{\infty}(\mathbb{R}^3 \times \mathbb{R}^3; \mathbb{C}^{8})$,
so all the more it cannot belong to 
$\mathcal{S}(\mathbb{R}^3; \mathbb{C}^4) \otimes \mathcal{S}(\mathbb{R}^3; \mathbb{C}^4) = E_1 \otimes E_1$
or to $\mathcal{S}^{0}(\mathbb{R}^3; \mathbb{C}^4) \otimes \mathcal{S}^{0}(\mathbb{R}^3; \mathbb{C}^4)
= E_2 \otimes E_2$ or to $E_1 \otimes E_2$ or finally to $E_2 \otimes E_1$.
In particular 
\begin{equation}\label{discon.2kappa.2kappa.map}
\phi \longmapsto 
{}^{{}^{2}}_{{}_{1}}\kappa_{1,0} \overset{\cdot}{\otimes} {}^{{}^{2}}_{{}_{2}}\kappa_{0,1}(\phi)
\end{equation}
cannot be continuous as a map 
\[
\mathscr{E}_i \longmapsto E_{{}_{n_1}} 
\otimes E_{{}_{n_2}}.
\]
From this it follows that 
\[
{}^{{}^{2}}_{{}_{1}}\kappa_{1,0} \overset{\cdot}{\otimes} {}^{{}^{2}}_{{}_{2}}\kappa_{0,1}
\]
cannot be extended to an element of 
\[
\mathscr{L}(E_{{}_{n_1}}^{*} \otimes  E_{{}_{n_2}}^{*}, \,\,
\mathscr{E}^{*}_{{}_{i}}).
\]

Of course from the last Lemma, part 3), it follows that the Wick product at the same point of 
any number of zero mass or massive fields is a well defined integral kernel operator belonging to 
\[
\mathscr{L}\big(\mathscr{E}, \, \mathscr{L}((\boldsymbol{E}), (\boldsymbol{E})^*)\big)
\cong \mathscr{L}\big((\boldsymbol{E}) \otimes \mathscr{E}, \, (\boldsymbol{E})^*\big)
\]
in the sense of Obata \cite{obataJFA} with vector-valued kernel.
We therefore have the following
\begin{prop*}
\begin{enumerate}
\item[1)]
For the Wick product at te same space-time point $x$
\begin{multline*}
\boldsymbol{}: \Xi_{l_1,m_1}\Big({}^{{}^{n_1}}_{{}_{1}}\kappa_{l_1,m_1}(x)\Big) \cdots 
\Xi_{l_M,m_M}\Big({}^{{}^{n_M}}_{{}_{M}}\kappa_{l_M,m_M}(x)\big) \boldsymbol{:} \\
= \Xi_{l,m}(\kappa_{lm}(x)), \,\,\, 
{}^{{}^{n_k}}_{{}_{k}}\kappa_{l_k,m_k} \in \mathfrak{K}_0 
\end{multline*}
of the integral kernel operators corresponding to the free fields of the theory
or their derivatives we have
\begin{multline*}
\kappa_{l,m} = \Big({}^{{}^{n_1}}_{{}_{1}}\kappa_{l_1,m_1} \Big) \overline{\dot{\otimes}} 
\cdots \overline{\dot{\otimes}} \,\, \Big( {}^{{}^{n_M}}_{{}_{M}}\kappa_{l_M,m_M} \Big) \\
\in
E^{*}_{{}_{n_1}} \otimes \cdots \otimes 
E^{*}_{{}_{n_M}} \otimes \mathscr{E}^{*}_{{}_{i}} 
\cong \mathscr{L}(E_{{}_{n_1}} \otimes \cdots \otimes  E_{{}_{n_M}}, \,\,
\mathscr{E}^{*}_{{}_{i}}), \,\,\, i=1,2.
\end{multline*}
Thus by (the generalization to tensor product of Fock spaces of) Thm. 3.9 of \cite{obataJFA}
\begin{multline*}
\boldsymbol{}: \Xi_{l_1,m_1}\Big({}^{{}^{n_1}}_{{}_{1}}\kappa_{l_1,m_1}\Big) \cdots 
\Xi_{l_M,m_M}\Big({}^{{}^{n_M}}_{{}_{M}}\kappa_{l_M,m_M}\Big) \boldsymbol{:} \\
= \Xi_{l,m}(\kappa_{lm}) \in 
\mathscr{L}\big((\boldsymbol{E}) \otimes \mathscr{E}_{{}_{i}}, \, (\boldsymbol{E})^* \big)
\cong \mathscr{L}\big(\mathscr{E}_{{}_{i}}, \, \mathscr{L}((\boldsymbol{E}), (\boldsymbol{E})^*) \big)
\end{multline*}
\item[2)]
If all $n_k = 1$, \emph{i.e.} among the factors 
\[
\Xi_{l_1,m_1}\Big({}^{{}^{n_k}}_{{}_{k}}\kappa_{l_k,m_k}(x)\Big)
\]
there are no integral kernel operators corresponding to mass less free fields
(electromagnetic potential field in case of QED) or their derivatives,  
then (by 4) of the preceding Lemma) the bilinear map
\begin{multline*}
\xi \times \eta \mapsto \kappa_{l,m}(\xi \otimes \eta), 
\\
\xi \in \overbrace{E_{i_1} \otimes \cdots 
\otimes E_{i_l}}^{\textrm{first $l$ terms $E_{i_j}$, $i_j \in \{1,2\}$}}, \\
\eta \in \overbrace{E_{i_{l+1}} \otimes \cdots 
\otimes E_{i_{l+m}}}^{\textrm{last $m$ terms $E_{i_j}$, $i_j\in \{1,2\}$}},
\end{multline*}
can be extended to a separately continuous bilinear map from
\[
\Big( \overbrace{E_{i_1} \otimes \cdots 
\otimes E_{i_l}}^{\textrm{first $l$ terms $E_{i_j}$}} \Big)^*
\times
\Big( \overbrace{E_{i_{l+1}} \otimes \cdots 
\otimes E_{i_{l+m}}}^{\textrm{last $m$ terms $E_{i_j}$}} \Big)
\,\,\, \textrm{into} \,\,\,\mathscr{L}(\mathscr{E}, \mathbb{C}) = \mathscr{E}^*.
\]
Thus by Thm. \ref{obataJFA.Thm.3.13}, Subsection \ref{psiBerezin-Hida}
\begin{multline*}
\boldsymbol{}: \Xi_{l_1,m_1}\Big({}^{{}^{n_1}}_{{}_{1}}\kappa_{l_1,m_1}\Big) \cdots 
\Xi_{l_M,m_M}\Big({}^{{}^{n_M}}_{{}_{M}}\kappa_{l_M,m_M}\Big) \boldsymbol{:} \\
= \Xi_{l,m}(\kappa_{lm}) \in 
\mathscr{L}\big((\boldsymbol{E}) \otimes \mathscr{E}_{{}_{i}}, \, (\boldsymbol{E}) \big)
\cong \mathscr{L}\big(\mathscr{E}_{{}_{i}}, \, \mathscr{L}((\boldsymbol{E}), (\boldsymbol{E})) \big)
\end{multline*}
\end{enumerate}
\end{prop*}

Now we pass to the operation of differentiation with respect to space-time coordinates. Suppose we have an
integral kernel operator $\Xi_{l,m}(\kappa_{l,m})$ with vector-valued 
kernel
\[
\kappa_{l,m} \in \mathscr{L} \big(\mathscr{E} , \,\, \big(E_{i_1} \otimes \cdots \otimes E_{i_{l+m}} \big)^* \, \big)
\cong \mathscr{L} \big(E_{i_1} \otimes \cdots \otimes E_{i_{l+m}}, \,\, \mathscr{E}^* \big)
\]
with the operator
\[
\Xi_{l,m}(\kappa_{l,m}) \in \mathscr{L}\big(\mathscr{E}, \, \mathscr{L}((\boldsymbol{E}), (\boldsymbol{E})^*)\big)
\cong \mathscr{L}\big((\boldsymbol{E}) \otimes \mathscr{E}, \, (\boldsymbol{E})^*\big)
\]
uniquely determined by
\begin{multline*}
\big\langle \big\langle \Xi_{l,m}(\kappa_{l,m})(\Phi \otimes \phi), \, \Psi  \big \rangle \big \rangle
=\big\langle \big\langle \Xi_{l,m}\big(\kappa_{l,m}(\phi)\big)\Phi, \, \Psi  \big \rangle \big \rangle \\
= \langle \kappa_{l,m}(\phi),  \eta_{\Phi, \Psi} \rangle
= \langle \kappa_{l,m}(\eta_{\Phi, \Psi}), \phi \rangle,
\,\,\,
\Phi, \Psi \in (\boldsymbol{E}), \phi \in \mathscr{E},
\end{multline*}
compare (\ref{VectValotimesXi=intKerOp'}) Subsection \ref{psiBerezin-Hida}. Suppose moreover
that 
\[
\begin{split}
\mathscr{E} = \mathscr{E}_{1} = \mathcal{S}_{H_{(4)}}(\mathbb{R}^4;\mathbb{C})=
\mathcal{S}(\mathbb{R}^4; \mathbb{C}) \,\,\, \textrm{or} \\
\mathscr{E} = \mathscr{E}_{2} = \mathcal{S}_{\mathscr{F}A^{(4)}\mathscr{F}^{-1}}(\mathbb{R}^4;\mathbb{C})=
\mathcal{S}^{00}(\mathbb{R}^4; \mathbb{C}).
\end{split}
\]
Let for $\kappa_{l,m}$ understood as an element of
\[
\mathscr{L} \big(E_{i_1} \otimes \cdots \otimes E_{i_{l+m}}, \,\, \mathscr{E}^* \big) \cong
\mathscr{L} \big(\mathscr{E} , \,\, \big(E_{i_1} \otimes \cdots \otimes E_{i_{l+m}} \big)^* \, \big)
\]
we have
\[
\kappa_{l,m}(\xi_1 \otimes \cdots \otimes \xi_{l+m}) \in \mathcal{O}_C(\mathbb{R}^4; \mathbb{C}), 
\,\,\,\
\xi_{k} \in E_{i_k},  i_k \in \{1, 2\}.
\]
We moreover include into consideration the special cases of integral kernel operators
\begin{equation}\label{FreeFieldXi-s}
\Xi_{0,1}({}^{1}\kappa_{0,1}), \Xi_{1,0}({}^{1}\kappa_{1,0}), 
\Xi_{0,1}({}^{2}\kappa_{0,1}), \Xi_{1,0}({}^{2}\kappa_{1,0}), 
\end{equation}
determined by the free fields of the theory with the integration in the general formula
(\ref{electron-positron-photon-Xi}) is restriced, respectively, only to fermi or only to bose
variables, and the Wick products of (\ref{FreeFieldXi-s}) at the same space-time point
(representing ordinary integral kernel operators (\ref{electron-positron-photon-Xi})
with vector-valued kernels and integration with integration in general ranging over both, bose and fermi, 
variables if the Wick product involves both, bose and fermi, field components). 

Then we can define the space-time derivative 
\[
\big(\frac{\partial}{\partial x^\mu} \Xi_{l,m}\big)(\kappa_{l,m})
\]
as the integral kernel operator uniquely determined by the condition
\begin{multline*}
\Big\langle \Big\langle \frac{\partial}{\partial x^\mu}\Xi_{l,m}\Big)(\kappa_{l,m})(\Phi \otimes \phi), 
\, \Psi  \Big \rangle \Big \rangle
=\Big\langle \Big\langle \Xi_{l,m}\Big(\Big(\frac{\partial}{\partial x^\mu}\kappa_{l,m}\big)(\phi)\Big)\Phi, \, \Psi  \Big \rangle \Big \rangle \\
= -\Big\langle \Big\langle \Xi_{l,m}\Big(\kappa_{l,m}\Big(\Big(\frac{\partial}{\partial x^\mu}\phi\big)\big)\Phi, \, \Psi  \Big \rangle \Big \rangle 
=  \Big\langle \Big(\frac{\partial}{\partial x^\mu}\kappa_{l,m}\Big)(\phi),  \eta_{\Phi, \Psi} \Big \rangle \\
= - \Big\langle \kappa_{l,m}\Big(\frac{\partial}{\partial x^\mu}\phi\Big),  \eta_{\Phi, \Psi} \Big \rangle
= - \langle \kappa_{l,m}(\eta_{\Phi, \Psi}), \frac{\partial}{\partial x^\mu}\phi \rangle,
\,\,\,
\Phi, \Psi \in (\boldsymbol{E}), \phi \in \mathscr{E}.
\end{multline*}

\begin{center}
{\small RULE III'}
\end{center}
\emph{We have the following computational rule}
\[
\Big(\frac{\partial}{\partial x^\mu} \Xi_{l,m}\Big)(\kappa_{l,m})
=
\Xi_{l,m}\Big(\frac{\partial}{\partial x^\mu}\kappa_{l,m}\Big)
\]
\emph{for $\kappa_{l,m}$ understood as an element of}
\[
\mathscr{L} \big(\mathscr{E} , \,\, \big(E_{i_1} \otimes \cdots \otimes E_{i_{l+m}} \big)^* \, \big)
\cong \mathscr{L} \big(E_{i_1} \otimes \cdots \otimes E_{i_{l+m}}, \,\, \mathscr{E}^* \big).
\]

Thus the operation of space-time differentiation performed on $\Xi(\kappa_{l,m})$
coresponds, via the Rule III', to the operation of differentiation performed upon the
vector-valued distributional kernel $\kappa_{l,m}$, undersdood as an
$\big(E_{i_1} \otimes \cdots \otimes E_{i_{l+m}} \big)^*$-valued distribution on the test function
space $\mathscr{E}$. Again  the Rule III' can be justified by utilizing the fact that
\begin{multline}\label{electron-positron-photon-Xi(x)}
\Xi_{l,m}\big(\kappa_{l,m}(x)\big) = 
\int \limits_{(\sqcup \mathbb{R}^3)^{(l+m)}}
\kappa_{l,m}(w_{i_1}, \ldots w_{i_l}, w_{i_{l+1}}, \ldots w_{i_{l+m}}; x) 
\, \\ \times
\partial_{w_{i_1}}^* \cdots \partial_{w_{i_l}}^* \partial_{w_{i_{l+1}}} \cdots \partial_{w_{i_{l+m}}}
\ud w_{i_1} \cdots \ud w_{i_l} \ud w_{i_{l+1}} \cdots \ud w_{i_{l+m}} = \\
= \int \limits_{(\sqcup \mathbb{R}^3)^{(l+m)}}
\kappa_{l,m}(w_{i_1}, \ldots w_{i_l}, u_{j_{1}}, \ldots u_{j_{m}}; x) \,\, \times \\ 
\times \,\, 
\partial_{w_{i_1}}^* \cdots \partial_{w_{i_l}}^* \partial_{u_{j_{1}}} \cdots \partial_{u_{j_{m}}}
\ud w_{i_1} \cdots \ud w_{i_l} \ud u_{j_{1}} \cdots \ud u_{j_{m}} 
\end{multline} 
exists pointwisely as a Pettis integral, just repeating the arguments
in constrution of space-time derivatives of the free electromagnetic potential field
during the proof of Bogoliubov-Shirkov Quantization Postulate, compare Subsection  5.9 of \cite{wawrzycki2018}.
Moreover during this proof we have given justification of the following Rules IV, V
and VI. 

For the integral kernel operator (\ref{electron-positron-photon-Xi(x)}) we have 

\begin{center}
{\small RULE IV'}
\end{center}
\[
\int \limits_{\mathbb{R}^4} \Xi_{l,m}\big(\kappa_{l,m}(x)\big) \, \ud^4 x
= \Xi_{l,m}\Bigg(\int \limits_{\mathbb{R}^4}\kappa_{l,m}(x) \ud^4 x \Bigg).
\]

\begin{center}
{\small RULE V'}
\end{center}
\[
\int \limits_{\mathbb{R}^4} \Xi_{l,m}\big(\kappa_{l,m}(\boldsymbol{\x}, x_0)\big) \, \ud^3 \boldsymbol{\x}
= \Xi_{l,m}\Bigg(\int \limits_{\mathbb{R}^4}\kappa_{l,m}(\boldsymbol{\x}, x_0) \, \ud^3 \boldsymbol{\x} \Bigg).
\]

Let $S \in \mathcal{S}(\mathbb{R}^4; \mathbb{C})^*$ then
\begin{center}
{\small RULE VI}
\end{center}
\begin{multline*}
S \ast \Xi_{l,m}(\kappa_{l,m})(x) = 
\int \limits_{\mathbb{R}^4} S(x-y) \Xi_{l,m}\big(\kappa_{l,m}(y)\big) \, \ud^4 y \\
= \Xi_{l,m}\Bigg(\int \limits_{\mathbb{R}^4} S(x-y)\kappa_{l,m}(y) \, \ud^4 y \,  \Bigg)
= \Xi_{l,m}\big( S \ast \kappa_{lm}(x) \big).
\end{multline*}
\emph{Here}
\begin{multline*}
S \ast \kappa_{lm}(\xi_1, \ldots, \xi_{l+m})(x)  \\
= \int \limits_{\mathbb{R}^4} S(x-y)\kappa_{l,m}(w_{i_1}, 
\ldots, w_{i_{l+m}}; y) \, \xi_{i_1}(w_{i_1}), 
\ldots, \xi_{i_{l+m}}(w_{i_{l+m}}) \ud^4 y,
\,\,\,\, \xi_{i_k} \in E_{i_k}
\end{multline*}
\emph{is well defined because}
\[
\kappa_{l,m}(\xi_1 \otimes \cdots \otimes \xi_{l+m}) \in \mathcal{O}_C(\mathbb{R}^{4}; \mathbb{C})
\subset \mathcal{O}'_C(\mathbb{R}^{4}; \mathbb{C}),
\]
\emph{and by definition is equal to the (kernel of the) distribution 
$S \ast (\kappa_{lm}(\xi_1, \ldots, \xi_{l+m}))$,
compare Appendix} \ref{convolutorsO'_C}.

The Rules III', IV', V', VI are also valid in case of more than just one space-time variable $x$. 
In order to see it we can repeat the proof replacing $\mathscr{E}$ (previously equal to $\mathscr{E}_1 = 
\mathcal{S}(\mathbb{R}^4; \mathbb{C})$ or $\mathscr{E}_2 = \mathcal{S}^{00}(\mathbb{R}^4; \mathbb{C})$)
by $\mathscr{E}$ equal to tensor product of several $\mathscr{E}_1$ or $\mathscr{E}_2$.
In this case we would obtain  more generally with
\[
\kappa_{l,m}(\xi_1 \otimes \cdots \otimes \xi_{l+m};x_1, \ldots, x_n) \in \mathcal{O}_C(\mathbb{R}^{4n}; \mathbb{C})
\]
the integral kernel operator
\begin{multline}\label{electron-positron-photon-Xi(x1...xn)}
\Xi_{l,m}\big(\kappa_{l,m}(x_1, \ldots, x_n)\big) = 
\int \limits_{(\sqcup \mathbb{R}^3)^{(l+m)}}
\kappa_{l,m}(w_{i_1}, \ldots w_{i_l}, w_{i_{l+1}}, \ldots w_{i_{l+m}}; x_1, \ldots, x_n) 
\, \times  \\ \times \,
\partial_{w_{i_1}}^* \cdots \partial_{w_{i_l}}^* \partial_{w_{i_{l+1}}} \cdots \partial_{w_{i_{l+m}}}
\ud w_{i_1} \cdots \ud w_{i_l} \ud w_{i_{l+1}} \cdots \ud w_{i_{l+m}} = \\
= \int \limits_{(\sqcup \mathbb{R}^3)^{(l+m)}}
\kappa_{l,m}(w_{i_1}, \ldots w_{i_l}, u_{j_{1}}, \ldots u_{j_{m}}; x_1, \ldots, x_n) \,\, \times \\
\times \, \partial_{w_{i_1}}^* \cdots \partial_{w_{i_l}}^* \partial_{u_{j_{1}}} \cdots \partial_{u_{j_{m}}}
\ud w_{i_1} \cdots \ud w_{i_l} \ud u_{j_{1}} \cdots \ud u_{j_{m}} 
\end{multline} 
existing pointwisely as a Pettis integral and with the following Rules:

\begin{center}
{\small RULE III}
\end{center}
\[
\Big(\frac{\partial^{n}}{\partial x_{1}^{\mu_1}\cdots \partial x_{n}^{\mu_n}} \Xi_{l,m}\Big)(\kappa_{l,m})
=
\Xi_{l,m}\Big(\frac{\partial^{n}}{\partial x_{1}^{\mu_1}\cdots \partial x_{n}^{\mu_n}}\kappa_{l,m}\Big)
\]
\emph{for $\kappa_{l,m}$ understood as an element of}
\[
\mathscr{L} \big(\mathscr{E} , \,\, \big(E_{i_1} \otimes \cdots \otimes E_{i_{l+m}} \big)^* \, \big)
\cong \mathscr{L} \big(E_{i_1} \otimes \cdots \otimes E_{i_{l+m}}, \,\, \mathscr{E}^* \big).
\]
\emph{with}
\[
\mathscr{E} = \mathscr{E}_{n_1} \otimes \cdots \otimes \mathscr{E}_{n_n}, \,\,\, n_k \in \{1,2\}.
\]

\begin{center}
{\small RULE IV}
\end{center}
\[
\int \limits_{\mathbb{R}^{4n}} \Xi_{l,m}\big(\kappa_{l,m}(x_1, \ldots, x_n)\big) \, \ud^4 x_1 \ldots \ud^4 x_n
= \Xi_{l,m}\Bigg(\int \limits_{\mathbb{R}^{4n}}\kappa_{l,m}(x_1, \ldots, x_n) \ud^4 x_1 \ldots \ud^4 x_n \Bigg).
\]

\begin{center}
{\small RULE V}
\end{center}
\begin{multline*}
\int \limits_{\mathbb{R}^{3n}} \Xi_{l,m}\big(\kappa_{l,m}(\boldsymbol{\x}_1, x_{10}, \ldots \boldsymbol{\x}_n, x_{n0})\big) \, \ud^3 \boldsymbol{\x}_1 \cdots \ud^3 \boldsymbol{\x}_n \\
= \Xi_{l,m}\Bigg(\int \limits_{\mathbb{R}^4}\kappa_{l,m}(\boldsymbol{\x}_1, x_{10}, \ldots \boldsymbol{\x}_n, x_{n0}) \, \ud^3 \boldsymbol{\x}_1 \ldots \ud^3 \boldsymbol{\x}_n \Bigg).
\end{multline*}

Now concerning the Rule VI for more space-time variables we can repeatedly combine the convolutions
of several distributions $S \in \mathcal{S}(\mathbb{R}^4; \mathbb{C})^*$ each in one space-time varible,
with the Wick product operation provided the correponding kernels $\kappa_{l,m}$ obtained in the intermediate steps 
are well defined elements of $\mathscr{L}(E_{i_1}\otimes \cdots \otimes, \,\, \mathscr{E}_{n_1}^{*} \otimes \cdots)$
with 
\[
\kappa_{l,m}(\xi_{i_1} \otimes \cdots )( x_{n_1}, \ldots) \in \mathcal{O}_C.
\]

Namely we have the following useful Lemma which allows us to operate with convolutions
of integral kernel operators with tempered distributions $S \in \mathcal{S}(\mathbb{R}^4; \mathbb{C})^*$:

\begin{lem}\label{S*Xi} 
Let $S \in \mathcal{S}(\mathbb{R}^4; \mathbb{C})^*$, and let 
\[
\kappa_{l,m} \in \mathscr{L} \big(\mathscr{E} , \,\, \big(E_{i_1} \otimes \cdots \otimes E_{i_{l+m}} \big)^* \, \big)
\cong \mathscr{L} \big(E_{i_1} \otimes \cdots \otimes E_{i_{l+m}}, \,\, \mathscr{E}^* \big)
\]
with
\[
\kappa_{l,m}(\xi_1 \otimes \cdots \otimes \xi_{l+m}) \in \mathcal{O}_C(\mathbb{R}^{4}; \mathbb{C}),
\,\,\,\,
\xi_k \in E_{i_k}, \,\,\, i_k \in \{1, 2\}.
\]
In particular this is the case (compare 1), 2), and 3) of Lemma \ref{kappaBarDotOtimeskappa}) for the kernel
\[
\kappa_{l,m} = \Big({}^{{}^{n_1}}_{{}_{1}}\kappa_{l_1,m_1} \Big) \overline{\dot{\otimes}} \cdots \overline{\dot{\otimes}} \,\, \Big( {}^{{}^{n_M}}_{{}_{M}}\kappa_{l_M,m_M} \Big)
\]
corresponding to the Wick product (at the same space-time point $x$) 
\[
\Xi_{l,m}(\kappa_{lm}(x)) =
\boldsymbol{}: \Xi_{l_1,m_1}\Big({}^{{}^{n_1}}_{{}_{1}}\kappa_{l_1,m_1}(x)\Big) \cdots 
\Xi_{l_M,m_M}\Big({}^{{}^{n_1}}_{{}_{M}}\kappa_{l_M,m_M}(x)\big) \boldsymbol{:} 
\]
of the integral kernel operators 
\[
\Xi_{l_k,m_k}\Big({}^{{}^{n_k}}_{{}_{k}}\kappa_{l_k,m_k}(x)\Big), \,\,\,\,
{}^{{}^{n_k}}_{{}_{k}}\kappa_{l_k,m_k} \in \mathfrak{K}_0.
\]
Let the integral kernel $S \ast \kappa_{l,m}$  be equal
\begin{multline*}
\langle S \ast \kappa_{l,m}(\xi_{i_1} \otimes \cdots \otimes \xi_{i_{l+m}}), \phi \rangle
=  \int \limits_{\mathbb{R}^4} S \ast \kappa_{lm}(\xi_1, \ldots, \xi_{l+m})(x) \, \phi(x) \, \ud^4 x \\
\int \limits_{\mathbb{R}^4 \times \mathbb{R}^4} S(x-y)\kappa_{l,m}(w_{i_1}, 
\ldots, w_{i_{l+m}}; y) \, \xi_{i_1}(w_{i_1}), 
\ldots, \xi_{i_{l+m}}(w_{i_{l+m}}) 
\ud w_{i_1} \cdots \ud w_{i_{l+m}}
\ud^4 y \ud^4 x, \\
\,\,\,\, \xi_{i_k} \in E_{i_k}, \, \phi \in \mathscr{E} = \mathcal{S}(\mathbb{R}^4; \mathbb{}C) 
\,\, \textrm{or} \,\, \mathscr{E} = \mathcal{S}^{00}(\mathbb{R}^4; \mathbb{C}). 
\end{multline*}
Then
\begin{enumerate}
\item[1)]
 the kernel
\[
S \ast \kappa_{l,m} \in \mathscr{L} \big(E_{i_1} \otimes \cdots \otimes E_{i_{l+m}}, \,\, \mathscr{E}^* \big);
\]
\item[2)]
and if 
\[
\kappa_{l,m} = \Big({}^{{}^{n_1}}_{{}_{1}}\kappa_{l_1,m_1} \Big) \overline{\dot{\otimes}} \cdots \overline{\dot{\otimes}} \,\, \Big( {}^{{}^{n_M}}_{{}_{M}}\kappa_{l_M,m_M} \Big), 
\,\,\,\,\,
{}^{{}^{n_k}}_{{}_{k}}\kappa_{l_k,m_k} \in \mathfrak{K}_0
\]
then
\[
S \ast \kappa_{lm}(\xi_1, \ldots, \xi_{l+m}) \in \mathcal{O}_C(\mathbb{R}^4; \mathbb{C}) 
\subset \mathcal{O}'_{C}(\mathbb{R}^4; \mathbb{C}).
\]
\end{enumerate}
\end{lem}

\qedsymbol \,
It is sufficient to consider the case $\mathscr{E} = \mathscr{E}_{1} = \mathcal{S}(\mathbb{R}^4; \mathbb{C})$,
because $\mathscr{E}_{1}^{*}$ is continously embedded into 
$\mathscr{E}_{2}^{*} = \mathcal{S}^{00}(\mathbb{R}^4; \mathbb{C})^{*}$, 
compare Subsection 5.5 of \cite{wawrzycki2018}.

Because the Schwartz' algebra $\mathcal{O}'_{C}(\mathbb{R}^4; \mathbb{C})$ of convolutors
of $\mathcal{S}(\mathbb{R}^4; \mathbb{C})^*$
(for definition of $\mathcal{O}'_{C}$ compare e.g. \cite{Schwartz} or Appendix \ref{convolutorsO'_C})
is dense in $\mathcal{S}(\mathbb{R}^4; \mathbb{C})^*$ in the strong dual topology, then for $\epsilon >0$
we can find $S_\epsilon \in \mathcal{O}'_{C}$ such that 
\[
\underset{\epsilon \rightarrow 0}{\textrm{lim}}S_\epsilon = S
\]
in the strong topology of the dual space $\mathcal{S}(\mathbb{R}^4; \mathbb{C})^*$ of tempered distributions.
Let $\xi$ be any element of
\[
E_{i_1} \otimes \cdots \otimes E_{i_{l+m}}.
\]
For $\epsilon >0$ we define the following linear operator $\Lambda_\epsilon$
\[
\Lambda_\epsilon(\xi) \overset{\textrm{df}}{=}
S_\epsilon \ast \kappa_{l,m}(\xi), \,\,\,
\xi \in E_{i_1} \otimes \cdots \otimes E_{i_{l+m}},
\]
on 
\[
E_{i_1} \otimes \cdots \otimes E_{i_{l+m}}.
\]
Because $S_\epsilon \in \mathcal{O}'_{C}$, $\epsilon>0$, and because 
\[
\kappa_{l,m} \in \mathscr{L} \big(E_{i_1} \otimes \cdots \otimes E_{i_{l+m}}, \,\, \mathscr{E}^* \big),
\]
then for each $\epsilon >0$ the operator 
\[
\Lambda_\epsilon: E_{i_1} \otimes \cdots \otimes E_{i_{l+m}} \longrightarrow
\mathscr{E}^*
\]
is continuous, i.e. 
\[
\Lambda_\epsilon \in \mathscr{L} \big(E_{i_1} \otimes \cdots \otimes E_{i_{l+m}}, \,\, \mathscr{E}^* \big).
\]

For each $\xi \in E_{i_1} \otimes \cdots \otimes E_{i_{l+m}}$
\[
\kappa_{l,m}(\xi) \in \mathcal{O}_C \subset \mathcal{O}'_C{}
\]
and 
\[
\underset{\epsilon \rightarrow 0}{\textrm{lim}} S_\epsilon = S \,\,\,
\textrm{in strong dual topology of} \,\, \mathcal{S}(\mathbb{R}^4)^* = \mathscr{E}^*
\]
so for each $\xi \in E_{i_1} \otimes \cdots \otimes E_{i_{l+m}}$
\[
\underset{\epsilon \rightarrow 0}{\textrm{lim}} \Lambda_\epsilon(\xi) = 
\underset{\epsilon \rightarrow 0}{\textrm{lim}} S_\epsilon \ast \kappa_{l,m}(\xi) \,\,\,
\]
in strong dual topology of $\mathscr{E}^*$
exists and is equal 
\[
\underset{\epsilon \rightarrow 0}{\textrm{lim}} \Lambda_\epsilon(\xi) = S \ast \kappa_{l,m}(\xi)
\]
(compare Appendix \ref{convolutorsO'_C} and references cited there).

Because  $E_{i_1} \otimes \cdots \otimes E_{i_{l+m}}$ is a complete Fr\'echet space then by the Banach-Steinhaus
theorem (e.g. Thm. 2.8 of \cite{Rudin}) it follows that
$S \ast \kappa_{l,m}$ is a continuous linear operator 
$E_{i_1} \otimes \cdots \otimes E_{i_{l+m}} \rightarrow \mathscr{E}^{*}$, i.e.
\[
S \ast \kappa_{l,m} \in \mathscr{L} \big(E_{i_1} \otimes \cdots \otimes E_{i_{l+m}}, \,\, \mathscr{E}^* \big).
\]
If $\mathscr{E} = \mathcal{S}^{00}(\mathbb{R}^4; \mathbb{C})$ then $S$ can be extended over to an element of
$\mathcal{S}^{00}(\mathbb{R}^4; \mathbb{C})^*$ (Hahn-Banach theorem), and the above proof can be repeated, 
because the algebra of convolutors of $\mathcal{S}^{00}(\mathbb{R}^4; \mathbb{C})^*$ is dense in 
$\mathcal{S}^{00}(\mathbb{R}^4; \mathbb{C})^*$ and contains $\mathcal{O}_{C}(\mathbb{R}^4; \mathbb{C})$
(compare Subsection 5.4, 5.5 of \cite{wawrzycki2018} and Appendix \ref{convolutorsO'_C}). 
This completes the proof of part 1). 

The assertion 2) follows by an explicit verification and essentially repeatition of the proof of the analogue assertion 
of Lemma \ref{kappa0,1,kappa1,0psi}, Subsection \ref{psiBerezin-Hida} or respectively Lemma \ref{kappa0,1,kappa1,0ForA},
Subsection \ref{A=Xi0,1+Xi1,0}. 
\qed

\begin{rem*}
We should emphasize here that the mere assumption
\[
\kappa_{l,m}(\xi_1 \otimes \cdots \otimes \xi_{l+m}) \in \mathcal{O}_C(\mathbb{R}^{4}; \mathbb{C}),
\,\,\,\,
\xi_k \in E_{i_k}, \,\,\, i_k \in \{1, 2\}
\]
would be insufficient for 

\[
S \ast \kappa_{lm}(\xi_1, \ldots, \xi_{l+m}) \overset{\textrm{df}}{=}
S \ast \big(\kappa_{lm}(\xi_1, \ldots, \xi_{l+m})\big) 
\]
to be an element of $\mathcal{O}_C \subset \mathcal{O}'_{C}$.
Indeed it is the special property of the plane wave distribution kernels defininig the free fields
which assures the validity of the assertion 2). Moreover the fact that the space $E_2$ is equal
\[
\mathcal{S}^{0}(\mathbb{R}^3; \mathbb{C}^4) \neq \mathcal{S}(\mathbb{R}^3; \mathbb{C}^4)
\]
intervenes here nontrivially. For the wrong space $\mathcal{S}(\mathbb{R}^3; \mathbb{C}^4)$ 
used for $E_2$ the assertion 2) would be false. But both parts, 1) and 2), 
are important for the construction of higher order contributions to interacting
fields understood as well defined integral kernel operators with vector-valued kernels. 
Analogue situation we encounter for any other zero mass field for which the corresponding space
$E_2$ must be equal $\mathcal{S}^{0}(\mathbb{R}^3; \mathbb{C}^r)$.
\end{rem*}

From the Rule VI and Lemma \ref{S*Xi} it folows the following

\begin{prop*}
If 
\[
\kappa_{l,m} \in \mathscr{L} \big(\mathscr{E} , \,\, \big(E_{i_1} \otimes \cdots \otimes E_{i_{l+m}} \big)^* \, \big)
\cong \mathscr{L} \big(E_{i_1} \otimes \cdots \otimes E_{i_{l+m}}, \,\, \mathscr{E}^* \big)
\]
with
\[
\kappa_{l,m}(\xi_1 \otimes \cdots \otimes \xi_{l+m};x) \in \mathcal{O}_C(\mathbb{R}^{4}; \mathbb{C}),
\,\,\,\,
\xi_k \in E_{i_k}, \,\,\, i_k \in \{1, 2\}.
\]
and $S \in \mathcal{S}(\mathbb{R}^4; \mathbb{C})^*$, then the operator
\begin{multline*}
S \ast \Xi_{l,m}(\kappa_{l,m})(x) = 
\int \limits_{\mathbb{R}^4} S(x-y) \Xi_{l,m}\big(\kappa_{l,m}(y)\big) \, \ud^4 y \\
= \Xi_{l,m}\Bigg(\int \limits_{\mathbb{R}^4} S(x-y)\kappa_{l,m}(y) \, \ud^4 y \,  \Bigg)
= \Xi_{l,m}\big( S \ast \kappa_{lm}(x) \big) 
\end{multline*}
defines integral kernel operator 
\[
\Xi_{l,m}\big( S \ast \kappa_{lm} \big) 
\in \mathscr{L}\big((\boldsymbol{E}) \otimes \mathscr{E}, \, (\boldsymbol{E})^*\big) \cong
\mathscr{L}\big(\mathscr{E}, \, \mathscr{L}((\boldsymbol{E}), (\boldsymbol{E})^*)\big)
\]
with the vector-valued kernel
\[
S \ast \kappa_{lm} \in \mathscr{L}\big(\mathscr{E}, \, E_{i_1} \otimes \cdots \otimes E_{i_{l+m}} \big)^* \, \big)
\cong \mathscr{L} \big(E_{i_1} \otimes \cdots \otimes E_{i_{l+m}}, \,\, \mathscr{E}^* \big).
\]
If moreover 
\[
\kappa_{l,m} = \Big({}^{{}^{n_1}}_{{}_{1}}\kappa_{l_1,m_1} \Big) \overline{\dot{\otimes}} \cdots \overline{\dot{\otimes}} \,\, \Big( {}^{{}^{n_M}}_{{}_{M}}\kappa_{l_M,m_M} \Big), 
\,\,\,\,\,
{}^{{}^{n_k}}_{{}_{k}}\kappa_{l_k,m_k} \in \mathfrak{K}_0
\]
then
\[
S \ast \kappa_{lm}(\xi_1, \ldots, \xi_{l+m}) \in \mathcal{O}_C(\mathbb{R}^4; \mathbb{C}) 
\subset \mathcal{O}'_{C}(\mathbb{R}^4; \mathbb{C}).
\]
\end{prop*}

\begin{twr}\label{g=1InteractingFieldsQED}
Let
\[
\boldsymbol{\psi}(x) = \Xi_{0,1}\big({}^{1}\kappa_{0,1}(x)\big) 
+ \Xi_{1,0}\big({}^{1}\kappa_{1,0}(x)\big), \,\,\,
A = \Xi_{0,1}\big({}^{2}\kappa_{0,1}(x)\big) + \Xi_{1,0}\big({}^{2}\kappa_{1,0}(x)\big),
\]
be the integral kernel operators defining the free fields of the spinor QED. 
Let
\[
\boldsymbol{\psi}_{{}_{\textrm{int}}}^{a}(g=1, x) =
\boldsymbol{\psi}^{a}(x) + \sum \limits_{n=1}^{\infty} \frac{1}{n!}
\int \limits_{\mathbb{R}^{4n}} \ud^4x_1 \cdots \ud^4 x_n \boldsymbol{\psi}^{a \, (n)}(x_1, \ldots, x_n; x), 
\]
with
\[
\boldsymbol{\psi}^{a \, (1)}(x_1; x) = 
e S_{{}_{\textrm{ret}}}^{aa_1}(x-x_1) \gamma^{\nu_1 \, a_1a_2} \boldsymbol{\psi}^{a_2}(x_1)A_{\nu_1}(x_1), 
\]
\begin{multline*}
\boldsymbol{\psi}^{a \, (2)}(x_1, x_2; x) = \\
e^2 \Bigg\{ S_{{}_{\textrm{ret}}}^{aa_1}(x-x_1) \gamma^{\nu_1 \, a_1a_2}S_{{}_{\textrm{ret}}}^{a_2a_3}(x_1-x_2)
\gamma^{\nu_2 \, a_3a_4} :\boldsymbol{\psi}^{a_4}(x_2)  A_{\nu_1}(x_1)A_{\nu_2}(x_2) : \\ 
- S_{{}_{\textrm{ret}}}^{aa_1}(x-x_1) \gamma^{\nu_1 \, a_1a_2} 
: \boldsymbol{\psi}^{a_2}(x_1) 
\overline{\boldsymbol{\psi}}^{a_3}(x_2) \gamma_{\nu_1}^{a_3a_4} \boldsymbol{\psi}^{a_4}(x_2): 
D^{{}^{\textrm{ret}}}_{0}(x_1-x_2) \\
+S_{{}_{\textrm{ret}}}^{aa_1}(x-x_1) \Sigma_{{}_{\textrm{ret}}}^{a_1a_2}(x_1-x_2)\boldsymbol{\psi}^{a_2}(x_2)
\Bigg\} \,\,\, +  \,\,\, \Bigg\{ x_1 \longleftrightarrow x_2 \Bigg\},
\end{multline*}
\[
\textrm{e. t. c.}
\]
and let 
\[
{A_{{}_{\textrm{int}}}}_{\mu}(g=1, x) =
A_{\mu}(x) + \sum \limits_{n=1}^{\infty} \frac{1}{n!}
\int \limits_{\mathbb{R}^{4n}} \ud^4x_1 \cdots \ud^4 x_n A_{\mu}^{\, (n)}(x_1, \ldots, x_n; x),
\]
with
\[
A_{\mu}^{\, (1)}(x_1;x) = -e D^{{}^{\textrm{av}}}_{0}(x_1-x) 
:\overline{\boldsymbol{\psi}}^{a_1}(x_1) \gamma_{\mu}^{a_1a_2} \boldsymbol{\psi}^{a_2}(x_1):,
\]
\begin{multline*}
A_{\mu}^{\, (2)}(x_1, x_2; x) = 
e^2 \Bigg\{ 
:\overline{\boldsymbol{\psi}}^{a_1}(x_1) 
\Big( 
 \gamma_{\mu}^{a_1a_2} S_{{}_{\textrm{ret}}}^{a_2a_3}(x_1-x_2) \gamma^{\nu_1 \, a_3a_4}
D^{{}^{\textrm{av}}}_{0}(x_1-x) A_{\nu_1}(x_2) \\
+ \gamma^{\nu_1 \, a_1a_2}S_{{}_{\textrm{av}}}^{a_2a_3}(x_1-x_2) \gamma_{\mu}^{a_3a_4}
D^{{}^{\textrm{av}}}_{0}(x_2-x)A_{\nu_1}(x_1)
\Big)  \boldsymbol{\psi}^{a_4}(x_2): \\
+ D^{{}^{\textrm{av}}}_{0}(x_1-x) {\Pi^{{}^{\textrm{av}}}}_{\mu}^{\nu_1}(x_2-x_1)A_{\nu_1}(x_2)
\Bigg\} \,\,\, + \,\,\, \Bigg\{ x_1 \longleftrightarrow x_2 \Bigg\}
\end{multline*}
\[
\textrm{e. t. c.}
\]
be equal to the formulas for (fixed components $a$ and $\mu$) of interacting Dirac and electromagnetic
fields $\boldsymbol{\psi}_{{}_{\textrm{int}}}$ and $A_{{}_{\textrm{int}}}$ in the causal 
St\"uckelberg-Bologoliubov spinor QED, \cite{DutFred}, \cite{DKS1} or \cite{Scharf}, 
in which the intensity-of-interaction function $g$
is put equal to the constant $1$. 

If the free fields $\boldsymbol{\psi}(x)$, $A(x)$ in these formulas for 
$\boldsymbol{\psi}_{{}_{\textrm{int}}}$ and $A_{{}_{\textrm{int}}}$
are understood as integral kernel operators
\[
\boldsymbol{\psi}(x) = \Xi_{0,1}\big({}^{1}\kappa_{0,1}(x)\big) 
+ \Xi_{1,0}\big({}^{1}\kappa_{1,0}(x)\big), \,\,\,
A = \Xi_{0,1}\big({}^{2}\kappa_{0,1}(x)\big) + \Xi_{1,0}\big({}^{2}\kappa_{1,0}(x)\big),
\]
and correspondingly the operations of Wick product $: \cdot :$ and integrations 
$\ud^4x_1, \ldots \ud^4x_n$
involved in the formulas for $\boldsymbol{\psi}_{{}_{\textrm{int}}}$ and $A_{{}_{\textrm{int}}}$
 are understood as Wick products and integrations of integral kernel operators with 
vector valued distributional kernels (which as we know have the properties expressed by the Rules
I-VI), then each $n$-th order term contribution
\[
\begin{split}
\boldsymbol{\psi}_{{}_{\textrm{int}}}^{a \,(n)}(g=1, x) =
\frac{1}{n!}
\int \limits_{\mathbb{R}^{4n}} \ud^4x_1 \cdots \ud^4 x_n \boldsymbol{\psi}^{a \, (n)}(x_1, \ldots, x_n; x), \\
{A_{{}_{\textrm{int}}}}_{\mu}^{\, (n)}(g=1, x) = 
\frac{1}{n!}
\int \limits_{\mathbb{R}^{4n}} \ud^4x_1 \cdots \ud^4 x_n A_{\mu}^{\, (n)}(x_1, \ldots, x_n; x),
\end{split}
\]  
respectively, to the interacting field $\boldsymbol{\psi}_{{}_{\textrm{int}}}^{a}(g=1, x)$
and ${A_{{}_{\textrm{int}}}}_{\mu}(g=1, x)$ is equal to a finite sum
\[
\sum \limits_{l,m} \Xi(\kappa_{l,m}(x)) \,\,\, \textrm{respectively} \,\,\,
\sum \limits_{l,m} \Xi(\kappa'_{l,m}(x))
\]
of integral kernel operators
\[
\Xi_{l,m}(\kappa_{lm}(x)), \,\,\, \textrm{respectively} \,\,\,
\Xi(\kappa'_{l,m}(x))
\]
which define integral kernel operators 
\[
\begin{split}
\Xi_{l,m}(\kappa_{lm}) \in \mathscr{L}\big((\boldsymbol{E}) \otimes \mathscr{E}_1, \,(\boldsymbol{E})^* \big)
\cong \mathscr{L}\big(\mathscr{E}_1, \, \mathscr{L}((\boldsymbol{E}), (\boldsymbol{E})^*) \big), \\
 \textrm{respectively} \\
\Xi_{l,m}(\kappa'_{lm}) \in \mathscr{L}\big((\boldsymbol{E}) \otimes \mathscr{E}_2, \,(\boldsymbol{E})^* \big)
\cong \mathscr{L}\big(\mathscr{E}_2, \, \mathscr{L}((\boldsymbol{E}), (\boldsymbol{E})^*) \big)
\end{split}
\]
with vector-valued distributional kernels
\[
\begin{split}
\kappa_{l,m} \in \mathscr{L} \big(E_{i_1} \otimes \cdots \otimes E_{i_{l+m}}, \,\, \mathscr{E}_{1}^* \big) \\
\kappa'_{l,m} \in \mathscr{L} \big(E_{i_1} \otimes \cdots \otimes E_{i_{l+m}}, \,\, \mathscr{E}_{2}^* \big). 
\end{split}
\]
Thus each $n$-th order term contribution $\boldsymbol{\psi}_{{}_{\textrm{int}}}^{a, \,(n)}(g=1)$
and ${A_{{}_{\textrm{int}}}}_{\mu}^{\, (n)}(g=1)$, respectively, to interacting fields
$\boldsymbol{\psi}_{{}_{\textrm{int}}}^{a}(g=1)$
and ${A_{{}_{\textrm{int}}}}_{\mu}(g=1)$ is equal
\[
\begin{split}
\boldsymbol{\psi}_{{}_{\textrm{int}}}^{a, \,(n)}(g=1) \,\, = \,\,
 \sum \limits_{l,m} \Xi(\kappa_{l,m}), \\
{A_{{}_{\textrm{int}}}}_{\mu}^{\, (n)}(g=1) \,\, = \,\,
\sum \limits_{l,m} \Xi(\kappa'_{l,m}),
\end{split}
\] 
to a finite sum of well defined integral kernel operators $\Xi(\kappa_{l,m}), \Xi(\kappa'_{l,m})$ 
with vector-valued distributional kernels $\kappa_{l,m}, \kappa'_{l,m}$ in the sense of Obata 
\cite{obataJFA} (compare Subsection \ref{psiBerezin-Hida}).  
\end{twr}

\qedsymbol \, 
The proof follows by induction and the repeated application of the Rules I-VI and the 
fundamental Lemma \ref{S*Xi}.
\qed

\begin{rem*}
Note that each $n$-th order contribution $\boldsymbol{\psi}_{{}_{\textrm{int}}}^{a, \,(n)}(g=1)$
and ${A_{{}_{\textrm{int}}}}_{\mu}^{\, (n)}(g=1)$ to interacting fields  
$\boldsymbol{\psi}_{{}_{\textrm{int}}}^{a}(g=1)$
and ${A_{{}_{\textrm{int}}}}_{\mu}(g=1)$ belongs to the same general class of (finite sums of) integral 
kernel operators (with vector-valued kernels) as the Wick products (at fixed space-time point)
of mass less fields. In fact some of the conributions to interacting fields are finite 
sums of integral kernel operators which even belong to a much better
behaved class of integral kernel operators, which belong to
\[
\begin{split}
\mathscr{L}\big((\boldsymbol{E}) \otimes \mathscr{E}_1, \,(\boldsymbol{E}) \big)
\cong \mathscr{L}\big(\mathscr{E}_1, \, \mathscr{L}((\boldsymbol{E}), (\boldsymbol{E})) \big), \\
 \textrm{respectively} \\
\mathscr{L}\big((\boldsymbol{E}) \otimes \mathscr{E}_2, \,(\boldsymbol{E}) \big)
\cong \mathscr{L}\big(\mathscr{E}_2, \, \mathscr{L}((\boldsymbol{E}), (\boldsymbol{E})) \big).
\end{split}
\]
In particular one can show that the first order contribution 
${A_{{}_{\textrm{int}}}}_{\mu}^{\, (1)}(g=1)$ to the interacting electromagnetic
potential field ${A_{{}_{\textrm{int}}}}_{\mu}(g=1)$ belogs to 
\[
\mathscr{L}\big((\boldsymbol{E}) \otimes \mathscr{E}_2, \,(\boldsymbol{E}) \big)
\cong \mathscr{L}\big(\mathscr{E}_2, \, \mathscr{L}((\boldsymbol{E}), (\boldsymbol{E})) \big).
\]
Let us emphasize here that the Wick product (at the the same space-time point)
of mass less free fields (or containig such among the factors) does not belong to
\[
\begin{split}
\mathscr{L}\big((\boldsymbol{E}) \otimes \mathscr{E}_1, \,(\boldsymbol{E}) \big)
\cong \mathscr{L}\big(\mathscr{E}_1, \, \mathscr{L}((\boldsymbol{E}), (\boldsymbol{E})) \big), \\
 \textrm{respectively} \\
\mathscr{L}\big((\boldsymbol{E}) \otimes \mathscr{E}_2, \,(\boldsymbol{E}) \big)
\cong \mathscr{L}\big(\mathscr{E}_2, \, \mathscr{L}((\boldsymbol{E}), (\boldsymbol{E})) \big).
\end{split}
\]
But we know that such product, as an integral kernel operator with vector-valued kernel,
belongs to
\[
\begin{split}
\mathscr{L}\big((\boldsymbol{E}) \otimes \mathscr{E}_1, \,(\boldsymbol{E})^* \big)
\cong \mathscr{L}\big(\mathscr{E}_1, \, \mathscr{L}((\boldsymbol{E}), (\boldsymbol{E})^*) \big), \\
 \textrm{respectively} \\
\mathscr{L}\big((\boldsymbol{E}) \otimes \mathscr{E}_2, \,(\boldsymbol{E})^* \big)
\cong \mathscr{L}\big(\mathscr{E}_2, \, \mathscr{L}((\boldsymbol{E}), (\boldsymbol{E})^*) \big).
\end{split}
\]
Similarly we know that each order term contribution to interacting fields is a finite sum of 
integral kernel operators which belong to  
\[
\begin{split}
\mathscr{L}\big((\boldsymbol{E}) \otimes \mathscr{E}_1, \,(\boldsymbol{E})^* \big)
\cong \mathscr{L}\big(\mathscr{E}_1, \, \mathscr{L}((\boldsymbol{E}), (\boldsymbol{E})^*) \big), \\
 \textrm{respectively} \\
\mathscr{L}\big((\boldsymbol{E}) \otimes \mathscr{E}_2, \,(\boldsymbol{E})^* \big)
\cong \mathscr{L}\big(\mathscr{E}_2, \, \mathscr{L}((\boldsymbol{E}), (\boldsymbol{E})^*) \big).
\end{split}
\]
But at least some of them, e.g. the first order contribution
$\boldsymbol{\psi}_{{}_{\textrm{int}}}^{a, \,(1)}(g=1)$ to the interacting Dirac field
$\boldsymbol{\psi}_{{}_{\textrm{int}}}^{a}(g=1)$, do not belong to
\[
\mathscr{L}\big((\boldsymbol{E}) \otimes \mathscr{E}_1, \,(\boldsymbol{E}) \big)
\cong \mathscr{L}\big(\mathscr{E}_1, \, \mathscr{L}((\boldsymbol{E}), (\boldsymbol{E})) \big).
\]
Nonetheless the contributions to interacting fields are finite sums
of integral kernel operators which belong to the same general class as the integral kernel operators
which are equal to Wick products (at the same space-time point) of mass less free fields.

One can even show that if the Wick products (at the same space-time point) of  
free fields (including mass less fields) were equal to finite sums of integral kernel 
operators belonging to 
\[
\begin{split}
\mathscr{L}\big((\boldsymbol{E}) \otimes \mathscr{E}_1, \,(\boldsymbol{E}) \big)
\cong \mathscr{L}\big(\mathscr{E}_1, \, \mathscr{L}((\boldsymbol{E}), (\boldsymbol{E})) \big), \\
 \textrm{respectively} \\
\mathscr{L}\big((\boldsymbol{E}) \otimes \mathscr{E}_2, \,(\boldsymbol{E}) \big)
\cong \mathscr{L}\big(\mathscr{E}_2, \, \mathscr{L}((\boldsymbol{E}), (\boldsymbol{E})) \big),
\end{split}
\]
then the same would be true of the contributions to interacting fields. 
But the assumption about the Wick product necessary to infer this conclusion
is however false (compare the corresponding Proposition of this Subsection).
\end{rem*}

The behaviour of each higher order term $S_n$ to the scattering operator, evaluated at $g\in \mathscr{E}$
\[
S_n\big(g \mathcal{L}\big) = + \sum \limits_{n=1}^{\infty} {\textstyle\frac{1}{n!}} 
\int \ud^4 x_1 \cdots \ud^4 x_n S_n(x_1, \ldots, x_n,x) \,
g(x_1) \cdots g(x_n), \,\,\, g\in \mathscr{E}
\]
need not be analysed separately,  and its begaviour can be inferred from the behaviour of the higher order contributions 
to the interacting fields
\[
\mathbb{A}_{{}_{\textrm{int}}}(x) = \mathbb{A}_{{}_{\textrm{int}}}(g,x), 
\]
where
\begin{multline*}
\mathbb{A}_{{}_{\textrm{int}}}(g,x) =  S^{-1}(g\mathcal{L})
\frac{\delta S(g\mathcal{L}+h\mathbb{A})}{\delta h(x)}\Bigg{|}_{{}_{h=0}}
\\
= \mathbb{A}(x) + \sum \limits_{n=1}^{\infty} {\textstyle\frac{1}{n!}} 
\int \ud^4 x_1 \cdots \ud^4 x_n \mathbb{A}^{(n)}(x_1, \ldots, x_n,x) \,
g(x_1) \cdots g(x_n),
\end{multline*}
and where $\mathbb{A}$ is the free electromagnetic field or the free Dirac field. Indeed the behaviour of the higher order contributions
to the scattering operator $S_n\big(g \mathcal{L}\big)$ can be obtained, e. g. by putting for the 
free electromagnetic potential operator $\mathbb{A}^\mu(x) = A^\mu(x) = \boldsymbol{1}$ into the formula for 
$\mathbb{A}_{{}_{\textrm{int}}}(g,x) = \mathbb{A}_{{}_{\textrm{int}}}(g,x)$, with arbitrary $g \in \mathscr{E}$. 
In particular from the last 
Theorem (or repeated application of Lemma \ref{S*Xi}) it follows the following
\begin{cor*}
For each fixed $g \in \mathscr{E}$, and $n \in \mathbb{N}$
\[
S_n\big(g \mathcal{L}\big) \in \mathscr{L}((\boldsymbol{E}), \, (\boldsymbol{E})^*)
\]
and the map
\[
\mathscr{E} \ni g \longmapsto S_n\big(g \mathcal{L}\big) \in \mathscr{L}((\boldsymbol{E}), \, (\boldsymbol{E})^*)
\]
is continuous.
\end{cor*}

The last Corollary is sufficient for the computation of the effective cross-sections for the \emph{in} and \emph{out}
states which are of the form of many particle plane-wave states, in the adiabatic limit $g=1$, as we have explained in Introduction,
without any need for handling infrared of ulra-violet infinities.
The reader is encouraged to consult the computation of the effective 
cross-section presented in \cite{Bogoliubov_Shirkov}, \S\S 24.5 and 25.

Note that $S_n(x_1, \ldots, x_n)$ in the last Corollary can be look upon as a definition of the 
``chronological products'' $S_n(x_1, \ldots, x_n) = T(i\mathcal{L}(x_1) \cdots i\mathcal{L}(x_1))$ 
of the Lagrange interaction density $\mathcal{L}(x)$, evaluated at $x_1, \ldots, x_n$. Even more,
each higher order therm $S_n(x_1, \ldots, x_n)$ in the causal perturbation series has the form of sums of normally
ordered products (finite sums of integral kernel operators), which can heuristically be looked at as if the 
``Wick theorem for chronological product'' had automatically been done in the causal construction of $S_n(x_1, \ldots, x_n)$.
Note that in the last Corollary the ``chronological products'' $S_n(x_1, \ldots, x_n) = T(i\mathcal{L}(x_1) \cdots i\mathcal{L}(x_1))$ 
depend in fact on the particular choices in the Epstein-Glaser splitting of the 
causally supported tempered distributions performed in each inductive step of the causal construction of the chronological product $S_n(x_1, \ldots, x_n)$,
and the last Theorem and its Corollary hold true for each of the particular choices in the splittings. 
This also shows that the axioms (I)-(IV) of Subsection \ref{psiBerezin-Hida} do not determine $S_n(x_1, \ldots, x_n)$
uniquely but only within the flexibility in decomposition of the causal tempered distributions into retarded and advanced parts, 
defined by the pairing functions of the theory in question. This arbitrariness can be further eliminated by the requirement posed on the 
interacting fields, which should respect the corresponding equations of motion, compare in particuar \cite{DKS1}. 
Also the computation of the splitting is involved into the analysis of the quasi-asymptotics, which unfortunately need to be 
performed separately at each order.

For this reason we give in Subsection  \ref{WickForChronological}  a construction of a particular  and ``natural'' example
of chronological product (which is equivalent to making particular choices in the splitting) and which respects (I)-(IV), 
and which is closely motivated by the  heuristic definition of the ``chronological product'' used in \cite{Bogoliubov_Shirkov}. 
This will allow us to avoid the analysis of quasi asymtotics, and provide a knew effective method for the computation 
of the perturbative series for $S_n(x_1, \ldots, x_n)$.
In Subsection \ref{WickForChronological} we also give another proof of the last Corollary for this particular
``naturally'' constructed chronological product.

\subsection{Wick's theorem for ``products''}\label{WickForProduct}

Finally let us return to the Wick product theorem for free fields. In the intermediate stage of the computations
of the scattering operator and interacting fields a so called Wick theorem (\cite{Bogoliubov_Shirkov}, \S 17.2) is used for 
decomposition of the ``product'' 
\begin{equation}\label{Wick(x)Wick(y)}
\boldsymbol{:} \mathbb{A}_{{}_{1}}^{a}(x) \ldots \mathbb{A}_{{}_{N}}^{a}(x) \boldsymbol{:} 
\boldsymbol{:} \mathbb{A}_{{}_{N+1}}^{b}(y) \ldots \mathbb{A}_{{}_{M}}^{b}(y)  \boldsymbol{:}
\end{equation}
of Wick product monomials 
\begin{equation}\label{WickMonomialsInGeneralFreeFields}
\boldsymbol{:} \mathbb{A}_{{}_{1}}^{a}(x) \ldots \mathbb{A}_{{}_{N}}^{a}(x) \boldsymbol{:} 
\,\,\,
\textrm{and}
\,\,\,
\boldsymbol{:} \mathbb{A}_{{}_{N+1}}^{b}(y) \ldots \mathbb{A}_{{}_{M}}^{b}(y)  \boldsymbol{:}
\end{equation}
in fixed components $\mathbb{A}_{{}_{k}}$ of free fields, each separately evaluated at the same space-time point $x$ or respectively, $y$,
into the sum of Wick monomials (each in the so-called ``normal order'').

The point lies in the correct definition of such ``product'', because each factor evaluated respectively at $x$ or $y$,
represetnts a generalized integral kernel operator transformning continously the Hida space $(\boldsymbol{E})$ into its strong dual
$(\boldsymbol{E})^*$, so that the product cannot be undersood as ordinary operarator composition, and therefore a correct definition is here required.
Recall that  $(\boldsymbol{E})$ is the Hida test space in the total Fock space of the free fileds involved in the product (in fact
in the total Fock space of all free fileds underlying the QFT in question).

The cruacial point is that the free fields and their Wick products define (finite sums of) integral kernel oerators with vector-valued
kernels in the sense of \cite{obataJFA}, as we have explained above, and the ``product'' can be given as a distributional
kernel operator. Indeed, from what we have already shown, it follows that each factor (\ref{WickMonomialsInGeneralFreeFields}) separately
represents an integral kernel operator which belongs to 
\[
\mathscr{L}\big(\mathscr{E}_i, \, \mathscr{L}((\boldsymbol{E}), \, (\boldsymbol{E})) \big), \,\,\, i =1  \, \textrm{or} \, 2,
\]
if among the factors $\mathbb{A}_{{}_{k}}$, $1 \leq k \leq M$, there are no mass-less fields (or their derivatives).
This means that the first factor in (\ref{WickMonomialsInGeneralFreeFields}) defines the corresponding continuous map
\[
\mathscr{E}_i \ni \phi \longmapsto \Xi(\phi) =
\sum\limits_{\substack{\ell,m \\ \ell+m=N}} 
\Xi_{\ell,m}\big(\kappa_{\ell,m}(\phi)\big)
\in \mathscr{L}\big( (E), (E)\big),  \,\,\, i =1  \, \textrm{or} \, 2,
\]
and similarily the second factor in (\ref{WickMonomialsInGeneralFreeFields}) defines continous map
\[
\mathscr{E}_j \ni \varphi \longmapsto \Xi'(\varphi) =
\sum\limits_{\substack{\ell,m \\ \ell'+m'=M-N}} 
\Xi_{\ell',m'}\big({\kappa'}_{\ell',m'}(\varphi)\big)
\in \mathscr{L}\big( (\boldsymbol{E}), (\boldsymbol{E})\big),  \,\,\, i =1  \, \textrm{or} \, 2,
\]
where $\mathscr{E}_i = \mathcal{S}_{{}_{B_{p_i}}}(\mathbb{R}^4; \mathbb{C}^4)$ and $E_j = \mathcal{S}_{{}_{B_{p_j}}}(\mathbb{R}^4; \mathbb{C}^4)$,
$i,j = 1,2$, $p_i, p_j= 1,2$ (compare Subsection \ref{psiBerezin-Hida} for the definition of the standard operatrs
$B_{p_i}$ on the correspondng standard Hilbert spaces).  Both factors $\Xi$ and $ \Xi'$
are equal to finite sums of integral kernel opertors vith $\mathscr{E}_{i}^{*}$- or $\mathscr{E}_{i}^{*}$-valued distributional kernels
$\kappa_{\ell,m}, \kappa'_{\ell',m'}$. In this case both factors $\Xi(\phi)$ and $ \Xi'(\varphi)$, when evaluated at the test functions $\phi, \varphi$,
are ordinary operators on the Fock space transforming continously the Hida space $(\boldsymbol{E})$ into itself,
and thus can be composed $\Xi(\phi) \circ  \Xi'(\varphi)$ as operators, giving the composition operator
\[
\Xi(\phi) \circ  \Xi'(\varphi) \in \mathscr{L}\big( (\boldsymbol{E}), (\boldsymbol{E})\big),
\]
defining the map
\[
\mathscr{E}_i \otimes \mathscr{E}_j \ni \phi \otimes \varphi \longmapsto \Xi(\phi) \circ  \Xi'(\varphi) \in \mathscr{L}\big( (\boldsymbol{E}), (\boldsymbol{E})\big),
\]
which by construction is separately continuous in the arguments $\phi \in \mathscr{E}_i$ and $\varphi \in \mathscr{E}_j$. Because
$\mathscr{E}_i, \mathscr{E}_j$ are complete Fr\'echet spaces, then by Proposition 1.3.11 of \cite{obataJFA} there exist
the corresponding operator-valued continuous map (say operator-valued distribution) 
\begin{multline*}
\phi \otimes \varphi \longmapsto 
 \Xi(\phi \otimes \varphi) \overset{\textrm{df}}{=} \\ \overset{\textrm{df}}{=}
\sum\limits_{a,b}
\int\limits_{\big[\mathbb{R}^4\big]^{\times \, 2}} 
\boldsymbol{:} \mathbb{A}_{{}_{1}}^{a}(x) \ldots \mathbb{A}_{{}_{N}}^{a}(x) \boldsymbol{:} 
\boldsymbol{:} \mathbb{A}_{{}_{N+1}}^{b}(y) \ldots \mathbb{A}_{{}_{M}}^{b}(y)  \boldsymbol{:} \,
\phi^a \otimes \varphi^b(x,y) \, \ud^4x \, \ud^4 y
\\
= \Xi(\phi) \circ  \Xi'(\varphi). 
\end{multline*}
In particular the operator map $\phi \otimes \varphi \mapsto \Xi(\phi \otimes \varphi )$ defines a 
generalized operator 
\[
\Xi \in \mathscr{L}\big(\mathscr{E}_i \otimes \mathscr{E}_j, \, \mathscr{L}((\boldsymbol{E}), \, (\boldsymbol{E}) \big)
\cong \mathscr{L}((\boldsymbol{E}) \otimes \mathscr{E}_i \otimes \mathscr{E}_j, \, (\boldsymbol{E})) 
\]
which by Theorem 4.8 of \cite{obataJFA} possesses unique (here finite) Fock expansiion
\[
\Xi = \sum\limits_{\ell'', m''} \Xi''_{\ell'', m''}({\kappa''}_{\ell'', m''}), \,\,\, \ell''+ m'' = M,
\]
into integral kernel operators
with $\mathscr{E}_{i}^{*} \otimes \mathscr{E}_{j}^{*} = \mathscr{L}(\mathscr{E}_i \otimes \mathscr{E}_j, \mathbb{C})$-valued
kernels ${\kappa''}_{\ell'', m''}$. This gives us the Wick theorem in case in which all factors
$\mathbb{A}_{{}_{k}}$ are massive free fields or their derivatives.  This form of Wick theorem is however
insufficient in realistic QFT, such as QED, because in the causal construction of the scattering operator or causal construction of interacting
fields from the scattering operator, the  Wick factors (\ref{WickMonomialsInGeneralFreeFields}) necessary include the Lagrange interaction density
\[
\mathcal{L}(x) = \boldsymbol{:} \boldsymbol{\psi}(x)^{+}\gamma_{0} \gamma^\mu \boldsymbol{\psi}(x) A_\mu(x) \boldsymbol{:} 
\]
and necessaty incude the mass-less electromagnetic potential field $A$ as one of the factors $\mathbb{A}_{{}_{k}}$
in the Wick products which have to be considered. In particulr as the first computational step in the causal perturbative construction of the 
scattering operator we need to consider the ``product'' (compare Section \ref{A(1)psi(1)} or \cite{Scharf}, \cite{Bogoliubov_Shirkov})
\begin{equation}\label{L(x)L(y)}
\mathcal{L}(x) \mathcal{L}(x)
\end{equation}
and apply Wick theorem of \cite{Bogoliubov_Shirkov} in order to write it in the form of ``normally ordered''
operators. 

In this situation, when among the factors $\mathbb{A}_{{}_{k}}$ in (\ref{Wick(x)Wick(y)}) there are present mass-less 
fields (or their derivatives), as in (\ref{L(x)L(y)}) -- a particular case of (\ref{Wick(x)Wick(y)}) -- 
then we replace the $\mathscr{E}_{2}^{*}$-valued kernels $\kappa_{0,1}, \kappa_{1,0}$ defining the mass-less factors by their massive counterparts.
In practice we just replace the zero mass energy functions
\[
p_{0}(\boldsymbol{\p}) = |\boldsymbol{\p}|
\]
in the mass-less kernels $\kappa_{0,1}, \kappa_{1,0}$ by the massive energy functions
\[
p_{0}(\boldsymbol{\p}) = \sqrt{|\boldsymbol{\p}|^2 + \epsilon^2}
\]
and obtain in this manner the massive kernels $\kappa_{\epsilon \,\, 0,1}, \kappa_{\epsilon \,\, 1,0}$, and the corresponding
``product'' (\ref{Wick(x)Wick(y)}):  $\Xi_{\epsilon}$, with all the mass-less kernels replaced with massive counterparts, as well as its 
Fock expansion
\[
\Xi_{\epsilon} = \sum\limits_{\ell'', m''} \Xi''_{\epsilon \,\, \ell'', m''}({\kappa''}_{\epsilon \,\,  \ell'', m''})
\in \mathscr{L}((\boldsymbol{E}) \otimes \mathscr{E}_i \otimes \mathscr{E}_j, \, (\boldsymbol{E})), \,\,\, \ell''+ m'' = M,
\]
into integral kernel operators
with $\mathscr{E}_{i}^{*} \otimes \mathscr{E}_{j}^{*} = \mathscr{L}(\mathscr{E}_i \otimes \mathscr{E}_j, \mathbb{C})$-valued
kernels ${\kappa''}_{\epsilon \,\, \ell'', m''}$, exactly as above for the massive fields.

It is easily checked that the distributional kernels   $\kappa_{\epsilon \,\, 0,1}, \kappa_{\epsilon \,\, 1,0}$
converge to $\kappa_{0,1}, \kappa_{1,0}$ in 
\[
\mathscr{L}(E_2, \mathscr{E}_{2}^{*})
\]
and from this it easily follows that ${\kappa''}_{\epsilon \,\, \ell'', m''}$ converge in
\[
\mathscr{L}(E_{{}_{n_1}} \otimes \ldots \otimes E_{{}_{n_M}}; \mathscr{E}_{i}^{*} \otimes  \mathscr{E}_{j}^{*}), \,\,\, \ell''+ m'' = M,
\]
when $\epsilon \rightarrow 0$. By Prop. 3.9 and Theorem 4.8 (or, respectively, by their Fock analogues, 
compare Subsection \ref{psiBerezin-Hida}) the operator ``product'' $\Xi_{\epsilon}$
converges to an operator 
\[
\Xi \in \mathscr{L}(\,\, (\boldsymbol{E}) \otimes \mathscr{E}_i \otimes \mathscr{E}_j \,\, , \,\,\, (\boldsymbol{E})^* \,\,) 
\] 
when $\epsilon \rightarrow 0$, which in general does not belong to 
\[
\mathscr{L}(\,\,(\boldsymbol{E}) \otimes \mathscr{E}_i \otimes \mathscr{E}_j \,\, , \,\,\, (\boldsymbol{E}) \,\,).
\]
This operator, when evaluated at fixed element $\phi \otimes \varphi \in \mathscr{E}_i \otimes \mathscr{E}_j$,
gives an operator in $\mathscr{L}((\boldsymbol{E}), \, (\boldsymbol{E})^*)$, and defines a continuous map
\begin{multline}\label{IntegralWick(x)Wick(y)phi(x)varphi(x)dxdy}
\mathscr{E}_i \otimes \mathscr{E}_j \ni \phi \otimes \varphi \longmapsto
 \Xi(\phi \otimes \varphi) \overset{\textrm{df}}{=} \\ \overset{\textrm{df}}{=}
\sum\limits_{a,b}
\int\limits_{\big[\mathbb{R}^4\big]^{\times \, 2}} 
\boldsymbol{:} \mathbb{A}_{{}_{1}}^{a}(x) \ldots \mathbb{A}_{{}_{N}}^{a}(x) \boldsymbol{:} 
\boldsymbol{:} \mathbb{A}_{{}_{N+1}}^{b}(y) \ldots \mathbb{A}_{{}_{M}}^{b}(y) \boldsymbol{:} \,
\phi^a \otimes \varphi^b(x,y) \, \ud^4x \, \ud^4 y
\\
\in \mathscr{L}((\boldsymbol{E}) \otimes \mathscr{E}_i \otimes \mathscr{E}_j, \, (\boldsymbol{E})^*),
\end{multline}
but this time, its value cannot be written as operator composition.
Again by Thm 4.8 of  \cite{obataJFA} or its fermionic analogue (compare Subsection \ref{psiBerezin-Hida}), 
which is applicable to general operators belonging also to
\[
\mathscr{L}(\,\, (\boldsymbol{E}) \otimes \mathscr{E}_i \otimes \mathscr{E}_j \,\, , \,\,\, (\boldsymbol{E})^* \,\, ),
\]
the generaliezed operator $\Xi$, defined by  (\ref{IntegralWick(x)Wick(y)phi(x)varphi(x)dxdy}), 
possesses unique (here finite) Fock expansion, which in fact gives the rigorous version of the Wick theorem stated in 
\cite{Bogoliubov_Shirkov}, \S 17, as ``The Wick's Theorem for Ordinary Products''\footnote{In the Enghish Edition we read there: ``Wick Theorem for 
Normal Products'', but ``The Wick's Theorem for Ordinary Products'' would be a better translation of the Russian original.}. 
Generalization of this theorem to the products
(\ref{Wick(x)Wick(y)}) containing a greather number of normally ordered Wick product factors (\ref{WickMonomialsInGeneralFreeFields}) 
is obvious.  This gives the mathematical justification for the Wick theorem stated
in \cite{Bogoliubov_Shirkov}, and shows that indeed the decomposition, or Fock expansion, can be easily
computed through the pairing functions, \emph{i.e.} commutation functions between the respective positive and negative frequency
parts of the free field factors $\mathbb{A}_{{}_{k}}$, because indeed it can be effectively computed through the operator
products (with mass-less factors replaced with the massive counterparts) and by the observation that in the zero-mass limit 
of the massive parings we indeed get the pairings of the mass-less fields.

\subsection{A natural chronological product. Wick's theorem for the natural chronological product}\label{WickForChronological}

Similarily as we did for the construction of the ``product'' of Wick ordered factors in Subsection \ref{WickForProduct}, 
we give here a similar construction of a `` natural'' chronological product, which is essentially based on the step theta
function $\theta$, and is immediately motivated by the heuristic definition used in \cite{Bogoliubov_Shirkov}:
\begin{multline}\label{thetaS_n}
S_n(x_1, \ldots, x_n) = \\ = i^{n}
\sum\limits_{\pi} \theta(t_{\pi(1)} -t_{\pi(2)}) \theta(t_{\pi(2)} -t_{\pi(3)}) \ldots \theta(t_{\pi(n-1)} -t_{\pi(n)})  \,
\mathcal{L}(x_{\pi(1)}) \ldots \mathcal{L}(x_{\pi(n)}), \\ 
\,\,\,\,\,\,\,\,\,\,\,\,\,\,\,\,\,\,\,\,\,\,\,\,\,\,\,\,\,\,\,\,
x_{\pi(k)} = (t_{\pi(k)}, \boldsymbol{\x}_{\pi(k)}), \,\,\, \pi \in \textrm{Permutations of} \,\{1, \ldots, n\}.
\end{multline}
We give here a strict meaning to the expression (\ref{thetaS_n}).
It is true that the ``regularized'' factors $\mathcal{L}_{\epsilon}(x) \in \mathscr{L}((\boldsymbol{E}), \, (\boldsymbol{E}))$ (with the kernels of the zero-mass electromagnetic field replaced by the 
kernels of the massive counterpart) and evaluated at 
fixed spacetime point $x$ can be multiplied by the value $\theta(t)$ of the theta function $\theta$. This is beacuse
$\mathcal{L}_{\epsilon}(x)$ for fixed $x$ is a well defined generalized operator, and it remanis to be so after being
multipled by ordinary number $\theta(t)$ (with fixed $x = (t,\boldsymbol{\x})$). It is however not the specific value
$\theta(t)\mathcal{L}_\epsilon(t,\boldsymbol{\x}) \in \mathscr{L}((\boldsymbol{E}), \, (\boldsymbol{E})^*)$ 
(with fixed $x = (t,\boldsymbol{\x})$) which is important here but the whole generalized operator $\theta \mathcal{L}_{\epsilon}$ with integral vector valued kernel, 
or the whole operator defined by the operator kernel $\theta(t)\mathcal{L}(t,\boldsymbol{\x})$. 
We know that the regularized
\[
\mathcal{L}_\epsilon \in \mathscr{L}( (\boldsymbol{E}) \otimes \mathscr{E} , (\boldsymbol{E}) )
\]
but after being multiplied by $\theta$ it becomes more singular, and
\[
\theta \mathcal{L}_\epsilon \in \mathscr{L}( (\boldsymbol{E}) \otimes \mathscr{E},  (\boldsymbol{E})^*).
\]
In this way the operation of mutiplication by $\theta$ converts the regular factors in 
\[
\sum\limits_{\pi} 
\mathcal{L}_\epsilon(x_{\pi(1)}) \ldots \mathcal{L}_\epsilon(x_{\pi(n)})
\]
into more singular. In other words we write explicitly
\[
\mathcal{L}_\epsilon \in \mathscr{L}( (\boldsymbol{E}) \otimes \mathscr{E} , (\boldsymbol{E}) )
\] 
in the form of Fock expansion
\[
\mathcal{L}_\epsilon =
\sum\limits_{\substack{\ell,m \\ \ell+m=3}} 
\Xi_{\ell,m}\big(\kappa_{\epsilon \,\, \ell,m}\big)
\in \mathscr{L}( (\boldsymbol{E}) \otimes \mathscr{E} , (\boldsymbol{E}) ),
\]
with 
\[
\kappa_{\epsilon \,\, \ell,m} \in \mathcal{L}(E_{1}^{*} \otimes E_{1}^{*} \otimes E_{2}^{*}, \mathscr{E}^*) 
\cong \mathcal{L}(\mathscr{E}, E_{1} \otimes E_{1} \otimes E_{2}), \,\,\, \ell+m = 3 
\]
with the kerels $\kappa_{\epsilon \,\,  \ell,m}$ constructed through the pontwise mutiplication and symmetrization or, respecively,
anti-symmetrization operations applied to the kernels defining the factor free fields in $\mathcal{L}$ and with the kernels corresponding to 
the free electromagnetic potential field replaced with their massive counterparts, as explained above. Then if we apply the operation 
of point wise mutiplication by $\theta$ function to the kernels $\kappa_{\epsilon \,\,  \ell,m}$, we will get the kernels
\[
\theta\kappa_{\epsilon \,\,  \ell,m}(s_1 \boldsymbol{\p}_1,s_2 \boldsymbol{\p}_2, \nu \boldsymbol{\p};t, \boldsymbol{\x}) \overset{\textrm{df}}{=}
\theta(t)\kappa_{\epsilon \,\,  \ell,m}(s_1 \boldsymbol{\p}_2,s_2 \boldsymbol{\p}_2, \nu \boldsymbol{\p};t, \boldsymbol{\x})
\]
of the generalized operator
\[
\theta\mathcal{L}_\epsilon =
\sum\limits_{\substack{\ell,m \\ \ell+m=3}} 
\Xi_{\ell,m}\big(\theta\kappa_{\epsilon \,\, \ell,m}\big)
\in \mathscr{L}( (\boldsymbol{E}) \otimes \mathscr{E} , (\boldsymbol{E}) ).
\]
But now the kernels
\[
\theta\kappa_{\epsilon \,\, \ell,m} \in \mathcal{L}(E_{1} \otimes E_{1} \otimes E_{2}, \mathscr{E}^*) 
\cong \mathcal{L}(\mathscr{E}, E_{1}^{*} \otimes E_{1}^{*} \otimes E_{2}^{*}),
\]
and
\[ 
\theta\kappa_{\epsilon \,\, \ell,m} \notin \mathcal{L}(E_{1}^{*} \otimes E_{1}^{*} \otimes E_{2}^{*}, \mathscr{E}^*) 
\cong \mathcal{L}(\mathscr{E}, E_{1} \otimes E_{1} \otimes E_{2}).
\]
Thus, by Thm. 3.13 of \cite{obataJFA} (or its fermionic analogue, Thm. \ref{obataJFA.Thm.3.13} of Subsection \ref{psiBerezin-Hida}) 
we arrive with the more singular $\theta \mathcal{L}_\epsilon$ transforming $(\boldsymbol{E})$ into $(\boldsymbol{E})^*$, and the construction used above for 
the definition of ``product'' of Wick-ordered factors cannot be applied for the ``chronological product'' through the intermediate operator composition
of the operators
\[
\theta_{{}_{\pi(k-1)}} \mathcal{L}_\epsilon(\phi_{\pi(k)}), \,\,\,\,\, \phi_{\pi(k)} \in \mathscr{E},
\]
where $\theta_{{}_{\pi(k+1)}}(t) = \theta(t- t_{\pi(k+1)})$, because 
\[
\theta_{{}_{\pi(k-1)}} \mathcal{L}_\epsilon(\phi_{\pi(k)}) \in \mathscr{L}((\boldsymbol{E}), \, (\boldsymbol{E})^*), \,\,\,\,\, 
\textrm{for} \,\,\, \phi_{\pi(k)} \in \mathscr{E},
\]
and the operators $\theta_{{}_{\pi(k-1)}} \mathcal{L}_\epsilon(\phi_{\pi(k)})$ cannot be composed as operators which transform Hida space into itself. 
A more sophisticated method is needed, involving a limit process not only with the regularization of the mass-less
kernels involved into the limit $\epsilon \rightarrow 0$, but likewise the theta function $\theta$ will have to  be
replaced with a smooth function $\theta_\varepsilon$, controlled by another parameter $\varepsilon$, 
and the final result achieved ony in the limit process.
Indeed we can always choose a one-parameter family $\{\theta_\varepsilon\}_{\varepsilon \in \mathbb{R}}$ of smooth functions 
$\theta_\varepsilon$ such that 
\[
\theta_\varepsilon \kappa_{\epsilon \,\, \ell,m} \in \mathcal{L}(E_{1}^{*} \otimes E_{1}^{*} \otimes E_{2}^{*}, \mathscr{E}^*) 
\cong \mathcal{L}(\mathscr{E}, E_{1} \otimes E_{1} \otimes E_{2}),
\]
\[
\theta_\varepsilon\mathcal{L}_\epsilon =
\sum\limits_{\substack{\ell,m \\ \ell+m=3}} 
\Xi_{\ell,m}\big(\theta_\varepsilon\kappa_{\epsilon \,\, \ell,m}\big)
\in \mathscr{L}( (\boldsymbol{E}) \otimes \mathscr{E} , (\boldsymbol{E}) ),
\]
and such that
\[
\theta_\varepsilon \kappa_{\epsilon \,\, \ell,m} \overset{\varepsilon, \epsilon \rightarrow 0}{\longrightarrow} \theta\kappa_{\ell,m}
\,\,\, \textrm{in} \,\,\, \mathcal{L}(E_{1} \otimes E_{1} \otimes E_{2}, \mathscr{E}^*) 
\cong \mathcal{L}(\mathscr{E}, E_{1}^{*} \otimes E_{1}^{*} \otimes E_{2}^{*}),
\]
and thus for each fixed $k\in \{1, \ldots,n\}$ and each fixed permutation $\pi$
\begin{multline*}
\theta_{{}_{\varepsilon \,\, \pi(k)}} \kappa_{\epsilon \,\, \ell,m} 
 \overset{\epsilon \rightarrow 0}{\longrightarrow}
\theta_{{}_{\pi(k)}} \kappa_{\ell,m} 
\,\,\,
\textrm{in} \,\,\, 
\mathcal{L}(E_{1} \otimes E_{1} \otimes E_{2}, \mathscr{E}^*) 
\cong \mathcal{L}(\mathscr{E}, E_{1}^{*} \otimes E_{1}^{*} \otimes E_{2}^{*}).
\end{multline*}
Therefore we can form the following continuous map
\begin{multline}\label{epsilonChronologicalProduct=OperatorProduct}
\mathscr{E}^{\otimes \, n}\ni \phi_1 \otimes \ldots \phi_n  \longmapsto \Xi_{\varepsilon, \epsilon}(\phi_1 \otimes \ldots \phi_n) \overset{\textrm{df}}{=} \\ \overset{\textrm{df}}{=} 
\sum\limits_{\pi}
\theta_{{}_{\varepsilon \,\, \pi(2)}} \mathcal{L}_\epsilon(\phi_{\pi(1)}) \circ 
\theta_{{}_{\varepsilon \,\, \pi(3)}} \mathcal{L}_\epsilon(\phi_{\pi(2)}) \circ \ldots \circ
\theta_{{}_{\varepsilon \,\, \pi(n)}} \mathcal{L}_\epsilon(\phi_{\pi(n-1)}) \circ
\mathcal{L}_\epsilon(\phi_{\pi(n)}) \\
\in \mathscr{L}\big( (\boldsymbol{E}), (\boldsymbol{E})\big),
\end{multline} 
defined by ordinary compositon $\circ$ of operators transforming continously the Hida space into itself.
Thus again by Proposition 3.9 (or, respectively, by its fermionic analogue, 
compare Subsection \ref{psiBerezin-Hida}) the operator product 
\[
\Xi_{\varepsilon,\epsilon}
\in \mathscr{L}( (\boldsymbol{E}) \otimes \mathscr{E}^{\otimes \, n} , (\boldsymbol{E}) )
\cong  \mathscr{L}\big(\mathscr{E}^{\otimes \, n}, \, \mathscr{L}( (\boldsymbol{E}), (\boldsymbol{E}) ) \, \big)
\]
converges to an operator 
\begin{multline*}
\Xi  \in \mathscr{L}( (\boldsymbol{E}) \otimes \mathscr{E}^{\otimes \, n} , (\boldsymbol{E})^* )
\cong  (\boldsymbol{E})^* \otimes \mathscr{E}^{* \, \otimes \, n} \otimes (\boldsymbol{E})^*
\supset \\ \supset
(\boldsymbol{E}) \otimes \mathscr{E}^{* \, \otimes \, n} \otimes (\boldsymbol{E})^* \cong
(\boldsymbol{E})^* \otimes \mathscr{E}^{* \, \otimes \, n} \otimes (\boldsymbol{E}) \cong
\mathscr{L}( (\boldsymbol{E}) \otimes \mathscr{E}^{\otimes \, n} , (\boldsymbol{E}) ),
\end{multline*}
when $\varepsilon, \epsilon \longrightarrow 0$.
This generalized operator $\Xi$ is to be interpreted as the rigorous definition of the ``chronological product'' (\ref{thetaS_n}).

Again by Thm. 4.8 of \cite{obataJFA} (resp. its fermionic version, compare Subsection \ref{psiBerezin-Hida})
this ``chronological product'' operator $\Xi$ possesees unique and finite Fock expansion
\[
\Xi = \sum\limits_{\substack{\ell,m \\ \ell+m=3n}} 
\Xi_{\ell,m}\big({\kappa''}_{\ell,m}\big)
\in \mathscr{L}( (\boldsymbol{E}) \otimes \mathscr{E} , (\boldsymbol{E})^* )
\]
into integral kernel operators $\Xi_{\ell,m}\big({\kappa''}_{\ell,m}\big)$ with $\mathscr{E}^{* \, \otimes \, n}=
\mathscr{L}(\mathscr{E}^{\otimes \, n}, \mathbb{C})$-valued kernels ${\kappa''}_{\ell,m}$. This Fock expansion
provides rigorous form of the so called Wick's theorem for the ``chronological product'' (\ref{thetaS_n}), \cite{Bogoliubov_Shirkov},
 \S 22.2, and provides at the same time the so-called normal form to the kernel $S_n(x_1, \ldots, x_n)$ of the $n$-th order contribution
$S_n$ to the scattering generalized operator. 

We shall emphasize here that the white noise analysis, in fact the theory of Fock expansions of generalized operators due to Hida, 
Obata and Sait\^o, not only gives unique kernels $S_n(x_1, \ldots, x_n)$ of the $n$-th order contributions $S_n$ to the scattering
operator but also gives the interpretation  to the $n$-th order contributions $S_n$
as particular cases of generalized operators 
\[
S_n = \Xi \in \mathscr{L}( (\boldsymbol{E}) \otimes \mathscr{E}^{\otimes \, n} , (\boldsymbol{E})^* )
\]
which, when evaluated at $\phi = \phi_1 \otimes \ldots \phi_n \in \mathscr{E}^{\otimes}$,  give
integral kernel operators
\[
\Xi(\phi) = \sum\limits_{\substack{\ell,m \\ \ell+m=3n}} 
\Xi_{\ell,m}\big({\kappa''}_{\ell,m}(\phi)\big)
\in \mathscr{L}( (\boldsymbol{E}) , (\boldsymbol{E})^* )
\]
wih scalar-valued kernels
\[
{\kappa''}_{\ell,m}(\phi) \in E_{{}_{n_1}}^{*} \otimes \ldots \otimes E_{{}_{n_{3n}}}^{*} 
\cong \mathscr{L}(E_{{}_{n_1}} \otimes \ldots \otimes E_{{}_{n_{3n}}}, \, \mathbb{C}).
\]
Because this $\Xi(\phi)$ gives the $n$-th order contribution $S_n\big(g\mathcal{L}\big)$ to the scattering operator, evaluated at $\phi
= g^{\otimes \, n}$,  then in particular we obtain the Corollary to Theorem \ref{g=1InteractingFieldsQED} of Subsection \ref{OperationsOnXi}. 
Thus we have just proven that each higher order contribution $S_n: \mathscr{E}^{\otimes \, n} \ni g^{\otimes \, n} \longmapsto S_n(g)$, 
restricted to the diagonal $\phi =\phi_1 \otimes \ldots \otimes \phi_n$, with $\phi_1 = \ldots = \phi_n =g$, defines an operator
\[
g \longmapsto S_n\big(g\mathcal{L}\big) \,\,\, \textrm{which belongs to} \,\, \mathscr{L}( (\boldsymbol{E}) \otimes \mathscr{E} , (\boldsymbol{E})^* );
\]
and thus we have proved Corollary of Thm. \ref{g=1InteractingFieldsQED} of Subsection \ref{OperationsOnXi} for the ``natural''
chronological product defined here.

Similarily, using the rigorous definition of the``chronological product'' stated above, we can give an idependent proof 
of Thm. \ref{g=1InteractingFieldsQED} of Subsection \ref{OperationsOnXi} that each higher order constribution to the interacting fields, evaluated at $g=1$, 
and based on the ``natural'' chronological product, is a generalized operator
equal to a finite sum of integral kernel operators with vector-valued kernels and belong to
\[
\mathscr{L}\big( \mathscr{E}, \mathscr{L}(\boldsymbol{E}) , (\boldsymbol{E})^* \big).
\]
Moreover, presented construction of the ``natural'' chronological product  is much more effective in comparison to the
method based on the Epstein-Glaser splitting of causally supported distributions. First of all, and contrary to the 
Epstein-Glasser splitting, it is unique. Moreover, it is computationally much more effective, because by the formula 
(\ref{epsilonChronologicalProduct=OperatorProduct}) it can be constructed in two independent steps, each being rather easily
adopted to practical computations. Namely in the first step the regularized chronological product for the 
regularized factor operators in the formula (\ref{epsilonChronologicalProduct=OperatorProduct}) can be reduced to the
compuatation of the ordinary commutation functions for the positive and negative frequency parts of the regularized factors.
Using the Rules of Subsection \ref{OperationsOnXi}, tis computation is reduced to  ordinary multiplication and symmetrization
of the ordinary functions defining the kernels of these factors.
This is is the case because the expression (\ref{epsilonChronologicalProduct=OperatorProduct}) is indeed equal to the ordinry 
product of operators. This stage of computation can be reduced to the algebraic
version of the so-called Wick theorem for the regularized chronological products of fields in which all mass-less factors are replaced by the 
massive counterparts. In the final step we pass to the
limit $\varepsilon, \epsilon \rightarrow 0$ in the scalar distributions (kernels) obtained in the first step. Both steps are essentially 
easer to be practically managed in comparision to the splitting problem for causally supported tempered distributions involved into the so-called
quasiasymtotic analysis, espesially for the higher order terms.  

The ``natural'' chronological product defined here is in agreement with the axioms (I)-(IV)
of Subsection \ref{MotivationForHida}.

However in the  last Theorem of Subsection \ref{OperationsOnXi} we have presented analysis of the higher order contributions
to interacting fields obtained independetly of the defintion of the ``chronological product'' given here, and instead construction 
of  the chronological product was based there on the causal splitting due to Epstein-Glaser together with the causality, symmetricity,
translational covariance, and unitarity (Krein isometricity for fields involving gauge freedom).
We did so because the discovery of Epstein-Glaser, that the splitting of causal distributions together with the axims
(I)-(IV) of Subsection \ref{MotivationForHida}, determines all higher order constributions
$S_n$ to the scattering operator presents an important contribution of its own value. First of all it shows that the aximos
(I)-(IV) cannot determine $S_n$ with the extend of uniqueness less than that involved in the splitting of causal distributions
into the advanced and retarded parts. An extra information is needed which cuts out the freedom in the particular choice of the splitting. 
In particular essential part of the freedom pertinent to the splitting is reduced by the equations of motion for the interacting fields, 
as indicated by the results of \cite{DKS1}. Therefore the fact that we have at our disposal a ``natural'' definition
of the ``chronological product'', which moreover avoids the rather laborious splitting analysis, seems to be important.

\subsection{Comparizon with the standard realization of the free Dirac field $\boldsymbol{\psi}$.
Bogoliubov-Shirkov quantization postulate}\label{StandardDiracPsiField}

In our formula  (\ref{psi(x)}) for the free Dirac field $\boldsymbol{\psi}(x)$:
\begin{multline}\label{psi(x)'}
\boldsymbol{\psi}(x)  =
\sum_{s=1}^{2} \, \int \limits_{\mathbb{R}^3} 
\frac{1}{2|p_0(\boldsymbol{\p})|}u_{s}(\boldsymbol{\p})e^{-ip\cdot x} \,\, b_{s}(\boldsymbol{\p}) \, 
\ud^3 \boldsymbol{\p} \\
+
\sum_{s=1}^{2} \, \int \limits_{\mathbb{R}^3} 
\frac{1}{2|p_0(\boldsymbol{\p})|}v_{s}(\boldsymbol{\p})e^{ip\cdot x} \,\, d_{s}(\boldsymbol{\p})^+ \, 
\ud^3 \boldsymbol{\p}.
\end{multline}
we have an additional weight $|2p_0(\boldsymbol{\p})|^{-1}$ in comparizon to the standard formula which can be 
found e.g. in \cite{Scharf} or \cite{Bogoliubov_Shirkov}, as well as in the classic works of Dirac. 
Of course this weight may be absorbed to the corresponding solutions $u_{s}(\boldsymbol{\p}), v_{s}(-\boldsymbol{\p})$,
$s=1,2$, constructed as in Appendix \ref{fundamental,u,v}. But this redefinition of 
$u_{s}(\boldsymbol{\p}), v_{s}(-\boldsymbol{\p})$ would have changed the orthonormality conditions
(\ref{u^+u=delta}) into the following conditions
\begin{equation}\label{u^+u=1/p-0^2delta}
\begin{split}
u_s(\boldsymbol{\p})^+ u_{s'}(\boldsymbol{\p}) = \frac{1}{(2|p_0(\boldsymbol{\p})|)^2} \, \delta_{ss'}, \,\,\,
v_s(\boldsymbol{\p})^+ v_{s'}(\boldsymbol{\p}) = \frac{1}{(2|p_0(\boldsymbol{\p})|)^2} \, \delta_{ss'}, \\
u_s(\boldsymbol{\p})^+ v_{s'}(-\boldsymbol{\p}) = 0.
\end{split}
\end{equation}
But because the same standard orthonormalization conditions (\ref{u^+u=delta}) are also assumed in 
\cite{Scharf}, pp. 38-41 (even exatly the same $u_{s}(\boldsymbol{\p}), v_{s}(-\boldsymbol{\p})$
are used there as we do for the standard representation of Dirac gamma matrices,
compare Appendix \ref{fundamental,u,v}), 
and the same we have in \cite{Bogoliubov_Shirkov}, formula (7.16) p. 67, 
(and the same is assumed in the classic works of the very founders of QED)
we see that the difference between our formula (\ref{psi(x)'}) and the standard formula: 
\begin{equation}\label{standardpsi(x)}
\boldsymbol{\psi}(x)  =
\sum_{s=1}^{2} \, \int \limits_{\mathbb{R}^3} 
u_{s}(\boldsymbol{\p})e^{-ip\cdot x} \,\, b_{s}(\boldsymbol{\p}) \, 
\ud^3 \boldsymbol{\p}
+
\sum_{s=1}^{2} \, \int \limits_{\mathbb{R}^3} 
v_{s}(\boldsymbol{\p})e^{ip\cdot x} \,\, d_{s}(\boldsymbol{\p})^+ \, 
\ud^3 \boldsymbol{\p}.
\end{equation}
of \cite{Bogoliubov_Shirkov} or \cite{Scharf},
cannot be explained by any redefinition of $u_{s}(\boldsymbol{\p}), v_{s}(-\boldsymbol{\p})$. 

Nonetheless the standard qunatum Dirac field $\boldsymbol{\psi}$ given by (\ref{standardpsi(x)}),
is unitarily isomorphic to the Dirac field $\boldsymbol{\psi}$ given by (\ref{psi(x)'}). 
Indeed the unitary equivalence between 
our $\boldsymbol{\psi}$ and (\ref{standardpsi(x)}) is realized by the lifting to the Fock space of the unitary
operator $\mathbb{U}$, and its inverse $\mathbb{U}^{-1}$, of point-wise multiplication by the function 
$\boldsymbol{\p} \mapsto |2p_0(\boldsymbol{\p})|^{-1}$ and respectively 
$\boldsymbol{\p} \mapsto |2p_0(\boldsymbol{\p})|$ regarded as unitary operators on the respecive single particle Hilbert spaces of the realizations of the field  $\boldsymbol{\psi}$: first is the space
$\mathcal{H}'= \mathcal{H}_{m,0}^{\oplus} \oplus \mathcal{H}_{-m,0}^{\ominus c}$ used by us and the secod 
$\mathbb{U}\mathcal{H}'$ is almost identical with ours, the only change is that we are using the ordinary measure 
$\ud^{3} \boldsymbol{\p}$ on the orbits 
$\mathscr{O}_{m,0,0,0}$, 
$\mathscr{O}_{-m,0,0,0}$ instead of $\frac{\ud^{3} \boldsymbol{\p}}{|2p_0(\boldsymbol{\p})|^{2}}$, 
in constructing Hilbert spaces of bispinors whose Fourier transforms 
are concentrated respectively on $\mathscr{O}_{m,0,0,0}$, $\mathscr{O}_{-m,0,0,0}$
and are component-wise square summable with respect to $\ud^{3}\boldsymbol{\p}$.
Therefore the corresponding function
$|2p_0(\boldsymbol{\p})|^{-1}$ is just equal to the square root of the Radon-Nikodym derivation of the measure 
$\frac{\ud^{3} \boldsymbol{\p}}{|2p_0(\boldsymbol{\p})|^{2}}$ on the orbits
$\mathscr{O}_{m,0,0,0}$, $\mathscr{O}_{-m,0,0,0}$ used by us (compare Subsection 2.1 of \cite{wawrzycki2018}) 
with respect to the knew one $\ud^{3} \boldsymbol{\p}$. 
Under this redefinition of measure on the orbits the formulas for 
$u_{s}(\boldsymbol{\p}), v_{s}(-\boldsymbol{\p})$ remain unchanged, similarily as the formulas for the 
projectors $P^\oplus, P^\oplus(p), P^\ominus, P^\ominus(p), E_\pm, E_\pm(p)$ (compare Appendix \ref{fundamental,u,v})
remain unchanged. The nuclear space $E$ in the corresponding Gelfand triples (\ref{SinglePartGelfandTriplesForPsi})
will remain unchanged with the single particle Hilber space $\mathcal{H}'$ replaced of course
by  $\mathbb{U}\mathcal{H}'$. The formula (\ref{isomorphismU}) for the unitary isomorphism $U$ jouning the 
Gelfand triple $E \subset \mathbb{U}\mathcal{H}' \subset E^*$ with the standard Gelfand triple
$\mathcal{S}_{A}(\mathbb{R}^3; \mathbb{C}^4) \subset L^2(\mathbb{R}^3; \mathbb{C}^4) \subset 
\mathcal{S}_{A}(\mathbb{R}^3; \mathbb{C}^4)^*$ will remain almost the same with the only difference that the additional
factor $1/|2p_0(\boldsymbol{\p})|$ will be absent in it, and accordingly the factor $2|p_0(\boldsymbol{\p})|$
will be absent in the formula for $U^{-1}$. It is readily seen now that the construction of Subsection  
\ref{psiBerezin-Hida},
with the mentionaed modification of the measure, will indeed produce the standard formula 
(\ref{standardpsi(x)}) for the Dirac field.

Note that the unitary operators $\mathbb{U}$, and $\Gamma(\mathbb{U})$, are well defined as unitary isomorphisms for fields
understood as integral kernel operators with vector-nalued kernels, because the operator 
$\mathbb{U}$ of multiplication by the function
$\boldsymbol{\p} \mapsto |2p_0(\boldsymbol{\p})|^{-1}$  transforms
$\mathcal{S}(\mathbb{R}^3; \mathbb{C})$ continously, and even isomorphically, into itself
and induces the isomorphism of the Gelfand triples
\[
\left. \begin{array}{ccccc}   & & \mathcal{H}_{m,0}^{\oplus} \oplus \mathcal{H}_{-m,0}^{\ominus c} & & \\
 & & \parallel & & \\
           E        & \subset & \mathcal{H}' & \subset & E^*        \\
                               \downarrow \uparrow &         & \mathbb{U}\downarrow \uparrow \mathbb{U}^{-1}      &         & \downarrow \uparrow  \\
                                         E   & \subset &  \mathcal{H}'' = \mathbb{U}\mathcal{H}' & \subset & E^*        \\ 
\end{array}\right..
\]
Let us denote the standard annihilation and creation operators 
over the Fock space $\Gamma(\mathbb{U}\mathcal{H}')$ by $a''(u \oplus v), a''(u \oplus v)^+$. They are constructed 
exactly as the operators $a'(u\oplus v), a'(u\oplus v)^+$
in Subsections \ref{electron}-\ref{electron+positron} with the only change that the weight 
$1/|2p_0(\boldsymbol{\p})|^2$ in the inner products will be absent, and analogousuly we extend them
over to $u\oplus v \in E^*$ using the corresponding isomorphism
\[
\left. \begin{array}{ccccc}   & & L^2(\sqcup \mathbb{R}^3; \mathbb{C}) & & \\
 & & \parallel & & \\
           \mathcal{S}_{A}(\mathbb{R}^3; \mathbb{C}^4)        & \subset & L^2(\mathbb{R}^3; \mathbb{C}^4) & \subset & \mathcal{S}_{A}(\mathbb{R}^3; \mathbb{C}^4)^*        \\
                               \downarrow \uparrow &         & U \downarrow \uparrow U^{-1}      &         & \downarrow \uparrow  \\
                                         E   & \subset &  \mathcal{H}'' = \mathbb{U}\mathcal{H}' & \subset & E^*        \\ 
\end{array}\right., 
\]
of the triple $E \subset \mathbb{U}\mathcal{H}'  \subset  E^*$ with the standard Gelfand triple,
and with $U,U^{-1}$ given by the formula (\ref{isomorphismU}) with the factors 
$1/|2p_0(\boldsymbol{\p})|$ (resp. $2|p_0(\boldsymbol{\p})$) removed. 
Then if $\boldsymbol{\psi}$ is the standard Dirac field (\ref{standardpsi(x)}) we have
\begin{equation}\label{standardpsi(x)''}
\boldsymbol{\psi}(f) = a''\big(P^\oplus\widetilde{f}|_{{}_{\mathscr{O}_{m,0,0,0}}} \oplus 0\big) + 
a''\Big( 0 \oplus \big(P^\ominus\widetilde{f}|_{{}_{\mathscr{O}_{-m,0,0,0}}}\big)^c \Big)^+,
\,\,\, f \in \mathcal{S}(\mathbb{R}^4; \mathbb{C}^4)
\end{equation}
correspondingly to the formula
\begin{equation}\label{psi(x)''}
\boldsymbol{\psi}(f) = a'\big(P^\oplus\widetilde{f}|_{{}_{\mathscr{O}_{m,0,0,0}}} \oplus 0\big) + 
a'\Big( 0 \oplus \big(P^\ominus\widetilde{f}|_{{}_{\mathscr{O}_{-m,0,0,0}}}\big)^c \Big)^+,
\,\,\, f \in \mathcal{S}(\mathbb{R}^4; \mathbb{C}^4)
\end{equation}
for the free Dirac field (\ref{psi(x)'}) constructed in Subsection \ref{psiBerezin-Hida}, and with the 
following isomorphism
\begin{multline}\label{G(bbU)^+a(bbU(u+v))G(bbU)=a'(u+v)}
 a'\big(\mathbb{U}^{+}(u\oplus v)\big) =  a''(u\oplus v), \\
 a'\big(\mathbb{U}^{+}(u\oplus v)\big)^+ =  a''(u\oplus v)^+, \\
u\oplus v \in E^*,
\end{multline} 
\begin{multline}\label{G(bbU)^+a(bbU(u+v))G(bbU)=a'(u+v)'}
 a'\big(\mathbb{U}^{-1}(u\oplus v)\big) =  a''(u\oplus v), \\
a'\big(\mathbb{U}^{-1}(u\oplus v)\big)^+ =  a''(u\oplus v)^+, \\
u\oplus v \in E \subset E^*.
\end{multline} 
joining the Hida operators $a'(u\oplus v)$ and $a''(u\oplus v)$.

Of course the plane waves defining the vector-valued distributional kernels
$\kappa_{0,1}, \kappa_{1,0}$ defining the standard Dirac field (\ref{standardpsi(x)})
as integral kernel operator
\[
\boldsymbol{\psi} = \Xi_{0,1}(\kappa_{0,1}) + \Xi_{1,0}(\kappa_{1,0})
\]
are equal
\begin{equation}\label{skappa_0,1}
\boxed{
\kappa_{0,1}(s, \boldsymbol{\p}; a,x) = \left\{ \begin{array}{ll}
u_{s}^{a}(\boldsymbol{\p})e^{-ip\cdot x} \,\,\, \textrm{with $p = (|p_0(\boldsymbol{\p})|, \boldsymbol{\p}) \in \mathscr{O}_{m,0,0,0}$} & \textrm{if $s=1,2$}
\\
0 & \textrm{if $s=3,4$}
\end{array} \right.,
}
\end{equation}
\begin{equation}\label{skappa_1,0}
\boxed{
\kappa_{1,0}(s, \boldsymbol{\p}; a,x) = \left\{ \begin{array}{ll}
0 & \textrm{if $s=1,2$}
\\
v_{s-2}^{a}(\boldsymbol{\p})e^{ip\cdot x} \,\,\, \textrm{with $p = (|p_0(\boldsymbol{\p})|, \boldsymbol{\p}) \in \mathscr{O}_{m,0,0,0}$} & \textrm{if $s=3,4$}
\end{array} \right.
}
\end{equation}

We claim that if the orthonormality conditions (\ref{u^+u=delta}) for 
$u_{s}(\boldsymbol{\p}), v_{s}(-\boldsymbol{\p})$,
$s=1,2$ (compare Appendix \ref{fundamental,u,v}) are to be preserved, then it is the formula
(\ref{psi(x)'}) for the free Dirac field $\boldsymbol{\psi}(x)$ which defines the 
Dirac field with the local and unitary transformation formula, as an immediate consequence of the locality of the 
transformation law (26) and (27) of Subsect. 2.1 of \cite{wawrzycki2018}. The locality of 
(26) and (27) of \cite{wawrzycki2018} is in turn an immediate consequence of the fact that there are no momentum 
dependent multipliers in the 
transformation law (24) and (25) of Subsect. 2.1 of \cite{wawrzycki2018}, acting on the Fourier transforms of 
bispinors concetrated respectively on $\mathscr{O}_{m,0,0,0}$ (elemets of $\mathcal{H}_{m,0}^{\oplus}$)
or on $\mathscr{O}_{-m,0,0,0}$ (elements of $\mathcal{H}_{-m,0}^{\ominus}$). 

Namely recall that that the representation $U(a,\alpha)$ of $(a,\alpha) \in T_4 \circledS SL(2, \mathbb{C})$ 
acts on the Fourier tramsform $\widetilde{\phi} \in \mathcal{H}_{m,0}^{\oplus}$ (concentrated on $\mathscr{O}_{m,0,0,0}$) of bispinor $\phi$ through the formulas (24) and (25) of \cite{wawrzycki2018},
and on $\phi$ through (26) and (27) of \cite{wawrzycki2018}. Similarily 
$U'(a,\alpha)^c$ act on $(\widetilde{\phi}')^c \in \mathcal{H}_{-m,0}^{\ominus c}$ by the conjugation
of the representation $U'(a,\alpha)$ acting on the bispinor $\widetilde{\phi}' \in \mathcal{H}_{-m,0}^{\ominus}$
by the same formula (24) and (25) of \cite{wawrzycki2018} and on $\phi'$
through the formula (26) and (27) of \cite{wawrzycki2018}, Subsect. 2.1. On writting
$\boldsymbol{U}(a,\alpha) = U(a,\alpha) \oplus U'(a,\alpha)^c$ for the representation of 
$(a,\alpha) \in T_4 \circledS SL(2, \mathbb{C})$ acting in the single particle Hilbert space 
$\mathcal{H}_{m,0}^{\oplus} \oplus \mathcal{H}_{-m,0}^{\ominus c}$ of
the field (\ref{psi(x)'}), we have
\begin{equation}\label{transformationGeneralLocalpsi}
\Gamma(\boldsymbol{U}(a,\alpha)) \boldsymbol{\psi}(f) \Gamma(\boldsymbol{U}(a,\alpha))^{-1}
= \boldsymbol{\psi}\big(U(a,\alpha)f\big)
\end{equation}
where $U(a, \alpha)$ acts on $f \in \mathcal{S}(\mathbb{R}^4; \mathbb{C}^4)$ and gives $U(a, \alpha)f$
in the same fashion as in  (26) and (27) of \cite{wawrzycki2018}. In particular\footnote{Recall that
here $\Lambda: \alpha \rightarrow \Lambda(\alpha)$ is an antihomomorphism.}
\begin{equation}\label{Ualphaxf}
U(\alpha) f(x) = 
\left( \begin{array}{cc}  \alpha & 0  \\
                                           
                                                   0              & {\alpha^*}^{-1}  \end{array}\right) 
 f(x\Lambda(\alpha^{-1}))
= \left( \begin{array}{cc}  \alpha & 0  \\
                                           
                                                   0              & {\alpha^*}^{-1}  \end{array}\right) 
 f(\Lambda(\alpha)x), 
\end{equation}
\begin{equation}\label{Uaxf}
T(a) f(x) = f(x - a).
\end{equation}
In particular the field (\ref{psi(x)'}) transforms locally, and in particular translations act on (\ref{psi(x)'})
in the standard fashion
\begin{equation}\label{transformationTranslationGeneralLocalpsi}
\Gamma(\boldsymbol{U}(a,0)) \boldsymbol{\psi}(f) \Gamma(\boldsymbol{U}(a,0))^{-1}
= \boldsymbol{\psi}\big(U(a,0)f\big) = \boldsymbol{\psi}\big(T(a) f\big) 
\end{equation}

It is easily seen that the operator of multiplication by the function 
$\boldsymbol{\p} \mapsto |p_0(\boldsymbol{\p})|^{-1}$  in action on 
$\mathcal{H}_{m,0}^{\oplus}$ and on $\mathcal{H}_{-m,0}^{\ominus}$ (compare Subsct. 2.1 of \cite{wawrzycki2018})
commutes with the translation operator (25) of Subsect. 2.1 of \cite{wawrzycki2018} and with the operators 
(24) of Subsect. 2.1 of \cite{wawrzycki2018} reperesenting spatial rotations (because $|p_0(\boldsymbol{\p})|
= \sqrt{|\boldsymbol{\p}|^2 + m^2}$ is invariant under rotations). Therefore both the free Dirac fields:
ours (\ref{psi(x)}) and the standard one (\ref{standardpsi(x)}), transform locally and identically under 
translations and spatial rotations. Namely for $(a,\alpha) = (a,0) \in T_4 \circledS SL(2, \mathbb{C})$
or for  $(a,\alpha) = (0, \alpha) \in T_4 \circledS SU(2, \mathbb{C}) \subset T_4 \circledS SL(2, \mathbb{C})$  
i.e. for translations or spatial rotations, we have
\[
\Gamma\big(\mathbb{U}\boldsymbol{U}(a,\alpha)\mathbb{U}^{-1}\big) \boldsymbol{\psi}(f) 
\Gamma\big(\mathbb{U}\boldsymbol{U}(a,0) \mathbb{U}^{-1}\big)^{-1}
= \boldsymbol{\psi}\big(U(a,\alpha)f\big) 
\]
with the standard local formula for the transformation formula (\ref{Ualphaxf}), (\ref{Uaxf})
for space-time transformed  bispinor $U(a,\alpha)f$, and for the standard Dirac quantum field (\ref{standardpsi(x)}) with the representation 
\[
\Gamma\big(\mathbb{U}\boldsymbol{U}(a,\alpha)\mathbb{U}^{-1}\big)
\]
acting in its Fock space
\[
\Gamma\big(\mathbb{U}\big) \big( \mathcal{H}_{m,0}^{\oplus} \oplus \mathcal{H}_{-m,0}^{\ominus c}\big)
= \Gamma\big(\mathbb{U}(\mathcal{H}_{m,0}^{\oplus} \oplus \mathcal{H}_{-m,0}^{\ominus c}) \big),
\]
 and with the representation
\[
\mathbb{U}\boldsymbol{U}(a,\alpha)\mathbb{U}^{-1}
\]
acting in its single particle Hilbert space
\[
\mathcal{H}'' = \mathbb{U}\big(\mathcal{H}_{m,0}^{\oplus} \oplus \mathcal{H}_{-m,0}^{\ominus c}\big)
= \mathbb{U} \mathcal{H}'.
\]

Note that for the bispinor $\underset{\circ}{\widetilde{\phi}} = \mathbb{U} \widetilde{\phi}$,
$\widetilde{\phi} \in \mathcal{H}_{m,0}^{\oplus}$, such that  
$\underset{\circ}{\widetilde{\phi}} \oplus 0 \in \mathcal{H}''$, concetrated on
$\mathscr{O}_{m,,0,0,0}$, or $0 \oplus {\underset{\circ}{\widetilde{\phi}}}^c  \in \mathcal{H}''$, 
$\underset{\circ}{\widetilde{\phi}} = \mathbb{U} \widetilde{\phi}$,
$\widetilde{\phi} \in \mathcal{H}_{-m,0}^{\ominus}$,
concentrated on $\mathscr{O}_{-m,0,0,0}$, 
we have
\[
\mathbb{U} U(\alpha)\mathbb{U}^{-1} \underset{\circ}{\widetilde{\phi}}(p) = \Bigg|\frac{p_0(\Lambda(\alpha)p)}{p_0(p)}\Bigg|
\left( \begin{array}{cc}  \alpha & 0  \\
                                           
                                                   0              & {\alpha^*}^{-1}  \end{array}\right) 
 \underset{\circ}{\widetilde{\phi}}(\Lambda(\alpha)p), 
\]
\[
\mathbb{U} T(a) \mathbb{U}^{-1} \underset{\circ}{\widetilde{\phi}}(p) 
= e^{i a \cdot p}\underset{\circ}{\widetilde{\phi}}(p).
\]

Therefore for the Lorentz transformations (24) of Subsect. 2.1 of \cite{wawrzycki2018} situation is
 different for the two mentioned 
realizations of the Dirac free field. Namely our field (\ref{psi(x)'}) by construction transforms locally as 
a bispinor field also under Lorentz transformations. But the operator $\mathbb{U}$ 
of point-wise multiplication by the function $\boldsymbol{\p} \mapsto |p_0(\boldsymbol{\p})|^{-1}$ does not commute with 
the operator $U(\alpha)$ for $\alpha \notin SU(2, \mathbb{C})$ given by (24) of \cite{wawrzycki2018}, 
and moreover it is immediately seen 
that transformation formula $\mathbb{U} U(\alpha) \mathbb{U}^{-1}$ gains non-trivial momentum dependend
multiplier 
\[
|p_0(\Lambda(\alpha)p)/p_0(p)| \neq 1
\]
for $\alpha \notin SU(2, \mathbb{C})$.
This additional multiplier means that $\mathbb{U}\boldsymbol{U}(a,\alpha)\mathbb{U}^{-1}$ in action on 
the elements of $\mathcal{H}''$, viewed as distributional Fourier 
transforms of positive (respectively conjugations of negative) energy solutions 
$\mathscr{F}^{-1}\widetilde{\underset{\circ}{\phi}}$ of Dirac equation, 
concentrated respectively on $\mathscr{O}_{m,0,0,0,0}$ or $\mathscr{O}_{-m,0,0,0}$, 
induce nonlocal transformation law on $\mathscr{F}^{-1}\widetilde{\underset{\circ}{\phi}}$. Aternatively
this additional multiplier, however, can be viewed as coming from the non-invariance of the ordinary euclidean measure 
$\ud^3 \boldsymbol{\p}$ under Lorentz transformation on the respective orbits $\mathscr{O}_{m,0,0,0}$ and 
$\mathscr{O}_{-m,0,0,0}$, which assures locality of Lorentz transformations not for the ordinary inverse Fourier transformed elements of 
$\mathcal{H}''$  but for the inverse Fourier transform of the elements 
$\mathbb{U}^{-1}\underset{\circ}{\widetilde{\phi}}$, $\underset{\circ}{\widetilde{\phi}} \in \mathcal{H}''$.
Namely consider the following formula 
\begin{multline*}
\phi(x) = \int \limits_{\mathscr{O}_{m,0,0,0}} \widetilde{\phi}(p) e^{-ip \cdot x} \, \ud \mu_{{}_{\mathscr{O}_{m,0,0,0}}}(p)
= \int \limits_{\mathbb{R}^3} \frac{\widetilde{\phi}(\boldsymbol{\p}, p_0(\boldsymbol{\p}))}{p_0(\boldsymbol{\p})} 
e^{-ip \cdot x} \, \ud^3 \boldsymbol{\p} \\
= \int \limits_{\mathbb{R}^3} \mathbb{U}\widetilde{\phi}(\boldsymbol{\p}) e^{-ip \cdot x}
 \, \ud^3 \boldsymbol{\p} 
= \int \limits_{\mathbb{R}^3} \underset{\circ}{\widetilde{\phi}}(\boldsymbol{\p}) e^{-ip \cdot x}
 \, \ud^3 \boldsymbol{\p},
\end{multline*}   
for the positive energy solutions. We have analogue formula for negative energy solutions.
Consider now the local transformation formula for $U(\alpha)\phi$ with $\phi$ expressed by the above formula.
We will get
\begin{multline*}
U(\alpha)\phi(x) = 
\left( \begin{array}{cc}  \alpha & 0  \\  
                                                   0              & {\alpha^*}^{-1}  \end{array}\right) 
\phi(\Lambda(\alpha)x) \\
=\left( \begin{array}{cc}  \alpha & 0  \\                                          
                                                   0              & {\alpha^*}^{-1}  \end{array}\right) 
\int \limits_{\mathbb{R}^3} \underset{\circ}{\widetilde{\phi}}(\boldsymbol{\p}) e^{-ip \cdot \Lambda x}
 \, \ud^3 \boldsymbol{\p} \\ =
\left( \begin{array}{cc}  \alpha & 0  \\                                           
                                                   0              & {\alpha^*}^{-1}  \end{array}\right) 
\int \limits_{\mathbb{R}^3} \underset{\circ}{\widetilde{\phi}}(\Lambda \boldsymbol{\p}) e^{-ip \cdot x}
 \, \ud^3 \Lambda \boldsymbol{\p} \\ =
\left( \begin{array}{cc}  \alpha & 0  \\                                           
                                                   0              & {\alpha^*}^{-1}  \end{array}\right) 
\int \limits_{\mathbb{R}^3} \underset{\circ}{\widetilde{\phi}}(\Lambda \boldsymbol{\p}) e^{-ip \cdot x}
 \, \Bigg|\frac{\ud^3 \Lambda \boldsymbol{\p}}{\ud^3 \boldsymbol{\p}} \Bigg| \, \ud^3 \boldsymbol{\p}.
\end{multline*}
Taking into account the invariance property
\[
\frac{\ud^3 \Lambda \boldsymbol{\p}}{|p_0(\Lambda \boldsymbol{\p})|} =
\frac{\ud^3 \boldsymbol{\p}}{|p_0(\boldsymbol{\p})|} \,\,\, \Longleftrightarrow \,\,\,
\Bigg| \frac{\ud^3 \Lambda \boldsymbol{\p}}{\ud^3 \boldsymbol{\p}} \Bigg| = 
\frac{|p_0(\Lambda \boldsymbol{\p})|}{|p_0(\boldsymbol{\p})|},
\]
we obtain 
\[
U(\alpha)\phi(x)  =
\left( \begin{array}{cc}  \alpha & 0  \\                                           
                                                   0              & {\alpha^*}^{-1}  \end{array}\right) 
\int \limits_{\mathbb{R}^3} \underset{\circ}{\widetilde{\phi}}(\Lambda \boldsymbol{\p}) e^{-ip \cdot x}
 \, \frac{|p_0(\Lambda \boldsymbol{\p})|}{|p_0(\boldsymbol{\p})|} \, \ud^3 \boldsymbol{\p},
\,\,\,\,\, p \in \mathcal{O}_{m,0,0,0},
\]
\emph{i.e.} again the assertion that the transformation $\mathbb{U} U(\alpha)\mathbb{U}^{-1} \underset{\circ}{\widetilde{\phi}}$ of $\underset{\circ}{\widetilde{\phi}} = \mathbb{U}\widetilde{\phi}$ is accompanied by the ordinary local bispinor transformation $U(\alpha) \phi$ of $\phi$, but not of $\mathscr{F}^{-1}\widetilde{\underset{\circ}{\phi}}$. Similar relation we obtain for the conjugations of the negative energy solutions whose Fourier transforms are concentrated on 
$\mathscr{O}_{-m,0,0,0}$. Therefore if 
$f \in \mathcal{S}(\mathbb{R}^4; \mathbb{C}^4)$
is a space-time test bispinor, then the transformation $\mathbb{U} U(\alpha)\mathbb{U}^{-1}$ (or its conjugation)
in action on 
\[
P^\oplus \mathbb{U}\widetilde{f}|_{{}_{\mathcal{O}_{m,0,0,0}}} \,\,\, \textrm{or resp.} \,\,\,
\big(P^\ominus \mathbb{U}\widetilde{f}|_{{}_{\mathcal{O}_{-m,0,0,0}}}\big)^c
\]
induces local bispinor transformation on $f$. This would be false for the action 
of $\mathbb{U} U(\alpha)\mathbb{U}^{-1}$ (or its conjugation) on 
\[
P^\oplus \widetilde{f}|_{{}_{\mathcal{O}_{m,0,0,0}}} \,\,\, \textrm{or resp.} \,\,\,
\big(P^\ominus \widetilde{f}|_{{}_{\mathcal{O}_{-m,0,0,0}}}\big)^c.
\]
Thus we see again that it is the field (\ref{psi(x)'}), or equivalently the field 
(\ref{psi(x)''}), which transforms locally as ordinary bispinor under the Fock lifting of
$U(\alpha)$ (summed up with its conjugation). The field (\ref{standardpsi(x)}), or equivalently the field
(\ref{standardpsi(x)''}), transforms non-locally under the Fock lifting
of the unitary representation $\mathbb{U} U(\alpha)\mathbb{U}^{-1}$ (summed up with its conjugation). 
 Correspondingly the standard Dirac quantum field
(\ref{standardpsi(x)}) transforms non-locally under Lorentz transformations if the unitarity of the transformation 
is to be preserved. 
Locality under proper Lorentz transformations of the standard field (\ref{standardpsi(x)})
can be restored, but then the unitarity of the Lorentz transformations will have to be abandoned.
Below in this Subsection we explain this fact together with its connection to the so called
Noether theorem for free fields.

Although the Dirac free fields (\ref{psi(x)'}) and (\ref{standardpsi(x)}) are unitarily isomorphic,
in the sense of the isomorphism (\ref{G(bbU)^+a(bbU(u+v))G(bbU)=a'(u+v)}) or (\ref{G(bbU)^+a(bbU(u+v))G(bbU)=a'(u+v)'}),
joining the corresponding Hida operators $a', a''$, 
there are some important differences between them. 

The first concerns locality
 under the proper Lorentz transformations, already explained. The field (\ref{psi(x)'}) is constructed from the 
direct sum of two (equivalent) irreducible represenations, giving the local transformation law for the elements 
of the single particle Hilbert 
space regarded as the space of (regular distributional) solutions of the Dirac equation, whose Fourier transforms
compose $\mathcal{H}'$ and are concetrated on the orbit $\mathscr{O}_{m,0,0,0}$ or eventually are equal to conjugations 
of bispinors concetrated on the orbit $\mathscr{O}_{-m,0,0,0}$. The standard field (\ref{standardpsi(x)})
is constructed from the slightly different representation, but unitary equivalent with it, which 
assures the local transformation law of the elements of the single particle space, uderstood as solutions of the Dirac equation, but only under the translation subgroup or spatial rotations.
It is a general paradigm that the locality of the transformation under the full $T_4 \circledS SL(2, \mathbb{C})$ is 
the fundamental assumption, and whenever we are able to construct a free field out of a representation
of $T_4 \circledS SL(2, \mathbb{C})$ it is customary to put the additional requirement of locality of the transformation
law induced by the representation. But it turns out that, at least in the realm of causal perturbatve approach to QFT, 
that it is the covariance under translations (with the standard local transformation formula) which plays the important role in the construction of the  causal perturbative series,
e.g. for interacting fields. The local Lorentz covariance and its unitarity turns out to be 
optional (which is of course a nontrivial fact).
Moreover it is known that also for determination of the commutation rules for free fields according to the classic
procedure due to Pauli-Bogoliubov-Shirkov, it is the the so-called Noether theorem for translations which is 
sufficient in derivation of these rules (compare \cite{Bogoliubov_Shirkov}, where it is understood as an example of the 
Bohr's \emph{correspondence principle}).
Therefore at least from the causal perturbative approach,  both 
(\ref{psi(x)'}) and (\ref{standardpsi(x)}) are equally well. 

Although (\ref{psi(x)}) and (\ref{standardpsi(x)}) are unitarily isomorphic, they
have \emph{different} ``commutation generalized functions'' as well as \emph{different} ``pairing functions'', 
which enter the causal perturbative series 
accordingly to different anti-commutation rules
\[
\begin{split}
\big\{a'(u \oplus v), a'(u'\oplus v')^+ \big\} = \big(u \oplus v, \, u' \oplus v' \big)_{{}_{\mathcal{H}'}}, \,\,\,\,
\,\,\,
u \oplus v \in E, \\
\big\{a''(u \oplus v), a''(u'\oplus v')^+ \big\} = \big(u \oplus v, \, u' \oplus v' \big)_{{}_{\mathbb{U}\mathcal{H}'}}, 
\,\,\,
u \oplus v \in E
\end{split}
\]
with different inner products: with the additional weight $|2p_0(\boldsymbol{\p})|^{-2}$
in the formula for $\big(\cdot, \cdot \big)_{{}_{\mathcal{H}'}}$ in comparison to 
$\big( \cdot, \cdot \big)_{{}_{\mathbb{U}\mathcal{H}'}}$, where the weight $|2p_0(\boldsymbol{\p})|^{-2}$
is absent. Because of the isomorphism between the Hida operators $a',a''$ defining respectively the fields
(\ref{psi(x)'}) and (\ref{standardpsi(x)}) we expect that both these fields should be physically equivalent,
in giving the same physical quantities, although it is still non trivial (nontriviality follows e.g. by the 
difference in commutation and pairing functions contributing to the perturbative series). At the present stage of the theory we should be carefull and keep in mind both possibilities (\ref{psi(x)'}) and (\ref{standardpsi(x)})
for the free Dirac field. 

That locality and unitarity under Lorentz transformations cannot be reconciled for the standard Dirac 
field (\ref{standardpsi(x)})
has so far been unnoticed, because of the rather heuristic approach in its construction, 
which either does not enter the theory of representations of $T_4 \circledS SL(2, \mathbb{C})$
at all or recalls to it, but in a rather disrespectful manner.
The lack of the adequate group theoretical construction of the Dirac field has been noted e.g. by Haag
\cite{Haag}, p. 48.

But there is also another difference between (\ref{psi(x)'}) and (\ref{standardpsi(x)}), which can be invariantly
expressed by recalling to the first Noether theorem applied to the free quantum fields. We devote the rest
part of this Subsection to the Noether theorem restricted to translations and Lorentz transformations and its 
relation to the fields
(\ref{psi(x)'}) and (\ref{standardpsi(x)}).

Let us recall the Noether theorem for free fields after \cite{Bogoliubov_Shirkov}, 
Chap. 2, \S 9.4 (in 1980 Ed.), where it is called 
the \emph{Quantization Postulate}: 

\emph{The operators for the energy-momentum four-vector
$\boldsymbol{P}$, and the angular momentum tensor $\boldsymbol{M}$, the charge
$\boldsymbol{Q}$, and so on, which are the generators of the corresponding 
symmetry transformations of state vectors, can be expressed in terms of the operator functions 
of the fields by the same relations as in classical field theory with the operators
arranged in the normal order}.
 
Let us start our analysis with translations.

Here we confine our attention to the Dirac field $\boldsymbol{\psi}$ given by 
(\ref{standardpsi(x)}) (and respectively (\ref{psi(x)'})).
Let $T^{0 \mu}$ be the $0-\mu$-components of the energy-momentum tensor for the free ``classic'' Dirac field
$\psi$ corresponding to translations via Emmy Noether theorem (compare \cite{Bogoliubov_Shirkov})
expressed in terms of $\psi(x)$ and of its derivatives $\partial_\nu \psi(x)$. According to this theorem the spatial 
integral
\[
\int T^{0\mu} \, \ud^3 \boldsymbol{\x}= \frac{i}{2}  \int  \Bigg(
\overline{\psi}(x)\gamma^0 \frac{\partial \psi}{\partial x_\mu}(x) 
- \frac{\partial \overline{\psi}}{\partial x_\mu}(x) \gamma^0 \psi(x)  \Bigg) \, \ud^3 \boldsymbol{\x}, 
\] 
is equal to the conserved integral corresponding to the translational symmetry, i.e. energy-momentum components
of the field $\psi$. Here $\overline{\psi}(x)$ stands for the Dirac adjoint $\psi(x)^+\gamma^0$,
and not for the complex conjugation, as usual. 
We replace the classical field $\psi$ in the above integral formally by the quantum field 
$\boldsymbol{\psi}$ with the counterpart of Dirac adjoint appropriately defined (see below) and with the product 
under the integral sign defined as the Wick product of the fields at the 
same space-time point (compare preceding Subsection \ref{OperationsOnXi}).  

Recall that in both cases, (\ref{psi(x)'}) and  (\ref{standardpsi(x)}), we realize the field operators as the integral kernel operators with the corresponding vector-valued distributions $\kappa_{0,1}, \kappa_{1,0}$,
over the standard Gelfand triple 
$E_1 = \mathcal{S}_{A}(\mathbb{R}^3; \mathbb{C}^4) \subset L^{2}(\mathbb{R}^3; \mathbb{C}^4) \subset E_{1}^{*}$ in 
both cases (\ref{psi(x)'}) and (\ref{standardpsi(x)}).

Thus we are going to check if 
\[
\int \boldsymbol{:} T^{0\mu} \boldsymbol{:} \, \ud^3 \boldsymbol{\x} = \boldsymbol{P}^\mu = d\Gamma(P^\mu),
\] 
where $P^\mu$, $\mu = 0,1,2,3$, are the translation generators of the represenation  
$U \boldsymbol{U}(a,\alpha)U^{-1}$, acting in $U\mathcal{H}' = L^2(\mathbb{R}^3; \mathbb{C}^4)$ 
(in the first case (\ref{psi(x)'}))
or $U \mathbb{U}\boldsymbol{U}(a,\alpha)\mathbb{U}^{-1}U^{-1}$ in the same $U\mathbb{U}\mathcal{H}'
= L^2(\mathbb{R}^3; \mathbb{C}^4)$ standard Hilbert space (in the second case 
(\ref{standardpsi(x)})),
and with $\boldsymbol{P}^\mu = d\Gamma(P^\mu)$, $\mu = 0,1,2,3$, equal to the generators of translations of the representation 
\[
\Gamma\Big(U\boldsymbol{U}(a, \alpha)U^{-1}\Big) \,\,\, \textrm{or resp.}, \,\,\,
\Gamma\Big(U \mathbb{U}\boldsymbol{U}(a,\alpha)\mathbb{U}^{-1}U^{-1}\Big)
\]
of $T_4 \circledS SL(2, \mathbb{C})$, both acting in the Fock space $\Gamma(U\mathcal{H}') = 
\Gamma(L^2(\mathbb{R}^3; \mathbb{C}^4))$
(in the second case corresponding to (\ref{standardpsi(x)}) we also have 
$\Gamma(U\mathbb{U}\mathcal{H}') = \Gamma(L^2(\mathbb{R}^3; \mathbb{C}^4))$ with the isomorphism 
$U$ given by the modification of (\ref{isomorphismU})
in which we remove the factor $1/p_0(\boldsymbol{\p})$, with the removal being compensated by the presence 
of $\mathbb{U}$). Note that in the first case (\ref{psi(x)'})
the unitary operator is given by the formula (\ref{isomorphismU}), and in the second case $U$
is given by the similar formula with the weight factor $1/p_0(\boldsymbol{\p})$ omitted.

Equivalently Bogoliubov-Shirkov Quantization Postulate for $\boldsymbol{\psi}$ demands the equality 

\begin{equation}\label{BSPpsi}
\frac{i}{2}  \int  \boldsymbol{:} \Bigg(
\overline{\boldsymbol{\psi}}(x)\gamma^0 \frac{\partial \boldsymbol{\psi}}{\partial x_\mu}(x)
- \frac{\partial \overline{\boldsymbol{\psi}}}{\partial x_\mu}(x) \gamma^0 \boldsymbol{\psi}(x)  \Bigg)
\boldsymbol{:}  \, \ud^3 \boldsymbol{\x}
=
d\Gamma(P^\mu), \,\,\, \textrm{in this order!}
\end{equation}
to hold.

The whole point about the Quantization Postulate (or Emmy Noether theorem for free fields) 
is that the operators $\boldsymbol{P}^\mu = d\Gamma(P^\mu)$ may be computed in 
therms of Wick polynomials in free fields -- integral kernel operators -- to which we know how to apply the 
perturbative series in the sense of Bogoliubov-Epstein-Glaser. 
In  checking its validity for the Dirac field we proceed in two steps. 
In the first step we show that for each $\mu = 0,1,2,3$, 
there exist a distribution $\kappa^\mu \in E_1 \otimes E_{1}^*$ 
such that the corresponding integral kernel operator  $\Xi_{1,1}(\kappa^\mu)$
  is equal to $\boldsymbol{P}^\mu = d\Gamma(P^\mu)$. Then according to the rule giving the Wick product of free fields at the same point as integral kernel operator with vector valued kernel as well as the rule giving its spatial integral
as an integral kernel operator with scalar kernel, given in the preceding Subsection, we show that the left hand side integral kernel operator is equal to the right hand side integral kernel operator $\Xi_{1,1}(\kappa^\mu)$ 
in (\ref{BSPpsi}) for the standard field (\ref{standardpsi(x)}). It turns out that 
(\ref{BSPpsi}) does not hold for the local field (\ref{psi(x)'}).

It is easily seen that the representors $U\boldsymbol{U}(a, \alpha)U^{-1}$ and respectively 
\[
U\mathbb{U}\boldsymbol{U}(a, \alpha)\mathbb{U}^{-1}U^{-1}
\]
are continuous as operators $E_1 \to E_1$, 
in case of both the representations of
$T_4\circledS SL(2, \mathbb{S})$: 
\begin{enumerate}
\item[1)]
for the representation $U\boldsymbol{U}(a, \alpha)U^{-1}$ acting in $U\mathcal{H}' = L^2(\mathbb{R}^3; \mathbb{C}^4)$,
with $U$ given by (\ref{isomorphismU}), corresponding to the field
(\ref{psi(x)'}), 
\item[2)]
for the representation $U\mathbb{U}\boldsymbol{U}(a, \alpha)\mathbb{U}^{-1}U^{-1}$, acting
in $U\mathbb{U}\mathcal{H}' = L^2(\mathbb{R}^3; \mathbb{C}^4)$, with $U$ 
given by (\ref{isomorphismU}) without the factor
$1/p_0(\boldsymbol{\p})$, which is compensated here by the operator $\mathbb{U}$, and
corresponding to the field (\ref{standardpsi(x)}).
\end{enumerate}
 
In particular this holds
for the translation subgroup representors.
And the translation representors in both of the representations are unitary and act identically on the common nuclear space $E_1 = \mathcal{S}_{A}(\mathbb{R}^3; \mathbb{C}^4)$.
Therefore the translation subgroup in both cases of representations 
compose the subgroup of the Yoshizawa group $U\big(E_1; L^2(\mathbb{R}^3; \mathbb{C}^4)\big)$.  
The Yoshizawa group $U\big(E_1; L^2(\mathbb{R}^3; \mathbb{C}^4)\big)$ is the group of unitary operators on 
$L^2(\mathbb{R}^3; \mathbb{C}^4)$ which induce homeomorphisms
of the test function space $E_1= \mathcal{S}_{A}(\mathbb{R}^3; \mathbb{C}^4)$ with respect to the nuclear topology of 
$E_1$. In other words the translation representors in both representations compose automorphisms of the Gelfand triple 
$E_1 \subset L^2(\mathbb{R}^3; \mathbb{C}^4) \subset E_{1}^*$. Moreover
any one parameter subgroup $\{T_\theta\}_{\theta \in \mathbb{R}}$ of translations in both considered representations 
is differentiable, i.e. $\lim_{\theta \to 0} (T_{\theta}\xi - \xi)/\theta = X\xi$ converges in $E_1$.  
Let us consider the one parameter 
subgroup of translations along the $\mu$-th axis and write in this case $X^\mu$ for $X$,
where in our case $X^\mu$ is the operator $M_{ip^\mu}$ of multiplication by the function 
$\boldsymbol{\p} \to ip^\mu(\boldsymbol{\p})$, and where 
$(p^0(\boldsymbol{\p}), \ldots p^3(\boldsymbol{\p}))
 = (\sqrt{\boldsymbol{\p} \cdot \boldsymbol{\p} + m^2}, \boldsymbol{\p}) \in \mathscr{O}_{(1,0,0,1)}$. 
Existence of the limit is equivalent to 
\begin{multline}\label{T-theta-differentiability-psi}
\lim \limits_{\theta \to 0} \bigg| \frac{T_{\theta}\xi - \xi}{\theta} - X^\mu \xi  \bigg|_k^2 \\ =
\lim \limits_{\theta \to 0}
\int \bigg( \frac{A^k \Big(e^{i\theta p^\mu} -1 
- i \theta p^\mu \Big)\xi(\boldsymbol{\p})}{\theta} \, , \,\, 
\frac{A^k \Big(e^{i\theta p^\mu} -1 
- i \theta p^\mu \Big)\xi(\boldsymbol{\p})}{\theta}
\bigg)_{{}_{\mathbb{C}^4}} \,\, 
\ud^3 \boldsymbol{\p} \,\, = 0, \\
 k = 0, 1, 2, \ldots, \,\,\, \xi \in E_1,
\end{multline}
where $p^\mu$, $\mu = 0,1,2,3$, in the exponent are the functions 
$\boldsymbol{\p} \mapsto (p^\mu(\boldsymbol{\p})) = (\sqrt{\boldsymbol{\p} \cdot \boldsymbol{\p}+m^2}, \boldsymbol{\p})$
and where $A$ is the standard operator (\ref{AinL^2(R^3;C^4)}) used in the construction of the standard
Gelfand triple $E_1 = \mathcal{S}_{A}(\mathbb{R}^3; \mathbb{C}^4) \subset L^2(\mathbb{R}^3; \mathbb{C}^4)
\subset E_{1}^*$. Explicit calculation shows that (\ref{T-theta-differentiability-psi}) is fulfilled. Therefore  
$\{T_\theta\}_{\theta \in \mathbb{R}}$ is differentiable subgroup and by the Banach-Steinhaus
theorem the linear operators $X^\mu$, $\mu = 0,1,2,3$, 
are continuous as operators $E_1 \to E_1$ and finally by Proposition 3.1 of \cite{hida}
every such subgroup is regular in the sense of \cite{hida}, \S  3.

For every operator $X$ which is continuous as the operator $E_1 \to E_1$  we define $\Gamma(X)$
and $d \Gamma(X)$ on $(E_1)$. Let $\Phi \in (E_1)$ be  
 be any element of the Hida space with decomposition (\ref{HidaPhi}) corresponding to the Gelfand triple
$E_1 = \mathcal{S}_{A}(\mathbb{R}^3; \mathbb{C}^4) \subset  L^2(\mathbb{R}^3; \mathbb{C}^4)
\subset E_{1}^*$, i.e. with the pairing 
$\langle \cdot, \cdot \rangle$  induced by the inner product
$(\cdot, \cdot)_{{}_{L^2(\mathbb{R}^3; \mathbb{C}^4)}}$ in $L^2(\mathbb{R}^3; \mathbb{C}^4)$. 
Then we define  
\[
\Gamma(X)\Phi
= \sum \limits_{n=0}^{\infty}  \, X^{\otimes n} \Phi_n;
\]
\[
d\Gamma(X)\Phi
= \sum \limits_{n=0}^{\infty} n \, (X \otimes I^{\otimes (n-1)}) \, \Phi_n.
\]
In this case it is easily seen that the Theorem 4.1 of \cite{hida} is easily adopted to our fermi case
and that $\{\Gamma(T_\theta ) \}_{\theta \in \mathbb{R}}$, with the generator $X^\mu$, is a regular 
one parameter subgroup with the generator $d\Gamma(X^\mu)$ which continuously maps $(E)$
into itself. 

In this situation it is not difficult to see that  for each $\mu =0,1,2,3$, the proof of Proposition 4.2 
and Theorem 4.3 of \cite{hida} is applicable in the fermi case to any of the one parameter translation subgroups
of the mentioned representations, in particular
for any of the traslation subgroup along the direction of the 
$\mu$-th axis, $\mu =0,1,2,3$, there exists  
a symmetric distribution $\kappa^\mu \in E_1 \otimes E_{1}^*$ such that 
\begin{equation}\label{dGammaX=Berexin-int-psi}
 d \Gamma(X^\mu) = \Xi_{1,1}(\kappa^\mu) = \sum_{s,s'} \, \int \limits_{\mathbb{R}^3 \times \mathbb{R}^3}
\kappa^\mu(\boldsymbol{\p}', s', \boldsymbol{\p},s) \,\, 
 \partial_{\boldsymbol{\p}',s'}^* 
\partial_{\boldsymbol{\p},s}  
\,\, \ud^3 \boldsymbol{\p}' \ud^3 \boldsymbol{\p},
\end{equation}
and $\kappa^\mu \in E_1 \otimes E_{1}^*$ fulfills
\begin{equation}\label{kappa-distribution-P-psi}
\langle \kappa^\mu, \zeta \otimes \xi \rangle =  \langle \zeta, X^\mu \xi \rangle,
\,\,\, \zeta, \xi \in E_1. 
\end{equation}
Because the pairings $\langle \cdot, \cdot \rangle$ in the formula are induced by 
the inner product $(\cdot, \cdot)_{{}_{L^2(\mathbb{R}^3; \mathbb{C}^4)}}$ in $L^2(\mathbb{R}^3; \mathbb{C}^4)$,
and because $X^\mu$ is the operator of multiplication by $ip^\mu(\boldsymbol{\p})$, we have
\[
(\overline{\zeta}, \, X^\mu \xi)_{{}_{\oplus L^2(\mathbb{R}^3)}} = \langle \zeta, X^\mu \xi \rangle
= \langle X^\mu \xi , \zeta \rangle = \langle \xi , X^\mu \zeta \rangle,
\,\,\, \zeta, \xi \in E, 
\]
so that
\[
\langle \kappa^\mu, \zeta \otimes \xi \rangle = \langle \kappa^\mu, \xi \otimes \zeta \rangle,
\,\,\, \zeta, \xi \in E, 
\] 
and $\kappa^\mu$ is indeed symmetric. 

On the other hand the pairing $\langle \cdot, \cdot \rangle$
on left hand side of (\ref{kappa-distribution-P-psi}) expressed in terms
of the kernel $\kappa^\mu (\boldsymbol{\p}', \boldsymbol{\p})$ is likewise induced by the inner
product $(\cdot, \cdot)_{{}_{\oplus L^2(\mathbb{R}^3)}}$ in $L^2(\mathbb{R}^3; \mathbb{C}^4)$. Therefore
we have
\[
\langle \kappa^\mu, \zeta \otimes \xi \rangle = \sum \limits_{s,s'}
\int \limits_{\mathbb{R}^3 \times \mathbb{R}^3}
\kappa^\mu(\boldsymbol{\p}', s', \boldsymbol{\p},s) \,\, 
\zeta(\boldsymbol{\p}',s') \xi(\boldsymbol{\p},s)
\,\, \ud^3 \boldsymbol{\p}' \ud^3 \boldsymbol{\p}.
\]  
Joining this with (\ref{kappa-distribution-P-psi}) we obtain
\[
\kappa^\mu(\boldsymbol{\p}', s' \boldsymbol{\p}, s) 
= i p^\mu(\boldsymbol{\p}) \delta_{s\,s'} \delta(\boldsymbol{\p}'- \boldsymbol{\p}).
\]

Therefore we get 
\begin{equation}\label{d(Gamma(P)-psi}
\boldsymbol{P}^\mu = d \Gamma(P^\mu) = \sum_{s,s'} \, \int \limits_{\mathbb{R}^3 \times \mathbb{R}^3}
p^\mu(\boldsymbol{\p}) \,\, \delta_{s  \, s'} \delta(\boldsymbol{\p}'- \boldsymbol{\p}) \,\,\, 
\partial_{\boldsymbol{\p}', s'}^{*} \partial_{\boldsymbol{\p}, s}
\,\,\, \ud^3 \boldsymbol{\p}' \ud^3 \boldsymbol{\p},
\end{equation}
which is customary to be written as
\begin{multline}\label{d(Gamma(P^0)-psi}
\boldsymbol{P}^0 = d \Gamma(P^0) = \sum \limits_s \int \limits_{\mathbb{R}^3}
|p^0(\boldsymbol{\p})| \,\, 
\partial_{\boldsymbol{\p}, s}^{*} \partial_{\boldsymbol{\p}, s}
\,\,\, \ud^3 \boldsymbol{\p} \\
= \sum \limits_{s=1,2} \int \limits_{\mathbb{R}^3}
|p^0(\boldsymbol{\p})| \,\, 
b_s(\boldsymbol{\p})^{+} b_s(\boldsymbol{\p})
\,\,\, \ud^3 \boldsymbol{\p}
+
\sum \limits_{s=1,2} \int \limits_{\mathbb{R}^3}
|p^0(\boldsymbol{\p})| \,\, 
d_s(\boldsymbol{\p})^{+} d_s(\boldsymbol{\p})
\,\,\, \ud^3 \boldsymbol{\p},
\end{multline}
\begin{multline}\label{d(Gamma(P^i)-psi}
\boldsymbol{P}^i = d \Gamma(P^i) = \sum \limits_s \int \limits_{\mathbb{R}^3}
p^i(\boldsymbol{\p}) \,\, 
\partial_{\boldsymbol{\p}, s}^{*} \partial_{\boldsymbol{\p}, s}
\,\,\, \ud^3 \boldsymbol{\p} \\
= \sum \limits_{s=1,2} \int \limits_{\mathbb{R}^3}
p^i(\boldsymbol{\p}) \,\, 
b_s(\boldsymbol{\p})^{+} b_s(\boldsymbol{\p})
\,\,\, \ud^3 \boldsymbol{\p}
+
\sum \limits_{s=1,2} \int \limits_{\mathbb{R}^3}
p^i(\boldsymbol{\p}) \,\, 
d_s(\boldsymbol{\p})^{+} d_s(\boldsymbol{\p})
\,\,\, \ud^3 \boldsymbol{\p}.
\end{multline}
Both operators $d \Gamma(P^\mu)$ and $\Xi_{1,1}(-i\kappa^\mu)$ transform (continuously)
the nuclear, and thus perfect, space $(E_1)$ into itself and both being equal and symmetric 
on $(E_1)$ have self-adjoint extension to self-adjoint operator
in the Fock space $\Gamma(L^2(\mathbb{R}^3; \mathbb{C}^4))$, again by the classical criterion of 
\cite{Riesz-Szokefalvy}
(p. 120 in Russian Ed. 1954). In general the criterion of Riesz-Sz\"okefalvy-Nagy
does not exclude existence of more than just one self-adjoint extension, but
in our case it is unique. Indeed because for each $\mu=0,1,2,3$, the one-parameter unitary
group generated by $d \Gamma(P^\mu)$ leaves invariant the dense nuclear space $(E_1)$, then by general
theory, e.g. Chap. 10.3., it follows that  $d \Gamma(P^\mu)$ with domain $(E_1)$ is essentially self adjoint
(admits unique self adjoint extension).

Now applying the Rules II and V' of Subsection \ref{OperationsOnXi} to the left hand side of  
(\ref{BSPpsi}) with $\boldsymbol{\psi}$ equal to the standard Dirac free field (\ref{standardpsi(x)}),
understood as an integral kernel operator
\[
\boldsymbol{\psi} = \Xi_{0,1}(\kappa_{0,1}) + \Xi_{1,0}(\kappa_{1,0})
\]
with the kernels $\kappa_{0,1}, \kappa_{1,0}$, (\ref{skappa_0,1}) and (\ref{skappa_1,0}), we immediately get 
the result equal to (\ref{d(Gamma(P)-psi}) or equivalently (\ref{d(Gamma(P^0)-psi}),
(\ref{d(Gamma(P^i)-psi}). Thus we arrive at the following
\begin{prop*}
The standard free Dirac field $\boldsymbol{\psi}$, equal (\ref{standardpsi(x)}), satisfies
the Bogoliubov-Shirkov Quantization Postulate (\ref{BSPpsi}) for translations: 
\[
\frac{i}{2}  \int  \boldsymbol{:} \Bigg(
\overline{\boldsymbol{\psi}}(x)\gamma^0 \frac{\partial \boldsymbol{\psi}}{\partial x_\mu}(x)
- \frac{\partial \overline{\boldsymbol{\psi}}}{\partial x_\mu}(x) \gamma^0 \boldsymbol{\psi}(x)  \Bigg)
\boldsymbol{:}  \, \ud^3 \boldsymbol{\x}
=
d\Gamma(P^\mu).
\]
\end{prop*}

On the other hand if we apply the Rules II and V' of Subsection \ref{OperationsOnXi} to the left hand side of  
(\ref{BSPpsi}) with $\boldsymbol{\psi}$ equal to the local Dirac free field (\ref{psi(x)'}),
understood as an integral kernel operator
\[
\boldsymbol{\psi} = \Xi_{0,1}(\kappa_{0,1}) + \Xi_{1,0}(\kappa_{1,0})
\]
with the kernels $\kappa_{0,1}, \kappa_{1,0}$, (\ref{kappa_0,1}) and (\ref{kappa_1,0}), 
Subsection \ref{psiBerezin-Hida}, we obtain an integral kernel operator not equal
to (\ref{d(Gamma(P)-psi}) or, equivalently, not equal to (\ref{d(Gamma(P^0)-psi}),
(\ref{d(Gamma(P^i)-psi}). Thus we arrive at the following
\begin{prop*}
The Bogoliubov-Shirkov Quantization Postulate (\ref{BSPpsi}) for translations is not satisfied
by the local Dirac field (\ref{psi(x)'}). 
\end{prop*}

Now let us consider Lorentz transformations. The Noether integral
generator corresponding to Lorentz transformations is equal
\begin{equation}\label{d(Gamma(M)-psi}
\frac{i}{2} \int  \boldsymbol{:} \Bigg(
\boldsymbol{\psi}(x)^{+}x^\mu\frac{\partial \boldsymbol{\psi}}{\partial x_\nu}(x)
- \boldsymbol{\psi}(x)^{+} x^\nu\frac{\partial \boldsymbol{\psi}}{\partial x_\mu}(x) 
+ \frac{1}{2} \boldsymbol{\psi}(x)^{+} \gamma^\mu \gamma^\nu \boldsymbol{\psi}(x)  \Bigg)
\boldsymbol{:}  \, \ud^3 \boldsymbol{\x} = \boldsymbol{M}^{\mu\nu}
\end{equation}
Again  applying the Rules II and V' of Subsection \ref{OperationsOnXi} we arrive at the follwing
(infinitisemal form of) local transformation formula
\[
i[\boldsymbol{M}^{\mu\nu}, \boldsymbol{\psi}^a] = \Sigma_{b}^{a \mu \nu}\boldsymbol{\psi}^b +
(x^\mu \partial^\nu - x^\nu \partial^\mu)\boldsymbol{\psi}^a
\]
for the standard Dirac free field (\ref{standardpsi(x)}) $\boldsymbol{\psi}$. It generates the ordinary
local bispinor transformation formula $\boldsymbol{U}(a, \alpha)$ in the single particle Hilbert space
$\mathcal{H}''$ of the standard Dirac field (\ref{standardpsi(x)}), which does not coincide with the 
unitary representation $\mathbb{U}\boldsymbol{U}(a, \alpha) \mathbb{U}^{-1}$, and which is not unitary
if regarded as representation in the single particle Hilbert space 
$\mathcal{H}'' = \mathbb{U}\mathcal{H}'$. In particular $\boldsymbol{M}^{\mu \nu} = d\Gamma(M^{\mu\nu})$, 
regarded as an operator in the Fock space
$\Gamma(\mathcal{H}'')$ of the standard Dirac free field (\ref{standardpsi(x)}), generates a nonunitary
transformation. Therefore the generator $\boldsymbol{M}^{\mu \nu} = d \Gamma(M^{\mu \nu})$ given by te Noether integral
 (\ref{d(Gamma(M)-psi}) coresponding to the Lorentz transformations, and computed for the standard Dirac
field (\ref{standardpsi(x)}) is not self-adjoint. 

We therfore have the following alternative: we can save locality of the transformation of the standard Dirac
field (\ref{standardpsi(x)}), with the generators of the local representation given by te Noether
integrals (with Wick ordered products), but unitarity of te Lorentz transformations have to be abandoned.
Alternatively we have the unitary representation 
$\Gamma(\mathbb{U}\boldsymbol{U}(a, \alpha) \mathbb{U}^{-1})$ in the Fock space $\Gamma(\mathcal{H}'')$ 
of the standard Dirac field (\ref{standardpsi(x)}), but locality of the Lorentz transformations is lost.

This alternative has not been discovered before. One reason lies in the fact that there are the white noise 
technics which allow us to construct equal time integrals of Wick products of free fields,
and to investigate their self-adjointness. As far as we know nobody has applied them before
to the realistic fields, and in particular to the analysis of Wick product fields and their Cauchy 
integrals. On the other hand the aproach more popular among mathematical physiscists, \emph{i. e.}
due to Wightman-G{\aa}rding, is not effective here, which was recognized by Segal \cite{Segal-NFWP.I},
p. 455. In particular non-self-adjointnes of the Lorentz transformations generator $\boldsymbol{M}^{\mu\nu}$
for the standard Dirac field (\ref{standardpsi(x)}) given by the Noether integral formula (\ref{d(Gamma(M)-psi}),
could have not been discovered by such founders of Quantum Field Theory like Pauli or Schwinger.       
This alternative explains, among other things, also the fact why we do not encounter
the standard Dirac field (\ref{standardpsi(x)}) among the free fields whose construction is based on the unitary 
and local representations. In particular it escaped the classification of free fields based on local unitary representations of the double covering of the Poincar\'e group given in \cite{lop1} 
or \cite{lop2}. This fact  was also recognized by 
Haag \cite{Haag}, p. 48. The local bispinor field (\ref{psi(x)'}) has the standard local
and unitary bispinor transformation formula, but it does not coincide with the standard Dirac field
(\ref{standardpsi(x)}). Note that the standard Dirac field (\ref{standardpsi(x)}) is a field which is 
obtainded through the canonical quantization, \emph{i.e.} it is uniquely determined by the condition
that it satisfies the Bogoliubov-Shirkov Quantization Postulate (\ref{BSPpsi}) for translations.
It seems that also the local bispinor field (\ref{psi(x)'}) has not been constructed before and apears 
here for the first time.

Note that the Wick product of the Dirac field components is skew-commutative, therefore the order is important 
in (\ref{BSPpsi}).

We end this Subsection with a remark on the Pauli theorem on spin-statistics relation. It
is based on the properties of the ``classical'', \emph{i.e.} before ``quantization'', fields.
Essentially it says that the energy component of the Noether energy-momentum tensor
is not positive definite for half-odd-integer free ``classical'' fields, compare e.g. \cite{Geland-Minlos-Shapiro}
and Pauli's book cited there. Technically speaking, 
generic half-odd-integer spin field (solution of equations of motion), when Fourier decomposed and inserded 
into te Noether energy integral, gives formally the expression (\ref{d(Gamma(P^0)-psi}),
but with operators $b_s(\boldsymbol{\p}), d_s(\boldsymbol{\p})$ replaced with
the Fourier coefficients and with the opposite sign at the second term in  
(\ref{d(Gamma(P^0)-psi}). Pauli then joined this result with the canonical quantization procedure,
equivalent to the Pauli-Bogoliubov-Shirkov Quantization Postulate (\ref{BSPpsi})
for translations. Because the Wick product of fermi fields in (\ref{d(Gamma(P)-psi})
repears the sign of the second term in the `classical'' counterpart of (\ref{d(Gamma(P^0)-psi}), 
Pauli arrived at the spin-statistics relation: 
half-odd-integer spin ``classical'' (free) fields should be quantized with the 
canonical anticommutation  relations. 

The so called ``spin-statistis theorem'' due to Wightman is different and in fact gives the relation
between the commutation relation of smeared out fields, within his axiomatic definition of 
a quantum field, and 
the representation defining a local transformation rule of the field. In Wightman's proof no relation with ``classical''
fields and with positivity of the energy-momentum of ``classical'' fields  intervenes. In this sense Pauli's spin-statistics theorem is different pointing out that such relation exists, and in this sense reveals what is untouched in the Wightman's version of spin-statistics theorem.

\subsection{The quantum electromagnetic potential field $A$ as an integral kernel operator
with vector-valued distributional kernel}\label{A=Xi0,1+Xi1,0}

Recall that the formula (294) of Subsection 5.9 of \cite{wawrzycki2018}: 
\begin{multline}\label{q-A-B'}
A^\mu(x) = \int \limits_{\mathbb{R}^3} \, \ud^3 p \, \bigg\{
\frac{1}{\sqrt{2 p^0(\boldsymbol{\p})}}\sqrt{B(\boldsymbol{\p}, p^0(\boldsymbol{\p}))}^{\mu}_{\lambda}
a^{\lambda} (\boldsymbol{\p}) e^{-ip\cdot x} \\
+  \frac{1}{\sqrt{2 p^0(\boldsymbol{\p})}}\sqrt{B(\boldsymbol{\p}, p^0(\boldsymbol{\p}))}^{\mu}_{\lambda} \,
\eta \, a^{\lambda}(\boldsymbol{\p})^+ \, \eta \, e^{ip\cdot x}  \bigg\} 
\end{multline} 
gives a well defined generalized operator transforming continously the Hida
space $(E)$ into its strong dual $(E)^*$, where $(E)$ is the Hida space of the Gelfand triple 
$(E) \subset \Gamma(\mathcal{H}') \subset (E)^*$
defining the electromagnetic potential field $A$ within the white noise setup. 
Recall that  $E = \mathcal{S}_{A}(\mathbb{R}^3; \mathbb{C}^4) = 
\mathcal{S}_{\oplus A^{(3)}}(\mathbb{R}^3; \mathbb{C}^4)$ is defined by the standard operator 
$A = \oplus_{0}^{3} A^{(3)}$ on the standard Hilbert space $L^2(\mathbb{R}^3; \mathbb{C}^4)$,
with the operator $A^{(3)}$ defined as in Subsection 5.3 of \cite{wawrzycki2018}. 
Recall that  the integral (\ref{q-A-B'}) exists pointwisely as the Pettis integral, compare 
(294), Subsection 5.9 of \cite{wawrzycki2018}. Nonetheless the potential field $A$ is naturally
a sum of two integral kernel operators 
\[
A = \Xi_{0,1}(\kappa_{0,1}) + \Xi_{1,0}(\kappa_{1,0})
 \in
\mathscr{L}\big( (E) \otimes \mathscr{E}, \, (E)^* \big) \cong
\mathscr{L}\big( \mathscr{E}, \,\, \mathscr{L}( (E), (E)^*) \big) 
\]
with vector valued kernels $\kappa_{0,1}, \kappa_{1,0} \in \mathscr{L}\big(E, \mathscr{E}^* \big)$
for
\[
\mathscr{E} = 
\mathcal{S}_{\mathscr{F}\oplus A^{(4)}\mathscr{F}^{-1}}(\mathbb{R}^4; \mathbb{C}^4) 
= \mathscr{F} \Big[\mathcal{S}_{\oplus A^{(4)}}(\mathbb{R}^4; \mathbb{C}^4) \Big] 
= \mathcal{S}^{00}(\mathbb{R}^4; \mathbb{C}^4),
\]
in the sense of Obata \cite{obataJFA} explained in Subsection \ref{psiBerezin-Hida}.
The vector valued distributions $\kappa_{0,1}, \kappa_{1,0}$ are defined by the following plane waves 
\[
\begin{split}
\kappa_{0,1}(\nu, \boldsymbol{\p}; \mu, x) =
\frac{\sqrt{B(\boldsymbol{\p}, p^0(\boldsymbol{\p}))}^{\mu}_{\nu}}{\sqrt{2 p^0(\boldsymbol{\p})}}
e^{-ip\cdot x}, \,\,\,\,\,\,
p = (| p^0(\boldsymbol{\p})|, \boldsymbol{\p}) 
\in \mathscr{O}_{1,0,0,1}, \\
\kappa_{1,0}(\nu, \boldsymbol{\p}; \mu, x) = (-1)^{(\mu)}
\frac{\sqrt{B(\boldsymbol{\p}, p^0(\boldsymbol{\p}))}^{\mu}_{\nu}}{\sqrt{2 p^0(\boldsymbol{\p})}}
e^{ip\cdot x},
\,\,\,\,\,\,
p = (| p^0(\boldsymbol{\p})|, \boldsymbol{\p}) 
\in \mathscr{O}_{1,0,0,1},
\end{split}
\]  
with
\[
(-1)^{(\mu)} \overset{\textrm{df}}{=}
\left\{ \begin{array}{ll}
-1 & \textrm{if $\mu = 0$}, \\
1 & \textrm{if $\mu = 1, 2, 3$}.
\end{array} \right., \,\,\,\,
p^0(\boldsymbol{\p}) = |\boldsymbol{\p}|.
\]
The above stated formulas for $\kappa_{0,1}, \kappa_{1,0}$ can be immediately read off from the
formula (\ref{q-A-B'}) and the commutation rules (219) of \cite{wawrzycki2018} 
for the Gupta-Bleuler operator $\eta$
and the Hida operators $\partial_{\mu, \boldsymbol{\p}} = a_{\mu}(\boldsymbol{\p})$:
\[
a_{0}(\boldsymbol{\p}) \eta = - \eta a_{0}(\boldsymbol{\p}), \,\,\,\,
a_{i}(\boldsymbol{\p}) \eta = \eta a_{i}(\boldsymbol{\p}), \,\,\,\, i= 1, 2, 3, \,\,\,\,
\eta^2 = \boldsymbol{1}.
\] 
Here we are using the standard convention of Subsection \ref{psiBerezin-Hida} that in the general integral kernel operator (\ref{electron-positron-photon-Xi}) in the tensor product of the Fock space of the Dirac field  $\boldsymbol{\psi}$ and of the electromagnetic potential field $A$ we have the ordinary Hida operators 
in the normal order with the ordinary adjoint (linear transpose)
$\partial_{\mu, \boldsymbol{\p}}^{*} = a_{\mu}(\boldsymbol{\p})^{+}$ corresponding to photon variables
$\mu, \boldsymbol{\p}$. This is the convention assumed in mathematical literature concerning integral 
kernel oprators. But physicict never use the ordinary
adjoint $\partial_{\mu, \boldsymbol{\p}}^{*} = a_{\mu}(\boldsymbol{\p})^{+}$ whenever usng expansions into normally ordered
creation-annihilation operators for the variables corresponding to the electromagnetic field, but instead they are using
the ``Krein-adjoined'' operators $\eta\partial_{\mu, \boldsymbol{\p}}^{*} \eta = \eta a_{\mu}(\boldsymbol{\p})^{+} \eta$
insted,  as in the formula  (\ref{q-A-B'}). Therefore it is more convenient, when adopting the integral kernel operators to QED (in Gupta-Bleuler gauge), to change slightly the convention of Subsection
\ref{psiBerezin-Hida} and use for $\partial_{w}^*$ in the general integral kernel operator 
(\ref{electron-positron-photon-Xi}),
on the tensor product of Fock spaces of the Dirac field $\boldsymbol{\psi}$ and the electromnagnetic potential field $A$,
the operators  $\eta\partial_{\mu, \boldsymbol{\p}}^{*} \eta$ whenever 
$w = (\mu, \boldsymbol{\p})$ corresponds to the photon variables $\mu, \boldsymbol{\p}$ 
in (\ref{electron-positron-photon-Xi}), 
insted of the ordinary transposed operators $\partial_{\mu, \boldsymbol{\p}}^{*}$. 
With this covention of physicists we will have the following formulas 
\begin{equation}\label{kappa_0,1kappa_1,0A}
\boxed{
\begin{split}
\kappa_{0,1}(\nu, \boldsymbol{\p}; \mu, x) =
\frac{\sqrt{B(\boldsymbol{\p}, p^0(\boldsymbol{\p}))}^{\mu}_{\nu}}{\sqrt{2 p^0(\boldsymbol{\p})}}
e^{-ip\cdot x}, \,\,\,\,\,\,
p \in \mathscr{O}_{1,0,0,1}, \\
\kappa_{1,0}(\nu, \boldsymbol{\p}; \mu, x) = 
\frac{\sqrt{B(\boldsymbol{\p}, p^0(\boldsymbol{\p}))}^{\mu}_{\nu}}{\sqrt{2 p^0(\boldsymbol{\p})}}
e^{ip\cdot x},
\,\,\,\,\,\,
p \in \mathscr{O}_{1,0,0,1},
\end{split}
}
\end{equation}
without the additional factor $(-1)^{(\mu)}$. In fact presence of the factors
\[ 
(-1)^{(\mu_1)} \cdots (-1)^{(\mu_l)}
\]
for the kernels of the corresponding integral kernel operators
is the only difference between the two conventions, and which are absorbed coincisely by the 
Gupta-Bleuler operator $\eta$.  

In other words: we will show that for the plane wave kernels (\ref{kappa_0,1kappa_1,0A}) we have
\begin{multline}\label{A=IntKerOpVectValKer}
A(\phi) = a'( \overline{\check{\widetilde{\phi}}}|_{{}_{\mathscr{O}_{1,0,0,1}}}) 
+ \eta a'(\widetilde{\phi}|_{{}_{\mathscr{O}_{1,0,0,1}}})^+ \eta \\ =
a\Big( U\big( \overline{\check{\widetilde{\phi}}}|_{{}_{\mathscr{O}_{1,0,0,1}}}\big) \Big) 
+ \eta a\Big( U\big(\widetilde{\phi}|_{{}_{\mathscr{O}_{1,0,0,1}}} \big) \Big)^+ \eta \\ =
a(\sqrt{B} \, \overline{\check{\widetilde{\phi}}}|_{{}_{\mathscr{O}_{1,0,0,1}}}) 
+ \eta a(\sqrt{B} \, \widetilde{\phi}|_{{}_{\mathscr{O}_{1,0,0,1}}})^+ \eta \\ =
\sum \limits_{\nu=0}^{3} \int \kappa_{0,1}(\phi)(\nu, \boldsymbol{\p}) \partial_{\nu, \boldsymbol{\p}} \, 
\ud^3 \boldsymbol{\p}
+ \sum \limits_{\nu=0}^{3} \int \kappa_{1,0}(\phi)(\nu, \boldsymbol{\p}) \eta \partial_{\nu, \boldsymbol{\p}}^{*}\eta
\, \ud^3 \boldsymbol{\p} \\ =
\Xi_{0,1}\big(\kappa_{0,1}(\phi)\big) + \Xi_{1,0}\big(\kappa_{1,0}(\phi)\big),
\,\,\,\,
\phi \in \mathscr{E}= \mathcal{S}^{00}(\mathbb{R}^4; \mathbb{C}^4).
\end{multline}
Moreover we will show that the kernels $\kappa_{0,1},\kappa_{1,0}$ defined 
by (\ref{kappa_0,1kappa_1,0A}) can be (uniquely) extended to the elements 
(and denoted by the same $\kappa_{0,1},\kappa_{1,0}$)
\[
\kappa_{0,1},\kappa_{1,0} \in \mathscr{L}(E^*, \mathscr{E}^*),
\]
so that by Thm 3.13 of \cite{obataJFA} (or Thm. \ref{obataJFA.Thm.3.13} of Subsection)
\ref{psiBerezin-Hida}
\[
A = \Xi_{0,1}(\kappa_{0,1}) + \Xi_{1,0}(\kappa_{1,0})
 \in
\mathscr{L}\big( (E) \otimes \mathscr{E}, \, (E) \big) \cong
\mathscr{L}\big( \mathscr{E}, \,\, \mathscr{L}( (E), (E)) \big) 
\] 
and $A$, understood as an integral kernel operator with vector-valued distributional kernels
(\ref{kappa_0,1kappa_1,0A}), determines a well defined operator-valued distribution
on the space-time nuclear test space 
\[
\mathscr{E} = \mathscr{F} \Big[\mathcal{S}_{\oplus A^{(4)}}(\mathbb{R}^4; \mathbb{C}^4)\Big] 
= \mathcal{S}^{00}(\mathbb{R}^4; \mathbb{C}^4).
\]
In the formula (\ref{A=IntKerOpVectValKer}) $\kappa_{0,1}(\phi), \kappa_{1,0}(\phi)$  denote the kernels 
representing distributions
in $E^* = \mathcal{S}_{A}(\mathbb{R}^3, \,\, \mathbb{C}^4)^*$ which are defined in the standard manner
\[
\kappa_{0,1}(\phi)(\nu, \boldsymbol{\p})
=  \sum_{\mu=0}^{3} \int \limits_{\mathbb{R}^3}
\kappa_{0,1}(\nu, \boldsymbol{\p}; \mu,x) \phi^{\mu}(x) \, \ud^4 x
\]
and analogously for $\kappa_{1,0}(\phi)$, where $\kappa_{0,1}, \kappa_{1,0}$
are understood as elements of 
\[
\mathscr{L}(\mathscr{E}, E^*) \cong \mathscr{L}\big( E, \, \mathscr{L}(\mathscr{E},\mathbb{C}) \big)
\cong \mathscr{L}\big( E, \, \mathscr{E}^* \big). 
\]
Similarily we have
\[
\kappa_{0,1}(\xi)(\mu, x) 
= \sum_{\nu=0}^{4}  \, \int \limits_{\mathbb{R}^3} 
\kappa_{0,1}(\nu, \boldsymbol{p}; \mu, x) \, \xi(\nu, \boldsymbol{\p}) \, \ud^3 \boldsymbol{\p},
\,\,\, \xi \in E,
\]
and analogously for $\kappa_{1,0}(\xi)(\mu, x)$, with
$\kappa_{0,1}, \kappa_{1,0}$
understood as elements of 
\[
\mathscr{L}\big( E, \, \mathscr{L}(\mathscr{E},\mathbb{C}) \big)
\cong \mathscr{L}\big( E, \, \mathscr{E}^* \big) \cong \mathscr{L}(\mathscr{E}, E^*); 
\]
with pairings
\begin{multline*}
\langle \kappa_{0,1}(\phi), \xi \rangle 
= \sum_{\nu=0}^{3} \, \int \limits_{ \mathbb{R}^4 \times \mathbb{R}^3} 
\kappa_{0,1}(\phi)(\nu, \boldsymbol{p}) \,\, \xi(\nu, \boldsymbol{\p}) \, \ud^3 \boldsymbol{\p} \\ 
= \sum_{\mu=0}^{3} \, \sum_{\nu=0}^{3} \, \int \limits_{\mathbb{R}^3} 
\kappa_{0,1}(\nu, \boldsymbol{p}; \mu, x) \, \phi^{\mu}(x) \,\, \xi(\nu, \boldsymbol{\p}) \, \ud^4 x \, \ud^3 \boldsymbol{\p}
= \langle \kappa_{0,1}(\xi), \phi \rangle,
\,\,\, \xi \in E, \phi \in  \mathscr{E},
\end{multline*}
defined through the ordinary Lebesgue integrals.

$U$ is the unitary isomorphism (and its inverse $U^{-1}$)
\[
\begin{split}
U: \mathcal{H}' \ni \xi \mapsto  \sqrt{B} \xi \in L^2(\mathbb{R}^3; \mathbb{C}^4), \\
U^{-1}: L^2(\mathbb{R}^3; \mathbb{C}^4) \ni \zeta \mapsto \sqrt{B}^{-1} \zeta \in \mathcal{H}',
\end{split}
\]
joining the Gelfand triples (272) of Subsection 2.1 of \cite{wawrzycki2018} defining the field $A$ 
through its Fock lifting,
and is defined as point-wise multiplication
\begin{multline*}
\sqrt{B}\xi(\boldsymbol{\p}) \overset{\textrm{df}}{=}
\frac{1}{\sqrt{2 p^0(\boldsymbol{\p})}}\sqrt{B(\boldsymbol{\p}, p^0(\boldsymbol{\p}))} \xi(\boldsymbol{\p}),
\\
\sqrt{B}^{-1}\zeta(\boldsymbol{\p}) \overset{\textrm{df}}{=}
\sqrt{2 p^0(\boldsymbol{\p})}\sqrt{B(\boldsymbol{\p}, p^0(\boldsymbol{\p}))}^{{}^{-1}} \zeta(\boldsymbol{\p})
\end{multline*}
by the matrix (and respectively its inverse)
\begin{equation}\label{sqrtB/sqrtp0}
\frac{1}{\sqrt{2 p^0(\boldsymbol{\p})}}\sqrt{B(\boldsymbol{\p}, p^0(\boldsymbol{\p}))},
\end{equation}
the same which is present in the fomula (\ref{q-A-B'}), with the matrix $\sqrt{B(p)}$, 
$p \in \mathscr{O}_{1,0,0,1}$ defined by (200) of Subsection 4.1 of \cite{wawrzycki2018}.

Note here that the Gelfand triples (272) of \cite{wawrzycki2018} with the joining unitary isomorphism
$U$ in (272) of \cite{wawrzycki2018} plays the same role in the construction of the field $A$ in Subsection 5.8 of \cite{wawrzycki2018} as does 
the triples (\ref{SinglePartGelfandTriplesForPsi})
joined by the unitary isomorphism (\ref{isomorphismU}) in the construction of the Dirac field
$\boldsymbol{\psi}$, Subsection \ref{psiBerezin-Hida}. 

Concerning the equality (\ref{A=IntKerOpVectValKer}) note that the first equality in 
(\ref{A=IntKerOpVectValKer}) follows by definition, second by the fact that $U$ is the unitary isomorphism
joninig the standard Gelfand triple 
\[
E = \mathcal{S}_{A}(\mathbb{R}^3; \mathbb{C}^4)
\subset L^2(\mathbb{R}^3; \mathbb{C}^4) \subset \mathcal{S}_{A}(\mathbb{R}^3; \mathbb{C}^4)^*
\]
with the triple 
\[
E \subset \mathcal{H}' \subset E^*
\]
over the single particle Hilbert space of the field $A$ (the analogue of the unitary isomorphism 
(\ref{isomorphismU}) of Subsection \ref{psiBerezin-Hida}) . The Fock lifting of the standard triple
serves to construct the standard Hida operators $a(\zeta)$, and the Fock lifting of the second triple serves to
construct the Hida operators $a'(\xi)$. Therefore we obtain the second equality (the analogue of the isomorphism
(\ref{a(U(u+v))=a'(u+v)})), compare also Subsection 5.8 of \cite{wawrzycki2018}. 
Third equality in (\ref{A=IntKerOpVectValKer}) follows by definition of the isomorphism
$U$. Finally note that it follows almost immediately from definition (\ref{kappa_0,1kappa_1,0A}) 
of $\kappa_{0,1}, \kappa_{1,0}$ that
\begin{equation}\label{kappa_0,1(varphi),kappa_1,0(varphi)}
\kappa_{0,1}(\phi) = \sqrt{B} \check{\widetilde{\phi}}|_{{}_{\mathscr{O}_{1,0,0,1}}}, \,\,\,\,\,\,\,
\kappa_{1,0}(\phi) = \sqrt{B} \widetilde{\phi}|_{{}_{\mathscr{O}_{1,0,0,1}}}.
\end{equation}
Thus the fourth equality in (\ref{A=IntKerOpVectValKer}) follows by 
Prop. 4.3.10 of \cite{obata-book} (compare also the fermi analogue of Prop. 4.3. 10 of  \cite{obata-book} -- 
the Corollary \ref{D_xi=int(xiPartial)}
of Subsection \ref{psiBerezin-Hida}).

Let $\mathcal{O}'_C, \mathcal{O}_M$ be the algebras of convolutors and multipliers of the ordinary Schwartz
algebra $\mathcal{S}(\mathbb{R}^4; \mathbb{C}^4)$, defined by Schwartz \cite{Schwartz},
compare also Appendix \ref{convolutorsO'_C}. 
If the elements of $\mathcal{O}'_C$ (resp.  of $\mathcal{O}_M$) are understood as continous linear operators
$\mathcal{S} \rightarrow \mathcal{S}$ of convolution
with distributions in $\mathcal{O}'_C$ (or respectively as continuous operators of multiplication by an element of 
$\mathcal{O}_M$) then we can endow $\mathcal{O}'_C, \mathcal{O}_M$ with the operator topolology of 
uniform convergence on bounded sets (after Schwartz). The Fourier exchange theorem of Schwartz then says
that the Fourier transorm becomes a topological isomorphism of $\mathcal{O}_M$ onto $\mathcal{O}'_C$,
which exchanges pointwise multiplication product defined by pointwise multiplication of functions in $\mathcal{O}_M$ (represeting the correponding tempered distributions) with the convolution product, defined through the composition
of the corresponding convolution operators in $\mathscr{L}(\mathcal{S, \mathcal{S}})$, compare \cite{Schwartz},
or Appendix \ref{convolutorsO'_C}.

Let $\mathcal{O}_C$ be the predual (a smooth function space determined explicitly by Horv\'ath) of 
the Schwartz convolution algebra $\mathcal{O}'_C$ endowed with the above 
Schwartz operator topology of uniform convergence on bounded sets on $\mathcal{O}'_C$ (strictly stronger than the topology inherited from the strong dual space $\mathcal{S}^*$ of tempered distributions), 
compare Appendix \ref{convolutorsO'_C}.  
 
Let $\mathcal{O}'_{CB_2}$ be the algebra of convolutors
of the algebra 
\[
\mathscr{E} = \mathcal{S}^{00}(\mathbb{R}^4; \mathbb{C}^4) = \mathscr{F}\Big[\mathcal{S}^{0}(\mathbb{R}^4; \mathbb{C}^4) \Big]
= \mathscr{F}\Big[\mathcal{S}_{\oplus A^{(4)}}(\mathbb{R}^4; \mathbb{C}^4) \Big]
= \mathcal{S}_{B_2}(\mathbb{R}^4; \mathbb{C}^4),
\]
where we have used the standard operator
\[
B_2 = \mathscr{F} \oplus_{0}^{3} A^{(4)} \mathscr{F}^{-1} \,\,\, \textrm{on} \,\,\,
\oplus_{0}^{3}L^2(\mathbb{R}^4; \mathbb{C}) = L^2(\mathbb{R}^4; \mathbb{C}^4),
\]
introduced in Subsection \ref{psiBerezin-Hida}, and further used in Subsection \ref{OperationsOnXi}.
Recall that the standard operators $A^{(n)}$ on $L^2(\mathbb{R}^n; \mathbb{C})$ have been constructed
in Subsection 5.3 of \cite{wawrzycki2018}.

Let $\mathcal{O}'_{MB_2}$ be the algebra of multipliers of the nuclear algebra
\[
\mathcal{S}^{0}(\mathbb{R}^4; \mathbb{C}^4) = \mathcal{S}_{\oplus A^{(4)}}(\mathbb{R}^4; \mathbb{C}^4)
= \mathcal{S}_{B_2}(\mathbb{R}^4; \mathbb{C}^4).
\]
All the spaces $\mathcal{O}_C, \mathcal{O}_M, \mathcal{O}_{MB_2}$ equipped with the Horv\'ath inductive limit or respectively Schwartz operator toplology of uniform convergence on bounded sets, and their strong duals
$\mathcal{O}'_C, \mathcal{O}'_M, \mathcal{O}'_{MB_2}$, equipped with the Schwartz operator toplology of uniform convergence
on bounded sets, are nuclear. 

We have:
\begin{equation}\label{O_C<O_CA,O_M<O_MA}
\begin{split}
\mathcal{O}_M \subset \mathcal{O}_{MB_2}, \\
\mathcal{O}'_C  \subset \mathcal{O}'_{CB_2}, \\
\mathcal{O}_C  \subset \mathcal{O}'_C  \subset \mathcal{O}'_{CB_2},
\end{split}
\end{equation}
by the results of Subsections 5.2-5.5 of \cite{wawrzycki2018}.

Recall that here $\mathcal{O}_M(\mathbb{R}^m; \mathbb{C}^n)$ is understood as 
the pointwise multiplication algebra of $\mathbb{C}^n$-valued functions on
$\mathbb{R}^3$ in $\mathcal{O}_M(\mathbb{R}^m; \mathbb{C}^n)$, with the elements of 
$\mathcal{O}_M(\mathbb{R}^m; \mathbb{C}^n)$, $\mathcal{S}(\mathbb{R}^m; \mathbb{C}^n)$
understood as $\mathbb{C}$-valued functions on the disjoint sum $\sqcup \mathbb{R}^m$ of $n$ 
copies of $\mathbb{R}^m$, compare Subsection \ref{psiBerezin-Hida}. The 
translation $T_b, b \in \mathbb{R}^m$ is understood as acting on $(a,x) \in \sqcup \mathbb{R}^m$, 
$a \in \{1,2, \ldots n\}$,
in the following manner $T_b(a,x) = (a, x+b)$. Equivalently $f \in \mathcal{O}_M(\mathbb{R}^m; \mathbb{C}^n)$
(or $f \in \mathcal{O}_C(\mathbb{R}^m; \mathbb{C}^n)$)
means that each component of $f$ belongs to $\mathcal{O}_M(\mathbb{R}^m; \mathbb{C})$
(or resp. to $\mathcal{O}_C(\mathbb{R}^m; \mathbb{C})$).

We need the following Lemma (analogously as in Subsection \ref{psiBerezin-Hida} for the Dirac field).
\begin{lem}\label{kappa0,1,kappa1,0ForA}
For the $\mathscr{L}(\mathscr{E},\mathbb{C})$-valued (or $\mathscr{E}^*$ -valued) distributions 
$\kappa_{0,1}, \kappa_{1,0}$, given by (\ref{kappa_0,1kappa_1,0A}),
in the equality (\ref{A=IntKerOpVectValKer}) defining the electromagnetic potential field $A$ we have
\begin{multline*}
\Bigg( \, (\mu,x) \mapsto \sum_{\nu} \, \int \limits_{\mathbb{R}^3}
\kappa_{0,1}(\nu, \boldsymbol{\p}; \mu, x)\, \xi(\nu,\boldsymbol{\p}) \,\ud^3\boldsymbol{\p} \,\, \Bigg) 
\in \mathcal{O}_C \subset \mathcal{O}_M \subset \mathscr{E}^*, \,\, 
\xi \in \mathcal{S}_{A}(\mathbb{R}^3, \mathbb{C}^4), \\
\Bigg( \, (\mu, x) \mapsto \sum_{\nu} \, \int \limits_{\mathbb{R}^3}
\kappa_{1,0}(\nu, \boldsymbol{\p}; \mu, x) \, \xi(s, \boldsymbol{\p}) \,\ud^3\boldsymbol{\p} \,\, \Bigg)  
\in \mathcal{O}_C \subset \mathcal{O}_M \subset \mathscr{E}^*, 
 \,\, \xi \in \mathcal{S}_{A}(\mathbb{R}^3, \mathbb{C}^4), \\
\Bigg( \, (\nu,\boldsymbol{\p}) \mapsto \sum_{\mu} \, \int \limits_{\mathbb{R}^4}
\kappa_{0,1}(\nu, \boldsymbol{\p}; \mu ,x) \, \varphi^\mu(x) \,\ud^4x \,\, \Bigg) \in 
\mathcal{S}_{A}(\mathbb{R}^3, \mathbb{C}^4), \,\, \varphi \in \mathscr{E}, \\
\Bigg( \, (\nu ,\boldsymbol{\p}) \mapsto \sum_{\mu} \, \int \limits_{\mathbb{R}^4}
\kappa_{1,0}(\nu, \boldsymbol{\p}; \mu, x) \, \varphi^\mu(x) \,\ud^4x \,\, \Bigg) \in 
\mathcal{S}_{A}(\mathbb{R}^3, \mathbb{C}^4), 
 \,\, \varphi \in \mathscr{E}. 
\end{multline*}
Moreover the maps 
\[
\begin{split}
\kappa_{0,1}: \mathscr{E} \ni \phi \longmapsto \kappa_{0,1}(\phi) 
\in \mathcal{S}_{A}(\mathbb{R}^3, \,\, \mathbb{C}^4), \\
\kappa_{1,0}: \mathscr{E} \ni \phi \longmapsto \kappa_{1,0}(\phi) 
\in \mathcal{S}_{A}(\mathbb{R}^3, \,\, \mathbb{C}^4)
\end{split}
\]
are continuous, with $\kappa_{0,1}, \kappa_{1,0}$ uderstood as maps 
in 
\[
\mathscr{L}\big( \mathscr{E}, \,\, \big(\mathcal{S}_{A}(\mathbb{R}^3, \mathbb{C}^4)^* 
\big) \cong \mathscr{L}\big( \mathcal{S}_{A}(\mathbb{R}^3, \mathbb{C}^4), \,\,
\mathscr{L}(\mathscr{E}, \mathbb{C})  \big)
\] 
and, equivalently,
the maps $\xi \longmapsto \kappa_{0,1}(\xi)$, $\xi \longmapsto \kappa_{1,0}(\xi)$ can be extended to
continuous maps
\[
\begin{split}
\kappa_{0,1}: \mathcal{S}_{A}(\mathbb{R}^3, \mathbb{C}^4)^* \ni \xi \longmapsto \kappa_{0,1}(\xi) 
\in \mathscr{E}^*, \\
\kappa_{1,0}: \mathcal{S}_{A}(\mathbb{R}^3, \mathbb{C}^4)^* \ni \xi \longmapsto \kappa_{1,0}(\xi) 
\in \mathscr{E}^*,
\end{split}
\]
(for $\kappa_{0,1}, \kappa_{1,0}$ uderstood as maps 
$\mathscr{L}\big( \mathcal{S}_{A}(\mathbb{R}^3, \mathbb{C}^4), \,\,
\mathscr{L}(\mathscr{E}, \mathbb{C})  \big) \cong 
\mathscr{L}\big( \mathcal{S}_{A}(\mathbb{R}^3, \mathbb{C}^4), \,\,
\mathscr{E}^*  \big)$). Therefore 
not only $\kappa_{0,1}, \kappa_{1,0}
\in \mathscr{L}\big( \mathcal{S}_{A}(\mathbb{R}^3, \mathbb{C}^4), \,\,
\mathscr{L}(\mathscr{E}, \mathbb{C})  \big)$, but both $\kappa_{0,1}, \kappa_{1,0}$
can be (uniquely) extended to elements of 
\[
\mathscr{L}\big( \mathcal{S}_{A}(\mathbb{R}^3, \mathbb{C}^4)^*, \,\,
\mathscr{L}(\mathscr{E}, \mathbb{C})  \big) \cong 
\mathscr{L}\big( \mathcal{S}_{A}(\mathbb{R}^3, \mathbb{C}^4)^*, \,\,
\mathscr{E}^*  \big)  \cong
\mathscr{L}\big( \mathscr{E}, \,\, 
\mathcal{S}_{A}(\mathbb{R}^3, \mathbb{C}^4)  \big).
\]
\end{lem}
\qedsymbol \,
That for each $\xi \in \mathcal{S}_{A}(\mathbb{R}^3, \mathbb{C}^4)$ the functions
$\kappa_{0,1}(\xi), \kappa_{1,0}(\xi)$ given by (here $x = (x_0, \boldsymbol{\x})$)
\begin{multline*}
(\mu,x) \mapsto \sum_{\nu=0}^{3} \, \int \limits_{\mathbb{R}^3}
\kappa_{0,1}(\nu, \boldsymbol{\p}; \mu,x)\, \xi(\nu,\boldsymbol{\p}) \,\ud^3\boldsymbol{\p} \\ =
\sum_{\nu =0}^{3} \, \int \limits_{\mathbb{R}^3}
\frac{\sqrt{B(\boldsymbol{\p}, p^0(\boldsymbol{\p}))}^{\mu}_{\nu}}{\sqrt{2p_0(\boldsymbol{\p})}} \, 
\xi(\nu,\boldsymbol{\p}) e^{-ip_0(\boldsymbol{\p})x_0 + i\boldsymbol{\p} \cdot \boldsymbol{\x}} \, \ud^3 \boldsymbol{\x}, \\
\end{multline*}
\begin{multline*}
(\mu, x) \mapsto \sum_{\nu=0}^{3} \, \int \limits_{\mathbb{R}^3}
\kappa_{1,0}(\nu, \boldsymbol{\p}; \mu, x)\, \xi(\nu,\boldsymbol{\p}) \,\ud^3\boldsymbol{\p} \\ =
\sum_{\nu=0}^{3} \, \int \limits_{\mathbb{R}^3}
\frac{\sqrt{B(\boldsymbol{\p}, p^0(\boldsymbol{\p}))}^{\mu}_{\nu}}{\sqrt{2p_0(\boldsymbol{\p})}} \, 
\xi(\nu, \boldsymbol{\p}) e^{i|p_0(\boldsymbol{\p})|x_0 - i\boldsymbol{\p} \cdot \boldsymbol{\x}} \, \ud^3 \boldsymbol{\x},
\end{multline*}
belong to $\mathcal{O}_C \subset \mathcal{O}_M \subset \mathscr{E}^*$ is immediate. Indeed, that they are smooth
is bovious, similarily as it is obvious the existence of such a natural $N$ (it is sufficient to take here $N=0$)
that for each multiindex $\alpha \in \mathbb{N}^4$ the functions
\[
(a,x) \mapsto (1 + |x|^2)^{-N} |D_{x^{\alpha}}^{\alpha}\kappa_{0,1}(\xi)(a,x)|, \,\,\,
(a,x) \mapsto (1 + |x|^2)^{-N} |D_{x^{\alpha}}^{\alpha}\kappa_{1,0}(\xi)(a,x)|
\]
are bounded (of course for fixed $\xi$). Here $D_{x^{\alpha}}^{\alpha}\kappa_{l,m}(\xi)$ denotes the ordinary derivative of 
the function $\kappa_{l,m}(\xi)$
of $|\alpha| = \alpha_0 + \alpha_1+\alpha_2+ \alpha_3$ order with respect to space-time 
coordinates $x= (x_0, x_1, x_2, x_3)$; and here 
$|x|^2= (x_{0})^2 + (x_{1})^2 + (x_{2})^2+ (x_{3})^2$. Recall that by the results of Subsections 5.4
and 5.5 of \cite{wawrzycki2018}, the operation of point-wise multiplication by the matrix (\ref{sqrtB/sqrtp0})
is a multiplier of the algebra  
$\mathcal{S}_{A}(\mathbb{R}^3, \mathbb{C}^4) = \mathcal{S}^{0}(\mathbb{R}^3; \mathbb{C}^4)$, similarily
multiplication by the function $|p_0(\boldsymbol{\p}|^{k} = |\boldsymbol{\p}|^k$, $k \in \mathbb{Z}$,
is a multiplier of this algebra, by the same Subsections. Thus the said integrals defining 
$\kappa_{0,1}(\xi), \kappa_{1,0}(\xi)$ are convergent, similarily as the integrals defining their space-time drivatives
with the obviously preserved mentioned above boundedness. 

Consider now the functions 
\[
\begin{split}
\phi \mapsto \kappa_{0,1}(\phi)
=  \sqrt{B} \check{\widetilde{\phi}}|_{{}_{\mathscr{O}_{1,0,0,1}}}, \\
\phi \mapsto \kappa_{1,0}(\phi)
= \sqrt{B} \widetilde{\phi}|_{{}_{\mathscr{O}_{1,0,0,1}}},
\end{split}
\] 
with $\phi \in \mathcal{S}^{00}(\mathbb{R}^4; \mathbb{C}^4)$. 
It is obvious that both functions
$\kappa_{0,1}(\varphi), \kappa_{1,0}(\phi)$ belong to 
$\mathcal{S}_{A}(\mathbb{R}^3, \mathbb{C}^4) = \mathcal{S}^{0}(\mathbb{R}^3; \mathbb{C}^4)$
whenever $\phi \in \mathcal{S}^{00}(\mathbb{R}^4; \mathbb{C}^4)$, by the results of Subsections
5.4 and 5.5 of \cite{wawrzycki2018}.
That both functions
$\kappa_{0,1}(\phi), \kappa_{1,0}(\phi)$ depend continously on $\phi$ as maps 
\[
\mathscr{E} = \mathcal{S}^{00}(\mathbb{R}^4; \mathbb{C}^4)
\longrightarrow \mathcal{S}_{A}(\mathbb{R}^3, \,\, \mathbb{C}^4) = \mathcal{S}^{0}(\mathbb{R}^3, \,\, \mathbb{C}^4)
\]
follows from:  1) the results of Subsection 5.5 of \cite{wawrzycki2018} and continuity of the Fourier 
transform as a map on the Schwartz space, 2) from the continuity of the restriction to the 
orbits $\mathscr{O}_{1,0,0,1}$ and $\mathscr{O}_{-1,0,0,1}$
 regarded as a map from 
\[
\mathcal{S}^{0}(\mathbb{R}^4;\mathbb{C}) =
\mathcal{S}_{\oplus A^{(4)}}(\mathbb{R}^4, \,\, \mathbb{C}^4)
\]
 into 
\[
\mathcal{S}^{0}(\mathbb{R}^3;\mathbb{C}) = \mathcal{S}_{\oplus A^{(3)}}(\mathbb{R}^3, \,\, \mathbb{C}^4),
\]
compare the second Proposition of Subsection 5.6 of \cite{wawrzycki2018}, and finally 3) from the fact 
that the operators of point-wise multiplication by the matrix (\ref{sqrtB/sqrtp0})
are multipliers of the nuclear algebra 
\[
\mathcal{S}_{A}(\mathbb{R}^3, \,\, \mathbb{C}^4) = 
\mathcal{S}_{\oplus A^{(3)}}(\mathbb{R}^3, \,\, \mathbb{C}^4) 
= \mathcal{S}^{0}(\mathbb{R}^3;\mathbb{C}),
\]
compare Subsections 5.4 and 5.5 of \cite{wawrzycki2018}. 
\qed

From the last Lemma \ref{kappa0,1,kappa1,0ForA} and from Thm. 3.13 of \cite{obataJFA} 
(or equivalently from Theorem \ref{obataJFA.Thm.3.13} of Subsection \ref{psiBerezin-Hida}) 
we obtain the following
\begin{cor}\label{A=intKerOpVectVal=OpValDistr}
Let $E = \mathcal{S}_{A}(\mathbb{R}^3; \mathbb{C}^4) = 
\mathcal{S}_{\oplus A^{(3)}}(\mathbb{R}^3; \mathbb{C}^4)$. 
Let
\[
A = \Xi_{0,1}(\kappa_{0,1}) + \Xi_{1,0}(\kappa_{1,0}) \in
\mathscr{L}\big( (E) \otimes \mathscr{E}, \, (E)^* \big) \cong
\mathscr{L}\big( \mathscr{E}, \,\, \mathscr{L}( (E), (E)^*) \big)
\]
be the free quantum electromagnetic potential field uderstood as an integral kernel operator with vector-valued kernels
\[
\kappa_{0,1}, \kappa_{1,0} \in \mathscr{L}\big( \mathcal{S}_{A}(\mathbb{R}^3, \mathbb{C}^4), \,\,
\mathscr{E}^*  \big) \cong \mathcal{S}_{A}(\mathbb{R}^3, \mathbb{C}^4)^* \otimes \mathscr{E}^*
= E^* \otimes \mathscr{E}^*,
\]
defined by (\ref{kappa_0,1kappa_1,0A}). Then the electromagnetic potential field operator
\[
A  = A^{(-)} + A^{(+)} = \Xi_{0,1}(\kappa_{0,1}) + \Xi_{1,0}(\kappa_{1,0}),
\]
belongs to $\mathscr{L}\big( (E) \otimes \mathscr{E}, \, (E) \big) \cong
\mathscr{L}\Big( \mathscr{E}, \,\, \mathscr{L}\big( (E), (E)\big) \, \Big)$, i.e.
\[
A = \Xi_{0,1}(\kappa_{0,1}) + \Xi_{1,0}(\kappa_{1,0}) \in
\mathscr{L}\big( (E) \otimes \mathscr{E}, \, (E) \big) \cong
\mathscr{L}\Big( \mathscr{E}, \,\, \mathscr{L}\big( (E), (E)\big) \, \Big),
\]
which means in particular that the electromagnetic potential field $A$, understood as a sum 
$A = \Xi_{0,1}(\kappa_{0,1}) + \Xi_{1,0}(\kappa_{1,0})$ of 
two integral kernel operators with vector-valued kernels, 
defines an operator valued distribution through the continuous map
\[
\mathscr{E} \ni \phi \longmapsto
\Xi_{0,1}\big(\kappa_{0,1}(\phi)\big) + \Xi_{1,0}\big(\kappa_{1,0}(\phi)\big)
\in \mathscr{L}\big( (E), (E)\big).
\]
\end{cor}

Note that the last Corollary likewise follows from: 
\begin{enumerate}
\item[1)]
the equality (\ref{A=IntKerOpVectValKer}),
\item[2)]
from Thm. 2.2 and 2.6 of \cite{hida}, 
\item[3)]
continuity of the Fourier transform as a map
on the Schwartz space, 
\item[4)]
continuity of the restriction to the orbit $\mathscr{O}_{1,0,0,1}$
regarded as a map $\mathcal{S}^{0}(\mathbb{R}^4) \longrightarrow \mathcal{S}^{0}(\mathbb{R}^3)$ and 
finally 
\item[5)]
from continuity of the multiplication by the matrix (\ref{sqrtB/sqrtp0}), regarded as a map
$\mathcal{S}^{0}(\mathbb{R}^3; \mathbb{C}^4) \longrightarrow \mathcal{S}^{0}(\mathbb{R}^3; \mathbb{C}^4)$.
\end{enumerate}

It is important to emphasize here that by the Thm. 3.13 of \cite{obataJFA}
(or Thm. \ref{obataJFA.Thm.3.13} of Subsection \ref{psiBerezin-Hida}) the continuity of the map
$\phi \longmapsto \kappa_{1,0}(\phi)$, regarded as a map
$\mathscr{E} \longrightarrow E = \mathcal{S}_{A}(\mathbb{R}^3; \mathbb{C}^4)$,
equivalent to the continuous unique extendibility of $\kappa_{1,0}$
to an element of $\mathscr{L}(E^*, \mathscr{E}^*)$, is a necessary and sufficient
condition for the operator $A = \Xi_{0,1}(\kappa_{0,1}) + \Xi_{1,0}(\kappa_{1,0})$
to be an element of 
\[
\mathscr{L}\big( (E) \otimes \mathscr{E}, \, (E) \big) \cong
\mathscr{L}\Big( \mathscr{E}, \,\, \mathscr{L}\big( (E), (E)\big) \, \Big),
\]
i.e. for $A$ being a sum of integral kernel operators with vector-valued kernels which defines 
an operator-valued distribution on $\mathscr{E}$.
On the other hand the continuity of the map
\[
\mathscr{E} \ni \phi \longmapsto \kappa_{1,0}(\phi) \in E = \mathcal{S}_{A}(\mathbb{R}^3; \mathbb{C}^4) 
\]
is equivalent, as we have seen, to the continuity of the restriction to the cone $\mathscr{O}_{1,0,0,1}$,
regarded as a map
\[
\widetilde{\mathscr{E}} \longrightarrow E = \mathcal{S}_{A}(\mathbb{R}^3; \mathbb{C}^4),
\]
followed by the multiplication by the matrix 
(\ref{sqrtB/sqrtp0}), and regarded as a map $E \rightarrow E$. 
From this it follows that 
\[
\widetilde{\mathscr{E}} \neq \mathcal{S}(\mathbb{R}^4), \,\,\, E \neq \mathcal{S}(\mathbb{R}^3)
\]
for the space-time test space of the zero mass field $A$ determined by a representation pertinent
to the cone orbit $\mathscr{O}_{1,0,0,1}$, because restriction to the cone
$\mathscr{O}_{1,0,0,1}$ is not continuous as a map $\mathcal{S}(\mathbb{R}^4) \rightarrow \mathcal{S}(\mathbb{R}^3)$,
nor the multiplication by the matrix (\ref{sqrtB/sqrtp0}) regarded as a map
$\mathcal{S}(\mathbb{R}^3) \rightarrow \mathcal{S}(\mathbb{R}^3)$.
This is in general the case for any zero mass (free) field. Namely we have the following
\begin{twr}\label{ZeromassTestspace}
For any zero mass field, pertinent to the cone orbit $\mathscr{O}_{1,0,0,1}$, such as the electromagnetic potential field, 
which can be regarded as an integral kernel operator
\[
\Xi_{0,1}(\kappa_{0,1}) + \Xi_{1,0}(\kappa_{1,0})
\]
with vector-valued kernels 
\[
\kappa_{0,1}, \kappa_{1,0} \in \mathscr{L}\big( \mathcal{S}_{A}(\mathbb{R}^3, \mathbb{C}^4), \,\,
\mathscr{E}^*  \big) \cong \mathcal{S}_{A}(\mathbb{R}^3, \mathbb{C}^4)^* \otimes \mathscr{E}^*
= E^* \otimes \mathscr{E}^*,
\]
extendible to 
\[
\kappa_{0,1}, \kappa_{1,0} \in \mathscr{L}\big( \mathcal{S}_{A}(\mathbb{R}^3, \mathbb{C}^4)^*, \,\,
\mathscr{E}^*  \big) \cong \mathcal{S}_{A}(\mathbb{R}^3, \mathbb{C}^4) \otimes \mathscr{E}^*
= E \otimes \mathscr{E}^*,
\]
and defined by plane waves
\[
\begin{split}
\kappa_{0,1}(s, \boldsymbol{\p}; a, x) = u^a(s, \boldsymbol{\p})\,e^{-ip\cdot x}, \,\,\,\,
p = (p_0(\boldsymbol{\p}), \boldsymbol{\p}) \in \mathscr{O}_{1,0,0,1}, \\
\kappa_{0,1}(s, \boldsymbol{\p}; a, x) = v^a(s, \boldsymbol{\p})\,e^{ip\cdot x}, \,\,\,\,
p = (p_0(\boldsymbol{\p}), \boldsymbol{\p}) \in \mathscr{O}_{1,0,0,1}, \\
s,a = 1,2, \ldots N
\end{split}
\]
the space-time test space $\mathscr{E}$ cannot be equal to the ordinary Schwartz space
 $\mathcal{S}(\mathbb{R}^4; \mathbb{C}^N)$ but instead it has to be equal
\[
\mathscr{E} = \mathcal{S}^{00}(\mathbb{R}^4; \mathbb{C}^N)
=  \mathscr{F} \Big[ \mathcal{S}^{0}(\mathbb{R}^4; \mathbb{C}^N) \Big]
= \mathscr{F} \Big[ \mathcal{S}_{\oplus A^{(4)}}(\mathbb{R}^4; \mathbb{C}^N) \Big],
\]
where $A^{(4)}$ is the standard operator on $L^2(\mathbb{R}^4; \mathbb{C})$
constructed in Subsection 5.3 of \cite{wawrzycki2018}, and $\oplus A^{(4)}$ denotes direct sum of
$N$ copies of the operator $A^{(4)}$ acting on 
\[
L^2(\mathbb{R}^4; \mathbb{C}^N) = \oplus_{1}^{N} L^2(\mathbb{R}^4; \mathbb{C}).
\]
\end{twr}
 
In particular this Theorem holds for all zero mass free gauge fields $A$ of the Standard Model. 

Let us stress once more that the conclusion of the last Theorem is inapplicable to 
zero-mass fields in the sense of Wightman, which allows the ordinary 
Schwartz space as the space-time test space. This follows immediately from the fact
that the integration of the restriction of the test function to the cone orbit 
$\mathscr{O}_{1,0,0,1}$ along $\mathscr{O}_{1,0,0,1}$ with respect to the measure 
induced by the ordinary measure of the ambient space $\mathbb{R}^4$, is a well defined
continuous functional on the ordinary Schwartz space $\mathcal{S}(\mathbb{R}^4; \mathbb{C})$.
We have also used this fact in extending the zero mass Pauli-Jordan function from
$\mathcal{S}^{00}(\mathbb{R}^4)$ over to a functional on $\mathcal{S}(\mathbb{R}^4)$,
with preservation of the homogeneity and its degree, compare Subsection 5.6 of \cite{wawrzycki2018}.

\subsection{Equivalent realizations of the free local electromagnetic
potential quantum field. Comparision with the realization used by other authors}\label{equivalentA-s}

Let $U^{*-1} = WU^{{}_{(1,0,0,1)}{\L}}W^{-1}$ and  $U = \big[WU^{{}_{(1,0,0,1)}{\L}}W^{-1}\big]^{*-1}$
be the {\L}opusza\'nski representation and its conjugation $U$ acting in the single
particle space of the quantum field $A$ realization of Sections 4 and  5 of \cite{wawrzycki2018}.
Both $U^{*-1}$, and $U$ transform continously the nuclear space $E_\mathbb{C}$ into itself (let us write
simply $E$ instead $E_\mathbb{C}$ for simplicity). Similarly the lifting $\Gamma(U)$ of $U$
acting in the Krein-Fock space $(\Gamma(\mathcal{H}'), \Gamma(\mathfrak{J}'))$ transforms
continously the nuclear Hida's test space $(E)$ onto itself, and is Krein isometric in the Krein-Fock
space of the field $A$. 

We can consider different such realizations of $A$, with the representations $U$ and $\Gamma(U)$ restricted to the translation subgroup commuting with the Krein fundamental symmetry $\mathfrak{J}'$, and resp. $\Gamma(U)$ commuting with the Gupta-Bleuler operator $\Gamma(\mathfrak{J}')$, and thus with translations being represented by unitary and Krein-unitary operators. The natural equivalence for such realizations is the existence of Krein isometric mapping transforming bi-uniquelly and bi-continously $E$, resp. $(E)$,
onto itself, and which intertwines the representations. It is easily seen that in case of ordinary non gauge fields with unitary representations, this equivalence reduces to the ordinary unitary equivalece
of the realizations of the fields. In case of gauge mass-less fields, such as electromagnetic potental field 
$A$, where $U$ and $\Gamma(U)$ are unbounded (and Krein-isometric) the equivalence is weaker,
although preserves the pairing functions of the field, the linear equation it fulfills and its local transformation formula. Nonetheless the analytic properties of the representation may be substantially different
for equivalent realizations of the field $A$, especially the behaviur of the restriction of the representation
$U$ or $\Gamma(U)$ of $T_4 \circledS SL(2, \mathbb{C})$ to the subgroup $SL(2, \mathbb{C})$, 
as is no very surprising as the representors of the Loretz hyperbolic rotations
are unbouded, contrary to the representors of translations, which are bounded (even unitary and Krein-unitary).

We illustrate this phenomena on a conctrete example of different equivalent realizations of the free field $A$. 
Although the example is concrete it can be shown 
that the construction encountered is generic, and that the general class of equivelnt realizations
may be constructed
without any substantial modification. The general construction of a realization of the free field $A$
is equivalent to the construction of the most general intertwining operator bi-uniquelly and bi-continously mapping the 
nuclear spaces, where the initial spaces and representations are these given in 
Sections 4 and 5 of \cite{wawrzycki2018} for the realization of $A$ given there.  
We give a concrete example of such an intertwining operator, in case where the nuclear spaces corresponding to different realizations are identical. Because this assumption is not relevant, and because the construction
of the general intretwining operator is general for the case where the nuclear spaces are identical, 
we prefer to give the concrete example
instead of going immediately into a general situation, which would be less transparent.

On the single particle space $(\mathcal{H}', \mathfrak{J}')$ of the realization of $A$ of Sect. 
4 and  5 of \cite{wawrzycki2018} there exists, besides $U, U^{*-1}$, the Krein-isometric
representation
\begin{equation}\label{Lop-rep--on-tildevarphi-ass}
\begin{split}
{}^{{}^{\textrm{ass}}}U(0,\alpha) \widetilde{\varphi} (p) = \sqrt{B(p)}^{-1}V(\alpha)\sqrt{B(p)} \widetilde{\varphi} (\Lambda(\alpha)p) = \sqrt{B(p)}^{-1}\Lambda(\alpha^{-1})\sqrt{B(p)} \widetilde{\varphi} (\Lambda(\alpha)p), \\
{}^{{}^{\textrm{ass}}}U(a,1) \widetilde{\varphi} (p) =  \sqrt{B(p)}^{-1}T(a) \sqrt{B(p)} \widetilde{\varphi}(p) 
= e^{i a \cdot p}\widetilde{\varphi}(p). 
\end{split}
\end{equation}
associated to the {\L}opusza\'nski representation $U^{*-1} = WU^{{}_{(1,0,0,1)}{\L}}W^{-1}$,
where $\sqrt{B(p)}$ is the (positive) square root of the (positive) matrix $B(p), p \in \mathscr{O}_{{}_{1,0,0,1}}$ 
(198) of \cite{wawrzycki2018}, equal  (200) of Subsection 4.1 of \cite{wawrzycki2018}. Recall that for each fixed point 
$p \in \mathscr{O}_{1,0,0,1}$, the matrices $\sqrt{B(p)}$, $B(p)$, $\mathfrak{J}'_{{}_{\bar{p}}}
=V(\beta(p))^{-1} \mathfrak{J}_{\bar{p}} V(\beta(p)) = 
\mathfrak{J}_{\bar{p}} B(p)$, are all Krein-unitary in the Krein space $(\mathbb{C}^4, \mathfrak{J}_{\bar{p}})$,
where $\mathfrak{J}_{\bar{p}}$ is the constant matrix (185) of Subsection 4.1 of \cite{wawrzycki2018}. 
In other words all the matrices 
$\sqrt{B(p)}$, $B(p)$, $\mathfrak{J}'_{{}_{\bar{p}}}$ are Lorentz matrices preserving the the Lorentz
metric $g^{\mu \nu} = \textrm{diag}(-1,1,1,1)$.

This representation is Krein-isometrically equivalent to the 
{\L}opusza\'nski representation $U^{*-1} = WU^{{}_{(1,0,0,1)}{\L}}W^{-1}$ given by (187)
of Subsection 4.1 of \cite{wawrzycki2018}.
(Analogously its conjugation is equivaelnt to the conjugation $U$ of the {\L}opusza\'nski representation
$U^{*-1}$). Indeed the intertwining operator $C$, understood as an operator
$(\mathcal{H}', \mathfrak{J}') \rightarrow (\mathcal{H}', \mathfrak{J}')$, 
acting in the single particle space is equal
\[
C \widetilde{\varphi}(p) = \sqrt{B(p)}^{-1} \widetilde{\varphi}(p), \,\,\,\,
C^{-1} \widetilde{\varphi}(p) = \sqrt{B(p)} \widetilde{\varphi}(p),
\]
and $C$ transforms bi-uniquelly and bi-continously the nuclear space $E$ onto itself (compare the first Proposition of Subsect. 5.6 of \cite{wawrzycki2018}) and the intertwining operator $\Gamma(C)$ transforms bi-uniquelly and bi-continously $(E)$ onto itself $(E)$, \cite{hida}, \cite{obata-book}. One easily checks that that $C$ indeed intertwines $U^{*-1}$ and 
${}^{{}^{\textrm{ass}}}U$:
\[
C \, U^{*-1} \, C^{-1} = {}^{{}^{\textrm{ass}}}U 
\] 
and thus that $\Gamma(C)$ intertwines $\Gamma(U^{*-1})$ and $\Gamma({}^{{}^{\textrm{ass}}}U)$. 

Let us introduce another operator $K$:
\[
K \widetilde{\varphi}(p) = \sqrt{B(p)} \widetilde{\varphi}(p), \,\,\,
K^{-1} \widetilde{\varphi}(p) = \sqrt{B(p)}^{-1} \widetilde{\varphi}(p),
\]
understood as a Krein-unitary operator mapping the Krein space $(\mathcal{H}', \mathfrak{J}')$
onto the Krein space $(K\mathcal{H}', K\mathfrak{J}'K^{-1}) 
= (K\mathcal{H}', \mathfrak{J}_{{}_{\bar{p}}})$, where the Krein fundamental symmetry
in the Krein space $(K \mathcal{H}', \mathfrak{J}_{{}_{\bar{p}}})$ is equal to the operator
of multiplication by the constant matrix $\mathfrak{J}_{{}_{\bar{p}}}$ equal (185)
of Subsection 4.1 of \cite{wawrzycki2018}.
Recall that the Krein fundamental symmetry operator $\mathfrak{J}'$ in the single paricle
Krein space $(\mathcal{H}', \mathfrak{J}')$ is equal to the operator of multiplication
by the matrix (193) of Subsection 4.1 of \cite{wawrzycki2018}:
\[
\mathfrak{J'}_{p} = V(\beta(p))^{-1} \mathfrak{J}_{\bar{p}} V(\beta(p)) = 
\mathfrak{J}_{\bar{p}} B(p),
\]
where $B(p)$ is equal to the matrix (198) of \cite{wawrzycki2018}. The operator $K$ gives a Krein-unitary equivalence
between the representation ${}^{{}^{\textrm{ass}}}U$ acting on the Krein space $(\mathcal{H}', \mathfrak{J}')$
and defined by the formula (\ref{Lop-rep--on-tildevarphi-ass}) with the dense nuclear domain
$(E)$, and the Krein-isometric representation given by formula (187) of \cite{wawrzycki2018} identical as for the
{\L}opusza\'nski representation $U^{*-1}$ on $(E)$, but on the Krein space  
$(K \mathcal{H}', \mathfrak{J}_{{}_{\bar{p}}})$ and with the nuclear domain $(E)$, which differs from 
the Krein space of Sections
4 and  5 of \cite{wawrzycki2018} by the replacement of the Lorentz matrices
$\sqrt{B(p)}$ and $B(p)$ everywhere with the constant unit matrix $\boldsymbol{1}$. Because on the other hand 
the {\L}opusza\'nski representation $U^{*-1}$, defined by (187) of \cite{wawrzycki2018},
and the representation ${}^{{}^{\textrm{ass}}}U$, both acting on the Krein space
$(\mathcal{H}', \mathfrak{J}')$ are Krein isometric equivalent (with $C$ defining the equivalence),
then it follows that the {\L}opusza\'nski representation, defined by (187) of \cite{wawrzycki2018},
with the nuclear domain $E$, on the Krein space $(\mathcal{H}', \mathfrak{J}')$ (with
the matrix $B(p) \neq \boldsymbol{1}$ and equal (198) of \cite{wawrzycki2018} is equivalent to the Krein isometric
represntation defined by the same formula (187) of \cite{wawrzycki2018} and the same nuclear domain $E$,
but on the Krein space in which the operators $B(p)$ and $\sqrt{B(p)}$ are everywhere replaced
by the constant unital matrices $\boldsymbol{1}$.
      
In this way we have obtained two equivalent realizations of the free quantum field $A$. The first one is obtained as in Sections 4 and  5 of \cite{wawrzycki2018}. The other is obtained exactly as in Sections
4 and 5 of \cite{wawrzycki2018} by the replacement everywhere in the formulas of the 
positive Lorentz matrices $B(p)$ and $\sqrt{B(p)}$ by the unit $4 \times 4$-matrix. A simple inspection shows that all proofs remain valid if we replace $B(p), \sqrt{B(p)}$ by $\boldsymbol{1}$ in Sections 
4 and 5 of \cite{wawrzycki2018}. In particular we obtain in this way a local
mass-less quantum four-vector field $A$, fulfilling d'Alembert equation with the pairing equal to the
zero mass Pauli-Jordan distribution function multilplied by the Minkowski metric components.
In particular this realization should be identified with the one used e.g. in \cite{Scharf},
\cite{DKS1}-\cite{DKS4}. In particular replacement of the matrix 
\[
\sqrt{B(\boldsymbol{\p}, p^0(\boldsymbol{\p}))}^{\mu}_{\lambda}
\]
by the unit $4\times 4$ matrix in the formula (294) of \cite{wawrzycki2018}: 
\begin{multline*}
A^\mu(x) = \int \limits_{\mathbb{R}^3} \, \ud^3 p \, \bigg\{
\frac{1}{\sqrt{2 p^0(\boldsymbol{\p})}}\sqrt{B(\boldsymbol{\p}, p^0(\boldsymbol{\p}))}^{\mu}_{\lambda}
a^{\lambda} (\boldsymbol{\p}) e^{-ip\cdot x} \\
+  \frac{1}{\sqrt{2 p^0(\boldsymbol{\p})}}\sqrt{B(\boldsymbol{\p}, p^0(\boldsymbol{\p}))}^{\mu}_{\lambda} \,
\eta \, a^{\lambda}(\boldsymbol{\p})^+ \, \eta \, e^{ip\cdot x}  \bigg\} 
\end{multline*} 
gives exactly the formula (2.11.45):
\begin{equation}\label{q-A'-B}
A^\mu(x) = \int \limits_{\mathbb{R}^3} \, \ud^3 p \, \bigg\{
\frac{1}{\sqrt{2 p^0(\boldsymbol{\p})}}
a^{\mu} (\boldsymbol{\p}) e^{-ip\cdot x} \\
+  \frac{1}{\sqrt{2 p^0(\boldsymbol{\p})}} \,
\eta \, a^{\mu}(\boldsymbol{\p})^+ \, \eta \, e^{ip\cdot x}  \bigg\} 
\end{equation} 
 of \cite{Scharf} (the lack of the additional constatnt factor $(2\pi)^{-3/2}$
in our formula comes from the fact that we have discarded the normalization factor for the measures in the Fourier transforms, in order to simplify notation). Similarly for other operator-valued distributions,
or ordinary operators, which we obtain by inserting the unit matrix for $\sqrt{B(p)}$. 

However the explicit formula for the Krein-isometric representation of
$T_4 \circledS SL(2, \mathbb{C})$ is lacking in the cited works as well as in other
works (as to the knowledge of the author) using the Gupta-Bleuler or BRST method. 
Moreover any analysis of the electromagnetic potential field in the Gupta-Bleuler approach, 
giving the linkage to 
the (generalized) induced representation theory of Mackey necessary uses the operator $\sqrt{B(p)}
\neq \boldsymbol{1}$. In particular no explicit construction of the representation of 
$T_4 \circledS SL(2, \mathbb{C})$ would be possible and its immediate linkage to the induced 
{\L}opusza\'nski representation, without the analysis using explicitly the
realization of the field $A$ with the matrix $B(p)$ equal (198) of Subsection 4.1 of \cite{wawrzycki2018}. 
We can pass to the (apparently)
simpler formulas only after using the intertwining operators, $C, K$, defined again with the hepl of 
$\sqrt{B(p)}$, and starting with the realization of $A$ presented in 4 and  
5 of \cite{wawrzycki2018}.

Perhaps we should emphasize that the two realizations of the free electromagnetic potential 
quantum field $A$: 1) the one with with $\sqrt{B(p)} \neq \boldsymbol{1}$ equal (200) of \cite{wawrzycki2018}
and presented in Sect. 4,  4 of \cite{wawrzycki2018} and 2) the one with 
$\sqrt{B(p)} = \boldsymbol{1}$, differ substantially. In particular we have the following
\begin{prop*}
 Consider the restriction
of the Krein-isometric representations of $T_4 \circledS SL(2, \mathbb{C})$ to the subgroup 
$SL(2, \mathbb{C})$, acting in the single particle Krein-Hilbert spaces in the two realizations, 1) and 2).
Then for the second realization 2)  (with $\sqrt{B(p)} = \boldsymbol{1}$) the restriction can be 
decomposed into
ordinary Hilbert space direct integral of subrepresentations $U^{{}^{\chi}}$ each acting in the
Hilbert space of generalized homogeneous of degree $\chi$ eigenstates $\in E^*$ (distributions) 
of the scaling operator $S_{\lambda}$:
\[
S_\lambda \widetilde{\varphi}(p) = \widetilde{\varphi}(\lambda p), \widetilde{\varphi} \in E,
\]       
where $\lambda$ is a fixed positive real number. 

No such decomposition is possible for the 1) realization of $A$ (with $\sqrt{B(p)} \neq \boldsymbol{1}$ 
and equal (200) in \cite{wawrzycki2018}.
\end{prop*}
{\bf REMARK}.
The statement of the last Proposition can be easily lifted to the Fock-Krein spaces of the realizations
1) and 2) of the 
field $A$, therefore we consider the statement and the proof only for the single particle 
Krein-Hilbert spaces. \qed

\qedsymbol \, (Proof of the Proposition. An outline.)
We consider the two versions of the {\L}opusza\'nski representation $U^{*-1}$ 
with $\sqrt{B(p)}$ equal respectively (200) of \cite{wawrzycki2018} or  $\boldsymbol{1}$ in case 1) or 2).
The results for its conjugation $U$ actually acting in the single particle space will follow
as a conseqence from the result for the {\L}opusza\'nski representation $U^{*-1}$ itself.  

Note that in both realizations the operator $S_\lambda$ (checking of which we leave as an easy exercise) 
has (unique) bounded extension to a normal operator, i.e. commuting with its adjoint $S_{\lambda}^{*}$ 
(with respect
to the ordinary Hilbert space inner product $(\cdot, \cdot)$, and not with respect to the Krein-inner product
$(\cdot, \mathfrak{J}' \cdot)$). 

The point is that the operators $S_\lambda, S_{\lambda}^{*}$, both commute 
with the {\L}opusza\'nski representation $U^{*-1}$ in the second realization 2) (with $\sqrt{B(p)} = \boldsymbol{1}$) and with the operator $\mathfrak{J}'$ (which in the realization 2) with $\sqrt{B(p)} = \boldsymbol{1}$
reduces to the constant matrix operator $\mathfrak{J}_{\bar{p}}$ equal to (185) of \cite{wawrzycki2018}. But in the first realization 1) (with $\sqrt{B(p)}$ equal (200) of \cite{wawrzycki2018}, although
$S_\lambda$ commutes with the {\L}opusza\'nski represntation $U^{*-1}$, the adjoint
operator $S_{\lambda}^{*}$ does not commute with the {\L}opusza\'nski representation $U^{*-1}$,
nor with the operator $\mathfrak{J}'$. Checking the commutation rules we again leave as an easy 
exercise to the reader. 

The proof of the statement of the Proposition can now be essentially reduced to the application of Theorems 1 and 2, \cite{Segal_dec_I}, with the commutative decomposition $*$-algebra $C$ of Thm. 2 in 
\cite{Segal_dec_I} equal to the one generated by the 
commuting operators $S_\lambda, S_{\lambda}^{*}$.

In both realizations, 1) and 2), the operators $S_\lambda, S_{\lambda}^{*}$ transform continously 
the nuclear space $E$ into itself, which follows easily by the results of Section 5
of \cite{wawrzycki2018} (compare the proof of the first Proposition of Subsection 5.6 of \cite{wawrzycki2018}). 
On the other hand $E$, the single particle Krein-Hilbert space $\mathcal{H}'$
and $E^*$, compose the Gefand triple $E \subset \mathcal{H}' \subset E^*$ (or a rigged Hilbert space). Thus  
the decomposition of $U$ (restricted to $SL(2, \mathbb{C})$) in the realization 2), is precisely the decomposition corresponding to the decomposition corresponding of the normal operator $S_\lambda$,
into the direct integral of subspaces of generalized eigen-subspaces of generalized eigenvectors in $E^*$
of $S_\lambda$, constructed as in Chap. I.4. of \cite{GelfandIV}.
\qed

Using the formula (\ref{q-A'-B}) for the electromagnetic potential field operator, regarded as 
the sum of integral kernel operators 
\[
A = \Xi_{0,1}(\kappa_{0,1}) + \Xi_{1,0}(\kappa_{1,0})
\]
with vector-valued distributional plane wave kernels 
\[
\kappa_{0,1}, \kappa_{1,0} \in \mathscr{L}\big( \mathcal{S}_{A}(\mathbb{R}^3, \mathbb{C}^4), \,\,
\mathscr{E}^*  \big) \cong \mathcal{S}_{A}(\mathbb{R}^3, \mathbb{C}^4)^* \otimes \mathscr{E}^*
= E^* \otimes \mathscr{E}^*,
\]
we will have the following formula for the plane wave kernels:
\begin{equation}\label{kappa_0,1kappa_1,0A'}
\boxed{
\begin{split}
\kappa_{0,1}(\nu, \boldsymbol{\p}; \mu, x) =
\frac{\delta_{\nu \mu}}{\sqrt{2 p^0(\boldsymbol{\p})}}
e^{-ip\cdot x}, \,\,\,\,\,\,
p \in \mathscr{O}_{1,0,0,1}, \\
\kappa_{1,0}(\nu, \boldsymbol{\p}; \mu, x) = 
\frac{\delta_{\nu \mu}}{\sqrt{2 p^0(\boldsymbol{\p})}}
e^{ip\cdot x},
\,\,\,\,\,\,
p \in \mathscr{O}_{1,0,0,1},
\end{split}
}
\end{equation}
defining the distributions $\kappa_{0,1}, \kappa_{1,0}$
instead of (\ref{kappa_0,1kappa_1,0A}). Proof that they can be (uniquely) extended
to elements
\[
\kappa_{0,1}, \kappa_{1,0} \in \mathscr{L}\big( \mathcal{S}_{A}(\mathbb{R}^3, \mathbb{C}^4)^*, \,\,
\mathscr{E}^*  \big) \cong \mathcal{S}_{A}(\mathbb{R}^3, \mathbb{C}^4) \otimes \mathscr{E}^*,
\] 
remains the same as for the kernels (\ref{kappa_0,1kappa_1,0A}) in Lemma \ref{kappa0,1,kappa1,0ForA}, 
Subsection \ref{A=Xi0,1+Xi1,0}. Thus by Thm. 3.13 of \cite{obataJFA} (or Thm. \ref{obataJFA.Thm.3.13} 
of Subsection \ref{psiBerezin-Hida}) we obtain the corollary that
\[
A = \Xi_{0,1}(\kappa_{0,1}) + \Xi_{1,0}(\kappa_{1,0})
\in
\mathscr{L}\big( (E) \otimes \mathscr{E}, \, (E) \big) \cong
\mathscr{L}\Big( \mathscr{E}, \,\, \mathscr{L}\big( (E), (E)\big) \, \Big),
\]
with $\kappa_{0,1}, \kappa_{1,0}$ defined by (\ref{kappa_0,1kappa_1,0A'}).
Thus the field $A = \Xi_{0,1}(\kappa_{0,1}) + \Xi_{1,0}(\kappa_{1,0})$, understood as integral kernel operator
defines an operator-valued distribution through the continuous map
\[
\mathscr{E} \ni \phi \longmapsto
\Xi_{0,1}\big(\kappa_{0,1}(\phi)\big) + \Xi_{1,0}\big(\kappa_{1,0}(\phi)\big)
\in \mathscr{L}\big( (E), (E)\big).
\]

\section{Higher order contributions $A_{\textrm{int}}^{\mu \,(n)}(g=1,x)$ and $\psi_{\textrm{int}}^{(n)}(g=1,x)$
to the interacting fields $A^{\mu}_{\textrm{int}}(g=1,x)$ and 
$\psi_{\textrm{int}}(g=1,x)$}\label{A(1)psi(1)}

The only modification which we introduce into the causal perturbative approach
to spinor QED, which goes back to St\"uckelberg and Bogoliubov is that we are using
the white noise construction of free fields of the theory.

This allows us to treat each free field at specified space-time point 
as a well defined generalized Hida operator, 
but moreover each free field gains the  mathematical interpretation
of an integral kernel operator with vector-valued kernel
in the sense of Obata \cite{obataJFA}. We have constructed the free Dirac and electromagnetic potential
fields as integral kernel operators with vector-valued kernels in the sense of Obata, respectively, 
in Subsections \ref{psiBerezin-Hida} and \ref{A=Xi0,1+Xi1,0}. 
The operations of Wick product, differentiation, integration, convolution with tempered distributions, 
which can be performed upon field operators understood
as integral kernel operators in the sense of Obata, have been described 
in Subsection \ref{OperationsOnXi}. Construction of the free fields as integral kernel 
operators opens us to 
the general and effective theory of integral kernel operators due to 
Hida-Obata-Sait\^o. In particular we can
treat the Wick product  (compare the so called ``Wick theorem''
in the book \cite{Bogoliubov_Shirkov}) in the rigorous mathematically controllable fashion, 
neccessary for the needs of the causal method (note here that in particular Wightman's definition is 
not effective here). The whole causal method is left completely untouched. We just put the free fields,
understood as integral kernel operators, into the formulas for the causal perturbative series
using the computational Rules for the Wick product, integration and convolution with tempered
distributions, which are given in Subsection \ref{OperationsOnXi}. The only nontrivial point is
the splitting of the causal distributions. Namely (if the free fields are understood as integral kernel operators) 
each contribution to the causal scattering matrix
is a finite sum 
\[
\sum \limits_{l,m} \Xi_{l,m}(\kappa_{l,m})
\]
of well defined integral kernel operators (which almost immediately follows from the 
our results summarized in Subsection \ref{OperationsOnXi})
\[
\Xi_{l,m}(\kappa_{l,m}) \in \mathscr{L}\big( (\boldsymbol{E}) \otimes \mathscr{E}, \, (\boldsymbol{E}^*) \big)
\cong \mathscr{L}\big(\mathscr{E}, \, \mathscr{L}((\boldsymbol{E}), (\boldsymbol{E})) \big)
\]
with vector-valued kernels
\[
\kappa_{l,m} \in \mathscr{L} \big(E_{i_1} \otimes \cdots \otimes E_{i_{l+m}}, \,\, \mathscr{E}^* \big) 
\cong E_{i_1}^{*} \otimes \cdots \otimes E_{i_{l+m}}^{*} \otimes \mathscr{E}^* 
\] 
in the sense of Obata, compare Subsections \ref{psiBerezin-Hida} and \ref{OperationsOnXi}, 
where the the Hida subspace $(\boldsymbol{E})$
in the tensor product of the Fock spaces of the Dirac fied and the electromagnetc potential field is constructed. 

Here 
\[
\mathscr{E} = \mathscr{E}_{n_1} \otimes \cdots \otimes \mathscr{E}_{n_M}, \,\,\,
n_k \in \{1,2\} 
\]
is equal to the tensor product of several space-time test function saces
\[
\mathscr{E}_{1} = \mathcal{S}(\mathbb{R}^4; \mathbb{C}) \,\,\, \textrm{or} \,\,\,
\mathscr{E}_{2} = \mathcal{S}^{00}(\mathbb{R}^4; \mathbb{C})
\]
correspondingly to the massive or mass less component field (compare Subsections \ref{psiBerezin-Hida} 
and \ref{OperationsOnXi}). The nontrivial task in construction is the splitting of vector valued
causal distribution kernels $\kappa_{l,m}$ into retarded and advanced parts, which in practical 
computation reduces to the slitting of causal distributions in
\[
\mathscr{E} = \mathscr{E}_{n_1}^{*} \otimes \cdots \otimes \mathscr{E}_{n_M}^{*}, \,\,\,
n_k \in \{1,2\} 
\]
causally supported into retarded and advanced parts. This problem has been solved by Epstein and 
Glaser \cite{Epstein-Glaser} but for the case where all factors $E_{n_k}$ are equal to the ordinary Schwartz space
$\mathcal{S}(\mathbb{R}^4; \mathbb{C})$. But, as we have  already explained in Subsection 
5.8 of \cite{wawrzycki2018} and in Subsections \ref{A=Xi0,1+Xi1,0}, \ref{OperationsOnXi} of this work, 
the modification of the space-time test space
into the space $\mathscr{E}_{2} = \mathcal{S}^{00}(\mathbb{R}^4; \mathbb{C})$
is necessary for the white noise construction of free mass less field to be possible. Moreover the white noise
construction allows
us to construct and controll the Wick product and allows rigorous formulation and proof of the ``Wick theorem'' of Bogoliubov-Shirkov \cite{Bogoliubov_Shirkov}, necessary for the causal method, compare Subsection
\ref{OperationsOnXi}. Therefore we need the splitting to be extended  over to causal elements of 
\[
\mathscr{E} = \mathscr{E}_{n_1}^{*} \otimes \cdots \otimes \mathscr{E}_{n_M}^{*}, \,\,\,
n_k \in \{1,2\} 
\] 
in which some of the factors $E_{n_k}^{*}$ are equal $\mathcal{S}^{00}(\mathbb{R}^4; \mathbb{C})^{*}$. The 
test space $\mathcal{S}^{00}(\mathbb{R}^4; \mathbb{C})$ in turn is much less flexible concerning localization, 
in particular it contains no non trivial elements with compact support. Fortunately the Pauli-Jordan functions
of mass less fields (e.g. of the free electromagnetic potential field) are by definition
homogeneous. This means that the causal distributions in    
\[
\mathscr{E} = \mathscr{E}_{n_1}^{*} \otimes \cdots \otimes \mathscr{E}_{n_M}^{*}, \,\,\,
n_k \in \{1,2\} 
\] 
which are to be split into retarded and advanced parts have the factors in 
$E_{n_k}^{*} = \mathcal{S}^{00}(\mathbb{R}^4; \mathbb{C})^{*}$ which are 
homogeneous and for homogeneous distributions we have enough elements in
$\mathcal{S}^{00}(\mathbb{R}^4; \mathbb{C})$ to realize the spliting of homogeneous and causal distributions,
compare Subsection 5.7 of \cite{wawrzycki2018}. Moreover all of the homogeneous factors in $E_{n_k}^{*} = \mathcal{S}^{00}(\mathbb{R}^4; \mathbb{C})^{*}$ which we encounter in practice can be extended over $\mathcal{S}(\mathbb{R}^4; \mathbb{C})^{*}$ with the preservation of homogeneity. Thus the splitting problem for causal distributions (homogeneous
over the factors $E_{n_k}^{*} = \mathcal{S}^{00}(\mathbb{R}^4; \mathbb{C})^{*}$) in 
\[
\mathscr{E} = \mathscr{E}_{n_1}^{*} \otimes \cdots \otimes \mathscr{E}_{n_M}^{*}, \,\,\,
n_k \in \{1,2\} 
\] 
can in fact be reduced to the splitting of Epstein-Glaser, compare Subsection 5.7 of \cite{wawrzycki2018}.

Summing up we can insert the free fields, undestood as integral kernel operators in the sense of Obata,
into the formulas for the causal perturbative series for interacting fields. The necessary
operations of Wick product, splitting, integrations, have a rigorous meaning as operations
perfomed upon integral kernel operations explained in  Subsection \ref{OperationsOnXi}. 
The formulas for the contributions are exactly the same as in the standrd perturbative causal spinor QED,
compare e.g. \cite{DKS1} or \cite{Scharf}, but with the Wick product and integration
in these formulas rigorously understood as performed upon integral kernel operators and expressed
by the Rules of  Subsection \ref{OperationsOnXi}. The computation being essetially simple can therefore 
be omitted. We give only the final formulas for the interacting fields (compare
\cite{DKS1}, \cite{Scharf}, \cite{DutFred}) 
\[
\boldsymbol{\psi}_{{}_{\textrm{int}}}^{a}(g, x) =
\boldsymbol{\psi}^{a}(x) + \sum \limits_{n=1}^{\infty} \frac{1}{n!}
\int \limits_{\mathbb{R}^{4n}} \ud^4x_1 \cdots \ud^4 x_n \boldsymbol{\psi}^{a \, (n)}(x_1, \ldots, x_n; x)
g(x_1) \cdots g(x_n), 
\]
with
\[
\boldsymbol{\psi}^{a \, (1)}(x_1; x) = 
e S_{{}_{\textrm{ret}}}^{aa_1}(x-x_1) \gamma^{\nu_1 \, a_1a_2} \boldsymbol{\psi}^{a_2}(x_1)A_{\nu_1}(x_1), 
\]
\begin{multline*}
\boldsymbol{\psi}^{a \, (2)}(x_1, x_2; x) = \\
e^2 \Bigg\{ S_{{}_{\textrm{ret}}}^{aa_1}(x-x_1) \gamma^{\nu_1 \, a_1a_2}S_{{}_{\textrm{ret}}}^{a_2a_3}(x_1-x_2)
\gamma^{\nu_2 \, a_3a_4} :\boldsymbol{\psi}^{a_4}(x_2)  A_{\nu_1}(x_1)A_{\nu_2}(x_2) : \\ 
- S_{{}_{\textrm{ret}}}^{aa_1}(x-x_1) \gamma^{\nu_1 \, a_1a_2} 
: \boldsymbol{\psi}^{a_2}(x_1) 
\overline{\boldsymbol{\psi}}^{a_3}(x_2) \gamma_{\nu_1}^{a_3a_4} \boldsymbol{\psi}^{a_4}(x_2): 
D^{{}^{\textrm{ret}}}_{0}(x_1-x_2) \\
+S_{{}_{\textrm{ret}}}^{aa_1}(x-x_1) \Sigma_{{}_{\textrm{ret}}}^{a_1a_2}(x_1-x_2)\boldsymbol{\psi}^{a_2}(x_2)
\Bigg\} \,\,\, +  \,\,\, \Bigg\{ x_1 \longleftrightarrow x_2 \Bigg\},
\end{multline*}
\[
\textrm{e. t. c.}
\]
and  
\[
{A_{{}_{\textrm{int}}}}_{\mu}(g, x) =
A_{\mu}(x) + \sum \limits_{n=1}^{\infty} \frac{1}{n!}
\int \limits_{\mathbb{R}^{4n}} \ud^4x_1 \cdots \ud^4 x_n A_{\mu}^{\, (n)}(x_1, \ldots, x_n; x)
g(x_1) \cdots g(x_n),
\]
with
\[
A_{\mu}^{\, (1)}(x_1;x) = -e D^{{}^{\textrm{av}}}_{0}(x_1-x) 
:\overline{\boldsymbol{\psi}}^{a_1}(x_1) \gamma_{\mu}^{a_1a_2} \boldsymbol{\psi}^{a_2}(x_1):,
\]
\begin{multline*}
A_{\mu}^{\, (2)}(x_1, x_2; x) = 
e^2 \Bigg\{ 
:\overline{\boldsymbol{\psi}}^{a_1}(x_1) 
\Big( 
 \gamma_{\mu}^{a_1a_2} S_{{}_{\textrm{ret}}}^{a_2a_3}(x_1-x_2) \gamma^{\nu_1 \, a_3a_4}
D^{{}^{\textrm{av}}}_{0}(x_1-x) A_{\nu_1}(x_2) \\
+ \gamma^{\nu_1 \, a_1a_2}S_{{}_{\textrm{av}}}^{a_2a_3}(x_1-x_2) \gamma_{\mu}^{a_3a_4}
D^{{}^{\textrm{av}}}_{0}(x_2-x)A_{\nu_1}(x_1)
\Big)  \boldsymbol{\psi}^{a_4}(x_2): \\
+ D^{{}^{\textrm{av}}}_{0}(x_1-x) {\Pi^{{}^{\textrm{av}}}}_{\mu}^{\nu_1}(x_2-x_1)A_{\nu_1}(x_2)
\Bigg\} \,\,\, + \,\,\, \Bigg\{ x_1 \longleftrightarrow x_2 \Bigg\}
\end{multline*}
\[
\textrm{e. t. c.}
\]
where $g$ is the intesity-of-interaction function over space-time which is assumed to be an element of
the ordinary Schwartz space $\mathcal{S}(\mathbb{R}^4; \mathbb{C})$, and which plays a technical role in 
realizing the causality condition in the form we have learned from Bogoliubov and Shirkov
\cite{Bogoliubov_Shirkov}, compare \cite{DKS1}, \cite{Scharf}, \cite{DutFred}.
This intensity function $g$ modifies the interaction into unphysical in the regions
which lie utside the domain on which $g$ is constant and equal to $1$. It is therefore important
problem to pass to a ``limit'' case of physical interaction with $g=1$ everywhere over the space-time.  

\[
\begin{split}
\boldsymbol{\psi}_{{}_{\textrm{int}}}^{a \,(n)}(g, x) =
\frac{1}{n!}
\int \limits_{\mathbb{R}^{4n}} \ud^4x_1 \cdots \ud^4 x_n \boldsymbol{\psi}^{a \, (n)}(x_1, \ldots, x_n; x), \\
{A_{{}_{\textrm{int}}}}_{\mu}^{\, (n)}(g, x) = 
\frac{1}{n!}
\int \limits_{\mathbb{R}^{4n}} \ud^4x_1 \cdots \ud^4 x_n A_{\mu}^{\, (n)}(x_1, \ldots, x_n; x),
\end{split}
\]  
are the repecitive $n$-th order contributions to the interacting Dirac and electromagnetic potential fields. 

Here in the above formulas for the $n$-th order contributions to interacting fields the free Dirac and electromagnetic fields $\boldsymbol{\psi}$ and $A$  we understood as integral kernel operators with vector-valued kernels as explained
in \ref{psiBerezin-Hida} and \ref{A=Xi0,1+Xi1,0}. Correspondingly the Wick product
and the integrations in these formulas are understood in a rigorous sense as operations performed 
upon integral kernel operators, and summarized in the Rules of Subsection \ref{OperationsOnXi}.
It turns out that each order contribution is equal
\[
\begin{split}
\boldsymbol{\psi}_{{}_{\textrm{int}}}^{a, \,(n)}(g) \,\, = \,\,
 \sum \limits_{l,m} \Xi(\kappa_{l,m}), \\
{A_{{}_{\textrm{int}}}}_{\mu}^{\, (n)}(g) \,\, = \,\,
\sum \limits_{l,m} \Xi(\kappa'_{l,m}),
\end{split}
\] 
to a finite sum of well defined integral kernel operators $\Xi(\kappa_{l,m}), \Xi(\kappa'_{l,m})$ 
with vector-valued distributional kernels $\kappa_{l,m}, \kappa'_{l,m}$ in the sense of Obata 
\cite{obataJFA} (compare Subsection \ref{OperationsOnXi}).

But the main and the whole point is that if the free fields are 
understood as integral kernel operators in the sense of Obata, then the above formulas for each 
$n$-th order contribution to interacting fields, preserve their rigorous mathematical meaning
even if we put $g=1$ everywhere:
namely for $g$ put everywhere equal to $1$ the formulas for each order contributions
to interacting fields represent well defined integral kernel operators in the sense of Obata. 
This we have proved as Theorem \ref{g=1InteractingFieldsQED}, Subsection \ref{OperationsOnXi}. 
Free fields are of course understood as integral kernel operators in the formulas
for contributions to interacting fields, 
and the respective operations of Wick product and integrations with pairing 
functions are understood as performed upon integral kernel operators according to the Rules
of Subsection \ref{OperationsOnXi}.   

Thus each order contribution to interacting fields in the adiabatic limit $g=1$ of physical
interaction is well defined inegral kernel operator and belongs to the same general class of integral 
kernel operators as the Wick product at the same space-time point of free mass less fields
(such as the free electromagnetic potential field). Thus the construction of the free fields within the white 
noise setup as integral kernel operators allows us to solve the adiabatic limit problem in the 
causal perturbative and spinor QED.

Presented method of solution of this problem is general enough to be applicable to other more general and realistic
QFT, provided they can be formulated within the causal perturbative approach, which is for example the case for the Standard Model with the Higgs field \cite{DKS2}, \cite{DKS3}.  

Moreover the interacting fields are given through Fock expansions
\[
\sum \limits_{l,m} \Xi(\kappa_{l,m})
\] 
into integral kernel operators in the sense of \cite{obataJFA} which can be subject to a precise and computable
convergence criteria, which utilize the symbol calculus of Obata, compare \cite{obataJFA},
\cite{obata}, \cite{obata-book}. This allows us to verify the convergence of the perturbative series 
with the tools which were beyong our reach before.

\subsection{Example 1: kernels $\kappa_{l,m}$ 
corresponding to $A_{{}_{\textrm{int}}}^{\mu \,(1)}(g=1,x)$}\label{analysis-of-klm-A(1)}

Here we give explicit formula for the (finite set of) kernels $\kappa'_{l,m}$ for which
\[
{A_{{}_{\textrm{int}}}}_{\mu}^{\, (1)}(g=1) \,\, = \,\,
\sum \limits_{l,m} \Xi(\kappa'_{l,m}),
\]
\emph{i. e.} which define (finite set of) integral kernel operators, (finite) 
sum of which gives the first order contribution to the interacting electromagnetic
potential field in the adiabatic limit $g=1$. 
More explicitly (using the notation of Subsections \ref{psiBerezin-Hida}
and \ref{A=Xi0,1+Xi1,0})
\begin{multline*}
{A_{{}_{\textrm{int}}}}_{\mu}^{\, (1)}(g=1, x) = \\
= \sum \limits_{s,s'=1}^{4} \int \limits_{\mathbb{R}^3\times \mathbb{R}^3}
\kappa'_{2,0}(\boldsymbol{\p}', s', \boldsymbol{\p}, s; \mu, x) \,
\partial_{s',\boldsymbol{\p}'}^{*} \partial_{s, \boldsymbol{\p}}^{*} \, \ud^3 \boldsymbol{\p}' \ud^3 \boldsymbol{\p} \\
+ \sum \limits_{s,s'=1}^{4} \int \limits_{\mathbb{R}^3\times \mathbb{R}^3}
\kappa'_{1,1}(\boldsymbol{\p}', s', \boldsymbol{\p}, s; \mu, x) \,
\partial_{s', \boldsymbol{\p}'}^{*} \partial_{s, \boldsymbol{\p}} \, \ud^3 \boldsymbol{\p}' \ud^3 \boldsymbol{\p} \\
+\sum \limits_{s,s'=1}^{4} \int \limits_{\mathbb{R}^3\times \mathbb{R}^3}
\kappa'_{0,2}(\boldsymbol{\p}', s', \boldsymbol{\p}, s; \mu, x) \,
\partial_{s', \boldsymbol{\p}'} \partial_{s, \boldsymbol{\p}} \, \ud^3 \boldsymbol{\p}' \ud^3 \boldsymbol{\p}
\end{multline*}
or otherwise (according to the notation for the Hida operators $\partial_{s, \boldsymbol{\p}}, 
\partial_{\nu, \boldsymbol{\p}}$
\emph{i. e.} the annihilation operators $a_{s}(\boldsymbol{\p}), a_{\mu}(\boldsymbol{\p})$
introduced in Subsection \ref{psiBerezin-Hida})
\begin{multline*}
{A_{{}_{\textrm{int}}}}_{\mu}^{\, (1)}(g=1, x) = \\
= \sum \limits_{s,s'=1}^{4} \int \limits_{\mathbb{R}^3\times \mathbb{R}^3}
\kappa'_{2,0}(\boldsymbol{\p}', s', \boldsymbol{\p}, s; \mu, x) \,
a_{s'}(\boldsymbol{\p}')^{+}a_{s}(\boldsymbol{\p})^{+} \, \ud^3 \boldsymbol{\p}' \ud^3 \boldsymbol{\p} \\
+ \sum \limits_{s,s'=1}^{4} \int \limits_{\mathbb{R}^3\times \mathbb{R}^3}
\kappa'_{1,1}(\boldsymbol{\p}', s', \boldsymbol{\p}, s; \mu, x) \,
a_{s'}(\boldsymbol{\p}')^{+}a_{s}(\boldsymbol{\p}) \, \ud^3 \boldsymbol{\p}' \ud^3 \boldsymbol{\p} \\
+\sum \limits_{s,s'=1}^{4} \int \limits_{\mathbb{R}^3\times \mathbb{R}^3}
\kappa'_{0,2}(\boldsymbol{\p}', s', \boldsymbol{\p}, s; \mu, x) \,
a_{s'}(\boldsymbol{\p}') a_{s}(\boldsymbol{\p}) \, \ud^3 \boldsymbol{\p}' \ud^3 \boldsymbol{\p}
\end{multline*}
or using still another notation for the annihilation and creation operators 
(used e.g. in \cite{Scharf}, compare Subsection \ref{psiBerezin-Hida})
\begin{multline*}
{A_{{}_{\textrm{int}}}}_{\mu}^{\, (1)}(g=1, x) = \\
= \sum \limits_{s,s'=1}^{2} \int \limits_{\mathbb{R}^3\times \mathbb{R}^3}
{\kappa'}_{2,0}^{++}(\boldsymbol{\p}', s', \boldsymbol{\p}, s; \mu, x) \,
b_{s'}(\boldsymbol{\p}')^{+}d_{s}(\boldsymbol{\p})^{+} \, \ud^3 \boldsymbol{\p}' \ud^3 \boldsymbol{\p} \\
+ \sum \limits_{s,s'=1}^{2} \int \limits_{\mathbb{R}^3\times \mathbb{R}^3}
{\kappa'}_{1,1}^{+-}(\boldsymbol{\p}', s', \boldsymbol{\p}, s; \mu, x) \,
b_{s'}(\boldsymbol{\p}')^{+}b_{s}(\boldsymbol{\p}) \, \ud^3 \boldsymbol{\p}' \ud^3 \boldsymbol{\p} \\
+\sum \limits_{s,s'=1}^{2} \int \limits_{\mathbb{R}^3\times \mathbb{R}^3}
{\kappa'}_{1,1}^{-+}(\boldsymbol{\p}', s', \boldsymbol{\p}, s; \mu, x) \,
d_{s'}(\boldsymbol{\p}')^{+}d_{s}(\boldsymbol{\p}) \, \ud^3 \boldsymbol{\p}' \ud^3 \boldsymbol{\p} \\
\sum \limits_{s,s'=1}^{2} \int \limits_{\mathbb{R}^3\times \mathbb{R}^3}
{\kappa'}_{0,2}^{--}(\boldsymbol{\p}', s', \boldsymbol{\p}, s; \mu, x) \,
d_{s'}(\boldsymbol{\p}')b_{s}(\boldsymbol{\p}) \, \ud^3 \boldsymbol{\p}' \ud^3 \boldsymbol{\p}
\end{multline*} 
where we have put
\[
\kappa'_{2,0}(\boldsymbol{\p}', s', \boldsymbol{\p}, s; \mu, x)
= 
\left\{\begin{array}{ll}
{\kappa'}_{2,0}^{++}(\boldsymbol{\p}', s', \boldsymbol{\p}, s-2; \mu, x) & 
s' = 1, 2, s = 3,4 \\
0 & \textrm{otherwise} 
\end{array} \right.,
\]
\[
\kappa'_{1,1}(\boldsymbol{\p}', s', \boldsymbol{\p}, s; \mu, x)
= 
\left\{\begin{array}{ll}
{\kappa'}_{1,1}^{+-}(\boldsymbol{\p}', s', \boldsymbol{\p}, s; \mu, x) & 
s' = 1, 2, s = 1,2 \\
{\kappa'}_{1,1}^{-+}(\boldsymbol{\p}', s'-2, \boldsymbol{\p}, s-2; \mu, x) & 
s' = 3,4, s = 3,4 \\
0 & \textrm{otherwise} 
\end{array} \right.,
\]
\[
\kappa'_{0,2}(\boldsymbol{\p}', s', \boldsymbol{\p}, s; \mu, x)
= 
\left\{\begin{array}{ll}
{\kappa'}_{0,2}^{--}(\boldsymbol{\p}', s'-2, \boldsymbol{\p}, s; \mu, x) & 
s' = 3, 4, s = 1,2 \\
0 & \textrm{otherwise} 
\end{array} \right..
\]

Let us assume the standard plane wave distribution kernels, $\kappa_{0,1}$ and $\kappa_{1,0}$, 
namely (\ref{skappa_0,1}),
(\ref{skappa_1,0}), Subsect. \ref{StandardDiracPsiField} and (\ref{kappa_0,1kappa_1,0A'}),
Subsection \ref{equivalentA-s}, which define, respectively, the free standard Dirac (\ref{standardpsi(x)}) 
and standard electromagnetic potential (\ref{q-A'-B}) 
fields as sums of two integral kernel operators with vector valued kernels $\kappa_{0,1}$ and $\kappa_{1,0}$.

Application of the Rules II, IV and VI immediately gives the following result
\begin{multline*}
\langle {\kappa'}_{2,0}^{++}(\zeta, \chi), \phi \rangle \overset{\textrm{df}}{=} \\
\overset{\textrm{df}}{=} 
\sum \limits_{s,s'=1}^{2} \int \limits_{\mathbb{R}^3 \times \mathbb{R}^3 \times \mathbb{R}^4} 
{\kappa'}_{2,0}^{++}(\boldsymbol{\p}', s', \boldsymbol{\p}, s; \mu, x) \, 
\zeta(s', \boldsymbol{\p}') \chi(s, \boldsymbol{\p})\phi(x)
\ud^3 \boldsymbol{\p}' \ud^3 \boldsymbol{\p} \ud^4 x \\ =
-e \sum \limits_{s,s'=1}^{2} \int \limits_{\mathbb{R}^3 \times \mathbb{R}^3} 
\ud^3 \boldsymbol{\p}' \ud^3 \boldsymbol{\p} u_{s'}(\boldsymbol{\p}')^{+}v_{s}(\boldsymbol{\p}) 
\frac{\widetilde{\phi}(\boldsymbol{\p} + \boldsymbol{\p}', E(\boldsymbol{\p}) + E'(\boldsymbol{\p}'))
\, \zeta(s', \boldsymbol{\p}') \chi(s', \boldsymbol{\p}')}
{|\boldsymbol{\p} + \boldsymbol{\p}'|^2 - (E(\boldsymbol{\p}) + E'(\boldsymbol{\p}'))^2}
\end{multline*}
\begin{multline*}
\langle {\kappa'}_{1,1}^{+-}(\zeta, \chi), \phi \rangle = \\ =
-e \sum \limits_{s,s'=1}^{2} \int \limits_{\mathbb{R}^3 \times \mathbb{R}^3} 
\ud^3 \boldsymbol{\p}' \ud^3 \boldsymbol{\p} u_{s'}(\boldsymbol{\p}')^{+}u_{s}(\boldsymbol{\p}) 
\frac{\widetilde{\phi}(\boldsymbol{\p}' - \boldsymbol{\p}, E'(\boldsymbol{\p}') - E(\boldsymbol{\p}))
\, \zeta(s', \boldsymbol{\p}') \chi(s', \boldsymbol{\p}')}
{|\boldsymbol{\p}' - \boldsymbol{\p}|^2 - (E'(\boldsymbol{\p}') - E(\boldsymbol{\p}))^2}
\end{multline*}
\begin{multline*}
\langle {\kappa'}_{1,1}^{-+}(\zeta, \chi), \phi \rangle = \\ =
-e \sum \limits_{s,s'=1}^{2} \int \limits_{\mathbb{R}^3 \times \mathbb{R}^3} 
\ud^3 \boldsymbol{\p}' \ud^3 \boldsymbol{\p} v_{s'}(\boldsymbol{\p}')^{+}v_{s}(\boldsymbol{\p}) 
\frac{\widetilde{\phi}(\boldsymbol{\p} - \boldsymbol{\p}', E(\boldsymbol{\p}) - E'(\boldsymbol{\p}'))
\, \zeta(s', \boldsymbol{\p}') \chi(s', \boldsymbol{\p}')}
{|\boldsymbol{\p} - \boldsymbol{\p}'|^2 - (E(\boldsymbol{\p}) - E'(\boldsymbol{\p}'))^2}
\end{multline*}
\begin{multline*}
\langle {\kappa'}_{0,2}^{--}(\zeta, \chi), \phi \rangle = \\ =
-e \sum \limits_{s,s'=1}^{2} \int \limits_{\mathbb{R}^3 \times \mathbb{R}^3} 
\ud^3 \boldsymbol{\p}' \ud^3 \boldsymbol{\p} v_{s'}(\boldsymbol{\p}')^{+}u_{s}(\boldsymbol{\p}) 
\frac{\widetilde{\phi}\big(-(\boldsymbol{\p} + \boldsymbol{\p}'), -(E(\boldsymbol{\p}) + E'(\boldsymbol{\p}'))\big)
\, \zeta(s', \boldsymbol{\p}') \chi(s', \boldsymbol{\p}')}
{|\boldsymbol{\p} + \boldsymbol{\p}'|^2 - (E(\boldsymbol{\p}) - E'(\boldsymbol{\p}'))^2}
\end{multline*}
with
\[
\zeta, \chi \in  \mathcal{S}(\mathbb{R}^3; \mathbb{C}^2), \,\,\,
\phi \in \mathscr{E}_2 = \mathcal{S}^{00}(\mathbb{R}^4; \mathbb{C}), \,\,\,
\widetilde{\phi} \in  \mathscr{F}\mathscr{E}_2 = \mathcal{S}^{0}(\mathbb{R}^4; \mathbb{C}),
\]
and with the convention that $\mathcal{S}(\mathbb{R}^3; \mathbb{C}^2) \subset \mathcal{S}(\mathbb{R}^3; \mathbb{C}^4)
= E_1$ with the convention that only two components of $\zeta$ or $\chi$ are non zero when 
$\xi, \chi$ are regarded as elements of $E_1$. Here
\[
E(\boldsymbol{\p}) = |\boldsymbol{\p}|, \,\,\, E(\boldsymbol{\p}') = |\boldsymbol{\p}'|.
\]

It follows from the general Theorem \ref{g=1InteractingFieldsQED} of 
Subsection \ref{OperationsOnXi} that    
\begin{equation}\label{kappaA^(1)}
\kappa'_{2,0}, \kappa'_{1,1}, \kappa'_{0,2} \in \mathscr{L} \big(E_1 \otimes E_2, \,\, \mathscr{E}_{2}^* \big),
\end{equation}
so that (compare generalization of Thm 3.9 of \cite{obataJFA}, and Subsection \ref{psiBerezin-Hida}) 
\[
\Xi_{l,m}(\kappa'_{l,m}) \in 
\mathscr{L}\big((\boldsymbol{E}) \otimes \mathscr{E}, (\boldsymbol{E})^*\big) \cong 
\mathscr{L}\Big(\mathscr{E}, \,\, \mathscr{L}\big((\boldsymbol{E}), \, (\boldsymbol{E})^*\big) \Big).
\]

But (\ref{kappaA^(1)}) can also be shown with the hepl of the explicit formulas for the kernels
$\kappa'_{l,m}$ by repeating the proof of Lemma \ref{Cont.Ofkappa.kappa}, 
Subsection \ref{OperationsOnXi}. 

Moreover we have the following 
\begin{prop*}
\begin{enumerate}
\item[1)]
The bilinear map
\[
\xi \times \eta \mapsto \kappa'_{1,1}(\xi \otimes \eta), 
\,\,\,\,
\xi , \eta \in E_1, 
\]
can be extended to a separately continuous bilinear map from
\[
E_{1}^{*} \times E_{1} \,\,\, \textrm{into} \,\,\,\mathscr{L}(\mathscr{E}, \mathbb{C}) = \mathscr{E}^*.
\]
\item[2)]
The bilinear map
\[
\xi \times \eta \mapsto \kappa'_{2,0}(\xi \otimes \eta), 
\,\,\,\,
\xi, \eta \in E_1, 
\]
can be extended to a continuous bilinear map from
\[
E_{1}^{*} \times E_{1}^{*} \,\,\, \textrm{into} \,\,\,\mathscr{L}(\mathscr{E}, \mathbb{C}) = \mathscr{E}^*.
\]
Therefore 
\[
\Xi_{l,m}(\kappa'_{l,m}) \in 
\mathscr{L}\big((\boldsymbol{E}) \otimes \mathscr{E}, (\boldsymbol{E})\big) \cong 
\mathscr{L}\Big(\mathscr{E}, \,\, \mathscr{L}\big((\boldsymbol{E}), \, (\boldsymbol{E})\big) \Big)
\]
and
\[
{A_{{}_{\textrm{int}}}}_{\mu}^{\, (1)}(g=1) \,\, = \,\,
\sum \limits_{l,m} \Xi(\kappa'_{l,m}) \in 
\mathscr{L}\big((\boldsymbol{E}) \otimes \mathscr{E}, (\boldsymbol{E})\big) \cong 
\mathscr{L}\Big(\mathscr{E}, \,\, \mathscr{L}\big((\boldsymbol{E}), \, (\boldsymbol{E})\big) \Big),
\]
by Thm. \ref{obataJFA.Thm.3.13}, Subsection \ref{psiBerezin-Hida}. 
\end{enumerate}
\end{prop*}

The same holds for all other possible choices, (\ref{kappa_0,1}), (\ref{kappa_1,0}),
Subsect. \ref{psiBerezin-Hida} and (\ref{kappa_0,1kappa_1,0A}), 
Subsection \ref{A=Xi0,1+Xi1,0}, of the plane wave distribution kernels 
$\kappa_{0,1}, \kappa_{1,0}$ defining the free fields $\boldsymbol{\psi}$, $A$ of the 
theory.

\subsection{Example 2: kernels $\kappa_{l,m}$ 
corresponding to $\boldsymbol{\psi}_{{}_{\textrm{int}}}^{\,(1)}(g=1,x)$}\label{analysis-of-klm-Psi(1)}

Here we give explicit formula for the (finite set of) kernels $\kappa'_{l,m}$ for which
\[
\boldsymbol{\psi}_{{}_{\textrm{int}}}^{a \,(1)}(g=1) \,\, = \,\,
 \sum \limits_{l,m} \Xi(\kappa_{l,m}).
\]
\emph{i. e.} which define (finite set of) integral kernel operators, (finite) 
sum of which gives the first order contribution to the interacting Dirac field in the adiabatic 
limit $g=1$. 
More explicitly (using the notation of Subsections \ref{psiBerezin-Hida}
and \ref{A=Xi0,1+Xi1,0})
\begin{multline*}
\boldsymbol{\psi}_{{}_{\textrm{int}}}^{a \,(1)}(g=1) = \\
= \sum \limits_{\nu'=0}^{3} \sum \limits_{s=1}^{4} \int \limits_{\mathbb{R}^3\times \mathbb{R}^3}
\kappa_{2,0}(\boldsymbol{\p}', \nu', \boldsymbol{\p}, s; a, x) \,
\eta\partial_{\nu',\boldsymbol{\p}'}^{*}\eta \partial_{s, \boldsymbol{\p}}^{*} \, 
\ud^3 \boldsymbol{\p}' \ud^3 \boldsymbol{\p} \\
+ \sum \limits_{\nu'=0}^{3} \sum \limits_{s=1}^{4} \int \limits_{\mathbb{R}^3\times \mathbb{R}^3}
\kappa_{1,1}(\boldsymbol{\p}', \nu', \boldsymbol{\p}, s; a, x) \,
\eta \partial_{\nu', \boldsymbol{\p}'}^{*} \eta \partial_{s, \boldsymbol{\p}} \, 
\ud^3 \boldsymbol{\p}' \ud^3 \boldsymbol{\p} \\
+ \sum \limits_{\nu=0}^{3} \sum \limits_{s'=1}^{4} \int \limits_{\mathbb{R}^3\times \mathbb{R}^3}
\kappa_{1,1}(\boldsymbol{\p}', s', \boldsymbol{\p}, \nu; a, x) \,
\partial_{s', \boldsymbol{\p}'}^{*} \partial_{\nu, \boldsymbol{\p}} \, 
\ud^3 \boldsymbol{\p}' \ud^3 \boldsymbol{\p} \\
+\sum \limits_{\nu'=0}^{3} \sum \limits_{s=1}^{4} \int \limits_{\mathbb{R}^3\times \mathbb{R}^3}
\kappa_{0,2}(\boldsymbol{\p}', \nu', \boldsymbol{\p}, s; a, x) \,
\partial_{\nu', \boldsymbol{\p}'} \partial_{s, \boldsymbol{\p}} \, \ud^3 \boldsymbol{\p}' \ud^3 \boldsymbol{\p}
\end{multline*}
or otherwise (according to the notation for the Hida operators $\partial_{s, \boldsymbol{\p}}, 
\partial_{\nu, \boldsymbol{\p}}$
\emph{i. e.} the annihilation operators $a_{s}(\boldsymbol{\p}), a_{\mu}(\boldsymbol{\p})$
introduced in Subsection \ref{psiBerezin-Hida})
\begin{multline*}
\boldsymbol{\psi}_{{}_{\textrm{int}}}^{a \,(1)}(g=1) = \\
= \sum \limits_{\nu'=0}^{3} \sum \limits_{s=1}^{4} \int \limits_{\mathbb{R}^3\times \mathbb{R}^3}
\kappa_{2,0}(\boldsymbol{\p}', \nu', \boldsymbol{\p}, s; a, x) \,
\eta a_{\nu'}(\boldsymbol{\p}')^{+}\eta a_{s}(\boldsymbol{\p})^{+} \, 
\ud^3 \boldsymbol{\p}' \ud^3 \boldsymbol{\p} \\
+ \sum \limits_{\nu'=0}^{3} \sum \limits_{s=1}^{4} \int \limits_{\mathbb{R}^3\times \mathbb{R}^3}
\kappa_{1,1}(\boldsymbol{\p}', \nu', \boldsymbol{\p}, s; a, x) \,
\eta a_{\nu'}(\boldsymbol{\p}')^{*} \eta a_{s}(\boldsymbol{\p}) \, 
\ud^3 \boldsymbol{\p}' \ud^3 \boldsymbol{\p} \\
+ \sum \limits_{\nu=0}^{3} \sum \limits_{s'=1}^{4} \int \limits_{\mathbb{R}^3\times \mathbb{R}^3}
\kappa_{1,1}(\boldsymbol{\p}', s', \boldsymbol{\p}, \nu; a, x) \,
a_{s'}(\boldsymbol{\p}')^{+} a_{\nu}(\boldsymbol{\p}) \, 
\ud^3 \boldsymbol{\p}' \ud^3 \boldsymbol{\p} \\
+\sum \limits_{\nu'=0}^{3} \sum \limits_{s=1}^{4} \int \limits_{\mathbb{R}^3\times \mathbb{R}^3}
\kappa_{0,2}(\boldsymbol{\p}', \nu', \boldsymbol{\p}, s; a, x) \,
a_{\nu'}(\boldsymbol{\p}') a_{s}(\boldsymbol{\p}) \, \ud^3 \boldsymbol{\p}' \ud^3 \boldsymbol{\p}
\end{multline*}
or using still another notation for the annihilation and creation operators 
(used e.g. in \cite{Scharf}, compare Subsection \ref{psiBerezin-Hida})
\begin{multline*}
\boldsymbol{\psi}_{{}_{\textrm{int}}}^{a \,(1)}(g=1) = \\
= \sum \limits_{\nu'=0}^{3} \sum \limits_{s=1}^{2} \int \limits_{\mathbb{R}^3\times \mathbb{R}^3}
\kappa_{2,0}^{++}(\boldsymbol{\p}', \nu', \boldsymbol{\p}, s; a, x) \,
\eta a_{\nu'}(\boldsymbol{\p}')^{+}\eta d_{s}(\boldsymbol{\p})^{+} \, 
\ud^3 \boldsymbol{\p}' \ud^3 \boldsymbol{\p} \\
+ \sum \limits_{\nu'=0}^{3} \sum \limits_{s=1}^{2} \int \limits_{\mathbb{R}^3\times \mathbb{R}^3}
\kappa_{1,1}^{+-}(\boldsymbol{\p}', \nu', \boldsymbol{\p}, s; a, x) \,
\eta a_{\nu'}(\boldsymbol{\p}')^{*} \eta b_{s}(\boldsymbol{\p}) \, 
\ud^3 \boldsymbol{\p}' \ud^3 \boldsymbol{\p} \\
+ \sum \limits_{\nu=0}^{3} \sum \limits_{s'=1}^{2} \int \limits_{\mathbb{R}^3\times \mathbb{R}^3}
\kappa_{1,1}^{-+}(\boldsymbol{\p}', s', \boldsymbol{\p}, \nu; a, x) \,
d_{s'}(\boldsymbol{\p}')^{+} a_{\nu}(\boldsymbol{\p}) \, 
\ud^3 \boldsymbol{\p}' \ud^3 \boldsymbol{\p} \\
+\sum \limits_{\nu'=0}^{3} \sum \limits_{s=1}^{2} \int \limits_{\mathbb{R}^3\times \mathbb{R}^3}
\kappa_{0,2}^{--}(\boldsymbol{\p}', \nu', \boldsymbol{\p}, s; a, x) \,
a_{\nu'}(\boldsymbol{\p}') b_{s}(\boldsymbol{\p}) \, \ud^3 \boldsymbol{\p}' \ud^3 \boldsymbol{\p}
\end{multline*}
where we have put
\[
\kappa_{2,0}(\boldsymbol{\p}', \nu', \boldsymbol{\p}, s; a, x)
= 
\left\{\begin{array}{ll}
\kappa_{2,0}^{++}(\boldsymbol{\p}', \nu', \boldsymbol{\p}, s-2; a, x) & 
s = 3,4 \\
0 & \textrm{otherwise} 
\end{array} \right.,
\]
\[
\kappa_{1,1}(\boldsymbol{\p}', \nu', \boldsymbol{\p}, s; a, x) 
= 
\left\{\begin{array}{ll}
\kappa_{1,1}^{+-}(\boldsymbol{\p}', \nu', \boldsymbol{\p}, s; a, x)  & 
s = 1,2 \\
0 & \textrm{otherwise} 
\end{array} \right.,
\]
\[
\kappa_{1,1}(\boldsymbol{\p}', s', \boldsymbol{\p}, \nu; a, x) 
= 
\left\{\begin{array}{ll}
\kappa_{1,1}^{-+}(\boldsymbol{\p}', s'-2, \boldsymbol{\p}, \nu; a, x) & 
s' = 3,4 \\
0 & \textrm{otherwise} 
\end{array} \right.,
\]
\[
\kappa_{0,2}(\boldsymbol{\p}', \nu', \boldsymbol{\p}, s; a, x) 
= 
\left\{\begin{array}{ll}
\kappa_{0,2}^{--}(\boldsymbol{\p}', \nu', \boldsymbol{\p}, s; a, x) & 
s = 1,2 \\
0 & \textrm{otherwise} \\
\end{array} \right..
\]

Let us assume the standard plane wave distribution kernels, $\kappa_{0,1}$ and $\kappa_{1,0}$, 
namely (\ref{skappa_0,1}),
(\ref{skappa_1,0}), Subsect. \ref{StandardDiracPsiField} and (\ref{kappa_0,1kappa_1,0A'}),
Subsection \ref{equivalentA-s}, which define, respectively, the free standard Dirac (\ref{standardpsi(x)}) 
and standard electromagnetic potential (\ref{q-A'-B}) 
fields as sums of two integral kernel operators with vector valued kernels $\kappa_{0,1}$ and $\kappa_{1,0}$.

Application of the Rules II, IV and VI immediately gives the following result
\begin{multline*}
\langle \kappa_{2,0}^{++}(\zeta, \chi), \phi \rangle \overset{\textrm{df}}{=} \\
\overset{\textrm{df}}{=} 
\sum \limits_{\nu'=0}^{3} \sum \limits_{s=1}^{2} \int \limits_{\mathbb{R}^3 \times \mathbb{R}^3 \times \mathbb{R}^4} 
\kappa_{2,0}^{++}(\boldsymbol{\p}', \nu', \boldsymbol{\p}, s; a, x) \, 
\zeta(s', \boldsymbol{\p}') \chi(s, \boldsymbol{\p})\phi(x)
\ud^3 \boldsymbol{\p}' \ud^3 \boldsymbol{\p} \ud^4 x \\ =
e \sum \limits_{\nu'=0}^{3} \sum \limits_{s=1}^{2} \int \limits_{\mathbb{R}^3 \times \mathbb{R}^3} 
\ud^3 \boldsymbol{\p}' \ud^3 \boldsymbol{\p} v_{s}^{c}(\boldsymbol{\p}) 
\big(
-(\boldsymbol{p}'+ \boldsymbol{p}) \cdot \vec{\boldsymbol{\gamma}}_{ab} 
+(E'(\boldsymbol{p}')+E(\boldsymbol{p}) \gamma^0 + \boldsymbol{1}_{ab}m 
\big)
\gamma^{\nu'}_{bc} \,\, \times \\ \times \,\, 
\frac{\widetilde{\phi}(\boldsymbol{\p} + \boldsymbol{\p}', E(\boldsymbol{\p}) + E'(\boldsymbol{\p}'))
\, \zeta(\nu', \boldsymbol{\p}') \chi(s, \boldsymbol{\p})}
{2|\boldsymbol{\p}'|\big( |\boldsymbol{\p}'| E(\boldsymbol{\p}) - \langle \boldsymbol{\p}'|\boldsymbol{\p}\rangle\big)}
\end{multline*}

\begin{multline*}
\langle \kappa_{1,1}^{+-}(\zeta, \chi), \phi \rangle = \\ =
e \sum \limits_{\nu'=0}^{3} \sum \limits_{s=1}^{2} \int \limits_{\mathbb{R}^3 \times \mathbb{R}^3} 
\ud^3 \boldsymbol{\p}' \ud^3 \boldsymbol{\p} u_{s}^{c}(\boldsymbol{\p}) 
\big(
-(\boldsymbol{p}'- \boldsymbol{p}) \cdot \vec{\boldsymbol{\gamma}}_{ab} 
+(E'(\boldsymbol{p}')-E(\boldsymbol{p}) \gamma^0 + \boldsymbol{1}_{ab}m 
\big)
\gamma^{\nu'}_{bc} \,\, \times \\ \times \,\, 
\frac{\widetilde{\phi}(\boldsymbol{\p}'- \boldsymbol{\p}, E'(\boldsymbol{\p}') - E(\boldsymbol{\p}))
\, \zeta(\nu', \boldsymbol{\p}') \chi(s, \boldsymbol{\p})}
{2|\boldsymbol{\p}'|\big(\langle \boldsymbol{\p}'|\boldsymbol{\p}\rangle - |\boldsymbol{\p}'| E(\boldsymbol{\p}) \big)}
\end{multline*}

\begin{multline*}
\langle \kappa_{1,1}^{-+}(\zeta, \chi), \phi \rangle = \\ =
e \sum \limits_{\nu'=0}^{3} \sum \limits_{s=1}^{2} \int \limits_{\mathbb{R}^3 \times \mathbb{R}^3} 
\ud^3 \boldsymbol{\p}' \ud^3 \boldsymbol{\p} v_{s}^{c}(\boldsymbol{\p}) 
\big(
(\boldsymbol{p}'- \boldsymbol{p}) \cdot \vec{\boldsymbol{\gamma}}_{ab} 
+(E'(\boldsymbol{p}')-E(\boldsymbol{p}) \gamma^0 + \boldsymbol{1}_{ab}m 
\big)
\gamma^{\nu'}_{bc} \,\, \times \\ \times \,\, 
\frac{\widetilde{\phi}(\boldsymbol{\p} - \boldsymbol{\p}', E(\boldsymbol{\p}) - E'(\boldsymbol{\p}'))
\, \zeta(\nu', \boldsymbol{\p}') \chi(s, \boldsymbol{\p})}
{2|\boldsymbol{\p}'|\big( \langle \boldsymbol{\p}'|\boldsymbol{\p}\rangle - |\boldsymbol{\p}'| E(\boldsymbol{\p})\big)}
\end{multline*}

\begin{multline*}
\langle \kappa_{0,2}^{--}(\zeta, \chi), \phi \rangle = \\ =
e \sum \limits_{\nu'=0}^{3} \sum \limits_{s=1}^{2} \int \limits_{\mathbb{R}^3 \times \mathbb{R}^3} 
\ud^3 \boldsymbol{\p}' \ud^3 \boldsymbol{\p} u_{s}^{c}(\boldsymbol{\p}) 
\big(
(\boldsymbol{p}'+ \boldsymbol{p}) \cdot \vec{\boldsymbol{\gamma}}_{ab} 
-(E'(\boldsymbol{p}')+E(\boldsymbol{p}) \gamma^0 + \boldsymbol{1}_{ab}m 
\big)
\gamma^{\nu'}_{bc} \,\, \times \\ \times \,\, 
\frac{\widetilde{\phi}\big(-(\boldsymbol{\p} + \boldsymbol{\p}'), -(E(\boldsymbol{\p}) + E'(\boldsymbol{\p}'))\big)
\, \zeta(\nu', \boldsymbol{\p}') \chi(s, \boldsymbol{\p})}
{2|\boldsymbol{\p}'|\big( |\boldsymbol{\p}'| E(\boldsymbol{\p}) - \langle \boldsymbol{\p}'|\boldsymbol{\p}\rangle\big)}
\end{multline*}
with summation over repeated spinor indices $b,c \{1,2,3,4\}$ and with
\[
\zeta \in  \mathcal{S}^{0}(\mathbb{R}^3; \mathbb{C}^4) = E_2, \,\,\,
\chi \in \mathcal{S}(\mathbb{R}^3; \mathbb{C}^2),
\phi \in \mathscr{E}_1 = \mathcal{S}(\mathbb{R}^4; \mathbb{C}), \,\,\,
\]
and with the convention that $\mathcal{S}(\mathbb{R}^3; \mathbb{C}^2) \subset \mathcal{S}(\mathbb{R}^3; \mathbb{C}^4)
= E_1$ with the convention that only two components of $\chi$ are non-zero when 
$\chi$ is regarded as an element of $E_1$.

It follows from the general Theorem \ref{g=1InteractingFieldsQED} of 
Subsection \ref{OperationsOnXi} that    
\begin{equation}\label{kappapsi^(1)}
\kappa_{2,0}, \kappa_{1,1}, \kappa_{0,2} \in \mathscr{L} \big(E_1 \otimes E_2, \,\, \mathscr{E}_{1}^* \big),
\end{equation}
so that (compare generalization of Thm 3.9 of \cite{obataJFA}, and Subsection \ref{psiBerezin-Hida}) 
\[
\Xi_{l,m}(\kappa'_{l,m}) \in 
\mathscr{L}\big((\boldsymbol{E}) \otimes \mathscr{E}, (\boldsymbol{E})^*\big) \cong 
\mathscr{L}\Big(\mathscr{E}, \,\, \mathscr{L}\big((\boldsymbol{E}), \, (\boldsymbol{E})^*\big) \Big).
\]

But (\ref{kappapsi^(1)}) can also be shown with the hepl of the explicit formulas for the kernels
$\kappa_{l,m}$ by repeating the proof of Lemma \ref{Cont.Ofkappa.kappa}, 
Subsection \ref{OperationsOnXi}. 

Thus the first order contribution to the interacting Dirac field is equal to a finite sum
\[
\boldsymbol{\psi}_{{}_{\textrm{int}}}^{a \,(1)}(g=1) \,\, = \,\,
 \sum \limits_{l,m} \Xi(\kappa_{l,m}) \in 
\mathscr{L}\big((\boldsymbol{E}) \otimes \mathscr{E}, (\boldsymbol{E})^*\big) \cong 
\mathscr{L}\Big(\mathscr{E}, \,\, \mathscr{L}\big((\boldsymbol{E}), \, (\boldsymbol{E})^*\big) \Big)
\]
of well defined integral kernel operators $\Xi(\kappa_{l,m})$ with vector-vaued distributional kernels
in the sense of Obata, compare \cite{obataJFA} or Subsections \ref{psiBerezin-Hida} and 
\ref{OperationsOnXi}.

However 
\[
\boldsymbol{\psi}_{{}_{\textrm{int}}}^{a \,(1)}(g=1) \,\, = \,\,
 \sum \limits_{l,m} \Xi(\kappa_{l,m}) 
\notin \mathscr{L}\big((\boldsymbol{E}) \otimes \mathscr{E}, (\boldsymbol{E})\big) \cong 
\mathscr{L}\Big(\mathscr{E}, \,\, \mathscr{L}\big((\boldsymbol{E}), \, (\boldsymbol{E})\big) \Big)
\]
similarily as for Wick products of free mass less fields (such as $A_\mu(x)$) at the same space-time point
$x$ which do belong to
\[
\mathscr{L}\big((\boldsymbol{E}) \otimes \mathscr{E}, (\boldsymbol{E})^*\big) \cong 
\mathscr{L}\Big(\mathscr{E}, \,\, \mathscr{L}\big((\boldsymbol{E}), \, (\boldsymbol{E})^*\big) \Big),
\] 
but do not belong to
\[
\mathscr{L}\big((\boldsymbol{E}) \otimes \mathscr{E}, (\boldsymbol{E})\big) \cong 
\mathscr{L}\Big(\mathscr{E}, \,\, \mathscr{L}\big((\boldsymbol{E}), \, (\boldsymbol{E})\big) \Big).
\] 

The same holds for all other possible choices, (\ref{kappa_0,1}), (\ref{kappa_1,0}),
Subsect. \ref{psiBerezin-Hida} and (\ref{kappa_0,1kappa_1,0A}), 
Subsection \ref{A=Xi0,1+Xi1,0}, of the plane wave distribution kernels 
$\kappa_{0,1}, \kappa_{1,0}$ defining the free fields $\boldsymbol{\psi}$, $A$ of the 
theory.

\section{APPENDIX: Fourier transforms $u_s(\boldsymbol{\p})$ and $v_s(-\boldsymbol{\p})$
of a complete system of distributional solutions
of the homogeneous Dirac equation}\label{fundamental,u,v}

As we have seen in Subsection 2.1 of \cite{wawrzycki2018} the Hilbert spaces 
$\mathcal{H}_{m,0}^{\oplus}$ and $\mathcal{H}_{-m,0}^{\ominus}$ of Fourier 
transforms of bispinor solutions of the Dirac equation, concentrated respectively 
on the orbit $\mathscr{O}_{m,0,0,0}$
and $\mathscr{O}_{-m,0,0,0}$, are equal to the images of the corresponding projection operators
$P^\oplus$ and $P^\ominus$ -- the multiplication operators by the corresponding
orthogonal projections $P^\oplus(p)$, $p \in \mathscr{O}_{m,0,0,0}$ and $P^\ominus(p), p \in \mathscr{O}_{-m,0,0,0}$
-- compare Subsection 2.1 of \cite{wawrzycki2018}. Recall that 
\[
\textrm{rank} P^\oplus(p) = 2, p \in \mathscr{O}_{m,0,0,0}, \,\,\,\,\,\,\,\,\,
\textrm{rank} P^\ominus(p) = 2, p \in \mathscr{O}_{-m,0,0,0}.
\]
It is therefore possible to choose at each pont $p = (\boldsymbol{\p}, p_0(\boldsymbol{\p}))
= (\boldsymbol{\p}, E(\boldsymbol{\p}) = \sqrt{|\boldsymbol{\p}|^2 +m^2})$ of the orbit $\mathscr{O}_{m,0,0,0}$
(specified uniquely by $\boldsymbol{\p} \in \mathbb{R}^3$) 
a pair of vectors $u_s(\boldsymbol{\p})$, $s=1,2$, which span the image $\textrm{Im} \, P^\oplus(\boldsymbol{\p}, p_0(\boldsymbol{\p})) = \textrm{Im} \, P^\oplus(\boldsymbol{\p}, E(\boldsymbol{\p}))$
of $P^\oplus(p) = P^\oplus(\boldsymbol{\p}, p_0(\boldsymbol{\p}))$.
Similarily for each point $p = (\boldsymbol{\p}, p_0(\boldsymbol{\p}))
= (\boldsymbol{\p}, -E(\boldsymbol{\p}) = -\sqrt{|\boldsymbol{\p}|^2 +m^2})$ of the orbit $\mathscr{O}_{-m,0,0,0}$
(specified by $\boldsymbol{\p} \in \mathbb{R}^3$) we can find a pair of two vectors
$v_s(\boldsymbol{\p})$, $s=1,2$, which span the image $\textrm{Im} \, P^\ominus(\boldsymbol{\p}, p_0(\boldsymbol{\p})) = \textrm{Im} \, P^\ominus(\boldsymbol{\p}, -E(\boldsymbol{\p}))$, 
$E(\boldsymbol{\p}) = \sqrt{|\boldsymbol{\p}|^2 +m^2}$ for 
$p = \big(\boldsymbol{\p}, -E(\boldsymbol{\p}) \big) = (\boldsymbol{\p}, -\sqrt{|\boldsymbol{\p}|^2 +m^2})
\in \mathscr{O}_{-m,0,0,0}$. We choose these vectors in such a manner that their components
depend smoothly on $\boldsymbol{\p}$ and are multipliers and even convolutors of the Schwartz nuclear
algebra $\mathcal{S}(\mathbb{R}^3; \mathbb{C})$. Moreover we choose them in such a manner that 
$\boldsymbol{\p} \mapsto u_s(\boldsymbol{\p})$ and $\boldsymbol{\p} \mapsto v_s(-\boldsymbol{\p})$
represent Fourier transforms of certain solutions of the free Dirac equation concentrated respectively on the orbits
$\mathscr{O}_{m,0,0,0}$ and $\mathscr{O}_{-m,0,0,0}$. That $\boldsymbol{\p} \mapsto v_s(-\boldsymbol{\p})$, $s=1,2$, represent the Fourier transforms of solutions of 
the Dirac equation and not simply $\boldsymbol{\p} \mapsto v_s(\boldsymbol{\p})$, $s=1,2$, is a matter
of tradition and does not have any dipper justification. Of course there is a whole infinity of different choices
for $u_s(\boldsymbol{\p})$ and  $v_s(\boldsymbol{\p})$, giving unitary equivalent constructions
of the Dirac field.

In this Appendix we construct one useful example of $u_s(\boldsymbol{\p})$ and  $v_s(\boldsymbol{\p})$, $s=1,2$
for the chiral representation of the Clifford algebra generators (Dirac matrices)
\begin{equation}\label{chiralgamma}
\gamma^0 = \left( \begin{array}{cc}   0 &  \bold{1}_2  \\
                                           
                                                   \bold{1}_2              & 0 \end{array}\right), \,\,\,\,
\gamma^k = \left( \begin{array}{cc}   0 &  -\sigma_k  \\
                                           
                                                   \sigma_k             & 0 \end{array}\right),
\end{equation}
which we have used in Subsection 2.1 of \cite{wawrzycki2018} as well as for the so called standard representation
\begin{multline}\label{standardgamma}
\gamma^0 = C \, \left( \begin{array}{cc}   0 &  \bold{1}_2  \\                    
                                                   \bold{1}_2              & 0 \end{array}\right)C^{-1}
= \left( \begin{array}{cc}  \bold{1}_2  &  0  \\
                                                          0       & -\bold{1}_2 \end{array}\right),
\\
\gamma^k = C \, \left( \begin{array}{cc}   0 &  -\sigma_k  \\
                                                   \sigma_k             & 0 \end{array}\right) C^{-1}
= \left( \begin{array}{cc}   0 &  \sigma_k  \\
                                                   -\sigma_k             & 0 \end{array}\right),
\end{multline}
of the Dirac matrices, where 
\[
C = \frac{1}{\sqrt{2}}\left( \begin{array}{cc}   \bold{1}_2 &  \bold{1}_2  \\
                                                   \bold{1}_2             & -\bold{1}_2 \end{array}\right)
= C^{+} = C^{-1}
\]
is unitary involutive $4 \times 4$ matrix.

\begin{center}
{\small THE SOLUTIONS  $u_s(\boldsymbol{\p})$ AND $v_s(\boldsymbol{\p})$ IN THE CHIRAL REPRESENTATION
(\ref{chiralgamma})}
\end{center}

Let us start with the chiral representation (used in Subsection 2.1 of \cite{wawrzycki2018}). Recall that 
\[
P^\oplus(p) = \frac{1}{2}
\left( \begin{array}{cc} 1 & \beta(p)^{-2}  \\
                                                  \beta(p)^2 &  1 \end{array}\right), \,\,\,\,
p \in \mathscr{O}_{m,0,0,0}
\]
with $\beta(p)$ (chosen correspondingly to the chiral representation, as there is infinitum of other possible choices
of $\beta(p)$, compare Subsect. 2.1 of \cite{wawrzycki2018}) corresponding to the orbit $ \mathscr{O}_{m,0,0,0}$, i.e.
\begin{equation}\label{betaO_m,0,0,0}
\begin{split}
\beta(p)^{-2} = \frac{1}{m} \big(p^0 \bold{1} + \vec{p} \cdot \vec{\sigma} \big), \,\,\,
p^0(\vec{p}) = \sqrt{\vec{p} \cdot \vec{p} + m^2} = E(\vec{p}), \\
\beta(p)^{2} = \frac{1}{m} \big(p^0 \bold{1} - \vec{p} \cdot \vec{\sigma} \big), \,\,\,
p^0(\vec{p}) = \sqrt{\vec{p} \cdot \vec{p} + m^2} = E(\vec{p}).
\end{split}
\end{equation}
Similarily recall that here
\[
P^\ominus(p) = \frac{1}{2}
\left( \begin{array}{cc} 1 & -\beta(p)^{-2}  \\
                                                  -\beta(p)^2 &  1 \end{array}\right), \,\,\,\,
p \in \mathscr{O}_{-m,0,0,0}
\]
with $\beta(p)$ corresponding to the orbit $ \mathscr{O}_{-m,0,0,0}$, i.e.
\begin{equation}\label{betaO_-m,0,0,0}
\begin{split}
\beta(p)^{-2} = \frac{1}{m} \big(-p^0 \bold{1} - \vec{p} \cdot \vec{\sigma} \big), \,\,\,
p^0(\vec{p}) = - \sqrt{\vec{p} \cdot \vec{p} + m^2} = -E(\vec{p}), \\
\beta(p)^{2} = \frac{1}{m} \big(-p^0 \bold{1} + \vec{p} \cdot \vec{\sigma} \big), \,\,\,
p^0(\vec{p}) = - \sqrt{\vec{p} \cdot \vec{p} + m^2} = - E(\vec{p}),
\end{split}
\end{equation}
compare Subsection 2.1 of \cite{wawrzycki2018}. In this case (of chiral representation (\ref{chiralgamma})) 
one can put 
\begin{multline}\label{chiral,u,v}
u_s(\boldsymbol{\p}) =  \frac{1}{\sqrt{2}} \sqrt{\frac{E(\boldsymbol{\p}) + m}{2 E(\boldsymbol{\p})}}
\left( \begin{array}{c}   \chi_s + \frac{\boldsymbol{\p} \cdot \boldsymbol{\sigma}}{E(\boldsymbol{\p}) + m} 
\\                                           
              \chi_s - \frac{\boldsymbol{\p} \cdot \boldsymbol{\sigma}}{E(\boldsymbol{\p}) + m} \chi_s                         \end{array}\right) \\ =
\frac{1}{\sqrt{2}} \sqrt{\frac{E(\boldsymbol{\p}) + m}{2 E(\boldsymbol{\p})}}
\left( \begin{array}{c}   \chi_s + \frac{\boldsymbol{\p} \cdot \boldsymbol{\sigma}}{E(\boldsymbol{\p}) + m} 
\\                                           
              \beta\big(p_0(\boldsymbol{\p}), \boldsymbol{\p}\big)^2 \, \big(\chi_s + \frac{\boldsymbol{\p} \cdot \boldsymbol{\sigma}}{E(\boldsymbol{\p}) + m} \chi_s\big)                         \end{array}\right),
\\
v_s(\boldsymbol{\p}) =  \frac{1}{\sqrt{2}} \sqrt{\frac{E(\boldsymbol{\p}) + m}{2 E(\boldsymbol{\p})}}
\left( \begin{array}{c}   \chi_s + \frac{\boldsymbol{\p} \cdot \boldsymbol{\sigma}}{E(\boldsymbol{\p}) + m} \chi_s
\\                                           
              -\big(\chi_s - \frac{\boldsymbol{\p} \cdot \boldsymbol{\sigma}}{E(\boldsymbol{\p}) + m}\chi_s \big)                          \end{array}\right) = \\
\frac{1}{\sqrt{2}} \sqrt{\frac{E(\boldsymbol{\p}) + m}{2 E(\boldsymbol{\p})}}
\left( \begin{array}{c}   \chi_s + \frac{\boldsymbol{\p} \cdot \boldsymbol{\sigma}}{E(\boldsymbol{\p}) + m} \chi_s
\\                                           
              -\beta(p_0\big(\boldsymbol{\p}), -\boldsymbol{\p}\big)^{2} \, \big(\chi_s + \frac{\boldsymbol{\p} \cdot \boldsymbol{\sigma}}{E(\boldsymbol{\p}) + m}\chi_s \big)                          \end{array}\right)
\end{multline}
where
\[
\chi_1 = \left( \begin{array}{c} 1  \\
                                                  0 \end{array}\right), \,\,\,\,\,
\chi_2 = \left( \begin{array}{c} 0  \\
                                                  1 \end{array}\right).
\]
Here $\beta(p)$ in the formula for $u_{s}(\boldsymbol{\p})$ is that (\ref{betaO_m,0,0,0}) corresponding
to the orbit $\mathscr{O}_{m,0,0,0}$ and in the formula for $v_{s}(\boldsymbol{\p})$ the matrix function
$\beta(p)$ equals 
(\ref{betaO_-m,0,0,0}) correspondingly
to the orbit $\mathscr{O}_{-m,0,0,0}$, so that by construction  the solutions $u_{s}(\boldsymbol{\p}), v_{s}(-\boldsymbol{\p})$ have the general form (with the respective $\beta(p)$ corresponding
to the respective orbit $\mathscr{O}_{\pm m,0,0,0}$)
\[
\begin{split}
u_{s}(\boldsymbol{\p}) \overset{\textrm{df}}{=}
u_s(p_0(\boldsymbol{\p}), \boldsymbol{\p}) =  \left( \begin{array}{c} \widetilde{\varphi}_{s +}(p)  \\
                                                  \beta(p)^2 \widetilde{\varphi}_{s+}(p) \end{array}\right), \,\,\,
p = (p_0(\boldsymbol{\p}), \boldsymbol{\p}) \in \mathscr{O}_{m,0,0,0}, \\
v_{s}(\boldsymbol{-\p}) \overset{\textrm{df}}{=}
v_s(p_0(\boldsymbol{\p}), -\boldsymbol{\p}) =  \left( \begin{array}{c} \widetilde{\varphi}_{s-}(p)  \\
                                                  -\beta(p)^2 \widetilde{\varphi}_{s-}(p) \end{array}\right), \,\,\,
p = (p_0(\boldsymbol{\p}), \boldsymbol{\p}) \in \mathscr{O}_{-m,0,0,0},
\end{split}
\]
with
\[
\begin{split}
\widetilde{\varphi}_{s+}\big(p = (p_0(\boldsymbol{\p}), \boldsymbol{\p}) \big) = 
\chi_s + \frac{\boldsymbol{\p} \cdot \boldsymbol{\sigma}}{E(\boldsymbol{\p}) + m} \chi_s, \,\,\,
p = (p_0(\boldsymbol{\p}), \boldsymbol{\p}) \in \mathscr{O}_{m,0,0,0}, \\
\widetilde{\varphi}_{s-}\big(p = (p_0(\boldsymbol{\p}), \boldsymbol{\p}) \big) = 
\chi_s - \frac{\boldsymbol{\p} \cdot \boldsymbol{\sigma}}{E(\boldsymbol{\p}) + m} \chi_s, \,\,\,
p = (p_0(\boldsymbol{\p}), \boldsymbol{\p}) \in \mathscr{O}_{-m,0,0,0}. 
\end{split}
\]
as expected by construction of $\mathcal{H}_{m,0}^{\oplus}$ and 
$\mathcal{H}_{-m,0}^{\ominus}$ in Subsection 2.1 of \cite{wawrzycki2018}.

The vectors $u_s(\boldsymbol{\p})$ and  $v_s(\boldsymbol{\p})$, $s=1,2$,
respect the following orthogonality relations:
\begin{equation}\label{u^+u=delta}
\begin{split}
u_s(\boldsymbol{\p})^+ u_{s'}(\boldsymbol{\p}) = \delta_{ss'}, \,\,\,
v_s(\boldsymbol{\p})^+ v_{s'}(\boldsymbol{\p}) = \delta_{ss'}, \,\,\,
u_s(\boldsymbol{\p})^+ v_{s'}(-\boldsymbol{\p}) = 0.
\end{split}
\end{equation}
By construction we have 
\begin{equation}\label{E+-(p)}
\begin{split}
E_+(\boldsymbol{\p}) 
= \sum_{s=1,2} u_{s}(\boldsymbol{\p}) u_s(\boldsymbol{\p})^+ =
\frac{1}{2E(\boldsymbol{\p})} \big( E(\boldsymbol{\p}) \boldsymbol{1} 
+\boldsymbol{\p} \cdot \boldsymbol{\alpha} + \beta m \big), \,\,\,\,
E(\boldsymbol{\p}) = \sqrt{|\boldsymbol{\p}|^2 + m^2} \\
E_-(\boldsymbol{\p}) 
= \sum_{s=1,2} v_{s}(\boldsymbol{\p}) v_s(\boldsymbol{\p})^+ =
\frac{1}{2E(\boldsymbol{\p})} \big( E(\boldsymbol{\p}) \boldsymbol{1} 
+\boldsymbol{\p} \cdot \boldsymbol{\alpha} - \beta m \big), \,\,\,\,
E(\boldsymbol{\p}) = \sqrt{|\boldsymbol{\p}|^2 + m^2}.
\end{split}
\end{equation}
Here
\[
\begin{split}
\boldsymbol{\sigma} = (\sigma_1, \sigma_2, \sigma_3), \,\,\,
\boldsymbol{\alpha} = (\alpha^1, \alpha^2, \alpha^3), \,\,\,\\
\boldsymbol{\p} \cdot \boldsymbol{\sigma} =
\sum_{i=1}^{3} p_i \sigma_i, \,\,\,
\boldsymbol{\p} \cdot \boldsymbol{\alpha} =
\sum_{i=1}^{3} p_i \alpha^i, \\
\alpha^i = \gamma^0 \gamma^i, \,\, \beta = \gamma^0.
\end{split}
\]
Note that $E_{+}(\boldsymbol{p})$ and $E_{-}(-\boldsymbol{p})$ are mutually orthogonal
projectors on $\mathbb{C}^4$ such that $E_{+}(\boldsymbol{p}) + E_{-}(-\boldsymbol{p}) = \boldsymbol{1}$ and
such that the operators $E_+$ and $E_-$ of Subsection \ref{FirstStepH} are equal to the operators of point-wise mutiplications by the matrices
$E_{\pm}(\pm \boldsymbol{p})$ on the Hilbert spaces  $\mathcal{H}_{m,0}^{\oplus}$ and $\mathcal{H}_{-m,0}^{\ominus}$
of bispinors concetrated respectively on $\mathscr{O}_{m,0,0,0}$
and $\mathscr{O}_{-m,0,0,0}$ (with the point $p= (p_0(\boldsymbol{\p}), \boldsymbol{\p})$ of the respective orbit identified with its cartesian coordinates $\boldsymbol{\p}$).

Moreover, recall that for any element $\widetilde{\phi} \in \mathcal{H}_{m,0}^{\oplus}$ 
the following algebraic relation holds (summation with respect to $i=1,2,3$)
\[
p_0 \gamma^0\widetilde{\phi}(p) =  \big[ p_i \gamma^i + m \boldsymbol{1} \big]\widetilde{\phi}(p), \,\,\,
p \in \mathscr{O}_{m,0,0,0},  
\]
compare Subsection 2.1 of \cite{wawrzycki2018}, so that
\[
E(\boldsymbol{\p}) \widetilde{\phi}(p) = \big[\boldsymbol{\p} \cdot \boldsymbol{\alpha} 
+ m \beta \big] \widetilde{\phi}(p), \,\,\, p = (p_0(\boldsymbol{\p}), \boldsymbol{\p}) \in \mathscr{O}_{m,0,0,0},
\]
for all $\widetilde{\phi} \in \mathcal{H}_{m,0}^{\oplus}$ and thus 
\begin{multline}\label{E_+Phi=Phi}
E_+(\boldsymbol{\p}) \widetilde{\phi}(p)
= \Bigg(\sum_{s=1,2} u_{s}(\boldsymbol{\p}) u_s(\boldsymbol{\p})^+\Bigg) \widetilde{\phi}(p)  \\ =
\frac{1}{2E(\boldsymbol{\p})} \big( E(\boldsymbol{\p}) \boldsymbol{1} 
+\boldsymbol{\p} \cdot \boldsymbol{\alpha} + \beta m \big) \widetilde{\phi}(p)
= \widetilde{\phi}(p), \\
p = (p_0(\boldsymbol{\p}), \boldsymbol{\p}) \in \mathscr{O}_{m,0,0,0},
\end{multline}
for each $\widetilde{\phi} \in \mathcal{H}_{m,0}^{\oplus}$.

Similarily for any element $\widetilde{\phi} \in \mathcal{H}_{-m,0}^{\ominus}$ 
the following algebraic relation holds (summation with respect to $i=1,2,3$)
\begin{multline*}
p_0 \gamma^0\widetilde{\phi}(p) =  \big[ p_i \gamma^i + m \boldsymbol{1} \big]\widetilde{\phi}(p), \\
p= (p_0(\boldsymbol{\p}, \boldsymbol{\p})) = (-E(\boldsymbol{\p}), \boldsymbol{\p}) \in \mathscr{O}_{-m,0,0,0},  
\end{multline*}
compare Subsection 2.1 of \cite{wawrzycki2018}, so that
\begin{multline*}
-E(\boldsymbol{\p}) \widetilde{\phi}(-E(\boldsymbol{\p}), \boldsymbol{\p}) = \big[\boldsymbol{\p} \cdot \boldsymbol{\alpha} 
+ m \beta \big] \widetilde{\phi}(-E(\boldsymbol{\p}), \boldsymbol{\p}), \\
p = (p_0(\boldsymbol{\p}), \boldsymbol{\p}) = (-E(\boldsymbol{\p}), \boldsymbol{\p}) \in \mathscr{O}_{-m,0,0,0},
\end{multline*}
for all $\widetilde{\phi} \in \mathcal{H}_{-m,0}^{\ominus}$ and thus 
\[
E(\boldsymbol{\p}) \widetilde{\phi}(-E(\boldsymbol{\p}), -\boldsymbol{\p}) =
\big(\boldsymbol{\p} \cdot \boldsymbol{\alpha} - \beta m \big)\widetilde{\phi}(-E(\boldsymbol{\p}), -\boldsymbol{\p}),
\,\,\, \widetilde{\phi} \in \mathcal{H}_{-m,0}^{\ominus}. 
\]
Therefore we have
\begin{multline}\label{E_-Phi=Phi}
E_-(\boldsymbol{\p}) \widetilde{\phi}(-E(\boldsymbol{\p}), -\boldsymbol{\p})
= \Bigg(\sum_{s=1,2} v_{s}(\boldsymbol{\p}) v_s(\boldsymbol{\p})^+\Bigg) 
\widetilde{\phi}(-E(\boldsymbol{\p}), -\boldsymbol{\p})  \\ =
\frac{1}{2E(\boldsymbol{\p})} \big( E(\boldsymbol{\p}) \boldsymbol{1} 
+\boldsymbol{\p} \cdot \boldsymbol{\alpha} - \beta m \big) \widetilde{\phi}(-E(\boldsymbol{\p}), -\boldsymbol{\p})
= \widetilde{\phi}(-E(\boldsymbol{\p}), -\boldsymbol{\p}), \\
p = (p_0(\boldsymbol{\p}), \boldsymbol{\p}) \in \mathscr{O}_{-m,0,0,0},
\end{multline}
for each $\widetilde{\phi} \in \mathcal{H}_{-m,0}^{\ominus}$. 

By construction we have
\begin{equation}\label{P^plusu=u,P^minusv=v}
P^\oplus \big(E(\boldsymbol{\p}), \boldsymbol{\p} \big) \, u_s(\boldsymbol{\p}) = u_s(\boldsymbol{\p}),
\,\,\,\,\,
P^\ominus \big(-E(\boldsymbol{\p}), \boldsymbol{\p} \big) \, v_s(-\boldsymbol{\p}) = v_s(-\boldsymbol{\p})
\end{equation}
or
\begin{equation}\label{P^minusv=v}
P^\ominus \big(-E(\boldsymbol{\p}), -\boldsymbol{\p} \big) \, v_s(\boldsymbol{\p}) = v_s(\boldsymbol{\p}),
\end{equation}
and 
\begin{equation}\label{P^plusPhi=Phi,P^minusPhi=Phi}
\begin{split}
P^\oplus \big(E(\boldsymbol{\p}), \boldsymbol{\p} \big) \, \widetilde{\phi}((E(\boldsymbol{\p}),\boldsymbol{\p}) 
= \widetilde{\phi}((E(\boldsymbol{\p}),\boldsymbol{\p}), \,\,\,\,\,\,
\widetilde{\phi} \in \mathcal{H}_{m,0}^{\oplus}, \\
P^\ominus \big(-E(\boldsymbol{\p}), \boldsymbol{\p} \big) \, \widetilde{\phi}(-E(\boldsymbol{\p}),\boldsymbol{\p}) 
= \widetilde{\phi}(-E(\boldsymbol{\p}),\boldsymbol{\p}), \,\,\,\,\,\,
\widetilde{\phi} \in \mathcal{H}_{-m,0}^{\ominus}.
\end{split}
\end{equation}
From the formulas (\ref{P^plusu=u,P^minusv=v}) or (\ref{P^minusv=v}) it follows in particular that
\begin{multline}\label{u^+P^plusPhi=u^+Phi}
u_s(\boldsymbol{\p})^+\widetilde{\phi}(E(\boldsymbol{\p}), \boldsymbol{\p}) =
\sum_{a=1}^{4} \overline{u_{s}^{a}(\boldsymbol{\p})}\widetilde{\phi}^{a}(E(\boldsymbol{\p}), \boldsymbol{\p}) =
\Big(u_{s}(\boldsymbol{\p}), \, \widetilde{\phi}(E(\boldsymbol{\p}), \boldsymbol{\p}) \Big)_{\mathbb{C}^4} \\
= \Big(P^\oplus(E(\boldsymbol{\p}), \,  \boldsymbol{\p}) u_{s}(\boldsymbol{\p}), \widetilde{\phi}(E(\boldsymbol{\p}), \boldsymbol{\p}) \Big)_{\mathbb{C}^4} =
\Big( u_{s}(\boldsymbol{\p}), \, P^\oplus(E(\boldsymbol{\p}), \boldsymbol{\p})\widetilde{\phi}(E(\boldsymbol{\p}), \boldsymbol{\p}) \Big)_{\mathbb{C}^4} \\ =
u_s(\boldsymbol{\p})^+ \big(P^\oplus(E(\boldsymbol{\p}), \boldsymbol{\p})\widetilde{\phi}(E(\boldsymbol{\p}), \boldsymbol{\p})\big) = u_s(\boldsymbol{\p})^+ \big(P^\oplus\widetilde{\phi}\big)(E(\boldsymbol{\p}), \boldsymbol{\p}),
\\
\,\,\, \textrm{for any smooth} \, \widetilde{\phi}
\end{multline}
and
\begin{multline}\label{v^+P^minusPhi=v^+Phi}
v_s(\boldsymbol{\p})^+\widetilde{\phi}(-E(\boldsymbol{\p}), -\boldsymbol{\p}) =
\sum_{a=1}^{4} \overline{v_{s}^{a}(\boldsymbol{\p})}\widetilde{\phi}^{a}(-E(\boldsymbol{\p}), -\boldsymbol{\p}) \\
\Big(v_{s}(\boldsymbol{\p}), \, \widetilde{\phi}(-E(\boldsymbol{\p}), -\boldsymbol{\p}) \Big)_{\mathbb{C}^4}
= \Big(P^\ominus(-E(\boldsymbol{\p}), \,  -\boldsymbol{\p}) v_{s}(\boldsymbol{\p}), \widetilde{\phi}(-E(\boldsymbol{\p}), -\boldsymbol{\p}) \Big)_{\mathbb{C}^4} = \\
\Big( v_{s}(\boldsymbol{\p}), \, P^\ominus(-E(\boldsymbol{\p}), -\boldsymbol{\p})\widetilde{\phi}(-E(\boldsymbol{\p}), -\boldsymbol{\p}) \Big)_{\mathbb{C}^4} \\ =
v_s(\boldsymbol{\p})^+ \big(P^\ominus(-E(\boldsymbol{\p}), -\boldsymbol{\p})\widetilde{\phi}(-E(\boldsymbol{\p}), -\boldsymbol{\p})\big) = v_s(\boldsymbol{\p})^+ \big(P^\ominus\widetilde{\phi}\big)(-E(\boldsymbol{\p}), -\boldsymbol{\p}),
\\
\,\,\, \textrm{for any smooth} \, \widetilde{\phi}.
\end{multline}
It should be stressed that the formulas (\ref{u^+P^plusPhi=u^+Phi}) and (\ref{v^+P^minusPhi=v^+Phi})
are valid for any $\widetilde{\phi}$ not necessary belonging to $\mathcal{H}_{m,0}^{\oplus}$ 
or $\mathcal{H}_{-m,0}^{\ominus}$. 

It is obvious that the projectors $P^\oplus(p)$, $p \in \mathscr{O}_{m,0,0,0}$ 
and $P^\ominus(p)$, $p \in \mathscr{O}_{-m,0,0,0}$, an be expressed in the following
manifestly covariant form
\begin{equation}\label{covariantPplusPminus}
\begin{split}
P^\oplus(p) = \frac{1}{2m} \big[g_{\nu \mu} p^\nu \gamma^\mu + m \boldsymbol{1}_{{}_{4}} \big]
=  \frac{1}{2m} \big[\slashed{p} + m  \big], \,\,\,
p \in \mathscr{O}_{-m,0,0,0}, \\
P^\ominus(p) = \frac{1}{2m} \big[g_{\nu \mu} p^\nu \gamma^\mu + m \boldsymbol{1}_{{}_{4}} \big]
=  \frac{1}{2m} \big[\slashed{p} + m  \big], \,\,\,
p \in \mathscr{O}_{-m,0,0,0}.
\end{split}
\end{equation} 

Finally  let us give the formulas useful in computation of the commutation functions 
and pairing functions for the Dirac field and its Dirac adjoined field. To this end 
let us recall that for a bispinor $u(\boldsymbol{\p})$ the Dirac adjoint $\overline{u}(\boldsymbol{\p})$
is defined to be equal $u(\boldsymbol{\p})^+ \gamma^0$. 
This (common) notation is somewhat unfortunate, because the Dirac adjoint may be mislead with the ordinary complex conjugaton, which we have already agreed to be denoted by overset bar (which also is a traditional notation
for complex conjugation). It must be explicitly stated 
what is meant in each case in working with bispinors. When working with quantum Dirac field $\boldsymbol{\psi}(x)$ the overset bar $\overline{\boldsymbol{\psi}}(x)$ will
always mean the Dirac adjoint. Denoting here $\overline{u}_s(\boldsymbol{\p}), \overline{v}_s(-\boldsymbol{\p})$ the
Dirac adjoints of the complete system of solutions $u_s(\boldsymbol{\p}), v_s(-\boldsymbol{\p})$, we get 
(summation with respect to $i= 1,2,3$)
\[
\begin{split}
\sum_{s=1,2} u_{s}(\boldsymbol{\p}) \overline{u}_s(\boldsymbol{\p}) =
\frac{1}{2E(\boldsymbol{\p})} \big( E(\boldsymbol{\p}) \gamma^0
-p_i \gamma^i + \boldsymbol{1} m \big), \,\,\,\,
E(\boldsymbol{\p}) = \sqrt{|\boldsymbol{\p}|^2 + m^2} \\
 \sum_{s=1,2} v_{s}(\boldsymbol{\p}) \overline{v}_s(\boldsymbol{\p}) =
\frac{1}{2E(\boldsymbol{\p})} \big( E(\boldsymbol{\p}) \gamma^0 
-p_i \gamma^i - \boldsymbol{1} m \big), \,\,\,\,
E(\boldsymbol{\p}) = \sqrt{|\boldsymbol{\p}|^2 + m^2},
\end{split}
\] 
on multiplying the formulas (\ref{E+-(p)}) for $E_{\pm}(\boldsymbol{\p})$ by $\gamma^0$
on the right, and which is frequently written as
\begin{equation}\label{covariantProj}
\begin{split}
\sum_{s=1,2} u_{s}(\boldsymbol{\p}) \overline{u}_s(\boldsymbol{\p}) =
\frac{\slashed{p} +m}{2E(\boldsymbol{\p})} = 
\frac{p_\mu \gamma^\mu +m}{2E(\boldsymbol{\p})}, \,\,\,\,
E(\boldsymbol{\p}) = \sqrt{|\boldsymbol{\p}|^2 + m^2} \\
 \sum_{s=1,2} v_{s}(\boldsymbol{\p}) \overline{v}_s(\boldsymbol{\p}) =
\frac{\slashed{p} -m}{2E(\boldsymbol{\p})} =
\frac{p_\mu \gamma^\mu - m}{2E(\boldsymbol{\p})}, \,\,\,\,
E(\boldsymbol{\p}) = \sqrt{|\boldsymbol{\p}|^2 + m^2}.
\end{split}
\end{equation}

\begin{center}
{\small THE SOLUTIONS  $u_s(\boldsymbol{\p})$ AND $v_s(\boldsymbol{\p})$ IN THE STANDARD REPRESENTATION
(\ref{standardgamma})}
\end{center}

Now let us give the formulas for the fundamental solutions $u_s(\boldsymbol{\p}), v_s(-\boldsymbol{\p})$, 
$s=1,2$, and projections $P^\oplus, P^\ominus$
$E_+, E_-$, in the so called standard representation (\ref{standardgamma}) of the Dirac gamma
matrices. It is not necessary to start the whole analysis with unitary Mackey's induced representations 
using the other choice of the functions $\beta(p)$ corresponding
to the orbits $\mathscr{O}_{m,0,0,0}$ and $\mathscr{O}_{-m,0,0,0}$, which determines
the Hilbert spaces of solutions of the Dirac equation with the standard Dirac matrices
(\ref{standardgamma}). Indeed in order to determine the corresponding projectors it is sufficient
to apply the adjoint homomorphism $C^{-1} (\cdot)C$, and in order to determine the corresponding solutions
$u_s(\boldsymbol{\p}), v_s(-\boldsymbol{\p})$ it is sufficient to apply the unitary operator of multiplication by $C$
\begin{multline}\label{standard,u,v}
u_s(\boldsymbol{\p}) = C \frac{1}{\sqrt{2}} \sqrt{\frac{E(\boldsymbol{\p}) + m}{2 E(\boldsymbol{\p})}}
\left( \begin{array}{c}   \chi_s + \frac{\boldsymbol{\p} \cdot \boldsymbol{\sigma}}{E(\boldsymbol{\p}) + m}\chi_s 
\\                                           
              \chi_s - \frac{\boldsymbol{\p} \cdot \boldsymbol{\sigma}}{E(\boldsymbol{\p}) + m}\chi_s                          \end{array}\right) \\ 
=  \sqrt{\frac{E(\boldsymbol{\p}) + m}{2 E(\boldsymbol{\p})}}
\left( \begin{array}{c}   \chi_s 
\\                                           
              \frac{\boldsymbol{\p} \cdot \boldsymbol{\sigma}}{E(\boldsymbol{\p}) + m} \chi_s                         \end{array}\right),
\\
v_s(\boldsymbol{\p}) =  C \frac{1}{\sqrt{2}} \sqrt{\frac{E(\boldsymbol{\p}) + m}{2 E(\boldsymbol{\p})}}
\left( \begin{array}{c}   \chi_s + \frac{\boldsymbol{\p} \cdot \boldsymbol{\sigma}}{E(\boldsymbol{\p}) + m} \chi_s
\\                                           
              -\big(\chi_s - \frac{\boldsymbol{\p} \cdot \boldsymbol{\sigma}}{E(\boldsymbol{\p}) + m}\chi_s \big)                          \end{array}\right)  \\ =
\sqrt{\frac{E(\boldsymbol{\p}) + m}{2 E(\boldsymbol{\p})}}
\left( \begin{array}{c}   \frac{\boldsymbol{\p} \cdot \boldsymbol{\sigma}}{E(\boldsymbol{\p}) + m} \chi_s
\\                                           
            \chi_s                            \end{array}\right)
\end{multline}
to the complete system of solutions in the chiral representation. For the corresponding
projectors in the standard representation (\ref{standardgamma}) we thus have
\begin{multline*}
P^\oplus(p) = C^{-1}\frac{1}{2}
\left( \begin{array}{cc} \boldsymbol{1}_2 & \beta(p)^{-2}  \\
                                                  \beta(p)^2 &  \boldsymbol{1}_2 \end{array}\right) C  \\ =
\frac{1}{2}
\left( \begin{array}{cc} \frac{m+ E(\boldsymbol{\p})}{m} \boldsymbol{1}_2 & 
-\frac{\boldsymbol{\p} \cdot \boldsymbol{\sigma}}{m}  \\
                           \frac{\boldsymbol{\p} \cdot \boldsymbol{\sigma}}{m} &  \frac{m- E(\boldsymbol{\p})}{m} \boldsymbol{1}_2 \end{array}\right), \,\,\,\,
p = (E(\boldsymbol{\p}), \boldsymbol{\p}) \in \mathscr{O}_{m,0,0,0},
\end{multline*}
(here with $\beta(p)$ equal (\ref{betaO_m,0,0,0})) and similarily for $P^\ominus(-E(\boldsymbol{\p}), \boldsymbol{\p})$
(with $\beta(p)$ equal (\ref{betaO_-m,0,0,0}) in the formula below)
\begin{multline*}
P^\ominus(p) = C^{-1}\frac{1}{2}
\left( \begin{array}{cc} \boldsymbol{1}_2 & -\beta(p)^{-2}  \\
                                                  -\beta(p)^2 &  \boldsymbol{1}_2 \end{array}\right) C  \\ =
\frac{1}{2}
\left( \begin{array}{cc} \frac{m - E(\boldsymbol{\p})}{m} \boldsymbol{1}_2 & 
-\frac{\boldsymbol{\p} \cdot \boldsymbol{\sigma}}{m}  \\
                           \frac{\boldsymbol{\p} \cdot \boldsymbol{\sigma}}{m} &  \frac{m+E(\boldsymbol{\p})}{m} \boldsymbol{1}_2 \end{array}\right), \,\,\,\,\,
p = (-E(\boldsymbol{\p}), \boldsymbol{\p}) \in \mathscr{O}_{-m,0,0,0}.
\end{multline*}
Of course we have the analogous formulas for $E_{\pm}(\boldsymbol{p})$ but we have to remember that 
with the corresponding matrices $\alpha^i = \gamma^0\gamma^i$ in the standard representation (\ref{standardgamma}).
By construction the (Fourier transforms) $u_s(\boldsymbol{\p}), v_s(-\boldsymbol{\p})$ of solutions 
in the standard representation (\ref{standardgamma}) respect the analogous relations 
(\ref{u^+u=delta})-(\ref{covariantProj}).

\begin{center}
{\small ON THE UNITARY ISOMORPHISM $U$ OF SUBSECTION \ref{psiBerezin-Hida} FOR THE DIRAC FIELD}
\end{center}

Note that the unitary isomorphism operator $U$, defined by (\ref{isomorphismU})
in Subsection \ref{psiBerezin-Hida},
can be regarded as the operator of pointwise multiplication by the matrix
\begingroup\makeatletter\def\f@size{5}\check@mathfonts
\def\maketag@@@#1{\hbox{\m@th\large\normalfont#1}}%
\[
U(\boldsymbol{\p}) =
\frac{1}{2|p_0(\boldsymbol{\p})|}
\left( \begin{array}{cccccccc}   \overline{u_{1}^{1}(\boldsymbol{\p})} &  \overline{u_{1}^{2}(\boldsymbol{\p})} &
\overline{u_{1}^{3}(\boldsymbol{\p})} & \overline{u_{1}^{4}(\boldsymbol{\p})} & & 0 & &   \\
\overline{u_{2}^{1}(\boldsymbol{\p})} &  \overline{u_{2}^{2}(\boldsymbol{\p})} &
\overline{u_{2}^{3}(\boldsymbol{\p})} & \overline{u_{2}^{4}(\boldsymbol{\p})} & & & &   \\
& & & & v_{1}^{1}(\boldsymbol{\p}) & v_{1}^{2}(\boldsymbol{\p}) & 
v_{1}^{3}(\boldsymbol{\p}) & v_{1}^{4}(\boldsymbol{\p}) \\
& & 0 & & v_{2}^{1}(\boldsymbol{\p}) & v_{2}^{2}(\boldsymbol{\p}) & 
v_{2}^{3}(\boldsymbol{\p}) & v_{2}^{4}(\boldsymbol{\p})  \end{array}\right)
\]
\endgroup
acting on the element  
$\widetilde{\phi} \oplus (\widetilde{\phi}')^c \in \mathcal{H}_{m,0}^{\oplus} \oplus \mathcal{H}_{-m,0}^{\ominus c}$;
where the value $\big(\widetilde{\phi} \oplus (\widetilde{\phi}')^c \big)(|p_0(\boldsymbol{\p})|, \boldsymbol{\p})$ 
at $p = ((|p_0(\boldsymbol{\p})|, \boldsymbol{\p})) \in \mathscr{O}_{m,0,0,0}$ of 
$\widetilde{\phi} \oplus (\widetilde{\phi}')^c$ is
written as a column vector
\[
\left( \begin{array}{c} 
\widetilde{\phi}(|p_0(\boldsymbol{\p})|, \boldsymbol{\p}) \\
\big[(\widetilde{\phi}')^c(|p_0(\boldsymbol{\p})|, \boldsymbol{\p})\big]^T
\end{array}\right).
\]
Similarily the inverse $U^{-1}$ of the isomorphism (\ref{isomorphismU}), Subsection \ref{psiBerezin-Hida}, 
can be regarded as the operator
of point wise multiplication by the matrix
\begingroup\makeatletter\def\f@size{5}\check@mathfonts
\def\maketag@@@#1{\hbox{\m@th\large\normalfont#1}}%
\[
U^{-1}(\boldsymbol{\p}) =
2|p_0(\boldsymbol{\p})|
\left( \begin{array}{cccc}   
u_{1}^{1}(\boldsymbol{\p}) &  u_{2}^{1}(\boldsymbol{\p}) & 0 & 0   \\
u_{1}^{2}(\boldsymbol{\p}) &  u_{2}^{2}(\boldsymbol{\p}) & 0 & 0   \\
u_{1}^{3}(\boldsymbol{\p}) &  u_{2}^{3}(\boldsymbol{\p}) & 0 & 0   \\
u_{1}^{4}(\boldsymbol{\p}) &  u_{2}^{4}(\boldsymbol{\p}) & 0 & 0   \\
0 & 0 & \overline{v_{1}^{1}(\boldsymbol{\p})} & \overline{v_{2}^{1}(\boldsymbol{\p})} \\
0 & 0 & \overline{v_{1}^{2}(\boldsymbol{\p})} & \overline{v_{2}^{2}(\boldsymbol{\p})} \\
0 & 0 & \overline{v_{1}^{3}(\boldsymbol{\p})} & \overline{v_{2}^{3}(\boldsymbol{\p})} \\
0 & 0 & \overline{v_{1}^{4}(\boldsymbol{\p})} & \overline{v_{2}^{4}(\boldsymbol{\p})} \\
  \end{array}\right)
\]
\endgroup
with the value 
$\Big((\widetilde{\phi})_1 \oplus (\widetilde{\phi})_2 \oplus (\widetilde{\phi})_3 \oplus (\widetilde{\phi})_4\Big)(\boldsymbol{\p}) $ of the elemet 
\[
(\widetilde{\phi})_1 \oplus (\widetilde{\phi})_2 \oplus (\widetilde{\phi})_3 \oplus (\widetilde{\phi})_4 
\in \oplus L^{2}(\mathbb{R}^3; \mathbb{C}) =  L^{2}(\mathbb{R}^3; \mathbb{C}^4)
\]
regarded as a column
\[
\left( \begin{array}{c} 
(\widetilde{\phi})_1(\boldsymbol{\p}) \\
(\widetilde{\phi})_2(\boldsymbol{\p}) \\
(\widetilde{\phi})_3(\boldsymbol{\p}) \\
(\widetilde{\phi})_4(\boldsymbol{\p})
 \end{array}\right).  
\]
Note that 
\[
U(\boldsymbol{\p})U^{-1}(\boldsymbol{\p}) = \boldsymbol{1}_4, \,\,\,\,
U^{-1}(\boldsymbol{\p}) U(\boldsymbol{\p}) = 
\left( \begin{array}{cc} 
E_+(\boldsymbol{\p}) & \boldsymbol{0}_4 \\
\boldsymbol{0}_4 & E_-(\boldsymbol{\p})^T
 \end{array}\right).
\]
Note also that 
\[
\left( \begin{array}{cc} 
E_+(\boldsymbol{\p}) & 0 \\
0 & E_-(\boldsymbol{\p})^T
 \end{array}\right) \left( \begin{array}{c} 
\widetilde{\phi}(|p_0(\boldsymbol{\p})|, \boldsymbol{\p}) \\
\big[(\widetilde{\phi}')^c(|p_0(\boldsymbol{\p})|, \boldsymbol{\p})\big]^T
\end{array}\right) = \left( \begin{array}{c} 
\widetilde{\phi}(|p_0(\boldsymbol{\p})|, \boldsymbol{\p}) \\
\big[(\widetilde{\phi}')^c(|p_0(\boldsymbol{\p})|, \boldsymbol{\p})\big]^T
\end{array}\right)
\]
for $\widetilde{\phi} \oplus 
(\widetilde{\phi}')^c \in \mathcal{H}' = \mathcal{H}_{m,0}^{\oplus} \oplus \mathcal{H}_{-m,0}^{\ominus c}$,
which follows from (\ref{E_+Phi=Phi}) and (\ref{E_-Phi=Phi}).

\section{APPENDIX: Schwartz' spaces of convolutors $\mathcal{O}'_{C}$ 
and multipliers $\mathcal{O}_M$ of $\mathcal{S}$}\label{convolutorsO'_C}

Schwartz \cite{Schwartz} introduced the following linear function spaces 
(in this Appendix we use notation of Schwartz including his notation $\mathcal{E}$
for $\mathscr{C}^\infty(\mathbb{R}^n,; \mathbb{C})$ and its strong dual space $\mathcal{E}'$ of distributions with compact support,
which should not be mislead with our notation $\mathscr{E}$ for a class of countably-Hilbert nuclear space-time test spaces
$\mathcal{S}(\mathbb{R}^4; \mathbb{C}^m)$ or $\mathcal{S}^{00}(\mathbb{R}^4; \mathbb{C}^m)$)
\begin{enumerate}
\item[]
$\mathcal{D} = \{\varphi \in \mathscr{C}^\infty(\mathbb{R}^n; \mathbb{C}), \,
\textrm{supp} \varphi \, \textrm{compact}  \}$,
\item[]
$\mathcal{S} = \mathcal{S}_{H_{(n)}}(\mathbb{R}^n; \mathbb{C}) = \mathcal{S}(\mathbb{R}^n; \mathbb{C}) 
= \{\varphi \in \mathscr{C}^\infty(\mathbb{R}^n; \mathbb{C}),
\forall \alpha, \beta \in \mathbb{N}_{0}^{n}: x^\alpha \partial^\beta \varphi \in \mathscr{C}_0  \}$,
\item[]
$\mathcal{D}_{L^p} = \{\varphi \in \mathscr{C}^\infty(\mathbb{R}^n; \mathbb{C}),
\forall \alpha \in \mathbb{N}_{0}^{n}: \partial^\alpha \varphi \in L^p  \}$ (Sobolev space $W^{\infty, p}$)
$1 \leq p < \infty$,
\item[]
$\mathcal{B} = \mathcal{D}_{L^{\infty}} = \{\varphi \in \mathscr{C}^\infty(\mathbb{R}^n; \mathbb{C}),
\forall \alpha \in \mathbb{N}_{0}^{n}: \partial^\alpha \varphi \in L^{\infty}  \}$,
\item[]
$\overset{\cdot}{\mathcal{B}} = \{\varphi \in \mathscr{C}^\infty(\mathbb{R}^n; \mathbb{C}),
\forall \alpha \in \mathbb{N}_{0}^{n}: \partial^\alpha \varphi \in \mathscr{C}_0  \}$,
\item[]
$\mathcal{O}_{C} = \{\varphi \in \mathscr{C}^\infty(\mathbb{R}^n; \mathbb{C}),
\exists k \in \mathbb{N}_0 \forall \alpha \in \mathbb{N}_{0}^{n}: \,  (1 + |x|^2)^{-k} \partial^\alpha 
\varphi \in \mathscr{C}_0 \}$ (very slowly increasing functions),
\item[]
$\mathcal{O}_{M} = \{\varphi \in \mathscr{C}^\infty(\mathbb{R}^n; \mathbb{C}),
\forall \alpha \in \mathbb{N}_{0}^{n} \exists k \in \mathbb{N}_0: \,  (1 + |x|^2)^{-k} \partial^\alpha 
\varphi \in \mathscr{C}_0 \}$ (slowly increasing functions),
\item[]
$\mathcal{E} = \mathscr{C}^\infty(\mathbb{R}^n; \mathbb{C})$;
\end{enumerate}
and their strong duals, which we will denote in this Appendix (after Schwartz \cite{Schwartz}) with the prime sign
$(\cdot)'$ 
\begin{enumerate}
\item[]
$\mathcal{D}'$ (distributions),
\item[]
$\mathcal{S}'$ (tempered distributions, denoted by us $\mathcal{S}(\mathbb{R}^n; \mathbb{C})^*$),
\item[]
$\mathcal{D}'_{L^p} = \{T \in \mathcal{D}',
\exists m \in \mathbb{N}_{0}: \,  T = \sum_{|\alpha| \leq m} \partial^\alpha f_\alpha \, \textrm{with} 
\, f_\alpha \in L^p  \}$,
\item[]
$\mathcal{O}'_{C} = \{T \in \mathcal{D}',
\forall k \in \mathbb{N}_{0} \exists m \in \mathbb{N}_{0}^{n}: \, 
(1 + |x|^2)^{k} T = \sum_{|\alpha| \leq m } \partial^\alpha f_{\alpha} \, \textrm{with} \, 
f_\alpha \in L^{\infty}  \}$ (rapidly decreasing distributions),
\item[]
$\mathcal{O}'_{M} = \{T \in \mathcal{D}',
\exists m \in \mathbb{N}_{0}^{n} \forall k \in \mathbb{N}_{0}: \,  
(1 + |x|^2)^{k} T = \sum_{|\alpha| \leq m } \partial^\alpha f_{\alpha} \, \textrm{with} \, 
f_\alpha \in L^{\infty}  \}$ (very rapidly decreasing distributions),
\item[]
$\mathcal{E}'$ (distributions with compact support).
\end{enumerate}
Here $\mathscr{C}_0$ is the space of continous $\mathbb{C}$-valued functions on $\mathbb{R}^n$,
tending to zero at infinity.  

All these linear topological spaces together with the topology were constructed in \cite{Schwartz}, except the
space $\mathcal{O}_{C}$ -- the predual of the Schwartz convolutor algebra $\mathcal{O}'_{C}$
of rapidly decreasing distributions. 
The function space $\mathcal{O}_{C}$ together with its inductive limit topoloy such that 
 $\mathcal{O}'_{C}$ with the Schwartz operator topology of uniform convergence on bounded sets, 
becomes the strong dual of $\mathcal{O}_{C}$, has been determined by Horv\'ath. Namely 
$\mathcal{O}'_{C} = \{T\in \mathcal{S}': T \, \textrm{extends uniquely to a continuous linear functional $\tilde{T}$
on  $\mathcal{O}_{C}$}  \}$, with the operator Schwartz topology of uniform convergence on bounded sets on
 $\mathcal{O}'_{C}$ coinciding with the strong dual topology on the space dual to  $\mathcal{O}_{C}$.

We have the following topological inclusions 
(with $E \subset F$ meaning that the topology of $E$ is finer than that of $F$):
\[
\left. \begin{array}{ccccccccccccccc}   & & 1 & \leq & p & \leq  & q & & & & & & & &  \\
 & & & & & & & & & & & & & &  \\
\mathcal{D} & \subset & \mathcal{S} & \subset & \mathcal{D}_{L^p} & \subset &  \mathcal{D}_{L^q} & \subset & \overset{\cdot}{\mathcal{B}} & \subset & \mathcal{B} & \subset & \mathcal{O}_{M} & \subset & \mathcal{E}     \\
\cap & \cap & \cap & \cap & \cap & \cap & \cap & \cap & \cap & \cap & \cap & \cap & \cap & \cap & \cap   \\
\mathcal{E}'& \subset & \mathcal{O}'_{C} & \subset & \mathcal{D}'_{L^p} & \subset & \mathcal{D}'_{L^q} & \subset & \overset{\cdot}{\mathcal{B}}'  & \subset & \mathcal{B}' & \subset & \mathcal{S}' & \subset & \mathcal{D}' 
\end{array}\right.,
\]
\[
\left. \begin{array}{ccccccccccccc}   
\mathcal{D} & \subset & \mathcal{S} & \subset & \mathcal{D}_{L^p} & \subset & \overset{\cdot}{\mathcal{B}} & \subset & \mathcal{O}_{C} & \subset & \mathcal{O}_{M} & \subset & \mathcal{E}     \\
\cap &  &  &  &  &  &  &  &  &  &  &  &   \cap   \\
\mathcal{E}'& \subset & \mathcal{O}'_{M} & \subset & \mathcal{O}'_{C} & \subset & \mathcal{D}'_{L^p} & \subset & \mathcal{D}'_{L^q} &  \subset & \mathcal{S}' & \subset & \mathcal{D}' \\
 & & & & & & & & & & & & \\
& & & & 1 & \leq & p & \leq  & q & & & & 
\end{array}\right.,
\]
\[
\mathcal{O}_{C}  \subset \mathcal{O}'_{C},
\]
compare \cite{Schwartz}, p. 420, or \cite{Horvath}, \cite{Larcher}, \cite{Kisynski}.

Therefore elements of all indicated spaces (except the whole of $\mathcal{E} = \mathscr{C}^\infty$ and $\mathcal{D}'$)
\[
\mathcal{D}, \mathcal{S}, \mathcal{D}_{L^p}, 
\mathcal{E}',
\mathcal{D}'_{L^p}, \mathcal{O}'_{M}, 
\mathcal{O}'_{C},
\]  
can be naturally regarded as tempered distributions, i.e. as elements of $\mathcal{S}'$. 
But we should empasize that the 
topology of each individual space is strictly stronger than the topology induced from the topology 
of the strong dual space $\mathcal{S}'$ of tempered distributions. 

Let us recall that the Fourier transform $\mathscr{F}$ maps isomporphically $\mathcal{S}$
onto $\mathcal{S}$. The Fourer transform is defined on the space of tempered distributions $\mathcal{S}'$
through the linear transpose (dual) of the Fourier transform on $\mathcal{S}$, which by the general
properties of the linear transpose \cite{treves} defines a continuous linear isomorphism 
$\mathcal{S}' \rightarrow \mathcal{S}'$ for the strong dual topology on $\mathcal{S}'$, and 
denoted by the same symbol $\mathscr{F}$. 

Because the elements of the linear spaces 
\[
\mathcal{D}, \mathcal{S}, \mathcal{D}_{L^p}, 
\mathcal{E}',
\mathcal{D}'_{L^p}, \mathcal{O}'_{M}, 
\mathcal{O}'_{C},
\]  
are naturally identified with elements of $\mathcal{S}'$ then
in particular the Fourier transform is a well defined
liner map on these spaces (although in general it leads us out of the particular space in question). 

Recall further that the operator $M_S$ of multiplication by any element $S$ of $\mathcal{O}_M$
maps isomorphically $\mathcal{S} \rightarrow \mathcal{S}$. Thus elements $S$ of $\mathcal{O}_M$
are naturally identified with contionous multiplication operators $M_S$ mapping continously 
$\mathcal{S}$ into $\mathcal{S}$,
i.e. with elements of $\mathscr{L}(\mathcal{S}, \mathcal{S})$. Therefore we can introduce 
on $\mathcal{O}_M$ after 
Schwartz \cite{Schwartz} the topology of uniform convergence on bounded sets induced from 
$\mathscr{L}(\mathcal{S}, \mathcal{S})$. 

Further recall that translation 
\[
T_b: \varphi \rightarrow T_b\varphi, \,\,\,\,\,\,\,
T_b\varphi(x) \overset{\textrm{df}}{=} \varphi(x-b)
\]
maps isomorphically $\mathcal{S} \rightarrow \mathcal{S}$. Again by duality we define 
\[
S \ast \varphi(x) \overset{\textrm{df}}{=} \langle S, T_x \varphi \rangle = S(T_x \varphi),
\]
where $\langle \cdot, \cdot \rangle$ stands for the canonical bilinear form on 
$\mathcal{S}' \times \mathcal{S} = \mathcal{S}^* \times \mathcal{S}$, i.e. the pairing
defined by taking the value of the functional. 
It turns out that if $S \in \mathcal{S}'$ then the operator 
\[
C_S: \varphi \mapsto S \ast \varphi = C_S(\varphi)
\] 
of convolution with $S \in \mathcal{S}'$ corresponding to $S$ maps continously 
$\mathcal{S} \rightarrow \mathcal{O}_C$, i.e. $C_S \in \mathscr{L}(\mathcal{S}, \mathcal{O}_C)$. 
Moreover $S \in \mathcal{O}'_{C}$ if and only if
the corresponding covloution operator $C_S \in \mathscr{L}(\mathcal{S}, \mathcal{S})$,
i.e. if and and only if $C_S$ maps (continously) the Schwartz space $\mathcal{S}$
into itself. Moreover if $S \in \mathcal{O}'_{C}$ then 
$C_{\tilde{S}} \in \mathscr{L}(\mathcal{O}_{C}, \mathcal{O}_{C})$, where $\tilde{S}$
is the unique extension of the fuctional $S$ on $\mathcal{S}$ over $\mathcal{O}_C$. 

Therefore we can, again after Schwartz \cite{Schwartz}, introduce the topology on
$\mathcal{O}'_{C}$ induced from the topology of uniform convergence on bounded sets 
on $\mathscr{L}(\mathcal{S}, \mathcal{S})$. 

These are the Schwartz operator topologies on 
$\mathcal{O}_M$ and $\mathcal{O}'_{C}$. These spaces become nuclear with these topologies,
(quasi-) complete and barreled. For their definitions as induced by systems of semi-norms we refer
the reader to the classic work \cite{Schwartz} or \cite{Horvath}, \cite{Larcher}, \cite{Kisynski}. 
In fact all indicated spaces are barreled,
although all of them are endowed with topology strictly stronger than the topology induced by the strong dual topology of
$\mathcal{S}'$ (for all of them except the whole of the space $\mathcal{E}$ and $\mathcal{D}'$
which cannot be naturally included into $\mathcal{S}'$). 

\begin{twr*}
Let $\mathcal{S}'$ be ednowed with the strong dual topology, and $\mathcal{O}_{M}$, $\mathcal{O}'_{C}$
with the Schwartz' operator topologies defined as above. 
On the space $\mathcal{S}'$ we can define the operation of multiplication by $S\in \mathcal{O}_M$ 
through the linear transpose of the map $M_S$, which maps continously 
$\mathcal{S}' \rightarrow \mathcal{S}'$ and defines a bilinear hypocontinuous multiplication map 
$\mathcal{S}' \times \mathcal{O}_{M} \rightarrow \mathcal{S}'$.
Similarily on the space $\mathcal{S}'$ we can define the operation of convolution by $S\in \mathcal{O}'_{C}$ 
through the linear transpose of the map $C_S$, which maps continously 
$\mathcal{S}' \rightarrow \mathcal{S}'$ and defines a bilinear hypocontinuous convolution map 
$\mathcal{S}' \times \mathcal{O}'_{C} \rightarrow \mathcal{S}'$.  
\end{twr*}
Compare \cite{Schwartz}, Thm. X and Thm. XI, Chap. VII, \S 5, pp. 245-248.

On the space $\mathcal{O}_M$ we can define the commutative  multiplication operation $S_1 \cdot S_2$:
\[
\mathcal{O}_{M}  \times  \mathcal{O}_{M} \ni S_1 \times S_2 \mapsto  S_1 \cdot S_2 \in \mathcal{O}_M
\]
through the composition
of the corresponding multiplication operators $M_{S_1} \circ M_{S_2} = M_{S_2} \circ M_{S_1} = M_{S_1 \cdot S_2}$,
which corresponds to the ordinary pointwise multiplication of functions $f_1, f_2 \in \mathcal{O}_M$
representing the corresponding tempered distributions $S_1, S_2 \in \mathcal{O}_M \subset \mathcal{S}'$.
Similarily we can define commutative convolution operation  $S_1 \ast S_2$:
\[
\mathcal{O}'_{C}  \times  \mathcal{O}'_{C} \ni S_1 \times S_2 \mapsto  S_1 \ast S_2 \in \mathcal{O}'_{C}
\]
through the composition
of the corresponding convolution operators $C_{S_1} \circ C_{S_2} = C_{S_2} \circ C_{S_1} = C_{S_1 \ast S_2}$,
which coincides with the ordinary convolution $f_1 \ast f_2$ of functions $f_1, f_2$
if the tempered distributions $S_1, S_2, S_1 \ast S_2  \in \mathcal{O}_M \subset \mathcal{S}'$
can be represented by ordinary functions $f_1, f_2, f_1 \ast f_2$.

\begin{twr*}
\begin{enumerate}
\item[1)]
The muliplication $S_1 \cdot S_2$ operation is not only hypocontinuous as a map 
$\mathcal{O}_{M}  \times  \mathcal{O}_{M} \rightarrow \mathcal{O}_{M}$,
but likewise (jointly) continuous.
\item[2)]
The convolution $S_1 \ast S_2$ operation is not only hypocontinuous as a map 
$\mathcal{O}'_{C}  \times  \mathcal{O}'_{C} \rightarrow \mathcal{O}'_{C}$,
but likewise (jointly) continuous.
\end{enumerate}
\end{twr*}
Compare \cite{Schwartz}, Remark on page 248, or \cite{Larcher}, Proposition 5. 

Similarily we define a function to be a multiplier (convolutor) of the indicated function space if the corresponding 
multiplication (convolution) operator maps the space continously into itself. Similarily we define by duality the
multipliers (convolutors) of the strong dual of the indicated function space.  

Recall the Schwartz' \emph{Fourier exchange Theorem} (\cite{Schwartz}, Chap. VII.8, Thm. XV)  
\begin{twr*}
If linear topological spaces $\mathcal{O}_{M}$ and $\mathcal{O}'_{C}$ are endowed with the 
Schwartz' operator topologies, defined as above, then the 
Fourier transform $\mathscr{F}$, regared as a map on $\mathcal{S}'$ restricted to $\mathcal{O}'_{C}$, 
transforms isomorphically $\mathcal{O}'_{C}$ onto $\mathcal{O}_{M}$, and the following formula
\[
\mathscr{F}(S \ast T) = \mathscr{F}S \cdot \mathscr{F}T,
\] 
is valid for any $S \in \mathcal{O}'_{C}$ and $T \in \mathcal{S}'$. 
\end{twr*}

All cited results in this Appendix are essentially contained in the classic work \cite{Schwartz}
of L. Schwartz. Some of the results are only remarked there or sometimes formulated without (detailed) proofs,
but the reader will find all details in the subsequent literateure on distribution theory. 
In paticular a topological supplement to the proof of the Fourier exchange Theorem XV (Chap. VII.8 \cite{Schwartz}) 
can be found e.g. in \cite{Kakita}, but a full and systematic treatement of this theorem can be found
in \cite{Kisynski}, where a detailed construction of the predual $\mathcal{O}_{C}$ of $\mathcal{O}'_{C}$
is also given. For further details on the indicated spaces and their multipliers and convolutors
compare \cite{Schwartz}, \cite{Yosida}, \cite{Larcher}, \cite{Larcher-Wengenroth}, \cite{Horvath}. 

\begin{rem*}
Note that the multiplication $\cdot$ map 
$\mathcal{O}_{M}  \times  \mathcal{O}_{M} \rightarrow \mathcal{O}_{M}$ (as well as the convolution $\ast$ map: 
$\mathcal{O}'_{C}  \times  \mathcal{O}'_{C} \rightarrow \mathcal{O}'_{C}$)
is not hypocontinuous with respect to the topology on $\mathcal{O}_{M}$ (resp. on $\mathcal{O}'_{C}$)
induced from the strong dual topology on $\mathcal{S}'$. Indeed if it was hypocontinuous then by the well known
extension theorem, compare  the Proposition of Chap. III, \S 5.4, p.90 in \cite{Schaefer}, a hypocontinuous extension
of the multiplication to a product $\mathcal{S}' \times \mathcal{S}' \rightarrow \mathcal{S}'$ 
(resp. extension of the convolution) 
could have been constructed, which coincides with the ordinary function point-wise multiplication (resp. convolution) product if the distributions can be represented by functions. Because $\mathcal{S'}$ is the strong dual of 
a reflexive Fr\'echet space $\mathcal{S}$, then by Thm. 41.1
of \cite{treves}, we could have obtained in this way a continuous extension of the product of distributions
respecting the natural algebraic laws under multiplication and differentiation and coinciding with the ordinary 
point-wise multiplication (resp. convolution) product of functions whenever the distributions coincide with ordinary functions. But this would be in contradiction to 
the classic result of Schwartz, which says that such extension is impossible, compare \cite{Schwart-mult.impossible}
or \cite{Schwartz}, Chap. V.1. 
Similarily we can show that the extension of the convolution product on the convolution algebra of 
$\mathcal{S}^{0}(\mathbb{R}^n; \mathbb{C})$
is not hypocontinuous with respect to the topology inherited from the strong dual 
$\mathcal{S}^{0}(\mathbb{R}^n; \mathbb{C})^*$, because of the topological inclusions
$\mathcal{S}^{0}(\mathbb{R}^n; \mathbb{C}) \subset \mathcal{S}(\mathbb{R}^n; \mathbb{C})$ and
$\mathcal{S}(\mathbb{R}^n; \mathbb{C})^* \subset \mathcal{S}^{0}(\mathbb{R}^n; \mathbb{C})^*$,
with the topology on $\mathcal{S}^{0}(\mathbb{R}^n; \mathbb{C})$ coinciding with that inherited fram
$\mathcal{S}(\mathbb{R}^n; \mathbb{C})$, compare Subsection 5.5 of \cite{wawrzycki2018}. Equivalently: 
the point-wise multiplication product defined on the multiplier algebra of 
$\mathcal{S}^{00}(\mathbb{R}^n; \mathbb{C})$
is not hypocontinuous with respect to the topology inherited from the strong dual 
$\mathcal{S}^{00}(\mathbb{R}^n; \mathbb{C})^*$.
\end{rem*}

\vspace*{0.5cm}

{\bf ACKNOWLEDGEMENTS}

\vspace*{0.3cm}

The author is indebted for inspiring discussions to prof. A. Staruszkiewicz.  
The author would especially like to thank prof. A. Staruszkiewicz and 
prof. M. Je\.zabek  
for the warm encouragement. He would like to thank prof. M. Je\.zabek 
for the excellent conditions for work at INP PAS where this work
has come into being. The author also would like to thank 
dr P. Duch and prof. A. Herdegen for helpful discussions.


\begin{thebibliography}{99}


\vspace*{.5cm}

\bibitem{Bargmann} Bargmann, V.: Ann. Math. 48, 568 (1947).
\bibitem{HBaum} Baum, H.: A remark on the spectrum of the Dirac operator on pseudo Riemannian spin 
manifolds. Preprint, 1996.
\bibitem{Berezin}  Berezin, F. A.: The method of second quantization. Acad. Press, New York, London, 1966.
\bibitem{BlaSen}  Blanchard, P., Seneor, R.: Annales de L' I. H. P. A23, 147 (1975). 
\bibitem{Bog}  Bognar, J.: Indefinite Inner Product Spaces. Springer, Berlin (1974).
\bibitem{Bogoliubov_Shirkov} Bogoliubov, N. N., Shirkov, D. V.: Introduction to the Theory of Quantized Fields.
New York (1959), second ed. John Wiley \& Sons, Inc., New York, Chichester, Brisbane, Toronto, 1980. 
\bibitem{Bratteli-Robinson} Bratteli, O, Robinson, D. W.: Operator algebras and quantum statistical mechanics,
Vol. II. Springer-Verlag, New York, Heidelberg, Berlin, 1981. 
\bibitem{duch} Duch, P.: Annales H. Poincare 19, 875 (2018); Preprint: math-ph/180110147.
\bibitem{DKS1} D\"utsch, M., Krahe, F., Scharf, G.: Nuovo Cimento A 103, 871 (1990).
\bibitem{DKS2} D\"utsch, M., Krahe, F., Scharf, G.: Nuovo Cimento A 1029, 871 (1993).
\bibitem{DKS3} D\"utsch, M., Krahe, F., Scharf, G.: Nuovo Cimento A 107, 375 (1994).
\bibitem{DKS4} D\"utsch, M., Krahe, F., Scharf, G.: Nuovo Cimento A 108, 737 (1995).
\bibitem{DutFred} D\"utsch, M., Fredenhagen, K.: Commun. Math. Phys. 203, 71 (1999). 
\bibitem{Epstein-Glaser} Epstein, H., Glaser, V.: Ann. Inst. H. Poincar\'e A19, 211 (1973).
\bibitem{epstein-glaser-al} Epstein, H., Glaser, V.: Contribution to the meeting on renormalization theory. C. N. R. S., Marseille, June 1971; C. E. R. N., preprint TH 1344; reprinted in: Renormalization Theory, G. Velo and A. S. Wightman (Eds.), D. Reider Publishing Company, Dordrecht-Holland 1976, pp. 193-254.
\bibitem{Geland-Minlos-Shapiro} Gelfand, I. M., Minlos, R. A,, Shapiro, Z. Ya.: Representations of the rotation 
and Lorentz groups and their applications. Pergamon Press Book, The Macmillan Company, New York, 1963.
\bibitem{GelfandI}  Gelfand, I. M., Shilov, G. E.: Generalized Functions. Vol I. Academic Press, New York, San Francisco, London, 1964.
\bibitem{GelfandIV} Gelfand, I. M. and Vilenkin, N. Ya.: Applications of Harmonic Analysis: Generalized functions. Vol. 4., Acad. Press, New York, 1964.
\bibitem{GelfandYaglom1} Gelfand, I. M., Yaglom, A. M.: Journal of Experimental and Theoretical 
Physics (in Russian ed.) 18, 703 (1948).
\bibitem{GelfandYaglom2} Gelfand, I. M., Yaglom, A. M.: Journal of Experimental and Theoretical Physics ((in Russian ed.) 18, 1096 (1948).
\bibitem{GelfandYaglom3} Gelfand, I. M., Yaglom, A. M.: Journal of Experimental and Theoretical Physics (in Russian ed.) 18, 1105 (1948).
\bibitem{Haag}  Haag, R.: Local Quantum Physics. Springer Verlag, 1996.
\bibitem{Hida1} Hida, T.: Brownian motion, Springer, Berlin, Heidelberg, New York, 1980.  
\bibitem{Hida2} Hida, T.: Causal analysis in terms of Brownian motion. In: Multivariate Analysis, Ed. P. R. Krishnaia. North-Holland, Amsterdam, 1980, pp. 111-118.
\bibitem{Hida3} Hida, T.: Causal analysis in terms of white noise. In: Quantum Fields-Algebras, Processes. 
Ed. Streit, L.. Springer, Berlin, Heidelberg, New York, 1980, pp. 1-19.
\bibitem{hida} Hida, T, Obata, N., Sait\^o, K.: Nagoya Math. J. 128, 65 (1992).
\bibitem{HKPS} Hida, T., Kuo, H.-H., Potthoff, J., Streit, L.: White noise. An infinite dimensional
calculus, Kluwer Academic Publishers, Dordrecht, Boston, London 1993.
\bibitem{Horvath} Horv\'ath, J.: Topological vector spaces and distributions, Vol. 1.  Addison-Wesley Publ. Comp.,
Mass- London-Don Mills 1966. Reed. Dover Publ. 2012.
\bibitem{Kakita} Kakita, T.: Proc. of the Japan Acad. 34 (No 1), (1958), 22.
\bibitem{Kisynski} Kisy\'nski, J.: On the exchange between convolution and multilplication via the Fourier 
transformation, Preprint IM PAN, 2017.
\bibitem{Larcher} Larcher, J.: Analysis 33 (No 4), (2013) 319.
\bibitem{Larcher-Wengenroth} Larcher, J and Wengenroth, J.: Bull. of the Belgian Math. Soc. 21 (2014), 887.
\bibitem{lop1} {\L}opusza\'nski, J.: Rachunek spinor\'ow. PWN, Warszawa 1985.
\bibitem{lop2}  {\L}opusza\'nski, J.: Fortschritte der Physik 26, 261 (1978).
\bibitem{luo} Luo, S.: J. Operator Theory 38, 367 (1997).
Chicago, London, 1976.
\bibitem{Murray_von_Neumann} Murray, F. J., von Neumann, J.: Ann. of Math. 37, 116 (1936).
\bibitem{obata} Obata, N.: J. Math. Soc. Japan 45, 421 (1993).
\bibitem{obataJFA} Obata, N.:  J. of Funct. Anal. 121, 185-232 (1994).
\bibitem{obata-book} Obata, N.: White noise calculus and Fock space, Lect. Notes in Math. Vol. 1577, 
Springer-Verlag (1994). 
\bibitem{PaneitzSegalI} Paneitz, S. M. and Segal, I. E.: J. Funct. Anal. {\bf 47}, 78 (1982).   
\bibitem{PaneitzSegalII} Paneitz, S. M. and Segal, I. E.: J. Funct. Anal. {\bf 49}, 335 (1982).
\bibitem{PaneitzSegalIII} Paneitz, S. M.: J. Funct. Anal. {\bf 54}, 18 (1983).
\bibitem{Riesz-Szokefalvy} Riesz, F. Sz\"okefalvy-Nagy, B.: Le{\c c}ons d'analyse fonctionnelle. Akad\'emiai
Kiad\'o, Budapest, 1952.
\bibitem{Rudin} Rudin, W.: Functional Analysis, McGraw-Hill, Inc. 1991.
\bibitem{Schaefer} Schaefer, H. H.: Topological vector spaces, Springer, 2nd ed. 
(rewritten with assistance of M.P. Wolff), New York 1999.
\bibitem{Scharf} Scharf, G: Finite Quantum electrodynamics, Dover Publications, Mineola, New York, 2014.
\bibitem{Schwartz} Schwartz, L.: Th\'eorie des distributions, Hermann, Paris, 1978.
\bibitem{Schwart-mult.impossible} Schwartz, L.: Comptes Rendus Ac. Sciences 239 (1954), 847.
\bibitem{Segal-NFWP.I} Segal, I.: Journal of Functional Analysis 4, 404 (1969).
\bibitem{Segal-ProcStone} Segal, I.: Local nonlinear functions of quantum fields. In: Proceedings
of the conference in honor of M. H. Stone, Chicago, May 1968. Ed. F. E. Browder. Springer 1970. Pages:
188-210.
\bibitem{Segal_dec_I} Segal, I. E.: Decomposition of Operator Algebras. I.  Memoirs of the American Mathematical 
Society. No. 9, 1951.
\bibitem{Segal-TnsorAlg-II} Segal, I. E.: Ann. Math. 63, 160 (1956).
\bibitem{SegalZhouPhi4} Segal, I. E. and Zhou, Z.: Ann. Phys. {\bf 218}, 279 (1992).
\bibitem{SegalZhouQED} Segal, I. E. and Zhou, Z.: Ann. Phys. {\bf 232}, 61 (1994).
\bibitem{Shimada} Shimada, Y.: White noise distribution theory for the fermion system.  arXiv: 0503051v3 [math-ph] (2005).
\bibitem{Simon} Simon, B.: J. Math. Phys. 12, 140 (1971).
\bibitem{Staruszkiewicz} Staruszkiewicz, A.: Ann. Phys. (N.Y.) 190, 354 (1989).
\bibitem{wig} Streater, R. F. and Wightman, A. S.: PCT, Spin and Statistics, and All
That, W. A. Benjamin, Inc., New York, 1964.  
\bibitem{Stro}  Strohmaier, A.: J. Geom. Phys. 56, 175 (2006).
\bibitem{treves} Treves, F.: Topological vector spaces, distributions and kernels. Academic Press, 1967.
\bibitem{wawrzycki2018}  Wawrzycki, J.: Preprint math-ph/180206719v3.

\bibitem{WightmanGarding} Wightman, A. S. and G{\aa}rding, L.: Arkiv Fysik. 28, 129 (1964).
\bibitem{Woronowicz} Woronowicz, S. L.: Studia Mathematica, 39, 217, (1971).
\bibitem{Yosida} Yosida, K.: Functional analysis, Springer, Berlin, Heidelberg, New York, 1988.




\end{thebibliography}
\end{document}